\documentclass[
headinclude=true, 
headsepline=true,
twoside=true,
BCOR=10mm, 
    fontsize=10 pt,
paper=17cm:24cm, 
pagesize, 
DIV=14, 
headings=normal, 
appendixprefix=true,
draft=false,
numbers=noendperiod, 
toc=bibliography, 
parskip=false, 
captions=bottombeside,
version=last 
]{scrbook}
 
\usepackage{ifdraft}
\ifdraft{
	\usepackage[notref,notcite]{showkeys}
	\usepackage[obeyspaces,hyphens,spaces]{url}
	\renewcommand*{\showkeyslabelformat}[1]{%
		\fbox{\parbox{1.15cm}{\normalfont\small\ttfamily\url{#1}}}}
}{}

\addtokomafont{caption}{\small}

\renewenvironment{thebibliography}[1]{
  \begin{oldthebibliography}{#1}
    \setlength{\itemsep}{.3em}
    \setlength{\parskip}{0em}
}
{
  \end{oldthebibliography}
}

\makeatletter
\DeclareRobustCommand*{\bfseries}{%
   \not@math@alphabet\bfseries\mathbf
   \fontseries\bfdefault\selectfont
   \boldmath
}
\makeatother

\usepackage{tabularx}

\makeatletter
\renewcommand{\@chapapp}{}

\makeatother

\usepackage{subfiles}
\usepackage[dutch,american]{babel}
\usepackage[T1]{fontenc}
\usepackage{lmodern, microtype}
\usepackage{wrapfig}
\setcapindent{0pt}
\usepackage[labelfont=bf]{caption}

\usepackage{mathrsfs}
\usepackage[hang, flushmargin]{footmisc}
\usepackage[pdftex]{hyperref}
\usepackage{graphicx}
\usepackage{amssymb}
\usepackage{amsmath}
\usepackage{verbatim}
\usepackage{bm}
\usepackage[separate-uncertainty=true,multi-part-units=single]{siunitx}
\usepackage{cancel}
\usepackage{cite}
\usepackage{makecell}
\usepackage{booktabs}
\usepackage{arydshln}
\usepackage{braket}
\usepackage{multirow}

\mathchardef\mhyphen="2D
\newcommand{\onlyinsubfile}[1]{#1}
\newcommand{\notinsubfile}[1]{}

\usepackage{footmisc} 
\usepackage{appendix}
\usepackage{chngcntr}
\usepackage{etoolbox}
\AtBeginEnvironment{subappendices}{
\section*{Appendices}
\counterwithin{figure}{chapter}
\counterwithin{table}{chapter}
}

\def\blfootnote{\xdef\@thefnmark{}\@footnotetext}
\long\def\symbolfootnote[#1]#2{\begingroup%
\def\thefootnote{\fnsymbol{footnote}}\footnote[#1]{#2}\endgroup}

\usepackage[dvipsnames,xcdraw]{xcolor}
\usepackage{tikz} 

\DeclareOldFontCommand{\rm}{\normalfont\rmfamily}{\mathrm}
\DeclareOldFontCommand{\sf}{\normalfont\sffamily}{\mathsf}
\DeclareOldFontCommand{\tt}{\normalfont\ttfamily}{\mathtt}
\DeclareOldFontCommand{\bf}{\normalfont\bfseries}{\mathbf}
\DeclareOldFontCommand{\it}{\normalfont\itshape}{\mathit}
\DeclareOldFontCommand{\sl}{\normalfont\slshape}{\@nomath\sl}
\DeclareOldFontCommand{\sc}{\normalfont\scshape}{\@nomath\sc}

\newcommand{\beq}{\begin{equation}}  \newcommand{\eeq}{\end{equation}}
\newcommand{\bal}{\begin{aligned}}   \newcommand{\eal}{\end{aligned}}
\newcommand{\bea}{\begin{eqnarray}}  \newcommand{\eea}{\end{eqnarray}}

\newcommand{\bmat}{\left(\begin{array}}
\newcommand{\emat}{\end{array}\right)}

\newcommand{\bbC}{\mathbb{C}}
\newcommand{\bbR}{\mathbb{R}}

\newcommand{\nn}{\nonumber}

\newcommand{\cO}{\mathcal{O}}

\newcommand{\cE}{\mathcal{E}}

\newcommand{\cC}{\mathcal{C}}

\newcommand{\cK}{\mathcal{K}}
\newcommand{\cN}{\mathcal{N}}

\newcommand{\cA}{\mathcal{A}}

\newcommand{\cF}{\mathcal{F}}
\newcommand{\cI}{\mathcal{I}}

\newcommand{\cR}{\mathcal{R}}

\newcommand{\cV}{\mathcal{V}}

\newcommand{\cM}{\mathcal M}
\newcommand{\cQ}{\mathcal Q}

\renewcommand{\Im}{\mathrm{Im}\,}
\renewcommand{\Re}{\mathrm{Re}\,}

\usepackage{cancel}

\newcommand{\be}{\begin{equation}}
\newcommand{\ee}{\end{equation}}

\DeclareMathOperator{\rk}{rk}
\usepackage{graphbox}
\usepackage{bbm}
\newcommand*\Bell{\ensuremath{\boldsymbol\ell}}
\newcommand*\Bdelta{\ensuremath{\boldsymbol\delta}}

\newcommand{\dd}{\mathrm{d}}

\DeclareMathOperator{\img}{img}  
\usepackage{ytableau}
\newcommand{\eq}[1]{\begin{equation}
                     \begin{split} #1 \end{split}
                     \end{equation}}
\newcommand{\op}{\hspace{1pt}}

\usepackage{dsfont}

\usepackage{tensor}

\usepackage[margin=0pt,font=small,labelfont=normalfont,skip=22pt]{subcaption}

\newcommand{\me}{\mathrm{e}}

\def\Im{\mathop{\mathrm{Im}}\nolimits}
\def\Re{\mathop{\mathrm{Re}}\nolimits}

\usepackage[framemethod=TikZ]{mdframed}

\begin{document}

\renewcommand{\onlyinsubfile}[1]{}
\renewcommand{\notinsubfile}[1]{#1}

\thispagestyle{empty}
\vspace*{4cm}

{\centering
{\huge \textbf{Asymptotic String Compactifications}}\vspace{.2cm} \\
{\large Periods, flux potentials, and the swampland} \vspace{1.45cm} \\
{\LARGE Damian van de Heisteeg}\\
}
\newpage~\thispagestyle{empty}
\vfill
\noindent 
PhD thesis, Utrecht University, March 2022\\
ISBN: 978-94-6423-845-7\\
\textbf{About the cover:} the cover of this booklet is a schematic depiction of the moduli space of a Calabi-Yau manifold. By varying the coordinates in this moduli space the Calabi-Yau manifold -- here depicted as a two-torus -- takes a different shape. Near boundaries of the moduli space the Calabi-Yau manifold degenerates, which corresponds to the pinching of the two-torus.

\newpage

\thispagestyle{empty}
\vspace*{0.2cm}

\begin{center}
{\huge \textbf{Asymptotic String Compactifications}}\vspace{.2cm} \\
{\large Periods, flux potentials, and the swampland} \vspace{1.45cm} \\

{\huge Asymptotische Snaarcompactificaties}\vspace{.2cm}  \\ 
{\large Periodes, fluxpotentialen, en het moerasland}\vspace{.35cm}\\

(met een samenvatting in het Nederlands)\vspace{2.2cm}\\
 
{\LARGE Proefschrift}\vspace{0.8cm}\\
\end{center}

{\large{{ \noindent ter verkrijging van de graad van doctor aan de Universiteit Utrecht op gezag van de rector magnificus, prof.~dr.~H.R.B.M. Kummeling, ingevolge het besluit van het college voor promoties in het openbaar te verdedigen op woensdag 29 juni 2022 des middags te 4.15 uur\\}}}\vspace{0.8cm}
{\centering
{\large door}\vspace{0.6cm}\\
{\LARGE Damian Theodorus Engelbert van de Heisteeg}\vspace{.6cm}\\
{\large geboren op 30 juli 1995\\ te Ede}\par}

{
 \newpage \thispagestyle{empty}
 {\raggedright
{\large Promotor: Prof.~dr.~T.W.~Grimm}\\
{\large Copromotor: Dr.~E.~Plauschinn}}
 \vfill



\frontmatter
\chapter*{Publications}
\textbf{Part I} of this thesis gives a review of asymptotic Hodge theory tailored to applications in string compactifications. It draws upon parts of \cite{Bastian:2021eom,Bastian:2020egp,Bastian:2021hpc,Grimm:2021ckh,Grimm:2019wtx,Grimm:2019bey,Grimm:2021ikg} listed below.
\\

\noindent\textbf{Part II} of this thesis is about geometrical applications of asymptotic Hodge theory in the study of period vectors. It is based on the publication:
\begin{itemize}

\item[\cite{Bastian:2021eom}] Brice Bastian, Thomas W. Grimm, Damian van de Heisteeg: \emph{Modeling General Asymptotic Calabi-Yau Periods}, \href{http://arxiv.org/abs/arXiv:2105.02232}{\textbf{[arXiv:2105.02232]}}.

\end{itemize}

\noindent\textbf{Part III} of this thesis is about applications of asymptotic Hodge theory in the swampland program and moduli stabilization. It is based on the publications:
\begin{itemize}

\item[\cite{Bastian:2020egp}] Brice Bastian, Thomas W. Grimm, Damian van de Heisteeg: \emph{Weak Gravity Bounds in Asymptotic String Compactifications}, \href{https://link.springer.com/article/10.1007/JHEP06(2021)162}{\textbf{JHEP 06 (2021) 162}}, \href{http://arxiv.org/abs/arXiv:2011.08854}{\textbf{[arXiv:2011.08854]}},

\item[\cite{Bastian:2021hpc}] Brice Bastian, Thomas W. Grimm, Damian van de Heisteeg: \emph{Engineering Small Flux Superpotentials and Mass Hierarchies}, \href{http://arxiv.org/abs/arXiv:2108.11962}{\textbf{[arXiv:2108.11962]}},

\item[\cite{Grimm:2021ckh}] Thomas W. Grimm, Erik Plauschinn, Damian van de Heisteeg: \textit{Moduli Stabilization in Asymptotic Flux Compactifications}, \href{https://link.springer.com/article/10.1007/JHEP03(2022)117}{\textbf{JHEP 03 (2022) 117}}, \href{http://arxiv.org/abs/arXiv:2110.05511}{\textbf{[arXiv:2110.05511]}}. 

\end{itemize}
\noindent Other publications concerning asymptotic Hodge theory applications to which the author has contributed, but not covered in detail in this thesis, are:
\begin{itemize}

\item[\cite{Grimm:2019wtx}] Thomas W. Grimm, Damian van de Heisteeg: \emph{Infinite Distances and the Axion Weak Gravity Conjecture}, \href{https://link.springer.com/article/10.1007/JHEP03(2020)020}{\textbf{JHEP 03 (2020) 020}}, \href{http://arxiv.org/abs/arXiv:1905.00901}{\textbf{[arXiv:1905.00901]}},

\item[\cite{Grimm:2019bey}] Thomas W. Grimm, Fabian Ruehle, Damian van de Heisteeg: \emph{Classifying Calabi\textendash{}Yau threefolds using infinite distance limits}, \href{https://link.springer.com/article/10.1007/s00220-021-03972-9}{\textbf{Commun. Math. Phys. 382, 239–275 (2021)}}, \href{http://arxiv.org/abs/arXiv:1910.02963t}{\textbf{[arXiv:1910.02963]}},

\item[\cite{Grimm:2021ikg}] Thomas W. Grimm, Jeroen Monnee, Damian van de Heisteeg: \emph{Bulk Reconstruction in Moduli Space Holography}, \href{https://link.springer.com/article/10.1007/JHEP05(2022)010}{\textbf{JHEP 05 (2022) 010}}, \href{http://arxiv.org/abs/arXiv:2103.12746}{\textbf{[arXiv:2103.12746]}}.

\end{itemize}
\noindent Finally, other publications to which the author contributed over the course of the PhD, but not included in this thesis, concerning near-horizon geometries of black holes in string theory are:
\begin{itemize}
\item[\cite{Couzens:2020jgx}] Christopher Couzens, Eric Marcus, Koen Stemerdink, Damian van de Heisteeg: \emph{The near-horizon geometry of supersymmetric rotating AdS$_{4}$ black holes in M-theory}, \href{https://link.springer.com/article/10.1007/JHEP05(2021)194}{\textbf{JHEP 05 (2021) 194}}, \href{http://arxiv.org/abs/arXiv:2011.07071}{\textbf{[arXiv:2011.07071]}},

\item[\cite{Couzens:2021rlk}] Christopher Couzens, Koen Stemerdink, Damian van de Heisteeg: \emph{M2-branes on Discs and Multi-Charged Spindles}, \href{https://link.springer.com/article/10.1007/JHEP04(2022)107}{\textbf{JHEP 04 (2022) 107}}, \href{http://arxiv.org/abs/arXiv:2110.00571}{\textbf{[arXiv:2110.00571]}}.

\end{itemize}


\tableofcontents

\mainmatter

\chapter{Introduction and preliminaries}\label{chap:intro}
Our goal in this first chapter is to set the stage for the remainder of this thesis. We begin with a heuristic overview of the present status of theoretical high-energy physics, gradually building up to string theory and the swampland program. This introduction includes a rudimentary outline of asymptotic Hodge theory, which forms the main arena of the work carried out in this thesis.
We conclude with a review on the four-dimensional supergravity theories that arise from string compactifications.


\section{The quest for a theory of quantum gravity}
In its current state, theoretical physics has already been highly successful in explaining a wide range of observed physical phenomena. At one of the smallest length scales probed so far (around $10^{-19}$ m) it is capable of describing the interactions detected by particle accelerators such as the Large Hadron Collider in Geneva. On the other end, at the largest scales (roughly $10^{27}$ m) our present understanding of cosmology gives a good account of the structures we see in the observable universe. Throughout this spectrum the main principles used by theoretical physicists follow the same approach: determine the dominant effects at play in a physical system, and build an effective theory around them. For particle physics this means we take three out of the four fundamental forces (the electromagnetic, weak and strong nuclear forces), while ignoring the effect of gravity in scattering processes. In contrast, for cosmology the gravitational force plays a central role, while the simplifying assumption typically takes matter to be distributed as some homogeneous and isotropic fluid.

These recent successes of theoretical physics can be traced back to two major breakthroughs that happened in the beginning of the 20th century: the development of quantum mechanics and the theory of relativity. First, Max Planck was able to describe the radiation emitted by a black body, interpolating from the low-frequency behavior in the infrared covered by Rayleigh-Jeans law up to Wien's law at high energies in the ultraviolet. This spectrum was explained by the realization that radiation is absorbed and emitted in discrete energy packets, sparking the underlying theory of quantum mechanics to be formalized in the following years. Roughly in the same timeframe the theory of relativity was developed. First special relativity abandoned the principle of an absolute time was abandoned, and it postulated that the speed of light should be the same in every reference frame. Subsequently Albert Einstein introduced the equivalence principle, culminating into the theory of general relativity. General relativity predicted physical phenomena such as gravitational waves (so-called ripples in spacetime) and black holes, which were recently confirmed with the detection of gravitational waves by LIGO in 2016 \cite{LIGOScientific:2016aoc} and the first image of a black hole by the Event Horizon Telescope in 2019 \cite{EventHorizonTelescope:2019dse}.

In its advancements, physics appears to have a tendency to move towards unified descriptions. To take a classical example, Maxwell was able to take the until then separate notions of electricity and magnetism and bring them into a combined framework, where his equations govern the theory of electrodynamics. Also general relativity itself can be viewed as how Newton's version of gravity was embedded into the setting of special relativity. This naturally leads one to ponder about the point of convergence for this sequence of events: is there a theory of everything at the end of our road? 

A first leap in this direction was taken with the unification of quantum mechanics and special relativity, leading in the end to quantum field theory. This coalescence began with the conception of quantum electrodynamics (QED), which brings Maxwell's laws of electrodynamics -- as described by special relativity -- together with the principles of quantum mechanics, and marks the first moment of conjunction between the two revelations of the early 1900s. QED has been tested to extreme lengths as of today, with predictions for quantities such as the anomalous magnetic moment of the electron matching to unreasonably great precision. In turn, by suitably generalizing this framework quantum field theory was developed, and in this setting two of the other fundamental forces -- the weak and strong nuclear force -- were encompassed within what we know now as the Standard Model of particle physics. This catalog contains all known elementary particles and describes how they can interact with each other. Its final piece and one of its most crucial components, the Higgs boson, was confirmed about 10 years ago in 2012 with the measurement of its mass by the LHC \cite{ATLAS:2012yve, CMS:2012qbp}. This particle plays an important role in the Standard Model, where by means of the Brout-Englert-Higgs mechanism it gives mass to some of its particles. 

It leaves us in the peculiar situation where we have two by now well-established theories -- quantum field theory and general relativity -- which together cover (most of) the broad spectrum of physical phenomena we observe. The trend in the past suggests that there should be some means to bring these two sides together into a unified framework. However, a direct approach to quantize gravity by using quantum field theory techniques fails spectacularly. The standard method is to apply so-called renormalization schemes: scattering amplitudes in particle physics are plagued by divergencies, and by introducing counterterms one is able to remedy these infinities. For the three forces in the Standard Model this approach has been used with great success, however, if one uses the same strategy for gravity this leads to an infinite number of counterterms, rendering the theory non-renormalizable. Roughly speaking, each counterterm would require the measurement of a new coupling constant for the theory, and from this point of view we would thus have an infinite number of parameters to fit with experiments.

A modern take on these issues is to describe physics by means of effective field theories (EFTs): these theories are valid up to a finite energy cutoff scale we denote by $\Lambda$, after which a more fundamental theory should take over. In the context of our Standard Model of particle physics this means we can view it as an EFT with $\Lambda_{\rm LHC} \sim 10^4$ GeV as cutoff scale, roughly the energy scale probed by colliders such as the LHC. Beyond this energy scale new effects could come into play -- such as heavier particles -- giving us a hint at how our Standard Model should be extended. From the birds eye view of a UV theory the low-energy EFT arises by integrating out all high-energy degrees of freedom with an energy $E > \Lambda$, as prescribed by the Wilson's renormalization group. In this top-down approach the couplings at low energies are modified by the UV physics through loop diagrams with high-energy states running in the cycle, resulting in the EFT with cutoff scale $\Lambda$. Conversely, in the bottom-up approach one writes down the most general low-energy theory compatible with the symmetries the system possesses, e.g.~the chiral symmetry of quantum chromodynamics, up to a satisfactory order. From this perspective it is not clear how the EFT can be embedded into a UV theory, or if it admits such a completion at all. Figuring out the criteria that distinguish EFTs with a UV completion into quantum gravity is one of the founding principle of the so-called \textit{swampland program}, which we review in section \ref{sec:swampland}. For gravity the expected cutoff scale is set by the Planck scale $\Lambda_{\rm Planck} \sim 10^{19}$ GeV (with $10^{-35}$ m as associated length scale), leaving us with a large window yet unexplored where quantum gravity effects could remedy its non-renormalizable nature. 

Before we speculate on what kind of new physics we expect, let us take a step back and ask ourselves why we need a theory of quantum gravity to begin with. Independently quantum field theory and general relativity have been highly successful in explaining physical phenomena, and at first sight they seem to work at the complete opposites of the spectrum: the former concerns itself with physics at the smallest length scales such as particle scatterings, while the latter describes gravitational effects between large objects in our universe. A challenging arena where these two meet is in the physics of black holes. These mysterious objects were found as solutions to the equations of general relativity by Schwarzschild almost immediately after its conception and have some bizarre properties, such as the presence of event horizons which even light itself cannot escape,\footnote{As an aside, let us note that semi-classically it was found by Hawking that black holes do in fact emit radiation, allowing them to evaporate slowly over time. This has led to the so-called information paradox, where information thrown into the black hole somehow gets lost over the course of its evaporation. Recently there has been much progress into understanding this process, where in \cite{Penington:2019npb,Almheiri:2019psf} it was shown how to reproduce the Page curve describing the black hole entropy.} hence their name. Moreover, the interior of a black hole contains a gravitational singularity, where the laws of general relativity completely break down. General relativity offers some sort of protection with principles such as cosmic censorship, which suggest that naked singularities never occur and are always shrouded behind such event horizons. Nevertheless, a complete description of these black holes necessitates a deeper understanding of quantum gravity.

There are more hints that our current Standard Model of particle physics is, as it stands now, not yet complete. A first clue follows from neutrino oscillation experiments: in the SM such particles are massless, but these measurements suggest a tiny mass instead, and it is not yet clear if there is an underlying mechanism that explains this phenomenon. Another recent suggestion comes from the measurement of the anomalous magnetic dipole moment of muons, where experiments by Fermilab indicate a value for $g-2$ that deviates by $4.2 \sigma$ from the Standard Model prediction -- almost sufficient to deem it strong evidence for new physics. Even larger discrepancies arise in cosmological estimates on the energy content of our universe: it is expected that only some 5\% of the physical universe is accounted for by the ordinary matter we know, while respectively about 26\% and 69\% are made up of dark matter and dark energy. For dark matter there have been speculations about new particles that could explain this large portion, although no conclusive evidence has been found so far. For the dark energy component the Standard Model prediction is off by even more dramatic proportions, known as the cosmological constant problem: the measured amount of dark energy suggests a value for the cosmological constant in the equations of general relativity, and the theoretical prediction for this value coming from the Standard Model is $10^{-120}$ orders smaller. 

Putting all of the above observations together, we realize there is yet much to learn about the physics underlying our universe. On the one hand, our quantum field theory understanding of particle physics is still lacking in some corners, where the precise nature of some experimental observations do not yet have a satisfactory explanation. On the other, general relativity does not yield to a direct attempt at quantization, and in extreme settings its description of gravity even breaks down. Faced by these adversities, it is remarkable to note that consistent formulations of quantum gravity do in fact exist at all. This thesis will consider string theory as its candidate for quantum gravity. In this setting for instance some aspects of black hole physics have been elucidated, such as a microscopic computation \cite{Strominger:1996sh} of their Bekenstein-Hawking entropy. However, let us already note that it presents us with numerous challenges as well: for instance, we observe an expanding universe, but it has proven to be difficult to construct vacua with the requisite de Sitter spacetime in string theory; moreover, string theory requires us to introduce extra dimensions for consistency reasons, which presents us with choices that affect the physics we observe from a four-dimensional point of view. In the following sections we lay out the framework underlying string theory, where we highlight key aspects that will play an important role later in this thesis.

\section{String theory, compactification and fluxes}\label{sec:stringintro}
Let us begin with a brief sketch of the main idea behind string theory. At its core it is a remarkably simple program, where we use extended one-dimensional objects -- strings -- instead of particles to describe our fundamental degrees of freedom. These strings can oscillate in different modes, where each mode replaces a kind of particle we observe, allowing us to use a single string to describe a variety of particles. Compared to particles which are localized in spacetime, also note that strings carry an intrinsic length scale with them. This length scale is typically given by the Regge slope $\alpha'$,\footnote{This terminology stems from the origins of string theory in the late 1960s, where it saw the light of day as a framework intended to describe the strong force, modeling interactions between hadrons such as protons and neutrons. It was quickly abandoned for quantum chromodynamics, the current description of the strong force in the Standard Model.} which has units of length squared. The tension of the string is then computed in natural units as
\begin{equation}
T=1/2\pi \alpha'\, .
\end{equation}
It is enticing to take this energy scale as a suggestion for the scale at which quantum gravity effects set in, i.e.~$\Lambda_{\rm Planck} \sim (\alpha')^{-1/2}$, although the string scale could in principle be anywhere between the energy scale of modern colliders and the Planck scale $\Lambda_{\rm LHC} < (\alpha')^{-1/2}  \lesssim \Lambda_{\rm Planck} $. Moreover, this small length would also give an explanation for why we have not yet detected their extended nature, since any scattering process with energy $E \ll (\alpha')^{-1/2}$ does not probe these characteristics.

\begin{figure}[h!]
\begin{center}
\includegraphics[width=7.5cm]{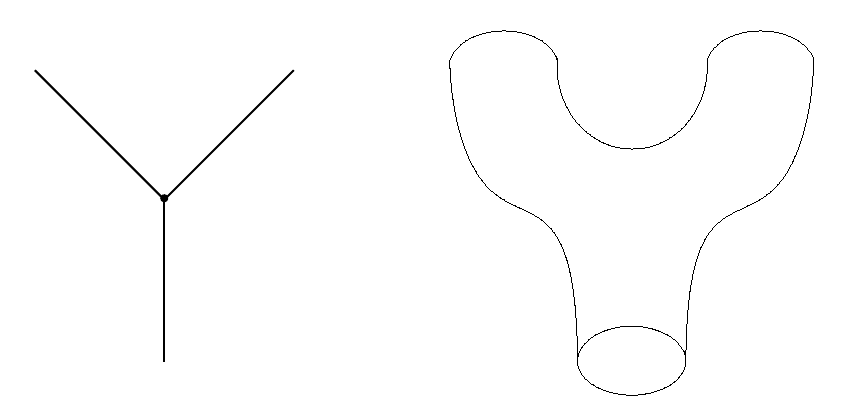}
\end{center}
\caption{\label{fig:interaction} Comparison of an interaction between three particles and three closed strings. The worldlines of the particles meet at an interaction point, while the worldsheets of the strings smoothen this process by tracing out a pair of pants.}
\end{figure}

One of the main advantages of working with string theory over standard particle physics becomes apparent when we look at interactions between states. Where the motion of a particle traces out a one-dimensional wordline in spacetime, strings trace out a two-dimensional worldsheet. This means that if two particles collide and combine into a new particle, there is a junction where these three wordlines meet. To contrast, string theory ameliorates this situation by smoothing out the process: two strings that form another string trace out the shape of a pair of pants, see figure \ref{fig:interaction}. From a more computational point of view, recall that quantum field theory amplitudes suffer from divergencies that have to be cured by renormalization schemes. These divergencies can be traced back to integrals over the high-momentum regimes, or equivalently, small length scales. Heuristically string theory puts a natural cutoff on these integration limits with the finite size of its strings, and hence it is by construction free of such infinities.

So far we have postponed specifying a shape for the string, but there is a choice in its topology: we can have closed strings with no endpoints and open strings with two endpoints. Depending on whether a string is open or closed we encounter different kinds of excitations in its oscillation modes. For open strings there are massless spin-one excitations, while for closed strings we find a massless spin-two excitation, both of which would fit naturally into our current picture of particle physics. Massless spin-one particles are gauge bosons such as the photon and gluon in our Standard Model. On the other hand, a massless spin-two particle would give rise to a graviton, the force carrier of gravity. String theory is therefore inherently a theory of gravity, and thus gives us a promising window into quantum gravity. Even if one were to start from a situation with only open strings, then these strings can merge to produce closed strings, thereby giving rise to gravity after all.

The endpoints of open strings deserve some further attention, because in contrast to closed strings propagating freely through spacetime, open strings are attached to objects known as D-branes. These D-branes are stable objects in their own right, and in fact can be thought of as the fundamental degrees of freedom of string theory in appropriate regimes of the physical couplings. We can even study the physical theory living on the worldvolume of these D-branes, which are gauge theories arising from the spin-one modes of the open strings, and look remarkably similar to their counterparts in our Standard Model. Crucially, note that this also makes string theory into not just a theory describing strings, but rather a theory that encompasses extended objects of any dimension.

The next step in setting up string theory is to quantize the theory living on the two-dimensional worldsheet of the string, in order to make a connection with quantum gravity. The position coordinates of the string in the ambient spacetime are then parametrized by scalar fields living on the string worldsheet. At a classical level this worldsheet theory possesses a conformal symmetry group, but this symmetry only persists at the quantum level for a particular number of scalar fields. In other words, since these scalars parametrize the position of the string, we obtain a requirement for the dimension of spacetime itself. For the bosonic strings this leads to a 26-dimensional spacetime, while for the superstring it is 10-dimensional.\footnote{Bosonic strings and supersymmetric strings differ in the field content of the worldsheet theory: bosonic strings only have scalar fields, while superstrings pair these up with fermions in order to make the spectrum supersymmetric.} In this thesis we will focus on the so-called Type II superstrings, so we have to deal with a ten-dimensional spacetime.

\begin{figure}[h!]
\begin{center}
\includegraphics[width=7.5cm]{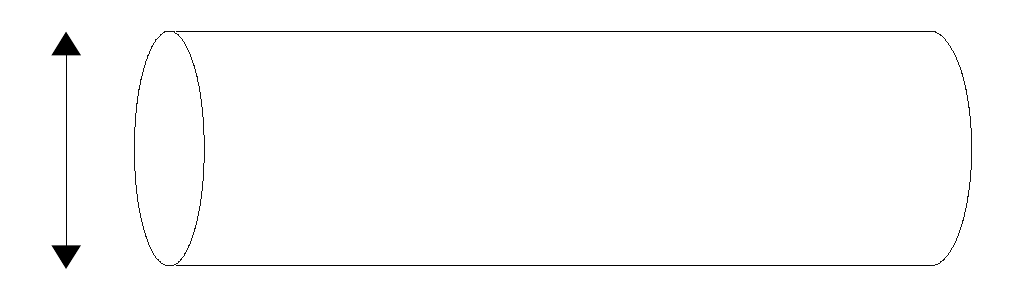}
\end{center}
\begin{picture}(0,0)
\put(37,55){\small $r \ll \Lambda^{-1}_{\rm LHC}$}
\end{picture}
\vspace*{-1cm}
\caption{\label{fig:cylinder} Depiction of a circle compactification.}
\end{figure}
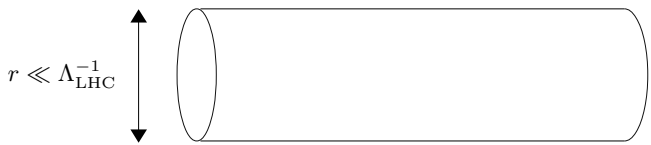

In order to connect with the four-dimensional spacetime we observe, we have to explain why these extra dimensions are absent from our viewpoint. In a similar spirit to the extended nature of the string, we can argue that as long as the length scale associated to the extra dimensions is small enough, particle accelerators do not detect these additional degrees of freedom. More concretely, this means we take the coordinates that parametrize the extra dimensions to lie on some compact space, whose volume is sufficiently small compared to the collider scale $\Lambda_{\rm LHC}\sim 10^4$ GeV. In order to make this more intuitive, let us consider an example and say we compactify one dimension on a circle as depicted in figure \ref{fig:cylinder}. A useful analogy here is to consider both a cord walker and an ant: while the cord walker can only move forwards or backwards, an ant the same size as the circle can also walk around it. In other words, the extra rolled up dimensions only become visible at the smallest length scales, or equivalently the highest energy scales.

\begin{figure}[h!]
\begin{center}
\includegraphics[width=9.5cm]{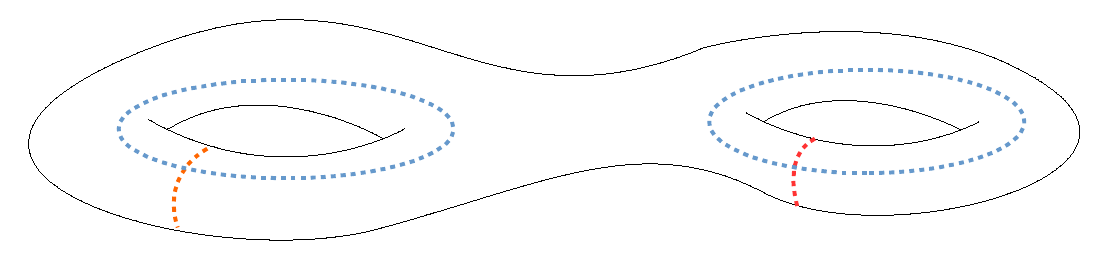}
\end{center}
\caption{\label{fig:riemannsurface} Depiction of a compactification manifold as a genus-two Riemann surface. The independent 1-cycles of this geometry have been colored in red and blue.}
\end{figure}
The presence of extra dimensions presents us with a choice of internal manifold. While in the one-dimensional case the only possible topologies are either homeomorphic to a line or a circle, in higher dimensions the number of options is much larger. For instance, in the two-dimensional case one could a priori pick a Riemann surface of any genus as a compact space -- any number of holes -- as depicted in figure \ref{fig:riemannsurface}. Requiring supersymmetry to remain unbroken in this process leads us to consider a particular class of compactification geometries known as Calabi-Yau manifolds. 

Furthermore, as we will make more precise later in section \ref{sec:CY}, the deformation parameters of the compactification space descend to neutral scalar fields in the four-dimensional effective theory. For our example at hand, this means the size of each of the 1-cycles in figure \eqref{fig:riemannsurface} would give rise to such a scalar field.\footnote{While the only Riemann surface that is Calabi-Yau has genus one, i.e.~a torus, the structure of cycles depicted in figure \ref{fig:riemannsurface} does give us an intuitive sketch of the nature of deformation parameters in higher-dimensional geometries.} In a compactification of string theory without any extra effects these scalar fields are massless, and are thus free to take any value, while we would for instance like the volume of the internal manifold to take a small value. Moreover, for a typical Calabi-Yau manifold there are $\mathcal{O}(100)$ so-called deformation moduli, while our current Standard Model contains only a single scalar particle, the Higgs particle, which is massive and charged under the weak interaction. Finally, scalars would mediate an attractive force between particles proportional to $e^{-mr}/r^2$, with $m$ the mass of the scalar field and $r$ the distance between them. In the massless case $m=0$ this gives rise to long-range interactions that would be picked up by fifth force experiments. To reconcile these issues, we have to set up a mechanism that generates a mass for these deformation moduli and thereby fixes them to suitable values. This program goes under the name of  \textit{moduli stabilization}.

In this thesis we will study one such approach to moduli stabilization: turning on background fluxes. We will sketch a heuristic picture here, and refer to section \ref{sec:CY} for how this works in practice. The  spectrum of the ten-dimensional superstring contains, among its massless modes, some higher-form field strengths. We can take the vacuum expectation values of these field strengths to vanish everywhere, but we can also choose to turn them on along certain cycles of the compactification geometry. The kinetic terms of these higher-form fields in the ten-dimensional action then generate a scalar potential for the moduli fields in four dimensions. For definiteness, let us take a two-form gauge potential $B_2$ as an example, and denote its three-form field strength by $H_3 = \dd B_2$. Schematically the scalar potential can then be written as an integral over the Calabi-Yau manifold $Y_3$ as\footnote{For ease of demonstration we chose to leave out a slightly complicated overall factor, and hence did not write a strict equal sign.}
\begin{equation}\label{eq:V4d}
V_{\rm 4d} \simeq \int_{Y_3} H_3 \wedge \ast H_3\, ,
\end{equation}
where the Hodge star operator $\ast$ on $Y_3$ induces a dependence on the Calabi-Yau moduli. This is seen most intuitively by writing the integral out in local coordinates, and in turn noting that deformations of the geometry affect the metric factors coming from the Hodge star.\footnote{We can write out \eqref{eq:V4d} in local coordinates as $\int_{Y_3}\dd^6x\, \sqrt{g} \, g^{kp} g^{lq} g^{mr} (H_3)_{klm} (H_3)_{pqr}$ up to some overall factor, with $g^{kp}$ denoting the Calabi-Yau metric and its indices run over $1, \ldots , 6$.} By picking a suitable choice of background flux $H_3$ one can then search for vacua where such a scalar potential is minimized.

Studying compactifications of string theory thus brings us into a mathematical arena where many aspects of the four-dimensional physics are encoded in the geometry of the internal manifold. This connection has led to many fruitful exchanges between research in physics and mathematics, with mirror symmetry between Calabi-Yau manifolds \cite{Candelas:1990rm} as one of the most prominent examples. More concretely, this setting requires us to develop a toolkit to explore the moduli spaces of these geometries -- the spaces parametrized by their deformation parameters. For simple geometries such as a circle this would just be the interval $[0,\infty)$ parametrized by its radius, but for more complicated cases such as Calabi-Yau manifolds one is compelled to look for a deeper underlying structure. In this thesis we will turn to the framework of \textit{asymptotic Hodge theory} \cite{Schmid, CKS}, which describes the asymptotic regimes in moduli space close to points where the Calabi-Yau manifold degenerates, see for instance figure \ref{fig:degeneration} for an illustration. It provides us with the necessary tools to probe these corners, and in particular gives us the asymptotic behavior of the Hodge star in the four-dimensional scalar potential \eqref{eq:V4d}.


\section{The swampland program}\label{sec:swampland}
Let us for the moment take a step back from string theory, and wonder what requirements an EFT should meet in order to admit a UV completion into a theory of quantum gravity. In other words, what distinguishes EFTs that can be coupled consistently to quantum gravity from those that cannot. This question was originally posed in \cite{Vafa:2005ui}, and the line of research intending to address it is known as the swampland program: the set of all EFTs that can be completed into a theory of quantum gravity are referred to as the \textit{landscape}, while those that cannot are said to be in the \textit{swampland}. We included a sketch of this decomposition of the set of effective theories in figure \ref{fig:venn}. For some recent reviews we refer the reader to \cite{Palti:2019pca,vanBeest:2021lhn}.

The criteria that aim to delineate the border between the swampland and the landscape go under the name of \textit{swampland conjectures}. These conjectures propose general properties that any EFT which can be embedded into quantum gravity should (or should not) have. By continually testing, refining and interconnecting these conditions this program attempts to uncover the fundamental structures that underlie a theory of quantum gravity. As a result, it also yields an increasingly stringent web of requirements that cause tension with bottom-up constructions of phenomenological models. In practice, these swampland criteria are often inspired either by semi-classical black hole arguments or observations based on string theory. Within the swampland program much effort has been put into examining the validity of these conjectures in Calabi-Yau compactifications of string theory \cite{Grimm:2018ohb,Blumenhagen:2018nts,Lee:2018urn,Lee:2018spm,Grimm:2018cpv,Corvilain:2018lgw,Lee:2019tst,Font:2019cxq,Marchesano:2019ifh,Lee:2019xtm,Grimm:2019wtx,Demirtas:2019lfi,Kehagias:2019akr,Lee:2019wij,Grimm:2019ixq,Baume:2019sry,Enriquez-Rojo:2020pqm,Andriot:2020lea,Cecotti:2020rjq,Gendler:2020dfp,Lanza:2020qmt,Heidenreich:2020ptx,Xu:2020nlh,Grimm:2020cda,Bastian:2020egp,Klaewer:2020lfg,Calderon-Infante:2020dhm,Grimm:2020ouv,Cota:2020zse,Brodie:2021ain,Cecotti:2021cvv,Lanza:2021qsu,Buratti:2021fiv,Castellano:2021yye,Palti:2021ubp, Bakker:2021uqw,Lee:2021qkx, Lee:2021usk,Grimm:2021vpn, Alvarez-Garcia:2021pxo,Long:2021jlv}. These setups yield lower-dimensional effective theories that feature for instance complicated scalar field spaces, thereby providing us with a landscape of challenging testing grounds. Below we introduce the swampland conjectures that are most relevant to the work carried out in this thesis. 

\begin{figure}[h!]
\begin{center}
\includegraphics[width=8.5cm]{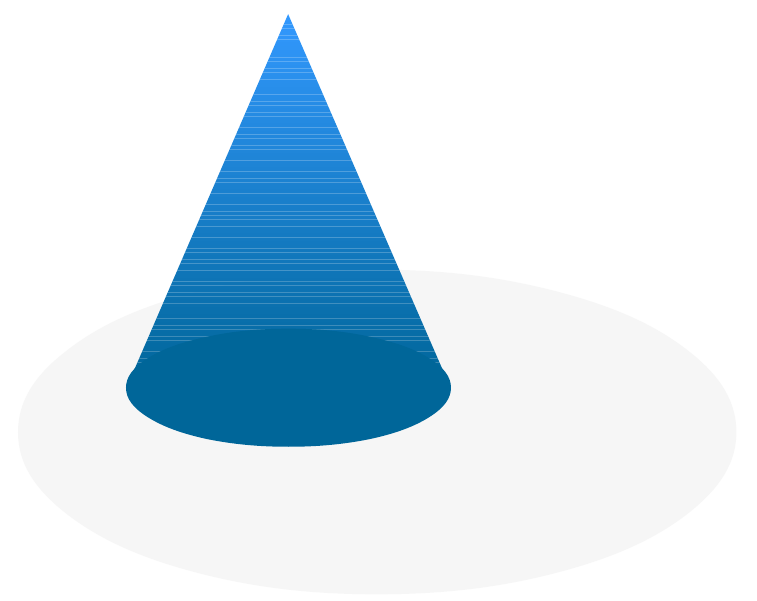}
\end{center}
\begin{picture}(0,0)
\put(155,206){\small quantum gravity}
\put(35,137){\tiny energy scale}
\put(70,130){ \rotatebox{90}{$\longrightarrow$}}
\put(170,60){\small swampland}
\put(130,88){\small \color{white} landscape}
\end{picture}\vspace*{-0.9cm}
\caption{\label{fig:venn} Set of all consistent (anomaly-free) effective field theories based on whether they admit a UV completion into quantum gravity (the landscape) or not (the swampland).}
\end{figure}
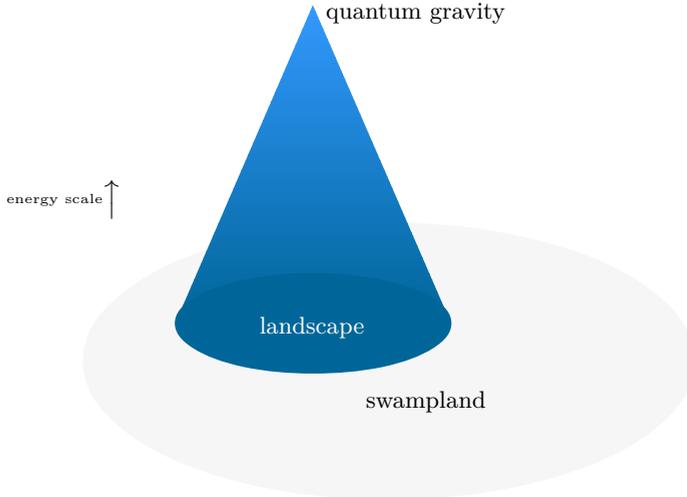

Perhaps the oldest swampland conjecture, even predating the establishment of the swampland program itself, stems from the expectation that a theory of quantum gravity does not have any global symmetries:
\begin{mdframed}[roundcorner=7pt]
\begin{minipage}{\textwidth}
\paragraph{No Global Symmetries Conjecture \cite{Banks:1988yz}:} any effective field theory coupled to quantum gravity cannot possess exact global symmetries.
\end{minipage}\end{mdframed}
In other words, the embedding of an EFT into quantum gravity must either break or gauge a global symmetry. The original motivation for this conjecture comes directly from string theory, where global symmetries on the worldsheet always appear as gauge symmetries in the target spacetime. Recently, this conjecture was also shown to hold true in the setting of AdS/CFT \cite{Harlow:2018jwu,Harlow:2018tng}, where it was argued that the presence of any global symmetry in the bulk, be it broken or discrete, lead to inconsistencies in the dual CFT.

The most intuitive argument for this absence of global symmetries in quantum gravity comes from black hole physics, see for instance \cite{Banks:2010zn} for a more detailed discussion. Say we have an effective field theory with a global U(1) symmetry, such as the B-L symmetry in the Standard Model. We can then imagine we create a black hole with the collapse of some baryonic matter, i.e.~we started from a state with a large global charge. Subsequently, we allow the black hole to evaporate through Hawking radiation, which is neutral with respect to the global charge. This leads us to a contradiction, where the initially large baryonic charge seems to have disappeared completely, and is thus in particular not conserved.\footnote{As an alternative, the black hole could also not decay completely but instead form some remnant, which are known to lead to other sorts of inconsistencies in the EFT.}  For our Standard Model this suggests the B-L symmetry must be broken at some higher energy scale, for instance by a scattering process.

Our next swampland criterion gives a prediction for the strength of gravity compared to the other forces. It suggests that gravity should always be the weakest force, and quantifies this into the following statement:
\begin{mdframed}[roundcorner=7pt]
\begin{minipage}{\textwidth}
\paragraph{Weak Gravity Conjecture (WGC) \cite{ArkaniHamed:2006dz}:} in any consistent theory of quantum gravity with a U(1) gauge field there should exist a superextremal particle of mass $m$ and charge $q$ such that
\begin{equation}\label{eq:WGC}
\frac{|q|}{m} \leq \frac{|Q|}{M}\bigg|_{\rm extr}\, ,
\end{equation}
with the charge-to-mass ratio of an extremal black hole on the right-hand side.
\end{minipage}\end{mdframed}
In other words, in appropriate units it requires the presence of a particle whose mass is lower than or equal to its charge. The interpretation of gravity as the weakest force becomes apparent by noting that if all particles were to violate the bound, then the gravitational attraction between two identical particles would overwhelm the repulsion under the gauge force; the WGC thus gives us a superextremal particle on which gravity acts weaker than the gauge force. Moreover, due to the different directions in which gauge forces and gravity act, the WGC particle is self-repulsive instead of self-attractive, so the WGC can also be stated as a demand for the presence of a self-repulsive particle.\footnote{This situation is slightly different when scalar fields are present, which also mediate an attractive force between identical particles. In this case a superextremal particle could still be self-attractive due to this scalar field interaction, and an appropriate refinement was proposed with the Scalar WGC \cite{Palti:2017elp}, also referred to as the Repulsive Force Conjecture.} 

In a similar fashion as the argument against global symmetries, the WGC can be motivated with a black hole argument. We now begin with an extremal black hole, replacing the baryonic matter content from before by a large amount of electric charge. The black hole again decays through Hawking radiation. This radiation can be electrically charged, but if it only contains subextremal particles then the black hole can never emit all of its charge. The WGC thus ensures the absence of remnants in the evaporation of these extremal black holes.


By applying electromagnetic duality we can obtain a magnetic version of the WGC. The analogous bound on the monopole mass $m_{\rm mon} \lesssim M_{\rm p}/g$ can then be used to obtain an upper bound on the cutoff scale of the effective theory itself as
\begin{mdframed}[roundcorner=7pt]
\begin{minipage}{\textwidth}
\paragraph{Magnetic WGC \cite{ArkaniHamed:2006dz}:} the cutoff scale of an EFT with a U(1) gauge field has to satisfy
\begin{equation}
\Lambda \lesssim g M_{\rm p}\, .
\end{equation}
\end{minipage}\end{mdframed}
Remarkable, this would suggest new physics to occur around $\Lambda \sim 10^{17}$ GeV -- well below the Planck scale -- by plugging in the value of the gauge couplings near the unification scale. Also note that in the limit $g \to 0$ where the gauge symmetry becomes global we recover the No Global Symmetries Conjecture: the EFT breaks down completely because its cutoff $\Lambda$ must go to zero. Let us note that the WGC also admits various generalizations: a convex hull condition \cite{Cheung:2014vva} when there are multiple U(1) gauge fields; stronger versions requiring sublattices \cite{Heidenreich:2015nta,Heidenreich:2016aqi,Montero:2016tif} and towers \cite{Andriolo:2018lvp} of WGC particles; and extensions from ordinary 1-form gauge symmetries to the case of $p$-form gauge fields.

Our next swampland conjecture addresses large excursions in scalar field spaces. It predicts an infinite tower of states to become massless at infinite distance points in field space. To be more precise, it relates the mass scale of the tower with the geodesic distance traveled as
\begin{mdframed}[roundcorner=7pt]
\begin{minipage}{\textwidth}
\paragraph{Swampland Distance Conjecture \cite{Ooguri:2006in}:} For two points $P,Q$ in field space there exists an infinite tower of states whose masses become exponentially small in the geodesic distance $d(P,Q)$ as
\begin{equation}\label{eq:distance}
m(P) = m(Q) e^{-\gamma d(P,Q)}\, ,
\end{equation}
where $\gamma$ denotes some $\cO(1)$ constant in units of $M_p$.
\end{minipage}\end{mdframed}
Here we assume that $Q$ is fixed as a reference point in the middle of field space with $m(Q)$ some associated mass scale, while $P$ is the point that is moved towards an infinite distance singularity in field space. The infinite tower of states becoming exponentially light signals a breakdown of the EFT near these points, because increasingly many light states have to be incorporated in the EFT below its fixed energy cutoff $\Lambda$. There have been proposed refinements \cite{Klaewer:2016kiy} and closely related conjectures \cite{Lee:2019wij,Lanza:2020qmt,Lanza:2021qsu}. 

The precise nature of the tower of states depends on the large field limit under consideration. Evidence for this distance conjecture stems from string theory, where the field space typically is the moduli space of some compactification manifold. A simple example, which we discuss in more detail in section \ref{sec:CY}, is a circle compactification: the tower at infinite radius is made up of Kaluza-Klein modes, while the tower at vanishing radius consists of winding modes of the string. Considering a more complicated setup: for Calabi-Yau compactifications of Type IIB string theory, near infinite distance boundaries in complex structure moduli space, infinite towers of BPS states arising from D3-brane wrapped on vanishing three-cycles have been constructed using asymptotic Hodge theory  in \cite{Grimm:2018ohb}. Another instructive example is the weak-coupling limit for the dilaton of the ten-dimensional Type II strings, where the massless states are the modes of the fundamental string itself. In fact, the Emergent String Conjecture of \cite{Lee:2019wij} proposes that any infinite distance limit is either a decompactification limit with a corresponding Kaluza-Klein tower, or a limit where a weakly coupled fundamental string (possibly in some dual frame) becomes tensionless. Another interesting perspective is given by the emergence proposal \cite{Grimm:2018ohb, Heidenreich:2018kpg,Corvilain:2018lgw}, which states that the infinite distance in the IR couplings arises from integrating out the infinite tower of massive states in the UV theory.

Finally we consider one of the more controversial swampland conjectures, which imposes bounds on scalar potentials by which it forbids any meta-stable de Sitter vacua. To be more precise, the conjecture states that
\begin{mdframed}[roundcorner=7pt]
\begin{minipage}{\textwidth}
\paragraph{de Sitter conjecture \cite{Obied:2018sgi,Ooguri:2018wrx}:} the scalar potential in any EFT compatible with quantum gravity should satisfy 
\begin{equation}\label{eq:dS1}
|\nabla V| \leq \frac{c}{M_p} V\, ,
\end{equation}
or
\begin{equation}\label{eq:dS2}
\min (\nabla_i \nabla_j V) \leq -\frac{c'}{M_p^2} V \, ,
\end{equation}
where $c,c'$ are some order one constants. The left-hand side of \eqref{eq:dS1} denotes the norm of the gradient $\nabla V$ with respect to the scalar fields computed with the field space metric, while for \eqref{eq:dS2} we consider the smallest eigenvalue of its Hessian in an orthonormal frame of this metric.
\end{minipage}\end{mdframed}
The first condition \eqref{eq:dS1} was proposed in the original conjecture \cite{Obied:2018sgi} and clearly prohibits any models of slow-roll inflation. The second condition was included later in the refinement \cite{Ooguri:2018wrx} to allow the scalar potential to have maxima as e.g.~occur in the Higgs potential, as was pointed out in \cite{Denef:2018etk}. 




\section{Superstrings, M-theory and dualities}\label{sec:TypeIIstrings}
Let us now switch gears a little bit and return to the setting of string theory. In this section we introduce the set of supersymmetric string theories in ten dimensions (and M-theory), and discuss in detail the low-energy limits of those theories most relevant to our work. In particular, we highlight the axio-dilaton field space of the Type IIB string and how it fits with asymptotic Hodge theory.


\subsection{Overview of supersymmetric string theories}
We begin with a brief overview of the set of possible string theories. Bosonic string theories do not give rise to fermions living in the ambient spacetime, so we focus our attention on the supersymmetric strings. To be precise, there are five ten-dimensional supersymmetric string theories,\footnote{We focus on string theories supersymmetric in the ten-dimensional target space. Note that there are more superstring theories which are supersymmetric on the worldsheet, but these do not give rise to supersymmetry in the target space.} which we will introduce here. A special role is played by a hypothetical eleven-dimensional called M-theory, which is expected to be related to each of these superstring theories in some limits via a web of dualities, see figure \ref{fig:Mtheorystar}

\paragraph{Type II strings.} There are two superstring theories in ten dimensions with 32 supercharges: the Type IIA and Type IIB superstrings. The IIA string has $\mathcal{N}=(1,1)$ supersymmetry and therefore is non-chiral, while the IIB string is chiral having $\mathcal{N}=(2,0)$ supersymmetry. The low-energy limits of these superstring theories can be matched precisely to the two ten-dimensional $\mathcal{N}=2$ supergravity theories according to their supersymmetries. Both Type II strings contain closed and open strings, where open strings in Type IIA end on D-branes of odd dimensions and in Type IIB of even dimensions. These objects play an important role in string phenomenology because they allow us to engineer gauge groups in the four-dimensional effective theories arising from compactifications.

\paragraph{Type I string.} The Type I string theory is a theory of open and closed unoriented strings, which has $\mathcal{N}=1$ supersymmetry and an SO(32) gauge group. In its low-energy limit it is described by ten-dimensional $\mathcal{N}=1$ supersymmetric Yang-Mills theories with an SO(32) gauge group coupled to the Type I supergravity. From a modern point of view the Type I string theory is understood as Type IIB string theory with 32 D9-branes and an orientifold projection.

\paragraph{Heterotic strings.} These strings have 16 supercharges, i.e.~minimal $\mathcal{N}=1$ supersymmetry, in ten dimensions. They are a hybrid between the bosonic string in the left-moving sector and the superstring in the right-moving sector. The heterotic string theories contain only closed strings, and have either an $E_8 \times E_8$ or SO(32) gauge group. In their low-energy limits these are described by ten-dimensional $\mathcal{N}=1$ supersymmetric Yang-Mills theories with the corresponding gauge groups, coupled to the Type I supergravity. Historically heterotic strings played an important role in the first superstring revolution due to their attractiveness for phenomenology: potential for building a grand unified theory (GUT) (SU(5), SO(10) and $E_6$ are subgroups of the ten-dimensional gauge groups) and a low amount of supersymmetry.

\paragraph{M-theory.} Finally, M-theory is a hypothetical eleven-dimensional theory introduced by Witten in 1995. In contrast to the above, it is a theory of membranes -- M2-branes and M5-branes -- rather than strings, which in its low-energy limit is described by the unique eleven-dimensional supergravity. It is expected that all five superstring theories arise as special limits of M-theory, and there is compelling evidence for this web of dualities. For instance, the Type IIA supergravity is recovered upon compactifying the eleven-dimensional supergravity on a circle, and conversely there is support for M-theory as the strong-coupling limit of Type IIA string theory.

\subsection{Supergravity actions}
We now discuss the low-energy limits of the Type II superstrings and M-theory. The spectrum of the strings can be divided into four sectors -- R-R, NS-NS, R-NS and NS-R -- of which the first two describe bosons in the ten-dimensional spacetime and the latter two fermions.\footnote{These sectors correspond to the boundary conditions for the left- and right-moving modes of the fermions living on the string worldsheet, where the Ramond (R) sector is periodic and the Neveu-Schwarz is anti-periodic.} For our purposes we will focus on the bosonic sectors, with the fermionic terms in the action fixed by supersymmetry. 

We begin with the NS-NS sector, which yields the same description for the Type IIA and IIB strings at low energies. Its field content consists of the spacetime metric, a dilaton $\phi$ and the Kalb-Ramond field $B_2$ with field strength $H_3 = \dd B_2$. The corresponding action can be written as
\begin{equation}\label{eq:NSNS}
S_{\rm (NS,NS)} = \frac{1}{2\kappa_{10}^2} \int_{\cM_{10}} \me^{-2\phi} \left( R \ast 1 - \frac{1}{2} H_3 \wedge \ast H_3 + 4 \dd\phi \wedge \ast \dd\phi \right)\, ,
\end{equation}
where $\kappa_{10}$ denotes the ten-dimensional gravitational coupling, related to the string length scale via $\kappa_{10}^{-2} = 4\pi \ell_s^{-8}$ with $\ell_s = 2\pi \sqrt{\alpha'}$. The two-form potential $B_2$ couples electrically to the worldsheet of the string, while its magnetic dual is a six-form sourcing an object known as the NS5-brane.

\subsubsection*{Type IIA}
For the R-R sector the low-energy spectrum does depend on the Type II string under consideration. For the Type IIA string the action is given by a sum of kinetic terms and Chern-Simons terms as
\begin{equation}\label{eq:RRIIA}
S_{\rm (R,R)} = -\frac{1}{4\kappa_{10}^2} \int_{\cM_{10}} \bigg( F_2 \wedge \ast F_2 +F_4 \wedge \ast F_4 + B_2 \wedge \dd C_3 \wedge \dd C_3 \bigg)\, ,
\end{equation}
with the field strenghts for the R-R potentials $C_1, C_3$ given by
\begin{equation}
F_2 = \dd C_1\, , \qquad F_4 = \dd C_3 + H_3 \, \wedge \, C_1\, .
\end{equation}
These potentials together with their magnetic duals $C_5$ and $C_7$ source the D0, D2, D4 and D6-branes on which open strings end. Including D8-branes requires us to introduce another non-dynamical field strength $F_{10} = F_0 \, \text{dvol}_{\rm 10}$ in the theory, where $F_0$ is known as Roman's mass.

\subsubsection*{Type IIB}
For the IIB string the low-energy field content of the R-R sector is made up of an axion $C_0$, a two-form potential $C_2$ and a four-form $C_4$. The action is given by
\begin{equation}\label{eq:RRIIB}
S_{\rm (R,R)} = -\frac{1}{4\kappa_{10}^2} \int_{\cM_{10}} \bigg(F_1 \wedge \ast F_1 +F_3 \wedge \ast F_3 + \frac{1}{2} F_5 \wedge \ast F_5 + C_4 \wedge H_3 \wedge F_3 \bigg) \, , 
\end{equation}
with field strengths
\begin{equation}
F_1 = \dd C_0\, , \quad F_3 = \dd C_2 - C_0 \, \dd B_2\, , \quad F_5 = \dd C_4 - \frac{1}{2} C_2 \, \wedge \, \dd B_2 +\frac{1}{2} B_2 \, \wedge \, \dd C_2\, . 
\end{equation}
Here D1 and D3-branes are charged electrically under $C_2$ and $C_4$ respectively, while these R-R potentials source D7 and D5-branes magnetically.

Let us also note that the action of the Type IIB supergravity -- the sum of \eqref{eq:NSNS} and \eqref{eq:RRIIB} -- is only a pseudo-action. In addition to the equations of motions coming from the action we have to impose a self-duality condition on the five-form field strength as
\begin{equation}\label{eq:selfdualF5}
F_5 = \ast F_5\, .
\end{equation}
By combining the equations of motions coming from \eqref{eq:NSNS} and \eqref{eq:RRIIB} with this self-duality condition we obtain the complete set of equations describing the IIB supergravity. Note in particular that \eqref{eq:selfdualF5} cannot be imposed at the level of the action, since the kinetic term in \eqref{eq:RRIIB} for $C_4$ would then vanish.

\subsubsection*{M-theory} 
In its low-energy limit M-theory is described by the unique eleven-dimensional supergravity. This theory contains a metric and a three-form potential $C_3$ described by the action
\begin{equation}\label{eq:M}
S_{\rm M} = \frac{1}{2\kappa_{11}^2} \int_{\cM_{11}} \left(R \ast 1 -\frac{1}{2} G_4 \wedge \star G_4 - \frac{1}{6} C_3 \wedge G_4 \wedge G_4 \right)\, ,
\end{equation}
with field strength $G_4 = \dd C_3$, and gravitational coupling $\kappa_{11}^{-2} = 4\pi \ell_{\rm M}^{-9}$ with $\ell_{\rm M}$ the eleven-dimensional Planck length. Three-dimensional objects known as M2-branes couple electrically to $C_3$, while this potential sources M5-branes magnetically.

\subsection{Dualities}
Let us now elaborate on how these ten- and eleven-dimensional theories are related under dualities. For simplicity we restrict ourselves to M-theory, Type IIA and Type IIB. We motivate our discussion from the perspective of the swampland program, with the distance conjecture \eqref{eq:distance} as our guide: moving to other corners of the M-theory star in figure \ref{fig:Mtheorystar} typically requires an infinite distance in field space, and hence we expect an infinite tower of states to arise. This infinite tower signals that a dual description takes over, requiring a reformulation in terms of the fundamental degrees of freedom at this other corner.

\begin{figure}[h!]
\begin{center}
\end{center}
\begin{tikzpicture}
\node[inner sep=0pt] (star) at (0,0) {\includegraphics[width=7cm]{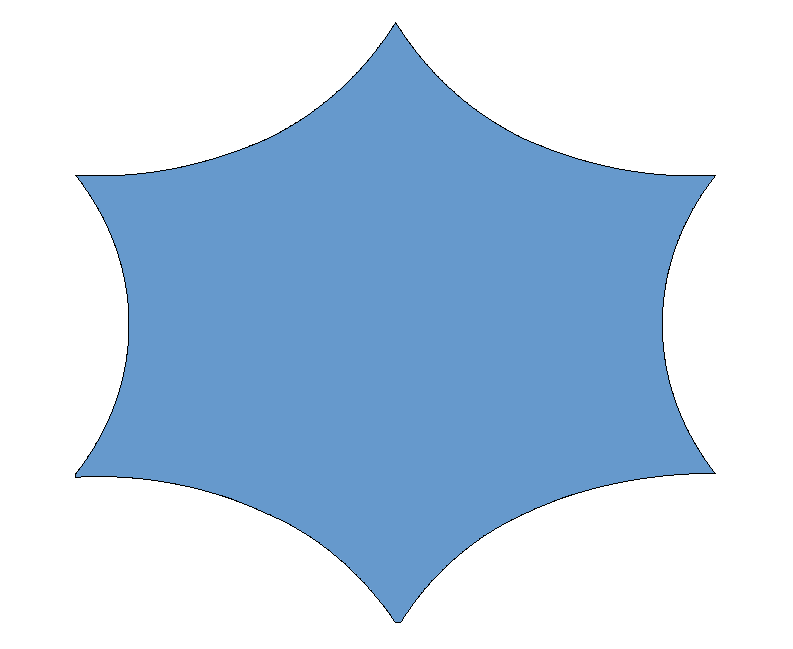}};
\node (a) at (0,3.1) {11d supergravity};
\node[right] (b) at (3,1.3) {heterotic $E_8 \times E_8$};
\node[right] (c) at (3,-1.3) {heterotic SO(32)};
\node (d) at (0,-3.1) {Type I};
\node (e) at (-3.9,-1.3) {Type IIB};
\node (f) at (-3.9,1.3) {Type IIA};

\node (m) at (0,0) {\Large \color{white} M-theory};

\draw [ ->] (a) to[out=0, in=120] node[pos=0.6, right] {\footnotesize on $S^1/\mathbb{Z}_2$} (b);
\draw [ <->] (b) to[out=285, in=75] node[midway, right] {\footnotesize T-duality} (c);
\draw [ <->] (c) to[out=240, in=0] node[pos=0.35, right] {\footnotesize S-duality} (d);
\draw [ <-] (d) to[out=180, in=300] node[midway, left] {\footnotesize orientifold} (e);
\draw [ ->] (e) to[out=240, in=180, looseness=4] node[midway, left] {\footnotesize S-duality} (e);
\draw [  <->] (e) to[out=105, in=255] node[midway, left] {\footnotesize T-duality} (f);
\draw [ ->] (a) to[out=180, in=60] node[midway, left] {\footnotesize on $S^1$} (f);
\end{tikzpicture}
\vspace*{-0.2cm}
\caption{\label{fig:Mtheorystar} M-theory star with the various ten- and eleven-dimensional theories as limits in special corners, with dualities among them indicated.}\end{figure}
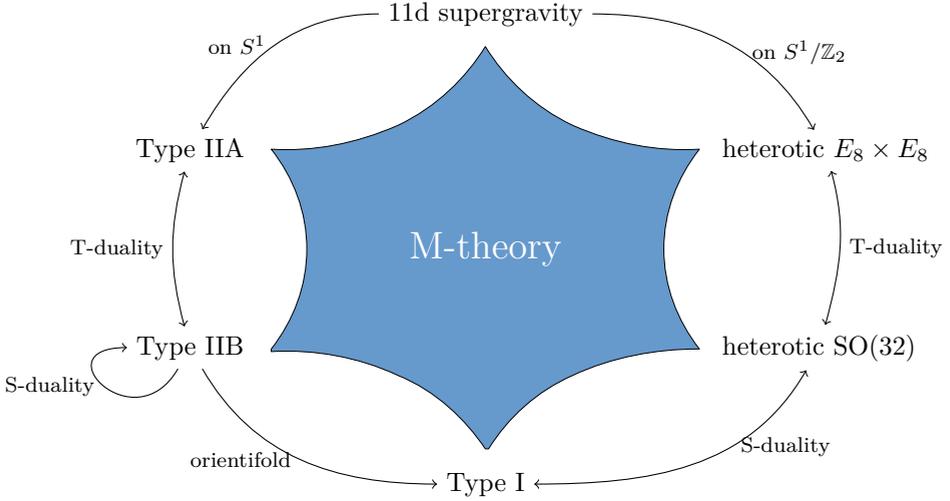

\subsubsection*{M-theory and Type IIA} 
Let us begin from the perspective of the Type IIA string. The supergravity action contains the dilaton $\phi$, for which the field space metric in the Einstein frame\footnote{Going to the Einstein frame corresponds to sending the ten-dimensional spacetime metric to $g_{\mu\nu} \to e^{\phi/2} g_{\mu \nu}$ such that the dilaton prefactor of the Ricci scalar in the action disappears.} is constant as can read off from \eqref{eq:NSNS}. This means we have two infinite distance points: weak-coupling at $\phi \to -\infty$ and strong-coupling at $\phi \to \infty$. The weak-coupling limit is easily explained as the limit where the modes of the Type IIA string become light. The strong-coupling requires us to change perspective and replace the modes of the Type IIA string as fundamental degrees of freedom by D0-branes: its mass is given by
\begin{equation}
m_{{\rm D0}} = \frac{2\pi}{\ell_s} e^{-\phi} \, ,
\end{equation}
which indeed becomes massless in the limit $\phi \to \infty$. By forming bound states of $n$ D0-branes we obtain a tower of states that becomes light. The spacing of the masses in this tower is given by $m_n^2 = m_{{\rm D0}}^2 n^2$, so we find a  quadratic scaling in $n$ rather than linear as occurs for a tower of string excitations. It hints that a decompactification limit occurs, and indeed this turns out to be the case: in the strong-coupling limit we decompactify to eleven-dimensional M-theory, with the tower of bound D0-branes as Kaluza-Klein modes. The radius of the circle as computed with the eleven-dimensional metric fixes the string coupling as $r=\ell_{\rm M} e^{2\phi/3}$. One can verify this more explicitly by dimensionally reducing the eleven-dimensional supergravity action \eqref{eq:M} on a circle and reproducing the Type IIA action \eqref{eq:NSNS} and \eqref{eq:RRIIA}.

\subsubsection*{T-duality between Type II strings}
T-duality is a duality between string theories compactified on circles, or more generally, on manifolds that admit continuous isometries. Let us consider the compactification of Type II strings on a circle of radius $r$. Closed strings can then carry a momentum $p$ along this circle direction, which must be quantized as $p=n/r$ with $n$ some integer. We can also allow strings to wind around the circle as 
\begin{equation}
X(\tau, \sigma+2\pi) = X(\tau, \sigma)+2\pi w r\, , 
\end{equation}
 with winding number $w$, and $X(\tau, \sigma)$ parametrizes the position on the circle as a function of the worldsheet  coordinates $\tau, \sigma$, with $\sigma \sim \sigma+2\pi$. The excitations of the closed string can thus be labeled by two quantum numbers $n,w$ indicating the momentum along and winding around the circle. Schematically the masses of these modes are given by
\begin{equation}\label{eq:Tdualmass}
m^2 = \left( \frac{n}{r}\right)^2 + \left( \frac{w r}{\alpha'} \right)^2 + \ldots\, ,
\end{equation}
where the ellipsis represents terms coming from oscillator modes irrelevant for our analysis here. The first term is the Kaluza-Klein mass due to the momentum along the circle, while the second is the mass of a string of tension $1/2\pi \alpha'$ winded $w$ times around a circle of length $2\pi r$. The idea of T-duality is to relate Type IIA string theory on a circle of radius $r$ to Type IIB string theory on a circle of radius $\alpha'/r$, where momentum and winding modes are exchanged $(n,w) \to (w,n)$. Notice that this indeed leaves the mass spectrum \eqref{eq:Tdualmass} unchanged.

It is instructive to investigate this duality from the perspective of the distance conjecture in the nine-dimensional effective theory. The radius $r$ can then be viewed as a dynamical field, and its field space $[0,\infty)$ has a metric proportional to $G_{rr} \sim 1/r^2$ as follows from a later computation \eqref{eq:radiusmetric}. This field space has an infinite distance point at both of its boundaries: 
\begin{itemize}
\item limit $r\to \infty$: the Kaluza-Klein modes $(n,w)=(n,0)$ become massless,
\item limit $r\to 0$: the winding modes $(n,w)=(0,w)$ become massless,
\end{itemize}
as can be seen from the mass formula \eqref{eq:Tdualmass}. T-duality then tells us that while the physics at these boundaries seems at first sight quite different -- Kaluza-Klein modes or winding modes -- they are actually two faces of a similar set of fundamental degrees of freedom. It is only a matter of perspective based on what theory we started from (Type IIA or IIB) and what regime of field space we consider (large or small compactification radius).

\subsubsection*{M-theory and Type IIB string theory}
We can now combine the previous two dualities -- M-theory to Type IIA and T-duality from Type IIA to IIB -- in order to establish a duality between M-theory and Type IIB. Throughout this chain of dualities we will keep track of the string couplings, because the Type IIB string coupling turns out to have an elegant geometric interpretation. In order to make the match between M-theory and Type IIB precise we compactify M-theory on a rectangular two-torus $T^2 = S^1_{\rm M} \times S^1_{\rm A}$ with radii $r_{\rm M}, r_{\rm A}$. The compactification of M-theory on the circle $S^1_{\rm M}$ yields Type IIA string theory, while subsequently T-dualizing along $S^1_{\rm A}$ brings us to Type IIB string theory. The string couplings are related as\footnote{Recall that the Type IIA string coupling is fixed in terms of the radius of the M-theory circle as $g_s^{\rm IIA} = (r_{\rm M}/\ell_{\rm M})^{3/2}$.} 
\begin{equation}\label{eq:gstotau}
g_s^{\rm IIB} = \sqrt{\frac{r_{\rm B}}{r_{\rm A}} }g_s^{\rm IIA} = \frac{r_{\rm M}}{r_{\rm A}} \, ,
\end{equation}
where we used that $r_{\rm B} = \alpha'/r_{\rm A}$. This allows us to interpret the string coupling of the Type IIB string as the imaginary part of the complex structure $\tau$ of the torus, i.e.~$g_s^{\rm IIB} = 1/ \Im \tau$. 

In order to recover ten-dimensional Type IIB string theory we decompactify by sending $r_{\rm B} \to \infty$, which on the Type IIA side corresponds to $r_{\rm A} = \alpha'/r_{\rm B} \to 0$. We would like to achieve this limit while keeping the string coupling $g_s^{\rm IIB}$ finite, so the radius $r_{\rm M}$ has to be sent to zero as well. From the perspective of the two-torus this means we send its volume -- which is proportional to $r_{\rm M} r_{\rm A}$ -- to zero while keeping the imaginary part of the complex structure parameter $\tau$ -- the ratio $r_{\rm M}/r_{\rm A}$ -- of the torus finite. In other words, there exists a duality between M-theory and Type IIB string theory given by
\begin{equation}\label{eq:MtoIIB}
\fbox{\parbox{8em}{\centering M-theory on $T^2$\\
vol$(T^2) \to 0$}} \qquad \longleftrightarrow \qquad \fbox{\parbox{10em}{\centering Type IIB string theory\\
with $g_s = \frac{1}{\Im \tau}$}}\, 
\end{equation}
While the sketch we gave here only used rectangular two-tori this duality holds more generally: we can use non-rectangular tori -- tori with non-vanishing $\Re \tau$ -- and even consider geometries with torus fibrations on the M-theory side, where $\tau$ is allowed to vary over (part of) the underlying nine-dimensional spacetime. This geometric interpretation of some of the dynamics of Type IIB string theory forms the basis for the framework of F-theory \cite{Vafa:1996xn}, where in particular D7-brane physics is encoded in the torus fibration. We will return to this matter later in light of the S-duality group of Type IIB when we have put all the pieces into place.

\subsubsection*{S-duality for Type IIB} 
We previously studied the strong-coupling limit for Type IIA, upon which we encountered a decompactification to eleven-dimensional M-theory, and we now intend to take the same limit for the Type IIB string. Rather than the emergence of another theory with a new set of fundamental degrees of freedom, Type IIB string theory appears to be dual to itself in the strong-coupling limit. In fact, it is expected that Type IIB string theory possesses an SL(2,$\mathbb{Z}$)-duality group, an element of which maps $g_s \to 1/g_s$ (when $C_0=0$), relating the weak and strong-coupling regimes.

In order to make this S-duality group for the Type IIB string theory become more apparent, let us first reformulate the terms in the IIB supergravity action given by \eqref{eq:NSNS} and \eqref{eq:RRIIB}. By performing a Weil rescaling to the Einstein frame $g^{\rm E}_{\mu \nu} = e^{-\phi/2} g_{\mu \nu}$ we can bring this action into the form
\begin{equation}\label{eq:IIBreformulated}
\begin{aligned}
S_{\rm IIB} = \frac{1}{2\kappa_{10}^2} \int_{\cM_{10}} \bigg( &R_{\rm E} \ast 1 - \frac{1}{2} \frac{\dd \tau \wedge \ast \dd \bar{\tau}}{(\tau_2)^2} - \frac{1}{2} \frac{G_3 \wedge \ast \bar{G}_3}{\tau_2} \\
&- \frac{1}{4} F_5 \wedge \ast F_5 - \frac{i}{4 \tau_2} C_4 \wedge G_3 \wedge \bar{G}_3 \bigg)\, ,
\end{aligned}
\end{equation}
with $R_{\rm E}$ the Einstein frame Ricci scalar, and we redefined our field content as
\begin{equation}\label{eq:IIBredef}
\tau = \tau_1+i \tau_2 = C_0 +ie^{-\phi}\, , \qquad G_3 = F_3 -ie^{-\phi} H_3 = \dd C_3- \tau \dd B_2\, .
\end{equation}
The action of the S-duality group takes a simple form in terms of these redefined fields. For an element of the symmetry group given by
\begin{equation}
\begin{pmatrix}
a & b \\
c & d 
\end{pmatrix} \in \mathrm{SL}(2, \mathbb{R})\, ,
\end{equation}
we find that the duality transformation acts as
\begin{equation}\label{eq:Sdualaction}
\tau \to \frac{a\tau+b}{c\tau+d}\, , \qquad \begin{pmatrix}
C_2\\
B_2
\end{pmatrix} \to \begin{pmatrix}
a & b \\
c & d 
\end{pmatrix}
\begin{pmatrix}
C_2 \\
B_2
\end{pmatrix}\, , 
\end{equation}
while all other fields are invariant. Non-perturbative effects coming from D($-1$)-instantons are expected to break this duality group down to the discrete subgroup SL(2,$\mathbb{Z}$); it is believed this remaining duality group persists as a symmetry of the full Type IIB spectrum, beyond the low-energy supergravity regime.

It is instructive to inspect the action of the S-duality transformations on the axio-dilaton $\tau$ a bit closer. We can consider the generators 
\begin{equation}\label{eq:Sgen}
S = \begin{pmatrix}
0 & 1 \\
-1 & 0 
\end{pmatrix}, \qquad T = \begin{pmatrix}
1 & 1 \\
0 & 1
\end{pmatrix} , \qquad S, T \in \mathrm{SL}(2, \mathbb{Z})\, .
\end{equation}
The duality between weak and strong-coupling regimes is implemented by the generator $S$, which for a background with $C_0=0$ indeed maps $g_s \to 1/g_s$. On the other hand, the axion shift symmetry $C_0 \to C_0 +1$ is generated by $T$. Together the elements $S,T$ can be used to generate any element in the discrete subgroup $\mathrm{SL}(2, \mathbb{Z})$. In fact, these can be used to bring the axio-dilaton $\tau$ from the upper-half plane -- where the string-coupling $g_s = 1/\tau_2$ is well-defined -- to the fundamental domain given in figure \ref{fig:funddomain}, sometimes loosely referred to as the keyhole domain. 

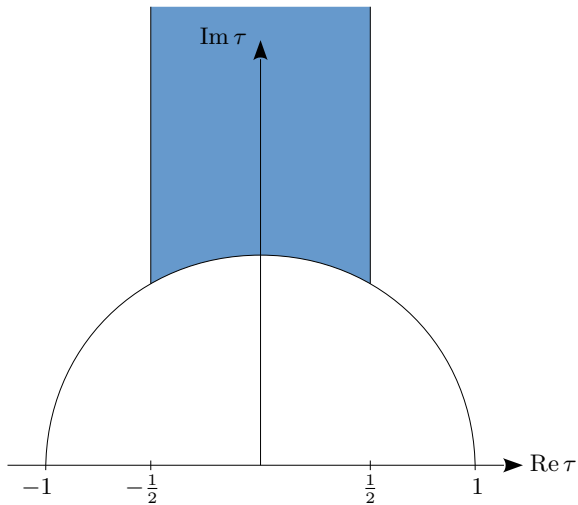
\begin{figure}[h!]
\begin{center}
\includegraphics[width=7.5cm]{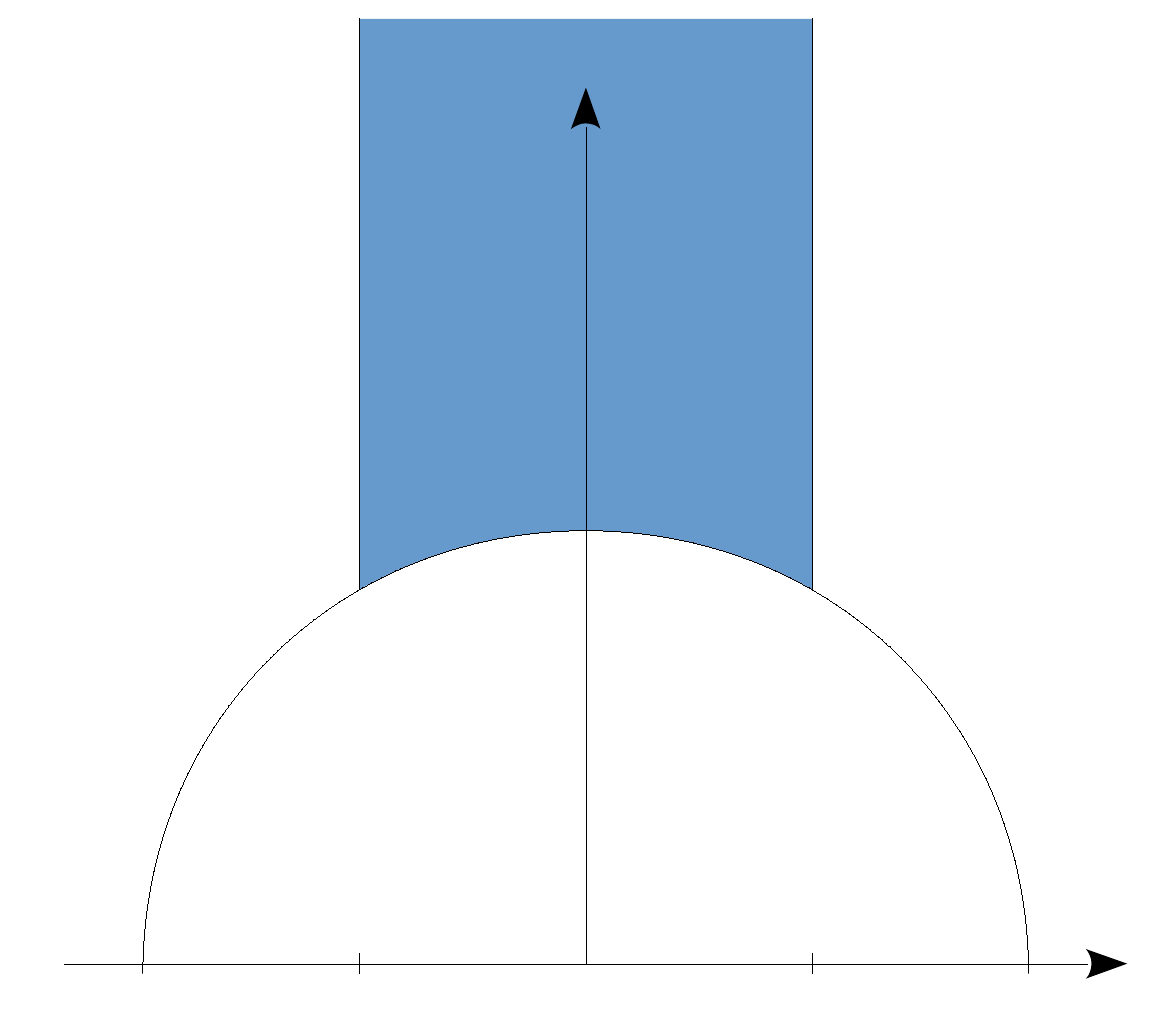}
\end{center}
\begin{picture}(0,0)
\put(280,30){\small $\Re \tau$}
\put(156,190){\small $\Im \tau$}
\put(89,21){\small $-1$}
\put(128,21){\small $-\tfrac{1}{2}$}
\put(258,21){\small $1$}
\put(217,21){\small $\tfrac{1}{2}$}
\end{picture}
\vspace*{-0.8cm}
\caption{\label{fig:funddomain} Depiction of the fundamental domain (colored in blue) of $\mathrm{SL}(2, \mathbb{Z})$.}
\end{figure}

Let us now consider S-duality from a swampland perspective with the distance conjecture \eqref{eq:distance}. The field space metric for the axio-dilaton $\tau$ is read off from its kinetic term in \eqref{eq:IIBreformulated} as $G_{\tau \bar{\tau}} =  1/(\tau_2)^2$. Taking $\tau$ to lie anywhere on the upper-half plane there are two infinite distance limits: strong-coupling $\tau_2 = 0$ and weak-coupling $\tau_2 = \infty$. Depending on the choice of limit either the fundamental F1-string of Type IIB or D1-branes become tensionless
\begin{equation}
T_{\rm F1} =\frac{2\pi}{\ell_s^2} \sqrt{g_s}\, , \qquad T_{\rm D1} = \frac{2\pi}{\ell_s^2} \frac{1}{\sqrt{g_s}}\, .
\end{equation}
In the weak-coupling limit the F1-string becomes tensionless and its excitations give rise the infinite tower of states, while for strong-coupling it is the D1-brane. According to \eqref{eq:Sdualaction} S-duality under the generator $S$ in \eqref{eq:Sgen} switches $B_2$ and $C_2$, so it precisely interchanges the F1-string and the D1-brane. The infinite tower of states arising in the strong-coupling limit can thus indeed be understood as coming from the weakly-coupled Type IIB string in a dual frame.

\paragraph{F-theory.} An insightful perspective on the above discussion is given by the geometrical framework of F-theory \cite{Vafa:1996xn}. The main idea behind this reformulation is to view the axio-dilaton $\tau$ of Type IIB string theory as the complex structure parameter of an auxiliary torus. This interpretation can be made more precise through the duality chain from M-theory to Type IIB \eqref{eq:MtoIIB}, where by compactifying M-theory on a torus of vanishing volume we recovered Type IIB string theory, with its string coupling set by the imaginary part of the complex structure parameter \eqref{eq:gstotau}. In a similar spirit the SL$(2,\mathbb{Z})$ symmetry group of S-duality we just encountered can be understood as the modular group acting on the complex structure parameter of the torus. Furthermore, this correspondence extends to Type IIB backgrounds with a varying axio-dilaton, which from the perspective of F-theory can be viewed as a geometry with a torus fibration over the underlying spacetime. Crucially, this allows us to encode D7-brane physics -- objects which induce a non-trivial axio-dilaton profile -- in terms of the compactification geometry, see e.g.~\cite{Denef:2008wq, Weigand:2018rez} for reviews. From a phenomenological point of view, this allows us to consider compactifications of F-theory on elliptically fibered Calabi-Yau fourfolds rather than Type IIB on threefolds, both of which give rise to four-dimensional effective theories. Phenomenologically the upshot of the F-theory perspective is that by the inclusion of torus fibrations we automatically have D7-branes in our setup, giving us a richer arena for model building.

\paragraph{Asymptotic Hodge theory.} Let us also note that the axio-dilaton field space -- or equivalently the complex structure moduli space of the torus -- we encountered here already gives us a glimpse into asymptotic Hodge theory \cite{Schmid, CKS}, a branch of mathematics that plays a central role in this thesis. This framework captures the structure that underlies limits in the moduli space of K\"ahler manifolds, i.e.~when we move towards points where the geometry degenerates. For our purposes, it can be used in the context of string compactifications on these manifolds, where asymptotic Hodge theory then tells us how physical couplings behave in these regimes. In the case at hand a torus is a simple (complex one-dimensional) example of such a K\"ahler manifold, with $\tau$ parametrizing its complex structure moduli space, i.e.~the keyhole domain in figure \ref{fig:funddomain}. The regime of interest to asymptotic Hodge theory is the weak-coupling region $\Im \tau \gg 1$, where the torus degenerates as pictured in figure \ref{fig:degeneration}. The dilaton-dependent couplings in the action \eqref{eq:IIBreformulated} (the kinetic terms of $\tau$ and $G_3$) can be viewed as determined by this underlying structure. Even the S-duality group SL(2,$\mathbb{Z}$) has a natural counterpart on the side of asymptotic Hodge theory, where the sl(2)-orbit theorem of \cite{CKS} associates an sl$(2,\mathbb{R})$-algebra to every limit. The upshot of asymptotic Hodge theory is that it associates these structures to any limit in any moduli space; while the complex structure moduli space of a torus is rather well-controlled, for more complicated geometries this need not be the case. In other words, we can thus think of asymptotic Hodge theory as a framework that describes asymptotic regimes in any complex structure moduli space by some generalized notion of a weak-coupling limit.

\section{Calabi-Yau manifolds and asymptotic moduli}
Having encountered some examples of circle compactifications, we next want to compactify the ten-dimensional string theories down to four dimensions. Previously we put one or two dimensions on a circle, but now we have to pick a real six-dimensional manifold in order to make contact with our four-dimensional universe. The choice of internal manifold here will have significant influence on the outcome of the effective theory. For instance, the straightforward generalization is to take a product of circles -- a six-torus $T^6 = (S^1)^6$ -- however, this would preserve all supercharges, leading to an effective theory with $\mathcal{N}=4$ or $\mathcal{N}=8$ supersymmetry in four dimensions.\footnote{This is correlated with whether we started from an $\mathcal{N}=1$ or $\mathcal{N}=2$ supersymmetric string theory in ten dimensions, i.e.~the Type I and heterotic strings or the Type II strings.} While as of yet we have not observed any supersymmetry with experiments at all, these extended versions do not even allow for the chiral fermions present in our Standard Model. On the other hand, it would be beneficial to keep at least some supersymmetry around: from a technical point of view it provides us with computational control over certain corrections, and phenomenologically it helps for instance in embedding the Standard Model into a GUT. A generic background would break all supersymmetry, so we have to find the right balance in selecting our internal manifold. This brings us to the class of Calabi-Yau manifolds, which are special in the sense that they lead us to four-dimensional theories with minimal $\mathcal{N}=1$ or $\mathcal{N}=2$ supersymmetry.

\paragraph{Holonomy group.} The amount of supersymmetry preserved by a compactification background can be traced back to the holonomy group of the manifold. This group keeps track of how objects can transform under parallel transport when we move along closed loops in the manifold. A simple example is the two-sphere, whose tangent vectors are rotated by SO(2). For our purposes we look at how supercharges -- which transform as spinors -- precisely behave along these loops: when they undergo a non-trivial rotation, this means the spinor cannot be defined globally, and thus some of the supersymmetry associated to this supercharge would be broken. In the case of a toroidal compactification on $T^6$ the holonomy group is trivial, so all supercharges are preserved. However, in general the holonomy group of a six-dimensional orientable Riemannian manifold is contained in SO(6), and a geometry with exact SO(6) holonomy would break all supersymmetry. The right balance is then provided by manifolds with exact $\text{SU(3)}\subset\text{SO(6)}$ holonomy -- Calabi-Yau threefolds -- which admit precisely one (covariantly constant and globally defined) spinor. This story extends naturally to Calabi-Yau manifolds of complex dimension $D$ with SU$(D)$ holonomy. It was famously proven by Yau that these manifolds admit a Ricci-flat metric.

\paragraph{Hodge decomposition.} One of the aspects of these Calabi-Yau manifolds that we will consider more closely are their cohomology groups. To be precise, recall that for any complex manifold these can be decomposed into so-called Dolbeault cohomology classes of differential forms with mixed holomorphic and anti-holomorphic indices. This splitting can be written out as
\begin{equation}\label{eq:Hodgedecomp}
H^k(Y_D) = \bigoplus_{p+q=k} H^{p,q}(Y_D)\, .
\end{equation}
We refer to the dimensions $h^{p,q} = \text{dim}_{\mathbb{C}} H^{p,q}(Y_D)$ of these subspaces as the Hodge numbers. For any K\"ahler manifold these have the symmetries
\begin{equation}
h^{p,q}=h^{q,p}\,, \quad h^{p,q} = h^{D-q,D-p}\, ,
\end{equation}
which follow from complex conjugation and duality under the Hodge star operator respectively. For Calabi-Yau manifolds we have the additional conditions that $h^{D,0}=1$ and $h^{p,0}=0$ for $0<p<D$: the first condition corresponds to the fact that a Calabi-Yau threefold has an (up to rescaling) unique holomorphic $(D,0)$-form $\Omega$; the second can be argued from the fact that a harmonic $(p,0)$-form transforms as a rank $p$ anti-symmetric tensor, while for exact SU$(D)$ holonomy it must transform as a singlet. Altogether we find that for a Calabi-Yau threefold the Hodge diamond reduces to
\begin{equation}\label{eq:hodgediamond}
\begin{tikzpicture}[baseline={([yshift=-.5ex]current bounding box.center)},scale=0.7,cm={cos(45),sin(45),-sin(45),cos(45),(15,0)}]
\draw (0,0) node {$h^{0,0}$};
\draw (1,0) node {$h^{0,1}$};
\draw (2,0) node {$h^{0,2}$};
\draw (3,0) node {$h^{0,3}$};

\draw (0,1) node {$h^{1,0}$};
\draw (1,1) node {$h^{1,1}$};
\draw (2,1) node {$h^{1,2}$};
\draw (3,1) node {$h^{1,3}$};

\draw (0,2) node {$h^{2,0}$};
\draw (1,2) node {$h^{2,1}$};
\draw (2,2) node {$h^{2,2}$};
\draw (3,2) node {$h^{2,3}$};

\draw (0,3) node {$h^{3,0}$};
\draw (1,3) node {$h^{3,1}$};
\draw (2,3) node {$h^{3,2}$};
\draw (3,3) node {$h^{3,3}$};
\end{tikzpicture} = \begin{tikzpicture}[baseline={([yshift=-.5ex]current bounding box.center)},scale=0.7,cm={cos(45),sin(45),-sin(45),cos(45),(15,0)}]
\draw (0,0) node {$1$};
\draw (1,0) node {$0$};
\draw (2,0) node {$0$};
\draw (3,0) node {$1$};

\draw (0,1) node {$0$};
\draw (1,1) node {$h^{1,1}$};
\draw (2,1) node {$h^{2,1}$};
\draw (3,1) node {$0$};

\draw (0,2) node {0};
\draw (1,2) node {$h^{2,1}$};
\draw (2,2) node {$h^{1,1}$};
\draw (3,2) node {$0$};

\draw (0,3) node {$1$};
\draw (1,3) node {$0$};
\draw (2,3) node {$0$};
\draw (3,3) node {$1$};
\end{tikzpicture}\,  .
\end{equation}

\subsection{Moduli spaces}
The moduli of Calabi-Yau manifolds are geometric deformations which preserve the Calabi-Yau condition. For definiteness we work with Calabi-Yau threefolds here, but these results can straightforwardly be extended to other dimensions. We can study the deformations most easily from the perspective of the metric, where by Yau's theorem one can then impose Ricci flatness on the perturbed metric $g+\delta g$. The moduli can then be divided into two classes: K\"ahler structure deformations corresponding to perturbations $\delta g_{\alpha\bar{\beta}}$, and complex structure deformations of the form $\delta g_{\bar{\alpha}\bar{\beta}}$. Infinitesimally one can expand the perturbations with mixed indices in $h^{1,1}$ harmonic $(1,1)$-forms spanning $H^{1,1}(Y_3)$, while the perturbations with purely anti-holomorphic indices can be written in terms of $h^{2,1}$ harmonic $(2,1)$-forms.\footnote{To be precise, these deformations in terms of (1,1)-forms $\omega_a$ and (2,1)-forms $\chi_i$ are given by
\begin{equation}
\delta g_{\bar{\alpha}\beta} = y^a (\omega_a)_{\bar{\alpha}  \beta}\, , \qquad \delta g_{\bar \alpha \bar \beta} = \frac{1}{2|\Omega|^2} z^i (\chi_{i})_{\bar{\alpha}\gamma \delta} \bar{\Omega}^{\gamma \delta}{}_{\bar \beta}\, ,
\end{equation}
with index ranges $a=1,\ldots , h^{1,1}$ and $i=1,\ldots , h^{2,1}$, and $| \Omega |^2 = \frac{1}{3!} \Omega_{\alpha \beta \gamma}\Omega^{\alpha \beta \gamma}$. } Moreover, it has been shown that these infinitesimal perturbations can be integrated into finite deformations. Hence we get a moduli space that factorizes into an $h^{1,1}$-dimensional and an $h^{2,1}$-dimensional part, which we refer to as the K\"ahler and complex structure moduli spaces $\cM^{\rm ks}(Y_3)$ and $\cM^{\rm cs}(Y_3)$ respectively. 

Both $\cM^{\rm ks}(Y_3)$ and $\cM^{\rm cs}(Y_3)$ can be viewed in their own regard as K\"ahler manifolds. This means these geometries come equipped with a K\"ahler potential from which we can compute the metrics on these moduli spaces by taking moduli derivatives. The K\"ahler potentials can be computed from geometrical data of the Calabi-Yau manifolds: the K\"ahler moduli space is encoded in the K\"ahler two-form $J$ of the Calabi-Yau manifold, while for the complex structure moduli space we take the holomorphic $(3,0)$-form $\Omega$. In the following we give a brief summary on both of these moduli spaces.

\paragraph{K\"ahler moduli space.} The deformations of the K\"ahler form give rise to $h^{1,1}$ real scalar fields $y^a$ by expanding $J = y^a \omega_a$ in a basis of real harmonic $(1,1)$-forms. The metric on this field space is computed from the volume $\cV$ of the Calabi-Yau manifold as
\begin{equation}
K_{a\bar{b}} = -\frac{1}{2} \frac{\partial}{\partial y^a}\frac{\partial}{\partial y^b} \log \cV \, , \qquad \cV =  \int_{Y_3} J \wedge J \wedge J = \cK_{abc} y^a y^b y^c\, ,
\end{equation}
where we defined intersection numbers
\begin{equation}
\cK_{abc} =  \int_{Y_3} \omega_a \wedge \omega_b \wedge \omega_c\, .
\end{equation}
The scalar field space spanned by the volume moduli $y^a$ is a real manifold, but in string compactifications these naturally combine with another set of real scalars into complexified K\"ahler moduli. These scalar fields arise from expanding the Kalb-Ramond field of the ten-dimensional string theories as $B_2 = x^a \omega_a$ in the same harmonic $(1,1)$-form basis, such that the complex coordinates are given by
\begin{equation}
t^a = x^a +i y^a\, .
\end{equation}
The complexified K\"ahler moduli space spanned by the $t^a$ is then a K\"ahler manifold, as becomes apparent by noting that we can define a K\"ahler potential as
\begin{equation}\label{eq:kahlerks}
K^{\rm ks}(t, \bar{t}) = -\log[\cK_{abc} y^a y^b y^c ]\, .
\end{equation}
Notice that the scalar fields $x^a$ coming from the Kalb-Ramond field do not appear here, and hence can be viewed as axions in the four-dimensional effective theory possessing a continuous shift symmetry. This shift symmetry is expected to be broken to a discrete version by worldsheet instanton corrections to the Calabi-Yau volume proportional to $e^{-2\pi y^a}$, arising from strings wrapped on 2-cycles.

\paragraph{Complex structure moduli space.} Deformations in the complex structure of a Calabi-Yau manifold can be detected in the Hodge decomposition \eqref{eq:Hodgedecomp} of the middle cohomology $H^3(Y_3)$, since such variations amount to changing what we call holomorphic and anti-holomorphic. Denoting the complex structure moduli by $z^i$, one finds for instance that under a derivative with respect to $z^i$ the holomorphic $(3,0)$-form $\Omega$ acquires a $(2,1)$-form piece $\chi_i$ as
\begin{equation}\label{eq:chi}
\frac{\partial}{\partial z^i} \Omega = \chi_i -   \frac{\partial K^{\rm cs} }{\partial z^i}  \Omega\, .
\end{equation}
The second term involves the K\"ahler potential for the complex structure moduli space, which is given by
\begin{equation}\label{eq:cskahler}
K^{\rm cs} (z, \bar{z}) = -\log\Big[i \int_{Y_3} \Omega \wedge \bar{\Omega} \Big]\, .
\end{equation}
The K\"ahler metric is then straightforwardly computed to be
\begin{equation}\label{eq:kahlermetriccs}
K_{i\bar{j}} = \frac{\partial}{\partial z^i}\frac{\partial}{\partial \bar{z}^j} K^{\rm cs} = - \frac{1}{\int_{Y_3} \Omega \wedge \bar{\Omega} } \int_{Y_3} \chi_i \wedge \bar{\chi}_j\, .
\end{equation}
Notice that the K\"ahler metric is invariant under so-called K\"ahler transformations $\Omega \to e^f \Omega$, with $f(z)$ a holomorphic function in the complex structure moduli. In this sense \eqref{eq:chi} gives us a K\"ahler covariant derivative $\chi_i = D_i \Omega \equiv \partial_i \Omega + (\partial_i K^{\rm cs} ) \Omega$, transforming as $D_i \Omega \to e^f D_i \Omega$ under this holomorphic rescaling.

\paragraph{Periods.} The dependence of the holomorphic $(3,0)$-form on the complex structure moduli in the above is rather hidden, but we can make this more explicit by expanding $\Omega$ in terms of a harmonic three-form basis $\gamma_\mathcal{I}  \in H^3(Y_3, \mathbb{Z})$. The coefficients in this expansion are then given by so-called period integrals $\Pi^I$, or periods for short, which are integrals of $\Omega$ over dual three-cycles $\Gamma_I \in H_3(Y_3, \mathbb{Z})$. To be more precise, this expansion reads
\begin{equation}\label{eq:omegaexpand}
\Omega = \Pi^\mathcal{I} \gamma_\mathcal{I}  \, , \qquad \Pi^\mathcal{I}  (z)= \int_{\Gamma_\mathcal{I} } \Omega \, .
\end{equation}
For later reference, a practical choice of basis for studying four-dimensional effective theories later is given by $\gamma_\mathcal{I}  = (\alpha_I, \beta^I)$ satisfying symplectic properties
\begin{equation}\label{eq:pairings}
 \int_{Y_3} \alpha_I \wedge \beta^J = \delta_I^{\ J}\, , \qquad \int_{Y_3} \alpha_I \wedge \alpha_J = 0\, , \qquad \int_{Y_3} \beta^I \wedge \beta^J = 0\, .
\end{equation}
The dependence of the periods $\Pi^\mathcal{I}$ on the moduli is, in general, given by complicated functions in the $z^i$ such as hypergeometrics. In practice one can compute these periods for examples by solving so-called Picard-Fuchs equations \cite{Hosono:1993qy,Hosono:1994ax,CoxKatz}. Near boundaries in complex structure moduli space where the Calabi-Yau manifold degenerates this behavior simplifies, and it can be described by means of asymptotic Hodge theory. This latter perspective plays a central role in this thesis, and in fact will be used to set up a method to construct general models for these asymptotic periods in chapter \ref{chap:models} near any one- or two-moduli boundary. For now, let us close off by noting that these periods allow us to write out the K\"ahler potential \eqref{eq:cskahler} as
\begin{equation}\label{eq:kahlercsperiods}
K^{\rm cs}(z, \bar{z}) = -\log\Big[i \, \Pi^I(z)  \eta_{IJ} \bar{\Pi}^J(\bar{z}) \Big]\, ,
\end{equation}
where we wrote $\eta_{IJ} = \int_{Y_3} \gamma_I \wedge \gamma_J$ for the intersection pairing.\footnote{For the choice of basis given in \eqref{eq:pairings} this takes the standard symplectic form $\eta = \left(\begin{array}{cc}
0 & \mathds 1 \\
- \mathds 1 & 0 
\end{array}\right)$.}
 This discussion of periods so far was rather abstract, but in a moment we will give an explicit example by using mirror symmetry.

\paragraph{Mirror symmetry.} Mirror symmetry \cite{Candelas:1990rm} relates two distinct Calabi-Yau manifolds by identifying the K\"ahler moduli space of one with the complex structure moduli space of another and vice versa. In the Hodge diamond \eqref{eq:hodgediamond} this corresponds to a reflection with respect to the diagonal, i.e.~interchanging $h^{2,1}$ and $h^{1,1}$. 
With the above descriptions of both moduli spaces we can make the action of mirror symmetry more explicit. Let us denote the mirror of our Calabi-Yau manifold by $\tilde{Y}_3$, and focus on the relation $\cM^{\rm cs}(Y_3) \simeq \cM^{\rm ks}(\tilde{Y}_3)$. The complex structure moduli space contains a special region known as the \textit{large complex structure regime}, which according to mirror symmetry we can identify with the \textit{large volume regime} in the K\"ahler moduli space. To be precise, we parametrize these points as
\begin{itemize}
\item large complex structure point in $\cM^{cs}(Y_3)$: $z^i=0$ for all $i =1, \ldots, h^{2,1}$,
\item large volume point in $\cM^{ks}(\tilde{Y}_3)$: $y^a=\infty$ for all $a =1, \ldots, \tilde{h}^{1,1}$,
\end{itemize}
where we wrote $\tilde{h}^{1,1} = h^{1,1}(\tilde{Y}_3)$. Using that $\tilde{h}^{1,1} =h^{2,1}$ (so we can relate the index ranges), mirror symmetry then asserts that these two patches are identified as\footnote{In general this mirror map contains an infinite series of higher-order terms in $z^i$ on the right-hand side, and one can then invert these relations order by order to switch between the two descriptions. For the purposes of our presentation here these details are not relevant, and hence we omitted such corrections.} 
\begin{equation}
 e^{2\pi i t^i} = z^i \, .
\end{equation}
Circling the singularity corresponds to sending $z^i \to e^{2\pi i}z^i$, which in the language of the K\"ahler moduli space is understood as shifting the axion by $x^i \to x^i+1$. The periods in this large complex structure regime take the form
\begin{equation}\label{eq:periodsexample}
\Pi = \big(1, \ t^i \ ,\frac{1}{6}\cK_{ijk} t^i t^j t^k +\frac{i \chi \zeta(3)}{8\pi^3}, \ -\frac{1}{2} \cK_{ijk} t^j t^k \big)\, .
\end{equation}
where the term with the Euler characteristic $\chi$ denotes an $\alpha'$-correction on the K\"ahler side we included for later reference. Notice that the K\"ahler potential of $\cM^{\rm cs}(Y_3)$ in \eqref{eq:kahlercsperiods} indeed reproduces the one of $\cM^{\rm ks}(\tilde{Y}_3)$ in \eqref{eq:kahlerks} upon plugging in these periods \eqref{eq:periodsexample} with pairing \eqref{eq:pairings}. The aforementioned worldsheet instanton corrections on the K\"ahler side appear in \eqref{eq:periodsexample} as exponentially suppressed corrections $e^{2\pi i t}$ to the periods. 

\paragraph{Hodge norm.} Finally we want to introduce a norm on the space of harmonic three-forms $H^3(Y_3)$. We can construct such a norm by introducing the Hodge star operator, whose action on the middle cohomology $H^3(Y_3)$ is defined via the spitting \eqref{eq:Hodgedecomp} as
\begin{equation}\label{eq:hodgedef}
\ast \omega_{p,q} = i^{p-q} \omega_{p,q}\, ,
\end{equation}
for an element $\omega_{p,q} \in H^{p,q}(Y_3)$. A natural choice of norm is then provided by integrating a three-form $\omega \in H^3(Y_3)$ and its complex conjugate over the Calabi-Yau threefold with the Hodge star operator as
\begin{equation}\label{eq:hodgenormintro}
\| \omega \|^2 = \int_{Y_3} \omega \wedge \ast\, \bar{\omega}\, .
\end{equation}
In string compactifications this norm computes for instance the scalar potential induced by fluxes as we wrote down before in \eqref{eq:V4d} by setting $\omega=H_3$, but also other physical quantities such as the physical charge of a BPS state \eqref{eq:charge}. It depends implicitly on the complex structure moduli via the Hodge star operator, which varies over $\cM^{\rm cs}(Y_3)$. We can make this dependence more explicit by decomposing in $(p,q)$-form pieces as
\begin{equation}\label{eq:hodgenormformbasis}
\| \omega \|^2 = e^{K^{\rm cs}} \Big(  \int_{Y_3} \Omega \wedge \omega \int_{Y_3} \bar{\Omega} \wedge \bar{\omega} +K^{i \bar{j}} \int_{Y_3} D_i \Omega \wedge \omega   \int_{Y_3} D_{\bar{j}} \bar{\Omega} \wedge \bar{\omega}  \Big)\, ,
\end{equation}
with $K^{i \bar{j}}$ the inverse of the K\"ahler metric \eqref{eq:kahlermetriccs}. The first term corresponds to the $(3,0)$ and $(0,3)$-form part of the three-form $\omega$, and the second its $(2,1)$ and $(1,2)$-form part. Via the expansion \eqref{eq:omegaexpand} this Hodge norm can be expressed purely in terms of the periods $\Pi$ and its derivatives, allowing for an explicit description in the dependence on the moduli.

\subsection{Asymptotic Hodge theory}
As mentioned, couplings appearing in the four-dimensional effective theories, such as the K\"ahler potential \eqref{eq:cskahler} and the Hodge norm \eqref{eq:hodgenormintro}, in general depend in a very complicated way on the moduli through the periods. In asymptotic regimes of the moduli space -- near points where the Calabi-Yau manifold degenerates -- this behavior simplifies, and can be made precise with the techniques of asymptotic Hodge theory. So far we have encountered two such examples in complex structure moduli space: the weak-coupling limit in Type IIB in section \ref{sec:TypeIIstrings} describing how a torus degenerates, and the large complex structure point for Calabi-Yau manifolds. A sketch of this degeneration has been depicted in figure \ref{fig:degeneration}. Asymptotic Hodge theory provides us with a general framework that describes the asymptotic behavior near any such boundary. Applying these techniques to learn lessons about the effective theories arising in string compactifications is one of the main goals of this thesis. Let us also mention that with \cite{cattani1994locus} it has provided strong evidence for the Hodge conjecture.

\begin{figure}[h!]
\begin{center}
\includegraphics[width=8cm]{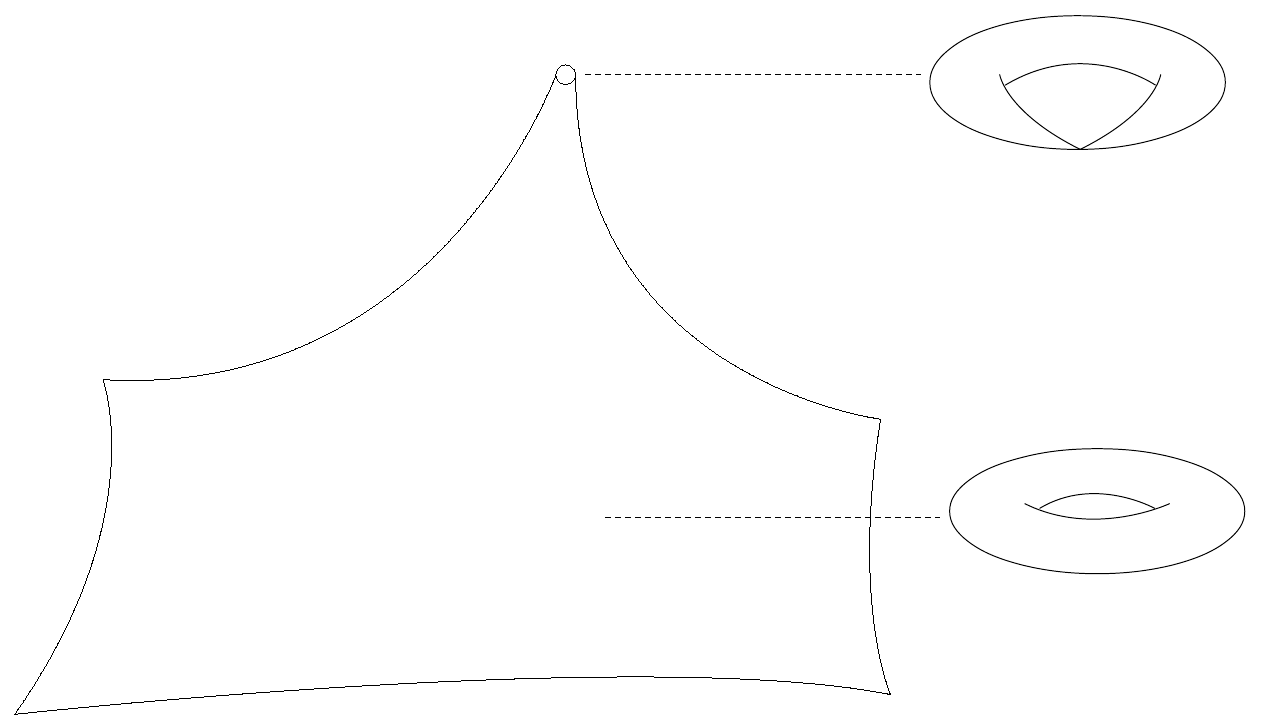}
\end{center}
\caption{\label{fig:degeneration} Example of how a K\"ahler manifold, here depicted as a two-torus, varies over its complex structure moduli space. In the bulk it takes a regular shape, while along infinite tails towards the boundary it degenerates by pinching.}
\end{figure}

\paragraph{Nilpotent orbit approximation.} One of the main pillars of asymptotic Hodge theory is the nilpotent orbit theorem by Schmid \cite{Schmid}. It gives us an asymptotic expansion for the periods of any harmonic $(p,q)$-form in $H^3(Y_3)$ near boundaries in complex structure moduli space. For the moment let us just focus on the holomorphic periods of the $(3,0)$-form $\Omega$. Parametrizing these boundaries as located at $t^i = x^i + i y^i = i \infty$ (analogous to the large complex structure/volume regime before) its period vector can then be written as
\begin{equation}\label{eq:periodsexpandintro}
\Pi(t^i)  = e^{t^i N_i} \big( a_0 + 
\cO(e^{2\pi i t^i})  \big)\, ,
\end{equation}
with $N_i$ nilpotent matrices. This form of the periods generalizes several observations we made previously for the large complex structure point to any boundary. Again we have a shift symmetry $x^i \to x^i +1$ when we circle the boundary, which for the period vector induces a so-called monodromy transformation $\Pi \to e^{N_i} \Pi$. Also there is a clear separation between leading polynomial terms in $t^i$ -- which arise from expanding the exponential in $N_i$ -- and exponentially suppressed corrections. For completeness, let us read off from \eqref{eq:periodsexample} the nilpotent orbit data of the large complex structure regime as
\begin{equation}\label{logN_001}
N_a = \begin{pmatrix}
0 & 0 & 0 & 0 \\
\delta_{ai} & 0 & 0 & 0 \\
0 &  0 & 0 & -\delta_{aj} \\
0 & -\kappa_{aij}  & 0 & 0
\end{pmatrix}\, , \hspace{30pt}
 a_0 = \left(1\, , \ 0 \, , \ 0\, , \ \frac{i \chi \zeta(3)}{8\pi^3} \right)^{\rm T}.
\end{equation}
Let us also mention that approximations similar to \eqref{eq:periodsexpandintro} can be made for elements of the other subspaces $H^{p,q}(Y_3)$. Their asymptotic forms retain the monodromy behavior under axion shifts and separation in polynomial and exponentially suppressed terms in $x,y$, losing only the holomorphicity which was special to the $(3,0)$-form. While we do not write these other (slightly more cumbersome) expansions explicitly here, we note that this leading polynomial behavior of the elements spanning the subspaces $H^{p,q}(Y_3)$ straightforwardly carries over to the Hodge star operator \eqref{eq:hodgedef}. In turn, this yields a similar behavior for the Hodge norm \eqref{eq:hodgenormintro}, and hence physical couplings such as scalar potentials are described to leading order as polynomials in the complex structure fields $x^i,y^i$ near the boundary. We give a more precise description of their behavior under the next approximation.

\paragraph{Sl(2)-orbit approximation.} The nilpotent orbit theorem was followed up by the multi-variable sl(2)-orbit theorem by Cattani, Kaplan and Schmid \cite{CKS}. This approximation implements a hierarchy between the fields $y^1 \gg y^2 \gg \ldots \gg y^n \gg 1$, with $n$ the number of moduli involved in this ordered limit towards the boundary. Each modulus is then associated to an sl$(2,\mathbb{R})$-triple $(N_i^{\pm}, N_i^0)$, where the nilpotent matrix $N_i$ is closely related to the lowering operator. Crucially, the weight operator $N_i^0$ captures the parametrical scaling in the $y^i$ of the Hodge norm \eqref{eq:hodgenormintro} as
\begin{equation}
 N_i^0 \omega_{\Bell} = \ell_i  \omega_{\Bell}\, : \qquad \| \omega_{\Bell} \|^2 \sim (y^1)^{\ell_1} \cdots (y^n)^{\ell_n}\, ,
\end{equation}
with $\Bell=(\ell_1, \ldots , \ell_n)$ the vector containing all weights. This approximation via the nilpotent and sl(2)-orbit brings the initially complicated, typically transcendental behavior of periods stepwise down to a manageable polynomial scaling in the complex structure moduli near the boundary.

\paragraph{Essential exponential corrections.} So far we have been mostly agnostic about the exponential terms in the periods \eqref{eq:periodsexpandintro}, but these corrections can be of essential importance near some boundaries \cite{Bastian:2021eom}. This can be traced back to the fact that one has to be careful with derivatives in limiting procedures. Recall from \eqref{eq:chi} that we obtain the periods of a $(2,1)$-form by taking a moduli space derivative of the $(3,0)$-form. However, one has to be careful with the order of taking limits and derivatives here: if one first plugs in the leading polynomial terms in \eqref{eq:periodsexpandintro} and then takes a derivative, the resulting $(2,1)$-form would for instance vanish if $N_i \mathbf{a}_0=0$ for all $i$, i.e.~to leading order the periods would be constant. This happens near certain boundaries such as a conifold point \cite{PhysRevLett.62.1956,Candelas:1989js}, and in order to resolve this conundrum one has to include essential exponential corrections in the periods when taking derivatives. A concrete situation where one has to be careful is for instance the computation of the K\"ahler metric \eqref{eq:kahlermetriccs}, which from a physical perspective is crucial in order to have well-defined kinetic terms for the scalars (see the discussion below \eqref{eq:example_potential} for an explicit example). We refer for more details to chapter \ref{chap:models}, where we explicitly construct asymptotic models for periods including essential exponential corrections. For now, we want to leave the reader with the cautionary remark that the approximations stated here are only applicable for the physical couplings themselves, and do not necessarily suffice to describe their derivatives.

\paragraph{Large complex structure lamppost.} With the above comments regarding essential exponential corrections in mind, let us return to the setting of the large complex structure point. Here one can always drop worldsheet instanton corrections consistently, meaning there are no essential exponential terms which have to be taken into account. Moreover, in this regime one has the advantage that it is relatively easy to describe the periods: one can bypass the computation of the periods with Picard-Fuchs methods by simply plugging in the topological data of the mirror Calabi-Yau manifold into \eqref{eq:periodsexample}. In fact, there exist large databases \cite{Candelas:1987kf, Kreuzer:2000xy} of Calabi-Yau threefolds from which this data can be computed using e.g.~toric geometry methods \cite{Greene:1996cy,Bouchard:2007ik}. In the physics literature this has led to the situation where much of the focus in studying string compactifications is set on these large complex structure regime. However, lessons learned from these effective theories do not necessarily carry over from this lamppost to other boundaries in moduli space, with essential exponential corrections that cannot be dropped as one of the prime examples. A complete overview of the string landscape thus requires us to shed light on the other corners of the complex structure moduli space, and asymptotic Hodge theory does so without any preference for particular regimes.

\section{String theory compactifications}\label{sec:CY}
In this section we lay out the details that go into the study of string compactifications. We already briefly touched upon this subject and its importance to string phenomenology in section \ref{sec:stringintro}, and now we take the time to explain how this procedure works in practice. As a warm-up we work out the simple case of a circle compactification first, setting the stage for more general string compactifications that make contact with the realm of four-dimensional effective theories. 

\subsection{Compactification on a circle}\label{sec:circle}
The study of circle compactifications goes back to the early days after the conception of general relativity. Kaluza in 1921 and later Klein in 1926 studied how general relativity in five spacetime dimensions could give rise to both Einstein gravity and electrodynamics in four dimensions by compactifying on a circle. While this attempt itself was not very successful, the main principles at play still go to the heart of the string phenomenology program today.

We first consider a $(d+1)$-dimensional theory describing only Einstein gravity. The action is then made up of just the Ricci scalar $R_{d+1}$, and can be written as
\begin{equation}\label{eq:d+1}
S_{{\rm EH}, d+1} = \frac{1}{2\kappa_{d+1}^2} \int_{\cM_{d+1}} R_{d+1} \ast 1\, .
\end{equation}
In turn, we want to compactify one dimension on a circle, meaning the spacetime topologically takes the form $\cM_{d+1} \simeq \cM_d \times S^1$. This allows us to decompose the higher-dimensional metric as
\begin{equation}\label{eq:d+1metric}
\dd s^2 = g_{mn} \dd x^m \dd x^n = g_{\mu \nu} \dd x^\mu \dd x^\nu + r^2 \dd y^2 \, , \qquad y \sim y+2\pi \, .
\end{equation}
where the indices $m=0,\ldots, d$ label the coordinates in $d+1$ dimensions, which we split as $x^m = (x^\mu, y)$ with $\mu=0,\ldots, d-1$. For simplicity we have set any mixed components $g_{\mu d},g_{d\nu}$ to zero, although in principle one could include these terms.\footnote{In fact, we can view $g_{\mu d}$ as a massless vector living in $d$ dimensions -- a U(1) gauge field -- and it was this insight that led Kaluza and Klein to use Einstein gravity in one dimension more as a means of unifying gravity and electrodynamics in the lower-dimensional setting. } We are mainly interested in $r$, which can be promoted to a scalar field in $d$ dimensions by allowing it to fluctuate over $\cM_d$. This parameter gives the radius of the compactification circle as measured with the $(d+1)$-dimensional metric
\begin{equation}
2\pi r = \int_0^{2\pi} \sqrt{g_{dd}} \, \dd y\, .
\end{equation}
Next we reduce the $(d+1)$-dimensional action \eqref{eq:d+1} to $d$ dimensions by plugging in the ansatz for the metric \eqref{eq:d+1metric} and integrating over the circle, resulting in
\begin{equation}
S_{{\rm EH},d} = \frac{1}{2\kappa_d^2} \int_{\cM_d} r \, R_d \ast 1 \, ,
\end{equation}
where we relate the gravitational couplings as $\kappa_d^2 = \kappa_{d+1}^2/(2\pi)$. Notice that this Einstein-Hilbert action is not written in the canonical form, but a dilaton-like coupling to the radius $r$ appears. We can get rid of this coupling by performing a Weyl rescaling of the metric as
\begin{equation}\label{eq:weylcircle}
g_{\mu \nu} \to r^{-\frac{2}{d-2}} g_{\mu \nu}\, ,
\end{equation} 
under which the action rewrites to
\begin{equation}
S_{{\rm EH},d} = \frac{1}{2\kappa_d^2} \int_{\cM_d} \left( R_d \ast 1 - \frac{d-1}{d-2} \dd \log r \wedge \ast \, \dd \log r \right)\, .
\end{equation}
The scalar field $r$ in this effective theory takes values in the interval $[0,\infty)$. We can read off the metric on this one-dimensional field space as the coefficient of the kinetic term $-\frac{1}{2} \dd r \wedge \ast \dd r$, meaning we find
\begin{equation}\label{eq:radiusmetric}
G_{rr} = \frac{2(d-1)}{d-2} \frac{1}{r^2}\, .
\end{equation} 
For this metric we find two infinite distance points on $[0, \infty )$: $r \to 0$ and $r\to \infty$. To gain some intuition for the dynamics that can appear in these two limits, it is instructive to add a massless scalar field $\Phi$ in the $(d+1)$-dimensional theory. Since it must be periodic on the circle, i.e.~$\Phi(x^d+2\pi)=\Phi(x^d)$, we can expand this scalar field as
\begin{equation}
\Phi(x^\mu, x^d) = \sum_{n=-\infty}^\infty \phi_n(x^\mu) e^{2\pi i n x^d}\, .
\end{equation} 
The modes $\phi_n$ are then typically referred to as the Kaluza-Klein modes, with $\phi_0$ being the zero-mode. The equation of motion for this massless scalar field $\partial^m \partial_m \Phi =0$ can then be written out individually for these modes
\begin{equation}
\partial^\mu \partial_\mu \phi_n =\frac{n^2}{r^2} \phi_n\, ,
\end{equation}
where for ease of computation we worked with the metric \eqref{eq:d+1metric} instead of the Weyl rescaled version \eqref{eq:weylcircle}. From these equations of motion we recognize the masses of the Kaluza-Klein modes as
\begin{equation}
M_n^2 = \frac{n^2}{r^2}\, .
\end{equation}
These Kaluza-Klein masses teach us multiple lessons about compactifications. First of all, the scaling in the radius is proportional to $M_n \sim 1/r$, so detection of the Kaluza-Klein tower -- a signal of an extra dimension -- would require energies at least of order $E \gtrsim 1/r$.
Secondly, in the large radius limit we see that this infinite tower becomes massless, confirming the distance conjecture expectations. This leaves us, as a final lesson, to wonder about the small radius limit and what kind of physics dominates in this regime. The answer is to consider a quantum theory of gravity that contains extended objects -- strings -- winding around the circle, thereby give rise to light states when the circle becomes small. This is precisely what happens in string theory as we saw in \eqref{eq:Tdualmass}: winding modes become light at small $r$, and are T-dual to KK modes of another string theory at large radius $\alpha'/r$.

\subsection{Type IIB compactifications on Calabi-Yau threefolds}\label{ssec:IIBN=2}
By compactifying Type IIB supergravity on Calabi-Yau threefolds we obtain four-dimensional $\mathcal{N}=2$ supergravities. Here we briefly summarize the relevant aspects of these theories, and refer to \cite{stromingerspecialkahler,Candelas:1990pi,Andrianopoli:1996cm, Craps:1997gp,Freed:1997dp,freedman_van_proeyen_2012} for more extensive reviews.

For this thesis it will suffice to focus on the gravity and vector multiplet sectors of these supergravity theories. The scalar field space is then spanned by the $h^{2,1}$ complex structure deformations $t^i$ of the Calabi-Yau threefold, while the $h^{2,1}+1$ vectors arise from the ten-dimensional R-R four-form potential. To be precise, by expanding $C_4$ in a symplectic three-form basis $\alpha_I, \beta^I \in H^3(Y_3, \mathbb{Z})$ we obtain
\begin{equation}\label{eq:C4expand}
C_4 = V^I \alpha_I - U_I \beta^I\, , 
\end{equation}
with intersections given by \eqref{eq:pairings}. The self-duality condition \eqref{eq:selfdualF5} eliminates half the degrees of freedom in $C_4$, allowing us to choose to use the one-forms $V^I$ over the $U_I$. Denoting the corresponding field strengths by $F^I = \dd V^I$ one obtains the effective $\mathcal{N}=2$ supergravity action
\begin{equation}\label{eq:action}
S^{(4)} \! =  \! \int_{\cM^4 }\! \bigg( \tfrac{1}{2} R \star  1- K_{i\bar \jmath}\, \dd t^i \wedge \star \, \dd \bar{t}^j + \tfrac{1}{4} \cI_{IJ} F^I \wedge \star \, F^{J} +  \tfrac{1}{4} \cR_{IJ} F^{I} \wedge \, F^{J} \bigg)\, , 
\end{equation}
where $\star$ denotes the 4d Hodge star. It is well known that these supergravity theories enjoy electro-magnetic duality for which Sp$(2(h_{2,1}+1),\mathbb{Z})$ is the relevant symmetry group. The coupling functions in this action are the K\"ahler metric $K_{i\bar \jmath}$ and the gauge kinetic functions $\cI_{IJ}, \cR_{IJ}$, all depending on the scalars $t^i,\bar t^i$. We first describe these couplings with the $\mathcal{N}=2$ supergravity formalism of special geometry, and only thereafter establish the geometrical connection with the underlying Calabi-Yau threefold. The K\"ahler potential that computes $K_{i\bar \jmath}$ is given by
\begin{equation}\label{eq:kahlersugra}
K = -\log\left[ i \bar{X}^I \cF_I - i X^I \bar{\cF}_I \right]\, .
\end{equation}
Here the $X^I, \cF_I$ are holomorphic functions in the complex scalar fields $t^i$. As explained in \cite{Craps:1997gp} one can always choose a symplectic frame such that the $\cF_I$ are obtained from a holomorphic prepotential $\cF(X^I)$ that is homogeneous of degree two. To be precise, one then finds that
\begin{equation}
\cF = \tfrac{1}{2} X^I \cF_I\, , \qquad \cF_I = \partial_I \cF \, , \qquad \cF_{IJ} = \partial_I \cF_J\, ,  \qquad \cF_{I} =  \cF_{IJ} X^J\, .
\end{equation}
The $\cI_{IJ}, \cR_{IJ}$ are straightforwardly computed from the prepotential as
\begin{equation}
\label{pm}
\mathcal N = \mathcal R + i\op \mathcal I\, ,  \qquad {\cal N}_{IJ}=\overline{\mathcal {F}}_{IJ}+2i \, \frac{
{\rm Im}(\mathcal F_{IM}) X^M \, {\rm Im}(\mathcal F_{JN}) X^N}{
           X^P \,{\rm Im}(\mathcal F_{PQ}) X^Q}  \,.
\end{equation}
Now let us make precise how these couplings are described geometrically on the side of the Calabi-Yau manifold. The holomorphic functions $X^I, \cF_I$ are the periods of the $(3,0)$-form, as can be seen by assembling them into a period vector as
\begin{equation}\label{eq:PeriodVector}
\Pi = \begin{pmatrix} X^I\\
-\cF_I
\end{pmatrix}\, , \qquad   \eta =  \left( \begin{array}{cc}
  0   & +\mathds 1
  \\
  - \mathds 1 & 0
  \end{array}\right) \, .
\end{equation}
By plugging these periods into \eqref{eq:kahlercsperiods} we indeed recover the K\"ahler potential written in \eqref{eq:kahlersugra}. On the other hand, recovering the gauge kinetic functions requires a bit more work. We can first express the Hodge star operator of the Calabi-Yau manifold in terms of the $\cR_{IJ}, \cI_{IJ}$ as
\begin{equation}\label{hodgestar01}
  C= \left( \begin{array}{cc}
  \mathcal R\op \mathcal I^{-1} &  -\mathcal I - \mathcal R\op \mathcal I^{-1} \mathcal R
  \\
  \mathcal I^{-1} & - \mathcal I^{-1} \mathcal R
  \end{array}\right) \,, 
\end{equation}
with $C$ the matrix representative of the Hodge star $\ast$ acting on the basis $(\alpha_I , \beta^J)$. In a similar fashion we can write down a matrix expression for the Hodge star norm \eqref{eq:hodgenormintro} as
\begin{equation}\label{def-cM}
  \mathcal M = \eta \, C =   
  \left( \begin{array}{cc}
   -\mathcal I - \mathcal R \mathcal I^{-1} \mathcal R
   &
  - \mathcal R \mathcal I^{-1} 
  \\
  - \mathcal I^{-1} \mathcal R & - \mathcal I^{-1} 
  \end{array}\right) .
\end{equation}
In turn, we can now read off the gauge kinetic functions explicitly as integrals
\beq \label{eq:normtogkfunctions}
\int_{Y_3} \alpha_I \wedge \ast \alpha_J   = - \cM_{IJ}\ , \quad \int_{Y_3} \beta^I \wedge \ast \beta^J  = - \cM^{IJ}\ , \quad 
\int_{Y_3} \alpha_I \wedge \ast \beta^J = - \cM_{I}^{\ J}\ . 
\eeq
which correspond respectively to the top left, bottom right and top right sub-blocks of $\cM$ in \eqref{def-cM}, as can be seen from the index positioning.

Finally, let us point out that the above gauge kinetic functions can also be computed outside of the prepotential frame. Assuming the periods are kept in the symplectic basis $\Pi=(X^I, -\cF_I)$, one then finds
\begin{equation}\label{eq:cNalt}
\cN_{IJ} = \left(\cF_I \, , \ D_{\bar{\imath}} \bar{\cF}_I \right) \left(X^J \, , \ D_{\bar{\imath}} \bar{X}^J \right)^{-1}
\end{equation}
For our work it is important to stress the presence of this identity. For instance, the asymptotic periods we derive in chapter \ref{chap:models} are not formulated in a prepotential frame, and in that case it is much easier to compute the coupling functions via \eqref{eq:cNalt} rather than deriving a prepotential first and subsequently using \eqref{pm}.

\paragraph{BPS states.} We will in particular be interested in extremal black hole solutions in these $\mathcal{N}=2$ supergravities that are BPS, see \cite{DallAgata:2011zkh} for a review. The electric and magnetic charges of these black holes are computed as
\begin{equation}
P^I = \int_{S_\infty} F^I  \, , \qquad Q^I = \frac{1}{4\pi} \int_{S_\infty} \star F^I \, .
\end{equation}
where $S_\infty$ denotes a two-sphere at spatial infinity. Geometrically these BPS states arise from D3-branes wrapping special Lagrangian three-cycles of the Calabi-Yau manifold. Quantized charges $q_I,p^I$ then specify a Poincar\'e dual three-form as $q= q^I \alpha_I - p_I \beta^I$, which are related to the black hole charges as
\begin{equation}
P^I = p^I\, , \qquad Q^I = - \big( \cI^{-1} \cdot q\big)^I +\big( \cI^{-1}\cdot  \cR\cdot  p\big)^I\, .
\end{equation}
The mass of this BPS state follows by definition from its central charge $M=|Z|$, which is computed as
\begin{equation}
\label{eq:centralcharge}
Z(q)= e^{K/2} q \,\eta \, \Pi \,   ,
\end{equation}
The physical charge of a BPS state is given by (see for instance \cite{Palti:2017elp})
\begin{equation}\label{eq:charge}
Q^2 = -\frac{1}{2}\, \mathbf{q}^T  \cM \mathbf{q} = \frac{1}{2} \|q \|^2\, ,
\end{equation}
with the charge matrix given by \eqref{def-cM}. Note that this is precisely the Hodge norm \eqref{eq:hodgenormintro} of the quantized charge $q$ of the BPS state, which can be controlled near boundaries of the moduli space by using asymptotic Hodge theory. There is a useful  identity in $\cN=2$ supergravity theories that relates this physical charge to the central charge via \cite{Ceresole_1996}\footnote{Let us note that there is a clever rewriting of \eqref{eq:N=2identity} by using the fact that $D_i(Z \bar{Z})=\partial_i (Z \bar{Z})$, i.e.~the squared norm of the central charge has zero K\"ahler weight. One obtains the form
\begin{align}\label{eq:N=2identityrewritten}
Q^2=|Z|^2 + 4  K^{i \bar{\jmath}} \partial_i |Z| \partial_{\bar{\jmath}} |Z|\, ,
\end{align}
which will be useful later in chapter \ref{chap:WGC}.}
\begin{equation}
\label{eq:N=2identity}
Q^2 = |Z|^2 + K^{i \bar{\jmath}} D_i Z D_{\bar{\jmath}} \bar{Z}\, ,
\end{equation}
where the K\"ahler covariant derivative acts on $Z$ as $D_i Z=\partial_i Z+ \frac{1}{2} (\partial_i K) Z$. Geometrically this is again just the splitting into (3,0) and (2,1)-form pieces as described by \eqref{eq:hodgenormformbasis}.

\subsection{Type IIB Calabi-Yau orientifold compactifications}\label{ssec:IIBN=1}
As a next step in the direction towards constructing phenomenologically compelling string vacua we consider Calabi-Yau orientifold compactifications with fluxes, see \cite{Grana:2005jc,Douglas:2006es,Giddings:2001yu,Grimm:2004uq} for some original literature and reviews. The orientifold projection brings down the supersymmetry to $\mathcal{N}=1$ supergravity, while turning on three-form background fluxes for $H_3$ and $F_3$ generates a potential for the axio-dilaton and the complex structure moduli. This discussion allows us to set the stage for detailed studies of moduli stabilization in chapter \ref{chap:modstab}. Let us note that we will again mostly ignore K\"ahler moduli in the following, although there are suitable approaches to induce a scalar potential for these as well.


\subsubsection*{Orientifold projection}
An orientifold of a Type II string theory is defined as a projection imposed on the string spectrum which involves the worldsheet parity operator $\Omega_p$. For a closed string $\Omega_p$ exchanges the left- and right-moving sectors, and for open strings it swaps the endpoints. There is a choice of parity operator on the string worldsheet here, which we fix by including a factor $(-1)^{F_L}$, with $F_L$ the left-moving fermion number. For the massless modes in the Type IIB spectrum one then finds that the fields split under $(-1)^{F_L} \Omega_p$ as
\begin{equation}
\text{even: } \ g_{\mu\nu}\, , \ \phi\, , \ C_0 \, , \ C_4 \, , \qquad \text{odd: } \ B_2\, , \ C_2\, .
\end{equation}
Subsequently we combine this worldsheet parity operator with a holomorphic and isometric involution $\sigma$ of the Calabi-Yau manifold, i.e.~$\sigma^2=1$. This involution then acts on the K\"ahler form and on the holomorphic three-form of $Y_3$ as
\begin{equation} 
\sigma^* J= + J\, , \qquad \sigma^*\Omega = -\Omega\, .
\end{equation}
Furthermore, the cohomology groups of $Y_3$ split into even and odd eigenspaces as 
\begin{equation}
H^{p,q}(Y_3) = H^{p,q}_+(Y_3) \oplus H^{p,q}_-(Y_3) \, .
\end{equation}
For the complex structure moduli this means $h^{2,1}_+$ fields are projected out, keeping only $h^{2,1}_-$ complex structure moduli. A scalar potential for these fields is generated by turning on three-form fluxes $H_3$ and $F_3$ along odd 3-cycles of the Calabi-Yau threefold, since both $B_2$ and $C_2$ are odd under $(-1)^{F_L} \Omega_p$. We expand these fluxes as
\eq{
\label{tad_02}
  H_3 = h^I \alpha_I - h_I \beta^I \,, \hspace{50pt}
  F_3 = f^I \alpha_I - f_I \beta^I \,, 
}
where $(\alpha_I, \beta^I) \in H^3_-(Y_3)$ denotes an odd symplectic basis with pairings given by \eqref{eq:pairings}. For the scope of this thesis we will not be concerned with these details of the orientifold projection and simply assume that $h^{2,1}_+=0$, meaning we set $h^{2,1}_-=h^{2,1}$ in the following. 

Fixed loci of the involution $\sigma$ in the Calabi-Yau manifold correspond to spacetime filling orientifold-planes, or O-planes for short. For our choice of orientifold projection these are O3- and O7-planes, which carry negative electric charge under $C_4$ and negative magnetic charge under $C_0$ respectively -- opposite to their counterpart D3-branes and D7-branes. When integrated over compact submanifolds the resulting total charges must cancel, leading to so-called tadpole cancelation conditions. For the O7-planes these are most easily canceled by stacking eight D7-branes on top of it. For the D3-brane tadpole the cancelation condition also involves a (positive) contribution coming from the fluxes and reads
\begin{equation}\label{tad_01}
N_{\rm flux} = \int_{Y_3} F_3 \wedge H_3 =  h_I f^I - h^If_I \, , \qquad    N_{\rm D3} - \frac{1}{4} N_{\rm O3} +\frac{1}{2} N_{\rm flux} = 0\, ,
\end{equation}
where for convenience we omitted induced charges from D7-branes and O7-planes.



\subsubsection*{Four-dimensional $\mathcal{N}=1$ supergravities}
By imposing the orientifold projection we obtain four-dimensional $\mathcal{N}=1$ supergravities. We will mainly be interested in the scalar field content of these theories, for which the K\"ahler potential reads
\eq{
  \label{kpot}
  K =  - \log\bigl[ -i(\tau-\bar \tau) \bigr] - \log \left[ +i\int_{Y_3} \Omega \wedge \bar \Omega \right]
  - 2\log \mathcal V\,,
}
where we recall that $\Omega$ depends on the complex structure moduli and the volume $\cV$ on the K\"ahler moduli. For ease in notation we will write out the axio-dilaton as $\tau = c+is$ from now onwards in this setting. 

The NS-NS and R-R three-form fluxes $H_3$ and $F_3$ induce a superpotential in this four-dimensional $\mathcal{N}=1$ effective theory, which can be written out in terms of the holomorphic $(3,0)$-form $\Omega$ as \cite{Gukov:1999ya}
\eq{
\label{eq:superpotential}
 W = \int_{Y_3} \Omega \wedge G_3  \,.
}
where we recall the complex combination $ G_3 = F_3 - \tau H_3 $. By means of standard $\mathcal{N}=1$ identities one can then write the scalar potential in terms of the superpotential. Using the no-scale property of the K\"ahler moduli we can write it out as
\begin{equation}\label{eq:potential}
V = e^{K}  K^{I \bar{J}} D_I W D_{\bar J} \bar{W} = \frac{1}{4 \cV^2 s} \big(\langle \bar{G}_3 , \ast G_3 \rangle - i \langle \bar{G}_3, G_3 \rangle \big)\, ,
\end{equation}
where $I$ runs over the complex structure moduli and axio-dilaton, with $K^{I\bar{J}}$ denotes the inverse of the K\"ahler metric and $D_I W = \partial_I W + K_I W$. For the second equal sign we recognized that the scalar potential corresponds to the (2,1)- and (1,2)-form part of the three-form flux, and hence can be rewritten in terms of the Hodge star norm as described by \eqref{eq:hodgenormformbasis}. 

One can now study the minima of this flux potential in two equivalent ways. The standard $\cN=1$ supergravity approach is to solve for vanishing F-terms, yielding as constraints
\begin{equation}\label{eq:Fterms}
D_I W = \partial_I W + K_I W =  0\, .
\end{equation}
Alternatively one considers the imaginary self-duality condition for the three-form flux $G_3$, which reads
\begin{equation}\label{eq:selfduality}
\ast G_3 = i G_3\, .
\end{equation}
From both approaches one sees that the scalar potential \eqref{eq:potential} vanishes at the minimum, giving rise to a Minkowski-type vacuum. A vanishing superpotential $W=0$ at the minimum corresponds to supersymmetric vacuum (including the K\"ahler moduli sector), while the second possibility is important for the KKLT and Large Volume Scenarios \cite{Kachru:2003aw,Balasubramanian:2005zx}. In particular, for KKLT it is needed that the vacuum superpotential takes an exponentially small value (see \cite{Demirtas:2019sip,Demirtas:2020ffz,Blumenhagen:2020ire,Honma:2021klo,Demirtas:2021nlu, Demirtas:2021ote,Broeckel:2021uty,Bastian:2021hpc,Carta:2021kpk,Carta:2022oex} for some recent constructions of such flux vacua), which we investigate further in chapter \ref{chap:modstab}.





\subsection{F-theory compactifications on Calabi-Yau fourfolds}\label{ssec:Ftheory}
As our final setup we consider Calabi-Yau fourfold compactifications of F-theory, and in particular the flux potential arising in the four-dimensional $\mathcal{N}=1$ supergravity theory. This can conveniently be studied by compactifying M-theory on the same Calabi-Yau fourfold to a three-dimensional $\mathcal{N}=1$ supergravity, where turning on $G_4$ flux induces a scalar potential for the complex-structure and K\"ahler structure moduli \cite{Haack:2001jz}. One can then lift this setting to a four-dimensional $\mathcal{N}=1$ supergravity theory by requiring $Y_4$ to be elliptically fibered and shrinking the volume of the torus fiber \cite{Denef:2008wq, Weigand:2018rez, Grimm:2010ks}. The scalar potential obtained in this way reads
\begin{equation}\label{eq:potentialFtheory}
V = \frac{1}{\cV_4^3} \Big( \int_{Y_4} G_4 \wedge \ast G_4 - \int_{Y_4} G_4 \wedge G_4 \Big)\, , 
\end{equation}
where $\cV_4$ denotes the volume of the Calabi-Yau fourfold $Y_4$ and $\ast$ is the corresponding Hodge-star operator. The scalar potential \eqref{eq:potentialFtheory} depends both on the complex-structure and K\"ahler moduli through the $\ast$ in the first term, and the overall volume factor $\cV_4$ gives an additional dependence on the K\"ahler moduli. We also note that the flux $G_4$ is constrained by the tadpole cancellation condition as \cite{Sethi:1996es}
\begin{equation}\label{eq:fourfoldtadpole}
\frac{1}{2} \int_{Y_4} G_4 \wedge G_4 +N_{\rm D3}= \frac{\chi(Y_4)}{24}\, .
\end{equation}
In this formulation the NS-NS and R-R fluxes $H_3, F_3$, as well as D7-brane fluxes in Type IIB are combined into a four-form field strength $G_4$. In particular, it generalizes the D3-brane tadpole condition written in \eqref{tad_01} to the F-theory setting. In \cite{Bena:2020xrh} it was conjectured that stabilization of a large number $h^{3,1}$ of complex structure moduli requires the flux contribution to grow as $h^{3,1}/3$. This tadpole conjecture thereby places vacua with all these complex structure moduli stabilized in the swampland, since the Euler characteristic scales as $\chi(Y_4) \sim 6 h^{3,1}$.

Let us now describe the dependence on the complex structure moduli more explicitly, where we (mostly) ignore the K\"ahler moduli in the following. This requires us to assume that $G_4$ is an element of the primitive cohomology $H^4_{\rm p}(Y, \mathbb{Z})$ \cite{Haack:2001jz}, which can be expressed as the condition $J \wedge G_4 = 0$ with $J$ being the K\"ahler two-form of $Y_4$. The K\"ahler potential $K=-3\log \cV_4 + K^{\rm cs}$ and superpotential giving rise to the scalar potential \eqref{eq:potentialFtheory} can then be written as \cite{Gukov:1999ya, Haack:2001jz}
\begin{equation}\label{fourfold_potentials}
K^{\rm cs} = - \log \int_{Y_4} \Omega \wedge \bar{\Omega}\, , \hspace{50pt} W  =  \int_{Y_4} \Omega \wedge G_4\, .
\end{equation}
Minima of the scalar potential \eqref{eq:potentialFtheory} are  found by either solving for vanishing F-terms or imposing a self-duality condition on $G_4$ \cite{Haack:2001jz}, and these constraints read 
\begin{equation}\label{fourfold_extremization}
D_I W = \partial_I W + K_I W = 0 \, ,  \hspace{50pt} \ast G_4 = G_4\, ,
\end{equation}
with $z^I$ the complex structure moduli of $Y_4$. One easily checks that this leads to a vanishing potential \eqref{eq:potentialFtheory} at the minimum, giving rise to a Minkowski vacuum. Remarkably, it has been proven recently in \cite{Grimm:2020cda, Bakker:2021uqw, Grimm:2021vpn} that the total number of vacua satisfying the tadpole bound \eqref{eq:fourfoldtadpole} in any complex structure moduli space is finite.

\RedeclareSectionCommand[ 
beforeskip=12ex,
            ]{part}
\setpartpreamble[u][\textwidth]{
\vspace*{1cm}
\hrulefill 
\vspace*{0.5cm}

The first part of this thesis is devoted to introducing the mathematical background underlying the bulk of our work: asymptotic Hodge theory. Our purpose is to familiarize the reader with the main ideas underlying this abstract framework, and explain how these techniques can be used in the study of string compactifications. This discussion is illustrated with examples that helped the author understand various parts of the story
. To give a more precise outline, chapter \ref{sec_hodge_gen} provides a quick sketch of the main principles, intended to give some guidance through later chapters. 
Chapter \ref{chap:asympHodge} establishes Hodge theory itself, describes limits in moduli space and how physical couplings behave in these asymptotic regimes. Chapter \ref{chap:mhs} covers the formal structure underlying these limits, introducing key concepts such as Deligne splittings. Finally, chapter \ref{sec:sl2splitting} reviews one of the most powerful results of asymptotic Hodge theory, the multi-variable sl(2)-orbit theorem \cite{CKS}, and explains both how this approximation works and how to construct it in examples.

\vspace*{0.5cm}
\hrulefill }
 
\part{Toolbox for Asymptotic Hodge Theory}\label{part1}
\chapter{Lightning review of asymptotic Hodge theory}
\label{sec_hodge_gen}
In this chapter we introduce the main ideas underlying asymptotic Hodge theory. In particular, we describe the behavior of the period vector \eqref{eq:omegaexpand} and the Hodge star \eqref{eq:hodgenormintro} near boundaries in complex structure moduli space (see also \eqref{eq:PeriodVector} and \eqref{def-cM} for their supergravity formulations). For simplicity, we send only one modulus to the boundary here, and cover the general case in the subsequent chapters.



\subsubsection*{Boundaries and monodromy symmetries}

In complex structure moduli space one can naturally associate to 
each boundary a discrete symmetry, known as a monodromy symmetry. 
Sending only a single modulus to the boundary we can  choose 
local coordinates such that $ z=0$ corresponds to the boundary locus, but
a more useful parametrization is given by
\eq{\label{coordinates}
   t =  x + i\op  y = \frac{1}{2\pi \op i} \log  z \,,
}
where the boundary corresponds to the limit $ t\to i\op \infty$.  
The monodromy symmetry is realized by 
encircling the boundary 
as $z\to e^{2\pi \op i} z$, which corresponds to a shift of the coordinate $x$ of the form $x\to x+ 1$. But, even though the effective theory is invariant under this shift, 
certain quantities transform non-trivially. A prominent example is the 
period vector $\Pi$ of the holomorphic three-form $\Omega$ shown in \eqref{eq:omegaexpand}, which behaves as
\eq{
  \label{monodromy_002}
  \Pi
  \hspace{5pt} \xrightarrow{\hspace{5pt}x\to x+1\hspace{5pt}}  \hspace{5pt}
  \Pi' = T\, \Pi\,,
}
where matrix notation is understood. Here, $T$ denotes an integer-valued monodromy matrix which in 
order for the K\"ahler potential \eqref{eq:cskahler} to be invariant has to satisfy 
\eq{
  T^T \eta \, T =  \eta\,,
  \hspace{50pt} T \in \mbox{Sp}(2h^{2,1}+2,\mathbb Z) \,.
}
Although not obvious, it turns out that for Calabi-Yau threefolds $Y_3$ the monodromy matrices
$T$ can always be made unipotent \cite{Landman}, that is $(T-\mathds 1)^{m+1}= 0$ for some $m\geq 0$.\footnote{This might require 
sending $z \rightarrow z^n$ and amounts to removing a possible semi-simple part of a general monodromy matrix $T$. } 
Furthermore, to each $T$ we can associate  a so-called log-monodromy matrix defined as 
$N = \log T$, which is an element of the Lie algebra $\mathfrak{sp}(2h^{2,1}+2,\mathbb R)$
and therefore satisfies
\eq{
  \label{def_log_mo}
  N = \log T\,,
  \hspace{40pt}
  ( \eta \op N) - ( \eta \op N)^T = 0 \,, 
  \hspace{40pt} N \in \mathfrak{sp}(2h^{2,1} +2,\mathbb R) \,.
}
Since the monodromy matrices $T$ are unipotent, it follows that the log-monodromy matrices
$N$ are nilpotent: $N^{m+1}=0$ for some $m\geq 0$. 
Note that 
the symmetry $N$ might induce an approximate 
continuous shift symmetry of the moduli space metric near certain boundaries. This is 
familiar, for example, from the large complex structure regime. In order to simplify the naming 
we will refer to $x$ as being the axion and $y$ the saxion, even if no continuous symmetry 
is restored in the limit.


\subsubsection*{Nilpotent orbit theorem}

The nilpotent orbit theorem \cite{Schmid} allows us to describe the moduli dependence of 
differential forms close to the boundary. 
An example is
the 
holomorphic three-form $\Omega$, and one finds that in the limit  $t\to i\op\infty$
the period vector $\Pi$ can be expressed as
\eq{
  \label{npot_001}
  \Pi(t) = e^{t\op N} e^{\Gamma(z)} \op \mathbf{a}_0 \,,
}
where $\Gamma(z)\in\mathfrak{sp}(2h^{2,1} +2,\mathbb C)$ (with $\Gamma(0)=0$) is a matrix varying
holomorphically in $z = e^{2\pi \op i t}$. The $(2h^{2,1} +2)$-component 
vector $\mathbf a_0$ is a reference point independent
of $t$, but in general depends holomorphically on the other moduli kept finite.\footnote{Let us note one can also choose to encode the dependence on these spectator moduli in the map $\Gamma$, see for instance \cite{Grimm:2020ouv} for more details.} We have therefore expressed the dependence of $\Pi$ on $t$ near the boundary 
in a simple form. Expanding the second exponential in \eqref{npot_001}  we find a natural 
split of $\Pi(t)$ as 
\beq \label{Pi-exp}
   \Pi(t) = \Pi_{\rm poly} +  \Pi_{\rm exp} = e^{t\op N}  \op \mathbf{a}_0 +  \cO(e^{2\pi i t})
\eeq
where we have collected all polynomial terms in $ \Pi_{\rm poly} =e^{t\op N}  \op \mathbf{a}_0 $ while 
all exponentially suppressed terms reside in $ \Pi_{\rm exp}$. It will be crucial below that 
the nilpotent orbit theorem implies that an expansion of the form \eqref{Pi-exp} 
also occurs for all its holomorphic derivatives.

Let us furthermore recall that for a Calabi-Yau threefold $\Omega$ is an element of 
$H^{3,0}(  Y_3)$. Taking a holomorphic (covariant) derivative of $\Omega$
lowers its holomorphic degree by one, and hence the fourth (covariant) derivative of $\Omega$ has to vanish. 
Combining this observation with \eqref{npot_001} and ignoring some technical subtleties, 
we find the necessary condition $N^4 = 0$. The highest non-vanishing power $m$ of $N$ depends 
on the boundary under consideration. We find the four choices
\eq{
  \label{npot_021}
  N^{m}\neq 0\, , \ N^{m+1}=0 \ , \hspace{50pt}\mbox{for}\quad  0\leq m \leq 3 \,,
}  
where for $m=0$ there is no unipotent monodromy associated to the boundary. 

\subsubsection*{Essential exponential corrections}

As stressed in \eqref{npot_021} the nilpotency order of $N$ does not have to be 
maximal, i.e.~$m=3$. This implies that the polynomial $\Pi_{\rm poly} =e^{t\op N}  \op \mathbf{a}_0 $ 
appearing in the expansion \eqref{Pi-exp} can be of any degree smaller or equal three and is given 
by the highest $k$ with non-vanishing $N^k a_0$. For a Calabi-Yau threefold the full middle cohomology 
can be obtained by taking holomorphic derivatives of $\Omega$. However, if 
$k<3$ the derivative of $\Pi_{\rm poly}$ with respect to $t$ cannot generate all three-forms. 
This implies that some of the exponential corrections in $ \Pi_{\rm exp}$ have to be present. Such corrections 
are thus essential near any boundary with $k<3$ and were termed  essential exponential corrections or essential instantons in \cite{Bastian:2021eom}.
A more precise statement can be made by introducing the expansion 
\beq \label{Pi-exp2}
    \Pi(t) =  e^{t\op N}  \op \big(\mathbf{a}_0 + e^{2\pi i t} \mathbf{a}_1 + e^{4\pi i t} \mathbf{a}_2 + \ldots \big)\ .
\eeq
Denoting by $k_i$ the lowest integer
such that $N^{k_i} \mathbf{a}_i = 0$, one finds that the term $\mathbf{a}_{i+1}$ has to be included whenever 
$k_0+ \ldots + k_i < 3 $.\footnote{Technically this only applies to one-modulus limits in one-dimensional moduli spaces, i.e.~$h^{2,1}=1$. Otherwise one also has to consider derivatives with respect to moduli not sent to the boundary, which can generate other three-forms as well. For the moment we ignore these subtleties, and refer to chapter \ref{chap:models} for a more detailed discussion.}  Essential instanton corrections are thus needed at almost all boundaries. While they can be constructed systematically as shown in \cite{Bastian:2021eom} (see chapter \ref{chap:models}), we will use their presence in a more indirect way in the following.


\subsubsection*{Nilpotent matrices and sl(2)}

As we have seen above, $N$ is a nilpotent matrix which belongs to a 
symplectic Lie algebra $\mathfrak{sp}(2h^{2,1}+2,\mathbb C)$. 
The classification of nilpotent elements of semi-simple Lie algebras is a
well-studied mathematical problem, and in the following we want  
to outline the main ideas of this classification. The classifications for $N \in \mathfrak{sp}(2h^{2,1}+2,\mathbb R)$ 
is slightly more involved and we will only quote the result in the following. 
For a clearer presentation, let us for this paragraph denote the nilpotent Lie algebra element 
by $N^-$ and let us use $m$ instead of $h^{2,1}+1$. 
\begin{itemize}

\item Let $N^-\in\mathfrak{sp}(2m,\mathbb K)$  be a nilpotent element of the Lie algebra $\mathfrak{sp}(2m,\mathbb C)$. Here 
 $\mathbb{K}$ can be either $\bbR$ or $\bbC$.  
A theorem by Jacobson and Morozov  states that one can always find elements 
$N^0,N^+\in \mathfrak{sp}(2m,\mathbb K)$ 
such that the algebra generated by $\{ N^-, N^+, N^0\}$ is isomorphic to $\mathfrak{sl}(2,\mathbb K)$. 
Recall that  $\mathfrak{sl}(2,\mathbb K)$ is generated by a triple $\{ n^-, n^+, n^0\}$ that satisfies the commutation relations
\eq{
\label{comm}
  [n^0,n^+]=+2\op n^+\,, \hspace{40pt}
  [n^0,n^- ]=-2\op n^- \,, \hspace{40pt}
  [n^+,n^-]= n^0\,.
}
Here, $n^+$ is a raising operator, $n^-$ is a lowering operator and $n^0$ is the weight operator, and we 
note that $\mathfrak{sl}(2,\mathbb K)$ is closely related to the Lie algebra $\mathfrak{su}(2)$. 

\item The triple $\{ n^-, n^+, n^0\} \in \mathfrak{sl}(2,\mathbb K)$ is represented by $2m\times 2m$ dimensional matrices 
$\{N^-,N^+,N^0\}$ acting on a $2m$-dimensional vector space $V$ over $\mathbb{K}$. 
However, this representation of  $\mathfrak{sl}(2,\mathbb K)$
is in general reducible and can therefore be expressed as a direct sum of irreducible 
representations of  $\mathfrak{sl}(2,\mathbb K)$. 
Concretely, this means that up to conjugation we can write each $N^-,N^+,N^0$ as
\eq{
  \label{nil_001}
  N^-,N^+,N^0=\left[\hspace{4pt}
  \arraycolsep0pt
  \begin{array}{ccc@{\hspace{6pt}}c@{\hspace{4pt}}}
  \setlength{\fboxsep}{6pt}
  \fbox{\hspace{0pt}$\ast$\hspace{0pt}}
  \\  
  &\setlength{\fboxsep}{14pt}
  \fbox{\hspace{0pt}$\ast$\hspace{0pt}}
\\
  &&\setlength{\fboxsep}{6pt}
  \fbox{\hspace{0pt}$\ast$\hspace{0pt}}
  \\
  &&& \ddots
  \end{array}
  \right] ,
}
where each block corresponds to an irreducible $\mathfrak{sl}(2,\mathbb K)$ representation
of $\{ n^-,n^+,n^0\}$ of a certain dimension. Note that typically some of these blocks 
correspond to the one-dimensional representation of $\mathfrak{sl}(2,\mathbb K)$ which 
is simply a zero. 
Also, the dimensions of the blocks 
have to add up to $2m$.

\item The classification of all possible nilpotent matrices $N^-$ becomes a combinatorial problem of how these can 
be decomposed in irreducible pieces.  Considering $N^- \in \mathfrak{sp}(2m,\mathbb C)$ it is well-known that 
Young diagrams classify irreducible representations. In the case of $N^- \in \mathfrak{sp}(2m,\mathbb R)$ this  problem is solved using 
so-called signed Young diagrams as explained, for example, in \cite{BrosnanPearlsteinRobles, Kerr2017,Grimm:2018cpv}. 

\end{itemize}
Let us now apply the above discussion to our situation. The nilpotent 
$\mathfrak{sp}(2h^{2,1} +2)$ matrix $N$ introduced in \eqref{def_log_mo}
can be identified with the representation $N^-$ of $n^-$ acting on the vector space $V = H^{3}(Y_3,\mathbb K)$.
Therefore, there exists a matrix $S\in \mbox{Sp}(2h^{2,1} +2)$
such that $S^{-1} N \op S$ has a block-diagonal form where
each block $\nu$ is a matrix representation $\sf N^-_{[\nu]}$ of the sl(2)-lowering operator $n^-$,
that is
\eq{
  S^{-1} N \op S = \left[\hspace{4pt}
  \arraycolsep0pt
  \begin{array}{ccc@{\hspace{6pt}}c@{\hspace{4pt}}}
  \setlength{\fboxsep}{6pt}
  \fbox{$\sf N^-_{[1]}$\hspace{-1pt}}
  \\  
  &\setlength{\fboxsep}{14pt}
  \fbox{$\sf N^-_{[2]}$\hspace{-2pt}}
\\
  &&\setlength{\fboxsep}{6pt}
  \fbox{$\sf N^-_{[3]}$\hspace{-1pt}}
  \\
  &&& \ddots
  \end{array}
  \right] .
}
Let us finally note that the weight operators $\sf N^0_{[\nu]}$ in each irreducible block 
are determined only up to conjugation. This freedom will be fixed in
asymptotic Hodge theory by picking a certain  Deligne splitting to be discussed in chapter \ref{chap:mhs}. 
Furthermore, we have suppressed in this discussion that the nilpotent matrix $N$ and the associated triple $\{N^-=N,N^+,N^0\}$ 
has to be compatible with the Hodge decomposition in the limit $t \rightarrow i \infty$. This  
     aspect will be also relevant in the discussion of the Hodge star and will be more central in chapter \ref{sec:sl2splitting}.


\subsubsection*{Weight-space decomposition}

We now want to get a better understanding of how a nilpotent matrix 
acts on  vectors, for instance how in \eqref{npot_001} the matrix $N$ acts  on $\mathbf a_0$.
This leads us to the  weight-space decomposition
under the action of $\mathfrak{sl}(2,\mathbb K)$, where again we can consider $\mathbb{K}$ being either $\bbR$ 
or $\bbC$. We start with a \textit{single irreducible} $\sf n$-dimensional 
representation $\{\sf N^-,\sf N^+,\sf N^0\}$
of $\mathfrak{sl}(2,\mathbb K)$, which corresponds to one particular block in \eqref{nil_001}. 
(We suppress the subscript $[\nu]$ for now.)
As is known from Lie-algebra representation theory, the vector space $\sf V$ on which 
the $\sf n\times \sf n$-dimensional 
matrices $\{\sf N^-,\sf N^+,\sf N^0\}$ are acting can be decomposed into one-dimensional weight spaces as
\eq{
  \label{wd_001}
  \sf V = \sf V_{d} \oplus \sf V_{d-2} \oplus \ldots \oplus \sf V_{-d}\,,
  \hspace{50pt}
  \sf V_{\ell} = \{ v\in \sf V : \sf N^0 \op v = \ell \op v \} \,,
}
where the eigenvalues $\ell$ of $\sf N^0$ are the weights and $d=\sf n-1$ is the highest weight. 
The raising and lowering operators $\sf N^+$ and $\sf N^-$ then map between these 
spaces as $\sf N^+: \sf V_{\ell} \to \sf V_{\ell+2}$ and $\sf N^-: \sf V_{\ell} \to \sf V_{\ell-2}$.
From this decomposition we find that $\sf N^-$ satisfies
\eq{
   (\sf N^-)^{d+1} = (\sf N^+)^{d+1}  = 0 \,.
}
So far we have focused on a single block of the decomposition \eqref{nil_001}, but 
we can now combine these building blocks as follows. The triple of 
$2m\times 2m$ dimensional matrices $\{N^-,N^+,N^0\}$ is a  sum of 
triples $\{\sf N^-_{[\nu ]},\sf N^+_{[\nu ]},\sf N^0_{[\nu]}\}$, which 
is therefore acting on a $2m$-dimensional vector space that can be decomposed as
\eq{
  \label{wd_005}
  V = \sf V_{[1]} \oplus \sf V_{[2]} \oplus \ldots\,.
}
The nilpotent matrix $N=N^-$ satisfies \eqref{npot_021}, and therefore
the largest allowed highest weight of the subspaces satisfies $d^{(i)}\leq 3$. In other words, 
in the decomposition \eqref{nil_001} at most four-dimensional irreducible representations of 
$\mathfrak{sl}(2,\mathbb K)$ can appear.


\subsubsection*{Hodge star}

The nilpotent orbit theorem can be used to determine the periods \eqref{npot_001} of the 
holomorphic three-form near the 
boundary. From this expression one can, in principle, determine the Hodge-star operator in that limit
for every asymptotic regime. 
However,  this approximation can be still too involved to 
be of practical use for moduli stabilization, since it will generally 
contain many sub-leading polynomial and exponential corrections.  
We are therefore going to perform 
other approximations as follows:
\begin{itemize}

\item Let us focus again on one particular block in the decomposition \eqref{nil_001}, and consider the 
weight-space decomposition shown in \eqref{wd_001}. For a given triple $\{\sf N^-,\sf N^+,\sf N^0\}$ we now introduce a real operator $\sf C_{\infty} $
satisfying the relations $\sf C^{-1}_{\infty} \sf N^+  \sf C_{\infty}=\sf N^-$, 
$\sf C^{-1}_{\infty} \sf N^0\op \sf C_{\infty}=-\sf N^0$ and the requirement $\sf C_{\infty}^2=-\mathds 1$.
This operator maps the subspaces $\sf V_{\ell} $ as
\eq{
  \label{wd_304}
  \sf C_{\infty} : \sf V_{\ell} \to \sf V_{-\ell} \,.
}
While these conditions significantly constrain $\sf C_\infty$ they do not 
fix it completely. In order to fix $\sf C_\infty$ we require that it corresponds to 
the Hodge star operator acting on this representation after being appropriately extended to the boundary.
To make this more precise, we combine the  $\sf C_\infty$
of the individual irreducible representations of 
$\mathfrak{sl}(2,\mathbb C)$ into a $C_\infty \in \mathfrak{sp}(2m,\bbR)$ acting 
on the full vector space $V$ given in \eqref{wd_005}.
$C_{\infty}$ is then obtained from the full Hodge star operator $C$ 
via the limiting procedure 
\begin{equation}\label{Cinfty_limit}
   C_\infty = \lim_{y \to \infty} e \, C \, e^{-1} \,,
   \hspace{70pt}
   e= \exp \left[\tfrac{1}{2} \log y \, N^0 \right] ,
\end{equation}
where $y=\mbox{Im}\, t$ is send to the boundary and where $N^0$ denotes the weight operator 
in the triple $\{ N, N^+,N^0\}$. Let us note that the simple expression \eqref{Cinfty_limit} is 
somewhat deceiving, since the construction of an appropriate $N^0$ requires to check for compatibility 
of the choice with the Hodge decomposition. In practice, as we will see in section \ref{sec_tech_details}, we will construct 
$C_\infty$ and the triple $\{ N, N^+,N^0\}$ at the same time. 

\item The operator $C_{\infty}$ in \eqref{wd_304} does not contain any dependence on 
the complex structure variable 
$y=\mbox{Im}\, t$ which is send to the boundary. We therefore introduce 
a so-called sl(2)-approximated Hodge star operator by
\eq{
  \label{wd_305}
  C_{\rm sl(2)} : \sf V_{\ell} \to  \: \sf V_{-\ell} \,,\qquad C_{\rm sl(2)} v  = y^\ell C_{\infty} v \ ,
}
where $v  \in V_{\ell}$ and the power in the variable $y$ corresponds to the weight of the weight-space $\sf V_{\ell}$. 
Using $e$ defined in \eqref{Cinfty_limit} we can make \eqref{wd_305} more precise as follows
\eq{
  \label{wd_306}
   C_{\rm sl(2)} =  e^{-1} \op C_{\infty} \, e \,, 
}
which produces precisely the type of mapping shown in \eqref{wd_305}. Again, the action 
on the full vector space $V$ is obtained by combining the action in each subspace. 
Note that in \eqref{wd_306} we set the axion  to zero, which can be 
re-installed by replacing $C_{\rm sl(2)} \to e^{+ x N} C_{\rm sl(2)} e^{- x N}$. 
Finally, the relation to the Hodge-star matrix \eqref{def-cM} is
\eq{
  \mathcal M_{\rm sl(2)} = \eta\op    C_{\rm sl(2)}  \,.
}

\item So far much of the above discussion was possible on the real vector space $V=H^3(Y_3,\bbR)$. 
As soon as one aims to talk about the Hodge decomposition and the compatibility with 
the construction of the $\mathfrak{sl}(2,\mathbb C)$ representation one is forced to work over $\bbC$.
Let us denote by  $Q$ the 
operator acting on elements of $H^{q,3-q}$ with eigenvalue $q-\frac{3}{2}$. Since $\overline{H^{p,q}} = H^{q,p}$ the operator 
$Q$ is imaginary and we have $Q \in i\, \mathfrak{sp}(2m,\bbR)$. It follows from \eqref{eq:hodgedef} that the Hodge star operator $C$
can be written in terms of $Q$ as $C = e^{\pi i Q} = (-1)^{Q}$. In analogy
to \eqref{Cinfty_limit} one can then extract the information about 
the boundary Hodge decomposition by evaluating 
\beq
     Q_\infty  = \lim_{y \to \infty} e \, Q \, e^{-1} \,.
\eeq
As we will explain in subsection \ref{sec_tech_details}, also $Q_\infty$ should actually be constructed together with the triple $\{N,N^+,N^0 \}$, since these operators are linked through non-trivial compatibility conditions. To display these compatibility relations 
it is useful to introduce a complex triple $\{L_{-1},L_0,L_{+1} \}$ by setting 
\beq
   L_{\pm 1} = \frac{1}{2}(N^+ + N^- \mp i N^0)\ , \qquad L_0 = i(N^- - N^+)\ .  
\eeq 
The algebra satisfied by $\{L_{-1},L_0,L_{+1},Q_\infty \}$ then reads
\beq
   [L_0,L_{\pm 1}] = \pm 2 L_{\pm 1}\ , \qquad [L_1,L_{-1}] = L_0\ , \qquad [Q_\infty, L_\alpha] = \alpha L_\alpha\ . 
\eeq
Note that this is the algebra of $\mathfrak{sl}(2,\bbC) \oplus \mathfrak{u}(1)$ if one introduces 
the operator $\hat Q = Q_\infty - \frac{1}{2} L_0$ as the generator of the $\mathfrak{u}(1)$.

\end{itemize}


\subsubsection*{Summary of main steps}

Let us finally summarize the necessary steps to construct the  sl(2)-approximated 
Hodge-star operator shown in equation \eqref{wd_306}:
\begin{enumerate}

\item One has to choose a modulus $y = \mbox{Im}\,t$ which approaches the boundary of moduli space. 
Associated to this boundary, the corresponding axion $x=\mbox{Re}\,t$ admits a 
discrete symmetry corresponding to the monodromy transformation of the period vector shown in \eqref{monodromy_002}.

\item The associated monodromy matrix $T$ can be made unipotent, and induces a 
nilpotent log-monodromy matrix $N=\log T$. For Calabi-Yau threefolds each boundary 
has an $0\leq m\leq 3$ with $N^m \neq 0$ and $ N^{m+1}=0$.

\item The log-monodromy matrix $N$ can be interpreted as a lowering operator in an 
$\mathfrak{sl}(2,\mathbb C)$ triple $\{ N^- = N, N^+, N^0\}$. 
Then, the weight operator $N^0$ needs to be constructed which is used in the definition of the 
sl(2)-approximated Hodge-star operator shown in \eqref{wd_306}. 

\item The $\mathfrak{sl}(2,\mathbb C)$ triple $\{ N^- = N, N^+, N^0\}$ has 
to be compatible with the Hodge decomposition extended to the boundary. 
The latter can be encoded by an operator $Q_\infty$ that is constructed 
jointly with the triple. 

\end{enumerate}
Let us emphasize that the above steps apply when sending a single modulus to the boundary. 
Additional complications arise when two or more moduli $t^i$ are considered. 
More concretely, even though the corresponding log-monodromy matrices 
$N_i$ can be shown to commute, when including the associated weight operators $N^0_i$ these operators generically do not all commute 
with each other and hence one cannot construct a consistent weight-space decomposition immediately. 
How to deal with this situation will be explained in section \ref{sec_tech_details}.

\chapter{Hodge theory and asymptotic regimes}\label{chap:asympHodge}
In this chapter we set the premise for asymptotic Hodge theory. First we specify the regimes of interest and describe how to parametrize asymptotic regimes near boundaries in the moduli space of Calabi-Yau manifolds. We then introduce the Hodge structures varying over these spaces, and explain how to switch from a physical formulation using period integrals to mathematical formulations with Hodge filtrations and period mappings. Combining these two stories, we explain how the behavior of Hodge structures near boundaries is captured by the nilpotent orbit theorem \cite{Schmid}. In particular, we show how this nilpotent orbit approximation can be used to describe the asymptotic form of physical couplings appearing in string compactifications.


\section{Limits in complex structure moduli space}
Asymptotic Hodge theory describes the behavior of Hodge structures near singular loci in the moduli space of K\"ahler manifolds. In the setting of this thesis we consider the complex structure moduli space of Calabi-Yau manifolds, with as Hodge structure the Hodge decomposition of their middle cohomology groups. Let us start by giving the reader some feeling for these moduli spaces, and show how to parametrize loci where the geometry becomes singular as boundaries.

\subsubsection*{One-modulus limits}

It is instructive to first recall the complex structure moduli space of a torus, given by the fundamental domain of SL$(2,\mathbb{Z})$ in figure \ref{fig:funddomain}. Here the complex structure parameter $\tau$ specifies the shape of the torus, which is simply the ratio of its two radii when $\Re \tau = 0$. The torus becomes singular when we take the limit $\tau \to i \infty$ -- it pinches -- and we can circle this locus by sending $\tau \to \tau+1$. To put it in simpler terms, the moduli space is shaped like an infinite tail towards this singular point, which is located at the very tip. This intuitive picture has been sketched in figure \ref{fig:degeneration}, and extends straightforwardly to Calabi-Yau manifolds of higher dimension. From a practical point of view, it allows us to divide the moduli space into a compact subregion in the middle where the manifold is smooth, with on the outside infinite tails along which the manifold degenerates. It is along these tails, when we move close to a singular point, where asymptotic Hodge theory comes into play.



To take a slightly more complicated example, we can consider the complex structure moduli space of the (mirror) quintic \cite{Candelas:1990rm}. This moduli space is parametrized by a single complex coordinate $z$ lying on $\mathbb{P}^1\backslash \{0,1,\infty\}$. In other words, it can be viewed as a sphere with three punctures. These punctures correspond to special points in moduli space where the Calabi-Yau manifold degenerates or takes some special shape. In this particular case the degenerations are
\begin{itemize}
\item $z=0$: large complex structure point,
\item $z=1$: conifold point,
\item $z=\infty$: Landau-Ginzburg point.
\end{itemize}
These three points differ in their singular nature -- e.g.~the large complex structure point lies at infinite distance, while the other two points are at finite distance with respect to the moduli space metric. Nevertheless, this help us in making our earlier sketch more explicit. We see that we can separate the moduli space into tails towards these special points, merged together by a compact bulk in the middle. Throughout this thesis we will loosely refer to these singularities as \textit{boundaries} of the moduli space, and moving towards these boundaries as \textit{limits}.

\begin{figure}[h!]
	\begin{center}
		\includegraphics[width=0.8\textwidth]{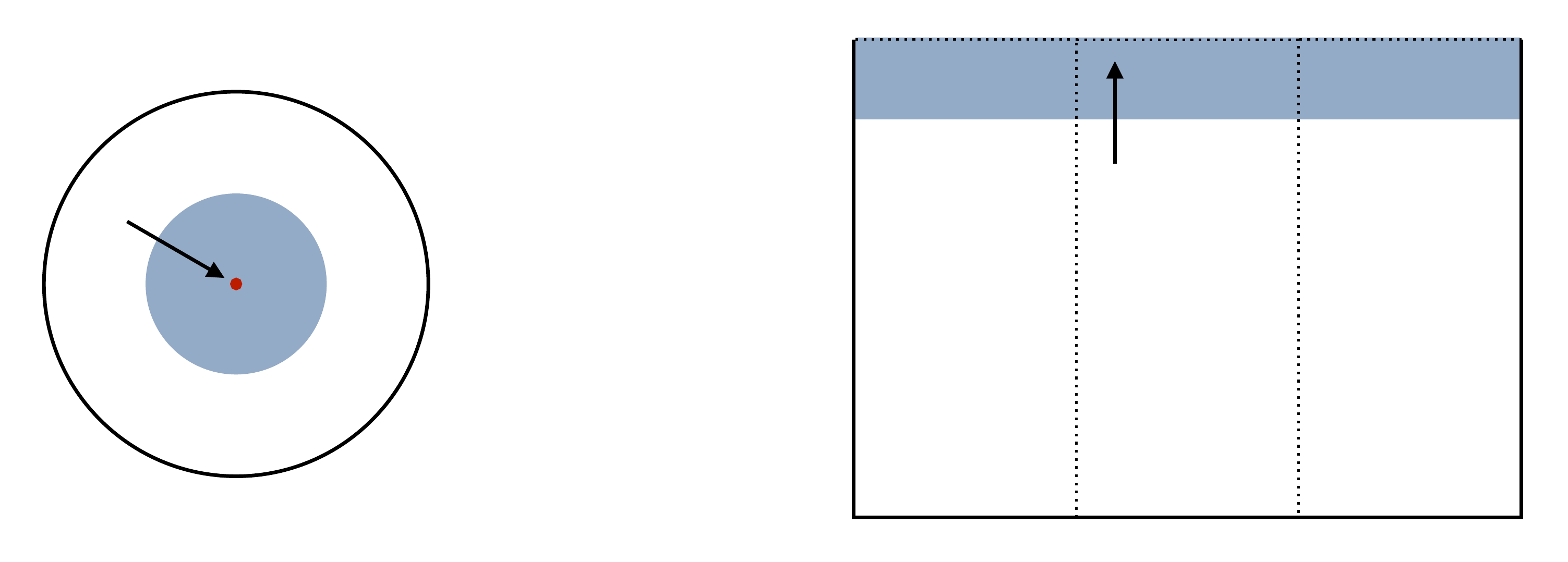}
		\vspace*{-1cm}
	\end{center}
	\begin{picture}(0,0)
	\put(54,64){\rotatebox{-31}{\small $z \rightarrow 0$}}
	\put(241,68){\small $y \rightarrow \infty$}
	\put(17,85){(a)}
	\put(165,85){(b)}
	\end{picture}
	\caption{Local patch in the space $\cM$ represented as Poincar\'e disc, figure (a), and upper half-plane, figure (b). The regime near the boundary at $z=0$ in the disc is mapped to $t = i \infty$ on the upper half plane by \eqref{eq:onemoduluscoordinates}. \label{fig:disc}}
	
\end{figure}
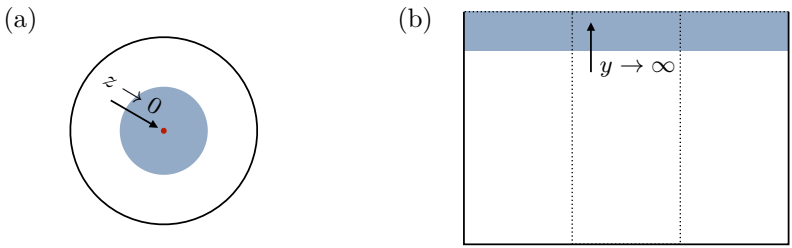

Let us now give coordinate patches to make the infinite tails in the setting of the quintic more explicit. For the coordinate $z$ above we can straightforwardly bring the location of the singularity to the form $z=0$ -- for the large complex structure point this is already the case, while for the conifold point we can shift $z\to z-1$, and for the Landau-Ginzburg point we map $z \to 1/z$. Circling the singularity in these patches then corresponds to $z \to e^{2\pi i } z$. However, from a physical point of view it is more convenient to work with a slightly different coordinate by moving to the so-called universal cover of the boundary. This simply amounts to reparametrizing as
\begin{equation}\label{eq:onemoduluscoordinates}
z = e^{2\pi i t} \, , \qquad t = x+i y\, .
\end{equation}
Similar to the fundamental domain of the torus, the boundary is then located at $t = i \infty$, while circling around the boundary corresponds to $t \to t+1$. The real and imaginary parts of this complex coordinate $t$ have a natural physical interpretation: $x$ can be thought of as an \textit{axion} field, since close to the boundary typically a continuous shift symmetry emerges when circling the boundary; its counterpart $y$ we refer to as a \textit{saxion}, which has to be large in order to be close to the boundary. The two different coordinate patches have been pictured in figure \ref{fig:disc}. 

\subsection{Multi-moduli limits}
Having discussed the one-modulus case, let us now move to the general multi-moduli setting. Similar to the one-modulus case we can describe singular components of co-dimension one as some locus of the form $z^k = 0$. Intersecting $n$ such boundaries can then be locally described as some normal intersection $z^1 = \ldots = z^n = 0$ (after appropriately reordering the coordinates). A two-modulus example of such an intersection has been depicted in figure \ref{fig:multimoduli}. The remaining coordinates $\zeta^a$ ($a = n+1,\ldots, h^{D-1,1}$) we keep away from any loci corresponding to further intersections of boundaries, and shall therefore be referred to as \textit{spectator moduli}.

\begin{figure}[h!]
\begin{center}
\includegraphics[width=7.5cm]{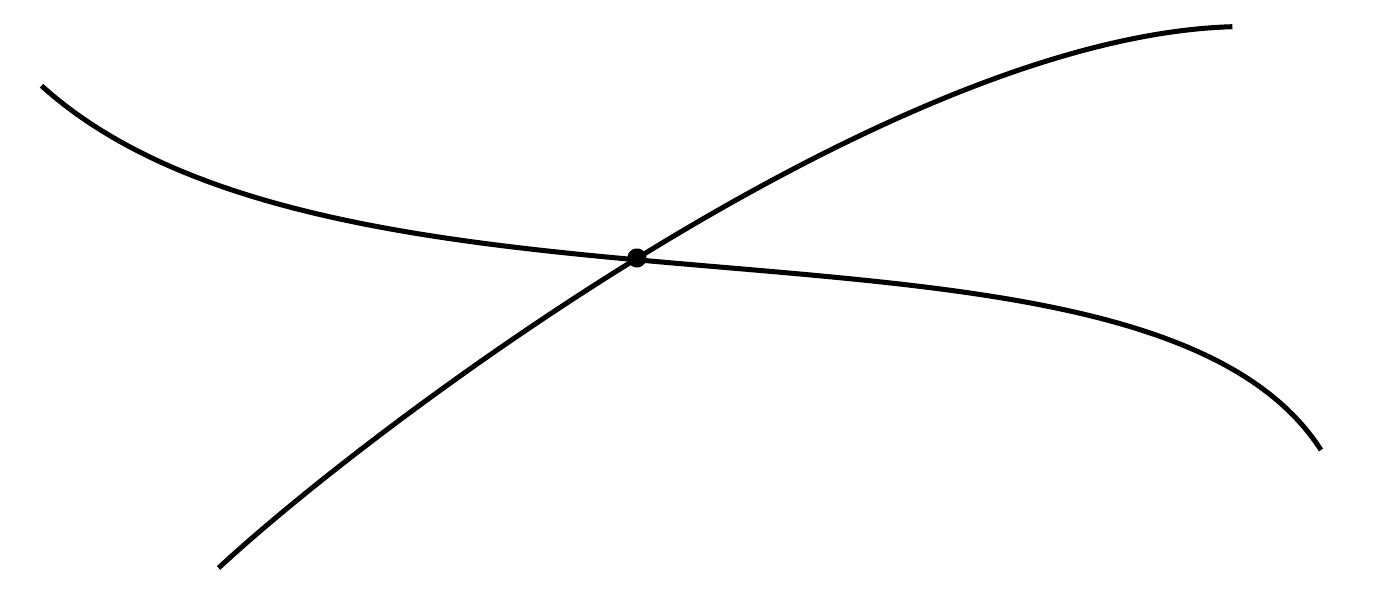}
\end{center}
\begin{picture}(0,0)
\put(270,110) {$z^1=0$}
\put(280,50) {$z^2=0$}
\end{picture}
\vspace*{-1cm}
\caption{\label{fig:multimoduli} Two boundaries $z^1=0$ and $z^2=0$ in $\cM^{\rm cs}(Y_D)$ intersecting at $z^1=z^2=0$.}
\end{figure}
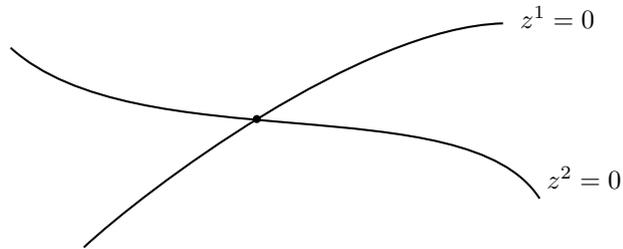
We can make the same switch of coordinates as in the one-modulus case \eqref{eq:onemoduluscoordinates}, which reads
\begin{align}
t^i = x^i + i y^i= \frac{1}{2 \pi i } \log[z^i] \, , \qquad i=1, \ldots, n\, , \label{eq:Coordinates}
\end{align}
where again $x^i$ and $y^i$ have the physical interpretation of axion and saxion fields, with $x^i \to x^i +1$ a rotation around the boundary and $y^i \to \infty$ the limit to the boundary. The relation \eqref{eq:Coordinates} should be kept in mind as we will use these two coordinate patches interchangeably throughout this work.

\section{Hodge structures, periods and period mappings}\label{sec:hodgestructures}
The complex structure moduli space of Calabi-Yau manifolds is accompanied by an underlying structure that varies with changes in the moduli. There are three equivalent formulations -- some more physical and some more mathematical -- given by Hodge structures, period integrals and period mappings. Below we introduce each of these descriptions and explain how one can switch between them.


\subsection{Hodge structures}
We begin with \textit{pure Hodge structures}. This perspective describes a decomposition of a complex vector space $H$ into $D+1$ complex subspaces as
\begin{equation}\label{eq:HodgeDecomp}
H = H^{D,0} \oplus \ldots \oplus H^{0,D} = \bigoplus_{p+q=D} H^{p,q}\, ,
\end{equation}
where $\overline{H^{p,q}}= H^{q,p}$. The integer $D$ we refer to as the \textit{weight} of the Hodge structure. Geometrically this splitting corresponds to the Hodge decomposition of the middle cohomology $H(Y_D,\mathbb{C})$ of the Calabi-Yau manifold into $(p,q)$-forms. To be precise, the vector spaces of interest are
\begin{itemize}
\item Calabi-Yau threefolds: the three-form cohomology $H=H^3(Y_3,\mathbb{C})$,
\item Calabi-Yau fourfolds: the primitive\footnote{This refers to taking four-forms $\omega \in H^4(Y_4, \mathbb{C})$ that satisfy the primitivity condition $\omega \wedge J = 0$ under the K\"ahler two-form $J$ of $Y_4$.} four-form cohomology $H=H^4_{\rm p}(Y_4, \mathbb{C})$.
\end{itemize}
The variation of these Hodge structures with the complex structure moduli can be understood as follows. Moving through complex structure moduli space amounts to changing what we call holomorphic and anti-holomorphic. This means that while the total space $H$ remains unchanged the orientation of the subspaces $H^{p,q}$ inside can vary. As an example, the complex vector space $H^{D,0}$ is one-dimensional and can be represented by a line passing through the origin in $H$, and by varying the complex structure moduli we rotate this line.

\paragraph{Charge operator and Hodge star.} We can naturally introduce a charge operator $Q$ which encodes this splitting into $H^{p,q}$. We define it by the action
\begin{equation}
\omega_{p,q} \in H^{p,q}: \qquad Q \omega_{p,q} = (p-D/2) \omega_{p,q} \, .
\end{equation}
We can use this $Q$ to define a Hodge star $C$ for the above Hodge decomposition, also referred to as the Weil operator in the math literature.\footnote{Technically speaking, the Hodge star operator acts on any differential form, while the Weil operator only acts on cohomology classes in $H^{p,q}$. This means the Weil operator does not see shifts by exact pieces under which cohomology classes are invariant. Throughout this thesis we will be more loose with our formulation, and refer to the operator acting on elements of $H$ simply as the Hodge star operator.} To be precise, we write $C = e^{\pi i Q}$ such that
\begin{equation}
\omega_{p,q} \in H^{p,q}: \qquad C\omega_{p,q} = i^{p-q} \omega_{p,q}\, .
\end{equation}
Note that since the splitting $H^{p,q}$ varies over the complex structure moduli space, the Hodge star operator must do so as well.

\paragraph{Hodge filtration.} We can repackage the splitting $H^{p,q}$ in terms of a so-called \textit{Hodge filtration} $F^p$. The idea is to collect all $H^{r,s}$ with $r \geq p$ into a vector space $F^p$. In other words, we take differential three-forms with at least $p$ $\dd z$'s when writing them out locally. We recover the old splitting $H^{p,q}$ from the Hodge filtration by intersecting $F^p$ with $\bar{F}^q$, i.e.~we take three-forms with at least $p$ $\dd z$'s and $q$ $\dd \bar{z}$'s. The two formulations are thus related as
\begin{align}\label{eq:Hpqdef}
F^p= \bigoplus_{k=p}^{D} H^{k,D-k} \, , \qquad F^p \cap \bar{F}^q = H^{p,q}\, , \quad p+q=D\, .
\end{align}
Note that the set of vector spaces $F^p$ forms a decreasing filtration of $H$ as 
\begin{align}
0 \subset F^D \subset F^{D-1} \subset \ldots \subset F^1 \subset F^0 = H\, .
\end{align}
The upshot of working with the vector spaces $F^p$ of the Hodge filtration is that these vary in a holomorphic manner with the complex structure moduli space -- in contrast to the $H^{p,q}$, where only $H^{3,0}$ varies holomorphically. This holomorphic dependence can be made explicit by the following rules for holomorphic and anti-holomorphic derivatives
\begin{align}
\omega^p \in F^p: \qquad \frac{\partial \omega^p}{\partial z} \in F^{p-1} \, , \quad \frac{\partial \omega^p}{\partial \bar{z}} \in F^p \,. \label{eq:Transversality}
\end{align}
We thus see that by taking derivatives with respect to the holomorphic complex structure moduli we move down the filtration but only by one degree at a time. This property is generally referred to as \textit{horizontality}. 

\paragraph{Polarization.} Next we move on to \textit{polarized} Hodge structures. By a polarization we here mean that the vector space $H$ is equipped with a bilinear pairing $\langle \, \cdot \, , \, \cdot \, \rangle$, under which the Hodge structure $H^{p,q}$ satisfies certain polarization conditions. In practice this bilinear pairing arises from integration over the Calabi-Yau manifold as
\begin{equation}\label{def-<>}
\langle u, v \rangle = \int_{Y_D} u \wedge v \, , \qquad \eta_{\mathcal{I}\mathcal{J}} = \langle \gamma_{\mathcal{I}} , \gamma_{\mathcal{J}} \rangle \, .
\end{equation}
where $u,v \in H$, and $\eta_{\mathcal I \mathcal J}$ denotes a matrix representative in terms of a basis $\gamma_{\mathcal{I}} \in H$.\footnote{For Calabi-Yau threefolds one can always choose a harmonic three-form basis that brings this pairing to the standard symplectic form, see \eqref{eq:pairings}.} Note that the pairing satisfies $\langle u, v \rangle = (-1)^{D} \langle v , u \rangle$,~i.e. it is symmetric (skew-symmetric) for D even (odd). Requiring a Hodge structure to be polarized now amounts to
\begin{equation}
\begin{aligned}
 \langle H^{p,q}, H^{r,s} \rangle &=0  \, , \quad \text{if $(p,q) \neq (s,r)$}\, ,\\
\langle \omega, C\bar{\omega} \rangle &> 0 \, ,\quad \text{for } \omega \in H^{p,q} \text{ and } \omega \neq 0 \,. \label{eq:HRrelations}
\end{aligned}
\end{equation}
These two conditions are referred to as the Hodge-Riemann bilinear relations. The first condition asserts that only integrals with integrands of Hodge type $(D,D)$ are non-vanishing. The second condition can be understood as a positivity condition for the Hodge norm \eqref{def-Hodgenorm} we will define shortly. Notice that we can reformulate the first condition in terms of the Hodge filtration \eqref{eq:Hpqdef} as $\langle F^p, \ F^{D+1-p} \rangle =0$.\footnote{A cute observation is that the vector space $F^p$ is then completely determined as the orthogonal complement $F^p = (F^{D+1-p})^\perp$ with respect to the bilinear pairing (assuming $p\neq (D+1)/2$).}

\paragraph{Hodge norm and product.} We can combine the Hodge star operator $C$ and the bilinear pairing $\langle \cdot, \cdot \rangle$ to construct a positive-definite inner product and norm on $H$. To be precise, these are given by
\begin{equation}\label{def-Hodgenorm}
(u,v) = \langle \bar{u}, C v \rangle \, , \qquad \| u \|^2 = (u, u)\, .
\end{equation}
where $u,v \in H$. Note that in the setting of Calabi-Yau threefolds these objects -- when written as matrices for a symplectic three-form basis  -- correspond to the charge matrix \eqref{def-cM} and physical charge \eqref{eq:charge} in the supergravity language.

\paragraph{Symmetry groups.} The above bilinear pairing \eqref{def-<>} has a symmetry group of real transformations which we shall denote by $G$, and we write $\mathfrak{g}$ for its corresponding Lie algebra. To be concrete, these are defined by
\begin{equation}
\begin{aligned}
g \in G &: \qquad &\langle gu, \, v \rangle &= \langle u, \, g^{-1} v)\, , \\
X \in \mathfrak{g} &: \qquad &\langle Xu, \, v \rangle &= -\langle u, \, Xv \rangle \, ,
\end{aligned}
\end{equation}
for $u,v \in H$. Depending on the dimension $D$ of the Calabi-Yau manifold this group changes, since for instance the bilinear pairing can be symmetric or skew-symmetric. For Calabi-Yau three- and fourfolds the symmetry groups are
\begin{equation}\label{eq:groups}
G = \begin{cases}
\mathrm{Sp}(2h^{2,1}+2)\, , \quad &D=3\, , \\
\mathrm{SO}(h^{2,2}_{\rm p}+2, 2h^{3,1})\, , \quad &D=4\, . \\
\end{cases}
\end{equation}
Likewise the Hodge product $(\cdot, \cdot)$ in \eqref{def-Hodgenorm} has a symmetry group $K$, with $\mathfrak{k}$ its corresponding algebra. This group is given by
\begin{equation}
K = \begin{cases}
\mathrm{U}(2h^{2,1}+2)\, , \quad &D=3\, , \\
\mathrm{SO}(2h^{3,1}+h^{2,2}_{\rm p}+2) \cap \big(\mathrm{O}(h^{2,2}_{\rm p}+2)\times \mathrm{O}(2h^{3,1}) \big)\, , \quad &D=4\, . \\
\end{cases}
\end{equation}
Let us note that $K$ is a maximal compact subgroup of $G$, and in particular $C \in K$ and $Q \in i \mathfrak{k}$.


\subsection{Periods}
Much of our discussion on asymptotic Hodge theory will concern vector spaces such as $H^{p,q}$, but in the physics literature we are more used to working with the differential forms represented by elements of these vector spaces. For instance, the vector space $H^{D,0}$ is spanned by the holomorphic $(D,0)$-form $\Omega$ of the Calabi-Yau manifold. The way to pass between these two descriptions is to represent a differential form by its period integrals, or periods for short. By integrating $\Omega$ over a basis of three-cycles $\Gamma_\mathcal{I} \in H_{\mathbb{Z}} $ we can expand it as
\begin{equation}\label{eq:omega-expand}
\Omega = \Pi^\mathcal{I} \gamma_\mathcal{I}\, , \qquad \Pi^\mathcal{I} =  \int_{\Gamma_\mathcal{I}} \Omega\, ,
\end{equation}
where $\gamma_I \in H^3(Y_3, \mathbb{Z})$ denotes the Poincar\'e dual three-form basis. The components $\Pi^I(z)$ then are the periods, which for $\Omega$ are holomorphic functions in the moduli. One can straightforwardly extend \eqref{eq:omega-expand} to any of the other subspaces $H^{p,q}$, however, holomorphicity will in general not be retained. 

\paragraph{Completeness principle.} The horizontality property \eqref{eq:Transversality} provides us with a useful method to derive the whole middle cohomology $H$ from just the period vector of $\Omega$. For Calabi-Yau threefolds there is a completeness principle, which states that we can obtain any $(p,q)$-form in the middle cohomology by taking derivatives of the $(3,0)$-form periods.\footnote{For Calabi-Yau fourfolds one can only recover the horizontal part of the middle cohomology $H^4_{\rm H}(Y_4)$ with derivatives of the $(4,0)$-forms \cite{stromingerspecialkahler,Greene:1993vm}. The embedding into the primitive cohomology $H^4_{\rm H}(Y_4) \subset H^4_{\rm p}(Y_4)$ is in general much more involved \cite{Braun:2014xka}.} To be precise, this means we can reproduce the Hodge filtration $F^p$ as
\begin{equation}\label{eq:Fpperiods}
F^p = \text{span}\left[\partial_{i_1} \cdots \partial_{i_{m}} \Pi \, |  \ 0 \leq m \leq 3-p \right]\, .
\end{equation}
In other words, the vector space $F^p$ is spanned by the period vector $\Pi$ and its first $3-p$ derivatives. In this basis we can recast the orthogonality identities in the polarization conditions \eqref{eq:HRrelations} as
\begin{align}\label{eq:horizontality}
\langle \Pi, \, \partial_i \Pi \rangle =0 \, , \qquad  \langle \Pi , \, \partial_i \partial_j \Pi \rangle =0 \,.
\end{align}
The pairing with the third order derivative does not vanish in general, and is related to the so-called Yukawa coupling.

\subsection{Period mappings}
As final formulation, we briefly want to mention the so-called \textit{period mapping}. While in our review of asymptotic Hodge theory here we will not utilize this perspective often, these object play an important role in the holographic description of the moduli space \cite{Grimm:2020cda, Grimm:2021ikg, Grimm:2021idu} (for related work see \cite{Cecotti:2020rjq,Cecotti:2020uek}). The period mapping can then be viewed as a set of matter fields living in the bulk, and in fact in \cite{Grimm:2021idu} it was shown to be described by a deformed Wess-Zumino-Witten model. 

The idea of the period mapping is to introduce a group-valued operator $h(t,\bar{t})$ that interpolates between a reference Hodge structure at the boundary and the Hodge structure in the interior of the moduli space. For this boundary Hodge structure $H^{p,q}_\infty$ we introduce a charge operator
\begin{equation}\label{eq:Qdef}
\omega_{p,q} \in H^{p,q}_\infty: \quad Q_\infty \omega_{p,q} = (p-D/2) \omega_{p,q}\, ,
\end{equation}
with corresponding boundary Hodge star $C_\infty = e^{\pi i Q_\infty}$. We denote the subgroup of $G$ that stabilizes the subspaces $H^{p,q}_\infty$ by $K$ as\footnote{We can equivalently view $V$ as the subgroup generated by operators which commute with $Q_\infty$, implying that $V$ is contained in $K$.}
\begin{equation}\label{Vdef}
V = \left\{ g \in G \, | \ g H^{p,q}_{\infty} = H^{p,q}_{\infty} \right\} = \begin{cases}
\mathrm{U}(1) \times \mathrm{U}(h^{2,1})\, , \quad &D=3\, , \\
\mathrm{U}(h^{3,1}) \times \mathrm{SO}(h^{2,2})  \, , \quad &D=4\, . \\
\end{cases}
\end{equation}
This brings us in the position to define $h(t,\bar{t})$ as
\begin{equation}\label{hdef}
H^{p,q}(t,\bar{t}) = h(t, \bar{t})\,  H^{p,q}_{\infty} \, .
\end{equation}
It assigns to each point in the moduli space an element in the coset space $G/V$,\footnote{To be more precise, when the moduli space is not simply connected -- there are singularities forbidding the contraction of loops to a point -- we have to take the monodromy group $\Gamma$ around these loci into account. This restricts $h(t,\bar{t})$ to the double coset $\Gamma \backslash G /V$.} which determines the periods of the $(p,q)$-forms through this relation. The stabilizer subgroup $V$ accounts for the local right multiplication of $h(t,\bar{t})$ that leaves the boundary Hodge structure $H^{p,q}_\infty$ in \eqref{hdef} invariant. We can use this period mapping to conveniently rewrite the Hodge star operator as
\begin{equation}\label{eq:Cperiodmapping}
C(t,\bar{t}) = h^{-1}(t, \bar{t}) \, C_\infty  \, h(t, \bar{t})\, .
\end{equation}
Note in particular that its dependence on the complex structure moduli is captured completely by the period mapping $h(t, \bar{t})$. In fact, the theorems underlying asymptotic Hodge theory are built around finding suitable approximations for this matrix-valued function $h(t,\bar{t})$ near the boundary.

\section{Approximation by the nilpotent orbit}
Let us now put the previous two sections on limits and Hodge structures together, and investigate how Hodge structures behave near boundaries in complex structure moduli space. This behavior is described by the nilpotent orbit theorem of Schmid \cite{Schmid}, which we introduce here. We will also use this opportunity to give a near-boundary example of the three equivalent formulations of Hodge structures introduced in the previous section.

\paragraph{Monodromies.} To prepare for this discussion, we first study what happens to the vector spaces $F^p$ when we circle around a boundary. Circling the boundary at $t^k = i \infty $ corresponds to shifting $t^k \to t^k+1$ and induces a monodromy transformation on elements of the Hodge filtration\footnote{The matrix action on a form is understood as an action on the vector of coefficients when expanded in a given three-form basis.}
\begin{align}\label{monodromy}
\omega^p \in F^p: \quad \omega^p(t^k + 1) = T_k\, \omega^p (t^k) \, ,
\end{align}
where the $T_k$ is the matrix generator of the monodromy that lies in the symmetry group $G$ of the bilinear pairing \eqref{eq:groups}. It follows from \cite{Landman} that these generators are unipotent,\footnote{In general, we could have quasi-unipotent matrices, i.e. $(T^q- \mathbb{I})^{m+1}=0$ for some integers $q,m$. However, we can always make them unipotent, i.e. setting $q=1$ by coordinates redefinitions of the form $z \to z^q$. } i.e. $(T-\mathds{1})^m=0$ for some positive integer $0 \leq m \leq  D$. From these generators we can define the so-called log-monodromy matrices as
\begin{align}
T_i = e^{N_i}: \qquad [N_i,N_j]=0\, , \quad N_i^T \eta = - \eta N_i \, , 
\end{align}
The first condition assures that monodromies around two boundaries $t^i=t^j=i\infty$ commute with each other, while the second condition guarantees that the log-monodromies are elements of the Lie algebra $N_i \in \mathfrak{g}$ of the bilinear pairing $\eta$. Given the unipotency of the monodromy generators $T_i$ one checks that the log-monodromy matrices are nilpotent of degree $0 \leq m \leq D$. 

\paragraph{Nilpotent orbit approximation of periods.} We are now in the position to state the nilpotent orbit theorem of \cite{Schmid}.\footnote{The name for this theorem comes from the period mapping $h(t,\bar{t})$ in \eqref{hdef}, which is approximated by an orbit of a nilpotent group in $G/V$.} Let us first write it out for the periods $\Pi$ of the holomorphic $(D,0)$-form before discussing the general behavior for the Hodge filtration $F^p$. It tells us that the period vector admits an expansion
\begin{equation}\label{eq:periodsexpansion}
\Pi(t, \zeta)  = e^{t^i N_i} \big( a_0(\zeta) + \sum_{r_i \geq 0} e^{2\pi i r_i t^i} a_{r_1 \cdots r_n}(\zeta)  \big)\, ,
\end{equation}
where $r_k > 0$ for at least some $k$, and the terms $a_{r_1\ldots r_n}$ are independent of the coordinates $t^1, \ldots, t^n$ taken close to the boundary, but in principle can depend on the spectator moduli $\zeta^k$. Note that the monodromy symmetry under $t^i \to t^i+1$ has now been made manifest by the exponential factor of log-monodromy matrices. Moreover, due to the order of the nilpotent log-monodromy matrices being bounded from above by $D$, we find that the leading term reduces to a degree $D$ polynomial in $t^i$ at most.

\paragraph{Essential exponential corrections.} We also want to point out that there is a clear separation in the periods between polynomial terms in $t^i$ arising from $a_0$ and exponentially suppressed terms associated with $a_{r_1\ldots r_n}$. Borrowing the nomenclature familiar from large complex structure and applying it to any boundary in complex structure moduli space, we loosely refer to $a_0$ as the perturbative term, while the other exponentially suppressed terms resemble the instanton expansion. However, let us already stress that near other boundaries in moduli space one can have \textit{essential exponential corrections} which cannot be dropped no matter how close one moves to the boundary. For instance, these corrections can be needed in certain physical couplings such as the  K\"ahler metric to obtain well-defined kinetic terms, in which case we refer to them as \textit{metric-essential}. From the perspective of the Hodge filtration these corrections are needed to reproduce the lower-lying elements $F^{D-1}, \ldots, F^{0}$ via \eqref{eq:Fpperiods}. Such behavior is rather surprising from the point of view of the large complex structure point, and we will therefore review some enlightening examples later in this chapter.

\paragraph{Nilpotent orbit approximation of Hodge filtrations.} Having discussed the asymptotic expansion of the $(D,0)$-form periods, let us now discuss the nilpotent orbit theorem for the Hodge filtration $F^p$. It follows that these can be approximated as
\begin{equation}\label{NilpOrbitGen}
F^p_{\rm nil}(t, \zeta) = e^{t^i N_i} F_0^p (\zeta)\, ,
\end{equation}
where the \textit{limiting filtration} $F^p_0$ is independent of $t^i$ and only varies in the spectator moduli $\zeta^k$. The key insight of Schmid was that the filtration $F^p_{\rm nil}$ still defines a polarized Hodge structure on $H$ provided that the saxions are sufficiently large $y^i \gg 0$. Moreover, \cite{Schmid} also shows that the distance between $F^p_{\rm nil}$ and $F^p$ with respect to the Hodge norm decays at an exponential rate in $y^i$, hence the form of the corrections in \eqref{eq:periodsexpansion}. 

\paragraph{Limiting filtration.} The vector spaces $F_0^p$ can be obtained by considering the Hodge filtration $F^p$ that lives in the bulk of the moduli space and taking the limit to the boundary. To be precise, we rotate out the divergent part involving the log-monodromy matrices such that
\begin{align}
F^p_0  = \lim_{t^i \to i \infty} e^{-t^i N_i} F^p(t)  \,. \label{eq:Limiting filtration}
\end{align}
In practice it can be quite involved to work out these limits explicitly -- they are taken at the level of vector spaces, so one has to be careful with overall factors that are exponentially small when working with the periods. From the horizontality property \eqref{eq:Transversality} one can easily show that the nilpotent matrices $N_i$ have to act on the limiting filtration as
\begin{equation}\label{eq:NFp0}
N_i F_0^p \subseteq F_0^{p-1}\, .
\end{equation}
Previously we saw that taking derivatives can only take us down by one degree at a time, and now it follows that this applies to the log-monodromy matrices as well.

\paragraph{Hodge star.} We can use the Hodge filtration $F^p_{\rm nil}$ to obtain a nilpotent orbit approximation for the Hodge star operator. Let us first write down the Hodge structure in this approximation as
\begin{equation}\label{eq:Hnil}
H^{p,q}_{\rm nil} = F^p_{\rm nil} \cap \overline{F_{\rm nil}^q} \, , \quad p+q=D\, .
\end{equation}
The polynomial structure of the vector spaces $F^p_{\rm nil}$ in $x^i, y^i$ carries over to the Hodge structure $H^{p,q}_{\rm nil}$. Consequently the corresponding Hodge star $C_{\rm nil}$ is made up such polynomial functions as well, while all exponential corrections are dropped. Moreover, the filtrations $F^p_{\rm nil}$ and $\overline{F}_{\rm nil}^q$ share a common axion-dependent factor $e^{x^k N_k}$ as can be seen from \eqref{NilpOrbitGen}, which we can rotate out of $C_{\rm nil}$ as   
\begin{equation}\label{eq:Cnilx}
C_{\rm nil}(x,y) = e^{x^k N_k} \hat{C}_{\rm nil}(y) e^{-x^k N_k}\, ,
\end{equation}
where clearly $C_{\rm nil}(0,y) = \hat{C}_{\rm nil}(y)$.

\subsection*{An example}
To illustrate this discussion on the nilpotent orbit approximation, let us consider a one-modulus example in explicit detail: the conifold point for a Calabi-Yau threefold. We begin with the periods of the $(3,0)$-form, which are given by (see appendix \ref{app:embedding} for the relation to the prepotential formulation)
\begin{equation}
\label{eq:periods_conifold}
\Pi = \big(
1, \ b\, e^{2\pi i t} , \ b \, e^{2\pi i t}( t  - \frac{1}{2\pi i}), \ i +\frac{b^2}{4\pi i}e^{4\pi it}\big) ,
\end{equation}
where $b \in \mathbb{C}$ is some model-dependent coefficient. 

\paragraph{Hodge filtration.} Let us now show how to derive the Hodge filtration $F^p$ from these periods according to \eqref{eq:Fpperiods}. By taking derivatives of this period vector we can generate a basis for the spaces $F^p$ as
\begin{equation}\label{eq:conifold_derivatives}
\begin{aligned}
\frac{1}{2\pi i } \partial_t \Pi &= b \, e^{2\pi i t}  \big(
 0 \, ,\  1\, , \  t\, , \  \frac{b}{2\pi i} e^{2\pi it} \big)\, , \\
\frac{1}{2\pi i} e^{2\pi i t}  \partial_t  (e^{-2\pi i t} \partial_t \Pi) &= b \, e^{2\pi i t}  \big(
 0 \, ,\  0\, , \  1\, , \  b \, e^{2\pi it}
\big)\, , \\
\frac{1}{(2\pi i)^2} e^{4\pi i t} \partial^2_t  (e^{-2\pi i t} \partial_t \Pi) &= b \, e^{4\pi i t}  \big(
 0 \, ,\  0 \, ,\  0 \, ,\  1
\big)\, ,
\end{aligned}
\end{equation}
where we made some clever choices of differential operators involving exponential factors to simplify the resulting expressions.\footnote{To highlight the technicalities with exponential factors in extracting the spaces $F^p_0$, we chose to reinstate those overall factors afterwards.} The vector spaces $F^p$ are spanned by the first $3-p$ derivatives of $\Pi$, e.g.~$F^2$ is spanned by $\Pi$ and $\partial_t \Pi$ above. 

\paragraph{Nilpotent orbit approximation.} In turn, we want to take the limit $t \to i \infty $ as prescribed by \eqref{eq:Limiting filtration} in order to obtain the nilpotent orbit approximation. Taking this limit directly for the elements in \eqref{eq:conifold_derivatives} does not work due to the exponentially suppressing factors, but we can rescale them since we are working with vector spaces. Collecting the vectors that span the $F^p_0$ in a period matrix we write
\begin{equation}\label{eq:periodmatrix_conifold}
\Pi_{\rm nil} = e^{tN^-} \scalebox{0.85}{$\begin{pmatrix}
1 & 0 & 0 & 0\\
0 & 1 & 0 & 0\\
0 & 0  & 1 & 0 \\
i & 0 & 0 & 1\\
\end{pmatrix}$} \, ,
\end{equation}
where the first $k$ columns span the vector space $F^{3-k}_{\rm pol}$. The vectors spanning the spaces $F^{3-k}_0$ can be read off as the first $k$ columns in the matrix on the right-hand side. Note that while the period vector in \eqref{eq:periods_conifold} contained exponential factors, this data has been mapped into a purely polynomial form by considering the nilpotent orbit formulation in \eqref{eq:periodmatrix_conifold}. Moreover, exponentially suppressed terms in the periods of the $(3,0)$-form were essential in deriving the vectors spanning the other $F^p_0$ with $p<3$. This is precisely the necessity we foreshadowed earlier: exponential corrections are needed in the $(3,0)$-form periods in order to obtain all vector spaces $F^p_0$ of the Hodge filtration.\footnote{By a slight reformulation of the nilpotent orbit in chapter \ref{chap:models} we will make this notion more precise as a rank condition.}

\paragraph{Hodge decomposition and Hodge star.} For the example at hand the Hodge decomposition associated to \eqref{eq:periodmatrix_conifold} can be computed via \eqref{eq:Hnil}. The resulting vector subspaces are spanned by
\begin{equation}\label{eq:conifold_periods}
\begin{aligned}
H^{3,0}_{\rm nil} &: \ \big( 1 , \ 0 , \ 0 , \ i \big), \quad &H^{2,1}_{\rm nil} &: \ \big( 0 , \ 1 , \ t , \ 0 \big), \\
\end{aligned}
\end{equation}
with the other two fixed by complex conjugation. The corresponding Hodge star is given by
\begin{equation}\label{eq:Cnilconifold}
C_{\rm nil} = \begin{pmatrix}
 0 & 0 & 0 & -1 \\
 0 & x & y & 0 \\
 0 & -\frac{x^2}{y}-\frac{1}{y} & -x & 0 \\
 1 & 0 & 0 & 0 
\end{pmatrix}\, .
\end{equation}
Note that this Hodge star operator could have equivalently followed from the period mapping in \eqref{hdef}. Below we give this description, where we primarily give the form of the various boundary and bulk objects, and refer to \cite{Grimm:2021ikg} for their explicit computation. 

\paragraph{Boundary data.} The boundary Hodge structure for the conifold point of the Calabi-Yau threefold takes the form
\begin{equation}
\begin{aligned}
H^{3,0}_{\infty} &: \ \big( 1 , \ 0 , \ 0 , \ i \big), \quad &H^{2,1}_{\infty} &: \ \big( 0 , \ 1 , \ i, \ 0 \big), \\
\end{aligned}
\end{equation}
with again the other two fixed by complex conjugation. The corresponding charge operator \eqref{eq:Qdef} and boundary Hodge star are
\begin{equation}
Q_\infty = \scalebox{0.85}{$\left(
\begin{array}{cccc}
 0 & 0 & 0 & -\frac{3 i}{2} \\
 0 & 0 & -\frac{i}{2} & 0 \\
 0 & \frac{i}{2} & 0 & 0 \\
 \frac{3 i}{2} & 0 & 0 & 0 \\
\end{array}
\right)$}, \qquad C_\infty = e^{\pi i Q_\infty} = \scalebox{0.85}{$\left(
\begin{array}{cccc}
 0 & 0 & 0 & -1 \\
 0 & 0 & 1 & 0 \\
 0 & -1 & 0 & 0 \\
 1 & 0 & 0 & 0 \\
\end{array}
\right)$}\, .
\end{equation}
\paragraph{Period mapping.} Now we can finally lift this boundary Hodge structure to the Hodge structure of the nilpotent orbit living in the asymptotic regime near the conifold point. The appropriate period mapping is given by
\begin{equation}
h(x,y) = \scalebox{0.85}{$\begin{pmatrix}
1 & 0 & 0 & 0\\
0 & \frac{1}{\sqrt{y}} & 0 & 0\\
0 & \frac{x}{\sqrt{y}} &  \sqrt{y} & 0\\
0 & 0 & 0 & 1
\end{pmatrix}$}\, ,
\end{equation}
which indeed reproduces the Hodge star \eqref{eq:Cnilconifold} via the relation \eqref{eq:Cperiodmapping}.

\section{Physical couplings in asymptotic regimes}
To illustrate the above discussion on asymptotic regimes and nilpotent orbits, let us use this technology to describe couplings arising in string compactifications. We will both use the nilpotent orbit approximation to give their general asymptotic behavior, as well as consider some explicit examples to show how these couplings behave in practice. We will be especially careful with our treatment of exponentially suppressed terms appearing in the periods \eqref{eq:Limiting filtration} and, in particular, demonstrate how essential corrections can become of importance from a physical point of view.

\paragraph{K\"ahler potential.} Let us begin with K\"ahler potential. By inserting the period expansion \eqref{eq:periodsexpansion} back into \eqref{eq:kahlercsperiods}, we can split the contributions into two parts
\beq \label{eq:kahlerexpansion}
K^{\rm cs} = - \log[\cK_{\rm pol} + \cK_{\rm inst} ] \, , 
\eeq
with
\beq
\begin{aligned}
\cK_{\rm pol} &=i \langle \bar a_0 , e^{2i y^i N_i} a_0 \rangle  \, , \nn \\
 \cK_{\rm inst} &=  \sum_{r_i, s_i \geq 0}  e^{-2\pi [  y^i (s_i +r_i)+i x^i( s_i - r_i)]}i \langle \bar  a_{r_1 \ldots r_n}  , e^{2i y^i N_i} a_{s_1 \ldots s_n}  \rangle  \, , \nn 
\end{aligned}
\eeq
where the sum runs over integers with either $r_i \neq 0$ or $s_i \neq 0$ for some index $i$. This partition separates the polynomial terms from the exponentially suppressed terms. Note in particular that the axions only appear in $\cK_{\rm inst}$ through sines and cosines.

\paragraph{Infinite distance example.} Let us momentarily consider an example boundary with a non-trivial $K_{\rm pol}$ and ignore exponential corrections. We take a simple one-modulus K\"ahler potential
\begin{equation}
K^{\rm cs}_{\rm pol} = -\log y^d\, , \qquad K_{t \bar{t}} = \frac{d}{y^2}\, ,
\end{equation}
where $d=1,\ldots, D$ is some integer that depends on the choice of boundary. For the distance in the metric we then find that
\begin{equation}
d(y_f, y_i) = \int_{y_i}^{y_f} \sqrt{K_{yy}} \,  \dd y = \int_{y_i}^{y_f} \frac{\sqrt{d}}{y} \,  \dd y = \sqrt{d} \log [y_f/y_i]\, ,
\end{equation}
where $y_i, y_f$ denote the starting point and ending point of the path respectively. Clearly this distance diverges as we move the endpoint to the boundary $y_f \to \infty$.\footnote{It has been proven in \cite{wang1} that one-modulus boundaries with $N a_0 \neq 0$ are at infinite distance.} 

\paragraph{Finite distance example.} To contrast, let us also consider a one-modulus point at finite distance. The conifold point we discussed above is an example of such a boundary, and by computing its K\"ahler potential from the periods \eqref{eq:periods_conifold} we find
\begin{equation}
K = -\log[ 2 -2 a^2\, y \,e^{-4\pi y} ] \, , \qquad \cK_{\rm pol}=2\, , \quad \cK_{\rm inst} = -2 a^2\, y \,e^{-4\pi y}\, .
\end{equation}
Here the $\cK_{\rm pol}$ is simply a constant, so the exponential corrections $\cK_{\rm inst}$ are essential to obtain a non-degenerate K\"ahler metric
\begin{equation}
K_{t \bar{t}} = 2 \pi a^2 e^{-4 \pi y} (2\pi y -1) + \cO(e^{-8\pi y})\, .
\end{equation}
Thus we see that the exponential corrections near the conifold point are not just essential as commented below \eqref{eq:periodmatrix_conifold}, but even metric-essential. By inspection of this metric one can straightforwardly show that the limit $y\to \infty$ is at finite distance for a conifold boundary.\footnote{The integral for this distance can be bounded as
\begin{equation}
\int^\infty \sqrt{K_{yy}} \, \dd y \sim  \int^\infty e^{-2\pi y} \sqrt{y} <  \int^\infty y^{-p}< \infty \, ,
\end{equation}
where we used that exponential decay is faster than any power law behavior, and by taking $p>1$ the last integral is finite. } 

From these examples we learn that infinite distance boundaries have a non-trivial $\cK_{\rm pol}$, while for finite distance boundaries the dependence on the saxions enters only through $\cK_{\rm inst}$. For some infinite distance boundaries such as large complex structure one finds that $\cK_{\rm inst}$ can be dropped consistently, but away from this lamppost that is not necessarily the case. The perturbative metric resulting from the polynomial part $\cK_{\rm pol} $ can be degenerate near certain boundaries, i.e.~it has a vanishing eigenvalue, resulting in ill-defined kinetic terms. 

\paragraph{Two-moduli example. } To demonstrate the above point, let us consider a two-moduli infinite distance boundary. We borrow the K\"ahler potential from \eqref{eq:II1kp}, which will be studied in more detail later. Its polynomial part reads
\begin{equation}\label{eq:example_potential}
K^{\rm cs}_{\rm pol} = - \log \cK_{\rm pol} = -\log[y_1+n_2 y_2]\, ,
\end{equation}
where $n_2 \geq 0$ is some model-dependent integer. Picking an example with $n_2>0$ leads to a boundary where $y_1 \to \infty$ or $y_2 \to \infty$ (or any combination) produces an infinite distance limit. However, by a holomorphic change of variables $(t_1',t_2')= (t_1+n_2 t_2, n_2 t_1- t_2)$ one easily checks that the dependence on $y_2'$ drops out. We can equivalently see this degeneracy by explicitly computing the K\"ahler metric
\begin{equation}\label{eq:example_metric}
K_{i\bar{j}} = \frac{1}{(y_1+n_2 y_2)^2} \begin{pmatrix} 1 & n_2 \\
n_2 & (n_2)^2
\end{pmatrix}\, .
\end{equation}
Its determinant vanishes, so taking just \eqref{eq:example_potential} as K\"ahler potential leads to ill-defined kinetic terms for the complex structure moduli in this asymptotic regime. To be more precise, the eigenvector $(1,n_2)$ has a polynomial eigenvalue, while $(n_2, -1)$ has a vanishing eigenvalue. By requiring the presence of exponential corrections to \eqref{eq:example_potential} we can cure this degeneracy. The relevant part of the exponential corrections takes the form
\begin{equation}\label{eq:examplecorrection}
\cK_{\rm inst} = -2 a^2 e^{-4\pi y_2}(n_1 y_1 + y_2)
\end{equation}
where $a \in \cR$ and an integer $n_1 \geq 0$ are some model-dependent coefficients. By including these terms one can check that the eigenvalue for $(n_2, -1)$ takes an exponentially small value proportional to $e^{-4\pi y_2}$ instead, so these metric-essential exponential corrections indeed cure the degenerate K\"ahler metric.

One important sidenote we should make is that not every essential instanton is also metric-essential. In the above example we can verify this explicitly: there are further essential corrections in \eqref{eq:II1kp} for the K\"ahler potential. We can understand this difference by following where derivatives of the $(3,0)$-form end up in the cohomology. For the K\"ahler metric we took only a single holomorphic derivative, so the necessary corrections are those needed to span $H^{2,1}$. In contrast, in order to span the middle cohomology $H$ via \eqref{eq:Fpperiods} we also look at higher-order derivatives, which might require further corrections to span the other spaces $H^{1,2}, H^{0,3}$. Therefore the mathematical notion of completeness that requires the derivatives of $\Omega$ to span the middle cohomology leads to a larger set of essential instantons than just those required by the K\"ahler metric.

\paragraph{Flux superpotential.} Having investigated the K\"ahler metric in detail, we next turn to the flux superpotential \eqref{eq:superpotential}. Similar to the K\"ahler potential we can use the expansion \eqref{eq:periodsexpansion} for the periods, and separate the terms in the superpotential into two parts
\begin{equation}\label{eq:Wexpansion}
\begin{aligned}
W &= W_{\rm pol} + W_{\rm inst} \, , \\
W_{\rm pol} &= \langle G_3\, , \ e^{t^i N_i} a_0 \rangle\, , \qquad  &W_{\rm inst} &=\sum_{r_i } e^{2\pi i r_i t^i}  \langle G_3\, ,  \  e^{t^i N_i} a_{r_1 \cdots r_n}  \rangle  \, ,
\end{aligned}
\end{equation}
where the sum over $r_i$ runs over at least one $r_i \neq 0$. Similar to the expansion of the K\"ahler potential \eqref{eq:kahlerexpansion}, essential instantons in the periods can play an important role for the superpotential, meaning one cannot drop all terms in $W_{\rm inst}$ near every boundary. For instance, one can find that some flux quanta only enter the superpotential through $W_{\rm inst}$, as is the case for the two-moduli example above we borrowed from section \ref{ssec:II1model}. 

\paragraph{Scalar potential.} We now want to study the behavior of the scalar potential \eqref{eq:potential}. Formulating it via the Hodge star tells us that in the asymptotic regime we can approximate it by algebraic functions in the moduli
\begin{equation}
 V =  \frac{1}{4 \cV^2 s} \big( \langle \bar{G}_3 , C_{\rm nil}(x,y) \, G_3 \rangle - i \langle \bar{G}_3, G_3 \rangle \big)\, .
\end{equation}
where $s$ denotes the dilaton of Type IIB and $\cV$ the K\"ahler moduli-dependent volume of the Calabi-Yau threefold. The algebraic dependence of $V$ on the complex structure moduli $t^i$ follows directly from how the Hodge star operator $C_{\rm nil}$ of the nilpotent orbit varies in the moduli as described below \eqref{eq:Hnil}. Moreover, we can rotate out the axion-dependence via \eqref{eq:Cnilx} as
\begin{equation}
V = \frac{1}{4 \cV^2 s} \big( \langle \bar{\rho}(x) , \hat{C}_{\rm nil}(y) \, \rho(x) \rangle - i \langle \bar{G}_3, G_3 \rangle \big)\, ,
\end{equation}
where we defined flux axion polynomials\footnote{Such redefinitions have also been used extensively in \cite{Bielleman:2015ina,Herraez:2018vae,Grimm:2019ixq,Marchesano:2019hfb,Grimm:2020ouv,Grimm:2021ckh} in studying moduli stabilization.}
\begin{equation}
\rho(x) = e^{-x^k N_k} G_3\, .
\end{equation}
Note that $\rho$ is invariant under monodromy transformations $x^i \to x^i +1$ when we shift the fluxes according to $G_3 \to e^{N_i} G_3$. Let us also mention that by going to F-theory one could straightforwardly include the axion $c$ into a complex structure modulus of the Calabi-Yau fourfold, while its counterpart $s$ would be part of its $C_{\rm nil}$. Finally, we want to remark that $C_{\rm nil}$ as a matrix is already non-degenerate when taking just its polynomial part, so in contrast to the K\"ahler metric we could safely drop exponential corrections to the scalar potential without losing essential information. 

\paragraph{Two-moduli example.} To see how this last observation fit with a flux superpotential that contains essential instantons, let us return to our two-moduli example. Turning on a flux for a metric-essential instanton in \eqref{eq:superpotentialII1} we find as superpotential
\begin{equation}
W = W_{\rm inst} = a \, e^{2\pi i t_2}\, .
\end{equation}
The F-term along the eigenvector with an exponential eigenvalue reads
\begin{equation}
n_2 D_{t_1} W_{\rm inst} - D_{t_2} W_{\rm inst} = 2\pi i \, a \, e^{2\pi i t_2}\, .
\end{equation}
By plugging this F-term into \eqref{eq:potential} we find that the exponential scaling of the inverse K\"ahler metric cancels off the scaling of the F-terms as
\begin{equation}
V = \frac{1}{n_1 y_1 + y_2} + \cO(e^{-2\pi y})\, .
\end{equation}
The lesson we learn is that metric-essential instantons have to be included in the superpotential as well:  these exponential terms in the superpotential produce polynomial order terms in the scalar potential via cancellations with the inverse K\"ahler metric, so they \textit{cannot} be dropped at the leading perturbative level. We refer to section \ref{ssec:strategy} for a discussion in the general setting.

\chapter{Mixed Hodge structures and boundary classifications}\label{chap:mhs}
In this chapter we discuss the formal structures underlying the asymptotic regimes of the nilpotent orbit approximation. To be precise, by using the nilpotent orbit data -- the log-monodromy matrices $N_i$ and the limiting filtration $F^p_0$ -- we can construct a finer splitting of the middle cohomology compared the standard Hodge decomposition. We explain how to determine this so-called Deligne splitting, and illustrate how it can be used as a classification device for boundaries in the complex structure moduli space of Calabi-Yau threefolds. We also describe how singularities can enhance when additional moduli are send to a limit by using this formalism.

\section{Mixed Hodge structures: Deligne splittings}
So far we have focused on \textit{pure} Hodge structures: the Hodge filtration $F^p$ gave rise to a decomposition of the middle cohomology via intersections of the form $H^{p,q} = F^p \cap \bar{F}^q$ with $p+q=D$. A natural question to ask is then whether a similar structure underlies the limiting filtration $F^p_0$ -- and indeed, the relevant 
framework in this case is that of a \textit{mixed} Hodge structure. 
Concretely, this means we consider another splitting of the middle cohomology $H$ known as the Deligne splitting -- a set of vector spaces $I^{p,q}$ with $0 \leq p,q \leq D$. In this section we introduce this finer splitting that characterizes boundaries in moduli space.

\paragraph{Monodromy weight filtration.} The Deligne splitting requires us to introduce another set of vector spaces based on the log-monodromy matrices $N_i$. 
For definiteness, say we are interested in a limit involving the first $k$ saxions $y^1,\ldots, y^k \to \infty$. The so-called \textit{monodromy weight filtration} for $N$ is then given by\footnote{In fact, one can consider any element $N=c_1 N_1 + \ldots +c_k N_k$ in the linear cone $c_i>0$. The resulting vector spaces $W_\ell(N)$ are independent of the choice of $c_i$.}
\begin{equation}\label{Wfiltr}
N =N_{(k)} = N_1 + \ldots + N_k: \ \ \  W_{\ell}(N)= \!\!\!\! \sum_{j \geq \max(-1,\ell-D)} \ker N^{j+1} \cap \img N^{j-\ell+D}\, .
\end{equation}
The idea of this monodromy weight filtration is to gather elements of $H$ based on the action of the nilpotent matrix $N$. To be more precise, one can check that the log-monodromy matrix $N$ acts on this filtration by moving down two steps as\footnote{This follows by using that $N \ker N^{j+1} \subseteq \ker N^j$ and $N \img N^{j-\ell+D} = \img N^{j-\ell+D+1}$ -- the power of $N$ is decreased by one in the kernel and increased by one in the image. By relabeling terms $j \to j+1$ to reinstate the kernel as $\ker N^{j+1}$ this can be understood as a shift $\ell \to \ell-2$.}
\begin{equation}\label{eq:NW}
N W_\ell \subseteq W_{\ell-2}\, ,
\end{equation}
which in fact applies to any $N_i$ ($1 \leq i \leq k$). It can sometimes also be helpful to introduce the vector space quotients $Gr_{\ell}=W_\ell/W_{\ell-1}$: the nilpotent matrix $N$ then defines an isomorphism
\begin{equation}\label{eq:NGr}
N^{\ell} : \ Gr_{D+\ell} \to Gr_{D-\ell}\, ,
\end{equation}
assuming $\ell \geq 0$. Intuitively, it gives us an identification between the upper part and the lower part of the weight filtration. To pre-empt our discussion of the sl(2)-approximation in chapter \ref{sec:sl2splitting}, it is instructive to compare these vector spaces with the weight decomposition \eqref{wd_001} under an sl$(2,\mathbb{R})$-triple. We can then write out $W_\ell$ by collecting all sl(2)-eigenspaces $V_m$ up to weight $\ell-D$ as
\begin{equation}
W_\ell = \bigoplus_m^{\ell-D} V_m\, .
\end{equation}
The relation \eqref{eq:NW} is then explained by identifying $N$ as the lowering operator of the sl$(2,\mathbb{R})$-triple, which lowers the weight by two. Meanwhile, for the graded subspaces $Gr_{\ell}$ we can simply take the sl(2)-eigenspaces $V_{\ell-D}$ as representatives, and \eqref{eq:NGr} then just relates $V_\ell$ and $V_{-\ell}$ under the action of $N^\ell$.

\paragraph{Deligne splitting.} The Deligne splitting $I^{p,q}$ is constructed out of the monodromy weight filtration $W_\ell$ and limiting filtration $F^p_0$ as the following intersection of vector spaces\footnote{\label{partial_limit}Let us stress that $e^{t^{k+1}N_{k+1} + \ldots + t^n N_n} F^p_0$ should be considered rather than $F^p_0$ for the Deligne splitting associated to $N_{(k)}$: the coordinates $y_{k+1}, \ldots, y_n$ have then not yet been taken to the boundary and are therefore technically spectator moduli. For ease of notation we assume that all $n$ moduli are send to the limit for the remainder of this section. In later sections we will explicitly indicate the coordinates taken to the boundary by writing $F^p_{(k)}$ instead of $F^p_0$.}  
\begin{equation}\label{Ipq}
I^{p,q} = F_{0}^{p} \cap W_{p+q} \cap \bigg( \bar{F}_{0}^{q}\cap W_{p+q}+\sum_{j\geq 1} \bar{F}_{0}^{q-j}\cap W_{p+q-j-1} \bigg)\, ,
\end{equation}
where we have $p,q=0,\ldots,D$. These vector spaces 
can be arranged into a diagram, which we have shown in figure~\ref{fig_deligne} for a Calabi-Yau threefold. The limiting filtration $F^p_0$ as well as the monodromy weight filtration $W_\ell$ can be recovered from the Deligne splitting 
via the relations
\begin{equation}\label{eq:ItoFW}
F^p_0 = \bigoplus_{r \geq p} \bigoplus_s I^{r,s}\, , \hspace{50pt} 
W_\ell = \bigoplus_{p+q = \ell} I^{p,q}\,.
\end{equation}
The spaces $W_\ell$ then have the straightforward interpretation of the sum of the first $\ell$ horizontal rows in figure~\ref{fig_deligne}, while the $F^p_0$ corresponds to the first $D-p+1$ diagonal columns when counting from the top-left.
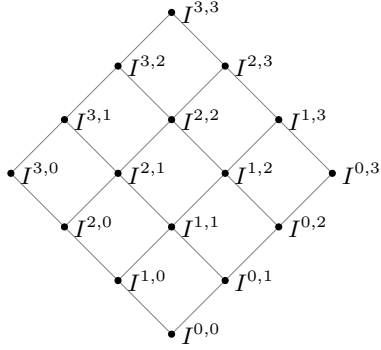
\begin{figure}[h!]
\centering
\begin{tikzpicture}[baseline={([yshift=-.5ex]current bounding box.center)},scale=1,cm={cos(45),sin(45),-sin(45),cos(45),(15,0)}]
  \draw[step = 1, gray, ultra thin] (0, 0) grid (3, 3);
  \draw[fill] (3, 0) circle[radius=0.04] node[right]{\small $I^{0,3}$};
  \draw[fill] (2, 0) circle[radius=0.04] node[right]{\small $I^{0,2}$};
  \draw[fill] (1, 0) circle[radius=0.04] node[right]{\small $I^{0,1}$};
  \draw[fill] (0, 0) circle[radius=0.04] node[right]{\small $I^{0,0}$};
  \draw[fill] (3, 1) circle[radius=0.04] node[right]{\small $I^{1,3}$};
  \draw[fill] (2, 1) circle[radius=0.04] node[right]{\small $I^{1,2}$};
  \draw[fill] (1, 1) circle[radius=0.04] node[right]{\small $I^{1,1}$};
  \draw[fill] (0, 1) circle[radius=0.04] node[right]{\small $I^{1,0}$};
  \draw[fill] (3, 2) circle[radius=0.04] node[right]{\small $I^{2,3}$};
  \draw[fill] (2, 2) circle[radius=0.04] node[right]{\small $I^{2,2}$};
  \draw[fill] (1, 2) circle[radius=0.04] node[right]{\small $I^{2,1}$};
  \draw[fill] (0, 2) circle[radius=0.04] node[right]{\small $I^{2,0}$};
  \draw[fill] (3, 3) circle[radius=0.04] node[right]{\small $I^{3,3}$};
  \draw[fill] (2, 3) circle[radius=0.04] node[right]{\small $I^{3,2}$};
  \draw[fill] (1, 3) circle[radius=0.04] node[right]{\small $I^{3,1}$};
  \draw[fill] (0, 3) circle[radius=0.04] node[right]{\small $I^{3,0}$};
\end{tikzpicture}
\caption{Arrangement of the Deligne splitting $I^{p,q}$ for Calabi-Yau threefolds.
\label{fig_deligne}}
\end{figure}
\subsubsection*{An example}
The expression \eqref{Ipq} for the Deligne splitting is rather involved, so let us pause our discussion for a moment and look at a simple example with $D=1$. Geometrically we can then think of this Deligne splitting as characterizing the degeneration of a two-torus $T^2$. As nilpotent matrix we take
\begin{equation}
N = \scalebox{0.85}{$\begin{pmatrix}
0 & 0 \\
1 & 0 
\end{pmatrix}$}\, ,
\end{equation}
while for the limiting filtration we consider
\begin{equation}
F^1_0 = \mathrm{span}\bigl[  (1,0)\bigr]\, , \quad F^0_0 = \mathrm{span}\bigl[  (1,0)\, , (0,1)\bigr]\, .
\end{equation}
We can straightforwardly compute the monodromy weight filtration \eqref{Wfiltr} from the kernels and images of $N$ to be
\begin{equation}
\begin{aligned}
W_2 &= H = \mathrm{span}\bigl[  (1,0)\, , (0,1)\bigr]\, , \\
W_1 &=  \ker N + \img N =  \mathrm{span}\bigl[  (0,1)\bigr]\, , \\
W_0 &= \ker N \cap \img N  = \mathrm{span}\bigl[  (0,1)\bigr]\, .\\
\end{aligned}
\end{equation}
Here it is instructive to verify the two properties of the weight filtration described below \eqref{Wfiltr}. As a first check we easily see that $N W_2 = W_0$. On the other hand, for the vector space quotients we find that $Gr_2=W_2/W_1$ is spanned by the equivalence class of $(1,0)$, while $Gr_0=W_0/W_{-1}$ is spanned by $(0,1)$ -- and indeed the nilpotent matrix $N$ is an isomorphism between these two.

Let us now return to the task at hand and compute the Deligne splitting according to \eqref{Ipq} as
\begin{equation}
I^{1,1} = F^1_0 \cap \bar{F}^1_0 \cap W_2 = \mathrm{span}\bigl[  (1,0)\, \bigr]\, , \quad I^{0,0} = F^0_0 \cap \bar{F}^0_0 \cap W_0 = \mathrm{span}\bigl[  (0,1)\, \bigr]\, ,
\end{equation}
while the vector spaces $I^{1,0} = \overline{I^{0,1}}$ are empty since $F^1_0 \cap W_1=0$. We can collect our findings nicely as
\begin{equation}\label{Ipq_example}
I^{p,q} = \begin{tikzpicture}[baseline={([yshift=-.5ex]current bounding box.center)},scale=1,cm={cos(45),sin(45),-sin(45),cos(45),(15,0)}]
  \draw[step = 1, gray, ultra thin] (0, 0) grid (1, 1);
  \draw[fill] (1, 1) circle[radius=0.05] node[right]{\small $(1,0)$ };
  \draw[fill] (0, 0) circle[radius=0.05] node[right]{\small $(0,1)$};
\end{tikzpicture}\, .
\end{equation}

\subsubsection*{Symmetries of the Deligne splitting}
After this brief detour, let us return to the general setting and try to make some further helpful remarks regarding the Deligne splitting. A first question is to wonder how these subspaces $I^{p,q}$ are related to each other under complex conjugation. For a pure Hodge structure we found that we simply had to interchange indices as $\overline{H^{p,q}} = H^{q,p}$, but the form of the Deligne splitting \eqref{Ipq} already hints that this is slightly more subtle for mixed Hodge structures. More precisely, from \eqref{Ipq} one can argue that $I^{p,q}$ and $I^{q,p}$ are related to each other under complex conjugation modulo lower-positioned elements as
\begin{equation}\label{Ipqbar}
\bar{I}^{p,q} = I^{q,p} \mod \bigoplus_{r<q,s<p} I^{r,s}\, .
\end{equation}
The case where the vector spaces are related by a simple interchange of indices as $\bar{I}^{p,q} = I^{q,p}$ is referred to as $\mathbb{R}$-split. In fact, there always exists a complex rotation of the limiting filtration $F^p_0$ to an $\mathbb{R}$-split, due to Deligne. Moreover, there is a further rotation to a special $\mathbb{R}$-split known as the sl(2)-split. Both these rotations play an important role in the sl(2)-approximation, and we elaborate on their construction in section \ref{sec_tech_details}. In the $\mathbb{R}$-split case \eqref{Ipq} reduces to the much simpler form
\begin{equation}\label{Ipq_Rsplit}
I^{p,q}_{\text{$\mathbb{R}$-split}}=   F_{0}^{p} \cap \bar{F}_{0}^{q} \cap W_{p+q} \, .
\end{equation}
In a similar spirit, a simple check we can carry out is to see how pure Hodge structures \eqref{eq:Hpqdef} arise as special cases of mixed Hodge structures. We recover a pure Hodge structure when the log-monodromy matrix is trivial $N=0$: the monodromy around the boundary has no unipotent part, only a finite order semi-simple piece such as e.g.~the Landau-Ginzburg point of the quintic. For this case the monodromy weight filtration \eqref{Wfiltr} is given by $W_\ell = 0$ for $\ell < D$ and $W_\ell = H$ for $\ell \geq D$. In turn, the Deligne splitting reduces to $I^{p,q} = F_{0}^{p} \cap \bar{F}_{0}^{q}$ for $p+q=D$ while all others are empty -- we indeed find the splitting of a pure Hodge structure.

Another feature to wonder about is whether the dimensions $i^{p, q} = \dim_\bbC I^{p, q}$ of the Deligne splitting admit any symmetries or other constraints. For instance, from the Hodge diamond of a Calabi-Yau manifold (see figure \ref{fig_deligne}) we are used to identifications under certain reflections. One such symmetry we have already observed: following \eqref{Ipqbar} the dimensions admit a symmetry $i^{p,q}=i^{q,p}$, even though the spaces themselves are only related modulo lower-positioned elements. The full symmetry pattern of the Deligne splitting is given by
\begin{equation}\label{eq:symmetries}
i^{p,q}=i^{q,p} = i^{D-q,D-p}\, ,
\end{equation}
where the horizontal reflection follows from the fact that $N^{p+q-D}$ is an isomorphism between $I^{p,q}$ and $I^{D-q,D-q}$ (assuming $p+q > D$), as follows from \eqref{eq:NGr}. These symmetries are complemented by the inequalities
\begin{equation}
i^{p-1,q-1} \leq i^{p,q} \, , \qquad p+q \leq D\, ,
\end{equation}
which follows from the fact that $N I^{p,q} \subseteq I^{p-1, q-1}$.\footnote{This can be seen from the definition of the Deligne splitting \eqref{Ipq} by using that $NF^p_0 \subseteq F^{p-1}_0$, $N\bar{F}^q_0 \subseteq \bar{F}^{q-1}_0$ and $NW_{\ell}\subseteq W_{\ell-2}$, which follow from \eqref{eq:NFp0} and \eqref{eq:NW}.} Finally, it is also insightful to point out that the dimensions $i^{p,q}$ of the Deligne splitting can be related to the Hodge numbers $h^{p,q} = \dim_\mathbb{C} H^{p,q}$ of the underlying pure Hodge structure by
\begin{equation}\label{eq:hformula}
h^{p,D-p} = \sum_{q=0}^D i^{p,q}\, .
\end{equation}
One can see this from the relation between the limiting filtration $F^p_0$ -- whose dimensions are determined by the $h^{p,q}$ -- and the Deligne splitting in \eqref{eq:ItoFW}.

\subsubsection*{Operator decompositions}
It will also be convenient to introduce decompositions for operators based on their action on the Deligne splitting $I^{p,q}$, i.e.~raising or lowering the indices $p,q$. Prime examples of such operators are the log-monodromy matrices. Namely, recall that
\begin{equation}\label{eq:NIpq}
N_i I^{p,q} \subseteq I^{p-1, q-1}\, .
\end{equation} 
In other words, application of log-monodromy matrices lowers elements by two rows in the Hodge-Deligne diamond in figure \eqref{fig_deligne}. This motivates us to set up a more general notation for operator decompositions as
\begin{equation}\label{eq:lambdapq}
\cO_{p,q} \in \Lambda_{p,q} : \qquad  \cO_{p,q} I^{r,s} \subseteq I^{r+p,s+q}\, .
\end{equation}
where $\Lambda_{p,q}$ denotes the space of operators that increase the indices of an element by $(p,q)$. The $N_i$ we can then refer to as $(-1,-1)$-maps with respect to the Deligne splitting, located in the operator subspace 
\begin{equation}\label{eq:Nmin1min1}
N_i \in \Lambda_{-1,-1}\, .
\end{equation}
It is now interesting to reflect on the fact that $N$ gives an isomorphism between the quotients $Gr_{D+\ell}$ and $Gr_{D-\ell}$ described below \eqref{Wfiltr}. On the level of the Deligne splitting this property can be understood in simpler terms as an identification between the sum of vector spaces $I^{p,q}$ with $p+q=D+\ell$ and $p+q=D-\ell$. Combining this with $N$ being a $(-1,-1)$-map, it means that $N$ gives an isomorphism between $I^{p,q}$ and $I^{D-q, D-p}$. In particular, note that this explains precisely the symmetry $i^{p,q}=i^{D-q,D-p}$ we pointed out before in \eqref{eq:symmetries}.

\subsubsection*{Primitive subspaces and polarization conditions}
Finally, we would like to introduce a helpful decomposition of the Deligne splitting: we can separate the vector spaces $I^{p,q}$ into primitive and non-primitive components with respect to the nilpotent matrix $N$. This allows us to focus our attention just on primitive elements and their descendants under $N$ when necessary, instead of working with more generic elements of $I^{p,q}$. Let us also briefly point out an analogy here with the decomposition of differential forms on K\"ahler manifolds, where the nilpotent matrix $N$ here replaces the role of the K\"ahler form $J$ there. We define the primitive component of $I^{p,q}$ as
\begin{equation}\label{eq:primsubspaces}
P^{p,q}(N)=I^{p,q}(N) \cap \ker N^{p+q+1-D}\, ,
\end{equation}
with $p+q \geq D$. The other components of $I^{p,q}$ can be recovered by acting with $N$ on the higher-lying primitive spaces as
\begin{equation}\label{Ipqdecomp}
I^{p,q}(N) = \oplus_k N^k P^{p+k,q+k}(N)\, .
\end{equation}
Recalling our example in \eqref{Ipq_example}: the vector $(1,0)$ would span $P^{1,1} = I^{1,1} \cap \ker N^2$ (since $N^2=0$) while $(0,1)$ is an element of $N P^{1,1}$.

We can now use this decomposition into primitive subspaces $P^{p,q}$ to state the \textit{polarization conditions} for a mixed Hodge structure in a concise way. For a pure Hodge structure we previously wrote these conditions as \eqref{eq:HRrelations}. The positivity condition is translated to
\begin{equation}\label{eq:pol}
v \in P^{p,q}: \quad i^{p-q} \langle N^{p+q-D} \bar{v}  \, , \  v \rangle >0 \, .
\end{equation}
It is instructive to see in what subspaces these elements are located: on the right we have simply $v \in I^{p,q}$, while on the left we find $N^{p+q-D} \bar{v} \in I^{p-D, q-D}$ -- in other words, the bilinear pairing is non-vanishing when the indices add up to $(D,D)$. This suggests an analogy with how only integrals of top-degree forms are non-vanishing, and indeed from conditions such as $\langle F^p_0, F^{D+1-p}_0 \rangle = 0$ one can show that\footnote{By writing out the limiting filtration via \eqref{eq:ItoFW} we find that $\langle F^p_0, F^{D+1-p}_0 \rangle=0$ implies that $\langle I^{p,q}, I^{r,s} \rangle =0$ for $p+r>D$. The case $p+r<D$ is slightly more non-trivial, but can be argued for via for instance the SL(2)-orbit theorem. }
\begin{equation}
\langle I^{p,q}, I^{r,s} \rangle =0\, ,  \quad (r,s)\neq (D-p, D-q)\, .
\end{equation}

\section{Boundary classification for Calabi-Yau threefolds}\label{sec:classification}
An interesting take on asymptotic Hodge theory is that it can be used to classify what boundaries can occur in complex structure moduli space \cite{robles_2015, Kerr2017}. Moreover, we can make a classification of allowed splittings $I^{p,q}$ of the middle cohomology $H$ of the Calabi-Yau manifold. 
In this section we discuss the application to Calabi-Yau threefolds, although many observations straightforwardly generalize to Hodge structures of any weight. For explicit work on the extension to Calabi-Yau fourfolds we refer the reader to \cite{Grimm:2019ixq}.

\subsubsection*{Singularity types}
We first work out what splittings $I^{p,q}$ are allowed -- in other words, what shapes can the Hodge-Deligne diamond in figure \ref{fig_deligne} take. Let us begin with the outer edge of the diamond. For a Calabi-Yau threefold we know that $h^{3,0}=1$, since the holomorphic $(3,0)$-form is unique up to rescalings. The relation \eqref{eq:hformula} between the Hodge numbers and the Deligne numbers then implies that only one of the numbers $i^{3,d}$ can be non-vanishing. Exploiting the symmetries \eqref{eq:symmetries} of the Hodge-Deligne diamond one then finds for the outer edge 
\begin{equation}
i^{3,d}=i^{d,3}=i^{0,3-d}=i^{3-d,0}=1\, ,
\end{equation}
with all other $i^{3,q}=i^{p,3}=i^{0,q}=i^{q,0}=0$. The integer $d$ for which $i^{3,d}=1$ can be read off straightforwardly from the leading behavior of the $(3,0)$-form periods \eqref{eq:periodsexpansion}: the leading term $a_0$ spans $I^{3,d}$, and $N$ can act at most $d$ times on it (otherwise we end up below $I^{3-d,0}$).\footnote{From the relation between the periods and the limiting filtration we know that $a_0$ spans $F^3_0$. Since $F^3_0$ is one-dimensional, one finds by inspecting \eqref{Ipq} that $I^{3,d}=F^3_0$, so $a_0$ spans $I^{3,d}$. The action of $N$ on $a_0$ then follows from the fact that $N^d$ is an isomorphism between $I^{3,d}$ and $I^{3-d,0}$.} Thus we find that
\begin{equation}\label{eq:ddef}
N^{d} \mathbf{a}_0 \neq 0\, , \qquad N^{d+1} \mathbf{a}_0 = 0\,,
\end{equation}
which tells us that the polynomial part of the periods in \eqref{eq:periodsexpansion} is of degree $d$. This motivates us to define principal types of singularities as 
\begin{equation}\label{all_limits}
\begin{aligned}
{\rm Type \ I}&: \ N a_0 =0\,, \quad &{\rm Type \ II}&: \ N^2 a_0 =0\,, \\
\quad {\rm Type \ III}&: \ N^3 a_0 =0\,, \quad &{\rm Type \ IV}&: \ N^4 a_0 =0  \,.
\end{aligned} 
\end{equation}
The remainder of the Hodge-Deligne diamond is then made up by the middle components $i^{p,q}$ with $1\leq p,q \leq 2$ as
\begin{equation}\label{Ipq_example}
I^{p,q} = \begin{tikzpicture}[baseline={([yshift=-.5ex]current bounding box.center)},scale=1,cm={cos(45),sin(45),-sin(45),cos(45),(15,0)}]
  \draw[step = 1, gray, ultra thin] (0, 0) grid (1, 1);
  \draw[fill] (1, 0) circle[radius=0.05] node[right]{\small $i^{2,1}$ };
  \draw[fill] (0, 1) circle[radius=0.05] node[right]{\small $i^{2,1}$};
  \draw[fill] (1, 1) circle[radius=0.05] node[right]{\small $i^{2,2}$ };
  \draw[fill] (0, 0) circle[radius=0.05] node[right]{\small $i^{2,2}$};
\end{tikzpicture}\, ,
\end{equation}
where we already used the symmetries in \eqref{eq:symmetries} which yield $i^{2,2}=i^{1,1}$ and $i^{2,1}=i^{1,2}$. Invoking the sum of all $i^{p,q}$ to be equal to the total dimension $2h^{2,1}+2$, only $i^{2,2}$ remains as a free index for the singularity type, and $i^{2,1}$ is fixed via
\begin{align}
{\rm Type \ I\, , \, IV}:& \quad 2 i^{2,1}+2i^{2,2}+2  =2h^{2,1}+2\,, \nonumber \\
{\rm Type \ II\, , \, III}:& \quad 2 i^{2,1}+2i^{2,2}+4  =2h^{2,1}+2 \,. \nonumber 
\end{align} 
Attaching the dimension $i^{2,2}$ as a subscript to the principal types, we end up with $4h^{2,1}$ possible types of singularities. This classification has been summarized in table \ref{table:HDclass}. Note that there are non-trivial restrictions on the index range $i^{2,2}$ depending on the type of singularity. For instance, for a IV$_d$ singularity we see that $d\geq 1$: since $N I^{3,3} \subseteq I^{2,2}$ and $N I^{3,3} \not = \emptyset$, the dimension of $I^{2,2}$ should be at least one.

\begin{table}[h!]
\centering
\scalebox{0.77}{
\renewcommand*{\arraystretch}{2.0}
\begin{tabular}{| c| c | c | c | c |}
\hline singularity & $\mathrm{I}_a$ & $\mathrm{II}_b$ & $\mathrm{III}_c$ & $\mathrm{IV}_d$ \\ \hline \hline 
\begin{minipage}{0.17\textwidth}
\vspace{-1.3cm}
HD diamond
\vspace{1.3cm}
\end{minipage} &
\rule[-0.25cm]{.0cm}{3.5cm} \begin{tikzpicture}[scale=0.65,cm={cos(45),sin(45),-sin(45),cos(45),(15,0)}]
  \draw[step = 1, gray, ultra thin] (0, 0) grid (3, 3);

  \draw[fill] (0, 3) circle[radius=0.05];
  \draw[fill] (1, 2) circle[radius=0.05] node[above]{$a'$};
  \draw[fill] (2, 1) circle[radius=0.05] node[above]{$a'$};
  \draw[fill] (1, 1) circle[radius=0.05] node[above]{$a$};
  \draw[fill] (2, 2) circle[radius=0.05] node[above]{$a$};
  \draw[fill] (3, 0) circle[radius=0.05];
\end{tikzpicture} &
\begin{tikzpicture}[scale=0.65,cm={cos(45),sin(45),-sin(45),cos(45),(15,0)}]
  \draw[step = 1, gray, ultra thin] (0, 0) grid (3, 3);

  \draw[fill] (0, 2) circle[radius=0.05];
  \draw[fill] (1, 3) circle[radius=0.05];
  \draw[fill] (1, 2) circle[radius=0.05] node[above]{$b'$};
  \draw[fill] (1, 1) circle[radius=0.05] node[above]{$b$};
  \draw[fill] (2, 1) circle[radius=0.05] node[above]{$b'$};
  \draw[fill] (2, 2) circle[radius=0.05] node[above]{$b$};
  \draw[fill] (2, 0) circle[radius=0.05];
  \draw[fill] (3, 1) circle[radius=0.05];
\end{tikzpicture} &
\begin{tikzpicture}[scale=0.65,cm={cos(45),sin(45),-sin(45),cos(45),(15,0)}]
  \draw[step = 1, gray, ultra thin] (0, 0) grid (3, 3);
  \draw[fill] (0, 1) circle[radius=0.05];
  \draw[fill] (1, 0) circle[radius=0.05];
  \draw[fill] (1, 2) circle[radius=0.05] node[above]{$c'$};
  \draw[fill] (2, 1) circle[radius=0.05] node[above]{$c'$};
  \draw[fill] (2, 3) circle[radius=0.05];
  \draw[fill] (1, 1) circle[radius=0.05] node[above]{$c$};
  \draw[fill] (2, 2) circle[radius=0.05] node[above]{$c$};
  \draw[fill] (3, 2) circle[radius=0.05];
\end{tikzpicture} &
\begin{tikzpicture}[scale=0.65,cm={cos(45),sin(45),-sin(45),cos(45),(15,0)}]
  \draw[step = 1, gray, ultra thin] (0, 0) grid (3, 3);

  \draw[fill] (0, 0) circle[radius=0.05];
  \draw[fill] (1, 1) circle[radius=0.05] node[above]{$d$};
  \draw[fill] (1, 2) circle[radius=0.05] node[above]{$d'$};
  \draw[fill] (2, 1) circle[radius=0.05] node[above]{$d'$};
  \draw[fill] (2, 2) circle[radius=0.05] node[above]{$d$};
  \draw[fill] (3, 3) circle[radius=0.05];
\end{tikzpicture} \\ \hline
index &  \begin{minipage}{.15\textwidth}\centering\vspace*{-0.cm} \begin{equation*}\begin{aligned} a+a'&=h^{2,1} \\ 0\leq a &\leq h^{2,1} \end{aligned}\end{equation*}  \vspace*{-0.cm} \end{minipage} & \begin{minipage}{.18\textwidth}\centering\vspace*{-0.cm} \begin{equation*}\begin{aligned}b+b'&=h^{2,1}-1 \\ 0\leq b &\leq h^{2,1}-1 \end{aligned}\end{equation*}  \vspace*{-0.cm} \end{minipage}& \begin{minipage}{.18\textwidth}\centering\vspace*{-0.cm} \begin{equation*}\begin{aligned} c+c'&=h^{2,1}-1 \\ 0 \leq c &\leq h^{2,1}-2 \end{aligned}\end{equation*}  \vspace*{-0.cm}\end{minipage}&\begin{minipage}{.15\textwidth}\centering\vspace*{-0.cm} \begin{equation*}\begin{aligned} d+d'&=h^{2,1} \\ 1 \leq d &\leq h^{2,1} \end{aligned}\end{equation*}  \vspace*{-0.cm} \end{minipage}\\ \hline
\begin{minipage}{0.18\textwidth}
\vspace{-1.1cm} 
\ \ \ \  (signed) \\
Young diagram 
\end{minipage}&   \begin{minipage}{0.156\textwidth}
\vspace{-1.1cm} \begin{tikzpicture}[scale=0.4]
    \draw (0, 0) rectangle (1, -1);
    \draw (.5, -.5) node {$+$};
    \draw (1, 0) rectangle (2, -1);
    \draw (1.5, -.5) node {$-$};
    \draw (2, -.5) node[right] {\small $\otimes \, a$};
    \draw (0, -1) rectangle (1, -2);
    \draw (2, -1.5) node[right] {\small $\otimes \, 2a' + 2$};
  \end{tikzpicture}
  \end{minipage} & 
   \rule[-0.4cm]{.0cm}{2.1cm}   \begin{tikzpicture}[scale=0.4]
    \draw (0, 0) rectangle (1, -1);
    \draw (.5, -.5) node {$+$};
    \draw (1, 0) rectangle (2, -1);
    \draw (1.5, -.5) node {$-$};
    \draw (2, -.5) node[right] {\small $\otimes \, b$};
    \draw (0, -1) rectangle (1, -2);
    \draw (.5, -1.5) node {$-$};
    \draw (1, -1) rectangle (2, -2);
    \draw (1.5, -1.5) node {$+$};
    \draw (2, -1.5) node[right] {\small $\otimes \, 2$};
    \draw (0, -2) rectangle (1, -3);
    \draw (2, -2.5) node[right] {\small $\otimes \, 2b'$};
  \end{tikzpicture} &
    \begin{tikzpicture}[scale=0.4]
    \draw[step = 1] (0, 0) grid (3, -1);
    \draw (3, -.5) node[right] {\small $\otimes \, 2$};
    \draw (0, -1) rectangle (1, -2);
    \draw (.5, -1.5) node {$+$};
    \draw (1, -1) rectangle (2, -2);
    \draw (1.5, -1.5) node {$-$};
    \draw (3, -1.5) node[right] {\small $\otimes \, c$};
    \draw (0, -2) rectangle (1, -3);
    \draw (3, -2.5) node[right] {\small $\otimes \, 2c' - 2$};
  \end{tikzpicture} &
    \begin{tikzpicture}[scale=0.4]
    \draw[step = 1] (0, 0) grid (4, -1);
    \draw (.5, -.5) node {$-$};
    \draw (1.5, -.5) node {$+$};
    \draw (2.5, -.5) node {$-$};
    \draw (3.5, -.5) node {$+$};
    \draw (4, -.5) node[right] {\small $\otimes \, 1$};
    \draw (0, -1) rectangle (1, -2);
    \draw (.5, -1.5) node {$+$};
    \draw (1, -1) rectangle (2, -2);
    \draw (1.5, -1.5) node {$-$};
    \draw (4, -1.5) node[right] { \small $\otimes \, d - 1$};
    \draw (0, -2) rectangle (1, -3);
    \draw (4, -2.5) node[right] {\small $\otimes \, 2d'$};
  \end{tikzpicture} \\ \hline
$\text{rk}(N,N^2,N^3)$ & $(a,\, 0,\, 0)$ & $(2+b,\, 0,\, 0)$ & $(4+c,\, 2,\, 0)$ & $(2+d,\, 2,\, 1)$ \\ \hline
eigvals $\eta N$ & $a$ negative & \begin{minipage}{0.15\textwidth}\centering\vspace*{0.2cm}
$b$ negative \\
2 positive  \vspace*{0.15cm}
\end{minipage} & not needed & not needed \\ \hline
\end{tabular}}
\caption{\label{table:HDclass} Classification of singularity types in complex structure moduli space based on the $4h^{2,1}$ possible different Hodge-Deligne diamonds. In each Hodge-Deligne diamond we indicated non-vanishing $i^{p,q}$ by a dot on the roster, where the dimension has been given explicitly when $i^{p,q} >1$. In the last two rows we listed the characteristic properties of the log-monodromy matrix $N$ and the symplectic pairing $\eta$ that are sufficient to make a distinction between the types.}
\end{table}

\subsubsection*{Geometrical interpretation}
To give some intuition for these singularities, let us briefly sketch what they correspond to geometrically by using mirror symmetry (we refer to \cite{Grimm:2019bey} for a more detailed study).  This perspective allows us to give a precise meaning to these degenerations based on the fiber structure of the mirror manifold. Note in particular that the K\"ahler potential \eqref{eq:kahlerexpansion} -- which now describes the (negative logarithm of) the volume of the geometry -- scales with power $d$ in the volume modulus. From this dimensional analysis one can then infer the following rule of thumb: the integer $d$ gives the dimension of the base of the fibration, and the degeneration limit corresponds to sending the base volume to infinity. 

We can make this rule of thumb more explicit by discussing geometric realizations of these degenerations. The type $\rm I$ singularities are finite distance boundaries, while the others are an infinite distance away when measuring in the K\"ahler metric. The most famous example of a type $\rm I$ boundary is the conifold point \cite{Candelas:1990rm} of the mirror quintic. The type $\rm II$ boundaries signal the presence of a K3 fibration in the mirror geometry \cite{doran2016mirror} -- which are relevant for the geometric engineering of Seiberg-Witten theories \cite{Kachru:1995fv,Eguchi:2007iw} from a physical point of view -- or a $T^4$ fibration. Type $\rm III$ boundaries signal the presence of an elliptic fibration at large complex structure. Finally, type $\rm IV$ boundaries corresponds to sending the overall volume of the mirror Calabi-Yau manifold itself to infinity, and are not related to any underlying fiber structure. 

In the large complex structure regime these observations can be made more precise with the intersection numbers $\cK_{ijk}$ of the mirror Calabi-Yau threefold. Recall the nilpotent orbit data from \eqref{logN_001} with the log-monodromy matrices given in terms of the $\cK_{ijk}$. Let us define sums of the intersection numbers as
\begin{equation}\label{notation}
\cK_{ij}^{(k)} \equiv \sum_{a =1}^k \cK_{aij}\; , \qquad \cK_{i}^{(k)}  \equiv \sum_{a,b =1}^k \cK_{abi}  \qquad \text{and} \qquad \cK_{}^{(k)}  \equiv \sum_{a,b,c =1}^k \cK_{abc}\;.
\end{equation}
The powers of $N_{(k)}=N_1 + \ldots + N_k$ are then computed to be 
\begin{equation}
\left(N_{(k)}\right)^2 = \scalebox{0.85}{$\left( \begin{array}{cccc} 0     &     0     & 0 & 0 \\
0     &     0     & 0 & 0 \\
0  &    \cK_{j}^{(k)}      & 0 & 0 \\
-\cK_{i}^{(k)}     & 0 & 0 & 0
\end{array}\right)$}\, , \qquad 
\left(N_{(k)}\right)^3  = \scalebox{0.85}{$\left(
\begin{array}{cccc}
0      & 0 & 0 & 0 \\
0      & 0 & 0 & 0 \\
 \cK^{(k)}       & 0 & 0 & 0 \\
0 & 0 & 0 & 0
\end{array}\right)$} \, .
\end{equation}
It is now straightforward to use table~\ref{table:HDclass} and translate the rank conditions into conditions on the intersection numbers. The results are presented in table~\ref{Type_Table2}.

\begin{table}[!ht]
	\centering
	\begin{tabular}{@{}lccr@{}} \toprule
		Type &~$\rk \cK_{}^{(k)}$ &~$\rk \cK_{i}^{(k)}$ &~$\rk \cK_{ij}^{(k)}$ \\ \midrule 
		II$_b$ & 0 & 0 &~$b$ \\  
		III$_c$ & 0 & 1 &~$c+2$ \\
		IV$_d$ & 1 & 1 &~$d$ \\ \bottomrule
	\end{tabular}
	\caption{List of types in the large complex structure regime in the limit $t^{i_1},...,t^{i_k}  \rightarrow i \infty$.
	        For numbers and vectors, we define the ranks~$\text{rk}(K_{}^{(k)})$ and~$\text{rk}(\cK_{i}^{(k)})$ to be either~$0$ or~$1$, depending on
		whether $K_{}^{(k)}=0$ and $\cK_{i}^{(k)}=0$ for all $i$.}
	\label{Type_Table2}
\end{table}

\subsubsection*{Minimal form of boundary data}

As a complementary perspective to the above scheme with Deligne splittings, it is useful to point out that nilpotent elements $N \in \mathfrak{g}$ -- infinitesimal isometries of the bilinear pairing -- admit a classification as well \cite{Djokovic1982} (see \cite{Collingwood1993} for an introductory review). This allows us to understand many aspects of the classification by Deligne splittings in terms of this representation theoretic language \cite{BrosnanPearlsteinRobles}. The idea is to use \textit{signed Young diagrams} to classify conjugacy classes of nilpotent elements under the adjoint action of the isometry group 
\begin{equation}
g \in G: \quad N \to g N g^{-1}\, .
\end{equation}
In table \ref{table:HDclass} we listed what signed Young diagrams correspond to each of the Hodge-Deligne diamonds. The upshot of mapping our classification to signed Young diagrams is that these can be represented by simple matrices in a Jordan normal form: we have given all the required building blocks for the Calabi-Yau threefold case in table \ref{table:buildingblocks} for completeness.

\ytableausetup
 {mathmode, boxsize=1.2em}
\newcolumntype{A}{>{\centering\arraybackslash} m{.3\linewidth} }

\begin{table}[h!]

\begin{center}
\scalebox{0.90}{
{\small
\begin{tabular}{|c|c|c|c|}
\hline
    \rule[-.1cm]{0cm}{0.5cm} signed Young diagram &  $N^-$  & $N^0$ \\
   \hline
   \hline
  \begin{ytableau}
    \ \\
    \ 
  \end{ytableau}
  &    \rule[-.5cm]{0cm}{1.2cm} $\left( \begin{array}{cc} 0 & 0 \\ 0 & 0\end{array} \right)$  & $\left( \begin{array}{cc} 0 & 0 \\ 0 & 0\end{array} \right)$  
  \\
   \hline
  \begin{ytableau}
    + & -
  \end{ytableau}
  & \rule[-.5cm]{0cm}{1.2cm}  $\left( \begin{array}{cc} 0 & 0 \\ -1 & 0\end{array} \right)$  &$\left( \begin{array}{cc} 1 & 0 \\ 0 & -1\end{array} \right)$  
  \\
    \hline
  \begin{ytableau}
    - & +
  \end{ytableau}
  &  \rule[-.5cm]{0cm}{1.2cm}  $\left( \begin{array}{cc} 0 & 0 \\ 1 & 0\end{array} \right)$& $\left( \begin{array}{cc} 1 & 0 \\ 0 & -1\end{array} \right)$  
  \\    
       \hline
  \begin{ytableau}
    \ & \ & \ \\
    \ & \ & \
  \end{ytableau}
  &  \rule[-1.3cm]{0cm}{2.8cm}   $\left( \begin{array}{cccccc} 0 & 0 & 0 & 0 & 0 & 0 \\ \ 1 & 0 & 0 & 0 &0 & 0\\0 & \ 1 & 0 &0 & 0 & 0\\ 0 & 0 &0 & 0 & -1 & 0\\ 0 &0 & 0 & 0 & 0 & -1\\ 0 & 0 & 0 & 0 & 0 & 0 \end{array} \right)$
    & $\left( \begin{array}{cccccc} 2 & 0 & 0 & 0 & 0 & 0 \\ \ 0 & 0 & 0 & 0 &0 & 0\\0 &  0 & -2 & 0 & 0 & 0\\ 0 & 0 &0 & -2 & 0 & 0\\ 0 &0 & 0 & 0 & 0 & 0\\ 0 & 0 & 0 & 0 & 0 & 2 \end{array} \right)$ 
    \\
    \hline
  \begin{ytableau}
    - & + & - & +
  \end{ytableau}
    & \rule[-0.9cm]{0cm}{2.0cm}  $\left( \begin{array}{cccc} 0 & 0 & 0 & 0  \\ \ 1 & 0 & 0 & 0 \\0 & 0 & 0 &-1 \\ 0 & 1 &0 & 0 
     \end{array} \right)$
    &  $\left( \begin{array}{cccc}    3 & 0 & 0 &0  \\   0 & 1 & 0 & 0 \\  0 & 0 & -3 & 0  \\   0 & 0 & 0 & -1 \end{array} \right)$    \\    
    \hline
\end{tabular}
}}
\caption{Building blocks for the lowering and weight operators $N^-,N^0$ for all relevant signed Young diagrams, where we assume the pairing matrix always takes the standard form \eqref{eq:pairings}. We can obtain simple normal forms for these matrices by combining the building blocks into the complete signed Young diagrams given in table \ref{table:HDclass}.} \label{table:buildingblocks}
\end{center}
\end{table}

Let us now elaborate on this correspondence between Deligne splittings and signed Young diagrams. For the moment we forget about the signing of these diagrams, and study conjugacy classes of $N$ under $G_{\mathbb{C}}$.\footnote{The difference between working with signed or unsigned Young diagrams amounts to whether we work over respectively $\mathbb{R}$ or $\mathbb{C}$, i.e.~with conjugacy classes under $G_{\mathbb{R}}$ or its complexification $G_{\mathbb{C}}$. } By Jacobson-Morosov we know that we can think of the nilpotent matrix $N$ as a lowering operator $N^-=N$, and complete it into an sl$(2,\mathbb{C})$-triple with a weight operator $N^0$ (the raising operator $N^+$ is then fixed by commutation relations \eqref{comm}). Under this sl$(2,\mathbb{C})$-triple we can perform a decomposition of the vector space $H$ into highest-weight components
\begin{equation}
H = \bigoplus_\ell \bigoplus_{k=0}^{\ell} N^k P_\ell \, ,
\end{equation}
where $P_\ell$ is spanned by highest-weight state as
\begin{equation}
v \in P_\ell : \qquad N^0 v= \ell v \, ,\quad (N^-)^{\ell+1}v = 0 \, .
\end{equation}
This information can now be encoded in terms of a Young diagram: irreducible representations are represented by rows, which are made up of $\ell+1$ blocks corresponding to the elements $N^k v$, $k=0,\ldots, \ell$. As an example, for a decomposition with $\dim P_1=1$ and $\dim P_0=2$ we write
\begin{equation}\ytableausetup{mathmode, boxsize=1.8em} 
\vspace*{0.2cm}\begin{ytableau}
    \footnotesize u & \footnotesize Nu  \\
     \footnotesize v \\
    \footnotesize w
\end{ytableau} 
\end{equation} \ytableausetup{mathmode, boxsize=1.2em}
where $u  \in P_1$ and $v, w \in P_0$ are linearly independent. In order to keep notation concise, we drop the labels of the blocks and can simply use the empty Young diagram to classify the nilpotent element $N$. Taking $v, u, w, -Nu$ as a basis, we see that we can write a simple representative for the conjugacy class of $N$ as
\begin{equation}\label{Nsimp}
N = \scalebox{0.85}{$\begin{pmatrix}
0 & 0 & 0 & 0 \\
0 & 0 & 0 & 0 \\
0 & 0 & 0 & 0 \\
0 & -1 & 0 & 0 
\end{pmatrix}$}\, .
\end{equation}
To see how signs in the Young diagrams come in the game, we have to look at the signature of the bilinear pairing. To be more precise, we introduce a non-degenerate bilinear on the vector space $P_\ell$ as
\begin{equation}\label{eq:<>l}
u,v \in P_\ell: \quad \langle u, v \rangle_\ell \equiv \langle N^\ell u,  v \rangle \, .
\end{equation}
Depending on the parity of $D+\ell$ we find that
\begin{itemize}
\item $D+\ell$ odd: the bilinear form \eqref{eq:<>l} is skew-symmetric. The dimension $\dim P_\ell$ must therefore be even, and in turn the Young diagram has an even number of rows of length $\ell$. Also note that this means we do not have to keep track of a signature.
\item $D+\ell$ even: the bilinear form \eqref{eq:<>l} is symmetric. Now there is a choice in signature for $\langle \cdot, \cdot \rangle_\ell $. We can encode this signature in the Young diagram as signs on its blocks: we sign the first column by the signature of $\langle \cdot, \cdot \rangle_\ell$ on $P_\ell$, after which we alternate for the subsequent blocks in these rows.
\end{itemize}
\subsubsection*{Minimal form of boundary data: an example}
Here we illustrate the above discussion of representation theory with an example. In particular, we show how to obtain the signed Young diagram from the Deligne splitting. We consider a one-modulus $\mathrm{I}_1$ limit, which for simplicity we take to be $\mathbb{R}$-split. The Deligne splitting of this singularity can then be spanned by
\begin{equation}\label{eq:IpqI1ex}
I^{p,q} = \begin{tikzpicture}[baseline={([yshift=-.5ex]current bounding box.center)},scale=0.65,cm={cos(45),sin(45),-sin(45),cos(45),(15,0)}]
  \draw[step = 1, gray, ultra thin] (0, 0) grid (3, 3);
  
  \draw[fill] (0, 3) circle[radius=0.05] node[above]{$e_1+ie_3$};
  \draw[fill] (1, 1) circle[radius=0.05] node[above]{$e_4$};
  \draw[fill] (2, 2) circle[radius=0.05] node[above]{$e_2$};
  \draw[fill] (3, 0) circle[radius=0.05] node[above]{$e_1-ie_3$};
\end{tikzpicture} 
\end{equation}
where $e_1,\ldots, e_4$ denote a real basis for $H$. We can take $e_4 = -Ne_2$, since $N$ is an isomorphism between $I^{2,2}$ and $I^{1,1}$, and furthermore $Ne_1=Ne_3=Ne_4=0$. As primitive spaces we identify $P_0 = \mathrm{span}\bigl[  e_1, e_3\bigr]$ and $P_1 = \mathrm{span}\bigl[  e_2 \bigr]$. The polarization conditions \eqref{eq:pol} on these spaces imply that
\begin{equation}
\begin{aligned}
i^3 \langle e_1-i e_3, e_1+ie_3 \rangle &>0   \quad \implies \quad \langle e_1, e_3 \rangle >0\, , \\
\langle N e_2 , e_2 \rangle &>0\, . 
\end{aligned}
\end{equation}
In other words, we find two one-dimensional sl$(2,\mathbb{R})$ states $e_1, e_3$ in $P_0$ with a skew-symmetric pairing, while on $P_1$ we find that \eqref{eq:<>l} is positive-definite. The signed Young diagram therefore is made up of three rows: two rows with one block, and one row with two blocks, starting with a plus sign. This indeed matches with the signed Young diagram given in table \ref{table:HDclass}
\begin{equation}
\begin{ytableau}
    + & -  \\
     \ \\
    \
\end{ytableau} \, .
\end{equation}
The nilpotent matrix corresponding to the basis $e_1,\ldots, e_4$ matches precisely with \eqref{Nsimp} given before. To conclude, let us write down the remaining data associated to this $\mathrm{I}_1$ boundary: the bilinear pairing $\eta$ and weight operator $N^0$ are given by
\begin{equation}
\eta = \scalebox{0.85}{$\begin{pmatrix}
0 & 0 & 1 & 0 \\
0 & 0 & 0 & 1 \\
-1 & 0 & 0 & 0 \\
0 & -1 & 0 & 0 
\end{pmatrix}$}\,, \qquad N^0 = \scalebox{0.85}{$\begin{pmatrix}
0 & 0 & 0 & 0 \\
0 & 1 & 0 & 0 \\
0 & 0 & 0 & 0 \\
0 & 0 & 0 & -1 
\end{pmatrix}$}\,,
\end{equation}
where the weight operator acts as $N^0 \omega_{p,q} = (p+q-3) \omega_{p,q}$ for $\omega_{p,q} \in I^{p,q}$ in \eqref{eq:IpqI1ex}.

\section{Enhancements of singularities}\label{sec:enhancements}
In this section we review how singularity types can change when we send additional moduli to the boundary. We first explain from a general perspective how Deligne splittings can enhance, and summarize the corresponding rules for Calabi-Yau threefolds. This includes an instructive example where we show how these enhancement rules work in practice. We then close off our discussion with chains of enhancements and restrictions on boundary types in the two-moduli case.

\subsubsection*{Enhancement rules}
It turns out that singularity types cannot jump arbitrarily, but there is an underlying ordering for what can happen. In table \ref{allowed_enhancements} we have summarized the restrictions for how singularity types of Calabi-Yau threefolds can enhance. In particular, note that we have a partial order on the principal types $\mathrm{I},\mathrm{II},\mathrm{III}, \mathrm{IV}$, which can only stay the same or increase via enhancements. On the other hand, when we look more closely we see that we could have $\mathrm{II}_0 \to \mathrm{II}_1 \to \mathrm{IV}_2$ while $\mathrm{II}_0 \not\to \mathrm{IV}_2$ is not allowed, so the enhancement rules are not transitive when we include subindices.  

\begin{table}[h!]
\centering
\scalebox{0.9}{
\begin{tabular}{cl}
\toprule
    \rule[-.16cm]{0cm}{0.6cm} \qquad starting  type \qquad\qquad& enhanced  type \qquad \qquad \\
   \midrule 
    \multirow{4}{*}{I$_a$} 
    & \rule[-.2cm]{0cm}{0.8cm} I$_{\hat a}$ for $a\leq \hat a$ \\
    &\rule[-.2cm]{0cm}{0.6cm} II$_{\hat b}$ for $a \leq \hat b $, $a < m$ \hspace*{.4cm}\\
    &\rule[-.2cm]{0cm}{0.6cm} III$_{\hat c}$ for $a \leq \hat c$, $a < m$\\
    &\rule[-.4cm]{0cm}{0.8cm} IV$_{\hat d}$ for $a < \hat d$, $a < m$ \\
     \multirow{3}{*}{II$_b$} 
    &\rule[-.2cm]{0cm}{0.8cm} II$_{\hat b}$ for $b\leq \hat b$\\
    &\rule[-.2cm]{0cm}{0.6cm} III$_{\hat c}$ for $2 \leq b \leq \hat c + 2$\\
    &\rule[-.4cm]{0cm}{0.8cm} IV$_{\hat d}$ for $1 \leq b \leq \hat d-1$ \\      
     \multirow{2}{*}{III$_c$} 
    &\rule[-.2cm]{0cm}{0.8cm} III$_{\hat c}$ for $c\leq \hat c$ \\
    &\rule[-.4cm]{0cm}{0.8cm} IV$_{\hat d}$ for $c + 2 \leq \hat d$ \\ 
       IV$_d$ & \rule[-.4cm]{0cm}{1cm} IV$_{\hat d}$ for $d\leq \hat d$  \\
    \bottomrule
\end{tabular}
\begin{picture}(0,0)
\put(-171,65){\begin{tikzpicture}
\draw [->] (0,0) -- (2,.7);
\end{tikzpicture}}
\put(-171,65){\begin{tikzpicture}
\draw [->] (0,0) -- (2,.1);
\end{tikzpicture}}
\put(-171,52){\begin{tikzpicture}
\draw [->] (0,-.2) -- (2,-.66);
\end{tikzpicture}}
\put(-171,38){\begin{tikzpicture}
\draw [->] (0,-.2) -- (2,-1.15);
\end{tikzpicture}}
\put(-171,-7){\begin{tikzpicture}
\draw [->] (0,0) -- (2,.5);
\end{tikzpicture}}
\put(-171,-10){\begin{tikzpicture}
\draw [->] (0,0) -- (2,-.1);
\end{tikzpicture}}
\put(-171,-26){\begin{tikzpicture}
\draw [->] (0,0) -- (2,-.66);
\end{tikzpicture}}
\put(-171,-64){\begin{tikzpicture}
\draw [->] (0,0) -- (2,.25);
\end{tikzpicture}}
\put(-171,-73){\begin{tikzpicture}
\draw [->] (0,0) -- (2,-.32);
\end{tikzpicture}}
\put(-169,-102){\begin{tikzpicture}
\draw [->] (0,0) -- (1.9,0);
\end{tikzpicture}}
\end{picture}}
\caption{List of all allowed enhancements of degeneration types \cite{Kerr2017}, with $m=2h^{2,1}+2$.}
\label{allowed_enhancements}
\end{table}

The underlying structure of these singularity enhancements can be described via the Deligne splittings.\footnote{This can also be described from the representation theoretic perspective discussed previously, where there are similar rules for how signed Young diagrams can enhance. For our discussion it is more natural to work with the Deligne splittings, so let us simply refer to \cite{BrosnanPearlsteinRobles} for more details.} Without loss of generality, let us assume that one saxion $y_1$ has been sent to the boundary, and we denote the corresponding Deligne splitting by $I^{p,q}_{(1)}$. We then want to investigate how the log-monodromy matrix $N_{2}$ can enhance this Deligne splitting as
\begin{equation}
I^{p,q}_{(1)} \quad \to \quad I^{p,q}_{(2)}\, .
\end{equation}
It suffices to restrict our attention to the primitive subspaces $P^{p,q}_{(1)}$ of the first Deligne splitting (see \eqref{eq:primsubspaces}), since the other components descend via $N_1$ by \eqref{Ipqdecomp}. Moreover, we can single out a particular row $p+q=D+\ell$: the primitive subspaces on this row define a pure Hodge structure of weight $D+\ell$ on
\begin{equation}
P^{D+\ell}_{(1)} = \bigoplus_{p+q = D+\ell} P^{p,q}_{(1)}\, ,
\end{equation}
polarized by the bilinear pairing
\begin{equation}
u, v \in P^{D+\ell}_{(1)}: \quad   \langle u, v \rangle_\ell = \langle N^\ell u, v \rangle\, .
\end{equation}
Note that this indeed provides a non-degenerate bilinear pairing by using \eqref{eq:pol}. The enhancement of a Deligne splitting can then be understood as follows. The nilpotent operator $N_2$ induces a mixed Hodge structure on each row of pure Hodge structures $P^{p,q}_{(1)}$, i.e.~at each weight $p+q=D+\ell$. In turn, these decompose as
\begin{equation}
P^{p,q}_{(1)} \quad \to \quad P^{p,q}_{(2)}, \ N_2 P^{p,q}_{(2)},\  \ldots ,\  (N_2)^D P^{p,q}_{(2)}\, ,
\end{equation}
i.e.~primitive components under $N_2$ and their descendants. The set of $I^{p,q}_{(2)}$ is then completed by applying $N_1$ on these vector spaces, by which we recover the descendants we momentarily ignored. 

To be slightly more concrete, we can also describe how dots in the Deligne diamond can be moved by enhancements. From the classification of the threefolds we already learn some lessons here: the dot corresponding to the perturbative term $a_0$ can only be moved to the top-right, since $i^{3,d}=1$ can only be replaced by some other $i^{3,d'}=1$ with $d \leq d'$. This rule of thumb can be used in general, where dots at $(p,q)$ (assuming $p+q >D$ and $p>q$) can only move up to some other $(p,q')$ with $q\leq q'$ and $ q' \leq p$. In simpler terms, the dots can move along diagonal steps to the top-right, and never across the middle. More explicitly, we can draw it for a Calabi-Yau threefold as
  \begin{equation}
 \begin{tikzpicture}[baseline={([yshift=-.5ex]current bounding box.center)},scale=0.75,cm={cos(45),sin(45),-sin(45),cos(45),(15,0)}]
  \draw[step = 1, gray, ultra thin] (0, 0) grid (3, 3);
  
\draw[fill] (0,3) circle[radius=0.05];
\draw[fill] (1,3) circle[radius=0.05];
\draw[fill] (2,3) circle[radius=0.05];
\draw[fill] (3,3) circle[radius=0.05];

\draw[fill] (1,2) circle[radius=0.05];
\draw[fill] (2,2) circle[radius=0.05];

\draw[->] (0.1,3) -- (0.9 ,3);
\draw[->] (1.1,3) -- (1.9 ,3);
\draw[->] (2.1,3) -- (2.9 ,3);

\draw[->] (1.1,2) -- (1.9 ,2);
\end{tikzpicture} \, .
 \end{equation}
 In these enhancements one is also allowed to move up multiple steps at the same time, even though we did not show this. Let us also stress that, while this rule gives us some intuition, one is not able to move up dots freely. The descendants under $N_2$ also must be present, i.e.~for each dot we move up another one must go down, so there are some further restrictions at play.

\subsubsection*{Singularity enhancement: an example}
Let us illustrate the above discussion on enhancements by considering an example. We will investigate how the Deligne splitting can enhance from a $\mathrm{II}_b$ singularity. For a detailed study of these and other enhancements in the literature we refer the reader to \cite{Grimm:2018cpv}. Let us begin by decomposing the vector space into primitive subspaces under $N_1$ for a $\mathrm{II}_b$ singularity as
\begin{equation}
H^3 = {\color{blue}P^3_{(1)}} \oplus {\color{red}P^4_{(1)}} \oplus {\color{ForestGreen}N_1 P^4_{(1)}}\, .
\end{equation}
We can draw this decomposition in the Deligne diamond as
\begin{equation}
I^{p,q}_{(1)}=\begin{tikzpicture}[baseline={([yshift=-.5ex]current bounding box.center)},scale=0.75,cm={cos(45),sin(45),-sin(45),cos(45),(15,0)}]
  \draw[step = 1, gray, ultra thin] (0, 0) grid (3, 3);

  \draw[fill, ForestGreen] (0, 2) circle[radius=0.05];
  \draw[fill, red] (1, 3) circle[radius=0.05];
  \draw[fill, blue] (1, 2) circle[radius=0.05] node[above]{$b'$};
  \draw[fill, ForestGreen] (1, 1) circle[radius=0.05] node[below]{$b$};
  \draw[fill, blue] (2, 1) circle[radius=0.05] node[above]{$b'$};
  \draw[fill, red] (2, 2) circle[radius=0.05] node[above]{$b$};
  \draw[fill, ForestGreen] (2, 0) circle[radius=0.05];
  \draw[fill, red] (3, 1) circle[radius=0.05];
  
  \draw[->] (0.95, 2.95) -- (0.05, 2.05);
  \draw[->] (1.95, 1.95) -- (1.05, 1.05);
  \draw[->] (2.95, 0.95) -- (2.05, 0.05);
\end{tikzpicture}\, ,
\end{equation}
where we explicitly depicted the action of the log-monodromy matrix $N_1$. To figure out how this Deligne splitting can enhance, we then study how $N_2$ can induce a mixed Hodge structure on the primitive components $P^3_{(1)}$ and $P^4_{(1)}$. 

We begin with $P^3_{(1)}$ since it is slightly simpler. We can decompose $P^{2,1}_{(1)}$ such that elements either stay in this position $(2,1)$, or produce together with its conjugate at $(1,2)$ new elements in $(2,2)$ and $(1,1)$. This means the primitive component enhances as
\begin{equation}\label{P3enhancement}
\begin{tikzpicture}[baseline={([yshift=-.5ex]current bounding box.center)},scale=0.8,cm={cos(45),sin(45),-sin(45),cos(45),(15,0)}]
  \draw[step = 1, gray, ultra thin] (0, 0) grid (3, 3);

  \draw[fill, blue] (1, 2) circle[radius=0.05] node[above]{\footnotesize $b'$};
  \draw[fill, blue] (2, 1) circle[radius=0.05] node[above]{\footnotesize $b'$};

\end{tikzpicture} \implies 
\begin{tikzpicture}[baseline={([yshift=-.5ex]current bounding box.center)},scale=0.8,cm={cos(45),sin(45),-sin(45),cos(45),(15,0)}]
  \draw[step = 1, gray, ultra thin] (0, 0) grid (3, 3);

  \draw[fill, blue] (1, 2) circle[radius=0.05] node[left]{\footnotesize $b'-k$};
  \draw[fill, blue] (2, 1) circle[radius=0.05] node[right]{\footnotesize $b'-k$};

  \draw[fill, blue] (2, 2) circle[radius=0.05] node[above]{\footnotesize $k$};
  \draw[fill, blue] (1, 1) circle[radius=0.05] node[below]{\footnotesize $k$};
  
  \draw[->, black] (1.9, 1.9) -- (1.1, 1.1) node [right, midway] {\footnotesize $N_2$};
  
\end{tikzpicture}\, ,
\end{equation}
where $0\leq k \leq b'$ keeps track of the number of elements that move up to $(2,2)$. It is interesting to point out that the $k$ elements that move down to $(1,1)$ -- previously primitive with respect to $N_1$ -- are no longer primitive with respect to $N_2$, since they lie in $N_2 P_{(2)}^{2,2}$.

Next we consider the primitive component $P^4_{(1)}$. Here we have a choice: the dot at $(3,1)$ can either stay in the same position, or move to one of the positions $(3,2)$ or $(3,3)$. Note in particular that it cannot move towards $(3,0)$: we have a pure Hodge structure of weight $4$ with non-vanishing $h^{p,q}$ for $1\leq p,q \leq 3$, and one can check that the resulting Deligne splitting should have the same index range. We thus find that it either stays the same, or enhances as
\begin{equation}\label{P4enhancement}
\begin{tikzpicture}[baseline={([yshift=-.5ex]current bounding box.center)},scale=0.77,cm={cos(45),sin(45),-sin(45),cos(45),(15,0)}]
  \draw[step = 1, gray, ultra thin] (0, 0) grid (3, 3);

  \draw[fill, red] (3, 1) circle[radius=0.05];
  \draw[fill, red] (2, 2) circle[radius=0.05] node[above]{\footnotesize $b$};
  \draw[fill, red] (1, 3) circle[radius=0.05];

\end{tikzpicture} \implies 
\begin{tikzpicture}[baseline={([yshift=-.5ex]current bounding box.center)},scale=0.77,cm={cos(45),sin(45),-sin(45),cos(45),(15,0)}]
  \draw[step = 1, gray, ultra thin] (0, 0) grid (3, 3);

  \draw[fill, red] (2, 2) circle[radius=0.05] node[above]{\footnotesize $b-2$};
  \draw[fill, red] (3, 2) circle[radius=0.05] ;
  \draw[fill, red] (2, 3) circle[radius=0.05] ;
  \draw[fill, red] (2, 1) circle[radius=0.05] ;
  \draw[fill, red] (1, 2) circle[radius=0.05] ;
  
  \draw[->, black] (2.9, 1.9) -- (2.1, 1.1) node [right, midway] {\footnotesize $N_2$};
  \draw[->, black] (1.9, 2.9) -- (1.1, 2.1) node [left, midway] {\footnotesize $N_2$};
  
\end{tikzpicture}\, , \
\begin{tikzpicture}[baseline={([yshift=-.5ex]current bounding box.center)},scale=0.77,cm={cos(45),sin(45),-sin(45),cos(45),(15,0)}]
  \draw[step = 1, gray, ultra thin] (0, 0) grid (3, 3);

  \draw[fill, red] (2, 2) circle[radius=0.05] node[right]{\footnotesize $b$};
  \draw[fill, red] (3, 3) circle[radius=0.05] ;
  \draw[fill, red] (1, 1) circle[radius=0.05] ;
  
  \draw[->, black] (2.9, 2.9) -- (2.1, 2.1) node [left, midway] {\footnotesize $N_2$};
  \draw[->, black] (1.9, 1.9) -- (1.1, 1.1) node [left, midway] {\footnotesize $N_2$};
  
\end{tikzpicture}\, ,
\end{equation}
Let us also point out that some of the elements in $(2,2)$ had to combine with the $(3,1)$ and $(1,3)$ elements to make this possible: for the first enhancement we had to produce two new elements at $(2,1)$ and $(1,2)$, while for the second the primitive state at $(3,3)$ has a descendant at $(2,2)$ and $(1,1)$. This in particular implies that we should have $b>2$ for the first enhancement and $b>1$ for the second.

Finally, we can reassemble the Deligne splitting $I^{p,q}_{(2)}$. We first put the above findings \eqref{P3enhancement},  \eqref{P4enhancement} for the primitive components together. In turn, we obtain the bottom half of the Deligne diamond by using the symmetry under $i^{p,q}=i^{3-q,3-p}$. In the end we find that $\mathrm{II}_b$ can enhance in one of the following two ways
\begin{equation}
\begin{tikzpicture}[baseline={([yshift=-.5ex]current bounding box.center)},scale=0.75,cm={cos(45),sin(45),-sin(45),cos(45),(15,0)}]
  \draw[step = 1, gray, ultra thin] (0, 0) grid (3, 3);

  \draw[fill] (0, 2) circle[radius=0.05];
  \draw[fill] (1, 3) circle[radius=0.05];
  \draw[fill] (1, 2) circle[radius=0.05] node[above]{$\hat{b}'$};
  \draw[fill] (1, 1) circle[radius=0.05] node[above]{$\hat{b}$};
  \draw[fill] (2, 1) circle[radius=0.05] node[above]{$\hat{b}'$};
  \draw[fill] (2, 2) circle[radius=0.05] node[above]{$\hat{b}$};
  \draw[fill] (2, 0) circle[radius=0.05];
  \draw[fill] (3, 1) circle[radius=0.05];

\end{tikzpicture} \, , \
\begin{tikzpicture}[baseline={([yshift=-.5ex]current bounding box.center)},scale=0.75,cm={cos(45),sin(45),-sin(45),cos(45),(15,0)}]
  \draw[step = 1, gray, ultra thin] (0, 0) grid (3, 3);

  \draw[fill] (0, 1) circle[radius=0.05];
  \draw[fill] (2, 3) circle[radius=0.05];
  \draw[fill] (1, 2) circle[radius=0.05] node[above]{$\hat{c}'$};
  \draw[fill] (1, 1) circle[radius=0.05] node[above]{$\hat{c}$};
  \draw[fill] (2, 1) circle[radius=0.05] node[above]{$\hat{c}'$};
  \draw[fill] (2, 2) circle[radius=0.05] node[above]{$\hat{c}$};
  \draw[fill] (1, 0) circle[radius=0.05];
  \draw[fill] (3, 2) circle[radius=0.05];
  
\end{tikzpicture}\, , \
\begin{tikzpicture}[baseline={([yshift=-.5ex]current bounding box.center)},scale=0.75,cm={cos(45),sin(45),-sin(45),cos(45),(15,0)}]
  \draw[step = 1, gray, ultra thin] (0, 0) grid (3, 3);

  \draw[fill] (0, 0) circle[radius=0.05];
  \draw[fill] (3, 3) circle[radius=0.05];
  \draw[fill] (1, 2) circle[radius=0.05] node[above]{$\hat{d}'$};
  \draw[fill] (1, 1) circle[radius=0.05] node[above]{$\hat{d}$};
  \draw[fill] (2, 1) circle[radius=0.05] node[above]{$\hat{d}'$};
  \draw[fill] (2, 2) circle[radius=0.05] node[above]{$\hat{d}$};
  
\end{tikzpicture}\, .
\end{equation}
where we defined integers $\hat{b}=b+k$, $\hat{c}=b-2+k$ and $\hat{d}=b+k$. From the index range $0 \leq k \leq b'$ we recover the conditions from table \ref{allowed_enhancements} as 
\begin{itemize}
\item $\hat{b} \geq b$ for a $\mathrm{II}_{\hat{b}}$ enhancement,
\item $2 \leq b \leq \hat{c}+2$ for a $\mathrm{III}_{\hat{c}}$ enhancement,
\item $1 \leq b \leq \hat{d}-1$ for a $\mathrm{IV}_{\hat{d}}$ enhancement,
\end{itemize}
where the lower bounds on the $\mathrm{III}_{\hat{c}}$ and $\mathrm{IV}_{\hat{d}}$ enhancements follows from the discussion below \eqref{P4enhancement}. Finally, let us point out that in section \ref{sec_aht_example} we consider an explicit $\mathrm{II}_1 \to \mathrm{IV}_2$ enhancement: it is an instructive for the reader to verify how the spaces in the Deligne splitting \eqref{sl2Ipq(1)} of the $\mathrm{II}_1 $ combine into the spaces \eqref{sl2Ipq(2)} of the $\mathrm{IV}_2$.

\paragraph{Enhancement chains.} We can now generalize the above discussion on one-step enhancements to \textit{enhancement chains} when several moduli are sent to the boundary in a sequence. For each saxion we get a singularity type, from $y^1 \to \infty$, $y^2 \to \infty$, up to $y^n \to \infty$. In other words, we can think of this limit as sending $y^1 , \ldots , y^n \to \infty$ while implementing some hierarchy $y^1 \gg y^2 \gg \ldots \gg y^n$, where the ratios between successive saxions also diverge asymptotically as $y^i/y^{i+1} \to \infty$.  At each step one then classifies the limit involving $y^1  , \, \ldots , \, y^k$ by the type associated with the sum of log-monodromy matrices $N_{(k)}$. Sending additional coordinates to their limit enhances the type, so we can combine the types of an ordered limit as
\begin{small}
\begin{equation}\label{eq:enhancementchain}
\mathrm{I}_0\xrightarrow{\ y^{1} \rightarrow \infty\ }\  {{\sf Type\ A}_{(1)}}\ \xrightarrow{\ y^{2} \rightarrow \infty\ }\  {\sf Type\ A}_{(2)} \ \xrightarrow{\ y^{3} \rightarrow \infty\ }\  
\ldots \ \xrightarrow{\ y^{n} \rightarrow \infty\ }\  {\sf Type\ A}_{(n)} \, .
\end{equation}
\end{small}
We always start in the non-degenerate case $\text{I}_0$  in the interior of the moduli space, and the types ${{\sf  A}_{(k)}}$ in the subsequent steps indicate the singularities associated with taking the limit $y^1  , \, \ldots , \, y^k \to \infty$.

\paragraph{Two-cubes.} Let us close our discussion by pointing out a complete classification of two-moduli boundaries for Calabi-Yau threefolds \cite{Kerr2017}. For such a singularity in moduli space there are two enhancement chains at play for $y^1 , y^2 \to \infty$, one for $y^1 \gg y^2$ and another for $y^2 \gg y^1$. The insight of \cite{Kerr2017} was that one cannot simply take any two singularity types $A_1 , A_2$ that both can enhance to some $A_{(2)}$ for $y^1,y^2 \to \infty$ according to \eqref{allowed_enhancements}, but there are instead some restrictions on this pattern coming from asymptotic Hodge theory. They combined these singularity types into a so-called \textit{2-cube} as $\langle \mathrm{A}_1 | \mathrm{A}_{(2)} | \mathrm{A}_2 \rangle$, and enumerated all possible 2-cubes. We do not review the details of this classification here, but will simply present the exhaustive set of 2-cubes that was obtained
\begin{align}\label{eq:cubesintro}
\text{$\mathrm{I}_2$ class} &: \quad \langle \mathrm{I}_1 | \mathrm{I}_2 | \mathrm{I}_1 \rangle\, , \ \langle \mathrm{I}_2 | \mathrm{I}_2 | \mathrm{I}_1 \rangle \, , \  \langle \mathrm{I}_2 | \mathrm{I}_2 | \mathrm{I}_2 \rangle    \, ,\nn \\
\text{Coni-LCS class} &: \quad \langle \mathrm{I}_1 | \mathrm{IV}_2 | \mathrm{IV}_1 \rangle \, , \   \langle \mathrm{I}_1 | \mathrm{IV}_2 | \mathrm{IV}_2 \rangle  \, , \nn \\
\text{$\mathrm{II}_1$ class} &: \quad \langle \mathrm{II}_0 | \mathrm{II}_1 | \mathrm{I}_1 \rangle \, , \  \langle \mathrm{II}_1 | \mathrm{II}_1 | \mathrm{I}_1 \rangle \, , \   \langle \mathrm{II}_0 | \mathrm{II}_1 | \mathrm{II}_1 \rangle     \, , \ \langle \mathrm{II}_1 | \mathrm{II}_1 | \mathrm{II}_1 \rangle  \, ,\\
\text{LCS class} &: \quad \langle \mathrm{II}_1 | \mathrm{IV}_2 | \mathrm{III}_0 \rangle \, , \   \langle \mathrm{II}_1 | \mathrm{IV}_2 | \mathrm{IV}_2 \rangle  \, , \ \langle \mathrm{III}_0 | \mathrm{IV}_2 | \mathrm{III}_0 \rangle \, , \ \langle \mathrm{III}_0 | \mathrm{IV}_2 | \mathrm{IV}_1 \rangle\, , \nn \\
& \qquad \langle \mathrm{III}_0 | \mathrm{IV}_2 | \mathrm{IV}_2 \rangle \, , \ \langle \mathrm{IV}_1 | \mathrm{IV}_2 | \mathrm{IV}_2 \rangle \, , \ \langle \mathrm{IV}_2 | \mathrm{IV}_2 | \mathrm{IV}_2 \rangle \, ,  \nn 
\end{align}
where we chose to sort 2-cubes with similar geometric characteristics together. The $\mathrm{I}_2$ class corresponds to the intersection of two finite distance divisors; an example of such a singularity has been considered in \cite{Curio:2000sc} as the collision of two conifold divisors. The boundaries of the coni-LCS class arise when one starts at the large complex structure point and sends one modulus away to a conifold locus; examples of such singularities have recently been considered in the physics literature in \cite{Demirtas:2020ffz,Blumenhagen:2020ire}. The $\mathrm{II}_1$ class includes the Seiberg-Witten point that played an important role in the geometrical engineering of field theories \cite{Kachru:1995fv}. Finally, each of the 2-cubes in the LCS class can be realized for particular values of the intersection numbers $\cK_{ijk}$ of some mirror Calabi-Yau manifolds. For illustration, we have depicted how a 2-cube characterizes the intersection of two boundary divisors in figure \ref{fig:2cube}.

\begin{figure}[h!]
\begin{center}
\includegraphics[width=6cm]{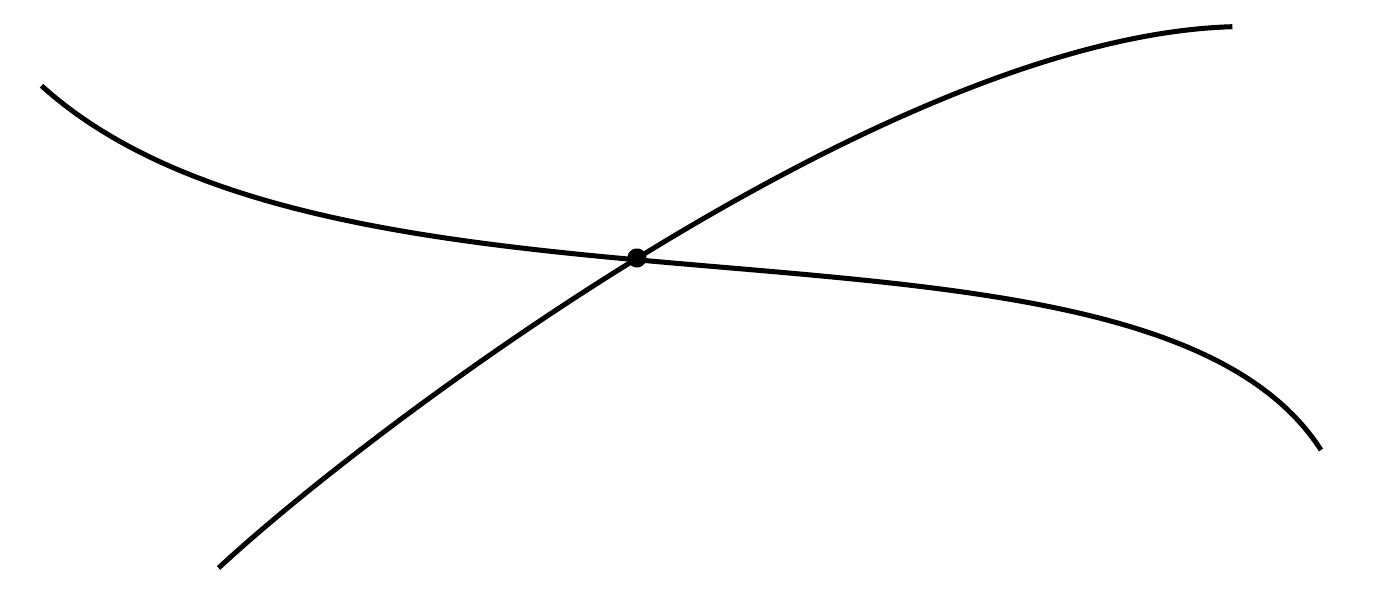}
\end{center}
\begin{picture}(0,0)\vspace*{-0.8cm}
\put(90,83){\small $\mathrm{I}_1$}
\put(250,91){\small $\mathrm{IV}_1$}
\put(166,73){\small $\mathrm{IV}_2$}
\end{picture}\vspace*{-1.0cm}
\caption{\label{fig:2cube}Example of the 2-cube $\langle \mathrm{I}_1 | \mathrm{IV}_2 | \mathrm{IV}_1 \rangle$ that characterizes a two-moduli coni-LCS boundary: A $\mathrm{I}_1$ and $\mathrm{IV}_1$ divisor in moduli space intersect at a $\mathrm{IV}_2$ singularity.}
\end{figure}

\chapter{Hodge theory in strict asymptotic regimes}\label{sec:sl2splitting}

In this section we discuss the final component of asymptotic Hodge theory we need: the sl(2)-orbit theorem of Cattani, Kaplan and Schmid \cite{CKS}. It provides us with a powerful framework to describe how couplings behave near the boundary in strict asymptotic regimes -- a refinement over the asymptotic regimes $y^i\gg 1$ used in the nilpotent orbit approximation before. The proof of this theorem is rather involved, so let us simply present its results here first and explain how to put them to use. The procedure to construct the sl(2)-orbit approximation in the multi-moduli case is postponed to section \ref{sec_tech_details}. We give a reformulated version of the one-modulus proof in a physical language in \cite{Grimm:2021ikg}. For our purposes, the main structures underlying the sl(2)-orbit theorem are
\begin{enumerate}
\item a set of $n$ commuting sl$(2,\mathbb{R})$-algebras, 
\item a pure Hodge structure $H^{p,q}_\infty$ on $H$ associated to the boundary.
\end{enumerate}
The sl$(2,\mathbb{R})$-algebras allow us to pick a special real basis that decomposes $H$ such that elements grouped in the same subspace are characterized by similar asymptotic behavior as we approach the boundary. On the other hand, the Hodge decomposition into spaces $H^{p,q}_\infty$ associated to the boundary allows us to fix the order one coefficients that appear with these parametrical scalings. 

\section{sl$(2,\mathbb{R})$-triples and boundary Hodge structures}\label{sec:boundarydata}
In this section we describe the algebraic structures associated to the boundary under consideration. We introduce the sl$(2,\mathbb{R})$-triples and boundary Hodge structure, and explain how this data arises from the Deligne splittings associated to the boundary. We also include several useful decompositions and identities using these structures, which will prove to be useful in later chapters of this thesis.

\subsubsection*{Strict asymptotic regimes}
Let us begin by introducing the regime of validity for this sl(2)-orbit approximation, or sl(2)-approximation for short. In addition to taking large saxion values $y^i\gg1$, we implement a hierarchy among the individual saxions as $y^1 \gg y^2 \gg \ldots \gg y^n$, which we refer to as the \textit{strict asymptotic regime}. We can formalize this limit into the notion of a growth sector near the singularity, given by
\begin{equation}\label{eq:growthsector}
\cR_{12\cdots n} = \big\{t^i = x^i+iy^i\,\Big|\, \frac{y^1}{y^2} > \lambda\, ,  \, \, \ldots\, , \ \frac{y^{n-1}}{y^n} > \lambda \, , \ y^n > \lambda, \, | x^i | <1 \big\}\, ,
\end{equation}
where we restricted the axions $x^i$ to a finite interval. Within these sectors the sl(2)-approximation describes how physical couplings grow parametrically; corrections appear at order $1/\lambda$ or higher, so by taking $\lambda \gg 1$ we ensure that it gives a good estimate in the strict asymptotic regime. Let us also point out the similarity between this growth sector and the sequential limits we took in \eqref{eq:enhancementchain}, allowing us to naturally associate an enhancement chain to each strict asymptotic regime. For a different ordering of the saxions this chain changes, and accordingly the boundary data changes as well. We therefore associate the algebraic data to the boundary of the strict asymptotic regime \eqref{eq:growthsector}, with the understanding that other orderings produce different results even though the limits all end up at $t^1,\ldots,t^n=i\infty$.

\subsubsection*{sl$(2,\mathbb{R})$-triples}
Given a growth sector, we next describe the structures we can associate to its asymptotic boundary. The first structure is provided by sl$(2,\mathbb{R})$-algebras. Taking the defining data of the nilpotent orbit as input -- the log-monodromy matrices $N_i$ and the limiting filtration $F^p_0$ -- it was shown in \cite{CKS} that one can construct
\begin{equation}\label{eq:sl2s}
\text{$n$ commuting sl$(2,\mathbb{R})$-triples:}\qquad (N_i^{-}, N_i^{+}, N^0_i) \, , \quad i=1,\ldots,n\, .
\end{equation}
These $sl(2,\mathbb{R})$-triples satisfy the standard commutation relations 
\begin{equation}\label{eq:sl2comm}
[N_i^0, N_i ^{\pm}]= \pm 2 N_i^{\pm}\, , \quad [N_i^+,N_i^-]=Y_i\, ,
\end{equation}
with $N_i^0$ the weight operator, and $N_i^\pm$ the raising and lowering operators. The procedure to obtain these sl$(2,\mathbb{R})$-triples is rather non-trivial, and is reviewed in detail in section \ref{sec_tech_details} including an illustrative example in section \ref{sec_aht_example}.

Let us nevertheless give a brief sketch of the origin of this algebraic structure. Recall that the log-monodromy matrices $N_i$ act on the Deligne splitting $I^{p,q}$ by lowering the row $p+q$ by two as in \eqref{eq:NIpq}. This suggests one can think of the $N_i$ as lowering operators, with the row $p+q$ as the weight under the sl$(2,\mathbb{R})$-algebras. Although this intuition is not far off, we have to take care of some subtleties first. For instance, recall that the Deligne splitting is not necessarily $\mathbb{R}$-split \eqref{Ipqbar}, so the resulting weight operator would not be real. The crucial insight is that we can always rotate a Deligne splitting to a special $\mathbb{R}$-split known as the sl(2)-split. This is accomplished by introducing two algebra-valued matrices $\zeta, \delta$ that rotate the filtration $F^p_0$ as
\begin{equation}\label{eq:Ftilde}
\tilde{F}^p_0 = e^{\zeta} e^{-i\delta} F^p_0\, .
\end{equation}
The purpose of the phase operator $\delta$ is to rotate the Deligne splitting to an $\mathbb{R}$-split \eqref{Ipq_Rsplit}. However, such an $\mathbb{R}$-split is not unique. This choice is made by $\zeta$, which selects a particular $\mathbb{R}$-split known as the sl(2)-split. We can fix this $\zeta$ componentwise in terms of $\delta$ via \eqref{zeta}. The idea is now that we get an sl(2)-split Deligne splitting for each singularity in the enhancement chain \eqref{eq:enhancementchain} as
\begin{equation}\label{eq:sl2Ipqs}
\tilde{I}^{p,q}_{(k)}\, , \quad k=1,\ldots, n\, .
\end{equation}
The weight operators  $N^0_{(k)}=N^0_{1}+\ldots N^0_{k} $ are then indeed defined by
\begin{equation}\label{eq:defN0}
\omega_{p,q} \in \tilde{I}^{p,q}_{(k)}: \quad N^0_{(k)} \omega_{p,q} = (p+q-D)\omega_{p,q}\, .
\end{equation}
In practice one has to start from the last Deligne splitting $\tilde{I}^{p,q}_{(n)}$ and go iteratively through the Deligne splittings at the previous steps and make them sl(2)-split. Let us stress that the rotation by the $\zeta$ to the sl(2)-splits here is crucial to ensure that the resulting sl$(2,\mathbb{R})$-triples commute. For now let us simply take a shortcut and assume that the sl(2)-data has been constructed: the sl(2)-split Deligne splittings are then related via their limiting filtrations as\footnote{Note that knowing the limiting filtrations $\tilde{F}^p_{(k)} $ is sufficient to fix the Deligne splittings, since the monodromy weight filtration $W(N_{(k)})$ is already fixed by the $N_i$. Moreover, we point out that the monodromy weight filtrations of the log-monodromy matrices and lowering operators agree: $W(N_{(k)}) =W(N_{(k)}^-)$. }
\begin{equation}\label{eq:Frecursion}
\tilde{F}^p_{(k)} = \bigoplus_{r \geq p} \bigoplus_s \tilde{I}^{r,s}_{(k)}: \qquad \tilde{F}^p_{(k)} = e^{i N_{k+1}^-}\tilde{F}^p_{(k+1)} \, .
\end{equation}
To gain some intuition for the above rotation, let us replace $N_{k+1}^-$ by the log-monodromy matrix $N_{k+1}$ for a moment. We can then think of this recursion as taking the nilpotent orbit \eqref{NilpOrbitGen} and setting a saxion to one $y^{k+1}=1$. In other words, we can interpret it as a limit for the first $k$ saxions $y^1 , \ldots , y^k$ while the others are kept finite, for which we should indeed recover the Deligne splitting at step $k$, see also footnote \ref{partial_limit} in chapter \ref{chap:mhs}. The precise relation between this exponential factor $e^{i N_{k+1}^-}$ and the saxion dependence will be given later with \eqref{eq:eyNey}, when we describe the parametrical behavior in the strict asymptotic regime.

Finally, let us justify this identification between the lowering operators $N_{i}^-$ and the log-monodromy matrices $N_{i}$. For the first element one indeed has $N_1^-=N_1$, while for the others this relation is slightly more intricate. We can decompose $N_i$ into eigenstates $N_{i,-\ell}$ of the previous weight operator $N_{(i-1)}^0$ as
\begin{equation}\label{eq:Ndecompose}
N_i = N_i^- + \sum_{\ell \geq 2} N_{i, -\ell}\, ,  \qquad [N_{(i-1)}^0, N_{i,-\ell}]=- \ell N_{i,-\ell}\, ,
\end{equation}
with $N_i^- =N_{i,0}$. This allowed us to remove precisely the pieces of $N_i$ that do not commute with this weight operator, such that the remainder $N_i^-$ does commute. And as a simple check, the identity for $N_1$ is then understood by noting we have a trivial weight operator $N_{(0)}^0 = 0$. 

\subsubsection*{Decomposition into sl(2)-eigenspaces}
We can now use these sl$(2,\mathbb{R})$-algebras to decompose $H$ into real eigenspaces of their weight operators $N_i^0$. This decomposition can be written out as\footnote{This decomposition is independent of $t^i$ but can vary with changes in the spectator moduli $\zeta^{k}$. }
\begin{equation}\label{eq:Sl2Decomp} 
H_{\mathbb{R}}= \bigoplus_{\Bell \in \cE} V_{\Bell}\, , \qquad \cE = \{ \Bell=(\ell_1, \ldots, \ell_n) : \ -D \leq \ell_i \leq D \}\, ,
\end{equation}
where we introduced an index set $\cE$ to denote the allowed values for the weights $\ell_i$. The eigenspaces $V_{\Bell}$ are then defined by the action of the weight operator as
\begin{equation}\label{eq:LevelOp}
v_{\Bell} \in V_{\Bell}:\qquad N^0_{(i)} v_{\Bell} = \ell_i v_{\Bell}\, .
\end{equation}
The range for the weights in \eqref{eq:Sl2Decomp} follows from the fact that the weight operators are constructed with the sl(2)-split Deligne splittings via \eqref{eq:defN0}, for which $0\leq p+q\leq 2D$. One can refine this index set by noting that the ranges ranges can be correlated with the singularity types occurring in the enhancement chain \eqref{eq:enhancementchain}. Concretely, for the Calabi-Yau threefold classification in table \ref{table:HDclass} we have that a singularity type at step $k$ yields
\begin{equation}
\text{I, II}: \ -1 \leq \ell_{(k)} \leq 1\, , \quad \text{III}: \ -2 \leq \ell_{(k)} \leq 2\, , \quad \text{IV}: \ -3 \leq\ell_{(k)} \leq 3\, ,
\end{equation}
where we wrote $\ell_{(k)}=\ell_1+\ldots+\ell_k$. Let us also write down how the lowering operators $N_i^-$ act on these sl(2)-eigenspaces $V_{\Bell}$. From the commutation relations of the sl$(2,\mathbb{R})$-triples \eqref{eq:sl2comm} one can straightforwardly show that
\begin{equation}\label{eq:lowering}
N^-_i V_{\Bell} \subseteq V_{\Bell-2\Bdelta_i} \, ,
\end{equation}
where $\Bdelta_i =(0, \ldots, 0,1,0,\ldots, 0)$ denotes the $n$-dimensional unit vector in the $i$-th direction. And as a final observation, we point out that we can write down orthogonality conditions between the subspaces $V_{\Bell}$ under the bilinear pairing. For two elements $v_{\Bell}\in V_{\Bell}$ and $v_{\mathbf{r}} \in V_{\mathbf{r}}$ we find that
\begin{equation}\label{eq:orthogonality}
\langle v_{\Bell} , v_{\mathbf{r}} \rangle = 0\, , \quad \text{unless $\Bell = -\mathbf{r}$}\, ,
\end{equation}
which can be shown by using that the weight operator is an infinitesimal isometry, i.e.~it obeys $\langle \cdot, N_i^{(0)} \cdot \rangle  = - \langle N_i^{(0)} \cdot, \cdot \rangle$.

\subsubsection*{Boundary Hodge structure}
Having discussed the sl$(2,\mathbb{R})$-triples and the corresponding sl(2)-decomposition, we next introduce the pure Hodge structure associated to the boundary. It is given by a Hodge decomposition of the middle cohomology as\footnote{ This decomposition is independent of the moduli $t^i$ sent to their limit, but can still vary with changes in the moduli $\zeta^k$ that are kept finite, similar to \eqref{eq:Sl2Decomp}.}
\begin{equation}\label{eq:boundaryHodge}
H= \bigoplus_{p+q=D} H^{p,q}_\infty\, ,
\end{equation}
where $\overline{H^{p,q}_\infty}=H^{q,p}_\infty$. For later reference it is convenient to introduce a boundary charge operator $Q_\infty$ that acts on these subspaces as
\begin{equation}\label{eq:chargeop}
v^{p,q} \in H^{p,q}_\infty: \quad Q_{\infty}\,  v^{p,q} = \big(p-D/2\big) \, v^{p,q}\, .
\end{equation}
The boundary Hodge star $C_\infty$ associated to \eqref{eq:boundaryHodge} is then obtained by exponentiating
\begin{equation}\label{eq:Cinf}
C_\infty = e^{\pi i Q_\infty}\, ,
\end{equation}
such that it acts on the vector spaces $H^{p,q}_\infty$ as
\begin{equation}\label{eq:Cinftyaction}
v^{p,q} \in H^{p,q}_\infty: \quad C_{\infty} v^{p,q} = i^{p-q} v^{p,q}\, .
\end{equation}
The origin of this pure Hodge structure can be traced back to the sl(2)-split Deligne splittings \eqref{eq:sl2Ipqs} we encountered before with the sl$(2,\mathbb{R})$-triples. We already noticed we can move between singularities in the enhancement chain \eqref{eq:enhancementchain} via application of the lowering operators $N_i^-$ as described by \eqref{eq:Frecursion}. In particular, it allows us to consider the Deligne splitting associated with the first type $\mathrm{I}_0$, which is a pure Hodge structure. This leads us to single out a particular element in this recursion
\begin{equation}\label{eq:boundaryfiltration}
F^p_\infty = e^{iN_{(n)}^-} \tilde{F}^p_0\,  ,
\end{equation}
with $\tilde{F}^p_0$ the sl(2)-split limiting filtration at step $n$.\footnote{In fact, note that one can write down \eqref{eq:boundaryfiltration} more generally as $F^p_\infty = e^{iN_{(k)}^-} \tilde{F}^p_{(k)}$ for any step $k$.} It determines the pure Hodge structure associated to the boundary of the strict asymptotic regime via the usual intersection $H^{p,q}_\infty = F^p_\infty \cap \bar{F}^q_\infty$.

\subsubsection*{Boundary Hodge structure and the sl(2)-decomposition}
Next let us look at how the sl$(2,\mathbb{R})$-algebras and the boundary Hodge structure cooperate. We can summarize this succinctly by writing down the commutation relations between the charge operator \eqref{eq:chargeop} and the elements of the sl$(2,\mathbb{R})$-triples \eqref{eq:sl2s} as
\begin{equation}\label{eq:QNcomm}
[Q_\infty, N^0_i] = i(N_i^+ + N_i^-) \, , \qquad [Q_\infty, N_i^{\pm}] = -\frac{i}{2} N^0_i \, .
\end{equation}
For our purposes, it will be particularly useful to know how the operator $C_\infty$ acts on the sl(2)-eigenspaces $V_{\Bell}$ introduced in \eqref{eq:Sl2Decomp}. To this end, we can show from \eqref{eq:QNcomm} by using the Baker-Campbell-Hausdorff formula that
\begin{equation}
N_i^0 C_\infty = - C_\infty N_i^0\, , \quad N_i^+ C_\infty = - C_\infty N_i^-\, .
\end{equation}
From the first relation one then immediately obtains
\begin{equation}\label{eq:Cv}
v_{\Bell} \in V_{\Bell} : \quad C_\infty v_{\Bell} \in V_{-\Bell}\, .
\end{equation}
We can also use this boundary Hodge star operator $C_\infty$ to define an inner product on $H$ as $\langle \cdot, C_\infty \cdot \rangle$. From the orthogonality conditions on the bilinear pairing \eqref{eq:orthogonality} and the way $C_\infty$ acts on the sl(2)-eigenspaces we then find that
\begin{equation}\label{eq:Cinftyorthogonality}
\langle C_{\infty} v_{\Bell},  v_{\mathbf{r}} \rangle = 0\, , \quad \text{unless $\Bell = \mathbf{r}$}\, .
\end{equation}
with $v_{\Bell} \in V_{\Bell}$ and $v_{\mathbf{r}} \in V_{\mathbf{r}}$.

\subsubsection*{Primitive subspaces}
It will also prove to be useful to break down the sl(2)-decomposition \eqref{eq:Sl2Decomp} further into primitive components and their descendants under the sl$(2,\mathbb{R})$-triples. The highest-weight states with respect to $N_1^-, \ldots, N_n^-$ are then defined as
 \begin{equation}\label{eq:primitive}
 P_{\Bell} = V_{\Bell}  \cap \ker[ (N_1^-)^{
\ell_1+1}] \cap \ldots \cap \ker [(N_n^-)^{
\ell_n+1}]\, .
 \end{equation}
On a given primitive subspace $P_{\Bell}$ the lowering operators $N_i^-$ can thus act at most $\ell_i$ times. This allows us to write out the sl(2)-decomposition \eqref{eq:Sl2Decomp} in terms of these primitive subspaces and their images under $N_i^-$ as
 \begin{equation}\label{eq:primitivedecomp}
  H_{\mathbb{R}} = \bigoplus_{\Bell \in \cE_{\rm prim}}\,  \bigoplus_{k_1=0}^{\ell_1} \cdots \bigoplus_{k_n=0}^{\ell_n}  (N^-_1)^{k_1} \cdots (N^-_n)^{k_n} P_{\Bell}\, ,
 \end{equation}
 where $\cE_{\rm prim}$ denotes the index set for the primitive states
 \begin{equation}\label{Eprim}
  \cE_{\rm prim} = \{\Bell = (\ell_1, \ldots, \ell_n):  \sum \ell_i \leq D\,, \ \ell_i \geq 0\,  \}\, .
 \end{equation}
In \eqref{Eprim} the condition on the sum of the $\ell_i$'s follows from the relation \eqref{eq:defN0} between the Deligne splitting and the weights under the sl$(2,\mathbb{R})$-triples, similar to the index set $\cE$ in \eqref{eq:Sl2Decomp}. Note that by going to the decomposition into primitive components and descendants we see a much more restricted pattern on the allowed weights. As an example, in the case $D=2$ of a K3 surface we find as descendants for the possible highest-weight states given in table \ref{table:K3}. We see that only two weights can be non-vanishing at most: when one weight is non-vanishing it can take any value $-2 \leq \ell_i \leq 2$, while if two weights are non-vanishing we can only have $-1 \leq \ell_i \leq 1$. Similar restrictions on the pattern of weights carry over to higher-dimensional Calabi-Yau manifold, only the set of allowed highest-weight states becomes significantly larger.

\begin{table}[htb]
\centering
\renewcommand{\arraystretch}{1.3}
\scalebox{0.85}{
\begin{tabular}{|c||c|c|}
\hline
primitive subspace & descendants & sl(2)-eigenspaces  \\
\hline \hline
$P_{\mathbf{0}}$ & & $V_{\mathbf{0}}$ \\ \hline
$P_{\Bdelta_i}$ & {\color{red} $N_i^- P_{\Bdelta_i} $ }& $V_{\Bdelta_i}$\,,\ {\color{red} $V_{-\Bdelta_i}$} \\ \hline
$P_{2\Bdelta_i}$ & {\color{red}$N_i^- P_{\Bdelta_i}$}\,,\  {\color{blue}$(N_i^-)^2 P_{\Bdelta_i}$}  & $V_{2\Bdelta_i}$\,,\ ${\color{red}V_{\mathbf{0}}}$\,,\  ${\color{blue}V_{-2\Bdelta_i}}$\\ \hline
$P_{\Bdelta_i+\Bdelta_j}$ & ${\color{red}N_i^- P_{\Bdelta_i+\Bdelta_j}}$\,,\ ${\color{blue}N_j^- P_{\Bdelta_i+\Bdelta_j}}$\,,\ ${\color{ForestGreen}N_i^- N_j^-  P_{\Bdelta_i+\Bdelta_j}}$ & ${\color{red}V_{-\Bdelta_i+\Bdelta_j}}\, ,$ ${\color{blue}V_{\Bdelta_i-\Bdelta_j} }$\,,\   ${\color{ForestGreen}V_{-\Bdelta_i-\Bdelta_j} }$ \\ \hline
\end{tabular}}
\renewcommand{\arraystretch}{1}
\caption{All possible primitive subspaces occurring for Deligne splittings of K3 surfaces. We indicated their descendants, and moreover cross-referenced the matching sl(2)-eigenspaces with colors by using \eqref{eq:lowering}. }\label{table:K3} 
\end{table}

\paragraph{Boundary Hodge star.} As next step, we can study the action of the boundary Hodge star operator $C_\infty$ on the highest-weight states. To this end, it is convenient to first introduce the splitting over the last Deligne diamond $I^{p,q}_{(n)}$ as
\begin{equation}
P_{\Bell} = \bigoplus_{p+q=\ell_n+D} P_{\Bell}^{p,q}\, .
\end{equation}
From the definition of the boundary Hodge filtration \eqref{eq:boundaryfiltration} one can then relate these primitive subspaces to the boundary Hodge structure $H^{p,q}_\infty$ as\footnote{The fact that this subspace lies in $F^p_\infty$ follows directly by using \eqref{eq:Frecursion}. On the other hand, it requires slightly more work to show that it lies in $\bar{F}^{D-p}_\infty$. To see this, note that descendants of $\bar{P}^{q,p}_{\Bell}=P^{p,q}_{\Bell}$ arise by acting at most $p+q-D$ times with the $N_i^-$. It means this highest-weight state and all its descendants lie in $\bar{F}^{D-p}_\infty$ (since we lower to $q-(p+q-D)=D-p$), and thus in particular the linear combination in \eqref{eq:eiNprim}.}
\begin{equation}\label{eq:eiNprim}
e^{iN^-_{(n)}} P_{\Bell}^{p,q} \subseteq H^{p, D-p}_{\infty}\, , \qquad p \geq q\, .
\end{equation} 
The action of $C_\infty$ on this primitive subspace is then described via \eqref{eq:Cinftyaction} as
\begin{equation}
v^{p,q}_{\Bell} \in P^{p,q}_{\Bell}: \quad C_\infty \, e^{iN^-_{(n)}} v_{\Bell}^{p,q} = i^{2p-D} \, e^{iN^-_{(n)}} v_{\Bell}^{p,q} \, .
\end{equation}
again assuming $p\geq q$. Combined with \eqref{eq:Cv} we can then show by expanding the exponentials and identifying sl(2)-eigenspaces that\footnote{This identity is derived by expanding the exponential in $e^{iN_{(n)}^-} v_{\Bell}^{p,q}$. The weights of the various terms with respect to $N^0_i$ follow by starting with $\Bell$ for the highest-weight state, and lowering according to \eqref{eq:lowering} under the action of $N_i^-$. By using that the Weil operator acts as $C_{\infty} V_{\Bell} \subseteq V_{-\Bell}$ one can then match terms with the same weights, resulting in the given identity.} 
\begin{equation}\label{eq:Cinftyprim}
v_\ell^{p,q}\in P^{p,q}_\ell: \quad C_\infty \prod_{i=1}^n \frac{(iN^-_i)^{k_i}}{k_i!} v_\ell^{p,q} = i^{2p-D}  \prod_{i=1}^n \frac{(iN^-_i)^{\ell_i-k_i}}{(\ell_i -k_i)!}  v_\ell^{p,q}\, .
\end{equation}
with again $p\geq q$. Thus the action of the boundary Hodge star $C_\infty$ can be described in a very precise way in a basis  of primitive subspaces and their descendants. To give some intuition, let us note the similarity of this relation with the K\"ahler form identity $\star J^k/k! = J^{D-k}/(D-k)!$ on a $D$-dimensional K\"ahler manifold. In fact, for Calabi-Yau manifolds this is precisely one of the manifestations of mirror symmetry between their K\"ahler and complex structure moduli space, where the operation of wedging with K\"ahler form $J$ is replaced by acting with the log-monodromy matrices.

\subsubsection*{Perturbative term $a_0$ in the periods}
To close off our discussion on the algebraic structures that underlie the boundary, let us specialize the above findings to the vector $a_0$. This vector determines the leading part of the periods \eqref{eq:periodsexpansion}, fixing for instance the leading polynomial behavior of the K\"ahler potential in \eqref{eq:kahlerexpansion}, so it is deserving of some special attention. We can locate it as the element that spans the one-dimensional vector space $F^D_0$. Going to the sl(2)-split with the matrices $\zeta,\delta$ means that
\begin{equation}
\tilde{a}_0 \equiv e^{\zeta}e^{i\delta} a_0 \in \tilde{F}^D_0\, .
\end{equation}
This $D$-form $\tilde{a}_0$ has a well-defined location in the sl(2)-eigenspaces as
\begin{equation}\label{eq:a0position}
\Re \tilde{a}_0\, , \, \Im \tilde{a}_0 \ \in P_{\mathbf{d}}^{D,d_n}\, , 
\end{equation}
where $ \mathbf{d}=(d_1, d_2-d_1, \dots , d_n-d_{n-1})$, with the $d_i$ defined by \eqref{eq:ddef} for each $N_i$. Following \eqref{eq:boundaryfiltration} we can apply lowering operators $N_i^-$ to construct another $D$-form out of $a_0$. This differential form can be placed in one of the subspaces $H^{p,q}_\infty$ of the boundary Hodge decomposition according to \eqref{eq:eiNprim} as
\begin{equation}\label{Sl2-orbit}
\Omega_\infty \equiv e^{iN_{(n)}^-} \tilde{a}_0 \in H^{D,0}_\infty\, .
\end{equation}
This means that according to \eqref{eq:Cinftyaction} the boundary Hodge star operator acts on this boundary $(D,0)$-form as $C_\infty \Omega_\infty = i^D \Omega_\infty$. Analogous to \eqref{eq:Cinftyprim} we can then fix how $C_\infty$ acts on the highest-weight state $\tilde{a}_0$ and its descendants. Its action is compactly summarized by the identity
\begin{equation}\label{eq:Cinftyidentity}
 C_{\infty}\,\prod_{i=1}^n \frac{i^{k_i}}{k_i!}(N_i^-)^{k_i} \, \tilde{a}_0= -i \prod_{i=1}^n \frac{i^{d_i-d_{i-1}-k_i}}{(d_i-d_{i-1}-k_i)!} (N_i^-)^{d_i-d_{i-1}-k_i}\, \tilde{ a}_0\, .
\end{equation}
This identity will prove to be very useful in fixing the charge-to-mass ratios of a particular set of BPS states in chapter \ref{chap:WGC}.

\section{Approximation by the sl(2)-orbit}
Having covered the intricacies of the boundary data, we are ready to utilize these algebraic structures to describe how couplings behave parametrically in the strict asymptotic regime with the sl(2)-orbit theorem of Cattani, Kaplan and Schmid \cite{CKS}. The sl(2)-decomposition \eqref{eq:Sl2Decomp} together with the boundary Hodge structure \eqref{eq:boundaryHodge} allows us to write down simple asymptotic behaviors. For our purposes here the statement of the sl(2)-orbit theorem can be formulated in two equivalent ways, either by using the Hodge star operator $C$ or the Hodge structure $H^{p,q}$ itself.

\subsubsection*{sl(2)-approximated Hodge star operator}
 Let us begin with the more practical result from a physical point of view, which gives us an approximation for the Hodge star in the strict asymptotic regime $\lambda \gg 1$. It then states that a good estimate is given by the sl(2)-approximated operator
\begin{align}
C_{\rm sl(2)}= e^{x^i N^-_i} e^{-1}(y) C_{\infty} e(y) e^{-x^i N^-_i}\,, \label{eq:Csl2ToCInf}
\end{align}
where we introduced a saxionic scaling operator
\begin{equation}\label{eq:ey}
e(y) = \exp\big[\,  \frac{1}{2}\sum_i \log[y^i] N^0_i\, \big]\, .
\end{equation}
This operator encodes the parametrical scaling in $y^i$ for the Hodge star. For instance, it acts on elements of one of the sl(2)-eigenspaces by multiplication as
\begin{equation}\label{eq:eyaction}
v_{\Bell} \in V_{\Bell} : \quad e(y) v_{\Bell} = (y^1)^{\ell_1/2} \cdots (y^n)^{\ell_n/2} v_{\Bell}\, .
\end{equation}
On the other hand, the dependence on the axions $x^i$ arises in a similar way compared to the nilpotent orbit approximation \eqref{eq:Cnilx}, only with the log-monodromy matrices $N_i$ replaced by the lowering operators $N_i^-$. 

To gain some intuition for the sl(2)-approximated Hodge star in \eqref{eq:Csl2ToCInf}, let us investigate how it acts on elements of sl(2)-eigenspaces, where we set the axions to zero $x^i=0$ for simplicity. We then find by a straightforward computation for the sl(2)-Hodge norm that
\begin{equation}\label{eq:growth}
v_{\Bell} \in V_{\Bell}: \quad \| v_{\Bell} \|_{\rm sl(2)}^2 \equiv \langle v_{\Bell}, C_{\rm sl(2)} v_{\Bell} \rangle =   (y^1)^{\ell_1} \cdots (y^n)^{\ell_n} \langle v_{\Bell}, C_{\infty} v_{\Bell} \rangle\, .
\end{equation} 
Here we clearly see the advantage of working with the sl(2)-approximation: (1) the scaling in the saxions is determined by the weights under the sl$(2,\mathbb{R})$-triples, and (2) the corresponding leading coefficient is fixed by the Hodge star norm with $C_\infty$ on the boundary. This scaling behavior suggests to make the following split in the weights of the sl(2)-eigenspaces
\begin{equation}
V_{\rm heavy} = \bigoplus_{\Bell \in \cE_{\rm heavy}} V_{\Bell}\, , \quad V_{\rm light} = \bigoplus_{\Bell \in \cE_{\rm light}} V_{\Bell} \, , \quad V_{\rm rest} = \bigoplus_{\Bell \in \cE_{\rm rest}} V_{\Bell}\, ,
\end{equation}
where we defined index sets
\begin{equation}
\begin{aligned}
\cE_{\rm heavy} &= \{ \Bell \in \cE: \, \ell_{(1)}\geq 0 \, , \ldots , \ell_{(n)}\geq 0, \text{ and at least one $\ell_{(i)} \geq 1$} \}\, , \\
\cE_{\rm light} &= \{ \Bell \in \cE: \, \ell_{(1)}\leq 0 \, , \ldots , \ell_{(n)}\leq 0, \text{ and at least one $\ell_{(i)} \leq -1$} \}\, ,
\end{aligned}
\end{equation}
where we wrote $\ell_{(k)} = \sum_{i=1}^k \ell_i$. The remainder of sl(2)-eigenspaces is contained in $\cE_{\rm rest}=\cE-(\cE_{\rm heavy}\cup \cE_{\rm light})$. The purpose of this split is that the integers $\ell_{(i)}$ describe precisely whether one of the ratios $y^i/y^{i-1} > \lambda \gg 1$ in the growth sector \eqref{eq:growthsector} appears with a positive or negative exponent in the growth estimate \eqref{eq:growth}. When we require all of them to be non-negative and one strictly positive, we are guaranteed to have a diverging Hodge norm for elements in $V_{\rm heavy}$ along any path within the growth sector with $\lambda \to \infty$. Similarly the vector space $V_{\rm light}$ contains all sl(2)-eigenspaces for which the Hodge norm vanishes asymptotically for any such path. The remainder $V_{\rm rest}$ is made up of elements whose Hodge norm increases or decreases asymptotically (or stays finite) depending on the choice of path in the strict asymptotic regime.

\subsubsection*{sl(2)-approximated Hodge structure}
Let us now back up the above observations by considering the limit $\lambda \to \infty$ of the strict asymptotic regime \eqref{eq:growthsector}, and argue that the nilpotent orbit approximation converges to the sl(2)-approximation given here. In order to do this, let us first write down the more formal formulation of the sl(2)-orbit theorem that typically appears in the math literature, in contrast to the Hodge star approximation in \eqref{eq:Csl2ToCInf}. It works from the perspective of the Hodge filtration, and states that we can approximate the nilpotent filtration $F^p_{\rm nil}$ in the strict asymptotic regime by
\begin{equation}\label{eq:sl2orbit}
F^{p}_{\rm sl(2)}(t) = e^{t^i N^{-}_{i}} \tilde{F}^{p}_{0}  \, .
\end{equation}
where $\tilde{F}^{p}_{0}$ is the sl(2)-split filtration defined in \eqref{eq:Ftilde}. The sl(2)-approximated Hodge star operator we introduced in \eqref{eq:Csl2ToCInf} is then precisely the Weil operator associated with the Hodge decomposition $H^{p,q}_{\rm sl(2)} = F^{p}_{\rm sl(2)} \cap \bar{F}^{q}_{\rm sl(2)}$. This can be made manifest by cleverly rewriting \eqref{eq:sl2orbit} with the saxionic scaling operator as
\begin{equation}\label{eq:sl2orbitrewritten}
F^{p}_{\rm sl(2)} =   e^{x^i N^{-}_{i}} e^{-1}(y) F^p_{\infty}\, ,
\end{equation}
where we rewrote in terms of the boundary Hodge filtration \eqref{eq:boundaryfiltration}. This can be seen by using the Baker-Campbell-Hausdorff formula to show that 
\begin{equation}\label{eq:eyNey}
e^{-1}(y) e^{iN_{(n)}^-} e(y) = e^{i y^i N_i^-}\, , 
\end{equation}
and the fact that $e(y)\tilde{F}^p_0 = \tilde{F}^p_0$.\footnote{It is instructive for the reader to check this last identity for the element $\tilde{a}_0$ spanning $\tilde{F}^D_0$. It has a well-defined location \eqref{eq:a0position} in the sl(2)-eigenspaces, and thus $e(y)$ simply acts by multiplication with an overall saxion-dependent factor as in \eqref{eq:eyaction}, which is irrelevant for vector spaces. } In \eqref{eq:sl2orbitrewritten} the axion- and saxion-dependent factors now appear explicitly. Moreover, since these are real they both appear in $F^p_{\rm sl(2)}$ and its conjugate, and can therefore be extracted as a common factor as
\begin{equation}
H^{p,q}_{\rm sl(2)} = e^{x^i N^{-}_{i}} e^{-1}(y) H^{p,q}_\infty\, .
\end{equation}
We can now note that when applying $C_{\rm sl(2)}$ on these spaces this common factor directly cancels off against its inverse factor appearing on the right in \eqref{eq:Csl2ToCInf}. The boundary Hodge star $C_\infty$ can then act on the boundary Hodge structure $H^{p,q}_\infty$ according to the standard multiplication rule, after which the common moduli-dependent factor is reinstated. A summary of the different regimes and the relevant operators and subspaces has been provided in table \ref{table:regimes}.

\begin{table}[htb]
\centering
\renewcommand{\arraystretch}{1.3}
\begin{tabular}{|c||c|c|c|}
\hline
regime & asymptotic & strict asymptotic & boundary \\
\hline \hline
validity & $e^{2\pi i t^i} \ll 1$ & $y^{1} \gg ... \gg y^n \gg 1$ &  $t^i = i \infty$ \\ \hline
filtration & $F^p_{\rm nil}$ & $F^p_{\rm sl(2)}$ & $F^p_{\infty}$ \\ \hline
Hodge star & $C_{\rm nil}$ & $C_{\rm sl(2)}$ & $C_{\infty}$ \\ \hline
\end{tabular}
\renewcommand{\arraystretch}{1}
\caption{Summary of the different regimes. We indicated the corrections that can be dropped in each of these regimes, as well as what Hodge star (\eqref{eq:Cnilx}, \eqref{eq:Csl2ToCInf} and \eqref{eq:Cinf}) and limiting subspaces (\eqref{NilpOrbitGen}, \eqref{eq:sl2orbit} and \eqref{eq:boundaryfiltration}) should be used to describe these regimes.}\label{table:regimes} 
\end{table}

\subsubsection*{The strict asymptotic limit}
Having established how the Hodge star \eqref{eq:Csl2ToCInf} and Hodge filtration formulation \eqref{eq:sl2orbit} of the sl(2)-approximation are related, we can now get to the limit $\lambda \to \infty$ in the strict asymptotic regime. The idea is that the saxionic scaling operator \eqref{eq:ey} encodes how periods diverge or vanish asymptotically as we approach the boundary of the strict asymptotic regime. In particular, by multiplying with this operator we can strip off this singular behavior. For the moment we will set the axions to zero again $x^i = 0$, which we will later justify with \eqref{eq:eyxNey}. Then we find in the limit $\lambda \to \infty$ that
\begin{equation}\label{eq:Flimit}
\lim_{\lambda \to \infty} e(y) F^{p}_{\rm nil}= F^{p}_{\infty}\, .
\end{equation}
For the details we refer to the original article \cite{CKS}, but let us nevertheless give a heuristic argument:
 \begin{itemize}
\item As a first check, we can verify that this identity is self-consistent for the sl(2)-orbit: indeed by replacing $F^{p}_{\rm nil}$ with $F^{p}_{\rm sl(2)}$ in \eqref{eq:sl2orbitrewritten} we see that the saxionic scaling operators cancel out, and hence the boundary Hodge filtration $F^{p}_{\infty}$ remains. In particular, taking the limit $\lambda \to \infty$ to the boundary of the strict asymptotic regime was not even needed here, as this equation holds identically for the sl(2)-orbit when the axions are dropped. 
\item The strict asymptotic limit becomes important when we consider corrections of the nilpotent orbit to the sl(2)-orbit approximation. One can work out a similar identity as \eqref{eq:eyNey}, but with the log-monodromy matrices $N_i$ instead: the lower-weight pieces in $N_i$ under sl$(2,\mathbb{R})$-triples given in \eqref{eq:Ndecompose} then produce corrections of the order $1/\lambda$ under application of the saxionic scaling operator. The limit $\lambda \to \infty $ therefore ensures that we can drop precisely these corrections between the sl(2)-orbit and the nilpotent orbit, hence resulting in \eqref{eq:Flimit}.
\end{itemize}
Let us next elaborate on the axion dependence in this limit, where for simplicity we work just with the sl(2)-orbit. Including the axion-dependent factor in \eqref{eq:sl2orbitrewritten} then amounts to considering its adjoint under the group element $e(y)$, since by moving this saxionic scaling operator to the right we reduce to our discussion above. For this adjoint we can show by a similar computation as in \eqref{eq:eyNey} that
\begin{equation}\label{eq:eyxNey}
e(y) e^{x^i N_i^-} e^{-1}(y) = e^{\frac{x^i}{y^i} N_i^-}\, .
\end{equation}
Since our axions are bounded in \eqref{eq:growth}, this factor clearly approaches the identity matrix in the limit $\lambda \to \infty$, which sends all $y^i\to \infty$. Hence the axion-dependent factor can be ignored in the strict asymptotic limit, justifying why we set $x^i=0$ earlier. Similar arguments can be made when one works with the log-monodromy matrices $N_i$ in the nilpotent orbit rather than the lowering operators $N_i^-$, see for instance \cite{Grimm:2020ouv} for a detailed discussion in the physics literature. The extra terms in the relation \eqref{eq:Ndecompose} between these operators then lead again to suppressed corrections, since they are by construction of a lower weight under the sl$(2,\mathbb{R})$-triples.

\subsubsection*{Leading behavior of periods and the K\"ahler potential}
The above discussion was rather abstract, so let us specialize to the periods of the $(D,0)$-form for a moment, which for instance specify the leading behavior of the K\"ahler potential via \eqref{eq:kahlerexpansion}. By taking the element $\Omega_\infty \in H^{D,0}_\infty$ we can lift it to its sl(2)-approximated version in the strict asymptotic regime by using \eqref{eq:sl2orbitrewritten} as
\begin{align}\label{eq:sl2orbit3form}
\Omega_{\rm sl(2)} = (y^1)^{\tfrac{d_1}{2}} (y^2)^{\tfrac{d_2-d_1}{2}} \cdots (y^n)^{\tfrac{d_n-d_{n-1}}{2}}  e^{x^i N_i^-} e^{-1}(y)  \Omega_{\infty}\, .
\end{align}
Here we included an overall saxion-dependent factor which normalizes the term $\tilde{a}_0$ appropriately, i.e.~it precisely cancels out the saxionic scaling of $e^{-1}(y) \tilde{a}_0$.

Next we can use this leading behavior to obtain an estimate for the parametrical scaling of the K\"ahler potential in the strict asymptotic regime. By a straightforward computation we find that
\begin{equation}\label{eq:Kahlerpotasymp}
K_{\rm sl(2)}=-\log\left[ i^{-D} \langle \Omega_{\infty} , \bar{\Omega}_{\infty} \rangle \, (y^1)^{d_1} (y^2)^{d_2-d_1} \cdots (y^n)^{d_n-d_{n-1}}\right] \, ,
\end{equation}
where the integers $d_{i}$ are defined by \eqref{eq:ddef}. For later reference, let us record the precise form of this coefficient $\langle \Omega_{\infty} , \bar{\Omega}_{\infty} \rangle$ as
\begin{equation}\label{eq:Kinf}
\langle \Omega_{\infty} , \bar{\Omega}_{\infty} \rangle =  (2i)^{d_n} \langle  \prod_i \frac{(N_i^-)^{d_i-d_{i-1}}}{(d_i-d_{i-1})!} \tilde{a}_0 ,  \bar{\tilde{a}}_0 \rangle \, .
\end{equation}
Notice that the polarization conditions \eqref{eq:pol} with $ \tilde{a}_0 \in P^{D, d_n}$ then assure us that the leading coefficient of the K\"ahler potential \eqref{eq:Kahlerpotasymp} is positive.

It is tempting to now try to compute the K\"ahler metric by taking derivatives of \eqref{eq:Kahlerpotasymp}. However, as we stressed before, one should be careful with interchanging the order of taking derivatives and taking limits. In fact, the K\"ahler metric provides us with a good example why this order matters. Namely, \eqref{eq:Kahlerpotasymp} does not depend on the coordinate $y^{i}$ whenever $d_{i}-d_{i-1}=0$. The integers $d_{i}$ are bounded by $0 \leq d_{i} \leq d_{i+1} \leq D$, so this already happens if one considers limits where $D+1$ or more moduli are scaled at different rates. In particular, note that $d_{i}-d_{i-1}=0$ is therefore not only specific to finite distance singularities, but can also occur for infinite distance singularities such as the large complex structure point. Thus simply taking \eqref{eq:Kahlerpotasymp} can result in a degenerate K\"ahler metric, and one should include corrections when necessary in order to resolve this issue. These corrections can have a polynomial nature, but can also be of an exponentially small order as occurred in the example in \eqref{eq:examplecorrection} we encountered before.







\section{Algorithmic approach to the sl(2)-approximation}\label{sec_tech_details}
In this section we give the algorithmic procedure for the construction of the sl(2)-approximation in explicit examples. It is built around the rotation of any Deligne splitting to a special $\mathbb{R}$-split version, the sl(2)-splitting. This complex rotation is iterated over the steps in the enhancement chain \eqref{eq:enhancementchain}, i.e.~at each hierarchy of the saxions. Here we explain how to run this program, and how the relevant boundary data can be read off in the end.

\subsubsection*{Algorithm I: rotation to the sl(2)-splitting}
A crucial observation relevant for the sl(2)-orbit theorem \cite{CKS} is that any Deligne splitting can be rotated to a special $\mathbb{R}$-split, known as the sl(2)-split. This rotation is implemented by rotating the limiting filtration $F^p_0$ and consists of two steps: first we use a real phase operator $\delta$ to rotate to an $\mathbb{R}$-split, and then we rotate to the sl(2)-split via another real operator $\zeta$
\eq{
  \label{notation_8492}
  \fbox{$I^{p,q}$\vphantom{sl(2)}} 
  \hspace{10pt}\xrightarrow{\hspace{10pt}\delta\hspace{10pt}}\hspace{10pt}
  \fbox{$\hat I^{p,q}$ $=$ $\mathbb R$-split $I^{p,q}$\vphantom{sl(2)}} 
  \hspace{10pt}\xrightarrow{\hspace{10pt}\zeta\hspace{10pt}}\hspace{10pt}
  \fbox{$\tilde I^{p,q}$ $=$ sl(2)-split $I^{p,q}$} 
}  
\paragraph{$\mathbb{R}$-split.} Let us begin by performing the rotation to the $\mathbb{R}$-split. We introduce a grading operator $\cN^0$ to ascertain how the complex conjugation rule $\bar{I}^{p,q} = I^{q,p}$ is modified by \eqref{Ipqbar}. This grading 
operator will serve as a starting point for the weight operator of an sl(2)-triple. To be precise, the grading operator $\cN^0$ acts on an element of $I^{p,q}$ as follows
\begin{equation}\label{Hcomplex}
\cN^0 \, \omega_{p,q} = (p+q-3)\,  \omega_{p,q}\, , \hspace{40pt} \omega_{p,q} \in I^{p,q}\, .
\end{equation}
Clearly this means that when the Deligne splitting is not $\mathbb{R}$-split,  i.e.~$\bar{I}^{p,q} \neq I^{q,p}$, then $\cN^0$ is not a real operator since $I^{p,q}$ and $I^{q,p}$ are part of the same eigenspace of $\cN^0$. In fact, we can use the way that $\cN^0$ transforms under complex conjugation to determine how the $F^p_0$ should be rotated. We can achieve this by writing the transformation rule conveniently as 
\begin{equation}\label{Hconjugate}
\bar{\cN}^0 = e^{-2 i \delta} \cN^0 \op e^{+2i \delta}\, , 
\hspace{50pt}
\delta \in \mathfrak{sp}(2h^{2,1} +2,\bbR)\,,
\end{equation}
where $\delta$ denotes the phase operator of the rotation. Let us also note that $\delta$ commutes with all 
log-monodromy matrices $N_i$, i.e.~$[\delta, N_i]=0$. This operator can be decomposed with respect to its action on $I^{p,q}$ as follows
\begin{equation}\label{delta_decomp}
\delta = \sum_{p,q \geq 1} \delta_{-p, -q}\, , \hspace{50pt} \delta_{-p,-q} \, I^{r,s} = I^{r-p, s-q}\, .
\end{equation}
The fact that $\delta$ only has components with $p,q \geq 1$ follows from the fact that under complex conjugation the $I^{p,q}$ are related only modulo lower-positioned elements, as can be seen from \eqref{Ipqbar}. One can then proceed and solve \eqref{Hconjugate} for the components $\delta_{-p,-q}$ of the phase operator for a weight 3 Hodge structure as\footnote{Here we used $e^{X} Y e^{-X} = Y + [X, Y] +\frac{1}{2!} [X, [X, Y] ]+ \ldots$ and $[H, \delta_{-p,-q} ] = -(p+q) \delta_{-p,-q}$.} 
\begin{equation}\label{delta}
\begin{aligned}
\delta_{-1, -1} &= \frac{i}{4} (\bar{\cN}^0- \cN^0)_{-1,-1} \, , \quad \delta_{-1, -2} = \frac{i}{6} (\bar{\cN}^0-\cN^0)_{-1,-2} \,, \\
 \delta_{-1, -3} &= \frac{i}{8} (\bar{\cN}^0-\cN^0)_{-1,-3} \, , \quad \delta_{-2,-2} = \frac{i}{8} (\bar{\cN}^0-\cN^0)_{-2,-2} \, , \\
 \delta_{-2, -3} &= \frac{i}{10} (\bar{\cN}^0-\cN^0)_{-2,-3} - \frac{i}{5} [\delta_{-1, -2}, \delta_{-1,-1}]\, , \\
\delta_{-3, -3} &= \frac{i}{12} (\bar{\cN}^0-\cN^0)_{-3,-3} - \frac{i}{3} [\delta_{-2,-2}, \delta_{-1,-1}]\, ,
\end{aligned}
\end{equation}
and the other components follow by complex conjugation.  The rotation of the limiting filtration $F^p_0$ to the $\mathbb{R}$-split is then straightforwardly given by
(as shown in \eqref{notation_8492}, we use a hat to distinguish the $\mathbb{R}$-split case from the non-$\mathbb{R}$-split case) 
\begin{equation}\label{FpR}
\hat{F}^p_0 \equiv e^{-i \delta} F^p_0\, .
\end{equation}
\paragraph{sl(2)-split.} Next we want to rotate from the $\mathbb{R}$-split to the sl(2)-split. We parametrize this rotation by an algebra element $\zeta \in \mathfrak{g}$, which is fixed componentwise in terms of $\delta$ as\footnote{These relations apply for Calabi-Yau threefolds, and their extended version for Calabi-Yau fourfolds is given in \eqref{fourfold_zeta}. Let us note that these $(p,q)$-decompositions are computed with respect to the $\mathbb{R}$-split $\hat{I}^{p,q}$ obtained from \eqref{FpR} and not $I^{p,q}$. In particular, this means that one cannot use the components $\delta_{-p,-q}$ from \eqref{delta} directly, since these were evaluated with respect to the starting Deligne splitting $I^{p,q}$. One should rather compute the $\mathbb{R}$-split $\hat{I}^{p,q}$ first explicitly by using \eqref{Ipq} for $\hat{F}^p_0$, and subsequently decompose $\delta$ with respect to $\hat{I}^{p,q}$ to determine $\zeta$.} (see also \cite{Kato})
\begin{equation}\label{zeta}
\begin{aligned}
\zeta_{-1,-2} &= -\frac{i}{2} \delta_{-1,-2}\, , \quad &\zeta_{-1,-3} &= - \frac{3i}{4} \delta_{-1,-3}\, , \\
 \zeta_{-2,-3} &= -\frac{3i}{8} \delta_{-2,-3} - \frac{1}{8} [ \delta_{-1,-1}, \delta_{-1,-2}]\, ,  \ \ &\zeta_{-3,-3} &= -\frac{1}{8} [ \delta_{-1,-1}, \delta_{-2,-2}]\, ,
\end{aligned}
\end{equation}
and all other components either vanish or are fixed by complex conjugation. The sl(2)-split is obtained by applying $\zeta$ on the limiting filtration in the following way
\begin{equation}\label{Fptilde}
\tilde{F}^p_0= e^{\zeta} \hat{F}^p_0= e^{\zeta}  e^{-i \delta} F^p_0\, ,
\end{equation}
where we used a tilde to distinguish the sl(2)-split from the other two cases. The sl(2)-split $\tilde{I}^{p,q}$ is then straightforwardly computed by using \eqref{Ipq} for $\tilde{F}^p_0$, similar to how the $\mathbb{R}$-split $\hat{I}^{p,q}$ was obtained.

\paragraph{Fixing $\zeta$.} Let us briefly elaborate on how the rotation by $\zeta$ to the sl(2)-split is precisely fixed, see for instance \cite{Grimm:2020cda} for a more extensive discussion.  The central identity of importance reads
\begin{equation}\label{delta-zeta-eta-review}
 e^{i \delta} = e^{\zeta} \big(1 + \sum_{k \geq 1} P_k(C_2, \ldots, C_{k+1}) \big) \, ,
\end{equation}
where the polynomials $P_k(C_2, \ldots, C_{k+1}) $ are defined recursively as
\begin{equation}
P_0 =1 \, , \qquad P_k = -\frac{1}{k} \sum_{j=1}^k P_{k-j} C_{j+1}\, ,
\end{equation}
and coefficients $C_k$ are given in terms of some operator $\eta \in \mathfrak{g}$ as
\begin{equation}
C_{k+1}(\eta) = \hspace*{-0.2cm} \sum_{p,q \geq 1,\,  
p+q \geq k+1} i \, b^{k-1}_{p-1, q-1}  \eta_{-p,-q}\, , \quad (1-x)^p (1+x)^q = \sum_{k=0}^{p+q} b^{k}_{p, q} x^k\, .
\end{equation}
The operators $\zeta$ and $\eta$ can then be determined in terms of components of $\delta$ by expanding the operators in \eqref{delta-zeta-eta-review} with respect to the real $\mathbb{R}$-splitting $\hat{I}^{p,q}$. Below we consider an explicit example where we show how this fixes these two operators.

\paragraph{K3 example.} The above set of conditions needed to derive $\zeta$ is rather abstract, so as an exercise let us work it out for a weight $D=2$ Hodge structure -- the middle cohomology of a K3 surface. The coefficients $C_{k+1}$ are then given by
\begin{equation}
\begin{aligned}
C_2 &= i \eta_{-1,-1} +i \eta_{-1,-2}+ i \eta_{-2, -1}  +i \eta_{-2,-2} \, , \\
C_3 &=i \eta_{-1,-2}  -i \eta_{-2,-1} \, , \qquad C_4 = -i \eta_{-2, -2}\, .
\end{aligned}
\end{equation}
For the polynomials we then get
\begin{equation}
\begin{aligned}
P_1 &= - C_2 =  -i \eta_{-1,-1} -i \eta_{-1,-2}- i \eta_{-2, -1}  -i \eta_{-2,-2}\, , \\
P_2 &= -\frac{1}{2} (P_1 C_2 +P_0 C_3) = -\frac{1}{2} \eta_{-1,-1} \eta_{-1,-1} +\frac{i}{2} \eta_{-2,-1} -\frac{i}{2} \eta_{-1,-2}\, , \\
P_3 &= -\frac{1}{2} (P_2 C_2 + P_1 C_3 + P_0 C_4) = \frac{1}{3} \eta_{-2,-2}\, , 
\end{aligned}
\end{equation}
where all higher-order polynomials vanish: these are made up of terms below degree $(-2,-2)$, which must vanish for $D=2$. We are now finally prepared to read off \eqref{delta-zeta-eta-review} componentwise as
\begin{align}
&\scalebox{0.95}{$ \delta_{-1,-1} = \zeta_{-1,-1} - i \eta_{-1, -1}\, , $}\nn \\
&\scalebox{0.95}{$i \delta_{-1, -2} = \zeta_{-1, -2} - \frac{3i}{2} \eta_{-1,-2}\, ,\qquad   i \delta_{-2, -1} = \zeta_{-2, -1} - \frac{i}{2} \eta_{-2,-1}\, , $} \\
&\scalebox{0.95}{$i \delta_{-2, -2} -\frac{1}{2} (\delta_{-1,-1})^2 = \zeta_{-2,-2} -\frac{2i}{3} \eta_{-2,-2} +\frac{1}{2}(\zeta_{-1,-1})^2 -i \zeta_{-1,-1} \eta_{-1,-1} -\frac{1}{2} (\eta_{-1,-1})^2\, .$} \nn
\end{align}
These conditions are solved for $\zeta$ by
\begin{equation}
\zeta_{-1,-2} = - \frac{i}{2} \delta_{-1,-2}\, , \qquad \zeta_{-2,-1} = \frac{i}{2} \delta_{-2,-1}\, ,
\end{equation}
and for $\eta$ by
\begin{equation}
\begin{aligned}
\eta_{-1,-1} &= - \delta_{-1,-1}\, , \quad &\eta_{-1,-2} &= - \delta_{-1,-2}\, , \\
 \qquad \eta_{-2,-1} &= -\delta_{-2,-1}\, ,\quad  &\eta_{-2,-2} &= -\frac{3}{2} \delta_{-2,-2}\, .
 \end{aligned}
\end{equation}


\subsubsection*{Algorithm II: iterating through the saxion hierarchies}
We now iterate the above procedure through all saxion hierarchies in order to obtain the sl(2)-approximation in the regime $y^1 \gg y^2 \gg \ldots \gg y^n \gg 1$ specified by \eqref{eq:growthsector}. We start from the lowest hierarchy where all saxions are taken to be large $y^1, \ldots, y^n \to \infty$, and move one saxion at a time up to the highest hierarchy $y^1 \to \infty$. A flowchart illustrating how this iteration runs has been provided in figure \ref{algorithm}. Our construction follows the same steps as \cite{CKS, Kato, Grimm:2018cpv}, in particular, we point out that \cite{Grimm:2018cpv} already contains some examples that have been worked out explicitly. 
\tikzstyle{block} = [draw,, rectangle, 
    minimum height=3em, minimum width=2em, align=center]
\begin{figure}[h!]
\centering
\scalebox{0.9}{
\begin{tikzpicture}
\node[block, text width=1.5cm] (aa) at (-3,3) {input: \\
$F^p_0, N_i$};
\node[block, text width=1.2cm] (a) at (0,3) {$I^{p,q}_{(n)}$};
\node[block, text width=1.2cm] (b) at (0,0) {$\tilde{I}^{p,q}_{(n)}$};
\node[block, text width=1.2cm] (c) at (4,3) {$I^{p,q}_{(n-1)}$};
\node[block, text width=1.2cm] (d) at (4,0) {$\tilde{I}^{p,q}_{(n-1)}$};

\node (e) at (6.5,1.5) { $\quad \ldots \quad $};

\node[block, text width=1.2cm] (f) at (9,3) {$I^{p,q}_{(1)}$};
\node[block, text width=1.2cm] (g) at (9,0) {$\tilde{I}^{p,q}_{(1)}$};

\draw[thick, ->] (aa) -- (a) ;
\draw[thick, ->] (a) -- (b) node[pos=0.65, left] {$e^{\zeta_n}e^{i\delta_n}$};
\draw[thick, ->] (b) -- (c) node[pos=0.65, left] {$e^{i N_n}$};
\draw[thick, ->] (c) -- (d) node[pos=0.65, left] {$e^{\zeta_{n-1}}e^{i\delta_{n-1}}$};
\draw[thick, ->] (d) -- (6.1, 1.4) node[pos=0.75, left] {$e^{i N_{n-1}}$};

\draw[thick, ->] (6.9,1.6) -- (f) node[pos=0.75, left] {$e^{i N_{2}}$};
\draw[thick, ->] (f) -- (g) node[pos=0.65, left]  {$e^{\zeta_{1}}e^{i\delta_{1}}$};
\end{tikzpicture}}
\caption{Flowchart illustrating  the algorithm for constructing the sl(2)-approximation. This figure focuses on the construction of the sl(2)-split Deligne splittings $\tilde{I}^{p,q}_{(k)}$.  We labelled each arrow by the rotation that has to be applied on the limiting filtration $F^p_0$ according to \eqref{recursion}. In particular, each downward arrow corresponds to an iteration of the sl(2)-splitting algorithm for the Deligne splitting $I^{p,q}_{(k)}$, where $\delta_k, \zeta_k$ are determined by \eqref{delta} and \eqref{zeta}. }\label{algorithm}
\end{figure}
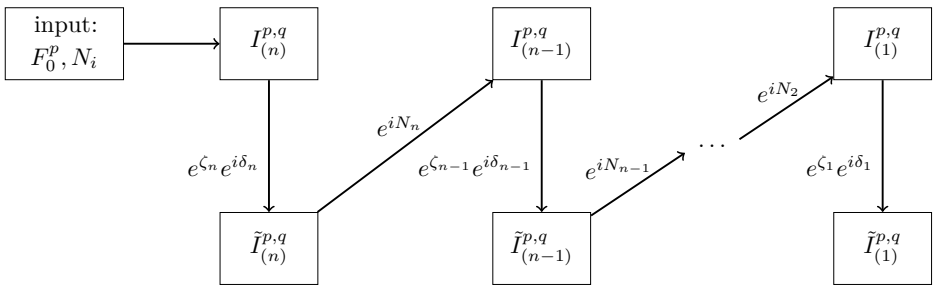

\begin{enumerate}

\item Our starting data from the periods is given by the limiting filtration $F^p_0$ obtained from \eqref{eq:Limiting filtration}, together with the log-monodromy matrices $N_i$. We begin from the lowest hierarchy where all saxions are taken to be large. This means we consider the monodromy weight filtration \eqref{Wfiltr} of $N_{(n)}=N_1 + \cdots + N_n$. By using \eqref{Ipq} we subsequently compute the Deligne splitting  $I^{p,q}_{(n)}$, where we included a subscript to indicate that it involves all $n$ limiting coordinates. We then apply the algorithm I above to compute rotation operators $\delta_n, \zeta_n$ by using \eqref{delta} and \eqref{zeta} in order to obtain the sl(2)-split $\tilde{I}^{p,q}_{(n)}$ through \eqref{Fptilde}.

\item As next step we consider the Deligne splitting for the limit $y^1, \ldots, y^{n-1} \to \infty$. Similar to the previous step we can compute the monodromy weight filtration \eqref{Wfiltr} for $N_{(n-1)}=N_1 + \cdots + N_{n-1}$. However, the limiting filtration we should consider requires slightly more work. Let us denote the limiting filtration of the sl(2)-split obtained in the previous step by $\tilde{F}^p_{(n)}$. Rather than taking this limiting filtration, we should apply $N_{n}$ in the following way
\begin{equation}
F_{(n-1)}^p = e^{i N_n} \tilde{F}^p_{(n)}\, .
\end{equation}
Roughly speaking this can be understood as fixing the saxion $y^n$ to a value -- $y^n = 1$ -- while keeping the other saxions large. We then take this limiting filtration and compute the Deligne splitting $I^{p,q}_{(n-1)}$ at hierarchy $n-1$. 

\item From here on the construction continues in the same manner as for the previous step, where we start from 
$I^{p,q}_{(n-1)}$.
We first determine the rotation matrices $\delta_{n-1}, \zeta_{n-1}$ to obtain the sl(2)-split $\tilde{I}^{p,q}_{(n-1)}$, and then move on to the next hierarchy by applying $ N_{n-1}$.

\end{enumerate}
To summarize, the full iterative process can therefore be described by the two recursive identities
\begin{equation}\label{recursion}
\tilde{F}^p_{(k)} = e^{\zeta_k} e^{i \delta_k} F^p_{(k)}\, , \hspace{50pt} F^p_{(k-1)} = e^{i N_{k}} \tilde{F}^{p}_{(k)}\, .
\end{equation}


\subsubsection*{Algorithm III: constructing the sl(2)-triples and boundary Hodge star
}

We now discuss the final steps in constructing the sl(2)-approximation. The above iteration of the sl(2)-splitting algorithm already provided us the necessary data in the form of the Deligne splittings $\tilde{I}^{p,q}_{(k)}$. The remaining task is now to read off the relevant boundary data.
The strategy is to first determine a set of mutually commuting sl(2)-triples 
\beq \label{commuting_triples}
    \{N^-_k,N^+_k,N^0_k \}\ , \qquad k=1,...,n\ ,
 \eeq 
one associated to each log-monodromy $N_k$. Subsequently we will read off the charge operator $Q_\infty$ and the boundary Hodge star $C_\infty$.

\paragraph{Weight operator.} We first determine the weight operators $N^{0}_{(k)}=N^0_1+\ldots +N^0_k $. Their action on the 
sl(2)-split $\tilde{I}^{p,q}_{(k)}$ is given as grading operators that multiply elements  as
\begin{equation}\label{def_weight}
N^0_{(k)}\,  \omega_{p,q} = (p+q-3)\,  \omega_{p,q} \, , \hspace{50pt} \omega_{p,q} \in \tilde{I}^{p,q}_{(k)} \, .
\end{equation}
Since the iteration of the sl(2)-splitting algorithm provides us with explicit expressions for the vector spaces $\tilde{I}^{p,q}_{(k)}$, this property suffices to write down the grading operators $N_{(k)}$ explicitly. The weight operators associated with the individual saxions are then determined via
\begin{equation}
N^0_k =N^0_{(k)} - N^0_{(k-1)}\, ,
\hspace{50pt}
N^0_{(0)}=0\,.
\end{equation}
\paragraph{Lowering operators.} Next we determine the lowering operators $N_k^-$ of the commuting sl(2)-triples. The idea is to construct these lowering operators out of the log-monodromy matrices $N_k$. However, while the $N_k$ commute with each other, generally they do not yet commute with the weight operators $N^0_{k'}$ of the other sl(2)-triples. This can be remedied by projecting each $N_k$ onto its weight-zero piece under $N^0_{(k-1)}$. We can write this projection out as
\begin{equation}\label{lowering}
N_k = N_k^- + \sum_{\ell \geq 2} N_{k, -\ell} \, , \hspace{50pt} [N^0_{(k-1)}, \,  N_{k, -\ell} ]  = -\ell \op N_{k,- \ell}\, ,
\end{equation}
where $\ell$ specifies the weight under $N^0_{(k-1)}$, and the lowering operator $N_{k}^- = N_{k, 0}$ is fixed as the weight-zero piece. By projecting out the other components $N_{k,-\ell}$ we ensure that the resulting sl(2)-triples are commuting. It is now straightforward to complete the sl(2)-triples \eqref{commuting_triples} by solving 
$[N_k^0,N_k^+]= 2 N_k^+$ and $[N_k^+,N_k^-]=N^0_k$ for the raising operators $N^+_i$. For our purposes this will not be necessary.

\paragraph{Boundary Hodge star.} Besides the sl(2)-triples we also need the boundary Hodge structure to construct the sl(2)-approximation. The defining Hodge filtration $F^p_\infty$ can be obtained from any of the sl(2)-split filtrations $\tilde{F}^p_{(k)}$. By applying $N^-_{(k)} = N_1^- + \ldots +N_k^-$ on this filtration in a similar manner as the second equation in \eqref{recursion} (also recall \eqref{eq:boundaryfiltration}) we obtain
\begin{equation}\label{Fpinfty}
F^p_\infty = e^{i N^-_{(k)}} \tilde{F}^p_{(k)}\, .
\end{equation}
The pure Hodge structure is then obtained straightforwardly through \eqref{eq:Hpqdef} and we can 
read off the operator $Q_\infty$ by 
\beq
 Q_{\infty} \omega = \frac{1}{2}\big(p- q\big) \omega\ , \qquad \omega \in H^{p,q}_\infty = F^p_\infty \cap \bar{F}^q_\infty\ .
\eeq 
Intuitively the appearance of this pure Hodge structure (rather than a mixed Hodge structure) follows from the fact that it corresponds to the Deligne splitting with a trivial nilpotent element $N_{(0)}^- = 0$, so it reduces to a pure Hodge structure. The boundary Hodge star operator can obtained through 
\begin{equation}\label{Cinfty}
C_\infty \op w_{p,q} = i^{p-q}\op w_{p,q}\, ,  \hspace{50pt} w_{p,q} \in H^{p,q}_\infty\,,
\end{equation}
or by recalling that $C_\infty  = e^{i \pi Q_\infty}$. Altogether the above boundary data suffices to write down the sl(2)-approximation, for instance for the Hodge star in \eqref{eq:Csl2ToCInf}.


\section{Example construction of the sl(2)-approximation}
\label{sec_aht_example}
In this subsection we work out in detail the sl(2)-approximation for a two-moduli example. We consider the large complex structure region for the mirror of the Calabi-Yau hypersurface inside $\mathbb{P}_{4}^{1,1,2,2,2}[8]$, 
which in this context was  studied originally in \cite{Candelas:1993dm, Hosono:1993qy, Hosono:1994ax}. 
A similar two-moduli model, $\mathbb{P}_4^{1,1,1,6,9}[18]$ of \cite{Hosono:1993qy, Candelas:1994hw}, has been worked out in 
\cite{Grimm:2018cpv} to which we refer for an additional example. The relevant topological data for our example can be found for instance in \cite{Candelas:1993dm}, which we recall as follows\footnote{
For a clearer presentation, we chose to interchange the complex structure moduli $t^1 \leftrightarrow t^2$
as compared to \cite{Candelas:1993dm}.
}
\begin{equation}
\label{data_ex_001}
\kappa_{122}  = 4\, , \hspace{40pt} \kappa_{222} = 8 \, , \hspace{40pt} \chi = -168\, ,
\end{equation}
and all other intersection numbers vanishing. The log-monodromy matrices \eqref{logN_001} are then given by 
\begin{equation}
N_1 = \scalebox{0.9}{$\begin{pmatrix}
 0 & 0 & 0 & 0 & 0 & 0 \\
 1 & 0 & 0 & 0 & 0 & 0 \\
 0 & 0 & 0 & 0 & 0 & 0 \\
 0 & 0 & 0 & 0 & -1 & 0 \\
 0 & 0 & 0 & 0 & 0 & 0 \\
 0 & 0 & -4 & 0 & 0 & 0 
\end{pmatrix}$} , \hspace{30pt}
N_2 = \scalebox{0.9}{$\begin{pmatrix}
 0 & 0 & 0 & 0 & 0 & 0 \\
 0 & 0 & 0 & 0 & 0 & 0 \\
 1 & 0 & 0 & 0 & 0 & 0 \\
 0 & 0 & 0 & 0 & 0 & -1 \\
 0 & 0 & -4 & 0 & 0 & 0 \\
 0 & -4 & -8 & 0 & 0 & 0 
\end{pmatrix}$} .
\end{equation}
In the following we work out the construction of the sl(2)-approximation in explicit detail for the asymptotic regime  $y^1 \gg y^2 \gg 1$.


\subsubsection*{Step 1: sl(2)-splitting of $I^{p,q}_{(2)}$}

We begin our analysis with  both saxions $y^1$ and $y^2$ being large, which corresponds to 
considering the sum of log-monodromy matrices 
\eq{
\label{example_001}
N_{(2)}=N_1+N_2 \,.
}
In order to write down the associated Deligne splitting $I^{p,q}_{(2)}$ 
we need to determine the monodromy weight filtration $W_{\ell}(N_{(2)})$ and the limiting filtration $F^p_0$. 
We discuss these structures in turn.
\paragraph{Monodromy weight filtration.} The monodromy weight filtration for the nilpotent operator \eqref{example_001} 
is  computed using  \eqref{Wfiltr}. 
Since $N_{(2)}$ is a finite-dimensional  and explicitly-known matrix this can easily be done using 
a computer-algebra program, and we find
\begin{small}
 \begin{equation}\label{W(2)}
\arraycolsep2pt
\renewcommand{\arraystretch}{1.3}
 \begin{array}{lclcl}
 W_0(N_{(2)}) &=&W_1(N_{(2)}) &=& \mathrm{span}\bigl[  (0,0,0,1,0,0)\bigr]\, , \\
 W_2(N_{(2)}) &=& W_3(N_{(2)})&=&  \mathrm{span}\bigl[ (0,0,0,0,1,0)\,, (0,0,0,0,0,1) \bigr] \oplus
 W_0(N_{(2)}) \, ,  \\
 W_4(N_{(2)}) &=& W_5(N_{(2)})&=& \mathrm{span}\bigl[  (0,1,0,0,0,0)\,,  (0,0,1,0,0,0) \bigr] \oplus 
  W_2(N_{(2)})\, ,\\ 
 &&W_6(N_{(2)}) &=&  \mathrm{span}\bigl[(1,0,0,0,0,0) \bigr] \oplus 
 W_4(N_{(2)}) 
 \, .
 \end{array}
 \end{equation}
 \end{small}
Note that this filtration corresponds  to the decomposition of the even homology into 
zero-, two-, four- and six-cycles on the mirror dual. To be precise, $W_{2p}(N_{(2)})$ is spanned by all 
even cycles of degree $2p$ or lower, which  turns out to be a general feature for the monodromy weight filtration of $N_{(h^{2,1}  )}$ at large complex structure.
  
\paragraph{Limiting filtration.} For the limiting filtration \eqref{eq:Limiting filtration} we recall that the vector space $F^p$ is spanned by the first $3-p$ derivatives of the period vector. At large complex structure the derivative with respect to $t^i$ simply lowers a log-monodromy matrix $N_i$ in \eqref{npot_001}, since we can ignore derivatives of $\Gamma(z)$ near this boundary. Subsequently multiplying by 
$e^{-t^i N_i}$ from the left we find we are left with the vectors $\mathbf{a}_0$, $N_i \mathbf{a}_0, \ldots$ to span the spaces $F^p_0$. In other words, the vector space $F^p_0$ is obtained by taking the span of up to $3-p$ log-monodromy matrices $N_i$ acting on $\mathbf{a}_0$. We can represent this information about the limiting filtration $F^p_0$ succinctly in terms of a period matrix as
\begin{equation}\label{F(2)}
\Pi_{(2)} =  \scalebox{0.85}{$\begin{pmatrix} 
1 & 0 & 0 & 0 & 0 & 0 \\
 0 & 1 & 0 & 0 & 0 & 0 \\
 0 & 0 & 1 & 0 & 0 & 0 \\
 -\frac{21 i \zeta (3)}{\pi ^3} & 0 & 0 & 0 & 0 & 1 \\
 0 & 0 & 0 & 1 & 0 & 0 \\
 0 & 0 & 0 & 0 & 1 & 0 
 \end{pmatrix}$}.
\end{equation}
Here the first column corresponds to $F^3_0$, the first three columns to $F^2_0$, the first five columns to $F^1_0$, and all six columns $F^0_0 = H^3(Y_3)$.

\paragraph{Deligne splitting.} Given the monodromy weight filtration \eqref{W(2)} and the limiting filtration \eqref{F(2)},
we can now compute the Deligne splitting via equation \eqref{Ipq}. Expressed in the 
diagrammatic form introduced in figure~\ref{fig_deligne},
with the associated vectors  understood as spanning the corresponding subspace, this 
yields
\begin{equation}\label{Ipq(2)}
I^{p,q}_{(2)}  = \begin{tikzpicture}[baseline={([yshift=-.5ex]current bounding box.center)},scale=0.8,cm={cos(45),sin(45),-sin(45),cos(45),(15,0)}]
  \draw[step = 1, gray, ultra thin] (0, 0) grid (3, 3);
  \draw[fill] (0, 0) circle[radius=0.04] node[right]{\small $(0,0,0,1, 0 , 0)$};
  \draw[fill] (1.05, 0.95) circle[radius=0.04] node[right]{\small $(0,0,0,0, 1 , 0)$};
  \draw[fill] (0.95, 1.05) circle[radius=0.04] node[left]{\small $(0,0,0,0, 0 , 1)$};
  \draw[fill] (1.95, 2.05) circle[radius=0.04] node[left]{\small $(0,1,0,0, 0 , 0)$};
  \draw[fill] (2.05, 1.95) circle[radius=0.04] node[right]{ \small $(0,0,1,0, 0 , 0)$};
  \draw[fill] (3, 3) circle[radius=0.04] node[right]{ \small $(1,0,0,-\frac{21 i \zeta(3)}{\pi^3}, 0 , 0)$};
\end{tikzpicture}
\end{equation}
However, note that that the Deligne splitting \eqref{Ipq(2)} is not $\mathbb{R}$-split. 
Indeed, under complex conjugation $I_{(2)}^{3,3}$ is shifted by a piece in $I_{(2)}^{0,0}$ as
\begin{align}\label{I33bar}
\bar{I}_{(2)}^{3,3} &= \mathrm{span} \bigl[ (1,0,0,+\tfrac{21 i \zeta(3)}{\pi^3}, 0 , 0)\bigr]\\
& = \mathrm{span} \bigl[ (1,0,0,-\tfrac{21 i \zeta(3)}{\pi^3}, 0 , 0)\bigr]\, \text{mod}\, \mathrm{span}\bigl[  \tfrac{42 i \zeta(3)}{\pi^3} (0,0,0,1, 0 , 0)\bigr]  = 
I_{(2)}^{3,3} \, \text{mod}\,  I_{(2)}^{0,0}\, , \nn
\end{align}
and hence the complex-conjugation rule shown in  \eqref{Ipqbar} follows. 
We now want to perform a complex rotation of the period matrix $\Pi_{(2)}$ to make the Deligne splitting $\mathbb{R}$-split. This procedure is outlined between equations \eqref{Hcomplex} and \eqref{FpR}. 

\paragraph{Grading operator.} To begin with, we determine the grading operator $\cN^0_{(2)}$ as defined by \eqref{Hcomplex} for the Deligne splitting \eqref{Ipq(2)}. Using for instance a computer algebra progam, this  operator is computed as
\begin{equation}\label{H2complex}
\cN^0_{(2)}= \scalebox{0.9}{$\begin{pmatrix}
 3 & 0 & 0 & 0 & 0 & 0 \\
 0 & 1 & 0 & 0 & 0 & 0 \\
 0 & 0 & 1 & 0 & 0 & 0 \\
 -\frac{126 i \zeta (3)}{\pi ^3} & 0 & 0 & -3 & 0 & 0 \\
 0 & 0 & 0 & 0 & -1 & 0 \\
 0 & 0 & 0 & 0 & 0 & -1 
\end{pmatrix}$}\, ,
\end{equation}
where the imaginary component of $\cN^0_{(2)}$ corresponds  to the breaking of the $\mathbb{R}$-split property of the Deligne splitting $I^{p,q}_{(2)}$ in \eqref{I33bar}.

\paragraph{Phase operator.} Next, we recall that the rotation is  implemented through a phase operator $\delta_2$, which  is fixed by the grading operator $\cN^0_{(2)}$ through \eqref{Hconjugate}, and $\delta_2$ can be computed explicitly by using \eqref{delta}. For the derived grading operator \eqref{H2complex} and Deligne splitting \eqref{Ipq(2)} we then  find the phase operator 
\begin{equation}
\delta_2 =  (\delta_2)_{-3,-3} = \scalebox{0.9}{$\begin{pmatrix}
 0 & 0 & 0 & 0 & 0 & 0 \\
 0 & 0 & 0 & 0 & 0 & 0 \\
 0 & 0 & 0 & 0 & 0 & 0 \\
- \frac{21 \zeta (3)}{\pi ^3} & 0 & 0 & 0 & 0 & 0 \\
 0 & 0 & 0 & 0 & 0 & 0 \\
 0 & 0 & 0 & 0 & 0 & 0 
\end{pmatrix}$}\, ,
\end{equation}
where we stressed that $\delta_2$ only has a $(-3,-3)$-component with respect to Deligne splitting $I^{p,q}_{(2)}$. By using \eqref{zeta} one therefore can already see that $\zeta_2=0$, so our rotation by $\delta$ will directly rotate us to the sl(2)-split. Following \eqref{recursion}, the period matrix of this limiting filtration $\tilde{F}^p_{(2)}$ is given by
\begin{equation}\label{sl2F2}
\tilde{\Pi}_{(2)} = e^{i \delta_2} \Pi_{(2)} =  \scalebox{0.9}{$\begin{pmatrix} 
1 & 0 & 0 & 0 & 0 & 0 \\
 0 & 1 & 0 & 0 & 0 & 0 \\
 0 & 0 & 1 & 0 & 0 & 0 \\
0 & 0 & 0 & 0 & 0 & 1 \\
 0 & 0 & 0 & 1 & 0 & 0 \\
 0 & 0 & 0 & 0 & 1 & 0 
 \end{pmatrix}$}.
\end{equation}

\paragraph{Sl(2)-splitting.} Combining this result  with the monodromy weight filtration \eqref{W(2)}, one straightforwardly shows that the sl(2)-splitting at the lowest hierarchy is spanned by
\begin{equation}\label{sl2Ipq(2)}
\tilde{I}_{(2)}^{p,q}  = \begin{tikzpicture}[baseline={([yshift=-.5ex]current bounding box.center)},scale=0.85,cm={cos(45),sin(45),-sin(45),cos(45),(15,0)}]
  \draw[step = 1, gray, ultra thin] (0, 0) grid (3, 3);

  \draw[fill] (0, 0) circle[radius=0.04] node[right]{\small $(0,0,0,1, 0 , 0)$};
  \draw[fill] (1.05, 0.95) circle[radius=0.04] node[right]{\small $(0,0,0,0, 1 , 0)$};
  \draw[fill] (0.95, 1.05) circle[radius=0.04] node[left]{\small $(0,0,0,0, 0 , 1)$};
  \draw[fill] (1.95, 2.05) circle[radius=0.04] node[left]{\small $(0,1,0,0, 0 , 0)$};
  \draw[fill] (2.05, 1.95) circle[radius=0.04] node[right]{ \small $(0,0,1,0, 0 , 0)$};
  \draw[fill] (3, 3) circle[radius=0.04] node[right]{ \small $(1,0,0,0, 0 , 0)$};
\end{tikzpicture}\, .
\end{equation}
Notice that the resulting sl(2)-split  $\tilde{I}_{(2)}^{p,p} $ can be interpreted precisely as the decomposition into $(p,p)$-forms on the mirror dual K\"ahler side, similar to our comment on the filtration $W(N_{(2)})$ below 
equation \eqref{W(2)}.


\subsubsection*{Step 2: sl(2)-splitting of $I^{p,q}_{(1)}$}

Following the algorithm outlined in figure~\ref{algorithm}, 
we next consider the hierarchy  set by $y^1$ with  $y^2 \ll y^1$. From a practical perspective this means we will be working with the Deligne splitting $I^{p,q}_{(1)}$ associated with the log-monodromy matrix $N_1$. 
We determine this splitting as follows.

\paragraph{Monodromy weight filtration.} Similarly as above, we first compute the monodromy weight filtration \eqref{Wfiltr} associated with $N_1$. 
It takes the following form
\begin{small}
\begin{align}\label{W(1)}
 W_0(N_1)=W_1(N_1) &=  0\, , \nn \\
 W_2(N_1) = W_3(N_1) &= \mathrm{span}\bigl[ (0,1,0,0,0,0)\,,  (0,0,0,0,0,1)\,, (0,0,0,1,0,0)\bigr]\, ,\\
 W_4(N_1) = W_5(N_1) &=
 \mathrm{span}\bigl[  (1,0,0,0,0,0)\,, (0,0,1,0,0,0)\, ,  (0,0,0,0,1,0) \bigr]
  \oplus W_2(N_1) \,,\nn
  \\
 W_6(N_1) &= W_4(N_1) \,.\nn
\end{align}
\end{small}
\paragraph{Limiting filtration.} Next, we determine $F^p_{(1)}$ from the sl(2)-split filtration $\tilde{F}^p_{(2)}$ 
of the previous hierarchy through equation \eqref{recursion}. 
At the level of the period matrix this means we rotate \eqref{sl2F2}  as
\begin{equation}\label{F(1)}
\Pi_{(1)} = e^{i N_2} \tilde{\Pi}_{(2)}=
e^{i N_2} e^{i \delta_2} \Pi_{(2)}
= \begin{pmatrix}
 1 & 0 & 0 & 0 & 0 & 0 \\
 0 & 1 & 0 & 0 & 0 & 0 \\
 -i & 0 & 1 & 0 & 0 & 0 \\
 \frac{4 i}{3} & -2 & -4 & 0 & - i & 1 \\
 2 & 0 & 4 i & 1 & 0 & 0 \\
 4 & 4 i & 8 i & 0 & 1 & 0 \\
 \end{pmatrix}.
 \end{equation}
\paragraph{Deligne splitting.} Using the definition of the Deligne splitting given in \eqref{Ipq} together with the monodromy weight filtration \eqref{W(1)} and the limiting filtration \eqref{F(1)}, we find that the Deligne splitting is spanned by
\begin{equation}\label{Ipq(1)}
I^{p,q}_{(1)} = 
\begin{tikzpicture}[baseline={([yshift=-.5ex]current bounding box.center)},scale=1,cm={cos(45),sin(45),-sin(45),cos(45),(15,0)}]
  \draw[step = 1, gray, ultra thin] (0, 0) grid (3, 3);
  \draw[fill] (0, 2) circle[radius=0.05] node[left]{\small $(0,1,0,-2,0,4i)$ };
  \draw[fill] (2, 0) circle[radius=0.05] node[right]{\small $(0,1,0,-2,0,-4i)$};
  \draw[fill] (1, 1) circle[radius=0.05] node[below]{\small $(0,1,0,2,0,0)$};
  \draw[fill] (2, 2) circle[radius=0.05] node[above]{\small $(1,-\frac{2i}{3},0,-\frac{4i}{3}, -2, -\frac{4}{3})$};
  \draw[fill] (1, 3) circle[radius=0.05] node[left]{\small $(1,0,-i,\frac{4i}{3},2,4)$ };
  \draw[fill] (3, 1) circle[radius=0.05] node[right]{\small  $(1,-\frac{4i}{3}, i, \frac{4i}{3}, 2, -\frac{4}{3})$};
\end{tikzpicture}\, .
\end{equation}
Note that none of the spaces in the upper part of the Deligne splitting \eqref{Ipq(1)} 
obey the $\mathbb{R}$-split conjugation rule. In  particular, under complex conjugation $I^{2,2}$ is related to itself up to a shift in $I^{1,1}$, while $I^{1,3}$ is related to $I^{3,1}$ up to a piece in $I^{2,0}$ and vice versa. To be precise, we find
\begin{small}
\begin{align}
\bar{I}_{(1)}^{2,2} &= 
\mathrm{span}\bigl[(1,+\tfrac{2i}{3},0,+\tfrac{4i}{3}, -2, -\tfrac{4}{3}) \bigr] \nn \\
&=
\mathrm{span}\bigl[ (1,-\tfrac{2i}{3},0,-\tfrac{4i}{3}, -2, -\tfrac{4}{3})\bigr]  \,  \text{mod} \, 
\mathrm{span}\bigl[ \tfrac{4i}{3} (0,1,0,2,0,0) \bigr] = I_{(1)}^{2,2} \,  \text{mod} \, I_{(1)}^{1,1}\, , \nn
\\
\bar{I}_{(1)}^{3,1} & =
\mathrm{span}\bigl[(1,+\tfrac{4i}{3}, -i, -\tfrac{4i}{3}, 2, -\tfrac{4}{3})\bigr]  \\
& = \mathrm{span}\bigl[ (1,0,-i,+\tfrac{4i}{3},2,4)\bigr] \,  \text{mod} \, 
\mathrm{span}\bigl[ \tfrac{4i}{3} (0,1,0,-2,0,4i)\bigr] = I_{(1)}^{3,1} \,  \text{mod} \, I_{(1)}^{2,0} \, , \nn
\end{align}
\end{small}
which is  in accordance with the complex conjugation rule \eqref{Ipqbar}.
We can now perform  the rotation to the sl(2)-split $\tilde{I}^{p,q}_{(1)}$.

\paragraph{Grading operator.} As before we first compute the grading operator defined in \eqref{Hcomplex} 
for the above Deligne splitting \eqref{Ipq(1)}. This grading operator is found to be
\begin{equation}
\cN^0_{(1)} = \scalebox{0.9}{$\begin{pmatrix}
 1 & 0 & 0 & 0 & 0 & 0 \\
 -\frac{4 i}{3} & -1 & -\frac{4}{3} & 0 & 0 & 0 \\
 0 & 0 & 1 & 0 & 0 & 0 \\
 0 & 0 & 0 & -1 & \frac{4 i}{3} & 0 \\
 0 & 0 & 0 & 0 & 1 & 0 \\
 0 & 0 & \frac{16 i}{3} & 0 & \frac{4}{3} & -1 \\
 \end{pmatrix}$}\, .
\end{equation}
\paragraph{Phase operator.}Next, using \eqref{delta}, $\delta$ is then computed from $\cN^0_{(1)}$ as
\begin{equation}
\delta_1 =(\delta_1 )_{-1,-1} \scalebox{0.9}{$\begin{pmatrix}
  0 & 0 & 0 & 0 & 0 & 0 \\
 -\frac{2}{3} & 0 & 0 & 0 & 0 & 0 \\
 0 & 0 & 0 & 0 & 0 & 0 \\
 0 & 0 & 0 & 0 & \frac{2}{3} & 0 \\
 0 & 0 & 0 & 0 & 0 & 0 \\
 0 & 0 & \frac{8}{3} & 0 & 0 & 0 \\
 \end{pmatrix}$}\, .
 \end{equation}
Note that similar to the Deligne splitting $I^{p,q}_{(2)}$ discussed above, we  find that $\zeta_1=0$ by using \eqref{zeta}, since $\delta_1$ only has a $(-1,-1)$-component. Therefore by rotating to the $\mathbb{R}$-split we again directly rotate to the sl(2)-split. At the level of the period matrix \eqref{F(1)} this means that $\tilde{F}^p_{(1)}$ can be represented as
\begin{equation}
\tilde{\Pi}_{(1)} = e^{i\delta_1} \Pi_{(1)} =  \scalebox{0.9}{$\begin{pmatrix}
 1 & 0 & 0 & 0 & 0 & 0 \\
 \frac{2 i}{3} & 1 & 0 & 0 & 0 & 0 \\
 -i & 0 & 1 & 0 & 0 & 0 \\
 0 & -2 & -\frac{4}{3} & -\frac{2 i}{3} & i & 1 \\
 2 & 0 & 4 i & 1 & 0 & 0 \\
 \frac{4}{3} & 4 i & \frac{16 i}{3} & 0 & 1 & 0 \\
 \end{pmatrix}$} \, .
\end{equation} 
\paragraph{Sl(2)-splitting.} Finally, using \eqref{Ipq} for the filtration $\tilde{F}^p_{(1)}$ we obtain
\begin{equation}\label{sl2Ipq(1)}
\tilde{I}^{p,q}_{(1)} = 
\begin{tikzpicture}[baseline={([yshift=-.5ex]current bounding box.center)},scale=0.9,cm={cos(45),sin(45),-sin(45),cos(45),(15,0)}]
  \draw[step = 1, gray, ultra thin] (0, 0) grid (3, 3);
  \draw[fill] (2, 0) circle[radius=0.05] node[right]{\small $(0,1,0,-2,0,-4i)$};
  \draw[fill] (0, 2) circle[radius=0.05] node[left]{\small $(0,1,0,-2,0,4i)$};
  \draw[fill] (1, 1) circle[radius=0.05] node[below]{\small $(0,1,0,2,0,0)$};
  \draw[fill] (2, 2) circle[radius=0.05] node[above]{\small $(1,0,0,0, -2, -\frac{4}{3})$};
  \draw[fill] (3, 1) circle[radius=0.05] node[right]{\small $(1,-\frac{2i}{3},i,0,2,\frac{4}{3})$};
  \draw[fill] (1, 3) circle[radius=0.05] node[left]{\small $(1,\frac{2i}{3},-i,0,2,\frac{4}{3})$};
\end{tikzpicture}\, .
\end{equation}
It is also worthwhile to reflect back on our earlier discussion of the $\mathrm{II}_b \to \mathrm{IV}_d$ enhancement in section \ref{sec:enhancements}, and in particular how the primitive component of weight four \eqref{P4enhancement} enhances. We can see precisely here how all three elements at the fourth row in \eqref{sl2Ipq(1)} are needed to make the linear combination $\tilde{a}_0=(1,0,0,0,0,0)$ spanning $I^{3,3}_{(2)}$ in \eqref{sl2Ipq(2)}. Similarly we can obtain its descendants $(0,-\tfrac{2}{3}, 1, 0,0,0)$ at $(2,2)$ and $ (0,0,0,0,1,\tfrac{2}{3})$ at $(1,1)$ from these elements.


\subsubsection*{Step 3: constructing the sl(2)-approximation}

We now construct the sl(2)-approximated Hodge star from the two sl(2)-splittings \eqref{sl2Ipq(1)} and \eqref{sl2Ipq(2)} determined above.

\paragraph{Sl(2)-triples.} The weight operators $N^0_{(i)}$ are fixed by the multiplication rule \eqref{def_weight} on the two sl(2)-splittings. For \eqref{sl2Ipq(1)} and \eqref{sl2Ipq(2)} these grading operators are respectively given by 
\begin{equation}
N^0_{(1)} = \scalebox{0.9}{$\begin{pmatrix} 
 1 & 0 & 0 & 0 & 0 & 0 \\
 0 & -1 & -\frac{4}{3} & 0 & 0 & 0 \\
 0 & 0 & 1 & 0 & 0 & 0 \\
 0 & 0 & 0 & -1 & 0 & 0 \\
 0 & 0 & 0 & 0 & 1 & 0 \\
 0 & 0 & 0 & 0 & \frac{4}{3} & -1 \\
 \end{pmatrix}$}\, , \quad N^0_{(2)} = \scalebox{0.9}{$\begin{pmatrix}
  3 & 0 & 0 & 0 & 0 & 0 \\
 0 & 1 & 0 & 0 & 0 & 0 \\
 0 & 0 & 1 & 0 & 0 & 0 \\
 0 & 0 & 0 & -3 & 0 & 0 \\
 0 & 0 & 0 & 0 & -1 & 0 \\
 0 & 0 & 0 & 0 & 0 & -1 \\
  \end{pmatrix}$}\, .
\end{equation}
In order to construct the lowering operators $N_i^-$ we need to decompose the log-monodromy matrices $N_a$ 
with respect to the weight operators as described in \eqref{lowering}. For the first lowering operator we find simply that $N_1 = N_1^-$, since $N^0_{(0)}=0$. On the other hand, for $N_2$ we find that it decomposes with respect to $N^0_{(1)}$ as
\eq{
  N_2 = (N_2)_0 + (N_2)_{-2} \,,
}
where
\begin{equation}
(N_2)_0 = \scalebox{0.85}{$\begin{pmatrix}
 0 & 0 & 0 & 0 & 0 & 0 \\
 -\frac{2}{3} & 0 & 0 & 0 & 0 & 0 \\
 1 & 0 & 0 & 0 & 0 & 0 \\
 0 & 0 & 0 & 0 & \frac{2}{3} & -1 \\
 0 & 0 & -4 & 0 & 0 & 0 \\
 0 & -4 & -\frac{16}{3} & 0 & 0 & 0 \\
\end{pmatrix}$} , \ \  
 \quad (N_2)_{-2} = \scalebox{0.85}{$\begin{pmatrix}
 0 & 0 & 0 & 0 & 0 & 0 \\
 \frac{2}{3} & 0 & 0 & 0 & 0 & 0 \\
 0 & 0 & 0 & 0 & 0 & 0 \\
 0 & 0 & 0 & 0 & -\frac{2}{3} & 0 \\
 0 & 0 & 0 & 0 & 0 & 0 \\
 0 & 0 & -\frac{8}{3} & 0 & 0 & 0 \\
 \end{pmatrix}$}.
\end{equation}
We then identify the lowering operator with the weight-zero piece as $N_2^- = (N_2)_0$.

\paragraph{Boundary Hodge star.} Next we construct the boundary Hodge star $C_\infty$. The Hodge filtration $F^p_\infty$ associated with this boundary Hodge structure follows from \eqref{Fpinfty}, and we can write the period matrix associated to these vector spaces $F^p_\infty$ as
\begin{equation}
\tilde{\Pi}_{(0)} = e^{i N_1^-} \tilde{\Pi}_{(1)} = e^{i (N_1^-+N_2^-)} \tilde{\Pi}_{(2)} = \begin{pmatrix}
 1 & 0 & 0 & 0 & 0 & 0 \\
 \frac{i}{3} & 1 & 0 & 0 & 0 & 0 \\
 i & 0 & 1 & 0 & 0 & 0 \\
 -2 i & -2 & -\frac{16}{3} & -\frac{i}{3} & -i & 1 \\
 2 & 0 & -4 i & 1 & 0 & 0 \\
 \frac{16}{3} & -4 i & -\frac{28 i}{3} & 0 & 1 & 0 \\
 \end{pmatrix}\, .
\end{equation}
From this boundary Hodge filtration one finds that the boundary $(p,q)$-spaces $H^{p,q}_\infty = F^p_\infty \cap F^q_\infty$ are spanned by
\begin{equation}
\begin{aligned}
H^{3,0}_\infty &= \mathrm{span}\bigl[ ( 1  ,   -\tfrac{i}{3}  ,  -i  ,  2 i  ,  2  ,  \tfrac{16}{3} ) \bigr]\, , 
\\
H^{2,1}_\infty &= \mathrm{span}\bigl[( 1  ,   0  ,   -\tfrac{3 i}{8}  ,   -2 i  ,   -\tfrac{1}{2}  ,   -\tfrac{11}{6})\, ,\,  ( 0  ,   1  ,   -\tfrac{3}{8}  ,   0  ,   -\tfrac{3 i}{2}  ,   \tfrac{i}{2}  )\bigr]\, , \\
\end{aligned}
\end{equation}
and the other spaces follow by complex conjugation. The corresponding Hodge star operator is then determined by the multiplication rule  \eqref{Cinfty} as
\begin{equation}
C_\infty = \scalebox{0.9}{$\begin{pmatrix}
 0 & 0 & 0 & \frac{1}{2} & 0 & 0 \\
 0 & 0 & 0 & 0 & \frac{11}{18} & -\frac{1}{6} \\
 0 & 0 & 0 & 0 & -\frac{1}{6} & \frac{1}{4} \\
 -2 & 0 & 0 & 0 & 0 & 0 \\
 0 & -2 & -\frac{4}{3} & 0 & 0 & 0 \\
 0 & -\frac{4}{3} & -\frac{44}{9} & 0 & 0 & 0 \\
\end{pmatrix}$} .
\end{equation}
Having constructed the necessary building blocks, we are now ready to put together the sl(2)-approximated Hodge star operator according to \eqref{eq:Csl2ToCInf}. Setting the axions to zero for simplicity, that is $x^i=0$, we find that 
\begin{equation}
\label{data_ex_csl2}
C_{\rm sl(2)} = \scalebox{0.85}{$\begin{pmatrix}
 0 & 0 & 0 & -\frac{1}{2 y_{1} y_{2}^2} & 0 & 0 \\
 0 & 0 & 0 & 0 & -\frac{y_{1}}{2 y_{2}^2}-\frac{1}{9 y_{1}} & \frac{1}{6 y_{1}} \\
 0 & 0 & 0 & 0 & \frac{1}{6 y_{1}} & -\frac{1}{4 y_{1}} \\
 2 y_{1} y_{2}^2 & 0 & 0 & 0 & 0 & 0 \\
 0 & \frac{2 y_{2}^2}{y_{1}} & \frac{4 y_{2}^2}{3 y_{1}} & 0 & 0 & 0 \\
 0 & \frac{4 y_{2}^2}{3 y_{1}} & \frac{8 y_{2}^2}{9 y_{1}}+4 y_{1} & 0 & 0 & 0 \\
\end{pmatrix}$},
\end{equation}
and the Hodge-star matrix $\mathcal M_{\rm sl(2)}$ is then again obtained as
$\mathcal M_{\rm sl(2)} = \eta\op C_{\rm sl(2)}$. 

It is instructive to compare the sl(2)-approximated Hodge star operator \eqref{data_ex_csl2} to the full Hodge star operator \eqref{hodgestar01} computed via the LCS periods \eqref{eq:periodsexample}. In particular, this allows us to see which terms are dropped in the sl(2)-approximation. We find that the full Hodge star operator is given by
\begin{align}\label{Cfull}
C &= \scalebox{0.8}{$\begin{pmatrix}
0 & 0 & \frac{6}{\cK} & 0 \\
0 & 0 & 0 & \frac{3}{2 \cK} K^{ij}\\
  - \frac{\cK}{6} & 0 & 0 & 0 \\
0 & -\frac{2\cK}{3} K_{ij}  & 0 & 0
\end{pmatrix}$} \\
&= \scalebox{0.8}{$\left(
\begin{array}{cccccc}
 0 & 0 & 0 & -\frac{3}{6 y_1 y_2^2+4 y_2^3} & 0 & 0 \\
 0 & 0 & 0 & 0 & -\frac{3 y_1^2+4 y_1 y_2+2 y_2^2}{6 y_1 y_2^2+4 y_2^3} &
   \frac{1}{6 y_1+4 y_2} \\
 0 & 0 & 0 & 0 & \frac{1}{6 y_1+4 y_2} & -\frac{3}{12 y_1+8 y_2} \\
 \frac{2}{3} y_2^2 (3 y_1+2 y_2) & 0 & 0 & 0 & 0 & 0 \\
 0 & \frac{6 y_2^2}{3 y_1+2 y_2} & \frac{4 y_2^2}{3 y_1+2 y_2} & 0 & 0 & 0 \\
 0 & \frac{4 y_2^2}{3 y_1+2 y_2} & \frac{4 \left(3 y_1^2+4 y_1 y_2+2
   y_2^2\right)}{3 y_1+2 y_2} & 0 & 0 & 0 \\
\end{array}
\right) $}\, , \nn
\end{align}
with $\cK = \cK_{ijk} y^i y^j y^k - \tfrac{3 \chi \zeta(3)}{16 \pi^3}   $ and K\"ahler metric
\begin{equation}
K_{ij} = \left(
\begin{array}{cc}
 \frac{144 y_2^4}{\left(8 y_2^2 (3 y_1+2 y_2)-3 \epsilon \right)^2} & \frac{96 y_2^4+36
   y_2 \epsilon }{\left(8 y_2^2 (3 y_1+2 y_2)-3 \epsilon \right)^2} \\
 \frac{96 y_2^4+36 y_2 \epsilon }{\left(8 y_2^2 (3 y_1+2 y_2)-3 \epsilon \right)^2} &
   \frac{96 y_2^2 \left(3 y_1^2+4 y_1 y_2+2 y_2^2\right)+36 \epsilon  (y_1+2
   y_2)}{\left(8 y_2^2 (3 y_1+2 y_2)-3 \epsilon \right)^2} \\
\end{array}
\right)\, ,
\end{equation} 
where in the second line of \eqref{Cfull} we dropped the $ \epsilon = \frac{ \chi\zeta(3)}{8 \pi^3}$ correction. Note by comparing \eqref{data_ex_csl2} and \eqref{Cfull}, we see that the sl(2)-approximation is more involved than simply dropping $\epsilon$, and in the limit $y_1 \gg y_2 \gg 1$ we exclude further polynomial corrections in $y_2/y_1$.

\begin{subappendices}

\section{Operator relations for Calabi-Yau fourfolds}
In this appendix we summarize the relations fixing the operators $\eta$ and $\zeta$ in terms of $\delta$ for Calabi-Yau fourfolds \cite{Grimm:2021ckh}. Recall that the central identity relating the three operators was given by \eqref{delta-zeta-eta-review}. In the mathematics literature \cite{Kato} this was solved for the components of $\zeta$ for threefolds.\footnote{Furthermore, general expressions for the components $\zeta_{-p,-q}$ and $\eta_{-p,-q}$ in terms of $\delta_{-p,-q}$ were derived, modulo commutators of $\delta_{-r,-s}$ ($r\leq p$ and $s\leq q$) left undetermined.} These results were extended in \cite{Grimm:2021ikg}, where the components of $\eta$ were given in the threefold case. Following the strategy laid out in section \ref{sec_tech_details}, we proceed and derive the componentwise expressions for $\eta,\zeta$ for fourfolds. We can express $\zeta$ componentwise in terms of $\delta$ as

\begin{footnotesize}
\begin{align}\label{fourfold_zeta}
 \zeta_{-1,-1}&= \zeta_{-2,-2}= 0\, , \quad \zeta_{-1,-2}= -\frac{i}{2} \delta_{-1,-2}\, , \quad \zeta_{-1,-3} = -\frac{3i}{4} \delta_{-1,-3}\, ,  \quad \zeta_{-1,-4} = -\frac{7i}{8} \delta_{-1,-4}\, ,\nn  \\
\zeta_{-2,-3} &=-\frac{3i}{8} \delta_{-2,-3}+ \frac{1}{8} [\delta_{-1,-2}, \delta_{-1,-1} ]\, , \qquad \zeta_{-2,-4} = -\frac{5i}{8} \delta_{-2,-4} + \frac{1}{4} [ \delta_{-1,-3}, \delta_{-1,-1} ]\, , \nn
\\
 \zeta_{-3,-3} &= \frac{1}{8}[\delta_{-2,-2}, \delta_{-1,-1}]\, ,  \\
\zeta_{-3,-4}  &= -\frac{5i}{16} \delta_{-3,-4} +\frac{3}{16} [\delta_{-2,-3} , \delta_{-1,-1} ] +\frac{3}{16} [\delta_{-1,-3}, \delta_{-2,-1} ] +\frac{i}{48} [\delta_{-1,-1}, [\delta_{-1,-1}, \delta_{-1,-2}]]\, ,  \nn \\
\zeta_{-4,-4} &= \frac{3}{16} [ \delta_{-3,-3}, \delta_{-1,-1} ] +\frac{3}{32} [\delta_{-3,-2} , \delta_{-1,-2}]+\frac{3}{32} [\delta_{-2,-3} , \delta_{-2,-1}] \nn \\ 
&\ \ \ +\frac{i}{32} [\delta_{-1,-1}, [\delta_{-2,-1}, \delta_{-1,-2} ]] \, , \nn
\end{align}
\end{footnotesize}
\noindent while $\eta$ is expressed componentwise as

\begin{footnotesize}
\begin{align}\label{fourfold_eta}
\eta_{-1,-1} &= - \delta_{-1,-1}\, , \qquad \eta_{-1,-2} = - \delta_{-1,-2}\, , \qquad \eta_{-1,-3} = - \frac{3}{4}\delta_{-1,-3}\, , \qquad \eta_{-1,-4} = - \frac{1}{2}\delta_{-1,-4}\, , \nn \\
 \eta_{-2,-2} &= - \frac{3}{2} \delta_{-2,-2}\, , \hspace{125.5pt} \eta_{-2,-3} = -\frac{3}{2} \delta_{-2,-3}+\frac{i}{2} [\delta_{-1,-1}, \delta_{-1,-2}]\, , \nn  \\
 \eta_{-2,-4} &= -\frac{5}{4} \delta_{-2,-4}+\frac{5i}{8}[\delta_{-1,-1}, \delta_{-1,-3}]\, , \hspace{30.5pt} \eta_{-3,-3} =  -\frac{15}{8}\delta_{-3,-3}+\frac{5i}{4}[ \delta_{-2,-1}, \delta_{-1,-2}] \, , \nn \\
 \eta_{-3,-4} &= -\frac{15}{8} \delta_{-3,-4}+\frac{3i}{8}[ \delta_{-1,-1},  \delta_{-2,-3}]  \\
 &\quad+\frac{3i}{2}[ \delta_{-2,-1},  \delta_{-1,-3}]+\frac{3i}{4}[ \delta_{-2,-2},  \delta_{-1,-2}] +\frac{1}{8} [  \delta_{-1,-1}, [ \delta_{-1,-1},  \delta_{-1,-2}]]\, , \nn  \\
   \eta_{-4,-4} &= -\frac{35}{16} \delta_{-4,-4} +\frac{63i}{32} [\delta_{-3,-1}, \delta_{-1,-3}]+\frac{21i}{16} [\delta_{-3,-2}, \delta_{-1,-2}]+\frac{21i}{16} [\delta_{-2,-1}, \delta_{-2,-3}] \nn \\
   & \quad +\frac{7}{48}[\delta_{-1,-1},  [\delta_{-1,-1}, \delta_{-2,-2}]]+\frac{7}{24}[\delta_{-2,-1},  [\delta_{-1,-1}, \delta_{-1,-2}]] \nn \\
  & \ \ \ +\frac{7}{24}[\delta_{-1,-2},  [\delta_{-1,-1}, \delta_{-2,-1}]]\, . \nn
\end{align}
\end{footnotesize}

\end{subappendices}

\setpartpreamble[u][\textwidth]{
\vspace*{1cm}
\hrulefill 
\vspace*{0.5cm}

In this second part we apply the machinery of asymptotic Hodge theory in a geometrical situation. In chapter \ref{chap:models} we study the $(3,0)$-form periods of Calabi-Yau threefolds near boundaries in complex structure moduli space. We explain that near most boundaries exponentially suppressed corrections must be present for consistency of e.g.~the K\"ahler metric. We develop a procedure to construct general models for these asymptotic periods including these essential terms, which we explicitly carry out for all possible one- and two-moduli boundaries.

\vspace*{0.5cm}
\hrulefill }

\part{Geometrical Applications}\label{part2}

\chapter{Constructing general models for asymptotic periods}\label{chap:models}

In this chapter our main focus will be the derivation of the period vector of the $(3,0)$-form of Calabi-Yau threefolds. It is well-known that these periods can be computed by using Picard-Fuchs equations \cite{Hosono:1993qy,Hosono:1994ax,CoxKatz}\footnote{See also \cite{ruddat2019period} for recently developed methods that take a different approach, which in particular does not require one to embed the Calabi-Yau manifold in some ambient space.} whose solutions are in general complicated functions of the complex structure moduli. While this strategy to study periods has been employed successfully for many Calabi-Yau examples in the past, we employ asymptotic Hodge theory here in order to construct general models for the asymptotic periods near boundaries in complex structure moduli space. For the large complex structure point such expressions are already available by using mirror symmetry, which writes the periods in terms of topological data of the mirror manifold. The goal of our work is to provide such lampposts near other boundaries in moduli space, which have received less attention so far in the string theory literature.

While it is straightforward to use the nilpotent orbit to determine the leading polynomial part of the period vector, we show that this polynomial approximation only suffices for boundaries of large complex structure type. For all other boundaries we find that exponential corrections to the period vector have to be included. This generalizes the observation already made in \cite{Grimm:2020cda} and nicely complements \cite{Palti:2020qlc,Cecotti:2020rjq}, in that we now note the importance of exponentially suppressed corrections near boundaries away from large complex structure. One way to see the necessity of these non-perturbative terms is that the K\"ahler metric derived from the polynomial periods can be singular. More generally, these issues can be traced back to a completeness principle: it is a fundamental result for Calabi-Yau threefolds $Y_3$ that it is possible to span the entire three-form cohomology $H^3(Y_3,\mathbb{C})$ from the period vector and its derivatives. The necessity of exponentially suppressed terms is expected from finite distance boundaries such as the conifold point, but we show that it extends to infinite distance boundaries as well. More precisely, we are able to systematically characterize which correction terms are needed depending on the boundary type by using the underlying structure provided by asymptotic Hodge theory. 

Our construction of the periods uses the techniques laid out in the mathematical works \cite{BrosnanPearlsteinRobles,KaplanPearlstein,BrosnanPearlstein,CattaniFernandez2000,CattaniFernandez2008}. We begin by constructing the most general nilpotent orbit compatible with the  sl$(2,\mathbb{R})^n$ boundary data, following the approach of \cite{BrosnanPearlsteinRobles,KaplanPearlstein,BrosnanPearlstein}. In particular, this procedure captures how to get the monodromy transformations from a given set of  sl$(2,\mathbb{R})^n$-data. We then turn to \cite{CattaniFernandez2000,CattaniFernandez2008} where the holomorphic expansion of the periods was encoded in terms of an analytic matrix-valued function $\Gamma$. This function generates the exponential corrections to the polynomial periods, hence we dub it the \textit{instanton map}. The form of this instanton map is constrained by imposing consistency with the boundary Hodge decomposition. Furthermore, its coefficients must obey certain differential equations following from orthogonality relations between derivatives of the periods. We then derive a rank condition on this instanton map which ensures that the entire three-form cohomology $H^3(Y_3, \mathbb{C})$ can be spanned by the period vector and its derivatives. This allows us to determine precisely which coefficients in $\Gamma$ are needed, consequently indicating the essential instanton terms required in the periods.  

We explicitly carry out this program for all possible one- and two-moduli boundaries in complex structure moduli space. In the one-modulus case boundaries are classified by their singularity type, while for two moduli one has to deal with intersections of divisors where the type can enhance. Nevertheless the two-moduli case can also be completely classified, as was worked out in \cite{Kerr2017}. At the one-modulus level we find two classes of boundaries that require instantons, which cover the conifold point of the (mirror) quintic \cite{Candelas:1990rm} and the so-called Tyurin degeneration \cite{Tyurin:2003}.   For two moduli we encounter three classes, among them the recently studied coni-LCS boundaries \cite{Demirtas:2020ffz,Blumenhagen:2020ire}. As a more involved example, we show how our results cover a degeneration for the Calabi-Yau threefold in $\mathbb{P}^{1,1,2,2,6}$ which played an important role in the geometrical engineering of Seiberg-Witten models \cite{Kachru:1995fv}. 

An interesting application is that the resulting periods can be used in the study of four-dimensional $\mathcal{N}=2$ and $\mathcal{N}=1$ supergravity theories arising from Calabi-Yau (orientifold) compactifications -- recall from sections \ref{ssec:IIBN=2} and \ref{ssec:IIBN=1} that the periods encode part of the K\"ahler potential and superpotential. For the K\"ahler potential we find the crucial instanton terms that remedy the singular behavior of the metric. Furthermore, we observe the emergence of a continuous axion shift symmetry near all boundaries we considered, only broken by terms subleading compared to these instantons. For the flux superpotential we find that these metric-essential terms are needed to couple all the fluxes, both for finite and even some infinite distance boundaries. In contrast, for the corresponding scalar potential all fluxes already appear at leading polynomial order.\footnote{Let us also mention that we have control over more instanton terms than just those required for the leading polynomial physical quantities. These subleading terms correct for instance the polynomial scalar potential at exponentially suppressed order.}

This chapter is organized as follows. In section \ref{sec:instantonperiods} we first argue that exponentially suppressed terms are required away from large complex structure, and give a lower bound on the required number based on the boundary data. We then lay out the procedure to construct the periods using the instanton map. In section \ref{sec:models} we provide general models for period vectors for all possible one- and two-moduli boundaries. In sections \ref{app:one-modulus} and \ref{sec:two-moduli} we go through the explicit construction of the one- and two-moduli models respectively. In appendix \ref{app:boundarydata} we construct the nilpotent orbit data for the two-moduli periods, and in appendix \ref{app:embedding} we embed some geometrical examples from the literature into our models for the periods.

\section{Instanton expansion of the periods}\label{sec:instantonperiods}
In this section we elucidate the structure behind exponentially suppressed corrections in the periods. We begin by explaining why these instanton terms are expected to be present from the perspective of asymptotic Hodge theory. To be concrete, we obtain in section \ref{sec:instantonsnes} a criterion \eqref{eq:instantonpresence} for the presence of instanton terms and a lower bound \eqref{eq:columncount} on the number required. We then turn to the techniques used to construct asymptotic expressions for the periods. In \ref{sec:construction} we explain how to write down the most general nilpotent orbit compatible with a given set of sl$(2,\mathbb{R})^n$-data, i.e.~how to construct the log-monodromy matrices $N_i$ and the filtration $F^p_0$. In \ref{sec:construction} we describe the instanton map $\Gamma(z)$ that encodes the instanton expansion the period vector \eqref{eq:periodsexpansion}. In particular, we discuss a rank condition \eqref{eq:rankGamma} for $\Gamma(z)$ indicating the essential instanton terms necessary for recovering the entire filtration $F^p_0$ from just the $(3,0)$-form periods. Finally, let us note that we focus on boundary components of codimension $h^{2,1}$ in this chapter, i.e.~points, so we set $n=h^{2,1}$ in the following.

\tikzstyle{block} = [draw,, rectangle, 
    minimum height=3em, minimum width=2em, align=center] 
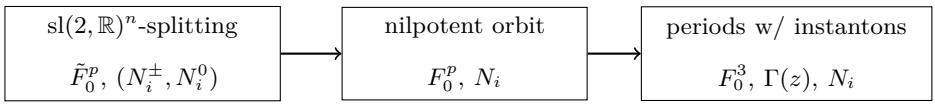
\begin{figure}[h!]
\centering
\begin{tikzpicture}
\node[block, text width=3.4cm] (a) at (0,0) {\small sl$(2,\mathbb{R})^n$-splitting\\ \vspace{0.2cm}
$\tilde{F}^p_{0}$, $(N_i^{\pm}, N_i^0)$ };
\node[block, text width=3.0cm] (b) at (4.25,0) {\small nilpotent orbit \\ \vspace{0.2cm}
$F^p_0$, $N_i$ };
\node[block, text width=3.6cm] (c) at (8.5,0) {\small periods w/ instantons \\ \vspace{0.2cm}
$F^3_0$, $\Gamma(z)$, $N_i$ };

\draw [ thick, ->] (a) -- (b);
\draw [ thick, ->] (b) -- (c);
\end{tikzpicture}
\caption{Flowchart illustrating the steps in constructing the periods. We start by writing down the sl$(2,\mathbb{R})^n$-data that characterizes a strict asymptotic regime near the boundary. We then extend from this strict asymptotic regime by constructing the most general nilpotent orbit compatible with this data. As final step we lift the data encoded in the nilpotent orbit $F^p_0$ into the leading terms of an instanton map $\Gamma(z)$ acting only on $F^3_0$, resulting in essential exponential corrections to the periods.}\label{fig:flowchart}
\end{figure}

\subsection{Presence of instanton terms} \label{sec:instantonsnes}
By using the classification of boundary types based on the dimensions of Deligne splittings in table \ref{table:HDclass}, we can already infer non-trivial information about the instanton terms $\mathbf{a}_{r_1 \ldots r_n}$ in the period vector expansion \eqref{eq:periodsexpansion}. We find that only one type of boundary region does not require the presence of instanton terms, while all other types need these terms in order to be able to recover the full mixed Hodge structure from the period vector. We can state this result as
\begin{small}
\begin{equation}\label{eq:instantonpresence}
\fbox{\rule[-0.5cm]{.0cm}{1.1cm} \
\text{\parbox{.85\textwidth}{\centering Every period vector $\mathbf{\Pi}$ near a co-dimension $h^{2,1}$ boundary component, \textit{not} of type $\mathrm{IV}_{h^{2,1}}$, \textit{must} contain 
instanton terms $\mathbf{a}_{r_1 \cdots r_n}$ in its expansion \eqref{eq:periodsexpansion}.  }} \ }
\end{equation}
\end{small}
Before we argue for this result, let us try to put it into a broader perspective. First, let us stress that such a simple statement cannot generally be formulated 
for periods near lower co-dimension boundary components. While we can apply a similar strategy near boundary components of lower co-dimensions, the necessity of 
instantons will depend on more details of the local period vector (for instance spectator moduli) that are not captured by simply stating the boundary type. Second, we recall that 
a well-known class of type $\mathrm{IV}_{h^{2,1}}$ boundaries are the large complex structure points. These points evade the above statement, and indeed do not require the presence of instanton terms for consistency. Mathematically this follows from the fact that the vector $\mathbf{a}_0$ and its descendants obtained by applying $N_i$ suffice to span the complete vector space $H^3(Y_3,\mathbb{C})$. Interestingly, there is another closely-related class of $\mathrm{IV}_{h^{2,1}}$ boundaries, which we dub coni-LCS points. These boundary components 
can be obtained, in certain examples, by considering a large complex structure point and then sending one modulus away to a conifold locus. 
In terms of the associated mixed Hodge structure one then finds that this one-modulus limit results in a $\mathrm{I}_a$ boundary component, whereas additionally sending the remaining moduli to the large complex structure regime can still yield a $\mathrm{IV}_{h^{2,1}}$ point at the intersection. While these coni-LCS points evade the above theorem as well, the type $\mathrm{I}_a$ mixed Hodge structure does require us to consider instanton terms for the conifold modulus. In fact, we will study these points later explicitly in section \ref{sec:coniLCS} as the intersection of $\mathrm{I}_1$ and $\mathrm{IV}_{1,2}$ divisors in two-dimensional moduli spaces. 

Let us now argue for the statement \eqref{eq:instantonpresence}. We know that the vector $\mathbf{a}_0$ spans the vector space $I_{(n)}^{3,d}$ in the Deligne splitting. Application of the log-monodromy matrices $N_i$ then lowers us within the same column according to \eqref{eq:Nmin1min1}. This implies that the dimension of the vector space spanned by $\mathbf{a}_0$ and its descendants is bounded from above by
\begin{equation}
\dim_\mathbb{C} \big(\text{span}_{\mathbb{C}}( N_{i_1} \cdots N_{i_k} \mathbf{a}_0 ) \big) \leq \sum_{k=0}^d i^{3-k,d-k}\, ,
\end{equation}
where the span runs over all values $k=0,1,2,3$ and $0 \leq i_1,i_2,i_3 \leq n$. In other words, the vector $\mathbf{a}_0$ and its descendants span at most the column of $I_{(n)}^{3,d}$ in the Deligne splitting. Looking at table \ref{table:HDclass}, this means that we can only generate the vector space $H^3(Y_3,\mathbb{C})$ in its entirety via $\mathbf{a}_0$ for type $\mathrm{IV}_{h^{2,1}}$ singularities. In order to span the other columns of the Deligne splitting, we need other elements to enter in the input $F^p_0$. These enter through the instanton terms $\mathbf{a}_{r_1 \ldots r_n}$ in the period vector expansion, thereby completing $H^3(Y_3,\mathbb{C})$. Thus we find that we must require the presence of instanton terms whenever a singularity is not of type  $\mathrm{IV}_{h^{2,1}}$. 

From \eqref{eq:instantonpresence} we do know whether instanton terms $\mathbf{a}_{r_1 \ldots r_n}$ must be present for a given boundary type, but let us now try to make the minimal number required more precise. We need additional elements in the boundary filtration $F^p_0$ in order to span the other columns of the Deligne splitting $I_{(n)}^{p,q}$, besides the column of $I_{(n)}^{3,d}$ corresponding to $\mathbf{a}_0$. Roughly speaking each instanton term $\mathbf{a}_{r_1 \ldots r_n}$ can only be identified with one column, since descendants via application of $N_i$ end up in the same column according to \eqref{eq:Nmin1min1}. Therefore we only need to count the number of columns in order to get a lower bound on the number of instanton terms required. Looking at table \ref{table:HDclass} we find that
\begin{equation}\label{eq:columncount}
\mathrm{I}_a : \ 2h^{2,1}-a+1\, , \quad \mathrm{II}_b :\ 2h^{2,1}-b-1\, , \quad \mathrm{III}_c : \ c+1\, , \quad \mathrm{IV}_d :\ 2(h^{2,1}-d)\, .
\end{equation}
Note that in this counting scheme we interpret $i^{p,q}>1$ at the top of a column as having $i^{p,q}$ columns. Namely, in this case we need at least $i^{p,q}$ instanton terms $\mathbf{a}_{r_1 \ldots r_n}$ in order to span the $i^{p,q}$-dimensional space $I^{p,q}$.

\subsection{Reconstructing the periods}\label{sec:construction} 
In this subsection we lay out how to reverse engineer asymptotic periods that include these essential instanton terms. We first summarize how to work out the first step in figure \ref{fig:flowchart}, i.e.~construct the nilpotent orbit. Subsequently we explain in detail how to lift this data into the essential instantons in the periods.

\paragraph{sl$(2,\mathbb{R})^n$-data.} Our construction starts from a given sl(2)-splitting. Recall from section \ref{sec:boundarydata} that this splitting is encoded in a set of commuting sl$(2,\mathbb{R})$-triples $(N_i^\pm, N_i^0)$ and an sl$(2,\mathbb{R})$-split Deligne splitting $\tilde{I}^{p,q}_{(n)}$. This data characterizes the boundary in a strict asymptotic regime of the form $y_1 \gg \ldots \gg y_n$. Crucially for us, the possible sl$(2,\mathbb{R})$-splittings that can arise are classified through the limiting mixed Hodge structures given in table \ref{table:HDclass}. Furthermore, there is a systematic procedure to write down simple expressions for the defining data of these sl$(2,\mathbb{R})$-splittings. This procedure translates the classifying Deligne splittings into signed Young diagrams as discussed in chapter \ref{sec:classification}. For our purposes here the details in this correspondence are not important, but we simply note that there exists a set of simple building blocks given in table \ref{table:buildingblocks} that can be used to assemble the defining elements of the sl$(2,\mathbb{R})$-splitting.

\paragraph{Nilpotent orbit data.} Our procedure to reconstruct the nilpotent orbit follows the approach taken in the study of Deligne systems in the mathematics literature, see for instance example 6.61 in \cite{BrosnanPearlsteinRobles} and also \cite{KaplanPearlstein,BrosnanPearlstein}. These systems formalize the structure behind the log-monodromy matrices $N_i$ and sl$(2,\mathbb{R})$-triples $(N_i^{\pm}, N_i^0)$ into a purely linear algebraic setup, without any reference to an underlying geometrical origin. The task of writing down the most general nilpotent orbit compatible with the sl$(2,\mathbb{R})$-splitting is then twofold:
\begin{itemize}
\item We want to construct the most general log-monodromy matrices $N_i$ that match with the lowering operators $N_i^-$ in the strict asymptotic regime: this is prescribed by the decomposition \eqref{lowering}, where one has to solve for the most general set of maps $N_{k,-\ell}$ to complete the log-monodromy matrices.\footnote{To be precise, one has to impose that we have isometries of the symplectic pairing and log-monodromies commute with each other, i.e.~$(N_{k,-\ell})^T\eta+\eta N_{k,-\ell}=0$ and $[N_k,N_r]=0$. Additionally, we have to require that $N_k$ is a $(-1,-1)$-map with respect to the sl(2)-split Deligne splitting $\tilde{I}^{p,q}_{(k)}$ as described by \eqref{eq:Nmin1min1}, for each $k=1,\ldots, n$.} 
\item We want to consider the most general rotation away from the sl(2)-split $\tilde{I}^{p,q}_{(n)}$ for the Deligne splitting: we can parametrize this by writing down the most general real phase operator $\delta$ that rotates the sl(2)-split Deligne splitting according to \eqref{eq:Ftilde}, with $\zeta$ fixed by \eqref{zeta}.\footnote{To be precise, $\delta$ has to admit an expansion \eqref{delta_decomp} with respect to the Deligne splitting $\tilde{I}^{p,q}_{(n)}$, and furthermore commute with the log-monodromy matrices.}
\end{itemize}

\subsubsection*{Instanton map}
We now turn to the instanton map $\Gamma(z)$. We use this map to describe the expansion in instanton terms $z=e^{2\pi i t}$ for the period vector. This map has originally been studied in great detail in \cite{CattaniFernandez2000,CattaniFernandez2008}, and we review the relevant aspects of their work here. Let us first state how we can recover the Hodge filtration $F^p$ from the boundary structure by using $\Gamma(z)$. We can write it in terms of the limiting filtration $\tilde{F}^p_0$ of the sl(2)-splitting $\tilde{I}^{p,q}_{(n)}$ as\footnote{In comparison to \cite{CattaniFernandez2000} we chose to expand $F^p_0 = e^{i\delta} e^{-\zeta} \tilde{F}^p_0$, and rewrite in terms of the filtration $\tilde{F}^p_0$ of the sl(2)-split mixed Hodge structure instead of $F^p_0$. Furthermore we commuted the exponentials involving $\delta$ and $\zeta$ to the left, which means that the instanton maps are related by $\Gamma |_{\rm here} =e^{\zeta} e^{-i\delta} \Gamma |_{\rm there}  e^{i\delta} e^{-\zeta}  $.  \label{fn:footnoteGamma}}
\begin{equation}\label{eq:FpGamma}
F^p = e^{i\delta} e^{-\zeta} e^{t^i N_i} e^{\Gamma(z)} \, \tilde{F}^p_0\, ,
\end{equation}
where $\Gamma(z)$ is a matrix-valued function holomorphic in $z=e^{2\pi it}$ with $\Gamma(0)=0$. Vanishing at $z=0$ ensures that the nilpotent orbit $e^{t^{i}N_{i}}F^{p}_{0}$ provides a good approximation for $F^{p}$ for $y^{i} \gg 1$. To be more precise, $\Gamma(z)$ is a map valued in the Lie algebra $\mathfrak{sp}(2h^{2,1}+2,\mathbb{C})$, located in 
\begin{equation}\label{eq:locGamma}
\Gamma(z) \in \Lambda_{-} = \bigoplus_{p<0}\bigoplus_q \Lambda_{p,q}\, ,
\end{equation}
where we consider the operator spaces $\Lambda_{p,q}$ with respect to the sl(2)-splitting $\tilde{I}^{p,q}_{(n)}$. From a practical perspective this means one needs to determine a basis for elements of $\Lambda_{-}$ that lie in the Lie algebra $\mathfrak{sp}(2h^{2,1}+2,\mathbb{C})$. One can then write out $\Gamma(z)$ by expanding in terms of this basis, where holomorphic functions vanishing at $z=0$ are taken as coefficients. Later we find that these holomorphic coefficients can be constrained by differential equations obtained from \eqref{eq:gammamin1} and \eqref{eq:recursiongamma}. 

For the purposes of this work we want to translate the vector space relation \eqref{eq:FpGamma} into an expression for the period vector. By taking a representative $\mathbf{\tilde{a}}_0$ of $\tilde{F}^{3}_{0}$, we find that we can write the period vector $\mathbf{\Pi}$ as
\begin{equation}\label{eq:PiGamma}
\boxed{
\rule[-.25cm]{.0cm}{0.8cm} \quad
\mathbf{\Pi}(t) = e^{i\delta} e^{-\zeta} e^{t^i N_i} e^{\Gamma(z)} \, \mathbf{\tilde{a}}_0\, . \quad}
\end{equation}

\paragraph{Horizontality.} In order to constrain the instanton map $\Gamma(z)$, we can now use the horizontality property of the Hodge filtration as described by \eqref{eq:Transversality}. The idea is that besides \eqref{eq:locGamma} the instanton map should satisfy certain differential conditions to produce a consistent period vector. To obtain these conditions, it is convenient to first combine the exponential maps acting on  $\tilde{F}^p_0$ in \eqref{eq:FpGamma} (respectively on $\mathbf{\tilde a}_{0}$ in \eqref{eq:PiGamma}) into a single Sp$(2h^{2,1}+2,\mathbb{C})$-valued matrix. We define this matrix as
\begin{equation}\label{eq:Edef}
E(t) =   \exp[X(t)]  \equiv e^{i\delta} e^{-\zeta} e^{t^i N_i} e^{\Gamma(z)}  \, ,
\end{equation}
where $X(t)$ is valued in $\mathfrak{sp}(2h^{2,1}+2,\mathbb{C})$ and $\Lambda_{-}$, since $\delta,\zeta,N_i,\Gamma(z)$ are all valued in these operator subspaces. For later reference let us write out the component in $\Lambda_{-1}=\bigoplus_{q} \Lambda_{-1,q}$ explicitly as
\begin{equation}\label{eq:Xmin1}
X_{-1}(t) = i\delta_{-1}-\zeta_{-1}+t^{i}N_{i}+\Gamma_{-1}(z)\, ,
\end{equation}
which follows simply from expanding the exponentials in \eqref{eq:Edef}. The horizontality property \eqref{eq:Transversality} can now be recast into a condition on $E(t)$ by rewriting the Hodge filtration $F^p$ with \eqref{eq:FpGamma}. This leads to a vector space relation that reads
\begin{equation}\label{eq:EdEsubset}
\big( E^{-1}\partial_{i} E \big) \tilde{F}^{p}_{0} \subseteq \tilde{F}^{p-1}_{0}\, .
\end{equation}
From writing out the exponentials in \eqref{eq:Edef} it already follows that $E^{-1}\partial_{i} E \in \Lambda_-$. However, we also know that the $\tilde{F}^p_0$ can be split into $\tilde{I}_{(n)}^{p,q}$ according to \eqref{eq:ItoFW}, so $E^{-1}\partial_{i} E$ can only be valued in the operator subspaces $\Lambda_{-1,q}$. Therefore we must impose
\begin{equation}\label{eq:EdE}
E^{-1}\partial_{i} E \in \Lambda_{-1}\, .
\end{equation}
By expanding the exponentials in \eqref{eq:Edef} we find that this implies
\begin{equation}\label{eq:EdEdX}
E^{-1}\partial_{i} E = \partial_i X_{-1} \, ,
\end{equation} 
since higher order terms are valued in the operator subspaces $\Lambda_{-2,q}$ or lower. Note in particular from \eqref{eq:Edef} and \eqref{eq:Xmin1} that the operators $\delta$ and $\zeta$ drop out of this relation, which can be seen immediately on the right-hand side since they are constant, while on the left-hand side they can be moved past the partial derivative.

\paragraph{Horizontality conditions.} The differential constraint \eqref{eq:EdEdX} on the instanton map $\Gamma(z)$ ensures that we can integrate the boundary data into a consistent period vector. However, imposing \eqref{eq:EdEdX} directly is not the most practical way to constrain this instanton map. In \cite{CattaniFernandez2000} a convenient approach was given to reduce \eqref{eq:EdEdX}. The idea is to first derive a necessary and sufficient condition \eqref{eq:gammamin1} on the component $\Gamma_{-1}(z)$ of the instanton map. Subsequently the lower-charged components $\Gamma_{-p}(z)$ with $p\geq 2$ can be fixed recursively through \eqref{eq:recursiongamma}. The differential condition on $\Gamma_{-1}(z)$ is obtained by taking another derivative $\partial_{j}$ of \eqref{eq:EdEdX} and antisymmetrizing in $i,j$, which yields\footnote{This condition is more naturally obtained by introducing an exterior derivative $d=\partial_{i} dt^{i}$ on the moduli space.}
\begin{equation}\label{eq:dXdX}
\partial_{[i} X_{-1} \partial_{j]}X_{-1} = 0\, .
\end{equation}
By using \eqref{eq:Xmin1} we can formulate this as a differential constraint on $\Gamma_{-1}(z)$ as
\begin{equation}\label{eq:gammamin1}
[N_{i}, \partial_{j} \Gamma_{-1}(z) ] + [\partial_i \Gamma_{-1}(z), N_j ] +[\partial_i \Gamma_{-1}(z), \partial_j \Gamma_{-1}(z) ] = 0 \, ,
\end{equation}
where we used that $[N_i, N_j] =0$ since the log-monodromy matrices commute. Next we need to obtain constraints on the lower-charged components $\Gamma_{-q}(z)$ with $q>2$ of the instanton map. First we write out \eqref{eq:EdEdX} by multiplying from the left with $\exp[\Gamma(z)]$ as
\begin{equation}
\partial_{i} \exp[\Gamma(z)] = [\exp[\Gamma(z)], N_{i}]+  \exp[\Gamma(z)]  \partial_{i} \Gamma_{-1}(z)\, .
\end{equation}
This condition can be translated into a constraint on the components in the subspaces $\Lambda_{-p} = \bigoplus_q \Lambda_{-p,q}$ as
\begin{equation}\label{eq:recursiongamma}
\partial_{i} \exp[\Gamma(z)]_{-p} = [\exp[\Gamma(z)]_{-p+1}, N_{i}]+  \exp[\Gamma(z)]_{-p+1}  \partial_{i} \Gamma_{-1}(z)\, .
\end{equation}
From  the left-hand side we obtain the term $\Gamma_{-p}(z)$ by expanding the exponential, while the other terms that appear in the equation are of charge $\Gamma_{-p+1}(z)$ or lower. This means we can fix $\Gamma_{-p}(z)$ uniquely in terms of the lower-charged components $\Gamma_{-1}(z),\ldots, \Gamma_{-p+1}(z)$. By induction we thus find that the entire map $\Gamma(z)$ is uniquely determined by its piece $\Gamma_{-1}(z)$, provided this piece solves the consistency requirement \eqref{eq:gammamin1}.

\paragraph{Coordinate changes.} It is worthwhile to check how coordinate redefinitions affect the instanton map $\Gamma(z)$ and $\delta$, since these transformations can later be used to reduce the number of arbitrary components for both. We can understand their effect most naturally by looking at $X_{-1}(t)$ given in \eqref{eq:Xmin1}. The most general divisor-preserving coordinate redefinition takes the form $z^{i} \to z^{i}f(z)$, where $f(z)$ is any holomorphic function with $f(0) \neq 0$. In terms of the coordinates $t^i$ defined in \eqref{eq:Coordinates} this amounts to shifting $2\pi i t^i \to 2\pi i t^i + \log f(z)$. Applying this shift to $t^i N_i$ produces two terms, a constant term involving $\log[f(0)]$ and a holomorphic term involving $\log[f(z)/f(0)]$ which vanishes at $z=0$. Taking $f(0)$ to be real we can absorb the former into the phase operator $\delta$, and the latter into $\Gamma(z)$. To be precise, from \eqref{eq:Xmin1} we find the following shifts
\begin{equation}\label{eq:shift}
\begin{aligned}
\delta_{-1,-1} &\to \delta_{-1,-1} - \frac{1}{2\pi }  \log[ f(0)] \,  N_i \, , \\
\Gamma_{-1}(z) &\to \Gamma_{-1}(z)+ \frac{1}{2\pi i}  \log\Big[ \frac{f(z^{i})}{f(0)} \Big] \, N_{j}\, .
\end{aligned}
\end{equation}
Later we will expand both of these maps into a basis for $\mathfrak{sp}(2h^{2,1}+2, \mathbb{K})$ (with $\mathbb{K}=\mathbb{R}, \mathbb{C}$ respectively) that is valued in the appropriate operator subspaces $\Lambda_{p,q}$. From these shifts we learn that we are free to set the components along the log-monodromy matrices $N_i$ to zero, effectively reducing the number of arbitrary coefficients that have to be dealt with. Let us also note that we did not yet exploit the full set of coordinate redefinitions: we can still rotate $f(0)$ by a complex phase, corresponding to a shift of the axion $x^i$ in $t^i=x^i+i y^i$. A shift $x^i \to x^i + c^i$ can then partially be absorbed by a basis transformation $e^{cN^i}$ for the Deligne splitting $\tilde{I}^{p,q}_{(n)}$, while it also rotates the complex phase of exponentially suppressed terms in the periods. The latter feature will prove to be useful in the explicit construction of the periods in one- and two-moduli settings in section \ref{sec:constructionperiods}, since it allows us to set the leading instanton coefficients to real values.

\paragraph{Completeness condition.} Finally, let us discuss the precise conditions that need to be imposed on $\Gamma(z)$ in order to realize the instanton terms required by \eqref{eq:instantonpresence}.  For Calabi-Yau threefolds we want that derivatives of the period vector together span the vector space $H^3(Y_3,\mathbb{C})$. In terms of the Hodge filtration $F^p$ this amounts to putting an equality sign in \eqref{eq:EdEsubset} when we take all possible linear combinations of the partial derivatives $E^{-1} \partial_i E$ into account on the left-hand side. Following \cite{CattaniFernandez2008} we can translate this statement into a more concrete condition involving the instanton map $\Gamma(z)$. The vector space $\tilde{F}^3_0 =\tilde I_{(n)}^{3,d}$ is one-dimensional, while the span of all $E^{-1} \partial_i E $ needs to be able to generate all lower lying spaces $I^{p,q}$ with $p<3$. The total dimension of these spaces is given by $2h^{2,1}+1$, so we find that
\begin{equation}\label{eq:rankGamma}
\dim \Big(\bigoplus_i  \img  (N_i+\partial_i \Gamma_{-1}) \Big)= 2h^{2,1}+1\, ,
\end{equation}
where we wrote out $E^{-1} \partial_i E $ in terms of $N_i$ and $\partial_i \Gamma_{-1}$ according to \eqref{eq:EdEdX} and \eqref{eq:Xmin1}. This condition can be understood intuitively by considering the Hodge-Deligne diamond in figure \ref{fig_deligne}. It implies that either a log-monodromy matrix $N_i$ or the instanton map $\Gamma_{-1}$ should map into every space $\tilde I_{(n)}^{p,q}$ in the Deligne splitting apart from $\tilde I_{(n)}^{3,d}$. Since the log-monodromy matrices $N_i$ are $(-1,-1)$-maps and therefore only act vertically on the diamond, this means we need $\Gamma_{-1}$ to generate the horizontally separated columns. In practice, we will use \eqref{eq:rankGamma} to determine which components of $\Gamma_{-1}$ are required to be non-vanishing. For the one-modulus setups discussed in section \ref{app:one-modulus} we find that $\Gamma_{-1}$ has only one functional degree of freedom, so \eqref{eq:rankGamma} dictates if this function must be non-vanishing. For the two-modulus setups studied in section \ref{sec:two-moduli} we find that $\Gamma_{-1}$ consists of several holomorphic functions, and \eqref{eq:rankGamma} will generically only indicate for some of these whether they must be non-vanishing.

\section{Models for one- and two-moduli periods}\label{sec:models}
Here we present general expressions for the periods near one- and two-moduli boundaries. We refer to sections \ref{app:one-modulus} and \ref{sec:two-moduli} for the construction of these periods to avoid distracting the reader by technical details. Crucially, we include the essential instanton terms for boundaries away from large complex structure in accordance with our discussion in section \ref{sec:instantonperiods}. This section is written such that it can be read without understanding the ingredients that go into this derivation. In particular, these periods can be used directly in studying four-dimensional supergravity theories, and to illustrate this point we readily compute the corresponding K\"ahler potentials, flux superpotentials and scalar potentials.

\subsection{Models for one-modulus periods} \label{sec:onemodels}
In this section we present general expressions for the periods near boundaries in one-dimensional moduli spaces, and refer to section \ref{app:one-modulus} for the details. Based on the classification reviewed in section \ref{sec:classification} there are three possible types of boundaries for $h^{2,1}=1$, given by
\begin{equation}
\mathrm{I}_1 : \  \text{conifold point}\, ,\quad \mathrm{II}_0 : \ \text{Tyurin degeneration}\, , \quad \mathrm{IV}_1 : \ \text{LCS point}\, .
\end{equation}
As indicated, each of these types of boundaries has a natural geometrical interpretation in the complex structure moduli space of Calabi--Yau threefolds. The type $\mathrm{I}_1$ characterizes conifold points, which arise for instance in the moduli space of the mirror quintic \cite{Candelas:1990rm}. Type $\mathrm{II}_0$ boundaries arise from so-called Tyurin degenerations \cite{Tyurin:2003}, see also \cite{Joshi:2019nzi} for a recent study of these periods in the context of the swampland program. Finally, $\mathrm{IV}_1$ boundaries correspond to large complex structure points, where the periods can be expressed in terms of the triple intersection numbers of the mirror Calabi--Yau manifold. By using \eqref{eq:columncount} we find that instanton terms play a crucial role in the asymptotic regime of $\mathrm{I}_1$ and $\mathrm{II}_0$ boundaries, so these provide us with an excellent setting to demonstrate how the formalism discussed in section \ref{sec:construction} describes periods. In contrast, instanton terms are insignificant in the asymptotic regime of $\mathrm{IV}_1$ boundaries, as follows from \eqref{eq:instantonpresence}. Since these periods are already well-understood from the study of mirror symmetry anyway, we do not include the periods at these boundaries in this chapter.

\subsubsection{Type $\text{I}_1$ boundaries}\label{ssec:I1}
We begin by writing down the periods for $\mathrm{I}_1$ boundaries. From the analysis of section \ref{ssec:constrconifold} we find that these periods can be expressed as
\begin{equation}\label{eq:I1periods}
\Pi = \begin{pmatrix}
1 +\frac{a^2}{8\pi} z^2, & a z, &i-\frac{i a^2}{8\pi} z^2, & \frac{i a}{2\pi} z \log[z] \end{pmatrix},
\end{equation}
where $a\in \mathbb{R}$ is a model-dependent coefficient. These periods contain two instanton terms, i.e.~terms exponentially suppressed in the saxion $y$ in $t=x+iy=\log[z]/2\pi i$. The periods depend on the complex structure modulus $t$ solely through these exponentially suppressed terms, so instanton terms clearly cannot be ignored for these boundaries. In fact, one can verify that $\Pi, \partial_z \Pi, \partial_z^2 \Pi, \partial_z^3 \Pi$ together span a four-dimensional space only when we include the terms at order $z^2$, so including just the terms at order $z$ does not suffice. This matches nicely with \eqref{eq:columncount}, which indicates two instanton terms for these $\mathrm{I}_1$ boundaries.  

For illustration, we compute the K\"ahler potential \eqref{eq:kahlercsperiods} from the above periods
\begin{align}\label{eq:kpI1}
e^{-K}=2-2a^2 e^{-4\pi y} y- \frac{a^4}{32\pi^2}e^{-8\pi y}\, ,
\end{align}
where we wrote $z=e^{2\pi i t}$ with $t=x+iy$. Let us now inspect this K\"ahler potential carefully. It depends exponentially on $y$, so by computing the K\"ahler metric one can straightforwardly verify that $\mathrm{I}_1$ boundaries are at finite distance. Also note that it does not depend on the axion $x$ even though these exponential terms are present, so close to the boundary a continuous shift symmetry $x\to x+c$ emerges for the K\"ahler metric.\footnote{Interestingly this differs from the usual K\"ahler potential one encounters through the prepotential \eqref{eq:I1prepotential}, where a cosine type term arises at order $|z|^2$. Compared to our formulation we have effectively removed this term by a K\"ahler transformation, so it does not make a difference at the level of the K\"ahler metric.}  Looking at the sign of the terms in the K\"ahler potential, we note that the subleading terms are fixed to be negative, which ensures the resulting K\"ahler metric is positive definite. From the perspective of asymptotic Hodge theory these signs follow from the polarization conditions \eqref{eq:pol} that the symplectic form satisfies.\footnote{To be precise, one finds that $a_0 \in P^{3,0}$, $a_1  \in P^{2,2}$ and $a_2 \in P^{0,3}$. The respective polarization conditions then imply that the coefficients of these terms satisfy $i\langle a_0\, , \ \bar{a}_0 \rangle >0$, $\langle a_1\, , \ N \bar{a}_1 \rangle < 0$ and $i \langle a_2\, , \  \bar{a}_2 \rangle <0$.}

Next we consider the flux superpotential \eqref{eq:superpotential}. From the above we obtain
\begin{equation}\label{eq:superpotentialI1}
W= ig_1-g_3-a e^{2\pi i t}\bigg(g_2 t +g_4 \bigg)-\frac{a^2 e^{4\pi i t} }{8\pi} (ig_1 +g_3)\, ,
\end{equation}
where we wrote out the fluxes as $G_3=(g_1,\ldots, g_4)$. In turn we find the leading polynomial scalar potential \eqref{eq:potential} to be
\begin{equation}\label{eq:potentialI1}
4 \cV^2 s  V_{\rm lead}= \bar{G}_3 \, e^{-x N^T}\begin{pmatrix} 1 & 0 & 0 & 0  \\
0 & y-\frac{1}{2\pi}& 0& 0 \\
0 & 0&1& 0  \\
0 & 0 & 0 & \frac{1}{y-\frac{1}{2\pi}} \\
\end{pmatrix} e^{-x N} G_3 \, ,
\end{equation}
where the log-monodromy matrix $N$ is given in \eqref{eq:I1N}. We dropped exponentially suppressed terms in $y$, and left out the $\langle G_3 , \bar{G}_3 \rangle$ term for convenience. The $1/2\pi$ is an artefact of setting the phase operator equal to $\delta=-N/2\pi$ to simplify the periods, and could in principle be removed by a coordinate redefinition as discussed above \eqref{eq:shift}. It is interesting to point out that all fluxes appear at polynomial order in the scalar potential, while  $ig_1+g_3,g_2,g_4$ were exponentially suppressed in the superpotential \eqref{eq:superpotentialI1}. In order to obtain \eqref{eq:potentialI1} it is therefore crucial to include the terms linear in $e^{-2\pi y}$ in the superpotential, while the terms at order $e^{-4\pi y}$ lead to exponentially suppressed corrections.

\subsubsection{Type $\text{II}_0$ boundaries}
Next we consider the periods near $\mathrm{II}_0$ boundaries. From the analysis of section \ref{ssec:constrII0} we find that these periods can be written as
\begin{equation}
\Pi = \Big(1+a z, \ i-iaz, \ \frac{\log[z]}{2 \pi i}+ \frac{az}{2\pi i} (\log[z]-2), \ \frac{\log[z]}{2 \pi } -\frac{az}{2\pi  } (\log[z]-2) \Big),
\end{equation}
where $a \in \mathbb{R}$ is a model-dependent coefficient. Note that these periods do have polynomial terms in $t=\log[z]/2\pi i$, but we also have a restricted form for the periods at order $z=e^{2\pi it}$. One needs this exponentially suppressed term in $t$ in order to span a four-dimensional space with $\Pi , \partial_z \Pi, \partial^2_z \Pi, \partial^3_z \Pi$. This matches nicely with \eqref{eq:columncount}, which indicates one instanton term for $\mathrm{II}_0$ boundaries.  

For illustration, let us again compute the K\"ahler potential \eqref{eq:kahlercsperiods} from the periods
\begin{align}\label{eq:kpII0}
e^{-K}=4  y + \frac{4  a^2(1+\pi y)}{\pi} e^{-4 \pi y} \, ,
\end{align}
where we wrote $z=e^{2\pi i t}$ with $t=x+iy$. Similar to $\mathrm{I}_1$ boundaries the K\"ahler potential does not depend on the axion $x$, both at leading and subleading order, so a continuous shift symmetry $x \to x+c$ emerges close to the boundary. Inspecting the sign of the terms in the K\"ahler potential, we note that both the leading polynomial term as the exponentially suppressed term are fixed to be positive. This ensures that the K\"ahler metric is positive definite, and these signs can again be traced back to the polarization conditions \eqref{eq:pol} of the symplectic form.\footnote{To be precise, one finds that $a_0 \in P^{3,1}$ and $(1+N/\pi i)a_1 \in P^{1,3}$, which implies that the coefficients satisfy $ \langle a_0 , N\bar{a}_0 \rangle >0$ and $\langle a_1, N \bar{a}_1 \rangle > 0$.} Finally, by computing the K\"ahler metric from \eqref{eq:kpI1} one finds that $\text{II}_0$ singularities are at infinite distance, as is expected from K\"ahler potentials that depend through polynomial terms on the saxion $y$ in the large field limit.

Next we consider the flux superpotential \eqref{eq:superpotential}. From the above we obtain
\begin{equation}\label{eq:superpotII0}
W=-g_3-ig_4+(g_1+ig_2)t +a e^{2\pi i t} \big(t-\frac{1}{\pi i}\big)(g_1-ig_2)-ae^{2\pi i t}(g_3-ig_4)\, .
\end{equation}
In turn, we find the leading polynomial scalar potential \eqref{eq:potential} to be
\begin{equation}\label{eq:potentialII0}
4 \cV^2 s V_{\rm lead}=\bar{G}_3e^{-x N^T}\begin{pmatrix} 
y & 0& 0& 0 \\
0 & y&0& 0  \\
0 & 0 & \frac{1}{y} &0\\
0 & 0 & 0 & \frac{1}{y}  \\
\end{pmatrix} e^{-x N} G_3 \, ,
\end{equation}
where the log-monodromy matrix $N$ is given in \eqref{eq:II0N}. We again dropped exponentially suppressed terms in $y$, and left out the $\langle G_3 , \bar{G}_3 \rangle$ term for convenience. Interestingly all fluxes now appear at polynomial order in \eqref{eq:potentialII0}, while before the linear combinations $g_1-ig_2$ and $g_3-ig_4$ were exponentially suppressed in the superpotential. However, $g_1+ig_2$ and $g_3+ig_4$ do appear at polynomial order in the superpotential, so one finds that the instanton terms do not contribute at leading order, but instead result in exponential corrections to the leading polynomial scalar potential. 

\subsection{Models for two-moduli periods}\label{sec:twomodels}
Having discussed the one-modulus periods, we now turn to periods in a two-moduli setting. We refer to section \ref{sec:two-moduli} for the construction of these periods, to avoid distracting the reader with technical details. Recall from section \ref{sec:classification} that two-moduli boundaries are characterized by three types of limiting mixed Hodge structures, written as a 2-cube $\langle \mathrm{A}_1 | \mathrm{A}_{(2)} | \mathrm{A}_2 \rangle$, two for the divisors $y_1 = \infty$ and $y_2=\infty$ and one for their intersection. Let us repeat the exhaustive set of \cite{Kerr2017}  given in \eqref{eq:cubesintro} as
\begin{align}\label{eq:cubes}
\text{$\mathrm{I}_2$ class} &: \quad \langle \mathrm{I}_1 | \mathrm{I}_2 | \mathrm{I}_1 \rangle\, , \ \langle \mathrm{I}_2 | \mathrm{I}_2 | \mathrm{I}_1 \rangle \, , \  \langle \mathrm{I}_2 | \mathrm{I}_2 | \mathrm{I}_2 \rangle    \, ,\nn \\
\text{Coni-LCS class} &: \quad \langle \mathrm{I}_1 | \mathrm{IV}_2 | \mathrm{IV}_1 \rangle \, , \   \langle \mathrm{I}_1 | \mathrm{IV}_2 | \mathrm{IV}_2 \rangle  \, , \nn \\
\text{$\mathrm{II}_1$ class} &: \quad \langle \mathrm{II}_0 | \mathrm{II}_1 | \mathrm{I}_1 \rangle \, , \  \langle \mathrm{II}_1 | \mathrm{II}_1 | \mathrm{I}_1 \rangle \, , \   \langle \mathrm{II}_0 | \mathrm{II}_1 | \mathrm{II}_1 \rangle     \, , \ \langle \mathrm{II}_1 | \mathrm{II}_1 | \mathrm{II}_1 \rangle  \, ,\\
\text{LCS class} &: \quad \langle \mathrm{II}_1 | \mathrm{IV}_2 | \mathrm{III}_0 \rangle \, , \   \langle \mathrm{II}_1 | \mathrm{IV}_2 | \mathrm{IV}_2 \rangle  \, , \ \langle \mathrm{III}_0 | \mathrm{IV}_2 | \mathrm{III}_0 \rangle \, , \ \langle \mathrm{III}_0 | \mathrm{IV}_2 | \mathrm{IV}_1 \rangle\, , \nn \\
& \qquad \langle \mathrm{III}_0 | \mathrm{IV}_2 | \mathrm{IV}_2 \rangle \, , \ \langle \mathrm{IV}_1 | \mathrm{IV}_2 | \mathrm{IV}_2 \rangle \, , \ \langle \mathrm{IV}_2 | \mathrm{IV}_2 | \mathrm{IV}_2 \rangle \, ,  \nn 
\end{align}
where we chose to sort 2-cubes with similar characteristics together. For the first three classes we find that instanton corrections are needed in $\mathbf{\Pi}$ in order to recover the information in the nilpotent orbit $F^p_{\rm nil}$. We determine the general models for the corresponding period vectors in section \ref{sec:two-moduli}, and give a summary of the obtained results here. The fourth subset of 2-cubes consists of cases that can be realized for particular values of the coefficients $\cK_{ijk}$ describing a large complex structure region and hence specify the intersection numbers of a candidate mirror Calabi-Yau threefold. At these boundaries we do not have any predictive capabilities regarding the instanton series with our machinery and we recover the usual K\"ahler cone restrictions, so we will not discuss the periods for these two-moduli setups later.

\subsubsection{Class $\text{I}_2$ boundaries}\label{ssec:I2}
Let us begin with the class of $\mathrm{I}_2$ boundaries. From the analysis in section \ref{ssec:I2construction} we found that we can write the periods near these boundaries as
\begin{equation}\label{eq:I2periods}
\Pi = \begin{pmatrix}
1 -\frac{a^2}{8 \pi  k_{2}}  z_{1}^{2 k_{1}} z_{2}^{2 k_{2}}-\frac{b^2}{8 \pi  m_{1}}  z_{1}^{2 m_{1}}
   z_{2}^{2 m_{2}} \\
 a z_{1}^{k_{1}} z_{2}^{k_{2}} \\
 b z_{1}^{m_{1}} z_{2}^{m_{2}} \\
i+ \frac{i a^2}{8 \pi  k_{2}} z_{1}^{2 k_{1}} z_{2}^{2 k_{2}}+\frac{i b^2 }{8 \pi  m_{1}} z_{1}^{2 m_{1}}
   z_{2}^{2 m_{2}} \\
-\frac{a}{2\pi i}  z_{1}^{k_{1}} z_{2}^{k_{2}}  \big( n_1 \log[z_1]+\log [z_{2}]-1/k_1\big)+i b \delta_{1}
   z_{1}^{m_{1}} z_{2}^{m_{2}} \\
 -\frac{b}{2\pi i}  z_{1}^{m_{1}}
   z_{2}^{m_{2}} \big(  \log [z_{1}]+n_2 \log[z_2]-1/m_2 \big)+ i a \delta_{1} z_{1}^{k_{1}} z_{2}^{k_{2}}
\end{pmatrix}\, .
\end{equation}
Let us briefly discuss the parameters that appear in these periods, whose properties have been summarized in table \ref{table:I2parameters}. The numbers $n_1,n_2 \in \mathbb{Q}_{\geq 0}$ parametrize the monodromy transformations under $z_i \to e^{2\pi i} z_i$. The integers $k_1,k_2 \in \mathbb{N}$ and $m_1,m_2 \in \mathbb{N}$ specify the order in the instanton expansion. These orders are fixed to be the (smallest) integers such that $n_1=k_1/k_2$ and $n_2=m_2/m_1$ (with $m_1,k_2 >0$), which follows from the horizontality property \eqref{eq:horizontality} of the periods. We must furthermore require $n_1 n_2 \neq 1$ to ensure that the derivatives of the periods together span a six-dimensional space, i.e.~the full three-form cohomology $H^3(Y_3, \mathbb{C})$. Finally, we have real coefficients $\delta_1 \in \mathbb{R}$ and $a,b\in \mathbb{R}$. The coefficient $\delta_1$ coming from the phase operator is always real, while the instanton coefficients $a,b$ have been rotated to real values using shifts of the axions $x^i$ in $x^i+i y^i = \log[z^i]/2\pi i$ as explained below \eqref{eq:shift}. 

\renewcommand*{\arraystretch}{2.0}
\begin{table}[ht!]
\centering
\scalebox{0.85}{
\begin{tabular}{| l || c | c | c | c |}
\hline parameters  & $ \langle \mathrm{I}_1 | \mathrm{I}_2 | \mathrm{I}_1 \rangle $ & $ \langle \mathrm{I}_2 | \mathrm{I}_2 | \mathrm{I}_1 \rangle $& $ \langle \mathrm{I}_2 | \mathrm{I}_2 | \mathrm{I}_2 \rangle $\\ \hline \hline 
log-monodromies $n_1, n_2$ & $n_1=n_2=0$ & $n_1 \in \mathbb{Q}_{>0}$, $n_2=0$ & $n_1,n_2 \in \mathbb{Q}_{>0}$, $n_1 n_2 \neq 1$  \\ \hline
instanton orders $k_1,k_2$ & $k_1=0,k_2=1$ & $k_1 = n_1 k_2$& $k_1 = n_1 k_2$ \\ \hline
instanton orders $m_1,m_2$ & $m_1=1,m_2=0$ & $m_1=1,m_2=0$ & $m_2 = n_2 m_1$  \\ \hline
instanton coefficients $a,b$ & \multicolumn{3}{c|}{$a,b \in \mathbb{R} -\{0\}$} \\ \hline
phase operator $\delta$ & \multicolumn{3}{c|}{$\delta_1 \in \mathbb{R}$} \\ \hline
\end{tabular}}
\caption{\label{table:I2parameters} Summary for the properties of the parameters in the periods \eqref{eq:I2periods} for each of the possible boundaries of class $\mathrm{I}_2$. }
\end{table}
Let us now compute the K\"ahler potential \eqref{eq:kahlercsperiods} from these periods. We find that
\begin{align}\label{eq:kpI2}
e^{-K} &= 2 -2 a^2 e^{-4\pi k_1 y_1-4\pi k_2 y_2} \big( n_1 y_1+ y_2+\frac{1}{2\pi k_2} \big) \nn\\
& \ \ \ -2b^2 e^{-4\pi m_1 y_1-4\pi m_2 y_2} \big(  y_1+n_2 y_2+\frac{1}{2\pi m_1} \big)  \\
& \ \ \ +4 \delta_1 ab e^{-4\pi (k_1+m_1)y_1-4\pi (k_2+m_2)y_2}   \cos[2\pi (k_1-m_1)x_1+2\pi(k_2-m_2)x_2]\nn  \, ,
\end{align}
where we only included terms up to square order in the two instanton expansions, i.e.~in $e^{-2\pi (k_1 y_1+ k_2 y_2)} $ and $e^{-2\pi (m_1 y_1+m_2 y_2)} $, and we used $2\pi i t_i=2\pi i(x_i+y_i) = \log z_i$ for convenience. Note that the sign of the first two non-constant terms is fixed to be negative similar to the one-modulus $\mathrm{I}_1$ boundaries, which again follows from the polarization conditions \eqref{eq:pol} that the symplectic form satisfies. The parameter $\delta_1$ of the phase operator controls the mixing between the two different instanton terms in the periods, i.e.~one coming from $a z_1^{k_1}z_2^{k_2}$ and the other from $ b z_{1}^{m_{1}} z_{2}^{m_{2}}$. This mixing term breaks the continuous shift symmetry for a particular linear combination of the axions $(k_1-m_1)x_1+(k_2-m_2)x_2$, while for the direction $(k_1-m_1)x_1=-(k_2-m_2)x_2$ we still find that a continuous shift symmetry emerges near the boundary for the K\"ahler potential. Finally, one can straightforwardly verify by computing the K\"ahler metric from \eqref{eq:kpI2} that class $\mathrm{I}_2$ boundaries are at finite distance for any large field limit in $y_1,y_2$ due to the exponential dependence. 

Next we consider the flux superpotential \eqref{eq:superpotential}. By using the above periods we find that
\begin{equation}
\begin{aligned}
W &=  ig_1-g_4-a \Big(g_2  \big( n_1 t_1+t_2 - \frac{1}{2 \pi  i k_1}\big)-i\delta_1 g_3 +g_5 \Big) e^{2\pi i (k_1 t_1+k_2 t_2)} \\
&\ \ \ -b \Big(g_3 \big( t_1 + n_2 t_2 -\frac{1}{2\pi i m_2} \big) -i\delta_1 g_2+g_6 \Big) e^{2\pi i (m_1 t_1+m_2 t_2)}\\
&\ \ \ +\frac{a^2}{8\pi k_2} (ig_1+g_4)e^{4\pi i (k_1 t_1+k_2 t_2)} +\frac{b^2}{8\pi m_1} (ig_1+g_4)e^{4\pi i (m_1 t_1+m_2 t_2)}   \, ,
\end{aligned}
\end{equation}
where we wrote out the fluxes as $G_3=(g_1,\ldots, g_6)$. In turn we find the leading polynomial scalar potential \eqref{eq:potential} to be
\begin{equation}
4 \cV^2 s  V_{\rm lead}= \bar{\rho}\, \scalebox{0.85}{$\begin{pmatrix} 1 & 0 & 0 & 0 & 0 & 0 \\
0 & n_1 y_1+y_2 & \delta_1 & 0 & 0 & 0 \\
0 & \delta_1 & y_1+n_2 y_2 & 0 & 0 & 0 \\
0 & 0 & 0 & 1 & 0 & 0 \\
0 & 0 & 0 & 0 & \frac{y_1+n_2 y_2}{\Delta} & \frac{\delta_1}{\Delta}  \\
0 & 0 & 0 & 0 & \frac{\delta_1}{\Delta}  & \frac{n_1y_1+ y_2}{\Delta}  \\
\end{pmatrix}$}\, \rho \, ,
\end{equation}
where we wrote $\rho = e^{-x^iN_i} G_3 $ (with the $N_i$ given in \eqref{eq:I2N}) and $\Delta=(n_1 y_1+y_2)(y_1+n_2 y_2)-\delta_1^2$. We dropped exponentially suppressed corrections in $y_1,y_2$ and left out the $\langle G_3 , \bar{G}_3 \rangle$ term. Note in particular that the linear combination of fluxes $ig_1+g_4$ as well as $g_2,g_3,g_5,g_6$ are exponentially suppressed in $y_1,y_2$ in the superpotential, while all fluxes appear at polynomial order in the scalar potential. We can trace these terms in the scalar potential back to the terms at orders $e^{-2\pi (k_1 y_1+ k_2 y_2)} $ and $e^{-2\pi (m_1 y_1+m_2 y_2)} $ in the superpotential, while the subleading corrections in the superpotential do produce exponential corrections in the scalar potential.

\subsubsection{Class $\text{II}_1$ boundaries}\label{ssec:II1model}
We continue with the class of $\text{II}_1$ boundaries. Within this class, it is interesting to point out that the periods near the boundary $ \langle \mathrm{II}_0 | \mathrm{II}_1 | \mathrm{II}_1 \rangle $ cover a well-studied degeneration for the K3-fibered Calabi-Yau threefold in $\mathbb{P}_4^{1,1,2,2,6}[12]$\cite{Hosono:1993qy, Candelas_1994, Kachru:1995fv, Curio:2000sc,Lee:2019wij}. The precise match between the two sets of periods is included in appendix \ref{app:Seiberg-Witten}. As outlined in section \ref{ssec:II1construction}, the period vector for boundaries of class $\mathrm{II}_1$ can be written as 
\begin{equation}\label{eq:II1periods}
\begin{aligned}
\mathbf{\Pi}&= \scalebox{0.805}{$\begin{pmatrix}
1+ c z_1^{m_1} z_2^{m_2} +\frac{1}{4}  \big(a^2 n_1 z_2^2+ 2 a b z_1 z_2\frac{1-n_1 n_2^2}{1-n_2}+b^2 n_2 z_1^2 \big)\\
i -i c z_1^{m_1} z_2^{m_2} - \frac{i}{4} \big(a^2 n_1 z_2^2+ 2 a b z_1 z_2\frac{1-n_1n_2^2}{1-n_2}+b^2 n_2 z_1^2 \big) \\
b n_2 z_1 + a z_2\\
   \frac{\log [z_1] + n_2 \log[z_2] }{2\pi i }\Big(1+ c z_1^{m_1} z_2^{m_2} +\frac{1}{4}  \big(a^2 n_1 z_2^2+ 2 a b z_1 z_2\frac{1-n_1 n_2^2}{1-n_2}+b^2 n_2 z_1^2 \big) \Big)  +f(z)\\
\frac{\log [z_1] + n_2 \log[z_2] }{2\pi  } \Big(1- c z_1^{m_1} z_2^{m_2} -\frac{1}{4}  \big(a^2 n_1 z_2^2+ 2 a b z_1 z_2\frac{1-n_1 n_2^2}{1-n_2}+b^2 n_2 z_1^2 \big)  \Big)-i f(z)\\
i ( b n_2 z_1 + a z_2) \frac{n_1 \log[z_1]+ \log[z_2]}{2\pi}- \frac{1-n_1 n_2}{2\pi i}(b z_1 - a z_2)\end{pmatrix}$} ,
\end{aligned}
\end{equation}
where we wrote
\begin{equation}
 f(z)= \frac{i c\, z_1^{m_1} z_2^{m_2} }{m_1 \pi } + \frac{1-n_1 n_2}{8\pi i} \Big(  a^2 z_2^2  +2 ab\, z_1 z_2 \frac{1+n_2^2}{1-n_2} + b^2 n_2 z_1^2\Big)\, .
\end{equation}
The information about the parameters in these periods has been summarized in table \ref{table:II1parameters}. In the construction it was assumed that the coefficients $a,b$ or $a,c$ are non-vanishing, which ensures the presence of essential instanton terms needed in order to span the entire space $H^3(Y_3,\mathbb{C})$. Furthermore $a,b$ have been rotated to real values using the residual axion shift symmetry as discussed below \eqref{eq:shift}. The parameters $n_1,n_2$ control the form of the log-monodromy matrices \eqref{eq:II1N} under $z_i \to e^{2\pi i} z_i$, and hence determine which member of the $\text{II}_1$ boundary class we are looking at. Also note that there is an interplay between the parameter $n_2$ and the orders of the instanton expansion $m_1,m_2$ similar to the class $\mathrm{I}_2$ boundaries, owing to the horizontality property of the periods \eqref{eq:horizontality}.

\begin{table}[h!]
\centering
\renewcommand*{\arraystretch}{2.0}
\scalebox{0.85}{
\begin{tabular}{| l || c | c | c | c | c |}
\hline parameters  & $ \langle \mathrm{II}_0 | \mathrm{II}_1 | \mathrm{I}_1 \rangle $ & $ \langle \mathrm{II}_1 | \mathrm{II}_1 | \mathrm{I}_1 \rangle $ & $ \langle \mathrm{II}_0 | \mathrm{II}_1 | \mathrm{II}_1 \rangle $&$ \langle \mathrm{II}_1 | \mathrm{II}_1 | \mathrm{II}_1 \rangle $ \\ \hline \hline 
log-mon.  & $n_1=n_2=0$ &$n_1 \in \mathbb{Q}_{>0}$, $n_2=0$ & $n_1=0,n_2 \in \mathbb{Q}_{>0}$ & $n_1,n_2 \in \mathbb{Q}_{>0}$, $n_1 n_2 \neq 1$    \\ \hline
inst. orders & \multicolumn{2}{c|}{$m_1 = 1, m_2=0$  } & \multicolumn{2}{c|}{$m_2 = n_2 m_1$  } \\ \hline
inst. coeff.  & \multicolumn{4}{c|}{$a,b \in \mathbb{R}, c \in \mathbb{C} : \ \  a,b \neq 0 \parallel  a,c \neq 0 $}  \\ \hline
\end{tabular}}
\caption{\label{table:II1parameters} Summary for the properties of the parameters in the periods \eqref{eq:II1periods} for each of the possible boundaries of class $\mathrm{II}_1$. }
\end{table}


Using these periods we compute the K\"ahler potential \eqref{eq:kahlercsperiods} for these boundaries
\begin{align}\label{eq:II1kp}
e^{-K}  &=  4(y_1+n_2 y_2)  -2 a^2 e^{-4\pi y_2} \Big(  n_1 y_1+y_2 + \tfrac{1-n_1 n_2}{2\pi} \Big)  \nn \\
 &\ \ \ -2 n_2 b^2 e^{-4\pi y_1} \Big( n_2(n_1y_1+y_2) -\tfrac{1-n_1 n_2}{2\pi} \Big) \\
 & \ \ \  +4 |c|^2 e^{-4\pi y_1} \Big( y_1+n_2 y_2 + \tfrac{1}{m_1 \pi} \Big)   \nn \\
&\ \ \ -4 ab e^{-2\pi y_1-2\pi y_2}\Big(  n_2 ( n_1 y_1+y_2) - \tfrac{(1-n_2)(1-n_1 n_2)}{4\pi} \Big) \cos(2\pi(x_1-x_2)) \, , \nn
\end{align}
where we used the coordinates $2 \pi t_i=2 \pi i (x_i + i y_i)= \log [z_i]$ for convenience, and dropped subleading corrections in the exponential expansion of order $e^{-6\pi y}$. For $n_2=0$ the coordinate dependence on $y_2$ enters only through exponentially suppressed terms as one would expect from the presence of an $\text{I}_1$ boundary associated with this coordinate. A noteworthy feature is also that the K\"ahler metric derived from the above potential does not require all instanton terms to become non-degenerate. The instanton term involving $a$ cures this degeneracy, while the one involving $b$ only does so for $n_2 \neq 0$. The instanton term involving $c$ does not suffice to fix the K\"ahler metric. This can be understood more precisely by looking at the derivatives $\partial_1 \Pi$ and $\partial_2 \Pi$ out of which the K\"ahler metric is constructed. For $a \neq 0$ or $b, n_2 \neq 0 $ these derivatives span a two-dimensional space, while if only $c \neq 0$ they are linearly dependent.  Finally, the signs of the first four leading terms in the K\"ahler potential are fixed by the polarization conditions \eqref{eq:pol}, similar to the examples we encountered previously. The remaining term breaks the continuous shift symmetry for the linear combination of axions $x_1-x_2$ at the level of the K\"ahler potential.

Next we consider the flux superpotential \eqref{eq:superpotential} and the corresponding scalar potential \eqref{eq:potential}. We find that the flux superpotential is given by
\begin{equation}\label{eq:superpotentialII1}
\begin{aligned}
W &=  ( g_1+ig_2) (t_1+n_{2}  t_2) - (g_4+i g_5)\\
& \ \ \ - a \, e^{2\pi i t_2}  \left(g_3 \Big( n_1 t_1 +t_2 - \tfrac{1-n_{1} n_{2}}{2\pi i} \Big)+    g_6 \right) \\
  & \ \ \ -  b  \, e^{2\pi i t_1} \left(g_3 \Big( n_2(n_1 t_1 +t_2) + \tfrac{1-n_{1} n_{2}}{2\pi i} \Big)  +
   n_2 g_6 \right) \\
  & \ \ \ +  c \, e^{2\pi i m_1 t_1} e^{2\pi i m_2 t_2} \left( \big(t_1+n_2 t_2 +\tfrac{i}{m_1 \pi}\big) (g_1-i g_2)-(g_4-ig_5) \right)  \,  ,
\end{aligned}
\end{equation}
where we expanded up to first order in the instanton expansion, dropping $e^{-4\pi y}$ and higher. In turn, we find as leading polynomial scalar potential
\begin{equation}\label{eq:potentialII1}
4 \cV^2 s  V_{\rm lead}= \bar{\rho} \scalebox{0.8}{ $\begin{pmatrix}  y_1+ n_2y_2 & 0 & 0 & 0 & 0 & 0 \\
0 &  y_1+ n_2y_2 & 0 & 0 & 0 & 0 \\
0 & 0 & n_1 y_1+ y_2 & 0 & 0 & 0 \\
0 & 0 & 0 & \frac{1}{ y_1+n_2y_2} & 0 & 0 \\
0 & 0 & 0 & 0 &\frac{1}{ y_1+n_2y_2} & 0\\
0 & 0 & 0 & 0 & 0 & \frac{1}{ n_1y_1+ y_2} \\
\end{pmatrix}$} \rho \, ,
\end{equation}
where we absorbed the axion-dependence as $\rho=e^{-x^i N_i}G_3$ with $N_i$ given in \eqref{eq:II1N}. We again dropped exponentially suppressed corrections in $y_1,y_2$ and left out the $\langle G_3 , \bar{G}_3 \rangle$ term. Note that only the linear combinations of fluxes $g_1+ig_2$ and $g_4+i g_5$ appear at polynomial order in $t_i = \log[z_i]/2\pi i$ in the superpotential \eqref{eq:superpotentialII1}. In particular the fluxes $g_3,g_6$ only appear through exponential corrections in the superpotential, while they appear at polynomial order in the scalar potential. In the computation of \eqref{eq:potentialII1} these exponential factors cancel out against factors in the K\"ahler metric, resulting in polynomial terms for the scalar potential. In other words, we find that class $\mathrm{II}_1$ boundaries require us to include essential exponential corrections in the superpotential, even though these are at infinite distance. To be more precise, the terms at first order in $e^{-2\pi y_i}$ in the superpotential contribute to the leading polynomial scalar potential, while the other instanton terms lead to exponential corrections.

\subsubsection{Coni-LCS class boundaries}\label{sec:coniLCS}
Finally we come to the class of coni-LCS boundaries. While these boundaries are characterized by a $\text{IV}_2$ singularity type similar to large complex structure points, one has to include essential instanton terms in the periods. Recently the periods near such boundaries have been considered in the context of small flux superpotentials in \cite{Demirtas:2020ffz,Blumenhagen:2020ire} (see also \cite{Demirtas:2019sip} for the original study at large complex structure). The period vector that we construct in section \ref{ssec:IV2construction} is given by
\begin{align} \label{eq:coniLCSperiods}
\Pi= \big(
&1  ,\  a z_1, \ \frac{ \log[z_2]}{2 \pi i}, \ -\frac{i \log[z_2]^3}{48 \pi^3}-\frac{ i a^2 n z_1^2 \log[z_2]}{4\pi }+   \frac{a^2}{4  \pi i} z_1^2+i \delta_2 +i \delta_1 a z_1, \nn \\ 
 &- a z_1\frac{ \log[z_1] + n \log[z_2]}{2 \pi i} +i\delta_1, \- \frac{\log[z_2]^2}{8 \pi^2}  -\frac{1}{2} a^2 n z_1^2 \big)\, .
 \end{align}
Note that the modulus $t_1=\log[z_1]/2\pi i$ only appears in terms with exponential factors $ e^{2\pi i t_1}$, while $t_2 = \log[z_2]/2\pi i$ appears polynomially. The former we typically attribute to conifold points in the moduli space, while the latter is familiar from large complex structure points, hence the term coni-LCS boundary.

The information about the different parameters is summarized in table \ref{table:IV2parameters}. It is assumed that the coefficient $a$ is non-vanishing as this is required in order to span the entire three-form cohomology $H^3(Y_3,\mathbb{C})$ from derivatives of the period vector. Furthermore, we have used the residual axion shift symmetry to set $a$ to a real value as discussed below \eqref{eq:shift}. The parameter $n$ controls which member of the coni-LCS class we are considering. 

\begin{table}[h!]
\centering
\renewcommand*{\arraystretch}{2.0}
\begin{tabular}{| l || c | c | c | }
\hline parameters  & $ \langle \mathrm{I}_1 | \mathrm{IV}_2 | \mathrm{IV}_1 \rangle $ & $ \langle \mathrm{I}_1 | \mathrm{IV}_2 | \mathrm{IV}_2 \rangle $\\ \hline \hline 
log-monodromies $n_1,n_2$  & $n=0$ & $n \in \mathbb{Q}_{>0}$  \\ \hline
instanton coefficient $a$ & \multicolumn{2}{c|}{$a\in \mathbb{R}- \{0 \}$} \\ \hline
phase operator $\delta$ & \multicolumn{2}{c|}{$\delta_1,\delta_2 \in \mathbb{R}$} \\ \hline
\end{tabular}
\caption{\label{table:IV2parameters} Summary for the properties of the parameters in the periods \eqref{eq:coniLCSperiods} for each of the possible boundaries of the coni-LCS class. }
\end{table}

Using these periods we calculate the K\"ahler potential \eqref{eq:kahlercsperiods} for coni-LCS class boundaries
\begin{equation}
\begin{aligned}
e^{-K} =\ & \frac{4y_2^3}{3} +2 \delta_2 +4a\delta_1 e^{-2\pi y_1} \cos[2\pi x_1] \\
&-2 a^2 e^{-4\pi y_1} \big( y_1+ny_2-(n y_2-1/4\pi ) \cos[4\pi x_1]\big)\, ,
\end{aligned}
\end{equation}
where we used the coordinates $2 \pi t_i=2 \pi i (x_i + i y_i)= \log [z_i]$ for convenience. The signs of the terms without parameters $\delta_i$ are fixed by the polarization conditions \eqref{eq:pol} similar to the previous examples. Note in particular that, as expected from the presence of a finite distance $\mathrm{I}_1$ divisor, the associated field $y_1$ only appears in exponentially suppressed terms. Furthermore, we can understand the role of the phase operator parameters $\delta_1,\delta_2$ by inspecting this K\"ahler potential. We find that $\delta_2$ gives rise to a constant term in the K\"ahler potential, similar to the Euler characteristic term at large complex structure. Interestingly, the parameter $\delta_1$ produces an axion-dependent term at order $e^{-2\pi y_1}$, which is leading compared to the usual term at order $e^{-4\pi y_2}$. See appendix \ref{app:coniLCS} for a more careful comparison with the standard large complex structure expressions.

We next compute the flux superpotential \eqref{eq:superpotential} by inserting the above periods as
\begin{align}
W &=  - g_4+ig_2 \delta_1 +ig_1 \delta_2 - g_6 t_2+ \frac{1}{2} g_3 t_2^2 -\frac{1}{6} g_1 t_2^3\\
& \ \ \ +ae^{2\pi i t_1} \big( -g_2 (t_1+n t_2) -g_5+ig_1 \delta_1 \big) -\frac{a^2 e^{4\pi i t_1}}{2} \big(g_3 n+g_1(n t_2+\tfrac{1}{2\pi i}) \big) \nn \, .
\end{align}
 We compute the corresponding scalar potential \eqref{eq:potential} to be
\begin{equation}\label{eq:potentialConiLCS}
4 \cV^2 s \,  V_{\rm lead}=  \bar{G}_3 e^{-x^iN_i^T}\scalebox{0.85}{$\begin{pmatrix} \frac{y_2^3}{6} & 0 & 0 & 0 & 0 & 0 \\
0 & \frac{y_2}{2} & 0 & 0 & 0 & 0 \\
0 & 0 & y_1+n y_2 & 0 & 0 & 0 \\
0 & 0 & 0 & \frac{6}{y_2^3} & 0 & 0 \\
0 & 0 & 0 & 0 & \frac{2}{y_2} & 0\\
0 & 0 & 0 & 0 & 0 & \frac{1}{y_1+n_2 y_2} \\
\end{pmatrix} $}e^{-x^iN_i} G_3 \, ,
\end{equation}
where the log-monodromy matrices $N_i$ are given in \eqref{eq:IV2N}. We again dropped exponentially suppressed corrections in $y_1$ and left out the $\langle G_3 , \bar{G}_3 \rangle$ term for convenience. Note that the fluxes $g_1,g_2,g_5$ as well as the linear combination $g_4-i\delta_1 g_2-i\delta_2 g_1$ appear at polynomial order in the superpotential, while the other fluxes are exponentially suppressed. In computing \eqref{eq:potentialConiLCS} the terms at order $e^{-2\pi y_1}$ in the superpotential are crucial to obtain the polynomial terms for the fluxes $g_3,g_6$ in the scalar potential, while the other instanton terms lead to exponential corrections. This is similar to the corrections for the one-modulus $\mathrm{I}_1$ boundary, and can be traced back to the fact that the divisor $y^1=\infty$ is at finite distance.

\section{Construction of one- and two-moduli periods}\label{sec:constructionperiods}
Here we construct the asymptotic periods for all possible boundaries in one- and two-dimensional complex structure moduli spaces. We begin by writing down the nilpotent orbit data that characterizes these boundaries. For the one-modulus case this data has been constructed in \cite{GreenGriffithsKerr}, and for the two-moduli case we refer to our analysis in appendix \ref{app:boundarydata}. From the given nilpotent orbits we then construct the most general compatible periods following the procedure laid out in section \ref{sec:construction}.

\subsection{Construction of one-modulus periods} \label{app:one-modulus}
In this section we explicitly construct general expressions for the periods near boundaries in one-dimensional moduli spaces. Recall from section \ref{sec:onemodels} that there are three possible singularity types for boundaries in complex structure moduli space when $h^{2,1}=1$: $\mathrm{I}_1$, $\mathrm{II}_0$ and $\mathrm{IV}_1$. Conveniently, we do not need to construct the boundary data from scratch as this was already done in \cite{GreenGriffithsKerr}, so we are simply going to record the results expressed in a different basis more suitable to us. With this information at hand, we write down the instanton map $\Gamma(z)$ as explained in section \ref{sec:construction}, and use it to construct the periods including the necessary instanton terms. These results are nothing inherently new in the sense that the periods for the one-modulus cases can also be systematically constructed by using the so-called Meijer G-functions, see e.g.~\cite{Garcia-Etxebarria:2014wla,Joshi:2019nzi}. However, we find the exercise of re-deriving these periods useful as it serves to illustrate the method we are also going to employ to tackle the two-moduli cases, where there is no systematic construction for periods away from the large complex structure lamppost. Furthermore, it allows us to fix our notation for these expressions as they are also used for computations in section \ref{sec:onemodels}.

\subsubsection{Type $\text{I}_1$ boundaries}\label{ssec:constrconifold}
Let us begin by studying $\mathrm{I}_1$ boundaries. The Hodge-Deligne diamond representing these boundaries has been depicted in figure \ref{fig:I1}. The nilpotent orbit data consists of the sl(2)-split Deligne splitting $\tilde{I}^{p,q}$, the log-monodromy matrix $N$ and a phase operator $\delta$.  The vector spaces $\tilde{I}^{p,q}$ of the Deligne splitting are spanned by
\begin{equation}\label{eq:DsplitI1}
\begin{aligned}
\tilde{I}^{3,0}&: \ \big( 1,\ 0,\ i,\ 0 \big),\qquad   &\tilde{I}^{2,2}&: \ \big( 0,\ 1,\ 0,\ 0 \big) ,  \\
\tilde{I}^{1,1}&: \ \big( 0, \ 0, \ 0, \ 1 \big) ,  \qquad   &\tilde{I}^{0,3}&: \ \big( 1, \ 0,\ -i,\ 0 \big), \\
\end{aligned}
\end{equation}
while the log-monodromy matrix and phase operator can be written as
\begin{align}\label{eq:I1N}
N= \scalebox{0.75}{$\begin{pmatrix}
0 & 0 & 0 & 0 \\
0 & 0 & 0 & 0 \\
0 & 0 & 0 & 0 \\
0 & -1 & 0 & 0
\end{pmatrix}$}\, , \qquad \delta = \delta_1 N\, .
\end{align}
Note that the phase operator is proportional to the log-monodromy matrix $N$. According to \eqref{eq:shift} we can therefore tune the parameter $\delta_1$ to simplify the periods later.

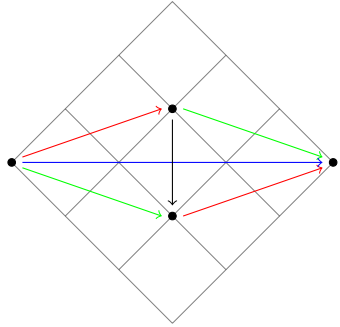
\begin{figure}[h!]
\centering
\begin{tikzpicture}[scale=1,cm={cos(45),sin(45),-sin(45),cos(45),(15,0)}]
  \draw[step = 1, gray, ultra thin] (0, 0) grid (3, 3);

  \draw[fill] (0, 3) circle[radius=0.05];
  \draw[fill] (1, 1) circle[radius=0.05];
  \draw[fill] (2, 2) circle[radius=0.05];
  \draw[fill] (3, 0) circle[radius=0.05];

\draw[->, red] (0.15,2.95) -- (1.9,2.1);
\draw[->, red] (1.1,0.9) -- (2.85,0.05);

\draw[->, green] (0.05,2.85) -- (0.9,1.1);
\draw[->, green] (2.1,1.9) -- (2.95,0.15);

\draw[->, blue] (0.1,2.9) -- (2.9,0.1);

\draw[->, black] (1.9,1.9) -- (1.1,1.1);
\end{tikzpicture}
\caption{\label{fig:I1} The Hodge-Deligne diamond that classifies $\mathrm{I}_1$ boundaries. We included colored arrows to denote the different components $\Gamma_{-1}, \Gamma_{-2}$ and $\Gamma_{-3}$ of the instanton map $\Gamma$ by red, green and blue respectively. We also used a black arrow to denote the action of the log-monodromy matrix $N$.}
\end{figure}

Now let us follow the procedure of section \ref{sec:construction} to construct the most general periods compatible with this boundary data. First we construct the instanton map $\Gamma(z)$. Let us write down the most general Lie algebra-valued map in $\Lambda_{-}$ with holomorphic coefficients, which reads
\begin{align}
\Gamma(z)= \tfrac{1}{2}\scalebox{0.8}{$\begin{pmatrix}
 c(z) & i b(z) & -i c(z) & -i a(z) \\
 a(z) & 0 & -i a(z) & 0 \\
 -i c(z) & b(z) & -c(z) & -a(z) \\
 b(z) & 0 & -i b(z) & 0 \\
\end{pmatrix}$}\, ,
\end{align}
where $a(z),b(z)$ and $c(z)$ make up the charge $\Gamma_{-1}, \Gamma_{-2}$ and $\Gamma_{-3}$ components respectively, with $a(0)=b(0)=c(0)=0$. Note that we set the piece proportional to the log-monodromy matrix $N$ to zero by using \eqref{eq:shift}. The periods can then be written in terms of these coefficients as
\begin{equation}
\Pi = \big(1+c(z), \  a(z) , \ i-ic(z), \ \frac{i a(z)}{2\pi} \log[z]+b[z]-i\delta_1\big)\, .
\end{equation}
The holomorphic functions $a(z)$, $b(z)$ and $c(z)$ that appear in these periods must satisfy the recursion relations \eqref{eq:recursiongamma}. We can write them out as differential constraints on the coefficients as
\begin{equation}\label{eq:I1diffs}
\begin{aligned}
z b(z)'   = \frac{1}{2\pi i} a(z)\, , \qquad c(z)' = \frac{i}{2} a(z)' b(z)\, .
\end{aligned}
\end{equation}
Since these coefficients are required to vanish at $z=0$, one finds that $b(z),c(z)$ are determined completely by $a(z)$, as can be verified by performing a holomorphic expansion in $z$. In order to obtain a more concrete model for the periods, let us include only the leading order term for $a(z)$ in this instanton expansion. We write as ansatz
\begin{equation}
a(z) = a z\, .
\end{equation}
Plugging this ansatz into the differential equations \eqref{eq:I1diffs} we can solve for the other two functions
\begin{equation}
b(z) = \frac{a}{2\pi i } z\, , \qquad c(z) = \frac{a^2}{8\pi} z^2\, .
\end{equation}
We then obtain the following expression for the asymptotic periods near $\mathrm{I}_1$ boundaries
\begin{align}
\Pi= \big(1 + \frac{ a^2}{8 \pi} z^2 , \ a z , \ i - \frac{i  a^2}{8 \pi} z^2 , \ \frac{ia}{2\pi} z \log[z] \big) \, ,
\end{align}
where we set $\delta_1=-1/2\pi$.

\subsubsection{Type $\text{II}_0$ boundaries}\label{ssec:constrII0}
Next we consider $\mathrm{II}_0$ boundaries. The Hodge-Deligne diamond representing these boundaries has been depicted in figure \ref{fig:II0}. Again let us begin by writing down the nilpotent orbit data. The spaces $\tilde{I}^{p,q}$ of the sl(2)-split Deligne splitting are spanned by
\begin{equation}
\begin{aligned}
\tilde{I}^{3,1}&: \ \big(1, \ i, \  0, \ 0 \big), \qquad &\tilde{I}^{2,0} &: \ \big(
0, \  0, \ 1, \ i
\big) ,\\
\tilde{I}^{1,3}&: \ \big(
1, \  -i, \   0, \  0 
\big), \qquad &\tilde{I}^{0,2} &: \ \big(
0, \   0, \ 1, \  -i
\big),
\end{aligned}
\end{equation} 
while the log-monodromy matrix can be written as
\begin{align}\label{eq:II0N}
N= \scalebox{0.75}{$\begin{pmatrix}
0 & 0 & 0 & 0 \\
0 & 0 & 0 & 0 \\
1 & 0 & 0 & 0 \\
0 & 1 & 0 & 0
\end{pmatrix}$},
\end{align}
and the phase operator $\delta = \delta_1 N$ has been set to zero by a coordinate shift \eqref{eq:shift}.

\begin{figure}[h!]
\centering
\begin{tikzpicture}[scale=1,cm={cos(45),sin(45),-sin(45),cos(45),(15,0)}]
  \draw[step = 1, gray, ultra thin] (0, 0) grid (3, 3);

  \draw[fill] (0, 2) circle[radius=0.05];
  \draw[fill] (1, 3) circle[radius=0.05];
  \draw[fill] (2, 0) circle[radius=0.05];
  \draw[fill] (3, 1) circle[radius=0.05];

\draw[->, red] (0.1,2) -- (2.9,1);

\draw[->, green] (1.1,2.9) -- (2.9,1.1);
\draw[->, green] (0.1,1.9) -- (1.9,0.1);

\draw[->, blue] (1,2.9) -- (2,0.1);

\draw[->] (0.9,2.9) -- (0.1,2.1);
\draw[->] (2.9,0.9) -- (2.1,0.1);
\end{tikzpicture}
\caption{\label{fig:II0} The Hodge-Deligne diamond that classifies $\mathrm{II}_0$ boundaries. We included colored arrows to denote the different components $\Gamma_{-1}, \Gamma_{-2}$ and $\Gamma_{-3}$ of the instanton map $\Gamma$ by red, green and blue respectively, and black arrows for the log-monodromy matrix $N$.}
\end{figure}
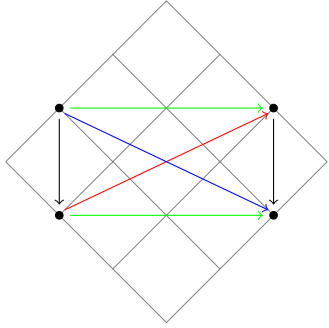

Using the procedure of section \ref{sec:construction} we now want to write down the most general periods compatible with this boundary data. We again begin by constructing the instanton map $\Gamma(z)$. In this case the most general Lie algebra-valued map in $\Lambda_-$ with holomorphic coefficients reads
\begin{align}
\Gamma=\tfrac{1}{2} \scalebox{0.8}{$\begin{pmatrix}
 b(z) & -i b(z) & i a(z) & a(z) \\
 -i b(z) & -b(z) & a(z) & -i a(z) \\
 c(z) & -i c(z) & -b(z) & i b(z) \\
 -i c(z) & -c(z) & i b(z) & b(z) \\
\end{pmatrix}$}\, ,
\end{align}
where $a(z),b(z)$ and $c(z)$ make up the charge $\Gamma_{-1}, \Gamma_{-2}$ and $\Gamma_{-3}$ components respectively, with $a(0)=b(0)=c(0)=0$. Note that we have again used coordinate transformations \eqref{eq:shift} to set the piece proportional to $N$ to zero. We can then write the periods in terms of these coefficients as
\begin{equation}\label{eq:II0genperiods}
\Pi = \big(1+b(z) , \ i-ib(z) , \ \big(1+b(z) \big) \frac{\log[z]}{2\pi i} +c(z) , \ i \big(1-b(z) \big) \frac{\log[z]}{2\pi i} +i c(z)\big)\, .
\end{equation}
The functions $a(z)$, $b(z)$ and $c(z)$ that appear in the periods must satisfy the recursion relations \eqref{eq:recursiongamma}. These can be written out as differential constraints
\begin{equation}\label{eq:II0diffs}
\begin{aligned}
z b'(z) = \frac{1}{2\pi} a(z)\, , \qquad z c'(z) = \frac{i}{\pi} b(z)\, .
\end{aligned}
\end{equation}
Note that while $a(z)$ did not appear in the periods directly, it does determine $b(z)$ and $c(z)$ uniquely similar to $\mathrm{I}_1$ boundaries. In order to obtain more concrete expressions for the periods near the boundary, let us include only the leading order term $a(z)$ in the holomorphic expansion. We write as ansatz
\begin{equation}
a(z) = 2\pi a z\, ,
\end{equation}
where we included a factor of $2\pi$ for later convenience. By using \eqref{eq:II0diffs} the other two functions are then found to be
\begin{equation}
b(z) = a z\, , \qquad c(z) = \frac{i}{\pi} a z\, .
\end{equation}
By plugging these expressions into \eqref{eq:II0genperiods} we find as asymptotic periods for $\mathrm{II}_0$ boundaries
\begin{align}
\Pi= \big(1+a z, \ i-iaz, \ \frac{\log[z]}{2 \pi i}+ \frac{az}{2\pi i} (\log[z]-2), \  \frac{\log[z]}{2 \pi } - \frac{az}{2\pi } (\log[z]-2) 
\big)\, .
\end{align}

\subsection{Construction of two-moduli periods}\label{sec:two-moduli}
In this section we derive general expressions for the periods near all possible two-moduli boundaries. Recall from section \ref{sec:twomodels} that there are three classes of boundaries we focus on: $\mathrm{I}_2, \mathrm{II}_1$ and coni-LCS class. The boundary data characterizing these classes has been constructed in appendix \ref{app:boundarydata}. We use the techniques discussed in section \ref{sec:construction} to construct the most general periods compatible with these sets of data. Before we begin, let us already note that we now find that \eqref{eq:dXdX} imposes non-trivial constraints on the coefficients of the component $\Gamma_{-1}(z)$ of the instanton map. In contrast to the one-modulus case, this means that one cannot consider any choice of holomorphic functions for these coefficients, but there will be some differential equations that have to be satisfied. For this reason we choose to make a simplified leading order ansatz for $\Gamma_{-1}(z)$, which allows us to illustrate the qualitative features of the models more easily.

\subsubsection{Class $\text{I}_{2}$ boundaries}\label{ssec:I2construction}
Let us begin by considering the class of $\mathrm{I}_2$ boundaries. The Hodge-Deligne diamond representing these boundaries has been depicted in figure \ref{fig:I2}. The nilpotent orbit data has been constructed in appendix \ref{app:I2data}, and is again given by the sl(2)-split Deligne splitting $\tilde{I}_{(2)}^{p,q}$ together with the log-monodromy matrices $N_i$ and the phase operator $\delta$. The Deligne splitting is spanned by
\begin{equation}\label{eq:DsplitI2}
\begin{aligned}
\tilde{I}_{(2)}^{3,0}&: \quad \big( 1, \  0,\ 0,\ i,\ 0,\ 0 \big) , \qquad \tilde{I}_{(2)}^{0,3}: \quad \big( 1, \  0,\ 0,\ -i,\ 0,\ 0\big) ,  \\
\tilde{I}_{(2)}^{2,2}&: \quad \big( 0, \  1,\ 0,\ 0,\ 0,\ 0 \big) , \  \big( 0, \  0,\ 1,\ 0,\ 0,\ 0 \big) ,  \\
\tilde{I}_{(2)}^{1,1}&: \quad \big( 0, \  0,\ 0,\ 0,\ 1,\ 0 \big) ,   \  \big( 0, \  0,\ 0,\ 0,\ 0,\ 1 \big) ,  \\
\end{aligned}
\end{equation}
while the log-monodromy matrices and phase operator are written as
\begin{footnotesize}
\begin{equation}\label{eq:I2N}
\begin{aligned}
N_1 &= -\scalebox{0.74}{$\begin{pmatrix}
 0 & 0 & 0 & 0 & 0 & 0 \\
 0 & 0 & 0 & 0 & 0 & 0 \\
 0 & 0 & 0 & 0 & 0 & 0 \\
 0 & 0 & 0 & 0 & 0 & 0 \\
 0 & 1 & 0 & 0 & 0 & 0 \\
 0 & 0 & n_1 & 0 & 0 & 0 \\
\end{pmatrix}$}, \ \ N_2 =- \scalebox{0.74}{$\begin{pmatrix}
 0 & 0 & 0 & 0 & 0 & 0 \\
 0 & 0 & 0 & 0 & 0 & 0 \\
 0 & 0 & 0 & 0 & 0 & 0 \\
 0 & 0 & 0 & 0 & 0 & 0 \\
 0 & n_2 & 0 & 0 & 0 & 0 \\
 0 & 0 & 1 & 0 & 0 & 0 
\end{pmatrix}$}, \ \  \delta = \scalebox{0.74}{$\begin{pmatrix}
 0 & 0 & 0 & 0 & 0 & 0 \\
 0 & 0 & 0 & 0 & 0 & 0 \\
 0 & 0 & 0 & 0 & 0 & 0 \\
 0 & 0 & 0 & 0 & 0 & 0 \\
 0 & 0 & \delta_1 & 0 & 0 & 0 \\
 0 & \delta_1 & 0 & 0 & 0 & 0 \\
\end{pmatrix}$}\, .
\end{aligned}
\end{equation}
\end{footnotesize}

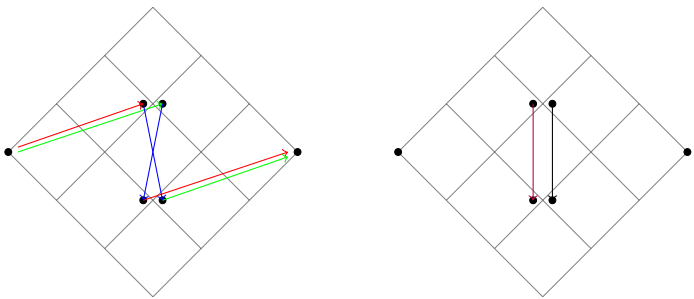
\begin{figure}[h!]
\centering
\begin{minipage}{0.4\textwidth}
\begin{tikzpicture}[scale=0.9,cm={cos(45),sin(45),-sin(45),cos(45),(15,0)}]
  \draw[step = 1, gray, ultra thin] (0, 0) grid (3, 3);

  \draw[fill] (0, 3) circle[radius=0.05];
  \draw[fill] (0.9, 1.1) circle[radius=0.05];
  \draw[fill] (1.9, 2.1) circle[radius=0.05];
  \draw[fill] (1.1, 0.9) circle[radius=0.05];
  \draw[fill] (2.1, 1.9) circle[radius=0.05];
  \draw[fill] (3, 0) circle[radius=0.05];

\draw[->, red] (0.15,2.95) -- (1.9,2.1);
\draw[->, red] (0.9,1.1) -- (2.9,0.1);

\draw[->, green] (0.1,2.9)  -- (2.1,1.9);
\draw[->, green] (1.1,0.9) -- (2.85,0.05);

\draw[->, blue] (1.9,2.1) -- (1.1,0.9);
\draw[->, blue] (2.1,1.9) -- (0.9,1.1);
\end{tikzpicture}
\end{minipage}
\begin{minipage}{0.4\textwidth}
\begin{tikzpicture}[scale=0.9,cm={cos(45),sin(45),-sin(45),cos(45),(15,0)}]
  \draw[step = 1, gray, ultra thin] (0, 0) grid (3, 3);

  \draw[fill] (0, 3) circle[radius=0.05];
  \draw[fill] (0.9, 1.1) circle[radius=0.05];
  \draw[fill] (1.9, 2.1) circle[radius=0.05];
  \draw[fill] (1.1, 0.9) circle[radius=0.05];
  \draw[fill] (2.1, 1.9) circle[radius=0.05];
  \draw[fill] (3, 0) circle[radius=0.05];

\draw[->, purple] (1.9,2.1) --(0.9,1.1);
\draw[->, black] (2.1,1.9) --(1.1,0.9);
\end{tikzpicture}
\end{minipage}
\centering
\caption{\label{fig:I2} The Hodge-Deligne diamond that classifies $\mathrm{I}_2$ boundaries. Note that we split up $\tilde{I}_{(2)}^{1,1},\tilde{I}_{(2)}^{2,2}$ according to the basis vectors in \eqref{eq:DsplitI2}, where left (right) vertices correspond to the first (last) two vectors. On the left we included colored arrows to denote the different components of $\Gamma_{-1}(z)$  of the instanton map $\Gamma(z)$, where we labeled $a(z)$, $b(z)$ and $c(z)$ by red, green and blue respectively. On the right the purple and black arrow denote the action of $N_1$ and $N_2$ respectively (setting $n_1,n_2=0$ for simplicity).}
\end{figure}
Now we want to write down the most general periods compatible with this boundary data. As explained in section \ref{sec:construction} we begin by considering the most general instanton map $\Gamma(z)$. For the above data the most general Lie algebra-valued map in $\Lambda_-$ with holomorphic coefficients is given by
\begin{equation}\label{eq:I2Gamma}
\Gamma(z_1,z_2) =\tfrac{1}{2} \scalebox{0.7}{$ \begin{pmatrix}
 f(z) & -i d(z) & -i e(z) & -i f(z) & -i
   a(z) & -i b(z) \\
 a(z) & 0 & 0 & -i a(z) & 0 & 0 \\
 b(z) & 0 & 0 & -i b(z) & 0 & 0 \\
 -i f(z) & -d(z) & -e(z) & -f(z) &
   -a(z) &  -b(z) \\
 -d(z) & 0 & -c(z) & i d(z) & 0 & 0 \\
 -e(z) & -c(z) & 0 & i e(z) & 0 & 0 \\
\end{pmatrix}$},
\end{equation}
where $a,b,c$ make up the charge component $\Gamma_{-1}$, $d,e$ correspond to $\Gamma_{-2}$ and $f$ to $\Gamma_{-3}$. Note that we set the coefficients proportional to the log-monodromy matrices $N_1,N_2$ to zero by using coordinate redefinitions \eqref{eq:shift}, where it is important that $n_1 n_2 \neq 1$. For illustration we depicted the action of the $\Gamma_{-1}$ coefficients on the Deligne splitting in figure \ref{fig:I2}. We can then use \eqref{eq:PiGamma} to write the periods in terms of these coefficients as
\begin{align}\label{eq:I2periodsfunctions}
\Pi = \big(&1+f(z)+ \tfrac{i}{12}   a(z) b(z) c(z) , \ a(z) , \ b(z) , \ i-i f(z)  +\tfrac{1}{12}   a(z) b(z) c(z) , \nn \\
& - a(z) \tfrac{ n_1 \log[z_1] + \log
   [z_2]}{2 \pi i}+i\delta_1
   b(z)-d(z)-\tfrac{1}{4} b(z) c(z), \nn \\
   & - b(z) \tfrac{  \log[z_1] + n_2 \log
  [z_2]}{2 \pi i}+i \delta_1
   a(z)-e(z)- \tfrac{1}{4} a(z) c(z)\big)\, .
   \end{align}
The holomorphic functions appearing in these periods are constrained by several sets of differential equations. Recall from section \ref{sec:construction} that we must first impose \eqref{eq:dXdX} on the coefficients $a,b,c$ of $\Gamma_{-1}$. Subsequently the coefficients $d,e,f$ of $\Gamma_{-2},\Gamma_{-3}$ are fixed uniquely by $a,b,c$ through \eqref{eq:recursiongamma}. Let us write out these equations explicitly in terms of the holomorphic coefficients. We find that \eqref{eq:dXdX} imposes
\begin{equation}\label{eq:dXdXI2}
\begin{aligned}
 z_{1} a^{(1,0)}- n_1 z_{2} a^{(0,1)}&=i \pi  z_{1} z_{2} \big(b^{(0,1)}c^{(1,0)}-b^{(1,0)} c^{(0,1)}\big)\, , \\
  z_{2} b^{(0,1)}-n_2 z_{1} b^{(1,0)}&=i\pi  z_{1} z_{2} \big(a^{(1,0)}c^{(0,1)}-a^{(0,1)}
   c^{(1,0)}\big)\, .
\end{aligned}
\end{equation}
Inspecting these differential equations carefully, we note that the right-hand side only contains mixed terms in $z_1,z_2$ after expanding the holomorphic functions around $z_1=z_2=0$. Therefore we obtain the following relations on their coefficients
\begin{equation}
a_{k0}=0\, ,\qquad b_{0l} = 0\, , \qquad n_1 a_{0l} =0\, , \qquad  n_2 b_{k0} = 0\, .
\end{equation}
For the remainder of the coefficients we have the relations
\begin{small}
\begin{equation}
\begin{aligned}
(n_1 l - k)a_{kl} &= \pi i \sum_{m,n} (k-m)n \, b_{k-m,l-n} c_{mn} -\pi i \sum_{m,n} (l-n)m \,  b_{k-m,l-n} c_{mn} \, ,\\
(n_2 k - l)b_{kl} &= \pi i \sum_{m,n} (k-m)n \, a_{k-m,l-n} c_{mn} - \pi i \sum_{m,n} (l-n)m \,  a_{k-m,l-n} c_{mn} \, .
\end{aligned}
\end{equation}
\end{small}
Note that coefficients $a_{kl}$ with $k=n_1 l$ and $b_{kl}$ with $l=n_2 k$ do not appear on the left-hand side of this equation, so they are unfixed by these differential constraints.

Next let us write down the differential equations that fix $d,e,f$ uniquely in terms of $a,b,c$. We find that \eqref{eq:recursiongamma} imposes the following set of constraints
\begin{small}
\begin{align}\label{eq:I2recursion}
z_1d^{(1,0)} &=- \frac{n_1}{2\pi i}a+\frac{1}{4} z_1 \big( b^{(1,0)}c-bc^{(1,0)}\big)\, , \quad z_1e^{(1,0)} =- \frac{1}{2\pi i} b+\frac{1}{4} z_1 \big( a^{(1,0)} c-ac^{(1,0)} \big)\, , \nn\\
z_2d^{(0,1)} &=- \frac{1}{2\pi i} a+\frac{1}{4} z_2 \big( b^{(0,1)}c-bc^{(0,1)} \big)\, , \quad z_2e^{(0,1)} =- \frac{n_2}{2\pi i} b+\frac{1}{4} z_2 \big( a^{(0,1)} c-ac^{(0,1)} \big)\, , \nn\\
i f^{(1,0)}&=-\frac{1}{24}  a^{(1,0)} b
   c+\frac{1}{2}   a^{(1,0)}
   d-\frac{1}{24}   a
   b^{(1,0)} c+\frac{1}{12} 
   ab c^{(1,0)}+\frac{1}{2}  
   b^{(1,0)} e\, , \\
if^{(0,1)}   &=-\frac{1}{24}  a^{(0,1)} b
   c+\frac{1}{2}   a^{(0,1)}
   d-\frac{1}{24}  a
   b^{(0,1)} c+\frac{1}{12} 
   a b c^{(0,1)}+\frac{1}{2}  
   b^{(0,1)} e\, . \nn
\end{align}
\end{small}
Now we want to find out what \eqref{eq:rankGamma} imposes on the functions $a,b,c$ that make up the $\Gamma_{-1}$ component of the instanton map. The dimension of the image of this set of matrices must be equal to 5. By inspecting \eqref{eq:I2Gamma} we find that the function $c$ is irrelevant, since they span the same part of the vector space as $N_1$ and $N_2$. On the other hand we can satisfy \eqref{eq:rankGamma} by turning on $a,b$. This can also be seen from figure \ref{fig:I2}, because in order to span $\tilde{I}_{(2)}^{2,2}$ and $\tilde{I}_{(2)}^{0,3}$ we need the components of $a,b$. Let us therefore take the following ansatz for the $\Gamma_{-1}$ coefficients
\begin{equation}
a(z)=a \, z_1^{k_1} z_2^{k_2}\, , \quad b(z)=b \, z_1^{m_1} z_2^{m_2}\, , \quad c(z)=0\, , 
\end{equation}
where $a,b \in \mathbb{C}$. Some comments are in order here. When $(k_1,k_2)\neq (m_1,m_2)$ we can use shifts of the axions $x^i$ to set $a,b \in \mathbb{R}$, while for $(k_1,k_2)= (m_1,m_2)$ this is generally only possible for one of the two. Also note that vanishing of $c(z)$ was not required by \eqref{eq:rankGamma}. Nevertheless $c$ only appear in products with $a(z),b(z)$ in the periods \eqref{eq:I2periodsfunctions}, so it would lead only to subleading corrections. Finally, we only wrote down one leading term for $a(z)$ and $b(z)$. We are however expanding with respect to two coordinates $z_1,z_2$, so there could in principle be two different leading terms for the two expansions. We will see shortly that \eqref{eq:dXdXI2} fixes the orders $k_1,k_2$ and $m_1,m_2$ for the leading terms, justifying the above expansion.  

Let us now solve the differential constraints that the functions of $\Gamma(z)$ must satisfy. We begin with \eqref{eq:dXdXI2}, which reduces to the following two conditions on our ansatz
\begin{equation}\label{eq:orders}
 n_1=k_1/k_2\, , \qquad n_2 = m_2/m_1\, .
\end{equation}
Thus the orders of the leading terms in the instanton expansion are fixed as the pairs of coprime integers $k_1,k_2 \in \mathbb{Z}$ and $m_1,m_2 \in \mathbb{Z}$ such that \eqref{eq:orders} holds. We can then continue and solve \eqref{eq:I2recursion}, which yields
\begin{equation}
\begin{aligned}
d(z) &=\frac{ia}{2\pi k_2} z_1^{k_1} z_2^{k_2} \, , \qquad e(z)=\frac{ib}{2\pi m_1} z_1^{m_1} z_2^{m_2} \, , \\
 f(z) &= \frac{a^2}{8\pi k_2} z_1^{2k_1} z_2^{2k_2} +\frac{b^2}{8\pi m_1} z_1^{2m_1} z_2^{2m_2}  \, .
\end{aligned}
\end{equation}
By inserting these expressions into \eqref{eq:I2periodsfunctions} we obtain the periods
\begin{align}
\Pi = \big(&1 +\tfrac{a^2 z_{1}^{2 k_{1}} z_{2}^{2 k_{2}}}{8 \pi  k_{1}}+\tfrac{b^2 z_{1}^{2 m_{1}}
   z_{2}^{2 m_{2}}}{8 \pi  m_{2}} , \  a z_{1}^{k_{1}} z_{2}^{k_{2}} , \  b z_{1}^{m_{1}} z_{2}^{m_{2}} , \ i- \tfrac{i a^2 z_{1}^{2 k_{1}} z_{2}^{2 k_{2}}}{8 \pi  k_{1}}-\tfrac{i b^2 z_{1}^{2 m_{1}}
   z_{2}^{2 m_{2}}}{8 \pi  m_{2}} \nn \\
&-a  z_{1}^{k_{1}} z_{2}^{k_{2}}  \tfrac{n_1 \log[z_1]+\log [z_{2}]-1/k_1}{2 \pi i}+i b \delta_{1}
   z_{1}^{m_{1}} z_{2}^{m_{2}} , \nn \\
   &- b  z_{1}^{m_{1}} z_{2}^{m_{2}} \tfrac{ \log [z_{1}]+ n_2\log[z_2]-1/m_2}{2 \pi i }+ i a \delta_{1} z_{1}^{k_{1}} z_{2}^{k_{2}}
   \big)\, .
\end{align}

\subsubsection{Class $\text{II}_1$ boundaries}
\label{ssec:II1construction}
Next we study class $\mathrm{II}_1$ boundaries. The Hodge-Deligne diamond representing such boundaries has been depicted in figure \ref{fig:II1}. The relevant boundary data has been constructed in appendix \ref{app:II1data}, and is again given by the sl(2)-split Deligne splitting $\tilde{I}_{(2)}^{p,q}$ together with the log-monodromy matrices $N_i$ and the phase operator. The Deligne splitting is spanned by 
\begin{equation}\label{eq:DsplitII1}
\begin{aligned}
\tilde{I}_{(2)}^{3,1}&: \quad \big( 1, \  i,\ 0,\ 0,\ 0,\ 0 \big) ,\quad  &\tilde{I}_{(2)}^{2,0}&: \quad\big( 0, \  0,\ 0,\ 1,\ i,\ 0 \big) , \\
\tilde{I}_{(2)}^{2,2}&: \quad \big( 0, \  0,\ 1,\ 0,\ 0,\ 0 \big) , \quad &\tilde{I}_{(2)}^{1,1}&: \quad \big( 0, \  0,\ 0,\ 0,\ 0,\ 1 \big) ,  \\
\tilde{I}_{(2)}^{1,3}&:\quad \big( 1, \  -i,\ 0,\ 0,\ 0,\ 0\big) , \quad &\tilde{I}_{(2)}^{0,2}&:\quad \big( 0, \  0,\ 0,\ 1,\ -i,\ 0\big) , 
\end{aligned}
\end{equation}
while the log-monodromy matrices are given by
\begin{footnotesize}
\begin{equation}\label{eq:II1N}
\begin{aligned}
N_1 &= \scalebox{0.8}{$\begin{pmatrix}
 0 & 0 & 0 & 0 & 0 & 0 \\
 0 & 0 & 0 & 0 & 0 & 0 \\
 0 & 0 & 0 & 0 & 0 & 0 \\
 1 & 0 & 0 & 0 & 0 & 0 \\
 0 & 1 & 0 & 0 & 0 & 0 \\
 0 & 0 & -n_1 & 0 & 0 & 0 \\
\end{pmatrix}$}, \qquad N_2 &= \scalebox{0.8}{$\begin{pmatrix}
 0 & 0 & 0 & 0 & 0 & 0 \\
 0 & 0 & 0 & 0 & 0 & 0 \\
 0 & 0 & 0 & 0 & 0 & 0 \\
 n_2 & 0 & 0 & 0 & 0 & 0 \\
 0 & n_2 & 0 & 0 & 0 & 0 \\
 0 & 0 & -1 & 0 & 0 & 0 
\end{pmatrix}$}, 
\end{aligned}
\end{equation}
\end{footnotesize}
where $n_1, n_2 \in \mathbb{Q}$ and $n_1,n_2 \geq 0$. The phase operator $\delta$ has been set to zero by using coordinate transformations \eqref{eq:shift}.

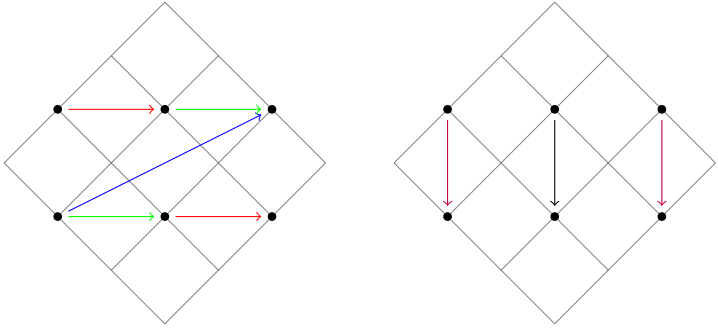
\begin{figure}[h!]
\centering
\begin{minipage}{0.4\textwidth}
\begin{tikzpicture}[scale=1,cm={cos(45),sin(45),-sin(45),cos(45),(15,0)}]
  \draw[step = 1, gray, ultra thin] (0, 0) grid (3, 3);

  \draw[fill] (0, 2) circle[radius=0.05];
  \draw[fill] (1, 3) circle[radius=0.05];
  \draw[fill] (2, 2) circle[radius=0.05];
  \draw[fill] (1, 1) circle[radius=0.05];
  \draw[fill] (2, 0) circle[radius=0.05];
  \draw[fill] (3, 1) circle[radius=0.05];

\draw[->, red] (1.1,2.9) -- (1.9,2.1);
\draw[->, red] (1.1,0.9) -- (1.9,0.1);

\draw[->, blue] (0.15,1.95) -- (2.85,1.05);

\draw[->, green]  (0.1,1.9) -- (0.9,1.1);
\draw[->, green]  (2.1,1.9) -- (2.9,1.1);
\end{tikzpicture}
\end{minipage}
\begin{minipage}{0.4\textwidth}
\begin{tikzpicture}[scale=1,cm={cos(45),sin(45),-sin(45),cos(45),(15,0)}]
  \draw[step = 1, gray, ultra thin] (0, 0) grid (3, 3);

  \draw[fill] (0, 2) circle[radius=0.05];
  \draw[fill] (1, 3) circle[radius=0.05];
  \draw[fill] (2, 2) circle[radius=0.05];
  \draw[fill] (1, 1) circle[radius=0.05];
  \draw[fill] (2, 0) circle[radius=0.05];
  \draw[fill] (3, 1) circle[radius=0.05];

\draw[->, purple] (0.9,2.9) -- (0.1,2.1);
\draw[->, purple] (2.9,0.9) -- (2.1,0.1);

\draw[->] (1.9,1.9) -- (1.1,1.1);

\end{tikzpicture}
\end{minipage}
\caption{\label{fig:II1} The Hodge-Deligne diamond that classifies $\mathrm{II}_1$ boundaries. On the left included colored arrows to denote the different components of $\Gamma_{-1}(z)$  of the instanton map $\Gamma(z)$, where we labeled $a(z)$, $b(z)$ and $c(z)$ by red, green and blue respectively. On the right the purple and black arrows denote the action of $N_1$ and $N_2$ respectively (setting $n_1,n_2=0$ for simplicity).}
\end{figure}

Now we want to write down the most general expressions for the periods compatible with this boundary data. Following section \ref{sec:construction} we begin by writing down the most general instanton map $\Gamma(z)$. It is the most general Lie algebra-valued map in $\Lambda_-$ with respect to \eqref{eq:DsplitII1} that has holomorphic coefficients
\begin{equation}
\Gamma(z) =\tfrac{1}{2} \scalebox{0.7}{$\begin{pmatrix}
 e(z) & -i e(z) & b(z) & c(z) & -i
   c(z) & 0 \\
 -i e(z) & -e(z) & -i b(z) & -i c(z) &
   -c(z) & 0 \\
 a(z) & -i a(z) & 0 & 0 & 0 & 0 \\
 f(z) & -i f(z) & -d(z) & -e(z) & i
   e(z) & -a(z) \\
 -i f(z) & -f(z) & i d(z) & i e(z) &
   e(z) & i a(z) \\
 -d(z) & i d(z) & 0 & -b(z) & i b(z) & 0 \\
\end{pmatrix}$},
\end{equation}
where $a(z),b(z),c(z)$ make up the $\Gamma_{-1}$ component of the instanton map, $d(z),e(z)$ the  $\Gamma_{-2}$ component and $f(z)$ the  $\Gamma_{-3}$ component. Note that we used coordinate redefinitions \eqref{eq:shift} to set the pieces along $N_1$ and $N_2$ to zero. The period vector then reads
\begin{align}
\Pi= \Big(&1+\tfrac{1}{4} a(z) b(z) + e(z) , \ i - \tfrac{i}{4} a(z) b(z) -i e(z) , \ a(z),  \nn \\
&\tfrac{\log [z_1] + n_2 \log[z_2] }{2\pi i } \big(1+\tfrac{1}{4}a(z)b(z)+e(z) \big)  +f(z) , \\
&\tfrac{\log [z_1] + n_2 \log[z_2] }{2\pi  } \big(1-\tfrac{1}{4}a(z)b(z)-e(z) \big)-if(z) , \ - a(z)\tfrac{ n_{1} \log [z_{1}] + \log[z_2] }{2\pi i }  -d(z) \Big)\,\nn .
\end{align}
Note in particular that the function $c(z)$ does not explicitly appear in the period vector. This can be attributed to the fact that we only read off how $\exp[\Gamma(z)]$ acts on $\mathbf{a}_0$, not the full matrix. Nevertheless, $c(z)$ enters indirectly in the periods through the recursion relations \eqref{eq:recursiongamma}, as we will see shortly.

From \eqref{eq:dXdX} we obtain two differential constraints on the $\Gamma_{-1}$ coefficients
\begin{equation}
\begin{aligned}
z_{2} n_1 a^{(0,1)}-z_{2} b^{(0,1)}(z) &=  z_{1} a^{(1,0)}(z)-n_{2} z_{1}
   b^{(1,0)}(z)\, ,\\
z_{2} c^{(0,1)}(z) -n_{2} z_{1}
   c^{(1,0)}(z)&= 2 i \pi  z_{1} z_{2}  \big(a^{(0,1)}(z) b^{(1,0)}(z) -a^{(1,0)}(z) b^{(0,1)}(z) \big)\, .  \label{eq:II1DiffConstraints}
\end{aligned}
\end{equation}
By expanding the holomorphic functions around $z_1=z_2=0$ we then obtain
\begin{align}
k(a_{kl}-n_2 b_{kl})&= l(n_1  a_{kl} -  b_{kl})  \, ,\\
l\, c_{kl}-n_2 k \, c_{kl} &=  2\pi i \sum_{m,n} (k-m)n \, a_{k-m,l-n} b_{mn}  - 2\pi i \sum_{m,n} (l-n)m \,  b_{k-m,l-n} a_{mn} \, . \nn
\end{align}
In particular, for $k=0$ or $l=0$ we find that
\begin{equation}
n_1 a_{0l}=b_{0l}\, ,\quad  a_{k0} =n_2 b_{k0}\, , \quad c_{0l}=0 \, , \quad n_2  c_{k0}= 0\, .
\end{equation}
Next we consider the component-wise constraints given in \eqref{eq:recursiongamma}. These result in the following set of relations
\begin{equation} \label{eq:II1DiffConstraints2}
\begin{aligned}
2 i \pi  z_{1} d^{(1,0)}(z) &= b(z)-n_1 a(z)\, , \qquad 2 i \pi  z_{2} d^{(0,1)}(z) = n_{2} b(z)- a(z)\, , \\
4 i \pi  z_{1} e^{(1,0)}(z) &= 2c(z)+i \pi  z_{1} \big(  a^{(1,0)}(z) b(z)-a(z)
   b^{(1,0)}(z) \big)\, , \\
4 i \pi  z_{2} e^{(0,1)}(z) &= 2 n_{2} c(z) +i \pi  z_{2}\big(  a^{(0,1)}(z) b(z)-a(z)
   b^{(0,1)}(z) \big) \, , \\
 i \pi  z_{1}
   f^{(1,0)}(z) &= -  e(z)+ i \pi  z_{1} a^{(1,0)}(z) d(z)\, , \\
 i \pi  z_{2}
   f^{(0,1)}(z) &= - n_2 e(z)+ i \pi  z_{2} a^{(0,1)}(z) d(z)\, .
\end{aligned}
\end{equation}
Similarly these relations can be cast into expressions for the series coefficients $e_{kl},f_{kl},g_{kl}$, but for brevity we do not write them down here. 

We then make a leading order ansatz for the components of $\Gamma_{-1}$. Turning on as many linear terms in $z_1,z_2$ as possible without violating \eqref{eq:II1DiffConstraints}, we write for these coefficients
\beq
\begin{aligned}
a(z) &= n_2 b\,  z_1+a\,  z_2\, ,  \qquad b(z)= b\, z_1 +n_1 a \, z_2 \, , \\
 c(z)&= 2\pi i m_1 c\,  z^{m_1}_1 z^{m_2}_2 +\pi i \, a b \frac{1-n_1 n_2}{1-n_2} z_1 z_2\,,
\end{aligned}
\eeq
where $a,b,c \in \mathbb{C}$, and we have non-negative integers $m_1, m_2$ with $m_1 > 0$ that satisfy
\begin{equation}
n_2 = m_2 /m_1 \, .
\end{equation}
When $n_1 n_2 \neq 1$ we can use axion shifts as discussed below \eqref{eq:shift} to set $a,b \in \mathbb{R}$, while for $n_1 n_2 = 1$ this is generally only possible for one of the two. The rank condition \eqref{eq:rankGamma} tells us that we must always require $a\neq 0$, and additionally $b \neq 0$ or $c \neq 0$ should hold. Note that the second term in $c(z)$ is ill-defined for $n_2 = 1$, meaning we should set $b=0$. In this special case the term at order $z_1 z_2$  in $c(z)$ then already follows from the first term by $m_1=m_2 = 1$. 

We can now solve the recursive differential equations  \eqref{eq:II1DiffConstraints2} for the components of $\Gamma_{-q}$ with $q< -1$, from which we obtain
\begin{equation}\label{eq:II1def}
\begin{aligned}
d(z)&=   \frac{1-n_1 n_2}{2\pi i} (b \, z_1-a \, z_2)\, ,\qquad  e(z) =c \, z_1^{m_1} z_2^{m_2} +ab\frac{(1+n_2)(1-n_1 n_2)}{4(1-n_2)} z_1 z_2\, ,\\
 f(z)&= \frac{i c\, z_1^{m_1} z_2^{m_2}}{\pi m_1 } + \frac{1-n_1 n_2}{8\pi i} \Big(  a^2 z_2^2 + b^2 n_2 z_1^2 +2 ab\, z_1 z_2 \frac{1+n_2^2}{1-n_2} \Big)\, .
 \end{aligned}
\end{equation}
Putting all this together we get as asymptotic period vector 
\begin{equation}
\mathbf{\Pi}=\scalebox{0.8}{$ \begin{pmatrix}
1+ c z_1^{m_1} z_2^{m_2} +\frac{1}{4}  \big(a^2 n_1 z_2^2+ 2 a b z_1 z_2\frac{1-n_1 n_2^2}{1-n_2}+b^2 n_2 z_1^2 \big)\\
i -i c z_1^{m_1} z_2^{m_2} - \frac{i}{4} \big(a^2 n_1 z_2^2+ 2 a b z_1 z_2\frac{1-n_1n_2^2}{1-n_2}+b^2 n_2 z_1^2 \big) \\
b n_2 z_1 + a z_2\\
   \frac{\log [z_1] + n_2 \log[z_2] }{2\pi i }\Big(1+ c z_1^{m_1} z_2^{m_2} +\frac{1}{4}  \big(a^2 n_1 z_2^2+ 2 a b z_1 z_2\frac{1-n_1 n_2^2}{1-n_2}+b^2 n_2 z_1^2 \big) \Big)  +f(z)\\
\frac{\log [z_1] + n_2 \log[z_2] }{2\pi  } \Big(1- c z_1^{m_1} z_2^{m_2} -\frac{1}{4}  \big(a^2 n_1 z_2^2+ 2 a b z_1 z_2\frac{1-n_1 n_2^2}{1-n_2}+b^2 n_2 z_1^2 \big)  \Big)-i f(z)\\
i ( b n_2 z_1 + a z_2) \frac{n_1 \log[z_1]+ \log[z_2]}{2\pi}- \frac{1-n_1 n_2}{2\pi i}(b z_1 - a z_2)\end{pmatrix}$}\, ,
\end{equation}
with $f(z)$ given in \eqref{eq:II1def}.

\subsubsection{Coni-LCS class boundaries}
\label{ssec:IV2construction}
Finally we study coni-LCS class boundaries. Such boundaries are characterized by a $\mathrm{IV}_2$ Hodge-Deligne diamond, which has been depicted in figure \ref{fig:IV2}. The nilpotent orbit data has been constructed in section \ref{app:coniLCSdata}, and is again given by the Deligne splitting $\tilde{I}_{(2)}^{p,q}$ together with the log-monodromy matrices $N_i$ and the phase operator $\delta$. Recall that the Deligne splitting is spanned by
\begin{equation}\label{eq:DsplitIV2}
\begin{aligned}
\tilde{I}_{(2)}^{3,3}&: \ \big( 1, \  0,\ 0,\ 0,\ 0,\ 0 \big) , \quad \tilde{I}_{(2)}^{0,0}: \ \big( 0, \  0,\ 0,\ 1,\ 0,\ 0 \big) ,\\
\tilde{I}_{(2)}^{2,2}&: \ \big( 0, \  1,\ 0,\ 0,\ 0,\ 0 \big) , \ \  \big( 0, \  0,\ 1,\ 0,\ 0,\ 0 \big) , \\
\tilde{I}_{(2)}^{1,1}&: \ \big( 0, \  0,\ 0,\ 0,\ 1,\ 0 \big) , \ \  \big( 0, \  0,\ 0,\ 0,\ 0,\ 1 \big) , \\
\end{aligned}
\end{equation}
while the log-monodromy matrices and phase operator are
\begin{footnotesize}
\begin{align}\label{eq:IV2N}
N_1= \scalebox{0.8}{$\begin{pmatrix}
0 & 0 & 0 & 0 & 0 & 0 \\
0 & 0 & 0 & 0 & 0 & 0 \\
0 & 0 & 0 & 0 & 0 & 0 \\
0 & 0 & 0 & 0 & 0 & 0 \\
0 & 1 & 0 & 0 & 0 & 0 \\
0 & 0 & 0 & 0 & 0 & 0 \\
\end{pmatrix}$} \, ,  \ 
N_2=\scalebox{0.8}{$\begin{pmatrix}
0 & 0 & 0 & 0 & 0 & 0 \\
0 & 0 & 0 & 0 & 0 & 0 \\
1 & 0 & 0 & 0 & 0 & 0 \\
0 & 0 & 0 & 0 & 0 & -1 \\
0 & -n & 0 & 0 & 0 & 0 \\
0 & 0 & 1 & 0 & 0 & 0 \\
\end{pmatrix}$},\ \delta = \scalebox{0.8}{$\begin{pmatrix} 0 & 0 & 0 & 0 & 0 & 0 \\
 0 & 0 & 0 & 0 & 0 & 0 \\
0 & 0 & 0 & 0 & 0 & 0 \\
 \delta_2 & \delta_1 & 0& 0 & 0 & 0 \\
 \delta_1 & \frac{1}{2\pi} & 0 & 0 & 0 & 0 \\
 0 & 0 &0 & 0 & 0 & 0 
\end{pmatrix}$}.
\end{align}
\end{footnotesize}
where we chose to set $\delta_3=1/2\pi$ and $\delta_4=0$.

\begin{figure}[h!]
\centering
\begin{minipage}{0.4\textwidth}
\begin{tikzpicture}[scale=0.85,cm={cos(45),sin(45),-sin(45),cos(45),(15,0)}]
  \draw[step = 1, gray, ultra thin] (0, 0) grid (3, 3);

  \draw[fill] (3, 3) circle[radius=0.05];
  \draw[fill] (0.9, 1.1) circle[radius=0.05];
  \draw[fill] (1.9, 2.1) circle[radius=0.05];
  \draw[fill] (1.1, 0.9) circle[radius=0.05];
  \draw[fill] (2.1, 1.9) circle[radius=0.05];
  \draw[fill] (0, 0) circle[radius=0.05];

\draw[->, red] (2.95,2.95) -- (1.95,2.15);
\draw[->, red] (0.85,1.05) -- (0.05,0.05);

\draw[->, blue] (2.1,1.9) -- (1.1,0.9);

\draw[->, green] (1.9,2.1) -- (1.1,0.9);
\draw[->, green] (2.1,1.9) -- (0.9,1.1);

\end{tikzpicture}
\end{minipage}
\begin{minipage}{0.4\textwidth}
\begin{tikzpicture}[scale=0.85,cm={cos(45),sin(45),-sin(45),cos(45),(15,0)}]
  \draw[step = 1, gray, ultra thin] (0, 0) grid (3, 3);

  \draw[fill] (3, 3) circle[radius=0.05];
  \draw[fill] (0.9, 1.1) circle[radius=0.05];
  \draw[fill] (1.9, 2.1) circle[radius=0.05];
  \draw[fill] (1.1, 0.9) circle[radius=0.05];
  \draw[fill] (2.1, 1.9) circle[radius=0.05];
  \draw[fill] (0, 0) circle[radius=0.05];

\draw[->] (2.95,2.95) -- (2.15,1.95);
\draw[->] (1.05,0.85) -- (0.05,0.05);

\draw[->] (2.1,1.9) -- (1.1,0.9);

\draw[->, purple] (1.9,2.1) -- (0.9,1.1);

\end{tikzpicture}
\end{minipage}
\caption{\label{fig:IV2} The Hodge-Deligne diamond that classifies $\mathrm{IV}_2$ boundaries of coni-LCS type. On the left we included colored arrows to denote the different components of $\Gamma_{-1}(z)$  of the instanton map $\Gamma(z)$, where we labeled $a(z)$, $b(z)$ and $c(z)$ by red, green and blue respectively. On the right we included purple and black arrows to indicate the action of the log-monodromy matrices $N_1,N_2$ respectively, where we considered $n=0$ for simplicity. Note that we split up $\tilde{I}_{(2)}^{1,1},\tilde{I}_{(2)}^{2,2}$ according to the vectors that span the spaces given in \eqref{eq:DsplitIV2}, where left (right) vertices correspond to the first (last) two vectors.}
\end{figure}
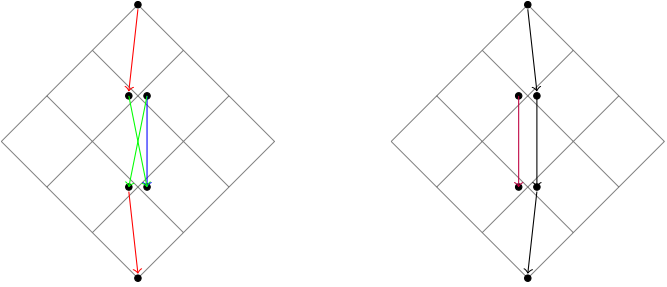

We now construct the most general periods compatible with this boundary data. Following section \ref{sec:construction} we first consider the most general instanton map $\Gamma(z)$. For the above boundary data the most general Lie algebra-valued map in $\Lambda_-$ is
\begin{small}
\begin{align}
\Gamma= \scalebox{0.72}{$\begin{pmatrix}
0 & 0 & 0 & 0 & 0 & 0 \\
a(z) & 0 & 0 & 0 & 0 & 0 \\
0 & 0 & 0 & 0 & 0 & 0 \\
f(z) & e(z) & d(z) & 0 & -a(z) & 0 \\
e(z) & 0 & b(z) & 0 & 0 & 0 \\
d(z) & b(z) & c(z) & 0 & 0 & 0   
\end{pmatrix}$}\, ,
\end{align}
\end{small}
where the holomorphic coefficients $a(z), b(z), c(z)$ specify the $\Gamma_{-1}$ component, while $d(z),e(z)$ and $f(z)$ correspond to $\Gamma_{-2}$ and $\Gamma_{-3}$ respectively. Note that we used coordinate shifts \eqref{eq:shift} to set coefficients associated with $N_1,N_2$ to zero. Next we can use \eqref{eq:PiGamma} to write the periods in terms of the holomorphic coefficients as
\begin{small}
\begin{align}\label{eq:periodsIV2general}
\Pi = \big(&1 , \ a(z) , \ \frac{ \log[z_2]}{2 \pi i} , \ i \delta_2 + i \delta_1 a(z) +f(z) + \frac{i (\frac{1}{2} a(z) c(z) + d(z)) \log[z_2]}{2 \pi } - \frac{i \log[z_2]^3}{48 \pi^3} \nn \\
&i \delta_1 +e(z)- \frac{ a(z)(\log[z_1]+n \log[z_2])}{2 \pi i} , \ \frac{1}{2} a(z) c(z) + d(z) - \frac{\log[z_2]^2}{8 \pi^2} \big) .
\end{align}
\end{small}
The coefficients of the instanton map $\Gamma$ now must satisfy several differential equations. The $\Gamma_{-1}$ coefficients obey \eqref{eq:dXdX}, while in turn the $\Gamma_{-2},\Gamma_{-3}$ coefficients are fixed uniquely by \eqref{eq:recursiongamma}. We can write out \eqref{eq:dXdX} as
\begin{equation}\label{eq:dXdXIV2}
\begin{aligned}
z_2 a^{(0,1)}(z)&=  z_1 (n a^{(1,0)}(z) b^{(1,0)}(z))\, , \\
 c^{(1,0)}(z) &= 2 \pi i z_2 (a^{(1,0)} (z)b^{(0,1)}(z)-a^{(0,1)} (z)b^{(1,0)}(z)) \, ,  
\end{aligned}
\end{equation}
while \eqref{eq:recursiongamma} imposes
\begin{equation}\label{eq:recursiongammaIV2}
\begin{aligned}
 2 d^{(1,0)}(z) &= a^{(1,0)}(z) b(z) - a(z) b^{(1,0)}(z) \, , \\
 2 \pi i z_2  d^{(0,1)}(z)&=c(z) +  \pi i z_2 (a^{(0,1)} (z)b(z) - a(z) b^{(0,1)}(z)) \, , \\
 2 \pi i z_1 e^{(1,0)}(z)&=  a(z) \, ,  \qquad \ \ \qquad 2 \pi i z_2 e^{(0,1)}(z)= b(z) + n a(z) \, ,\\
 f^{(1,0)}(z) &= a^{(1,0)}(z) e(z) \, ,  \qquad \ \ \, \pi f^{(0,1)}(z)= \pi a^{(0,1)}(z) e(z) -id(z)\, .
\end{aligned}
\end{equation}
We now want to construct asymptotic models for the periods by performing a leading order expansion for the $\Gamma_{-1}$ coefficients. First we consider the rank condition \eqref{eq:rankGamma}, which gives us an indication which coefficients have to be turned on. In terms of the Hodge-Deligne diamond it implies that there should be an ingoing arrow due to $\Gamma_{-1}$ or $N_i$ for every vertex apart from $\tilde{I}_{(2)}^{3,3}$. Looking at figure \ref{fig:IV2} this means that $a(z)$ must be turned on. In order to solve \eqref{eq:dXdX} we find that we must also turn on $b(z)$. Let us therefore make as ansatz
\begin{align}
a(z)= a z_1 \, , \qquad b(z)=b z_1 \, , \qquad c(z)=0\, ,
\end{align}
where $a \in \mathbb{R}$ has been rotated to a real value using the axion shift symmetry described below \eqref{eq:shift}. We then find that \eqref{eq:dXdXIV2} reduces on our ansatz to
\begin{equation}
b = -n a\, .
\end{equation}
Consequently we can solve \eqref{eq:recursiongammaIV2} for the coefficients of $\Gamma_{-2},\Gamma_{-3}$ as
\begin{align}
d(z)=0\, , \qquad e(z)= \frac{a}{2 \pi i} \, , \qquad f(z)= \frac{a^2}{4 \pi i } z_1^{2 } \, .
\end{align}
Inserting these leading order behaviors into \eqref{eq:periodsIV2general} we find as asymptotic model for the periods
\begin{align}\label{eq:IV2Period}
\Pi= \big(&1 , \ a z_1, \ \frac{ \log[z_2]}{2 \pi i} , \ -\frac{i \log[z_2]^3}{48 \pi^3}-\frac{ i a^2 n z_1^2 \log[z_2]}{4\pi }+   \frac{a^2}{4  \pi i} z_1^2+i \delta_2 +i \delta_1 a z_1, \nn \\
 &- a z_1\frac{ \log[z_1] + n \log[z_2]}{2 \pi i} +i\delta_1, \ - \frac{\log[z_2]^2}{8 \pi^2}  -\frac{1}{2} a^2 n z_1^2\big)\, .
\end{align}

\begin{subappendices}

\section{Nilpotent orbit data for two-moduli periods}\label{app:boundarydata}
In this section we construct the nilpotent orbits of the two-moduli periods. Following figure \ref{fig:flowchart}, we begin by writing down an sl(2)-splitting each boundary. Recall from section \ref{sec:sl2splitting} that the sl$(2,\mathbb{R})^n$-boundary data is encoded in a set of commuting sl$(2,\mathbb{R})$-triples $(N_i^{\pm},N_i^0)$ and the sl(2)-split Deligne splitting $\tilde{I}^{p,q}_{(2)}$ characterizing the intersection. By using the building blocks given in table \ref{table:buildingblocks} we can straightforwardly obtain simple expressions for this sl$(2,\mathbb{R})^n$-data. The main task in this appendix is then to complete this sl$(2,\mathbb{R})^n$-boundary data into the most general compatible nilpotent orbit, which comes in two parts. First we construct the most general log-monodromy matrices $N_i$ out of the lowering operators $N_i^-$. Secondly we determine the most general phase operator $\delta$ that rotates the sl(2)-split $\tilde{I}^{p,q}_{(2)}$ into a generic Deligne splitting. This construction follows the approach laid out in section \ref{sec:construction}.

\subsection{Class $\text{I}_2$ boundaries}\label{app:I2data}
Let us begin by considering boundaries of class $\mathrm{I}_2$, which consists of the 2-cubes $\langle \mathrm{I}_1|\mathrm{I}_2|\mathrm{I}_1 \rangle$, $\langle \mathrm{I}_2|\mathrm{I}_2|\mathrm{I}_1 \rangle$ and $\langle \mathrm{I}_2|\mathrm{I}_2|\mathrm{I}_2 \rangle$. For $\langle \mathrm{I}_1|\mathrm{I}_2|\mathrm{I}_1 \rangle$ we consider the enhancement chain $\mathrm{I}_1 \to \mathrm{I}_2$, while we consider $\mathrm{I}_2 \to \mathrm{I}_2$ for $\langle \mathrm{I}_2|\mathrm{I}_2|\mathrm{I}_1 \rangle$ and $\langle \mathrm{I}_2|\mathrm{I}_2|\mathrm{I}_2 \rangle$. Both these enhancement chains have the singularity type $\mathrm{I}_2$ for $y^1=y^2=\infty$ in common. We span the vector spaces $\tilde{I}_{(2)}^{p,q}$ of this sl(2)-split Deligne splitting by
\begin{equation}
\begin{aligned}
\tilde{I}_{(2)}^{3,0}&: \ \big( 1, \  0,\ 0,\ i,\ 0,\ 0 \big) , \qquad \tilde{I}_{(2)}^{0,3}: \ \big( 1, \  0,\ 0,\ -i,\ 0,\ 0\big) . \\
\tilde{I}_{(2)}^{2,2}&: \ \big( 0, \  1,\ 0,\ 0,\ 0,\ 0 \big) , \  \big( 0, \  0,\ 1,\ 0,\ 0,\ 0 \big) ,  \\
\tilde{I}_{(2)}^{1,1}&: \ \big( 0, \  0,\ 0,\ 0,\ 1,\ 0 \big) ,   \  \big( 0, \  0,\ 0,\ 0,\ 0,\ 1 \big) ,  \\
\end{aligned}
\end{equation}

\subsubsection*{Enhancement step $\text{I}_1 \to \text{I}_2$}
Here we construct the nilpotent orbit data for the 2-cube $\langle \mathrm{I}_1 | \mathrm{I}_2 | \mathrm{I}_1 \rangle$. Let us begin by writing down the commuting sl$(2,\mathbb{R})$-triples as
\begin{equation}
\begin{aligned}
 N_1 = N_1^- &=  \scalebox{0.7}{$\begin{pmatrix}
 0 & 0 & 0 & 0 & 0 & 0 \\
 0 & 0 & 0 & 0 & 0 & 0 \\
 0 & 0 & 0 & 0 & 0 & 0 \\
 0 & 0 & 0 & 0 & 0 & 0 \\
 0 & 0 & 0 & 0 & 0 & 0 \\
 0 & 0 & -1 & 0 & 0 & 0 \\
 \end{pmatrix}$}, \qquad &Y_1 &= \scalebox{0.7}{$\begin{pmatrix}
 0 & 0 & 0 & 0 & 0 & 0 \\
 0 & 0 & 0 & 0 & 0 & 0 \\
 0 & 0 & 1 & 0 & 0 & 0 \\
 0 & 0 & 0 & 0 & 0 & 0 \\
 0 & 0 & 0 & 0 & 0 & 0 \\
 0 & 0 & 0 & 0 & 0 & -1 \\
\end{pmatrix}$}\, , \\
 N_2^- &= \scalebox{0.7}{$\begin{pmatrix}
 0 & 0 & 0 & 0 & 0 & 0 \\
 0 & 0 & 0 & 0 & 0 & 0 \\
 0 & 0 & 0 & 0 & 0 & 0 \\
 0 & 0 & 0 & 0 & 0 & 0 \\
 0 & -1 & 0 & 0 & 0 & 0 \\
 0 & 0 & 0 & 0 & 0 & 0 \\
 \end{pmatrix}$}, \qquad &Y_2 &= \scalebox{0.7}{$\begin{pmatrix}
 0 & 0 & 0 & 0 & 0 & 0 \\
 0 & 1 & 0 & 0 & 0 & 0 \\
 0 & 0 & 0 & 0 & 0 & 0 \\
 0 & 0 & 0 & 0 & 0 & 0 \\
 0 & 0 & 0 & 0 & -1 & 0 \\
 0 & 0 & 0 & 0 & 0 & 0 \\
\end{pmatrix}$}\, ,
\end{aligned}
\end{equation}
where we did not include $N_i^+$ since the raising operators are irrelevant for our discussion. Let us point out that the sign of the coefficients in the lowering operators is fixed by \eqref{eq:pol}, which requires $\eta N_1^-$ and $\eta N_2^-$ to have negative eigenvalues.

Next we want to write down the most general log-monodromy matrix $N_2$ compatible with the above boundary data. From \eqref{lowering} we find that we must identify matrices with eigenvalue $\ell\leq -2$ under the adjoint action of $Y_1$. The only map that satisfies this property is proportional to $N_1$, so we find that
\begin{equation}
N_2 = N_2^- + n N_1 \, .
\end{equation}
Similar to the lowering operators we find that \eqref{eq:pol} requires $N_2$ to have two negative eigenvalues, so we must impose $n\geq 0$. For $n=0$ we find that $N_2$ produces a $\mathrm{I}_1$ singularity type for the $y^2=\infty$ divisor, while for $n>0$ it produces a $\mathrm{I}_2$ singularity. Thus for the 2-cube $\langle \mathrm{I}_1 | \mathrm{I}_2 | \mathrm{I}_1 \rangle$ one can simply take $N_1^-$ and $N_2^-$ as log-monodromy matrices.

\subsubsection*{Enhancement step $\text{I}_2 \to \text{I}_2$}
Here we construct the nilpotent orbit data for the 2-cubes $\langle \mathrm{I}_2|\mathrm{I}_2|\mathrm{I}_1 \rangle$ and $\langle \mathrm{I}_2|\mathrm{I}_2|\mathrm{I}_2 \rangle$. Let us begin by writing down the commuting sl$(2,\mathbb{R})$-triples as
\begin{equation}
 N_1 =N_1^-= - \scalebox{0.7}{$\begin{pmatrix}
 0 & 0 & 0 & 0 & 0 & 0 \\
 0 & 0 & 0 & 0 & 0 & 0 \\
 0 & 0 & 0 & 0 & 0 & 0 \\
 0 & 0 & 0 & 0 & 0 & 0 \\
 0 & 1 & 0 & 0 & 0 & 0 \\
 0 & 0 & 1 & 0 & 0 & 0 \\
 \end{pmatrix}$}, \qquad Y_1 = \scalebox{0.7}{$\begin{pmatrix}
 0 & 0 & 0 & 0 & 0 & 0 \\
 0 & 1 & 0 & 0 & 0 & 0 \\
 0 & 0 & 1 & 0 & 0 & 0 \\
 0 & 0 & 0 & 0 & 0 & 0 \\
 0 & 0 & 0 & 0 & -1 & 0 \\
 0 & 0 & 0 & 0 & 0 & -1 \\
\end{pmatrix}$},
\end{equation}
while the second sl$(2,\mathbb{R})$-triple is trivial, i.e.~$N_2^-=Y_2=0$. The sign of the coefficients in $N_1$ is fixed by the requirement of the polarization conditions \eqref{eq:pol} that $\eta N_1$ has two negative eigenvalues.

Next we want to determine the most general log-monodromy matrix $N_2$ compatible with the above boundary data. There are three matrices with eigenvalue $\ell=-2$ under the adjoint action of $Y_1$ that are infinitesimal isometries of the symplectic pairing $\langle \cdot, \cdot \rangle$. These matrices map from $\tilde{I}_{(2)}^{2,2}$ to $\tilde{I}_{(2)}^{1,1}$, and the most general linear combination is given by

\begin{equation}
N_2
=- \scalebox{0.7}{$\begin{pmatrix}
 0 & 0 & 0 & 0 & 0 & 0 \\
 0 & 0 & 0 & 0 & 0 & 0 \\
 0 & 0 & 0 & 0 & 0 & 0 \\
 0 & 0 & 0 & 0 & 0 & 0 \\
 0 & n_1 & n_3 & 0 & 0 & 0 \\
 0 & n_3 & n_2 & 0 & 0 & 0 \\
\end{pmatrix}$}\, .
\label{NilpI1}
\end{equation}
We now want to simplify this expression by considering a change of basis. First we want to diagonalize the $2\times 2$ block that appears in $N_2$ by a symplectic basis transformation (while keeping $N_1$ the same). This yields
\begin{equation}
N_2
= -\scalebox{0.7}{$\begin{pmatrix}
 0 & 0 & 0 & 0 & 0 & 0 \\
 0 & 0 & 0 & 0 & 0 & 0 \\
 0 & 0 & 0 & 0 & 0 & 0 \\
 0 & 0 & 0 & 0 & 0 & 0 \\
 0 & \lambda_1 & 0 & 0 & 0 & 0 \\
 0 & 0 & \lambda_2 & 0 & 0 & 0 \\
\end{pmatrix}$}\, ,
\end{equation}
where the $2\times 2$ matrix needs to have at least one non-zero eigenvalue, which we take to be $\lambda_1 \neq 0$. This allows us to rescale by the symplectic basis transformation $M=\text{diag}(1,\sqrt{\lambda_1},1,1,1/\sqrt{\lambda_1},1)$, after which our log-monodromy matrices become
\begin{equation}
 N_1 =-\scalebox{0.7}{$\begin{pmatrix}
 0 & 0 & 0 & 0 & 0 & 0 \\
 0 & 0 & 0 & 0 & 0 & 0 \\
 0 & 0 & 0 & 0 & 0 & 0 \\
 0 & 0 & 0 & 0 & 0 & 0 \\
 0 & n_1 & 0 & 0 & 0 & 0 \\
 0 & 0 & 1 & 0 & 0 & 0 \\
 \end{pmatrix}$}, \qquad N_2
= -\scalebox{0.7}{$\begin{pmatrix}
 0 & 0 & 0 & 0 & 0 & 0 \\
 0 & 0 & 0 & 0 & 0 & 0 \\
 0 & 0 & 0 & 0 & 0 & 0 \\
 0 & 0 & 0 & 0 & 0 & 0 \\
 0 & 1 & 0 & 0 & 0 & 0 \\
 0 & 0 & n_2 & 0 & 0 & 0 \\
\end{pmatrix}$}\, ,
\end{equation}
by relabeling $n_1=1/\lambda_1$ and $n_2=\lambda_2$. From the polarization conditions \eqref{eq:pol} we find that $\eta N_2$ needs to have one positive and one non-negative eigenvalue, so $n_2 \geq 0$. For $n_2=0$ we are dealing with a $\mathrm{I}_1$ divisor at $y^2=\infty$, while for $n_2>0$ it is a $\mathrm{I}_2$ divisor. Note that the log-monodromy matrices for the $\langle \mathrm{I}_1 | \mathrm{I}_2 | \mathrm{I}_1\rangle$ boundary in the $\mathrm{I}_1 \to \mathrm{I}_2$ enhancement step take the same form with $n_1=n_2=0$. This conveniently allows us to use the same form for the log-monodromy matrices for all boundaries of $\mathrm{I}_2$ class.

\subsubsection*{Construction of the phase operator}
Based on the above data let us write down the phase operator $\delta$. The only possible matrices that are valued in $\Lambda_{-p,-q}$ with $p,q>0$ map from $\tilde{I}_{(2)}^{2,2}$ to $\tilde{I}_{(2)}^{1,1}$, effectively reducing $\delta$ to a $2\times 2$ sub-block. We therefore find that the most general $\delta$ takes the form
\begin{equation}
\delta = \scalebox{0.7}{$\begin{pmatrix}
 0 & 0 & 0 & 0 & 0 & 0 \\
 0 & 0 & 0 & 0 & 0 & 0 \\
 0 & 0 & 0 & 0 & 0 & 0 \\
 0 & 0 & 0 & 0 & 0 & 0 \\
 0 & \delta_2 & \delta_1 & 0 & 0 & 0 \\
 0 & \delta_1 & \delta_3 & 0 & 0 & 0 \\
\end{pmatrix}$}\, ,
\end{equation}
where the off-diagonal components of the sub-block are equal because $\delta^T \eta+\eta \delta = 0$. One can reduce $\delta$ further by using coordinate shifts \eqref{eq:shift}, setting $\delta_2=\delta_3=0$.

\subsection{Class $\mathrm{II}_1$ boundaries} \label{app:II1data}
Next we consider boundaries of class $\mathrm{II}_1$, which consists of the 2-cubes $\langle \mathrm{II}_0 | \mathrm{II}_1 | \mathrm{I}_1 \rangle$, $\langle \mathrm{II}_0 | \mathrm{II}_1 | \mathrm{II}_1 \rangle$,  $\langle \mathrm{II}_1 | \mathrm{II}_1 | \mathrm{I}_1 \rangle$ and  $\langle \mathrm{II}_1 | \mathrm{II}_1 | \mathrm{II}_1 \rangle$. For the first two we reconstruct the nilpotent orbit data starting from the enhancement chain $\mathrm{II}_0 \to \mathrm{II}_1$, while for the latter two boundaries we consider $\mathrm{II}_1 \to \mathrm{II}_1$. In either case the enhancement chains have the singularity type $\mathrm{II}_1$ at $y^1=y^2=\infty$ in common. Let us therefore begin by writing down basis vectors for $\tilde{I}_{(2)}^{p,q}$ of this sl$(2,\mathbb{R})$-split Deligne splitting as
\begin{equation}\label{eq:appII1}
\begin{aligned}
\tilde{I}_{(2)}^{3,1}&: \quad \big( 1, \  i,\ 0,\ 0,\ 0,\ 0 \big) ,\qquad  &\tilde{I}_{(2)}^{2,0}&: \quad\big( 0, \  0,\ 0,\ 1,\ i,\ 0 \big) , \\
\tilde{I}_{(2)}^{2,2}&: \quad \big( 0, \  0,\ 1,\ 0,\ 0,\ 0 \big) , \qquad &\tilde{I}_{(2)}^{1,1}&: \quad \big( 0, \  0,\ 0,\ 0,\ 0,\ 1 \big) ,  \\
\tilde{I}_{(2)}^{1,3}&:\quad \big( 1, \  -i,\ 0,\ 0,\ 0,\ 0\big) , \qquad &\tilde{I}_{(2)}^{0,2}&:\quad \big( 0, \  0,\ 0,\ 1,\ -i,\ 0\big) .
\end{aligned}
\end{equation}

\subsubsection*{Enhancement step $\text{II}_0 \to \text{II}_1$}
Here we construct the nilpotent orbit data for the 2-cubes $\langle \mathrm{II}_0 | \mathrm{II}_1 | \mathrm{I}_1 \rangle$ and $\langle \mathrm{II}_0 | \mathrm{II}_1 | \mathrm{II}_1 \rangle$. Let us begin by writing down the sl$(2,\mathbb{R})$-triples as
\begin{small}
\begin{equation}
 N_1 = N_1^- =  \scalebox{0.7}{$\begin{pmatrix}
 0 & 0 & 0 & 0 & 0 & 0 \\
 0 & 0 & 0 & 0 & 0 & 0 \\
 0 & 0 & 0 & 0 & 0 & 0 \\
 1 & 0 & 0 & 0 & 0 & 0 \\
 0 & 1 & 0 & 0 & 0 & 0 \\
 0 & 0 & 0 & 0 & 0 & 0 \\
 \end{pmatrix}$}, \qquad Y_1 = \scalebox{0.7}{$\begin{pmatrix}
 1 & 0 & 0 & 0 & 0 & 0 \\
 0 & 1 & 0 & 0 & 0 & 0 \\
 0 & 0 & 0 & 0 & 0 & 0 \\
 0 & 0 & 0 & -1 & 0 & 0 \\
 0 & 0 & 0 & 0 & -1 & 0 \\
 0 & 0 & 0 & 0 & 0 & 0 \\
\end{pmatrix}$},
\end{equation}
\end{small}
\begin{small}
\begin{equation}
 N_2^- = \scalebox{0.7}{$\begin{pmatrix}
 0 & 0 & 0 & 0 & 0 & 0 \\
 0 & 0 & 0 & 0 & 0 & 0 \\
 0 & 0 & 0 & 0 & 0 & 0 \\
 0 & 0 & 0 & 0 & 0 & 0 \\
 0 & 0 & 0 & 0 & 0 & 0 \\
 0 & 0 & -1 & 0 & 0 & 0 \\
 \end{pmatrix}$}, \qquad Y_2 = \scalebox{0.7}{$\begin{pmatrix}
 0 & 0 & 0 & 0 & 0 & 0 \\
 0 & 0 & 0 & 0 & 0 & 0 \\
 0 & 0 & 1 & 0 & 0 & 0 \\
 0 & 0 & 0 & 0 & 0 & 0 \\
 0 & 0 & 0 & 0 & 0 & 0 \\
 0 & 0 & 0 & 0 & 0 & -1 \\
\end{pmatrix}$},
\end{equation}
\end{small}
where the signs in $N_1^-,N_2^-$ are fixed by the polarization conditions \eqref{eq:pol}.\footnote{To be precise,  $\eta N_1^-$ must have two positive eigenvalues and $\eta N_2^-$ one negative eigenvalue.}

Next we determine the most general log-monodromy $N_2$: there are three real maps with eigenvalue $\ell\leq -2$ under $Y_1$, which means we find
\begin{small}
 \begin{equation}
 N_2 = \scalebox{0.7}{$\begin{pmatrix}
 0 & 0 & 0 & 0 & 0 & 0 \\
 0 & 0 & 0 & 0 & 0 & 0 \\
 0 & 0 & 0 & 0 & 0 & 0 \\
 n_2 & n_4 & 0 & 0 & 0 & 0 \\
 n_4 & n_3 & 0 & 0 & 0 & 0 \\
 0 & 0 & -1 & 0 & 0 & 0 \\
 \end{pmatrix}$} \,.
\end{equation}
\end{small}
Also recall that  $N_2$ is a $(-1,-1)$-map with respect to the Deligne splitting \eqref{eq:appII1}, which requires $n_2=n_3$ and $n_4=0$. Polarization conditions \eqref{eq:pol} tell us that $\eta N_2$ should have two non-negative and one negative eigenvalue, which sets $n_2 \geq 0$. For $n_2=0$ we encounter a $\mathrm{I}_1$ divisor at $y^2=\infty$, while for $n_2>0$ a $\mathrm{II}_1$ divisor.

\subsubsection*{Enhancement step $\text{II}_1 \to \text{II}_1$}
Here we construct the nilpotent orbit data for the 2-cubes $\langle \mathrm{II}_1 | \mathrm{II}_1 | \mathrm{I}_1 \rangle$ and $\langle \mathrm{II}_1 | \mathrm{II}_1 | \mathrm{II}_1 \rangle$. Let us begin by writing down the sl$(2,\mathbb{R})$-triples as 
\begin{small}
\begin{equation}
 N_1 = N_1^- =  \scalebox{0.7}{$\begin{pmatrix}
 0 & 0 & 0 & 0 & 0 & 0 \\
 0 & 0 & 0 & 0 & 0 & 0 \\
 0 & 0 & 0 & 0 & 0 & 0 \\
 1 & 0 & 0 & 0 & 0 & 0 \\
 0 & 1 & 0 & 0 & 0 & 0 \\
 0 & 0 & -1 & 0 & 0 & 0 \\
 \end{pmatrix}$}, \qquad Y_1 = \scalebox{0.7}{$\begin{pmatrix}
 1 & 0 & 0 & 0 & 0 & 0 \\
 0 & 1 & 0 & 0 & 0 & 0 \\
 0 & 0 & 1 & 0 & 0 & 0 \\
 0 & 0 & 0 & -1 & 0 & 0 \\
 0 & 0 & 0 & 0 & -1 & 0 \\
 0 & 0 & 0 & 0 & 0 & -1 \\
\end{pmatrix}$},
\end{equation}
\end{small}
while the second $\mathfrak{sl}(2,\mathbb{R})$-triple is trivial, i.e.~$N_2^-=Y_2=0$. The signs of $N_1$ are fixed by the polarization conditions \eqref{eq:pol}, which requires  $\eta N_1^-$ to have two positive and one negative eigenvalue.

We now want to construct the most general log-monodromy matrix $N_2$ compatible with the above boundary data. The most general map with eigenvalue $\ell \leq -2$ under the adjoint action with $Y_1$ is given by
\begin{small}
\begin{equation}
 N_2 = \scalebox{0.7}{$\begin{pmatrix}
 0 & 0 & 0 & 0 & 0 & 0 \\
 0 & 0 & 0 & 0 & 0 & 0 \\
 0 & 0 & 0 & 0 & 0 & 0 \\
  n_1 & n_4 & n_5 & 0 & 0 & 0 \\
 n_4 & n_2 & n_6 & 0 & 0 & 0 \\
 n_5 & n_6 & -n_3 & 0 & 0 & 0 \\
 \end{pmatrix}$}  \,.
\end{equation}
\end{small}
where we required the $3\times 3$ block to be symmetric to ensure that $N_2^T \eta + \eta N_2 = 0$. Additionally we must require that $N_2$ is a $(-1,-1)$-map with respect to the Deligne splitting \eqref{eq:appII1}. This requires us to set $n_1=n_2$ and $n_4=n_5=n_6=0$. Polarization conditions \eqref{eq:pol} then require us to put $n_1>0$ and $n_3>0$.

Let us now try to bring these log-monodromy matrices into a similar form as we found for the enhancement chain $\mathrm{II}_0 \to \mathrm{II}_1$. We can apply a symplectic basis transformation $M = \text{diag}(1,1,\sqrt{n_3},1,1,1/\sqrt{n_3})$, which yields
\begin{equation}
N_1 = \scalebox{0.7}{$\begin{pmatrix}
 0 & 0 & 0 & 0 & 0 & 0 \\
 0 & 0 & 0 & 0 & 0 & 0 \\
 0 & 0 & 0 & 0 & 0 & 0 \\
 1 & 0 & 0 & 0 & 0 & 0 \\
 0 & 1 & 0 & 0 & 0 & 0 \\
 0 & 0 & -n_1 & 0 & 0 & 0 \\
 \end{pmatrix}$}, \qquad N_2 = \scalebox{0.7}{$\begin{pmatrix}
 0 & 0 & 0 & 0 & 0 & 0 \\
 0 & 0 & 0 & 0 & 0 & 0 \\
 0 & 0 & 0 & 0 & 0 & 0 \\
 n_2 & 0 & 0 & 0 & 0 & 0 \\
 0 & n_2 & 0 & 0 & 0 & 0 \\
 0 & 0 & -1 & 0 & 0 & 0 \\
\end{pmatrix}$} \,,
\end{equation}
where we relabeled $n_1=1/n_3$. 
\subsubsection*{Construction of the phase operator}
Based on the above data let us write down the most general phase operator $\delta$. For the above Deligne splitting there are two real maps $\delta_{-p,-q}$ with $p,q>0$ that satisfy $\delta_{-p,-q}^T \eta+\eta \delta_{-p,-q}=0$, one mapping $\tilde{I}_{(2)}^{3,1},\tilde{I}_{(2)}^{1,3}$ to $\tilde{I}_{(2)}^{2,0},\tilde{I}_{(2)}^{0,2}$ and another from $\tilde{I}_{(2)}^{2,2}$ to $\tilde{I}_{(2)}^{1,1}$. Taking $\delta_1,\delta_2  \in \mathbb{R}$ as proportionality constants for these two maps we find
\begin{equation}
\delta = \scalebox{0.7}{$ \begin{pmatrix}
 0 & 0 & 0 & 0 & 0 & 0 \\
 0 & 0 & 0 & 0 & 0 & 0 \\
 0 & 0 & 0 & 0 & 0 & 0 \\
 \delta_1 & 0 & 0 & 0 & 0 & 0 \\
 0 & \delta_1 & 0 & 0 & 0 & 0 \\
 0 & 0 & \delta_2& 0 & 0 & 0 \\
\end{pmatrix}\, .$}
\end{equation}
Note that this matrix is precisely of the same form as the log-monodromies $N_1,N_2$, so we can use coordinate shifts \eqref{eq:shift} to set $\delta_1=\delta_2=0$.

\subsection{Coni-LCS class boundaries}\label{app:coniLCSdata}
Finally we consider coni-LCS class boundaries, which consists of the 2-cubes $\langle \mathrm{I}_1 | \mathrm{IV}_2 | \mathrm{IV}_1 \rangle$ and $\langle \mathrm{I}_1 | \mathrm{IV}_2 | \mathrm{IV}_2 \rangle$. These boundaries share a $\mathrm{IV}_2$ singularity type at the intersection $y^1=y^2=\infty$. Let us begin by writing down a basis for the vector spaces $\tilde{I}_{(2)}^{p,q}$ of this sl$(2,\mathbb{R})$-split Deligne splitting as
\begin{equation}
\begin{aligned}
\tilde{I}_{(2)}^{3,3}&: \ \big( 1, \  0,\ 0,\ 0,\ 0,\ 0 \big) , \quad  \tilde{I}_{(2)}^{0,0}: \ \big( 0, \  0,\ 0,\ 1,\ 0,\ 0 \big) .\\
\tilde{I}_{(2)}^{2,2}&: \ \big( 0, \  1,\ 0,\ 0,\ 0,\ 0 \big) , \ \  \big( 0, \  0,\ 1,\ 0,\ 0,\ 0 \big) , \\
\tilde{I}_{(2)}^{1,1}&: \ \big( 0, \  0,\ 0,\ 0,\ 1,\ 0 \big) , \ \  \big( 0, \  0,\ 0,\ 0,\ 0,\ 1 \big) , \\
\end{aligned}
\end{equation}

\subsubsection*{Enhancement step $\text{I}_1 \to \text{IV}_2$}
Here we construct the log-monodromy matrices for the 2-cubes $\langle \mathrm{I}_1 | \mathrm{IV}_2 | \mathrm{IV}_1 \rangle$ and $\langle \mathrm{I}_1 | \mathrm{IV}_2 | \mathrm{IV}_2 \rangle$. Let us begin by writing down the sl$(2,\mathbb{R})$-triples as
\begin{small}
\begin{equation}
N_1 = N_1^- = \scalebox{0.7}{$\begin{pmatrix}
 0 & 0 & 0 & 0 & 0 & 0 \\
 0 & 0 & 0 & 0 & 0 & 0 \\
 0 & 0 & 0 & 0 & 0 & 0 \\
 0 & 0 & 0 & 0 & 0 & 0 \\
 0 & -1 & 0 & 0 & 0 & 0 \\
 0 & 0 & 0 & 0 & 0 & 0 
\end{pmatrix}$},  \quad Y_1 = \scalebox{0.7}{$\begin{pmatrix}
 0 & 0 & 0 & 0 & 0 & 0 \\
 0 & 1 & 0 & 0 & 0 & 0 \\
 0 & 0 & 0 & 0 & 0 & 0 \\
 0 & 0 & 0 & 0 & 0 & 0 \\
 0 & 0 & 0 & 0 & -1 & 0 \\
 0 & 0 & 0 & 0 & 0 & 0 
\end{pmatrix}$} ,
\end{equation}
\end{small}
\begin{equation}
\quad N_2^- = \scalebox{0.7}{$\begin{pmatrix}
 0 & 0 & 0 & 0 & 0 & 0 \\
 0 & 0 & 0 & 0 & 0 & 0 \\
 1 & 0 & 0 & 0 & 0 & 0 \\
 0 & 0 & 0 & 0 & 0 & -1 \\
 0 & 0 & 0 & 0 & 0 & 0 \\
 0 & 0 & 1 & 0 & 0 & 0 
\end{pmatrix}$}, \quad Y_2 = \scalebox{0.7}{$\begin{pmatrix}
 3 & 0 & 0 & 0 & 0 & 0 \\
 0 & 0 & 0 & 0 & 0 & 0 \\
 0 & 0 & 1 & 0 & 0 & 0 \\
 0 & 0 & 0 & -3 & 0 & 0 \\
 0 & 0 & 0 & 0 & 0 & 0 \\
 0 & 0 & 0 & 0 & 0 & -1 
\end{pmatrix}.$}
\end{equation}
The signs of the coefficients of the lowering operators $N^-_1,N^-_2$ are fixed by the polarization condition \eqref{eq:pol}. We must require $\eta N^-_1$ and $\eta (N^-_2)^3$ both to have one negative eigenvalue. In turn the condition that $N_2$ is an infinitesimal isomorphic of the symplectic product $(N^-_2)^T \eta + \eta N_2^-=0$ fixes the sign of the other two coefficients of $N_2$.

Next we construct the most general log-monodromy matrix $N_2$ compatible with the above boundary data. There is only one map with weight $\ell\leq -2$ under the adjoint action of $Y_1$, which is $N_1$. By using \eqref{lowering} we therefore find
\begin{equation}
N_2 = N_2^- + n N_1  \, ,
\label{NilpIV2}
\end{equation}
where polarization conditions require $n\geq 0$. For $n=0$ we encounter a $\mathrm{IV}_1$ divisor at $y^2=\infty$, while for $n>0$ it is a $\mathrm{IV}_2$ divisor.

\subsubsection*{Construction of the phase operator}
Based on the above data let us construct the most general phase operator $\delta$. For the given Deligne splitting there are four real maps $\delta_{-p,-q}$ with $p,q\geq 0$ we can write down that are infinitesimal isometries of $\langle \cdot, \cdot \rangle$. Taking their linear combination gives us
\begin{equation}
\delta =  \scalebox{0.7}{$\begin{pmatrix} 0 & 0 & 0 & 0 & 0 & 0 \\
 0 & 0 & 0 & 0 & 0 & 0 \\
 \delta_4 & 0 & 0 & 0 & 0 & 0 \\
 \delta_2 & \delta_1 & 0& 0 & 0 & -\delta_4 \\
 \delta_1 & \delta_3 & 0 & 0 & 0 & 0 \\
 0 & 0 & \delta_4 & 0 & 0 & 0 
\end{pmatrix}$}.
\end{equation}
The components related to the coefficients $\delta_3,\delta_4$ are proportional to $N_1$ and $N_2$, so these can be tuned by using \eqref{eq:shift}.

\section{Embedding periods for geometrical examples}\label{app:embedding}
In this appendix we show how the periods constructed in our work relate to some familiar geometrical examples. We rewrite the periods of the one-modulus $\mathrm{I}_1$ boundary and two-modulus coni-LCS class boundaries in terms of the prepotential formulation of the conifold and coni-LCS periods. Furthermore we show how the periods found for the 2-cube $\langle \mathrm{II}_0 | \mathrm{II}_1 | \mathrm{II}_1 \rangle$ cover the periods for the Calabi-Yau threefold in $\mathbb{P}_4^{1,1,2,2,6}[12]$ near a particular degeneration.

\subsection{Conifold point}
We begin by rewriting the periods \eqref{eq:I1periods} of $\mathrm{I}_1$ boundaries in terms of the prepotential formulation in e.g.~\cite{PhysRevLett.62.1956,Candelas:1989js,Strominger_1995}. In this frame the periods can be written as $\Pi=(X^0,X^1,\cF_0,\cF_1)$, where the $\cF_i = \partial_{X^i}\cF$ are obtained by taking derivatives of the prepotential $\cF(X^i)$. In order to bring our periods into this frame, one has to perform K\"ahler transformations and basis changes. We typically set $X^0=1$, so let us first rescale the periods by an overall factor $\Pi \to e^f \Pi$ with $f=1-\frac{a^2 z^2}{8\pi}$ to set the first entry equal to one. Next we want to set the second entry equal to the special coordinate $X^1 =z$, so we also apply a symplectic basis transformation $M=\text{diag}(1,1/a,1,a)$. Consequently the transformed period vector reads
\begin{equation}
\Pi = \big(1, \ z, \ i-\frac{ia^2}{4\pi}z^2, \ \frac{ia^2}{2\pi} z\log[z] \big),
\end{equation}
up to corrections in $z^3$. One can straightforwardly verify that these periods indeed match with the prepotential
\begin{equation}\label{eq:I1prepotential}
\cF= \frac{i}{2}(X^0)^2 +\frac{i a^2}{4\pi} (X^1)^2 \log\big[X^1/X^0\big] - \frac{i a^2}{8\pi} (X^1)^2 \, ,
\end{equation}
where afterwards we can set $X^0=1$ and $X^1=z=e^{2\pi i t}$. Note in particular that $a^2>0$ now fixes the sign of the second term in this prepotential.

\subsection{Coni-LCS point}\label{app:coniLCS}
Next we rewrite the periods \eqref{eq:coniLCSperiods} near coni-LCS boundaries in the prepotential formulation, see e.g.~\cite{Demirtas:2020ffz,Blumenhagen:2020ire} for recent constructions using different methods. Again we want to set one period proportional to the conifold modulus as $X^1=z_1$, so let perform a symplectic basis transformation $M = \text{diag}(1,1/a,1,1,a,1)$ analogous to the $\mathrm{I}_1$ boundaries. The periods then read
\begin{align}
\Pi = \big(&1, \ z_1, \ \frac{ \log[z_2]}{2 \pi i}, \ -\frac{i \log[z_2]^3}{48 \pi^3}-\frac{ i a^2 n z_1^2 \log[z_2]}{4\pi }+   \frac{a^2}{4  \pi i} z_1^2+i \delta_2 +i \delta_1 a z_1, \nn \\
&- a^2 z_1\frac{ \log[z_1] + n \log[z_2]}{2 \pi i} +i\delta_1 a, \ - \frac{\log[z_2]^2}{8 \pi^2}  -\frac{1}{2} a^2 n z_1^2\big)\, .
\end{align}
Equivalently these periods can be obtained from the prepotential
\begin{equation}
\cF = \frac{1}{6} \frac{\cK_{ijk} X^i X^j X^k}{X^0} - \frac{1}{2}A_{ij} X^i X^j+B_i X^0 X^i+C(X^0)^2 + D (X^1)^2 \log[X^1/X^0] \, ,   
\end{equation}
where we set $X^0=1$, $X^1=z_1 $ and $X^2=\frac{ \log[z_2]}{2 \pi i}$ afterwards. The coefficients in the periods are then related by
\begin{equation}
 \cK_{112}=-a^2n\, , \ \ \cK_{222}=1\, , \ \ A_{11} =\frac{ia^2}{4\pi}\, , \ \ B_1= i\delta_1 a \, , \ \ C=\frac{i\delta_2}{2} \, , \ \ D=\frac{ia^2}{4\pi}\, ,
\end{equation}
and all other coefficients vanish. Note that the sign of the imaginary piece of $D$ is again fixed by $a^2>0$, similar to the $\mathrm{I}_1$ boundary.

\subsection{Degeneration for the Calabi-Yau threefold in $\mathbb{P}_4^{1,1,2,2,6}[12]$}\label{app:Seiberg-Witten}
In this appendix we illustrate how the period vector near the Seiberg-Witten point of the K3-fibered Calabi-Yau threefold in $\mathbb{P}_4^{1,1,2,2,6}[12]$ can be embedded into our models. This geometry has been studied in detail in the literature, see e.g.~\cite{Hosono:1993qy, Candelas_1994, Kachru:1995fv, Curio:2000sc,Lee:2019wij}, and we will follow the analysis of the periods of \cite{Lee:2019wij} here. In our models this Seiberg-Witten point corresponds to the $\langle \mathrm{II}_0 | \mathrm{II}_1 | \mathrm{II}_1 \rangle$ boundary, whose periods have been constructed section \ref{ssec:II1construction}. The period vector as computed from the relevant Picard-Fuchs equations takes the form
\begin{align}
\Pi_{\rm P12}=\frac{1}{\pi} \Big(&1+ \frac{5}{36} z_1, \ z_1,\ - \sqrt{z_1}, \ \frac{i}{\pi} (\log[z_2]-6 \log[2]+7) \sqrt{z_1}  \\
&\frac{i}{2 \pi }  (5 + 2 \log[z_1]+\log[z_2] )(1+\frac{5}{36} z_1), \ \frac{i}{2 \pi }(1+2 \log[z_1] + \log[z_2] )z_1 \Big). \nn
\end{align}
One observation that we can immediately make is that the period vector depends on square roots of the coordinates, which results in monodromy transformations that are only quasi-unipotent. We remedy this by an appropriate coordinate transformation further below. As generic solutions to the Picard-Fuchs equations, these periods are not in a symplectic basis. To find the appropriate basis transformation one uses the fact that such a basis can naturally be found at the LCS point and then by analytic continuation one can compute the transition matrix. The latter is also given in \cite{Lee:2019wij}. Furthermore, we require another basis transformation to bring the periods into the symplectic basis we use in this work. The combined transition matrix is given by 
\begin{align}
M_{\rm P12}=\tfrac{\sqrt{2}}{X} \scalebox{0.75}{$ \begin{pmatrix}
1 &- \frac{5}{36}+ X^2 & 0 & 0 & 0 & 0 \\
i &  - i  (\frac{5}{36 }+X^2) & 0 & 0 & 0 & 0 \\
0 & 0 & X & 0 & 0 & 0 \\
\frac{5 i}{8 \pi } & \frac{5i (36 X^2-1)}{288 \pi }&0 &0 & \frac{-1}{4}& \frac{1}{144}(5-36 X^2) \\
-\frac{5}{8 \pi} & \frac{5(36X^2 +1)}{288 \pi} & 0 & 0 & - \frac{i}{4} & \frac{i}{144}(5 + 36 X^2) \\
0 & 0 & \frac{3 i X}{2 \pi}(\log[4] -1)& -\frac{X}{2} & 0 & 0
\end{pmatrix}$},
\end{align}
where $X=\frac{\Gamma(3/4)^4}{\sqrt{3} \pi^2}$. In addition, we perform the divisor preserving coordinate redefinition
\begin{align}
 z_1 \to z_1^2 \exp \Big( \frac{5}{16}z_2+\frac{131}{2048}z_2^2 \Big) \, ,  \qquad z_2 \to z_2 \exp \Big(-\frac{5}{8}z_2-\frac{131}{1024}z_2^2 \Big) \,,
\end{align}
that, among other things, makes the monodromies unipotent. After these transformations the period vector takes the form
\begin{small}
\begin{align}
\Pi_{\rm P12}=\Big(&1 + X^2 z_1^2, \ i - i  X^2 z_1^2 , \ -X z_1, \nn \\
& - \tfrac{i}{8 \pi} (4 \log[z_1] + \log[z_2]) + \tfrac{iX^2}{8 \pi} (4- 4 \log[z_1]-\log[z_2])  z_1^2 \, ,   \label{eq:P12Period} \\
 &\frac{1}{8 \pi} (4 \log[z_1] + \log[z_2]) + \tfrac{X^2}{8 \pi} (4- 4 \log[z_1]-\log[z_2])  z_1^2, \ -\tfrac{iX}{2 \pi} (4+\log[z_2] )z_1  \Big) \, . \nn 
\end{align}
\end{small}
Our model \eqref{eq:II1periods} reproduces the periods given in \eqref{eq:P12Period} upon identifying
\begin{equation}
n_1=0\, , \qquad n_2 = 1/4 \, , \qquad b = -4 X \, , \qquad a=c =0\, .
\end{equation}

\end{subappendices}



\setpartpreamble[u][\textwidth]{
\vspace*{1cm}
\hrulefill 
\vspace*{0.5cm}

In this third part we discuss two applications of asymptotic Hodge theory in string compactifications. In chapter \ref{chap:WGC} we investigate the bounds put by the WGC in Type IIB Calabi-Yau compactifications. We focus on the extremality region of electric BPS black holes, which is formed by an ellipsoid. We compute the values of its radii in strict asymptotic regimes, thereby giving us a non-trivial order one coefficient for the WGC in infinite distance limits. Chapter \ref{chap:modstab} studies moduli stabilization near boundaries in complex structure moduli space. We use the self-duality condition to set up a systematic procedure for finding flux vacua using the sl(2)-structures. On the other hand, for the F-terms we explain how to include essential exponential corrections in the extremization conditions. We also present a method for finding vacua with a small flux superpotential, including some explicit examples.

\vspace*{0.5cm}
\hrulefill }

\part{Applications in String Compactifications}\label{part3}
     

\chapter{Weak Gravity Conjecture in Type IIB Calabi-Yau compactifications}\label{chap:WGC}

The main focus of this chapter will be the WGC \eqref{eq:WGC} in Type IIB Calabi-Yau compactifications. To be precise, we want to determine the order one coefficients set by the extremality bound of black holes. Recall from section \ref{ssec:IIBN=2} that in this setting there are multiple U(1) gauge fields present, which arise from expanding the R-R four-form potential $C_4$ along harmonic three-forms of the Calabi-Yau threefold as described by \eqref{eq:C4expand}. Instead of requiring the existence of a single superextremal particle, the WGC then has to be satisfied for every direction in the charge lattice. In \cite{Cheung:2014vva} this observation was formalized into the statement that there should exist a set of electrically charged particles whose charge-to-mass vectors span a convex hull that contains the black hole extremality region. The BPS states charged under the R-R gauge fields arise from D3-branes wrapped on three-cycles of the Calabi-Yau manifold. It is then natural to pose the question whether one can in fact identify these D3-brane states, and moreover if charge lattice sites populated by BPS states suffice or if non-BPS states are also necessary in order to satisfy the convex hull condition. In this chapter we set ourselves a more modest goal, and we merely aim to make the bounds put by the extremality region of electrically charged BPS black holes as precise as possible.\footnote{For an enlightening discussion on subtleties regarding the charge-to-mass ratio of BPS states and the black hole extremality bound we refer the reader to \cite{Alim:2021vhs}.}

As a first step in approaching WGC bounds we single out a special set of candidate BPS states that are elementary with respect to the asymptotic sl$(2,\mathbb{R})^n$-structure, i.e.~they sit in a single eigenspace under the sl(2)-decomposition. If in addition these particles couple to the graviphoton asymptotically, we find that their asymptotic charge-to-mass ratio is given by
\begin{equation} \label{intro-Q/M}
 \lim_{\lambda \to \infty} \bigg( \frac{Q}{M} \bigg)^{-2} \bigg|_{q_{\rm G}}= 2^{1-d_n}  \prod_{i=1}^n {\Delta d_i\choose{(\Delta d_i -  \ell_i)/2}} \times \begin{cases}
1 \text{ for $d_n = 3$}\, ,\\
\frac{1}{2} \text{ for $d_n \neq 3$}\, . 
\end{cases}\, 
\end{equation}
In this formula the $d_i$ and $\ell_i$ correspond to discrete data characterizing the candidate BPS state and the type of limit: recall that $\Delta d_i = d_i - d_{i-1}$ specify the weights of the leading term $a_0$ of the periods (see \eqref{eq:a0position}) and similarly $ \ell_i$ the weights of the charge $q_{\rm G}$ under the $N_i^0$. The notation $q_{\rm G}$ stands for sl(2)-elementary charges that couple asymptotically to the graviphoton. Our results significantly extend the recent formula of \cite{Gendler:2020dfp}, which was derived using asymptotic Hodge theory for a large class of infinite distance limits. While the two formulas look rather different we find that they agree in most cases, with some particular exceptions where our formula contains additional terms.

The formula \eqref{intro-Q/M} will be essential in establishing actual bounds on the charge-to-mass spectrum of electric BPS states in four-dimensional $\mathcal{N}=2$ supergravities in the asymptotic regime. In order to do this we elaborate on a result of \cite{Gendler:2020dfp} that the charge-to-mass vectors of electric BPS states lie on an ellipsoid with two non-degenerate directions $\gamma_1,\gamma_2$. We will compute the asymptotic values of these radii using the above general formula for all limits in complex structure moduli space, both at finite and infinite distance. We find that there are only three possible sets of values for these radii corresponding to the three ellipsoids depicted in figure \ref{fig:ellipsoids}. We note that for finite distance singularities $\gamma_2^{-2}=0$, so another direction of the ellipsoid degenerates, and only a single non-degenerate direction remains. Besides specifying a structure for the charge-to-mass spectrum, the smallest radius of this ellipsoid also serves as a lower bound on the charge-to-mass ratio for electric BPS states. 
Inserting the numerical values for the radii, we find that for infinite distance singularities the asymptotic charge-to-mass ratio of electric BPS states is bounded from below by $2/\sqrt{3}$. 

\begin{figure}[h!]
\vspace*{-.2cm}
\centering
\hspace{-1.5cm} \subfloat[$\gamma_1=2/\sqrt{3},\gamma_2=2$]{\label{fig:ellipsoid1}\hspace{0.5cm} \includegraphics[height=2cm]{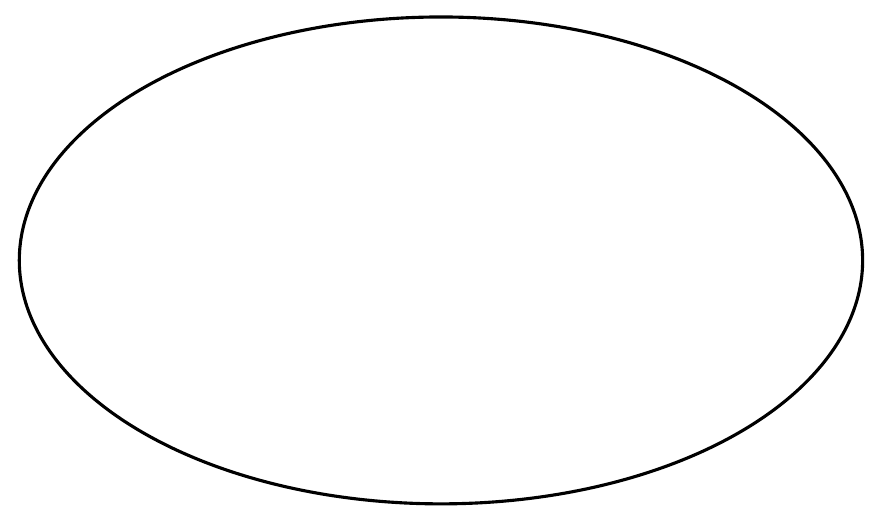}  \hspace{0.5cm}}
\hspace{0.2cm}
\subfloat[$\gamma^{-2}_1=\gamma^{-2}_2=1/2$]{\label{fig:ellipsoid2}\hspace{.3cm}\includegraphics[height=3cm]{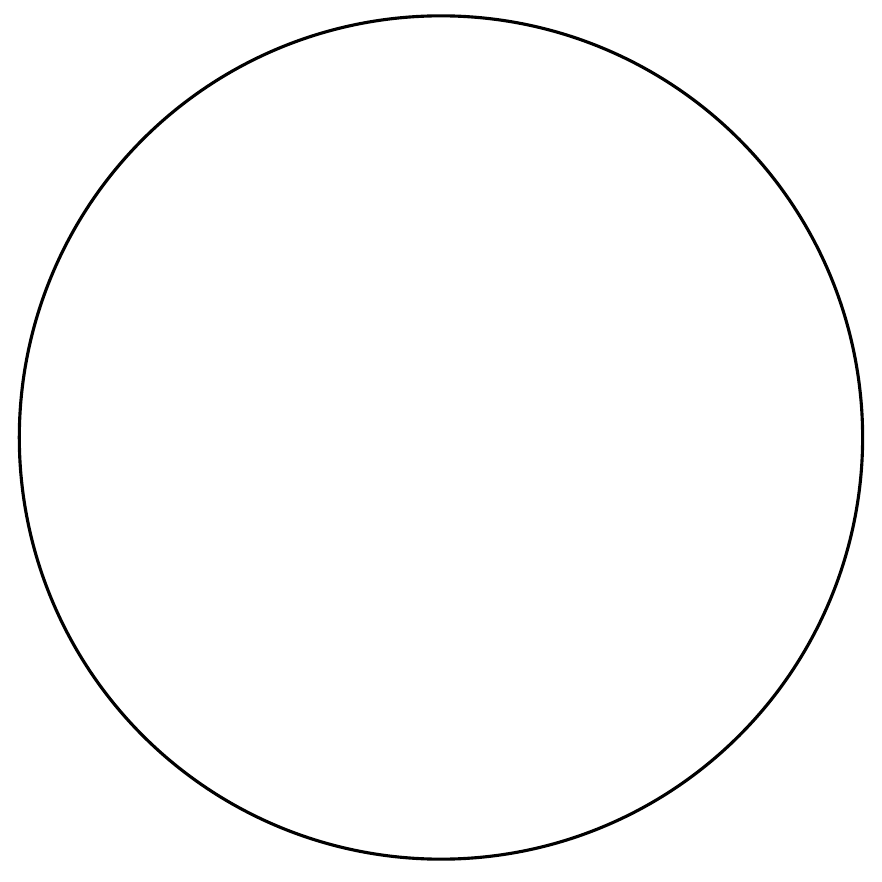}}
\hspace{0.5cm}
\subfloat[$\gamma^{2}_1=1,\gamma^{-2}_2=0$]{\label{fig:ellipsoid3}\hspace{0.2cm}\includegraphics[height=1.9cm]{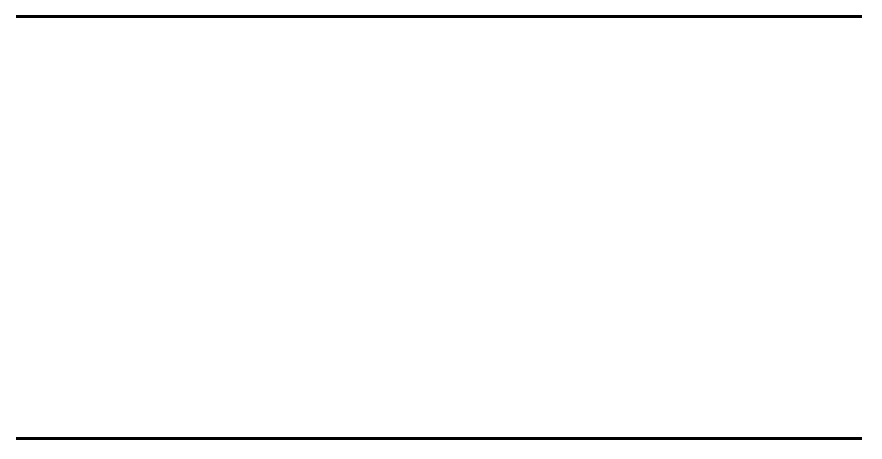}}
\caption{The three possible asymptotic shapes of the ellipsoid that forms the extremality region of electric BPS black holes. Figures \ref{fig:ellipsoid1} and \ref{fig:ellipsoid2} occur for asymptotic regions at infinite distance, whereas figure \ref{fig:ellipsoid3} occurs for finite distance limits. }\label{fig:ellipsoids}
\end{figure}

Having obtained a numerical bound for the charge-to-mass ratio, it is natural to investigate how this relates to order one coefficients in other swampland conjectures. In that sense, we are able to determine asymptotic values for the relevant order-one coefficients appearing in the asymptotic de Sitter \cite{Obied:2018sgi,Ooguri:2018wrx,Garg:2018reu} and Swampland Distance Conjectures \cite{Ooguri:2006in,Klaewer:2016kiy}. A relation with the de Sitter conjecture is established by making particular flux choices and rewriting the flux potential such that the order-one coefficient of the de Sitter conjecture can be evaluated by using the above general formula \eqref{intro-Q/M}. We find bounds that are known from the literature (see \cite{Andriot:2020lea,Lanza:2020qmt} and references therein) depending on which contributions to the flux potential are taken into account, i.e.~we separate the contributions coming from the axio-dilaton, complex structure and tree-level K\"ahler moduli. In particular, we are able to saturate the recently proposed Trans-Planckian Censorship Conjecture bound \cite{Bedroya:2019snp} $c \geq \sqrt{2/3}$, when we only consider complex structure moduli for infinite distance singularities. Furthermore, we use the relation of the Weak Gravity Conjecture with the Swampland Distance Conjecture suggested in \cite{Palti:2017elp,Grimm:2018ohb,Lee:2018spm, Gendler:2020dfp}. This allows us to match the order-one coefficients as outlined in \cite{Lee:2018spm, Gendler:2020dfp}. We find agreement with results from the literature \cite{Grimm:2018ohb, Andriot:2020lea,Gendler:2020dfp} and in particular with the lowest value of $\lambda = 1/\sqrt{6}$, which matches with the recently proposed relation $\lambda = c/2$ \cite{Andriot:2020lea}.

The chapter is structured as follows. In section \ref{sec:Review} we discuss the charge-to-mass spectrum of BPS states in 4d $\mathcal{N}=2$ supergravity theories, and we look at two examples to gain some intuition for the question at hand. Next we put the full machinery of asymptotic Hodge theory to use in section \ref{sec:generalanalysis} to perform a general analysis of the charge-to-mass spectrum at asymptotic regions in complex structure moduli space. Finally, in section \ref{sec:remarks} we discuss connections between the Weak Gravity Conjecture and the de Sitter and Distance Conjectures, and how bounds obtained for the former relate to those of the latter two.

\section{Charge-to-mass ratios and limits in moduli space}
\label{sec:Review}
In this section we discuss the structure of the charge-to-mass spectrum of BPS states in four-dimensional $\mathcal{N}=2$ supergravity theories. We outline a remarkable observation made by the authors of \cite{Gendler:2020dfp}, namely that the so-called charge-to-mass vectors (defined in \ref{CMVect}) of electric BPS states of these theories lie on a degenerate ellipsoid with exactly two finite radii. Furthermore, we provide two examples where these radii can be computed explicitly as the prepotential is known. Studying these examples gives a feeling for the task at hand and serves as a precursor to section \ref{sec:generalanalysis}, where we perform a general analysis that holds for any type of singularity in complex structure moduli space and does not rely on a prepotential formulation of the underlying supergravity theory.

\subsection{Charge-to-mass spectrum of BPS states}
\label{ssec:examples}
First let us clarify what we mean by charge-to-mass vectors. In our setting we compute the physical charge of a BPS state via \eqref{eq:charge}. For electric states $\mathbf{q}=(0, q_I)$ this physical charge then follows from the right-bottom block of the matrix $\cM$ in \eqref{def-cM}. In order to determine the individual electric charges of this state, we have to decompose the matrix $\cI_{IJ}$ that appears in this expression. By
introducing a symmetric matrix $G$ such that 
\beq
    - 2\, \cI_{IJ}=G_{I}^{K} \delta_{KL} G^{L}_{J}\ ,
\eeq  
we then define the charge-to-mass vectors as
\begin{equation} 
    \mathfrak{z}_I = \frac{|Q|}{M} \hat Q_I\, , \label{CMVect}
\end{equation}
where $Q_I=(G^{-1})^{J}_{I} q_J$ and $\hat Q_I$ denotes the unit vector in this direction. 

The interest for these charge-to-mass vectors $ \mathfrak{z}_I$ stems from the electric Weak Gravity Conjecture. 
In the case of considering only a single gauge field,  
this conjecture states that there should always exist an electric state whose charge-to-mass ratio is larger than the black hole extremality bound. 
In a setting with multiple gauge fields, e.g.~for the general theory that we consider in \eqref{eq:action}, one has to consider the ratio between multiple electric charges and the mass of a state. In \cite{Cheung:2014vva} a generalized version of the Weak Gravity Conjecture with multiple gauge fields is proposed. Motivated by black hole remnant arguments the authors formulate a convex hull condition, which states that there should exist a set of particles such that the convex hull spanned by their charge-to-mass vectors contains the black hole extremality region. 

It is a non-trivial task to obtain the form of the extremality region in a general 4d $\mathcal{N}=2$ supergravity. A first 
step is to restrict to electric BPS black holes and determine the charge-to-mass spectrum of electric BPS states. In this case the defining equation for the shape of the charge-to-mass spectrum is the BPS condition $|Z|^2=M^2$, which can be written as
\begin{equation}
e^K \frac{(\mathbf{q}^T \eta \mathbf{\Pi} )(\mathbf{\bar \Pi}^T \eta \mathbf{q}) }{M^2}=1\, .
\end{equation}
Specializing to electric states $\mathbf{q}=(0,q_I)$ and using \eqref{CMVect}, we can express the quantized 
charge vectors $q_I$ in terms of the charge-to-mass vectors $\mathfrak{z}_I$ via $q_I=M G^J_I \mathfrak{z}_J$. The BPS condition then tells us that the charge-to-mass vectors obey
\begin{equation}\label{eq:spectrumequation}
e^K \big( \mathfrak{z}_J G_I^J X^I \big) \big(\bar{X}^K  G_K^L \mathfrak{z}_L \big) = 1\, .
\end{equation}
where we plugged in the periods as \eqref{eq:PeriodVector}. This condition can now be interpreted as a matrix equation for the charge-to-mass vectors. It can be written as
\begin{equation}
\mathfrak{z}_I \cA^{IJ} \mathfrak{z}_J = 1\, ,
\end{equation}
with the matrix $\cA$ given by
\begin{equation}\label{eq:radiimatrix}
\cA^{IJ} =   e^K  G_K^I  \bigg(\Re X^K\, \Re X^L +\Im X^K\, \Im X^L \bigg) G_L^J \, .
\end{equation}
This tells us that the eigenvalues of this matrix specify the shape of the charge-to-mass spectrum. Looking at the form of $\cA$, we notice that this matrix has at most two non-zero eigenvalues, since there are only two independent vectors in its image. Denoting these eigenvalues by $\gamma_1^{-2}$ and $\gamma_2^{-2}$, and expanding the charge-to-mass vectors in terms of an eigenbasis for $\cA$
writing $ \mathfrak{\tilde z}_I$, we can write the BPS condition as
\begin{equation} \label{ellip1}
\gamma_1^{-2}  \mathfrak{\tilde z}_1^2 + \gamma_2^{-2} \mathfrak{ \tilde z}_2^2 =1\, ,
\end{equation}
where the components $\mathfrak{\tilde z}_i$ with $i\neq 1,2$ are unconstrained. Viewed as an equation constraining $\mathfrak{\tilde z}_I$ the condition \eqref{ellip1} describes an ellipsoid with two non-degenerate directions if $\gamma_1,\gamma_2 < \infty$. We can determine its radii $\gamma_{1,2}$ by computing the eigenvalues of the matrix $\cA$. This problem reduces to finding the eigenvalues of a $2\times 2$ matrix, by noting that only linear combinations of $ G_K^I  \Re X^K$ and $ G_K^I \Im X^K$ can be eigenvectors of $\cA$ with non-zero eigenvalues. Its eigenvalues are given by
\begin{equation}\label{eq:radiigeneral}
\gamma^{-2}_{1,2} = \frac{e^K}{R^2 I^2}  \bigg( (R^2+I^2)(R^2 I^2+P^2) \pm \sqrt{\Delta} \bigg) \, ,
\end{equation}
with
\begin{equation}
\Delta= 2R^2 I^2 \big(I^4+6R^2 I^2+R^4\big) P^2 +\big( R^4 I^4 + P^4 \big) (R^2- I^2)^2  \, ,
\end{equation}
and where we used short-hands
\begin{small}
\begin{equation}
R^2=-  \cI_{IJ} \Re X^I  \Re X^J \, ,\ \ \ I^2=- \cI_{IJ}   \Im X^I  \Im X^J\, , \ \ \ P=  \cI_{IJ} \Re X^I \Im X^J\, .
\end{equation}
\end{small}
In general, we have that $P \neq 0$. In order to set $P$ to zero one has to rescale\footnote{To be precise, this rescaling is given by $f=-(i/2) \arctan(2P/(I^2-R^2)) $. This can be verified by checking how this rescaling acts on $\cI_{IJ} X^I X^J = I^2-R^2+2i P$, since it cancels out its complex phase and therefore sets the imaginary part to zero. Let us point out that this $f$ need not be holomorphic for our purposes, since the structure of the charge-to-mass spectrum does not involve derivatives. In particular the rescaling that sets $P$ to zero is generically not a holomorphic rescaling, and is therefore not a K\"ahler transformation.} the period vector by $\mathbf{\Pi} \to e^f \mathbf{\Pi}$, which maps $X^I \to e^f X^I$. Note in particular that the defining equation for the shape of the charge-to-mass spectrum \eqref{eq:spectrumequation} is invariant under such rescalings, and therefore so are the formulas for the radii given in \eqref{eq:radiigeneral}. By using this rescaling to set $P=0$, the expressions for the radii \eqref{eq:radiigeneral} reduce to
\begin{equation}
\label{eq:radiiperiod}
\begin{aligned}
\gamma_1^{-2} &= -2e^K  \cI_{IJ} \, \Re X^I \Re X^J \, , \quad &\gamma_2^{-2} &= -2e^K  \cI_{IJ} \, \Im X^I  \Im X^J \, .   \\
\end{aligned}
\end{equation}
This provides us with a simple way to compute the radii of the charge-to-mass spectrum from the supergravity data, i.e.~the K\"ahler potential $K$, the gauge kinetic functions $\cI_{IJ}$ and the periods $X^I$. Note in particular that the general $\cN=2$ special geometry identity $- 2 e^K  \cI_{IJ} \,  X^I \bar X^J  = 1$ implies 
\beq \label{gamma_constr}
   \gamma_1^{-2}  +  \gamma_2^{-2} =1\ . 
\eeq
In other words the ellipsoid \eqref{ellip1} is not general but restricted by the $\cN=2$ condition \eqref{gamma_constr}.\footnote{Recently in \cite{Gonzalo:2018guu,Aalsma:2019ryi,Palti:2020qlc,Andriolo:2020lul,Loges:2020trf} supersymmetry and duality groups have been used as guiding principles in studying the Swampland.}

Let us close this subsection with two remarks. 
Firstly, the symplectic frame used to formulate this data is not necessarily the frame in which a prepotential formulation exists. Namely, we want to choose a symplectic frame in which we obtain a weakly-coupled description for the $U(1)$ gauge fields. In other words, we pick our electric charges based on the behavior of the physical charge \eqref{eq:charge}, since a small physical charge for electrically charged states indicates that the gauge kinetic functions $\cI_{IJ}$ in the action \eqref{eq:action} are large. For now we assume that this choice of electric charges or symplectic frame has already been made for us. How to make this choice will be discussed in more detail in section \ref{ssec:bounds}, where we make use of an alternative manner to compute these radii that follows from \eqref{eq:radiiperiod}, namely via the charge-to-mass ratios of a particular set of electric states \eqref{eq:asymptoticradii}.

\paragraph{BPS states with generic charge.} Secondly, we can also study the charge-to-mass spectrum of BPS states with generic charge. In order to define their charge-to-mass vectors we have to decompose the matrix $\cM$ that appears in the physical charge \eqref{eq:charge} via a symmetric matrix $G$ as $2\cM^{-1} = - Z Z^T$. The charge-to-mass vectors are then given by $\mathbf{z}=M Z^T \mathbf{q}$.\footnote{Note that we do not recover the electric charge-to-mass vectors $(0,\mathbf{z}_E)$ via $M Z^T (0,\mathbf{q}_E)$ due to the off-diagonal components of $\cM$. Namely, we find that $ Z^T (0,\mathbf{q}_E) \neq (0,  G^T \mathbf{q}_E)$ since application of $G^T$ on $(0,\mathbf{q}_E)$ generally results in a non-vanishing piece in the first component.} Following similar steps as in the analysis of electric states, the BPS condition can be rewritten as $\mathbf{z}^T \mathcal{Z} \mathbf{z} = 1$, 
where the matrix $\mathcal{Z}$ is given by
\begin{equation}\label{eq:radiimatrixgeneral}
\mathcal{Z} =   e^K \, Z \eta \Big(\Re \mathbf{\Pi}\  \Re \mathbf{\Pi}^T+\Im \mathbf{\Pi}\  \Im\mathbf{\Pi}^T \Big)\eta Z^T \, .
\end{equation}
Again we find a matrix with a two-dimensional image, in this case spanned by the vectors $Z\eta\Re \mathbf{\Pi}$ and $Z\eta\Im \mathbf{\Pi}$, so $\mathcal{Z}$ has only two non-vanishing eigenvalues. Let us denote these eigenvalues by $r_1^{-2}$ and $r_2^{-2}$ to avoid confusion with the eigenvalues $\gamma_1^{-2}$ and $\gamma_2^{-2}$ that were found for the electric charge-to-mass spectrum. We then obtain a similar relation for the charge-to-mass vectors by expanding in terms of an eigenbasis for $\mathcal{Z}$ as
\begin{equation}
r_1^{-2} \tilde z_1^2 + r_2^{-2} \tilde z_2^2 =1\, ,
\end{equation}
where the components $\tilde z_{\alpha}$ with $\alpha\neq 1,2$ are unconstrained. This means we are again dealing with an ellipsoid with two non-degenerate directions. We can determine the radii by computing the eigenvalues of the $2\times 2$ matrix by projecting onto the subspace spanned by $G \eta \Re \mathbf{\Pi}$ and $G \eta \Im \mathbf{\Pi}$. After some slightly involved computations, we find as radii $r_1 = r_2=1$. So we find that the ellipse forms a circle with unit radius at any point in moduli space. Note that this puts $Q/M \geq 1$ as lower bound on the charge-to-mass ratio of any BPS state, which is also expected from \eqref{eq:N=2identity}.

\subsection{Examples}
\label{ssec:examples2}
As promised, we now turn to two examples where we calculate the \eqref{eq:radiiperiod} explicitly using the known prepotential formulations. We will highlight some of the ingredients that will play a central role in the more sophisticated general analysis of section \ref{sec:generalanalysis}. Let us stress that the general approach is also essential to draw conclusions when a prepotential is hard to determine or unavailable. 

\subsubsection{Example 1: conifold point}
\label{ssec:conifold}
Here we study the behavior of the charge-to-mass spectrum for an example of a finite distance singularity, namely the one-modulus conifold point. Such a singularity is realized in e.g.~the complex structure moduli space of the quintic \cite{Candelas:1990rm}. Other than for the obvious reason, which is the knowledge of the prepotential, we chose this example because in \cite{Gendler:2020dfp} only infinite distance singularities were treated. We show that the charge-to-mass spectrum of electric BPS states consists of two parallel lines separated from each other by a distance of 2. For the general asymptotic analysis of finite distance singularities, we refer to section \ref{sec:generalanalysis} and appendix \ref{app:Ichain}.  

The conifold prepotential is given by (see for instance \cite{Strominger_1995})
\begin{equation}\label{eq:conifoldprepotential}
\cF(X^0,X^1) = -i c_1 (X^0)^2 -i c_2 (X^1)^2 \log \frac{X^1}{X^0} \, ,
\end{equation}
with $c_1,c_2$ real positive constants. Then we obtain from \eqref{eq:PeriodVector} the period vector
\begin{equation}
\mathbf{\Pi}  = \begin{pmatrix}
1 \\
e^{2\pi i t} \\
-2 i c_1+ic_2 e^{4\pi i t}\\
 -\frac{c_2}{2\pi} t e^{2\pi i t} - i c_2 e^{2\pi i t}
\end{pmatrix}
\end{equation} 
where we set $X^0=1$ and $X^1=e^{2\pi i t}$. Under $t \to t+1  $ the period vector undergoes a monodromy transformation $\mathbf{\Pi}(t+1) = M \mathbf{\Pi}(t)$, with monodromy matrix
\begin{equation}\label{eq:conifoldmonodromy}
M = \begin{pmatrix}
1 & 0 & 0 & 0 \\
0 & 1 & 0 & 0 \\
0 & 0 & 1 & 0 \\
0 & -\frac{c_2}{2\pi} & 0 & 1
\end{pmatrix}.
\end{equation}
The associated log-monodromy matrix is given by $N=\log M = M-\mathbb{I}$.

From now on we will write $t=+iy$ and set $b=0$ for simplicity, i.e.~the axions will not be relevant in what follows. By plugging the prepotential \eqref{eq:conifoldprepotential} into \eqref{pm} we find as polynomial part of the gauge kinetic functions
\begin{equation}
\cI_{IJ} = \begin{pmatrix}
-2 c_1 & 0 \\
0 & -4 \pi c_2  y
\end{pmatrix},
\end{equation}
while the $\cR_{IJ} = 0$ because we set the axion to zero. Then we find that the leading order part of \eqref{eq:charge} becomes
\begin{equation}\label{eq:conifoldchargematrix}
\cM = \begin{pmatrix}
-2 c_1 & 0 & 0 & 0\\
0 & -4 \pi c_2  y & 0 & 0 \\
0 & 0 & -\frac{1}{2c_1} & 0 \\
0 & 0 & 0 & -\frac{1}{4\pi c_2 y}
\end{pmatrix}.
\end{equation}
Let us examine the form of this matrix in detail in light of asymptotic Hodge theory. We observe that $\cM$ takes a diagonal form, and that each diagonal component scales as a power-law in the modulus $y$. This is precisely the behavior that is predicted by asymptotic Hodge theory in \eqref{eq:growth}, and it is one of the features that makes this formalism so powerful. Namely, one is able to control the asymptotic behavior of couplings without any reference to a prepotential, which allows for general statements instead of being restricted to a particular example.

In our current choice of symplectic frame we take the electric charges to be of the form $\mathbf{q}_E=(0,0,q_0,q_1)$. Note that these are charges for which the physical charge becomes small (or finite) at the conifold point according to \eqref{eq:conifoldchargematrix}, which ensures a weakly-coupled description for the $U(1)$ gauge fields.\footnote{More generally, one could pick $\mathbf{q}_E = (q_0 \sin \theta,0,q_0 \cos \theta, q_1)$ as electric charges: in the corresponding symplectic frame one finds $P=\cI_{IJ} \Re X^I \Im X^J \neq 0$. This means that \eqref{eq:radiiperiod} for the radii no longer hold, but one should use \eqref{eq:radiigeneral} instead. However, instead of computing the radii via this formula there is another way to see that the radii do not depend on $\theta$: one can rescale $\mathbf{\Pi} \to e^{i \theta} \mathbf{\Pi}$ -- setting $P=0$ -- which can be understood as a rotation back to $\mathbf{q}_E \to  (0,0,q_0, q_1)$.}  Then the electric part of the period vector that couples to these charges is given by
\begin{equation}\label{eq:conifoldelectricperiods}
X^I = \big( 1,e^{- 2\pi y}  \big)\ .
\end{equation}
By inserting \eqref{eq:conifoldchargematrix} and \eqref{eq:conifoldelectricperiods} into the matrix \eqref{eq:radiimatrix} we obtain
\begin{equation}\label{eq:conifoldmatrix}
\cA = \frac{1}{2c_1}\begin{pmatrix}
2c_1 & 0  \\
0 & 4 \pi c_2 y e^{-4\pi y} \\
\end{pmatrix}\, .
\end{equation}
The eigenvalues of this matrix give the radii of the ellipsoid, and we find as asymptotic values
\begin{equation}
\gamma_1^{-2} =  1 \, , \qquad \gamma_2^{-2} = 0\, . 
\end{equation}
This structure of the charge-to-mass spectrum could also have been expected from the charge-to-mass ratios of the two states $\mathbf{q}_0 = (0,0,1,0)$ and $\mathbf{q}_1=(0,0,0,1)$. From the perspective of emergence, note that the state $\mathbf{q}_1$ is precisely the state that has to be integrated out to produce the conifold singularity \cite{Strominger_1995}. It does not couple to the polynomial part of the period vector in $t$ but to one of the exponentially suppressed terms, and this state therefore becomes massless at the singularity. Furthermore the monodromy matrix acts trivially on the charge vector, so the log-monodromy matrix annihilates $\mathbf{q}_1$ as $N \mathbf{q}_1 = 0$. This indicates that we are only dealing with a single state that has to be integrated out, instead of an entire tower that can be generated via monodromy transformations as was found for infinite distance limits in \cite{Grimm:2018ohb,Grimm:2018cpv}. We find that the leading order behavior of the charge-to-mass ratios of the states $\mathbf{q}_0,\mathbf{q}_1$ is given by
\begin{equation}
\bigg(\frac{Q}{M}\bigg)^2 \bigg|_{\mathbf{q}_0} = 1 \,, \qquad
\bigg(\frac{Q}{M}\bigg)^2 \bigg|_{\mathbf{q}_1} = \frac{c_1}{2 \pi  c_2 y}e^{4\pi y}\, . 
\end{equation}
These ratios match nicely with the structure observed for the charge-to-mass spectrum from the radii of the ellipse. On the one hand, we found that there is a state that attains the lowest value possible value for its charge-to-mass ratio. Namely, as can be seen from \eqref{eq:N=2identity}, the charge-to-mass ratio of BPS states in 4d $\mathcal{N}=2$ supergravities is always bounded from below by 1. On the other hand, we found a state for which its charge-to-mass ratio diverges at the conifold point. Together these results combine into a compelling picture: one radius diverges, and the ellipsoid degenerates into two lines separated from each other by a distance of 2. This shape can also be inferred from the radii, since the asymptotic value of the second radii is given by $\gamma_2^{-2}=0$, which means that this radius must diverge at the conifold point. It turns out that this behavior is characteristic for finite distance singularities, and we find in section \ref{sec:generalanalysis} that the ellipsoid always degenerates in this manner at finite distance points.

\subsubsection{Example 2: large complex structure point}
\label{ssec:LCS}
For our next example we turn to the large complex structure point, considering an arbitrary number of moduli. This choice of singularity allows us to study a large class of infinite distance limits all at once, since the prepotential always takes a cubic form at this point. Before we begin we should note that, depending on the form of the intersection numbers $\cK_{ijk}$ and the choice of path, there arise some subtleties in the choice of electric charges. To avoid distraction from the main purpose of the examples, we give here only the calculation for one of the two kinds of paths explicitly. The other path involves more technical details and will therefore not be considered here, but it is covered by the general analysis in section \ref{sec:generalanalysis}.

The prepotential at the large complex structure point can be conveniently written as
\begin{equation}\label{eq:LCSprepotential}
\cF(X^I) = -\frac{\cK_{ijk}X^i X^j X^k}{6X^0}\, ,
\end{equation} 
with $X^I = (X^0,X^i)$, and $\cK_{ijk}$ the intersection numbers of the mirror dual of the Calabi-Yau threefold $Y_3$. We can then write the period vector \eqref{eq:PeriodVector} as
\begin{equation}
\mathbf{\Pi} = \begin{pmatrix}
1 \\
t^i \\
\frac{1}{6}\cK_{klm}t^k t^l t^m \\
-\frac{1}{2} \cK_{ikl}t^k t^l
\end{pmatrix}, 
\end{equation}
where we used special coordinates $X^I=(1,t^i)$. We again write out $t^i=x^i+iy^i$ and set $x^i=0$ for simplicity. The K\"ahler potential \eqref{eq:kahlersugra} then reads
\begin{equation}
K = -\log \Big( \frac{4}{3}\cK_{ijk}y^i y^j y^k \Big) \, .
\end{equation}
Now we want to study electric BPS states at large complex structure. We can distinguish electric charges from magnetic charges by asking for what charges BPS states become light. Note that this approach deviates slightly from the prescription that will be used in section \ref{ssec:bounds}, where we look directly at how the physical charge \eqref{eq:charge} behaves asymptotically. Looking at the mass of states can become a problem when exponentially suppressed contributions to the period vector are important. These contributions can cause the mass associated with magnetic charges to vanish asymptotically, while their physical charge does diverge. For the large complex structure point this is not an issue and the two methods agree, but it can be an issue at e.g.~the conifold point, which is why we motivated our choice of electric charges via the physical charge in the previous subsection. Taking a closer look at the mass of a BPS state \eqref{eq:centralcharge}, we find that
\begin{equation}
M^2 = \frac{3}{4\cK_{ijk}y^i y^j y^k}\big| q_0 + i q_i y^i + \frac{1}{6} p^0 i\cK_{ijk}y^i y^j y^k -\frac{1}{2} i \cK_{ikl}p^i y^j y^k \big|^2 \, ,
\end{equation}
where we wrote $\mathbf{q}=(q_0,q_i,p^0,p^i)$. The most natural choice of electric charges is given by $q_0,q_i$, and then $p^0,p^i$ form their dual magnetic charges. BPS states with these charges become light when asymptotically
\begin{equation}\label{eq:pathcondition}
\frac{y^i}{\sqrt{\cK_{ijk}y^i y^j y^k}} \to 0\, .
\end{equation}
However, note that if $\cK_{11i}=0$ for all $i$, then one can send the modulus $y^1$ to large complex structure at a rate much faster than all other moduli, say $y^1 \gg (y^i)^2$. In that case the charge $q_1$ is not electric, but one should consider $p^1$ as electric charge instead. One can then view \eqref{eq:pathcondition} as a constraint that specifies a certain sector of the moduli space around the large complex structure point. We only consider the charges $q_0,q_i$ to be electric in the following, and refer to the general analysis in section \ref{sec:generalanalysis} for other sectors around this singularity.

Having identified the electric charges, the electric periods that couple to these charges are simply the periods $X^I$. To compute the radii of the ellipsoid from \eqref{eq:radiiperiod}, we need to know the gauge kinetic functions. By plugging the cubic prepotential into \eqref{pm} we find that $\cR_{IJ}=0$ and
\begin{equation}\label{eq:gaugecouplings}
\cI  = -\frac{\cK}{6}\begin{pmatrix}
1 & 0 \\
0 & 4K_{ij}
\end{pmatrix}\,, \qquad K_{ij} = \partial_i \partial_{\bar{j}} K = -\frac{3}{2}\big(\frac{\cK_{ij}}{\cK}-\frac{3}{2} \frac{\cK_i \cK_j}{\cK^2} \big)\, ,
\end{equation}
where we wrote $\cK_{ij}=\cK_{ijk}y^k$, $\cK_{i}=\cK_{ijk}y^j y^k$ and $\cK=\cK_{ijk}y^i y^j y^k$. 

We next evaluate the expressions for the radii \eqref{eq:radiiperiod} by 
writing $X^I=(1,iy^i)$ and using \eqref{eq:gaugecouplings} to find
\begin{equation}\label{eq:radii}
\gamma_1^{-2} = \frac{3}{2\cK} \frac{\cK}{6} =\frac{1}{4}\, ,\qquad \gamma_2^{-2} = \frac{3}{2\cK}  \frac{\cK}{2} = \frac{3}{4}\, .
\end{equation}
Let us also note that if one looks at paths that do not lie in the sector given by \eqref{eq:pathcondition}, then the radii are found to be $\gamma_1=\gamma_2=\sqrt{2}$ instead, and we elaborate further on this matter in section \ref{ssec:bounds}.

Even though large complex structure points form only a subset of all possible infinite distance singularities, we can already draw some lessons from our study of this singularity. First, note that the lower bound put by \eqref{eq:N=2identity} ($Q^2/M^2 \geq 1$) cannot be saturated, but that the charge-to-mass ratio of an electric BPS state is bounded from below by $2/ \sqrt{3}$ or $\sqrt{2}$ instead for these infinite distance limits.  Secondly, although the large complex structure point provides us with a large variety of infinite distance limits, we only find two different shapes that the ellipsoid can take. Quite remarkably, we will find that many of the observations made for the conifold point and large complex structure point apply generally, as we will see in section \ref{sec:generalanalysis}. For instance, the three allowed shapes of the charge-to-mass spectrum found here constitute all possibilities that can arise in any infinite distance limit.

To conclude, we discuss how the large complex structure point provides us with infinite distance paths for which the analysis of \cite{Gendler:2020dfp} is not applicable. The reason for this was already stated by the authors of the latter work. For each modulus $y^i$ that is scaled at a different rate compared to the others, one introduces an integer $d_i$. Assuming an ordered limit by $y^i \gg y^j$ for $i > j$, these integers need to satisfy $d_i \geq d_j$ if $i > j$. Furthermore, these integers are bounded from below by $d_i \geq 0$, and from above by the complex dimension of the Calabi-Yau manifold, so here $d_i \leq 3$. In the asymptotic analysis of \cite{Gendler:2020dfp} it was crucial that these integers satisfy $d_i \neq d_{i-1}$. However, this condition can clearly not be realized if one takes a limit with four different scalings of the moduli, and might not even be realized for a lower number of scalings depending on the values that the integers $d_i$ take. From this perspective, one can always find limits that are not covered by this analysis in moduli spaces with dimension $h^{2,1} \geq 4$. It is then interesting to point out that our above analysis of the large complex structure point did not require this assumption, and one is free to pick any relative scalings for the moduli. Even for the simple computation presented here that requires limits to obey \eqref{eq:pathcondition}, one can scale as many moduli at different rates as one wants, only how much these rates can differ is constrained. In the study of higher-dimensional moduli spaces there is thus a large class of limits still left unexplored, which will be the subject of section \ref{sec:generalanalysis}.

\section{Asymptotic analysis of the charge-to-mass spectrum}\label{sec:generalanalysis}
Here we study the charge-to-mass spectrum of BPS states for \textit{any} limit in complex structure moduli space, both at finite and infinite distance. In order to perform this analysis we apply the machinery of asymptotic Hodge theory. First we derive a formula for the charge-to-mass ratio of sl(2)-elementary BPS states. 
We then apply this formula to give general bounds on the electric charge-to-mass spectrum. These bounds are obtained by computing the radii of the ellipsoid that is spanned by the charge-to-mass vectors of electric BPS states. The values found for these radii are listed in table \ref{table:WGCradii}. To illustrate these results, we conclude by considering some examples where we demonstrate how to use this formula for charge-to-mass ratios.

\subsection{Formula for asymptotic charge-to-mass ratios}
We want to put bounds on the charge-to-mass spectrum of BPS states. In general one finds that the charge-to-mass ratio of a state depends in a highly non-trivial manner on the complex structure moduli. This behavior simplifies when we move towards the boundary of moduli space, where we can precisely describe how the charges and masses of BPS states scale in the moduli via asymptotic Hodge theory. The aim of this section is thus to study BPS states in these limits in moduli space, and thereby obtain asymptotic bounds on their charge-to-mass ratios.

\paragraph{Sl(2)-elementary states.} For a BPS state with a generic set of charges, it is however still a rather complicated problem to give its asymptotic charge-to-mass ratio. In order to simplify this problem, we turn to the sl$(2)^n$-splitting \eqref{eq:Sl2Decomp} that plays a central role in asymptotic Hodge theory. This splitting decomposes the charge space $H^3(Y_3,\mathbb{R})$ into irreducible sl$(2)^n$ representations, where the weights $\ell_i$ ($i=1,\ldots,n$) of a charge fix the scaling in the moduli via equations such as \eqref{eq:growth}. We restrict our attention for now to states that can be specified by a single set of weights $\ell_i$, which were referred to as  single-charge states in \cite{Gendler:2020dfp}, but we will adopt the name \textit{sl(2)-elementary} states. We want to emphasize that these states do not need to be BPS, but rather they are a convenient basis in which any BPS state can be expanded. Let us denote the set of charges for the \textit{sl(2)-elementary} by
\begin{equation}
\cQ_{\rm sl(2)} = \{ q \in H^3(Y_3,\mathbb{R}) \, | \ q \in V_{\Bell} \text{ for some } \Bell \}\, .  \label{eq:SetElementary}
\end{equation}
The space $\cQ_{\rm sl(2)} $ is a union of vector spaces if we consider the charges to be continuous. We note that this 
split can also be performed over the rationals to accommodate quantized charges, but we will not address this issue any further in the following. 
At first, the restriction to this particular set of states limits the generality of our results. However, let us point out that a formula for the charge-to-mass ratio of these sl$(2)$-elementary  states suffices to obtain the 
asymptotic shape of the charge-to-mass spectrum of all electric BPS states, as we will see in section \ref{ssec:bounds}.

\paragraph{Asymptotic behavior of couplings.} Let us examine the asymptotic behavior of the charge-to-mass ratio for a candidate sl$(2)$-elementary  BPS state $q_{\Bell} \in V_{\Bell}$ piece by piece. By using the growth theorem \eqref{eq:growth} we find that the physical charge \eqref{eq:charge} of this state asymptotes in the strict asymptotic regime \eqref{eq:growthsector} to
\begin{equation}
\label{eq:piece1}
Q^2 = -\frac{1}{2} (y^1)^{\ell_1} (y^2)^{\ell_2} \cdots (y^n)^{\ell_n } \langle q_{\Bell},\, C_{\infty} q_{\Bell} \rangle \, ,
\end{equation}
where we remind the reader that $C_{\infty}$ is the Weil operator associated with the boundary. The mass of a BPS state $M(q_{\Bell})$, given in \eqref{eq:centralcharge}, consists of two factors. The first one involves the K\"ahler potential, and by using \eqref{eq:Kahlerpotasymp} we find that its leading term is given by
\begin{equation}
\label{eq:piece2}
e^{-K} = (y^1)^{d_1} (y^2)^{d_2-d_1}\cdots (y^n)^{d_n-d_{n-1}}\ i \langle \Omega_{\infty} , \bar{\Omega}_{\infty} \rangle \, .  
\end{equation}
The second factor asymptotes to
\begin{equation}
\label{eq:piece3}
| \langle q_{\Bell}, \Omega \rangle |^2 = (y^1)^{\ell_1+d_1} (y^2)^{\ell_2+d_2-d_1} \cdots (y^n)^{\ell_n+d_n-d_{n-1}}  \ |\langle q_{\Bell},  \Omega_\infty \rangle |^2 \, .
\end{equation}
This follows by using the sl$(2)$-orbit approximation for the $(3,0)$-form $\Omega$ given in \eqref{eq:sl2orbit3form} 
which is valid in the strict asymptotic regime \eqref{eq:growthsector}. The operator $e^{-1}(y)$ defined in \eqref{eq:ey} can be moved to the other side via $\langle e(y) \cdot, \cdot \rangle = \langle \cdot, e^{-1}(y) \cdot \rangle$, after which it can be applied on the charge $q_{\Bell}$ to obtain part of the parametrical scaling.

\paragraph{Asymptotic graviphoton.} Now we can put the pieces of the charge-to-mass ratio back together. When we compare their scalings in the moduli, we find that the factors of $y^i$ cancel out precisely. However, this relies crucially on the coefficients of these leading terms being non-zero. The coefficients of $|Q|^2$ and $e^{-K}$ are indeed non-zero, since both can be interpreted as a vector norm computed with the metric $\langle \cdot, C_{\infty} \bar{\cdot} \rangle$, where we note that $C_\infty \Omega_\infty = -i\Omega_\infty$. However, the coefficient in \eqref{eq:piece3} is trickier, and we require that
\begin{equation}
\label{eq:graviphotoncoupling}
\langle q_{\Bell} , \ \Omega_\infty  \rangle \neq 0\, .
\end{equation}
This quantity has the natural interpretation as the asymptotic coupling of the state to the graviphoton. We can see this by looking at the scaling of the charge-to-mass ratio for states for which \eqref{eq:graviphotoncoupling} vanishes. Namely, when this product is zero, a term subleading to \eqref{eq:piece3} sets the asymptotic behavior of the mass $M(q_{\Bell})$. Previously the scaling of the different pieces of the charge-to-mass ratio precisely matched, so now $|Q|$ grows parametrically compared to $M(q_{\Bell})$. This means that the charge-to-mass ratio of such states must diverge along the limit, which leads us to consider \eqref{eq:graviphotoncoupling} as the asymptotic coupling to the graviphoton.

\paragraph{Charge-to-mass ratio.} For sl$(2)$-elementary states with a non-vanishing coupling to the graviphoton, we find that the charge-to-mass ratio is given by
\begin{equation}
\label{eq:centralchargeasymptotically2}
\bigg( \frac{Q}{M}\bigg)^2 \bigg|_{q_{\Bell}} =  \frac{\langle q_{\Bell}, C_{\infty} q_{\Bell} \rangle \ i \langle \Omega_{\infty} , \bar{\Omega}_{\infty} \rangle }{2 | \langle q_{\Bell}, \Omega_{\infty} \rangle |^2} + \cO\Big(\frac{v^{i+1}}{v^i}\Big)  \, .
\end{equation}
To compute this ratio, one first has to identify the sl$(2)$-elementary  states that have a non-vanishing coupling to the graviphoton, i.e.~charges satisfying \eqref{eq:graviphotoncoupling}. However, this condition does not fix a unique set of charges. One is free to add any charges with a vanishing coupling to charges with a non-vanishing coupling to the asymptotic graviphoton, so we have to specify how we pick these charges. A natural choice is to consider charges that sit in the same irreducible sl$(2)^n$ representation as the asymptotic graviphoton. These charges can be obtained from $\tilde{a}_0$ by applying lowering operators $N_i^-$. We write this set of charges as
\begin{equation}\label{eq:gravitystates}
\cQ_{\rm G} = \{ q \in \cQ_{\rm sl(2)} \, | \, q = (N_1^-)^{k_1}  \cdots (N_n^-)^{k_n} \, (a \Re \tilde{ a}_0 + b  \Im \tilde{a}_0), \text{ $a,b \in \mathbb{R}$}   \} \, .
\end{equation}
For the remaining charges we define the set of so-called field theory states
\begin{equation}\label{eq:fieldstates}
\cQ_{\rm F} = \{ q \in \cQ_{\rm sl(2)} \, | \ \langle q , \, \Omega_{\infty} \rangle = 0   \} \, .
\end{equation}
Together $\cQ_{\rm G}$ and $\cQ_{\rm F} $ provide us with a complete basis for the charges of BPS states
and have been discussed in \cite{Grimm:2018ohb,Grimm:2018cpv} in the context of the distance conjecture. One can then apply identity \eqref{eq:Cinftyaction} to compute the charge-to-mass ratio for the states in $\cQ_{\rm G}$, the details of which have been moved to appendix \ref{app:charge-to-mass}. In the end, one finds that the charge-to-mass ratio of an sl(2)-elementary state with non-vanishing coupling to the graviphoton is given by the formula
\begin{equation}\label{eq:chargetomass}
\boxed{\quad \rule[-.5cm]{.0cm}{1.2cm} \lim_{\lambda \to \infty} \bigg( \frac{Q}{M} \bigg)^{-2} \bigg|_{q_{\rm G}}= 2^{1-d_n}  \prod_{i=1}^n {\Delta d_i\choose{(\Delta d_i -  \ell_i)/2}} \times \begin{cases}
1 \text{ for $d_n = 3$}\, ,\\
\frac{1}{2} \text{ for $d_n \neq 3$}\, ,
\end{cases}\, }
\end{equation}
where $\gamma$ denotes the constant involved in the definition of the strict asymptotic regimes \eqref{eq:growthsector} and we used the abbreviation $\Delta d_i = d_i - d_{i-1}$. We stress that sending $\lambda \rightarrow \infty$ can 
also be viewed as performing a consecutive limit sending $y^1 \rightarrow \infty$, then $y^2 \rightarrow \infty$, up to $
y^n \rightarrow \infty$.
A different order of limits requires to consider another choice of sector  \eqref{eq:growthsector} and will, in general, change the 
integers appearing in \eqref{eq:chargetomass}. The formula \eqref{eq:chargetomass} admits a straightforward generalization 
for any Calabi-Yau $D$-fold as we show in appendix \eqref{eq:Cinftyaction}. Explicitly, we find  
\beq
\lim_{\gamma \to \infty} \bigg( \frac{Q}{M} \bigg)^{-2} \bigg|_{q_{\rm G}}= 2^{1-d_n}  \prod_{i=1}^n {\Delta d_i\choose{(\Delta d_i -  \ell_i)/2}} \times \begin{cases}
1 \text{ for $d_n = D$}\, ,\\
\frac{1}{2} \text{ for $d_n \neq D$}\, ,
\end{cases}\, \label{eq:GeneralChargetomass}
\eeq
which trivially agrees with \eqref{eq:chargetomass} when setting $D=3$. While we will not use this formula in this generality any further, it is nice to see 
that the same general pattern arises in any dimension. 

\paragraph{Intermediate summary.} Before we continue, let us briefly summarize our findings. We studied the asymptotic behavior of the charge-to-mass ratio for sl(2)-elementary BPS states. We found that this behavior depends crucially on whether the charges of these states couple to the asymptotic graviphoton via \eqref{eq:graviphotoncoupling} or not. When this coupling vanishes the charge-to-mass ratio diverges, whereas if this coupling is non-vanishing the charge-to-mass ratio stays finite and is given by \eqref{eq:chargetomass}. This formula expresses the charge-to-mass ratio purely in terms of the discrete data $d_i,\ell_i$ that characterizes the limit and the choice of sl(2)-elementary state. In particular, note that these charge-to-mass ratios are independent of the spectator moduli $\zeta^k$ that are not taken to a limit and therefore remain constant to leading order, up to suppressed corrections in $y^{i+1}/y^{i}$ and $1/y^{n}$.

\paragraph{Electromagnetic duality.} Let us now take a closer look at these charge-to-mass ratios given in \eqref{eq:chargetomass}. First of all, it is interesting to point out that this formula even applies for limits with $\Delta d_i = 0$, since it involves binomial coefficients. This is the upshot of working with the boundary Hodge structure via identities such as \eqref{eq:Cinftyidentity}, instead of an asymptotic approximation for the K\"ahler metric that follows from \eqref{eq:Kahlerpotasymp}. Secondly, notice that the charge-to-mass ratios are symmetric under
\begin{equation}\label{eq:electromagneticduality}
\Delta \ell_i \to -\Delta \ell_i:\quad \frac{Q}{M} \to \frac{Q}{M}\, .
\end{equation}
This symmetry has a natural interpretation from a physics perspective, since it tells us that dual electric and magnetic states have the same charge-to-mass ratio. Namely, recall from the orthogonality condition \eqref{eq:orthogonality} that dual electric and magnetic charges are related by $\ell_i \to - \ell_i$, which is equivalent to sending $\Delta \ell_i \to -\Delta \ell_i$.

The formula we presented in \eqref{eq:chargetomass} is only applicable for the charge-to-mass ratios of sl(2)-elementary states, but one might wonder if it can be extended to apply for BPS states with generic charges. Ideally one could simply identify its elementary charge with the largest parametrical growth according to \eqref{eq:growth}, and argue that this elementary charge fixes its charge-to-mass ratio. However, when looking more carefully at the spaces $V_{\Bell}$ in which these elementary charges reside, one realizes that things can become more complicated. The first issue arises when the parametrical growth associated with two (or more) of these spaces via \eqref{eq:growth} is the same for a given path. In that case both charges contribute to the charge-to-mass ratio of the state, such that one ends up with some combination between their charge-to-mass ratios. Another issue arises when we try to add one of the charges in \eqref{eq:fieldstates} that does not couple to the asymptotic graviphoton to a charge in \eqref{eq:gravitystates} that does have a non-vanishing coupling. Assuming that both charges lie in the same eigenspace $V_{\Bell}$, we find that this added charge does contribute asymptotically to the physical charge of the state but not its mass, so the charge-to-mass ratio changes. It would be interesting to see what the generalized formula for the charge-to-mass ratio that resolves these issues looks like, but this lies beyond the scope of this work. We will, however, argue in the next subsection that our results for sl(2)-elementary states allows us to make statements about the asymptotic shape of the  general charge-to-mass spectrum of electric BPS states.

\subsection{Asymptotic shape of the electric charge-to-mass spectrum}\label{ssec:bounds}
Now that we have derived a formula for the charge-to-mass ratio of sl(2)-elementary BPS states with \eqref{eq:chargetomass}, we can put it to use to determine more properties of the charge-to-mass spectrum of electric BPS states. An elegant way to do so was given in \cite{Gendler:2020dfp}, where it was shown that the charge-to-mass vectors of electric BPS states lie on an ellipsoid with at most two nondegenerate directions. We will determine the asymptotic shape of this ellipsoid, by deriving the asymptotic values for its radii. It turns out that these radii can then be determined from the charge-to-mass ratios of electric sl(2)-elementary states via \eqref{eq:asymptoticradii}. Besides specifying a structure for the electric charge-to-mass spectrum, this also provides the limiting value for the smallest radius as lower bound on the asymptotic charge-to-mass ratio of any electric BPS state. 

\paragraph{Electric charges.} In order to study the charge-to-mass spectrum of electric BPS states, we first have to establish how we can identify electric charges. When we look at the prepotential formulation of 4d $\mathcal{N}=2$ supergravities, a natural choice is to pick the charges $q_I$ that couple to the periods $X^I$ in the central charge \eqref{eq:centralcharge}. However, this method is not suitable for our purposes. First of all, the general techniques that we borrow from asymptotic Hodge theory simply do not make use of a prepotential. Secondly, one wants the physical charge \eqref{eq:charge} of an electric BPS state to be small in order to provide a weakly-coupled description for the $U(1)$ gauge fields. For instance, if we recall our analysis of the large complex structure point in section \ref{ssec:LCS}, we found that some of the charges $q_I$ had to be replaced by $p^I$ as electric charges when considering limits outside the sector \eqref{eq:pathcondition}. This teaches us that we should study the asymptotic behavior of the physical charge \eqref{eq:charge} carefully in order to identify the electric charges correctly. As a first step let us therefore take  sl(2)-elementary states as basis for the electric charges, since the parametrical behavior of their physical charges is described by \eqref{eq:growth}.  A complication that can then arise is that the physical charge of a BPS state does not diverge or vanish asymptotically, but stays finite instead. In that case one can use that sl(2)-elementary states with finite physical charge come in pairs that are each others electro-magnetic dual, as can be seen by using \eqref{eq:orthogonality} and \eqref{eq:growth}. This allows us to pick the electric charge out of each pair by hand, which in particular means that our choice of electric charges is not necessarily unique.

\paragraph{Sector-dependence.} The task that remains is then to fix a sector in complex structure moduli space such that we know precisely what sl(2)-elementary charges are electric and magnetic. We find that we can specify these sectors simply by imposing constraints on the scalings of the moduli. To begin with we limit ourselves to considering strict asymptotic regimes, which already restricts the saxions $v^i$ via the constraints given in \eqref{eq:growthsector} with $\lambda \gg 1$. Subsequently we want electric states to have an asymptotically vanishing physical charge, which leads to additional conditions such as $(y^{1})^2\gg (y^i)^2$ by imposing \eqref{eq:growth} to decrease for a given set of $\ell_i$. In practice the allowed values for $\ell_i$ are fixed by the type of singularity under consideration, so one can systematically determine all possible subsectors. While we do not outline a procedure to construct these subsectors here, let us refer to appendix \ref{app:radii} where we have to make such divisions in the general analysis.

To be more precise, let us summarize the above conditions that specify the choice of electric charges in terms of an equation. We define the set of elementary electric charges in a given sector by 
\begin{align}\label{eq:defelectric}
\cQ_{\text{el}}= \{ q, q' \in \cQ_{\text{sl(2)}} \, \, | \, \, \|q \|^2 , \|q'\|^2 < \infty \text{ and } \langle q , q' \rangle =0 \}\, ,
\end{align}
where the mutual non-locality condition makes sure that among the electric charges we picked, none of the ones with finite physical charge are dual to each other. As stated above, there can of course be more than one possibility to make this choice, so this definition of $\cQ_{\text{el}}$ does not define a unique set. Having defined our space of elementary electric charges, we can define the dual magnetic charges by application of $C_{\infty}$ as follows
\begin{align}
\cQ_{\text{mag}}= C_{\infty} \cQ_{\text{el}} \,. \label{eq:EMduality}
\end{align}
Note that this choice of magnetic charges ensures that products between electric and magnetic charges computed with the asymptotic Hodge norm $\langle \cdot, C_{\infty} \cdot \rangle $ vanish, as can be shown by using that $C_{\infty}^{2}=-1$. In other words, the gauge kinetic functions $\cR_{IJ}$ that describe the coupling between electric and magnetic charges via \eqref{def-cM} vanish in the strict asymptotic regime. This follows from the asymptotic behavior of the Hodge norm \eqref{eq:growth}, and by expressing the Hodge norm in terms of the gauge kinetic functions via \eqref{eq:normtogkfunctions}.

\paragraph{Three-form basis.} In order to compute the radii of the electric charge-to-mass spectrum we now want to make a particular choice of symplectic 
basis $(\tilde \alpha_I, \tilde \beta^J)$, $I=1,...,h^{2,1}+1$. We first pick linearly independent $(\tilde \alpha_I,\tilde \beta^J)$ 
such that 
\beq
  \tilde \alpha_I \in \cQ_{\rm el}\ , \qquad \tilde \beta^I \in \cQ_{\rm mag}\ . 
\eeq 
Crucially, we make sure that these elements satisfy a number of further conditions that 
will be useful below.  
As a start we pick a basis that preserves the splitting in terms of the sets $\mathcal{Q}_{\rm G}$ and $\mathcal{Q}_{\rm F}$, the reasons for which are twofold. On the one hand, this splits the charges based on whether they couple to the asymptotic graviphoton or not, which provides us with a precise description of their charge-to-mass ratios via expressions such as \eqref{eq:chargetomass}. On the other hand, it proves to be useful to pick a basis that diagonalizes the gauge kinetic functions $\cI_{IJ}$ in the strict asymptotic regime.  The advantage of splitting our basis elements in $\mathcal{Q}_{\rm G}$ and $\mathcal{Q}_{\rm F}$ is then that mixed terms between these subsets vanish in this setting. This follows from expressing the gauge kinetic functions in terms of the Hodge norm via \eqref{eq:normtogkfunctions}, which in turn can be described by the boundary Hodge norm $\langle \cdot, C_{\infty} \cdot \rangle$ via the approximation \eqref{eq:growth}.\footnote{It can then be argued that mixed terms vanish from the fact that $C_{\infty}$ maps $\mathcal{Q}_{\rm G}$ back into $\mathcal{Q}_{\rm G}$ as follows from \eqref{eq:Cinftyidentity}, and that charges in $\mathcal{Q}_{\rm F}$ have a vanishing symplectic product with elements of $\mathcal{Q}_{\rm G}$ by construction.} 

\paragraph{Graviphoton coupling.} Let us now construct a particular basis for the charges in $\cQ_{\rm el}  \cap \mathcal{Q}_{\rm G}$ in detail. 
This basis will make up parts of the elements $\tilde \alpha_I$. 
From the expressions for the radii given in \eqref{eq:radiiperiod} we know that the coupling of these charges to the real and imaginary parts of the holomorphic $(3,0)$-form $\Omega$ plays an important role. This motivatives us to define our basis elements via
\begin{align}
\mathcal{Q}_{\Re}&= \{ q \in \cQ_{\rm el} \cap \cQ_{\rm G} \, \,  | \, \,  \text{linearly independent and }  \langle q , \Im (\Omega_{\infty}) \rangle =0 \}\, ,  \nonumber \\
\mathcal{Q}_{\Im}&= \{ q \in \cQ_{\rm el} \cap \cQ_{\rm G}  \, \,  | \, \,  \text{linearly independent and }  \langle q , \Re (\Omega_{\infty}) \rangle =0 \} \, . \label{eq:ReImBasisSets}
\end{align}
It can be argued that this choice of basis provides us with a diagonalization for the boundary Hodge norm $\langle \cdot, C_{\infty} \cdot \rangle$.\footnote{To be precise, this follows from the action of $C_{\infty}$ on elements of $Q_{\rm G}$ as given by \eqref{eq:Cinftyidentity}, where one also needs to use that $C_{\infty}$ is a real map together with the orthogonality conditions \eqref{eq:orthogonality} and polarization conditions \eqref{eq:polarization1} and \eqref{eq:polarization2}.} In principle one can then complete the basis $\tilde \alpha_I$ for $\mathcal{Q}_{\rm el}$ by picking linearly independent elements of $\cQ_{\rm el} \cap \cQ_{\rm F}$ that also diagonalize $\langle \cdot, C_{\infty} \cdot \rangle$, but for our purposes we do not need to derive more explicit expressions for these charges.

\paragraph{Diagonal gauge kinetic functions.} The above choice of basis $\tilde \alpha_I$ for the electric charges allows us to diagonalize the gauge kinetic functions $\cI_{ IJ}$ in the strict asymptotic regime, since they can be expressed in terms of the Hodge norm via \eqref{eq:normtogkfunctions}. The appropriate quantity to describe the Hodge norm in the strict asymptotic regime is the sl(2)-norm $\langle \cdot, C_{\rm sl(2)} \cdot \rangle $ introduced in \eqref{eq:growth}, so let us write the gauge kinetic functions as
\begin{align}
\langle \tilde \alpha_I , C_{\rm sl(2)} \tilde \alpha_I \rangle = (\tilde{\cI}_{II})^{-1} \, , \label{eq:Csl2DiagonalElements}
\end{align}
where the matrix $ \tilde{\cI}_{IJ}$ is (the imaginary part of) the gauge coupling functions associated to the sl(2)-orbit.
 
\paragraph{Radii computations.} We now express the radii \eqref{eq:radiigeneral} in terms of charge-to-mass ratios of sl(2)-elementary electric charges in the strict asymptotic regime (see table \ref{table:regimes} for a reminder of this notion). As the derivation is analogous for both radii we will only be explicit for $\gamma_1$. We start by rewriting the expression for $\gamma_1^{-2}$ in the sl(2)-basis outlined above as
\begin{equation}
\begin{aligned}
\gamma_1^{-2} &=-2 e^{K_{\rm sl(2)}} \Im \tilde{\cN}_{II} \Re \tilde{X}^I \Re \tilde{X}^I + \cO\Big(\frac{v^{i+1}}{v^i}\Big)  \, ,
\end{aligned}
\end{equation}
where the $\tilde{X}^I$ are obtained by expanding $\Omega_{\rm sl(2)}$ along $\tilde \alpha_I$, and the quantities $K_{\rm sl(2)}, \tilde{\cN}_{II}$
are the $\cN=2$ data associated to the sl(2)-orbit. We can now manipulate this strict asymptotic expression using the techniques introduced in section \ref{sec:sl2splitting}, which gives
\begin{align}
-2 e^{K_{\rm sl(2)}} \Im \tilde{\cN}_{II} \Re \tilde{X}^I \Re \tilde{X}^I  &=- 2 e^{K_{\rm sl(2)}} \sum_{I} \frac{\langle \tilde \alpha_I , \Re \Omega_{\rm sl(2)} \rangle^2 }{\langle \tilde \alpha_I , C_{\rm sl(2)} \tilde \alpha_I \rangle} \nonumber \\ 
&= - 2 e^{K_{\rm sl(2)}} \sum_{I} \frac{\langle e(y) \tilde \alpha_I , e(y)\Re \Omega_{\rm sl(2)} \rangle^2 }{ \langle e(y) \tilde \alpha_I , C_{\infty} e(y) \tilde \alpha_I \rangle} \nonumber \\
&=-2  \sum_{I} \frac{\langle \tilde \alpha_I , \Re \Omega_{\infty} \rangle^2 }{\langle \Omega_{\infty} , \bar{\Omega}_{\infty} \rangle \langle \tilde \alpha_I , C_{\infty} \tilde \alpha_I \rangle}  \nonumber \\
&=  -2   \sum_{  \tilde \alpha_I \in \mathcal{Q}_{\Re} } \frac{\langle \tilde \alpha_I , \Omega_{\infty} \rangle^2 }{\langle \Omega_{\infty} , \bar{\Omega}_{\infty} \rangle \langle \tilde \alpha_I , C_{\infty} \tilde \alpha_I \rangle} \, . \label{eq:radDerivation}
\end{align}
In the first step, we used \eqref{eq:Csl2DiagonalElements} and reformulated things in the language of forms. In the second step, we inserted the identity in the form of $e^{-1}(y) e(y)$ into the symplectic products and used that $\langle e^{-1}(y) \cdot , \cdot \rangle= \langle \cdot , e(y) \cdot \rangle$. For the third equality, we used \eqref{eq:sl2orbit3form}-\eqref{eq:Kinf} which makes clear that all the parametrical scaling cancels out. In the last step, we used the defining property of the set $\cQ_{\Re}$. We want to emphasize that the basis charges sitting in $\cQ_{\rm el} \cap \cQ_{F}$ can be safely ignored here as they give a vanishing contribution to the sum. We recognize that \eqref{eq:radDerivation} is a sum over the inversed of strict asymptotic charge-to-mass ratios \eqref{eq:centralchargeasymptotically2}. As the strict asymptotic expression for $\gamma_2^{-2}$ can be rewritten in a similar manner, we directly state the result from \cite{Gendler:2020dfp}
\begin{align} \label{eq:asymptoticradii} 
\gamma_1^{-2}= \sum_{\tilde \alpha_I \in \cQ_{\Re}} \bigg( \frac{Q}{M} \bigg)^{-2} \bigg|_{q=\tilde \alpha_I}\, , \quad \quad \gamma_2^{-2}=\sum_{\tilde \alpha_I \in \cQ_{\Im}} \bigg( \frac{Q}{M} \bigg)^{-2} \bigg|_{q=\tilde \alpha_I} \,.
\end{align}
where corrections in $y^{i+1}/y^i$ are dropped. We can provide a quick check of our formula for the charge-to-mass ratios of sl(2)-elementary states \eqref{eq:chargetomass} by verifying the $\mathcal{N}=2$ constraint \eqref{gamma_constr} on the radii. To derive this relation, we will make use of a well known identity for binomial coefficients, which in our specific setup reads
\begin{equation}
\sum_{\Delta \ell_i}  {{\Delta d_i}\choose{\frac{\Delta d_i-\Delta \ell_i}{2}}} = 2^{\Delta d_i}\, .
\end{equation}
The sum $\gamma_1^{-2}+ \gamma_2^{-2}$ amounts to adding up the inverse squares of charge-to-mass ratios for all electric states, as can be seen from \eqref{eq:asymptoticradii}. Since we found that the charge-to-mass ratios are the same for dual electric and magnetic charges, we can just as well sum over all charges and compensate by dividing by two, which yields
\begin{equation}
\begin{aligned}
\gamma_1^{-2}+ \gamma_2^{-2} &= \frac{1}{2} \sum_{\Bell} \bigg(\frac{Q}{M}\bigg)^{-2} \times \begin{cases}
1 \text{ for $d_n = 3$}\, ,\\
2 \text{ for $d_n \neq 3$}\, ,
\end{cases} \\
&= 2^{-d_n} \prod_{i=1}^n \sum_{\Delta\ell_i}  {{\Delta d_i}\choose{\frac{\Delta d_i-\Delta \ell_i}{2}}}  \\
&= 2^{-d_n} \prod_{i=1}^n 2^{\Delta d_i} = 1\, ,
\end{aligned}
\end{equation}
Here the extra factor of two in the first line for the case that $d_n \neq 3$ follows from the fact that each sl(2)-level site is populated by two states. Namely, one has both states coming from applying lowering operators $N_i^-$ on $\Re \tilde{a}_0$ and on $\Im \tilde{a}_0$, whereas for $d_n=3$ they only come from the real three-form $\tilde{a}_0$. In the next line this factor of two cancels against the factor of two that has to be included in the expression for the charge-to-mass ratio in \eqref{eq:chargetomass}.

This relation already gives us some insight into the bounds for the charge-to-mass ratio. As mentioned before, the smallest radius of the ellipsoid serves as lower asymptotic bound on the charge-to-mass spectrum via
\begin{equation}\label{eq:radiibound}
 \frac{Q}{M} \bigg|_{\rm asym} \gtrsim \min(\gamma_1,\gamma_2)\, .
\end{equation}
Let us briefly explain what we mean by this asymptotic bound. Firstly, we can consider the bound after taking the asymptotic limit, i.e.~to the boundary of the moduli space, by taking the limit $\lambda \rightarrow \infty$ in \eqref{eq:growthsector} or sending consecutively $y^1,...,y^n \rightarrow \infty$.  
In this limit \eqref{eq:radiibound} turns into a proper inequality and we can replace $\gtrsim$ with $\geq$. However, as soon as we go away from the boundary, there are corrections to the ratio of the order $y^{i+1}/y^i$. These are suppressed in the strict asymptotic regime \eqref{eq:growthsector}, but we did not infer any information about the signs of these corrections. Therefore, also any bound on the general expression for 
$Q/M$ for an electric  state can have already in the strict asymptotic regime small 
corrections that become increasingly irrelevant near the boundary. In the following 
we will compute the radii $\gamma_1,\gamma_2$ for all possible asymptotic limits. 
Before doing so let us briefly note that \eqref{gamma_constr} implies that the radii are bounded from below by $\gamma_{1,2} \geq 1$. This tells us that the charge-to-mass ratio of any electric BPS state is bounded by
\begin{equation}\label{eq:generallowerbound}
 \frac{Q}{M}  \geq 1\, .
\end{equation}
This nicely agrees with our knowledge from 4d $\mathcal{N}=2$ supergravities, since \eqref{eq:N=2identity} predicts the same lower bound.

\begin{table}[]
\centering
\begin{tabular}{|c|c|c|c|}
\hline
enhancement chain & subsector & $\gamma_1^{-2}$ & $\gamma_2^{-2}$ \\
\hline \hline
$\mathrm{I}$ & & 1 & 0 \\ \hline
$\mathrm{I} \to \mathrm{II}$ & & 1/2 & 1/2 \\ \hline
$\mathrm{I} \to \mathrm{III}$ & & 3/4 & 1/4 \\ \hline
$\mathrm{I} \to \mathrm{II} \to \mathrm{III}$ & & 1/2 & 1/2  \\ \hline
$\mathrm{I} \to \mathrm{IV}$ & & 3/4 & 1/4 \\ \hline
$\mathrm{I} \to \mathrm{II} \to \mathrm{IV}$ & \begin{minipage}{2.5cm}\vspace{0.1cm}\centering
$y^{\mathrm{II}} \gg (y^\mathrm{IV})^2 $ \\ 
$y^{\mathrm{II}} \ll (y^\mathrm{IV})^2 $ \vspace{0.1cm}
\end{minipage} & \begin{minipage}{1cm}\centering
1/2 \\
3/4
\end{minipage} & \begin{minipage}{1cm}\centering
1/2 \\
1/4
\end{minipage} \\ \hline
$\mathrm{I} \to \mathrm{III} \to \mathrm{IV}$ & & 3/4 & 1/4  \\ \hline
$\mathrm{I} \to \mathrm{II} \to \mathrm{III} \to \mathrm{IV}$ &\begin{minipage}{2.5cm}\vspace{0.1cm}\centering
$y^{\mathrm{II}} \gg y^\mathrm{III} y^\mathrm{IV} $ \\
$y^{\mathrm{II}} \ll y^\mathrm{III} y^\mathrm{IV} $ \vspace{0.1cm}
\end{minipage} & \begin{minipage}{1cm}\centering
1/2 \\
3/4
\end{minipage} & \begin{minipage}{1cm}\centering
1/2 \\
1/4
\end{minipage} \\ \hline
\end{tabular}
\caption{Asymptotic values for the radii of the ellipsoid that forms the charge-to-mass spectrum of electric BPS states, depending on the enhancement chain that characterizes the limit, and subsector of the strict asymptotic regime \eqref{eq:growthsector} considered. The latter we indicated the constraints on scaling of the saxions when relevant, where $y^{\mathrm{A}}$ corresponds to the saxion that sources an increase to roman numeral $\mathrm{A}$ for the enhancement chain.}\label{table:WGCradii} 
\end{table}

\paragraph{Radii classification.} Let us now make the bound \eqref{eq:radiibound} precise by explicitly computing the radii. In order 
to do that we apply \eqref{eq:asymptoticradii} to determine the radii from the charge-to-mass ratios of sl(2)-elementary states. We can then turn to our formula for charge-to-mass ratios \eqref{eq:chargetomass}, and go through all possible limits by considering all possible enhancement chains. These results are summarized in table \ref{table:WGCradii}, and the details are included in appendix \ref{app:radii}. We found only three different sets of values for the radii
\begin{equation}
(\gamma_1^{-2},\gamma_2^{-2})=\quad (1,0),\quad (\frac{3}{4},\frac{1}{4}),\quad (\frac{1}{2},\frac{1}{2})\, .
\end{equation}
Let us briefly elaborate on the sector-dependence of the results in table \ref{table:WGCradii}. The underlying reason is that depending on in what subsector of the strict asymptotic regime we are, we pick different electric charges. To be more precise, the constraints given in \ref{table:WGCradii} ensure that for each eigenspace $V_{\Bell}$ we can make a definite statement about whether its asymptotic Hodge norm given in \eqref{eq:growth} diverges, stays finite or vanishes asymptotically. We need this information because we require the physical charge of our electric states to be bounded as described by \eqref{eq:defelectric}.  In fact, the limits towards large complex structure that we ignored in section \ref{ssec:LCS} -- outside of the sector \eqref{eq:pathcondition} -- are precisely those that lie in $y^{\mathrm{II}} \gg (y^\mathrm{IV})^2 $ or $y^{\mathrm{II}} \gg y^\mathrm{III} y^\mathrm{IV} $, for which the radii according to table \ref{table:WGCradii} are $\gamma_{1}=\gamma_{2}=\sqrt{2}$.

Having determined the radii for all possible limits, let us first look at the differences between finite and infinite distance limits. Finite distance limits only involve $\mathrm{I}_a$ singularities, whereas infinite distance limits include one of the other types of singularities, i.e.~$\mathrm{II}_b$, $\mathrm{III}_c$ or $\mathrm{IV}_d$. Then we observe from table \ref{table:WGCradii} that one of the radii always diverges for finite distance singularities, resulting in an ellipse that degenerates into two lines, separated from each other by a distance of 2. On the other hand, for infinite distance limits both radii remain finite, and we find either an ellipse with radii $\gamma_1=2/\sqrt{3}$ and $\gamma_2=2$, or a circle with radius $\gamma_1=\gamma_2=\sqrt{2}$. 

\paragraph{WGC bounds.} We can now use these values for the radii to bound the charge-to-mass ratio of electric BPS states based on the singularity under consideration. For finite distance limits we find that the lower bound for charge-to-mass ratios given in \eqref{eq:generallowerbound} can be saturated, since the smallest radius is given by $\gamma_1=1$, so we do not obtain a new bound. For infinite distance limits we do find new bounds, and depending on the limit we obtain $Q/M \geq 2/\sqrt{3}$ or $Q/M \geq \sqrt{2}$ for the charge-to-mass ratio. In either case, the charge-to-mass ratio is bounded from below by
\begin{equation}
 \frac{Q}{M}  \geq \frac{2}{\sqrt{3}}\, .
\end{equation} 
It is interesting to point out that the state with minimal charge-to-mass ratio need not be a sl(2)-elementary state. Namely, when the sum over charge-to-mass ratios for one of the radii in \eqref{eq:asymptoticradii} runs over only one state, then the charge-to-mass ratio of this state is equal to the radius. But once the sum runs over multiple states, we find that for none of these states the charge-to-mass ratio can be equal to the radius. In fact, since there are $h^{2,1}+1$ electric charges, one finds for $h^{2,1}>1$ that for at least one of the radii multiple charges should contribute, so the state corresponding to this radius cannot be sl(2)-elementary. \footnote{As an example we want to point out the LCS point discussed in section \ref{ssec:LCS}. There the smallest radius $\gamma_2$ in equation \eqref{eq:radii} is obtained by summing over multiple electric periods, and thus the state with minimal charge-to-mass must be realized as a linear combination of sl(2)-elementary states.} It is quite remarkable that the charge-to-mass ratios of sl(2)-elementary states fix this minimal charge-to-mass ratio via \eqref{eq:asymptoticradii}, even though the formula for the charge-to-mass ratio \eqref{eq:chargetomass} only applies for sl(2)-elementary states.

\section{Remarks on other swampland conjectures}
\label{sec:remarks}
In this section we discuss connections between the order-one coefficients in various swampland conjectures. In the previous section we derived a formula for the charge-to-mass ratio of sl(2)-elementary BPS states that applies to any limit in complex structure moduli space $\cM^{\rm cs}(Y_3)$, which provides us with an order-one coefficient for the Weak Gravity Conjecture in the strict asymptotic regime. First, we point out that these charge-to-mass ratios also appear in the order-one coefficient in the asymptotic de Sitter conjecture for a particular class of flux potentials. Then we review the connection between the Weak Gravity Conjecture and the Swampland Distance Conjecture, and comment on the order-one coefficient that we obtain for the Swampland Distance Conjecture via this connection.

\subsection{Bounds for the de Sitter conjecture}\label{ssec:dSbounds}
We first study the order-one coefficients in the asymptotic de Sitter conjecture \cite{Obied:2018sgi} for flux potentials \eqref{eq:potential} arising in Type IIB Calabi-Yau orientifolds. To be precise, we focus on the parameter $c$ in \eqref{eq:dS1} in asymptotic regimes of the moduli space, and we do not investigate $c'$ in \eqref{eq:dS2} signalling the presence of an unstable direction of the potential. Our goal is to establish a link between the order-one coefficients that appear in the Weak Gravity Conjecture and the one appearing in the de Sitter conjecture. For this we will rewrite a specific class of flux potentials -- arising from picking just $F_3$ or just $H_3$ flux -- in terms of charge-to-mass ratios of some BPS state\footnote{Let us note that we use the term BPS state here very loosely, since for our argument to work we do not need to make sure that the charge lattice site we pick is actually populated by a physical BPS state. In fact, it would be more appropriate to view these potentials as sourced by domain walls, as recently considered in \cite{Lanza:2020qmt}.}. This will allow us to use our asymptotic expression for the charge-to-mass ratio \eqref{eq:chargetomass} in order to evaluate the order-one constant $c$ from \eqref{eq:dS1} numerically at the boundary of complex structure moduli space.

The calculation for both flux choices is similar and only differs slightly in the dilaton factor. So we will only be explicit for the case where $F_3=q$ and $H_3=0$. By using the expressions for the charge \eqref{eq:charge} and mass \eqref{eq:centralcharge} for a BPS state that would be associated with this charge $q$, we can suggestively rewrite the potential \eqref{eq:potential} as 
\begin{equation}
V = \frac{1}{2 \cV^2} \bigg( \frac{Q}{M}\bigg)^2 e^{\phi} M^2\, .
\end{equation}
We are now restricting to fluxes for which the above `charge-to-mass' ratio approaches a constant value along the limit. As explained in section \ref{sec:generalanalysis}, this can be realized by requiring the charge $q$ to belong to $\cQ_G$ defined in \eqref{eq:gravitystates}. In 
the following we will assume that
\begin{equation} \label{QMassumption}
\bigg| \nabla  \frac{Q}{M} \bigg|^2_{q \in \cQ_G} = 2 K^{i \bar{j}} \partial_i  \frac{Q}{M} \partial_{\bar{j}} \frac{Q}{M} \to 0\, ,
\end{equation}
along the limit. It should be noted, however, that we inferred this condition from studying a number of examples, and did not yet manage to show it rigorously in the framework of asymptotic Hodge theory. From there, we can see that the ratio that is of interest in the de Sitter conjecture reduces to\footnote{The Cauchy-Schwarz inequality tells us that we do not have to consider the mixed term between $\partial_I (Q/M)$ and $\partial_I( \cV e^{\phi} M^2)$.}
\begin{equation}
\frac{\big| \nabla  V \big|^2}{V^2}  = \frac{2K^{A \bar{B}} \partial_A V \partial_{\bar{B}}V }{V^2} =  \frac{2K^{A \bar{B}} \partial_A ( \cV^{-2} e^{\phi} M^2 ) \partial_{\bar{B}} (\cV^{-2}e^{\phi} M^2 ) }{\cV^{-4} e^{2\phi} M^4}\, ,
\end{equation}
with the indices $A,\bar{B}$ running over all the moduli, i.e.~the complex structure moduli, the K\"ahler structure moduli and the axio-dilaton. Furthermore, by making use of the identity \eqref{eq:N=2identityrewritten} we then find that asymptotically\footnote{By making the flux choice $F_3=0$ and $H_3=q $ instead, one would obtain the same numbers on the right-hand side.}
\begin{equation}
\lim_{\lambda \to \infty} \frac{\big| \nabla  V \big|^2}{V^2} \bigg|_{F_3 \in \cQ_G} = 2\bigg[ \bigg( \frac{Q}{M} \bigg)^2-1\bigg]+ 2 + 6\, . \label{eq:AsyRatioH}
\end{equation}
where the last two terms represent the positive contribution that arise from including the axio-dilaton and K\"ahler moduli respectively. We choose to neglect contributions from the K\"ahler moduli, as their stabilization would anyway require including quantum corrections to source a non-trivial profile. Depending on the type of singularity, we found different lowest values for the charge-to-mass ratios that we can now use. To keep things compact, we will only distinguish between the finite and infinite distance singularities, which gives 
\begin{equation}
\frac{\big| \nabla  V \big|}{V} \bigg|_{\rm asym}   \gtrsim   \begin{cases} \, \sqrt{2} &\text{ finite distance}  \\ 2 \sqrt{2/3}   &\text{ infinite distance} \end{cases}\, ,
\end{equation}
where the bound is to be understood as explained below \eqref{eq:radiibound}. A more refined analysis can of course be performed for the infinite distance case by considering the different enhancement chains. The bounds we obtain here coincide with bounds that were found recently in \cite{Andriot:2020lea,Lanza:2020qmt}, and also with previously established no-go theorems \cite{Blaback:2010sj, Andriot:2016xvq}. Ignoring contributions coming from the axio-dilaton, note that we recover the recently proposed Trans-Planckian Censorship Conjecture bound \cite{Bedroya:2019snp}, i.e.~$c \geq \sqrt{2/3}$ for infinite distance boundaries.

In our analysis, we also neglected D7-brane moduli which would also give a contribution to the superpotential \eqref{eq:superpotential} and enter in the K\"ahler potential at the next to leading order in the string coupling. A systematic way to include these moduli would be to look at F-theory flux vacua where they become -- together with the axio-dilaton -- complex structure moduli of the relevant Calabi-Yau fourfold \cite{Sen:1996vd}. In fact, such setups were already studied within the framework of asymptotic Hodge theory in \cite{Grimm:2019ixq}, and it would be interesting to revisit these flux potentials in the future. 

\subsection{Comments on the Swampland Distance Conjecture}
We next turn to the order-one coefficient of the Swampland Distance Conjecture \cite{Ooguri:2006in,Klaewer:2016kiy}, stated in \eqref{eq:distance}. For our purposes it is important to point out the towers of wrapped D3-brane states constructed in \cite{Grimm:2018ohb,Grimm:2018cpv}, since these form the infinite towers of states that become massless at infinite distance loci in complex structure moduli space for Type IIB Calabi-Yau compactifications. This construction starts from a particular sl(2)-elementary state that belongs to $\mathcal{Q}_{\rm G}$, i.e.~it couples to the asymptotic graviphoton. This state becomes light close to the singular loci, and the infinite tower of states is generated by acting with monodromy transformations on this `seed charge'. In studying the bounds put by the Swampland Distance Conjecture it then suffices to consider the mass of this sl(2)-elementary state, since it sets the parametrical behavior for the masses of all states in this infinite tower.


Our goal is now to relate the order-one coefficient we computed for the Weak Gravity Conjecture to its counterpart for the Swampland Distance Conjecture. The connection between these conjectures has already been studied before, and how to relate their order-one coefficients was spelled out in \cite{Lee:2018spm,Gendler:2020dfp}. Following \cite{Gendler:2020dfp}, we can express $\lambda$ in terms of the gradient of the mass of the sl(2)-elementary state as
\begin{equation}\label{eq:lambda}
\lambda = 2 \Big| K^{ij}\frac{\partial_i M}{M} u_j \Big|\, ,
\end{equation}
where $u_i$ denotes the unit vector that points along the geodesic. By making use of \eqref{eq:N=2identityrewritten} one can then bound the coefficient $\lambda$ via a Cauchy-Schwarz inequality. By picking a geodesic with $u_i =  \frac{\partial_i \log M}{| \nabla   \log M|}$ one can saturate this bound which yields
\begin{equation}\label{eq:SDCbound}
\lambda^2 = 2  \Big| K^{ij}\frac{\partial_i M \partial_j M}{M^2} \Big| = \frac{1}{2} \bigg( \Big(\frac{Q}{M}\Big)^2-1\bigg)\, .
\end{equation}
One can then try to study the coefficient $\lambda$ of the Swampland Distance Conjecture either directly from \eqref{eq:lambda}, or indirectly via the bound given in \eqref{eq:SDCbound}. For the former approach one needs to have control over the asymptotic behavior of the inverse K\"ahler metric $K^{ij}$, which is achieved to some extend by approximations such as \eqref{eq:Kahlerpotasymp}. However, as mentioned before this approximation does not necessarily provide the complete picture of the K\"ahler metric, and in particular it can lead to a mismatch for charge-to-mass ratios. We therefore take for the latter approach, and provide an upper bound for $\lambda$ via the charge-to-mass ratio of the sl(2)-elementary state. By using \eqref{eq:chargetomass} we obtain the bound
\begin{equation}
\lambda^2 =  \begin{cases} 2^{d_n-2}  \prod_{i=1}^n \frac{1}{ {\Delta d_i \choose{ (\Delta d_i - \ell_i)/2}}} -\frac{1}{2}
 \text{ for $d_n = 3$}\, ,\\
2^{d_n-1} \prod_{i=1}^n \frac{1}{ {\Delta d_i \choose{ (\Delta d_i - \ell_i)/2}}} -\frac{1}{2} \text{ for $d_n \neq 3$}\, .
\end{cases}
\end{equation}
In comparison to \cite{Gendler:2020dfp} this extends the bounds obtained for $\lambda$ to limits characterized by discrete data with $d_i=d_{i-1}$ for some $i$. Overall we find that the lowest value attained by $\lambda$ is still given by
\begin{equation}
\lambda \geq \frac{1}{\sqrt{6}}\, .
\end{equation}
Another way to obtain the order-one coefficient of the Swampland Distance Conjecture has been noted in \cite{Andriot:2020lea}, where it was conjectured that it can be related to the order-one coefficient of the de Sitter Conjecture via $\lambda = c/2$. In our setting this relation holds true when contributions from the K\"ahler moduli and axio-dilaton to the gradient of the Type IIB flux potential are ignored, cf.~\eqref{eq:AsyRatioH}. We only considered infinite distance limits that involved complex structure moduli for the SDC, so it would be interesting to see if limits that also involve the axio-dilaton and/or K\"ahler moduli lead to a matching value for $\lambda$.\footnote{Work in this direction has already been performed in \cite{Font:2019cxq}, where they studied the Swampland Distance Conjecture in the mirror Type IIA setup and found tensionless branes when the dilaton was also sent to a limit. Moreover in \cite{Lanza:2020qmt} connections between swampland conjectures were studied by looking at such extended objects, and it would be interesting to see if the approach taken in our work for computing order-one constants leads to new insights into this matter. }

\paragraph{Summary.} In this chapter we have studied the asymptotic charge-to-mass spectrum of BPS states in 4d $\mathcal{N}=2$ supergravity theories. Specifically we focused on Calabi--Yau threefold compactifications of Type IIB string theory, where these BPS states arise from D3-branes wrapped on three-cycles. Both the physical charges and the masses of such states vary with changes in the complex structure moduli. Using powerful tools from asymptotic Hodge theory we can make their leading behavior explicit when moving towards the boundary of the moduli space with the sl(2)-approximation. This description relies on the universal structure that emerges at every such limit and can be formulated without referring to specific examples. We used this structure to derive a general formula \eqref{eq:chargetomass} for the charge-to-mass ratios of a particular set of states, which we called sl(2)-elementary, at strict asymptotic regimes in complex structure moduli space. Given this formula we were then able to obtain numerical bounds for the Weak Gravity Conjecture, and also indirectly for the asymptotic de Sitter Conjecture and Swampland Distance Conjecture.

\begin{subappendices}

\section{Computations on charge-to-mass ratios}
In this appendix we include some computations on the charge-to-mass spectrum of BPS states. We first compute the charge-to-mass spectrum of sl(2)-elementary BPS states that couple to the asymptotic graviphoton in appendix \ref{app:charge-to-mass}. In appendix \ref{app:radii} we then use these results to determine the radii of the charge-to-mass spectrum of electric BPS states.

\subsection{Derivation of the formula for charge-to-mass ratios}
\label{app:charge-to-mass}
In this appendix we derive formula \eqref{eq:chargetomass} for the charge-to-mass ratio of sl(2)-elementary BPS states that couple asymptotically to the graviphoton. For the sake of generality we perform these computations for Calabi-Yau manifolds of arbitrary complex dimension $D$. In this work we consider $D=3$, but it turns out to be fairly simple to compute quantities such as \eqref{eq:centralchargeasymptotically2} for generic dimension $D$. We make a separation of cases based on the integer $d_{n}$ that characterizes the limit, since the value that this integer takes matters for the construction of our charges. Namely, when a limit ends with $d_{n}=D$ one can take $\tilde{a}_{0}$ to be real, whereas for $d_{n } \neq D$ we have that $\Re \tilde{a}_0$ and $\Im \tilde{a}_0$ are linearly independent. 
Recalling the definition of charges that couple to the asymptotic graviphoton from \eqref{eq:gravitystates}, we notice that we have to `double' the amount of charges we consider for $\mathcal{Q}_{\rm G}$ when $d_{n} \neq D$.

Before we make this separation of cases, let us make some general comments about computing the charge-to-mass ratios first. To begin we recall expression \eqref{eq:centralchargeasymptotically2} for the charge-to-mass ratio, which reads
\begin{equation}\label{eq:startingpoint}
\bigg( \frac{Q}{M}\bigg)^2 \bigg|_{q} =  \frac{\langle q, C_{\infty} q \rangle \ i^{D} \langle  \bar{\Omega}_\infty \, , \ \Omega_\infty \rangle }{2 | \langle q,\ \Omega_\infty \rangle |^2} \, ,
\end{equation}  
where we replaced the factor of $i^{3}$ by $i^{D}$. This expression serves as our starting point for computing the charge-to-mass ratios. Without knowledge of the form of the charges, we can already write out the second factor of the numerator as (see also \eqref{eq:Kinf})
\begin{equation}\label{eq:1}
 \langle \Omega_\infty , \ \bar{\Omega}_\infty \rangle = (-2i)^{d_n} \langle \tilde{a}_0 , \ \prod_i \frac{(N_i^-)^{d_i-d_{i-1}}}{(d_i-d_{i-1})!} \bar{\tilde a}_0 \rangle\, ,
\end{equation}
where we expanded the exponentials in $\Omega_\infty = e^{iN_{(n)}^-} \tilde{a}_0$ and its conjugate into lowering operators $N_i^-$. We also used the relation $\langle \cdot, N_i^- \cdot \rangle = -\langle N_i^- \cdot, \cdot \rangle$, and the powers of each $N_i^-$ are fixed by the orthogonality condition \eqref{eq:orthogonality}.

Then remain the other two factors that appear in the charge-to-mass ratio, both of which involve the charge $q$. For convenience in notation we write the charges as 
\begin{equation}\label{eq:sl2elementaries}
q^r_{\mathbf{k}} = (N_1^-)^{k_1} \ldots (N_n^-)^{k_n} \Re  \tilde{a}_0\, , \qquad q^i_{\mathbf{k}} = (N_1^-)^{k_1} \ldots (N_n^-)^{k_n} \Im  \tilde{a}_0\, .
\end{equation}
The integers $\mathbf{k}=(k_1,\ldots,k_n)$ that label the charges should not be confused with the eigenvalues of the level operators $N_i^0$, which follow from $\ell_i-\ell_{i-1} = d_i-d_{i-1}-2k_i$. In the case that $\tilde{a}_0$ is real only the charges $q^r_{\mathbf{k}}$ matter, which in turn allows us to ignore the superscript. 

In order to evaluate the remaining two factors in the charge-to-mass ratio, let us introduce some relations relevant for the above charges. It is useful to specialize the polarization conditions \eqref{eq:pol} to the leading polynomial term $\tilde{a}_0$ of the periods. To be precise, by using \eqref{eq:a0position} these read
\begin{equation}
\label{eq:polarization1}
-(-i)^{D+d_n}\langle \tilde{a}_0, \ (N_1^-)^{d_1} (N_2^-)^{d_2-d_1} \cdots (N_n^-)^{d_n-d_{n-1}}  \bar{\tilde{a}}_0 \rangle > 0\, .
\end{equation}
In the case that $d_n \neq D$ this positivity condition can be supplemented by the vanishing constraint
\begin{equation}
\label{eq:polarization2}
\langle \tilde{a}_0, \ (N_1^-)^{d_1} (N_2^-)^{d_2-d_1} \cdots (N_n^-)^{d_n-d_{n-1}}  \tilde{a}_0 \rangle = 0\, .
\end{equation}
Finally, also recall the action of $C_\infty$ on $\tilde{a}_0$ and its descendants as \eqref{eq:Cinftyidentity}. Together these relations suffice to evaluate products between charges of BPS states constructed out of $\Re \tilde{a}_0$, $\Im \tilde{a}_0$ and its descendants. In the following two subsections we now write out the remaining two factors of the charge-to-mass ratios in \eqref{eq:startingpoint} for the cases $d_n=D$ and $d_n \neq D$.

\subsubsection{Limits with $d_{n} = D$}
Let us first consider the case where the limit is characterized by an integer $d_{n}=D$. For a Calabi-Yau threefold this corresponds to an enhancement chain that ends with a $\mathrm{IV}$ singularity. In this case $\tilde{a}_0$ is real, so we only need to consider the charges $\mathbf{q}^r_{\mathbf{k}}$, and therefore we drop the superscript label $r$ in this subsection.

Let us begin with the factor appearing in the denominator in \eqref{eq:startingpoint}. By expanding $e^{i N^-_{(n)}}$ in terms of lowering operators $N_i^-$ we find that
\begin{equation}\label{eq:2D}
|\langle q_{\mathbf{k}},\ e^{i N^-_{(n)} } \tilde{a}_0 \rangle| = \prod_i \frac{1}{(d_i-d_{i-1}-k_i)!} \ |\langle \tilde{a}_0, \ \prod_i (N_i^-)^{d_i-d_{i-1} } \tilde{a}_0 \rangle| \, ,
\end{equation}
where we used that we needed $d_i-d_{i-1}-k_i$ factors of $N_i^-$ in this expansion to satisfy the orthogonality condition \eqref{eq:orthogonality}. Then remains the first factor in the numerator of \eqref{eq:startingpoint}. By using \eqref{eq:Cinftyidentity} for the action of $C_{\infty}$ on the charges, it reduces to
\begin{equation}\label{eq:3D}
\langle q_{\mathbf{k}} , \ C_{\infty} q_{\mathbf{k}} \rangle  =- \prod_i \frac{k_i!}{(d_i-d_{i-1}-k_i)!} \ |\langle \tilde{ a}_0, \ \prod_i (N_i^-)^{d_i-d_{i-1} } \tilde{a}_0 \rangle|\, ,
\end{equation}
Putting all factors together (\eqref{eq:1}, \eqref{eq:2D} and \eqref{eq:3D}), we find the charge-to-mass ratio to be 
\begin{equation}\label{eq:D}
\bigg( \frac{Q}{M} \bigg)^2 = 2^{d_n-1} \prod_i \frac{ (d_i-d_{i-1}-k_i)!k_i!}{(d_i-d_{i-1})!} \, .
\end{equation}

\subsubsection{Limits with $d_{n} \neq D$}
Now we consider the case where the limit is characterized by an integer $d_{n} \neq D$. For threefolds this corresponds to an enhancement chain that ends with $\mathrm{I}_a$, $\mathrm{II}_b$ or $\mathrm{III}_c$. Here the computations become slightly more involved, but in the end the factors only differ by some factors of two compared to the previous subsection.

First let us exploit the polarization conditions \eqref{eq:polarization1} and \eqref{eq:polarization2} to write down conditions for products involving the vectors $\Re \tilde{a}_0$ and $\Im \tilde{a}_0$. These identities will be useful for computing the charge-to-mass ratio. Depending on the choice of $d_n$, these identities look different. In the case that $D+d_n$ is odd we find from $\langle v, w \rangle = (-1)^D \langle w, v \rangle$  and $\langle v, N_i^- w \rangle = - \langle N_i^- v, w \rangle$ that
\begin{equation}
\langle \Re \tilde{a}_0, \ \prod_i (N_i^-)^{d_i-d_{i-1} } \Re \tilde{a}_0 \rangle =0\, , \quad \langle \Im \tilde{a}_0, \ \prod_i (N_i^-)^{d_i-d_{i-1} } \Im \tilde{a}_0 \rangle =0\, ,
\end{equation}
whilst from \eqref{eq:polarization1} we know that
\begin{equation}\label{eq:Dodd}
- i^{D+d_n+1}\langle \Re \tilde{a}_0, \ \prod_i (N_i^-)^{d_i-d_{i-1} } \Im \tilde{a}_0 \rangle >0\, .
\end{equation}
On the other hand when $D+d_n$ is even we find that
\begin{equation}
\langle \Re \tilde{a}_0, \ \prod_i (N_i^-)^{d_i-d_{i-1} } \Im \tilde{a}_0 \rangle = 0\, .
\end{equation}
which follows as a non-trivial constraint from the vanishing of the imaginary part of \eqref{eq:polarization1}. Meanwhile by combining the polarization conditions \eqref{eq:polarization1} and \eqref{eq:polarization2} we find for $D+d_n$ even that
\begin{equation}\label{eq:Deven}
\begin{aligned}
\langle \Re \tilde{a}_0, \ \prod_i (N_i^-)^{d_i-d_{i-1} } \Re \tilde{a}_0 \rangle &= \langle \Im \tilde{a}_0, \ \prod_i (N_i^-)^{d_i-d_{i-1} } \Im \tilde{a}_0 \rangle \, , \\
(-i)^{D+d_n}\langle \Re \tilde{a}_0, \ \prod_i (N_i^-)^{d_i-d_{i-1} } \Re \tilde{a}_0 \rangle &<0\, ,
\end{aligned}
\end{equation}
where \eqref{eq:polarization2} implied that the products involving $\Re \tilde{a}_0$ and $\Im \tilde{a}_0$ are equal to one another. For threefolds the case $D+d_n$ odd corresponds to a  $\mathrm{I}_a$ or $\mathrm{III}_c$ singularity, while the case $D+d_n$ even corresponds to a $\mathrm{II}_b$ singularity. Keeping these identities in mind, we now write out the remaining two factors of the charge-to-mass ratio below.

We begin with the factor that appears in the denominator of the charge-to-mass ratio in \eqref{eq:startingpoint}. By expanding $e^{i N^-_{(n)}}$ in terms of lowering operators $N_i^-$ we find that
\begin{equation}\label{eq:2!D}
|\langle q^{r,i}_{\mathbf{k}},\ e^{i N^-_{(n)} } \tilde{a}_0 \rangle| = \frac{1}{2} \prod_i \frac{1}{(d_i-d_{i-1}-k_i)!} \ |\langle \tilde{a}_0, \ \prod_i (N_i^-)^{d_i-d_{i-1} }  \bar{\tilde{a}}_0 \rangle |\, ,
\end{equation}
where we divide by two in comparison to \eqref{eq:2D}. This division by two was necessary because when writing out the right-hand side, either the product in \eqref{eq:Dodd} appears twice if $D+d_n$ is odd, or both products in \eqref{eq:Deven} appear when $D+d_n$ is even, whereas these products appear only once on the left-hand side.

Then remains the first factor that appears in the numerator in \eqref{eq:startingpoint}, and we find that
\begin{equation}\label{eq:3!D}
\langle q^{r,i}_{\mathbf{k}} , \ C_{\infty} q^{r,i}_{\mathbf{k}} \rangle  = \frac{1}{2} \prod_i \frac{k_i!}{(d_i-d_{i-1}-k_i)!} \ |\langle \tilde{a}_0, \ \prod_i (N_i^-)^{d_i-d_{i-1} }  \bar{\tilde{a}}_0 \rangle|\, ,
\end{equation}
where we take either both charges with superscript $r$ or both with superscript $i$. In deriving this expression we made use of \eqref{eq:Cinftyidentity} and that $C_{\infty}$ is a real map. Furthermore we divided by two in comparison to \eqref{eq:3D} for reasons similar to \eqref{eq:2!D}.

If we now put all the different factors together (\eqref{eq:1}, \eqref{eq:2!D} and \eqref{eq:3!D}), we find for the charge-to-mass ratio
\begin{equation}
\bigg( \frac{Q}{M} \bigg)^2 = 2^{d_n} \prod_i \frac{ (d_i-d_{i-1}-k_i)!k_i!}{(d_i-d_{i-1})!}\, .
\end{equation}
Note that in the end we picked up an additional factor of two compared to the case $d_n = D$ in \eqref{eq:D}.

\subsection{Computation of radii of the ellipsoid}
\label{app:radii}
In this appendix we determine the radii of the electric charge-to-mass spectrum for limits in complex structure moduli space $\cM^{\rm cs}(Y_3)$. We go through all possible enhancement chains that classify these limits, and compute the radii from the charge-to-mass ratios of sl(2)-elementary states via \eqref{eq:asymptoticradii}. The states relevant for this computation are the ones that couple to the asymptotic graviphoton as described by \eqref{eq:gravitystates}, since their charge-to-mass ratio stays finite. The discrete data $d_i$ associated with the enhancement chain suffices to characterize this subset of sl(2)-elementary states. This means that the relevant information about the enhancement chain is captured by just the presence or absence of the segments  $ \mathrm{II}$, $\mathrm{III}$ and $\mathrm{IV}$, so the problem reduces to considering eight different kinds of enhancement chains in total. Let us note that for boundaries ending on a $\mathrm{III}_c$ our formula for the charge-to-mass ratio differs from \cite{Gendler:2020dfp}.\footnote{The underlying reason is that $\mathrm{III}_c$ singularities have a non-empty space $I^{2,3}$ in the Deligne splitting (see table \ref{table:HDclass}). In \cite{Gendler:2020dfp} they took derivatives of only the leading polynomial periods of the $(3,0)$-form, while as we know from chapter \ref{chap:models} we need essential exponential corrections (or spectator moduli) to span this vector space. We take these extra terms into account by working with the Hodge star instead of period derivatives to compute the physical charge \eqref{eq:charge}. We refer to \cite{Bastian:2020egp} for more details, where we explain how this works with some examples.} 

Before we go through each of these kinds of enhancement chains, let us briefly summarize how one can obtain the relevant properties of the sl(2)-elementary states under consideration \eqref{eq:sl2elementaries}. The first thing we need to know are their eigenvalues under the weight operators $N_{i}^0$ of the sl$(2,\mathbb{R})$-algebras. These follow from the level of $\tilde{a}_{0}$ as indicated by \eqref{eq:a0position}, together with how the lowering operators $N_{i}^{-}$ lower these levels according to \eqref{eq:lowering}. The charge-to-mass ratios of these states can then be obtained simply by evaluating \eqref{eq:chargetomass} for their discrete data. We next need to determine whether charges are electric or magnetic. This distinction is based upon whether the physical charge of these states diverges or vanishes asymptotically, which in turn can be deduced from the sl(2)-data by making use of \eqref{eq:piece1}.\footnote{For some particular elementary charges this method does not lead to a definite statement, e.g.~when the physical charge is finite asymptotically. In these cases we clarify whether elementary charges are electric or magnetic on the spot.} The last thing we need for the computation of the radii is whether charges couple to the real or imaginary part of the asymptotic graviphoton $\Omega_{\infty}$, cf.~\eqref{eq:ReImBasisSets}. This follows from the polarization conditions \eqref{eq:polarization1} and \eqref{eq:polarization2}. To be more precise, one finds for even $d_{n}$ (odd $d_{n}+3$) that charges obtained from $\Re \tilde{a}_{0}$ couple to charges obtained from $\Im \tilde{a}_{0}$, whereas for odd $d_{n}$ (even $d_{n}+3$) charges obtained from $\Re \tilde{a}_{0}$ couple to other charges obtained from $\Re \tilde{a}_{0}$ and similarly for charges obtained from $\Im \tilde{a}_{0}$. Keeping in mind that each application of an $N_{i}^{-}$ on $\tilde{a}_{0}$ comes together with an $i$ for $\Omega_{\infty}$, one can then straightforwardly determine whether a charge couples to the real or imaginary part of the asymptotic graviphoton. Having gathered all this information on the sl(2)-elementary states in $\mathcal{Q}_{\rm G}$, one is then finally ready to compute the radii of the ellipsoid.

\subsubsection*{Enhancement chain $\mathrm{I}$}
\label{app:Ichain}
Here we consider limits characterized by enhancement chains of the form $\mathrm{I}$, i.e.~it consists only of $\mathrm{I}_{a}$ singularities. This sort of limit is a finite distance limit, and the discrete data of such a limit is given by $d_{1},\ldots,d_{n}=0$. From this discrete data we can infer that all lowering operators $N_{i}^{-}$ annihilate the 3-form $\tilde{a}_{0}$, so we only have to consider the sl(2)-elementary charges $\Re \tilde{a}_{0}$ and $\Im \tilde{a}_{0}$. Their properties have been summarized in table \ref{table:I}. From this information we can straightforwardly compute the radii via \eqref{eq:asymptoticradii} to be
\begin{equation}
\begin{aligned}
\gamma_1^{-2} = 0 \, ,  \qquad \gamma_2^{-2} = \Big( \frac{Q}{M} \Big)^{-2} \Big|_{\Re \tilde{a}_0} = 1\, .
\end{aligned}
\end{equation}

\begin{table}[htb]\centering
\renewcommand{\arraystretch}{1.3}
\scalebox{0.85}{
\begin{tabular}{|c|c|c|c|c|}
\hline
charges & sl(2)-level   & $Q/M$ & electric/magnetic &period \\ \hline
$\Re \tilde{a}_{0}$ & $3$  & 1 & electric & imaginary \\ \hline
$\Im \tilde{a}_{0}$ & $3$   & 1 & magnetic & real\\ \hline
\end{tabular}}
\renewcommand{\arraystretch}{1}
\caption{Properties of the charges that couple to the asymptotic graviphoton for $\mathrm{I}$. The choice of electric and magnetic charge was picked by hand since the physical charges of both states are finite asymptotically.}
\label{table:I}
\end{table}

\subsubsection*{Enhancement chain $\mathrm{I} \to \mathrm{II}$}
Here we consider limits characterized by enhancement chains of the form $\mathrm{I} \to \mathrm{II}$. The discrete data of such a limit is given by $d_{1},\ldots,d_{k-1} = 0$ and $d_{k},\ldots,d_{n}=1$. The enhancement to a $\mathrm{II}_{b}$ singularity occurs at step $k$, and we denote the lowering operator $N_{k}^{-}$ therefore by $N_{\rm II}$. From the discrete data we can infer that $N_{\rm II}$ can be applied once on $\tilde{a}_{0}$. Moreover, charges obtained from $\Re \tilde{a}_{0}$ and $\Im \tilde{a}_{0}$ are linearly independent, so there are four charges to consider in total. Their properties have been summarized in table \ref{table:II}. From this information we can straightforwardly compute the radii via \eqref{eq:asymptoticradii} to be
\begin{equation}
\begin{aligned}
\gamma_1^{-2} = \Big( \frac{Q}{M} \Big)^{-2} \Big|_{N \Re \tilde{a}_0} = \frac{1}{4}\, ,  \qquad \gamma_2^{-2} = \Big( \frac{Q}{M} \Big)^{-2} \Big|_{N \Im \tilde{a}_0} = \frac{1}{4}\, .
\end{aligned}
\end{equation}

\begin{table}[htb]\centering
\renewcommand{\arraystretch}{1.3}
\scalebox{0.85}{
\begin{tabular}{|c|c|c|c|c|}
\hline
charges & sl(2)-level   & $Q/M$ & electric/magnetic &period \\ \hline
$\Re \tilde{a}_{0}$ & $1$  & $\sqrt{2}$ & magnetic & imaginary \\ \hline
$\Im \tilde{a}_{0}$ & $1$   & $\sqrt{2}$ & magnetic & real\\ \hline
$N_{\rm II}\Re \tilde{a}_{0}$ & $-1$  & $\sqrt{2}$ & electric & real \\ \hline
$N_{\rm II}\Im \tilde{a}_{0}$ & $-1$   & $\sqrt{2}$ & electric & imaginary \\ \hline
\end{tabular}}
\renewcommand{\arraystretch}{1}
\caption{Properties of the charges that couple to the asymptotic graviphoton for $\mathrm{I} \to \mathrm{II}$. }
\label{table:II}
\end{table}

\subsubsection*{Enhancement chain $\mathrm{I} \to \mathrm{III}$}
Here we consider limits characterized by enhancement chains of the form $\mathrm{I} \to \mathrm{III}$. The discrete data of such a limit is given by $d_{1},\ldots,d_{k-1} = 0$ and $d_{k},\ldots,d_{n}=2$. The enhancement to a $\mathrm{III}_{c}$ singularity occurs at step $k$, and we denote the lowering operator $N_{k}^{-}$ therefore by $N_{\rm III}$. From the discrete data we can infer that $N_{\rm III}$ can be applied twice on $\tilde{a}_{0}$. Moreover, charges obtained from $\Re \tilde{a}_{0}$ and $\Im \tilde{a}_{0}$ are linearly independent, so there are six charges to consider in total. Their properties have been summarized in table \ref{table:III}. From this information we can straightforwardly compute the radii via \eqref{eq:asymptoticradii} to be
\begin{equation}
\begin{aligned}
\gamma_1^{-2} &= \Big( \frac{Q}{M} \Big)^{-2} \Big|_{N \Re \tilde{a}_0} + \Big( \frac{Q}{M} \Big)^{-2} \Big|_{N^2 \Im \tilde{a}_0}   = \frac{1}{2}+\frac{1}{4} = \frac{3}{4}\, ,  \\ \gamma_2^{-2} &= \Big( \frac{Q}{M} \Big)^{-2} \Big|_{N^2 \Re \tilde{a}_0} = \frac{1}{4}\, .
\end{aligned}
\end{equation}

\begin{table}[htb]\centering
\renewcommand{\arraystretch}{1.3}
\scalebox{0.85}{
\begin{tabular}{|c|c|c|c|c|}
\hline
charges & sl(2)-level  & $Q/M$ & electric/magnetic & period \\ \hline
$\Re \tilde{a}_{0}$ & $2$  & 2 & magnetic & imaginary \\ \hline
$\Im \tilde{a}_{0}$ & $2$& 2& magnetic & real\\ \hline
$N \Re \tilde{a}_{0}$ & $0$  &$\sqrt{2}$ & electric & real \\ \hline
$N \Im \tilde{a}_{0}$ & $0$ & $\sqrt{2}$ & magnetic & imaginary \\ \hline
$N^{2} \Re \tilde{a}_{0}$ &  $-2$ & 2 & electric & imaginary\\ \hline
$N^2 \Im \tilde{a}_{0}$ & $-2$ & 2 & electric & real \\ \hline
\end{tabular}}
\renewcommand{\arraystretch}{1}
\caption{Properties of the charges that couple to the asymptotic graviphoton for $\mathrm{I} \to \mathrm{III}$. The parametrical scaling of the physical charges of $N \Re \tilde{a}_{0}$ and $N \Im \tilde{a}_{0}$ allows us to pick the electric charge by hand, and we chose $N \Re \tilde{a}_{0}$ as electric charge and $N \Im \tilde{a}_{0}$ as dual magnetic charge.}
\label{table:III}
\end{table}

\subsubsection*{Enhancement chain $\mathrm{I} \to \mathrm{II} \to \mathrm{III}$}
Here we consider limits characterized by enhancement chains of the form $\mathrm{I} \to \mathrm{II} \to \mathrm{III}$. The discrete data of such a limit is given by $d_{1},\ldots,d_{k-1} = 0$, $d_{k},\ldots,d_{l-1}=1$ and $d_{l},\ldots,d_{n}=2$. The enhancements to $\mathrm{II}_{b}$ and $\mathrm{III}_{c}$ singularities occur at steps $k$ and $l$ respectively, and we denote the lowering operators $N_{k}^{-}$ and $N_{l}^{-}$ therefore by $N_{\rm II}$ and $N_{\rm III}$. From the discrete data we can infer that $N_{\rm II}$ and $N_{\rm III}$ can both be applied once on $\tilde{a}_{0}$. Moreover, charges obtained from $\Re \tilde{a}_{0}$ and $\Im \tilde{a}_{0}$ are linearly independent, so there are eight charges to consider in total. Their properties have been summarized in table \ref{table:II-III}. From this information we can straightforwardly compute the radii via \eqref{eq:asymptoticradii} to be
\begin{equation}
\begin{aligned}
\gamma_1^{-2} &= \Big( \frac{Q}{M} \Big)^{-2} \Big|_{N_{\rm II} \Re \tilde{a}_{0}} +\Big( \frac{Q}{M} \Big)^{-2} \Big|_{N_{\rm II} N_{\rm III} \Im \tilde{a}_{0}} = \frac{1}{4} + \frac{1}{4} = \frac{1}{2}\, ,  \\
\gamma_1^{-2} &= \Big( \frac{Q}{M} \Big)^{-2} \Big|_{N_{\rm II} \Im \tilde{a}_{0}} +\Big( \frac{Q}{M} \Big)^{-2} \Big|_{N_{\rm II} N_{\rm III} \Re \tilde{a}_{0}} = \frac{1}{4} + \frac{1}{4} = \frac{1}{2}\, ,  \\
\end{aligned}
\end{equation}

\begin{table}[htb]\centering
\renewcommand{\arraystretch}{1.3}
\scalebox{0.85}{
\begin{tabular}{|c|c|c|c|c|}
\hline
charges & sl(2)-level  & $Q/M$ & electric/magnetic & period \\ \hline
$\Re \tilde{a}_{0}$ & $(1,1)$  & 2 & magnetic & imaginary \\ \hline
$\Im \tilde{a}_{0}$ & $(1,1)$ & 2& magnetic & real\\ \hline
$N_{\rm II} \Re \tilde{a}_{0}$ & $(-1,1)$  & 2 & electric & real \\ \hline
$N_{\rm II} \Im \tilde{a}_{0}$ & $(-1,1)$ & 2 & electric & imaginary \\ \hline
$N_{\rm III} \Re \tilde{a}_{0}$ & $(1,-1)$  & 2 & magnetic & real \\ \hline
$N_{\rm III} \Im \tilde{a}_{0}$ & $(1,-1)$ & 2 & magnetic & imaginary \\ \hline
$N_{\rm II} N_{\rm III} \Re \tilde{a}_{0}$ &  $(-1,-1)$ & 2 & electric & imaginary\\ \hline
$N_{\rm II} N_{\rm III} \Im \tilde{a}_0$ & $(-1,-1)$ & 2 & electric & real \\ \hline
\end{tabular}}
\renewcommand{\arraystretch}{1}
\caption{Properties of the charges that couple to the asymptotic graviphoton for $\mathrm{I} \to \mathrm{II} \to \mathrm{III}$. }
\label{table:II-III}
\end{table}

\subsubsection*{Enhancement chain $\mathrm{I} \to \mathrm{IV}$ }
Here we consider limits characterized by enhancement chains of the form $\mathrm{I} \to \mathrm{IV}$. The discrete data of such a limit is given by $d_{1},\ldots,d_{k-1} = 0$ and $d_{k},\ldots,d_{n}=3$. The enhancement to  a $\mathrm{IV}_{d}$ singularity occurs at step $k$, and we denote the lowering operator $N_{k}^{-}$ therefore by $N_{\rm IV}$. From the discrete data we can infer that $N_{\rm IV}$ can be applied three times on $\tilde{a}_{0}$, so there are four charges to consider in total. Their properties have been summarized in table \ref{table:IV}. From this information we can straightforwardly compute the radii via \eqref{eq:asymptoticradii} to be
\begin{equation}
\begin{aligned}
\gamma_1^{-2} = \Big( \frac{Q}{M} \Big)^{-2} \Big|_{N^3 \tilde{a}_0} = \frac{3}{4}\, ,  \qquad \gamma_2^{-2} = \Big( \frac{Q}{M} \Big)^{-2} \Big|_{N^2 \tilde{a}_0} = \frac{1}{4}\, .
\end{aligned}
\end{equation}

\begin{table}[htb]\centering
\renewcommand{\arraystretch}{1.3}
\scalebox{0.85}{
\begin{tabular}{|c|c|c|c|c|}
\hline
charges & sl(2)-levels  & $Q/M$ & electric/magnetic & period \\ \hline
$\tilde{a}_{0}$ & $3$   & 2 & magnetic & imaginary \\ \hline
$N \tilde{a}_{0}$ & $1$  & $2/\sqrt{3}$ & magnetic & real\\ \hline
$ N^2 \tilde{a}_{0}$ & $-1$ & $2/\sqrt{3}$ & real & imaginary\\ \hline
$N^{3} \tilde{a}_{0}$ & $-3$ &  2 &electric & real\\ \hline
\end{tabular}}
\renewcommand{\arraystretch}{1}
\caption{Properties of the charges that couple to the asymptotic graviphoton for $\mathrm{I} \to \mathrm{IV}$.}
\label{table:IV}
\end{table}

\subsubsection*{Enhancement chain $\mathrm{I} \to \mathrm{II} \to \mathrm{IV}$}
Here we consider limits characterized by enhancement chains of the form $\mathrm{I} \to \mathrm{II} \to \mathrm{IV}$. The discrete data of such a limit is given by $d_{1},\ldots,d_{k-1} = 0$, $d_{k},\ldots,d_{l-1}=1$ and $d_{l},\ldots,d_{n}=3$. The enhancements to $\mathrm{II}_{b}$ and $\mathrm{IV}_{d}$ singularities occur at s $k$ and $l$ respectively, and we denote the lowering operators $N_{k}^{-}$ and $N_{l}^{-}$ therefore by $N_{\rm II}$ and $N_{\rm IV}$. From the discrete data we can infer that $N_{\rm II}$ can be applied once on $\tilde{a}_{0}$ and  $N_{\rm IV}$ twice, so there are six charges to consider in total. Their properties have been summarized in table \ref{table:II-IV}. The computation of the radii for these limits depends on the subsector of the growth sector \eqref{eq:growthsector} we consider, since moving between these sectors changes what charges we consider to be electric. Below we go through both sectors.

\begin{table}[htb]\centering
\renewcommand{\arraystretch}{1.3}
\scalebox{0.85}{
\begin{tabular}{|c|c|c|c|c|c|}
\hline
charges & sl(2)-levels  & scaling & $Q/M$ & electric/magnetic & period \\ \hline
$\tilde{a}_{0}$ & $(1,2)$   & $y_{\rm II} (y_{\rm IV})^2$ & 2 & magnetic & imaginary \\ \hline
$N_{\rm II} \tilde{a}_{0}$ & $(-1,2)$  & $\frac{(y_{\rm IV})^2}{y_{\rm II} }$ & $2$ & sector-dep. & real\\ \hline
$ N_{\rm IV} \tilde{a}_{0}$ & $(1,0)$ & $y_{\rm II} $ & $\sqrt{2}$ & magnetic & real\\ \hline
$N_{\rm II} N_{\rm IV} \tilde{a}_{0}$ & $(-1,0)$ & $\frac{1}{y_{\rm II}}$ &  $\sqrt{2}$ &electric & imaginary\\ \hline
$ N^2_{\rm IV} \tilde{a}_{0}$ &  $(1,-2)$  & $\frac{y_{\rm II} }{(y_{\rm IV})^2}$& 2 & sector-dep. & imaginary\\ \hline
$N_{\rm II} N_{\rm IV}^{2}\tilde{a}_{0}$ & $(-1,-2)$  & $ \frac{1}{y_{\rm II}(y_{\rm IV})^2}$ &  2 & electric & real \\ \hline
\end{tabular}}
\renewcommand{\arraystretch}{1}
\caption{Properties of the charges that couple to the asymptotic graviphoton for $\mathrm{I} \to \mathrm{II} \to \mathrm{IV}$.}
\label{table:II-IV}
\end{table}

\textbf{Subsector 1:} $y^{\rm II} \gg (y^{\rm IV})^2$. In this subsector we find that the charge $N_{\rm IV}^{2} \tilde{a}_{0}$ has a decreasing physical charge and is therefore electric, whereas the dual charge $N_{\rm II} \tilde{a}_{0}$ is magnetic. From the information in table \ref{table:II-IV} we then compute the radii via \eqref{eq:asymptoticradii} to be
\begin{equation}
\begin{aligned}
\gamma_1^{-2} &= \Big( \frac{Q}{M} \Big)^{-2} \Big|_{N_{\rm II} \tilde{a}_{0}} + \Big( \frac{Q}{M} \Big)^{-2} \Big|_{N_{\rm II} N_{\rm IV}^{2}\tilde{a}_{0}} = \frac{1}{4} + \frac{1}{4} =\frac{1}{2}\, ,  \\
 \gamma_2^{-2} &= \Big( \frac{Q}{M} \Big)^{-2} \Big|_{N_{\rm II} N_{\rm IV} \tilde{a}_{0}} = \frac{1}{4}\, .
\end{aligned}
\end{equation}
\textbf{Subsector 2:} $y^{\rm II} \ll (y^{\rm IV})^2$. In this subsector we find that the charge $N_{\rm II} \tilde{a}_{0}$ has a decreasing physical charge instead and is therefore electric, whereas the dual charge $N^{2}_{\rm IV} \tilde{a}_{0}$ is now magnetic. From the information in table \ref{table:II-IV} we then compute the radii via \eqref{eq:asymptoticradii} to be
\begin{equation}
\begin{aligned}
\gamma_1^{-2} &= \Big( \frac{Q}{M} \Big)^{-2} \Big|_{N_{\rm II} N_{\rm IV}^{2}\tilde{a}_{0}} = \frac{1}{4}\, ,  \\
 \gamma_2^{-2} &= \Big( \frac{Q}{M} \Big)^{-2} \Big|_{N_{\rm II} N_{\rm IV} \tilde{a}_{0}}+  \Big( \frac{Q}{M} \Big)^{-2} \Big|_{N^2_{\rm IV} \tilde{a}_{0}} = \frac{1}{4}+\frac{1}{2} = \frac{3}{4}\, .
\end{aligned}
\end{equation}

\subsubsection*{Enhancement chain $\mathrm{I} \to \mathrm{III} \to \mathrm{IV}$}
Here we consider limits characterized by enhancement chains of the form $\mathrm{I} \to \mathrm{III} \to \mathrm{IV}$. The discrete data of such a limit is given by $d_{1},\ldots,d_{k-1} = 0$, $d_{k},\ldots,d_{l-1}=2$ and $d_{l},\ldots,d_{n}=3$. The enhancements to $\mathrm{III}_{c}$ and $\mathrm{IV}_{d}$ singularities occur at steps $k$ and $l$ respectively, and we denote the lowering operators $N_{k}^{-}$ and $N_{l}^{-}$ therefore by $N_{\rm III}$ and $N_{\rm IV}$. From the discrete data we can infer that $N_{\rm III}$ can be applied twice on $\tilde{a}_{0}$ and  $N_{\rm IV}$ once, so there are six charges to consider in total. Their properties have been summarized in table \ref{table:III-IV}. From this information we can straightforwardly compute the radii via \eqref{eq:asymptoticradii} to be
\begin{equation}
\begin{aligned}
\gamma_1^{-2} &= \Big( \frac{Q}{M} \Big)^{-2} \Big|_{N_{\rm III} N_{\rm IV}^{2}\tilde{a}_{0}} = \frac{1}{4}\, ,  \\
 \gamma_2^{-2} &= \Big( \frac{Q}{M} \Big)^{-2} \Big|_{N^2_{\rm III} \tilde{a}_{0}} +\Big( \frac{Q}{M} \Big)^{-2} \Big|_{N_{\rm III} N_{\rm IV} \tilde{a}_{0}}= \frac{1}{4}+\frac{1}{2} = \frac{3}{4}\, .
\end{aligned}
\end{equation}

\begin{table}[htb]\centering
\renewcommand{\arraystretch}{1.3}
\scalebox{0.85}{
\begin{tabular}{|c|c|c|c|c|c|}
\hline
charges & sl(2)-levels & scaling & $Q/M$ & electric/magnetic & period \\ \hline
$\tilde{a}_{0}$ & $(2,1)$ & $(y_{\rm III})^{2} y_{\rm IV}$  & 2 & magnetic & imaginary \\ \hline
$N_{\rm III} \tilde{a}_{0}$ & $(0,1)$ &$y_{\rm IV}$ & $\sqrt{2}$ & magnetic & real\\ \hline
$ N_{\rm IV} \tilde{a}_{0}$ & $(2,-1)$&  $\frac{(y_{\rm III})^{2}}{y_{\rm IV}}$ & 2 &magnetic & real\\ \hline
$N_{\rm III}^{2} \tilde{a}_{0}$ & $(-2,1)$& $\frac{y_{\rm IV}}{(y_{\rm III})^{2}}$ &  2 &electric & imaginary\\ \hline
$N_{\rm III} N_{\rm IV} \tilde{a}_{0}$ &  $(0,-1)$ & $\frac{1}{y_{\rm IV}}$ & $\sqrt{2}$ & electric & imaginary\\ \hline
$N^2_{\rm III} N_{\rm IV} \tilde{a}_{0}$ & $(-2,-1)$ & $\frac{1}{(y_{\rm III})^{2}y_{\rm IV}}$ & 2 & electric & real \\ \hline
\end{tabular}}
\renewcommand{\arraystretch}{1}
\caption{Properties of the charges that couple to the asymptotic graviphoton for $\mathrm{I} \to \mathrm{III} \to \mathrm{IV}$.}
\label{table:III-IV}
\end{table}

\subsubsection*{Enhancement chain $\mathrm{I} \to \mathrm{II} \to \mathrm{III} \to \mathrm{IV}$}
Finally we consider limits characterized by enhancement chains of the form $\mathrm{I} \to \mathrm{II} \to \mathrm{III} \to \mathrm{IV}$. The discrete data of such a limit is given by $d_{1},\ldots,d_{k-1} = 0$, $d_{k},\ldots,d_{l-1}=1$, $d_{l},\ldots,d_{m-1}=2$ and $d_{m},\ldots,d_{n}=3$. The enhancements to $\mathrm{II}_{b}$, $\mathrm{III}_{c}$ and $\mathrm{IV}_{d}$ singularities occur at steps $k$, $l$ and $m$ respectively, and we denote the lowering operators $N_{k}^{-}$, $N_{l}^{-}$ and $N_{m}^{-}$ therefore by $N_{\rm II}$, $N_{\rm III}$ and $N_{\rm IV}$. From the discrete data we can infer that $N_{\rm II}$, $N_{\rm III}$ and $N_{\rm IV}$ each can be applied once on $\tilde{a}_{0}$, so there are six charges to consider in total. Their properties have been summarized in table \ref{table:II-III-IV}. The computation of the radii for these limits depends on the subsector of the growth sector \eqref{eq:growthsector} we consider, since moving between these sectors changes what charges we consider to be electric. Below we go through both sectors.

\begin{table}[htb]\centering
\renewcommand{\arraystretch}{1.3}
\scalebox{0.85}{
\begin{tabular}{|c|c|c|c|c|c|}
\hline
charges & sl(2)-levels  & scaling & $Q/M$ & electric/magnetic & period \\ \hline
$\tilde{a}_{0}$ & $(1,1,1)$  & $y_{\rm II}y_{\rm III}y_{\rm IV}$ & 2 & magnetic & imaginary \\ \hline
$N_{\rm II} \tilde{a}_{0}$ & $(-1,1,1)$ & $\frac{y_{\rm III}y_{\rm IV}}{y_{\rm II}}$ & 2 & sector-dep. & real\\ \hline
$N_{\rm III} \tilde{a}_{0}$ & $(1,-1,1)$ & $\frac{y_{\rm II}}{y_{\rm III}y_{\rm IV}}$  & 2 &magnetic & real\\ \hline
$ N_{\rm IV} \tilde{a}_{0}$ & $(1,1,-1)$ & $\frac{y_{\rm II}y_{\rm III}}{y_{\rm IV}}$ & 2 &magnetic & real\\ \hline
$N_{\rm II} N_{\rm III} \tilde{a}_{0}$ &  $(-1,-1,1)$  & $\frac{y_{\rm IV}}{y_{\rm II}y_{\rm III}}$ & 2 & electric & imaginary\\ \hline
$N_{\rm II} N_{\rm IV} \tilde{a}_{0}$ & $(-1,1,-1)$  & $\frac{y_{\rm III}}{y_{\rm II}y_{\rm IV}}$ &  2 &electric & imaginary\\ \hline
$N_{\rm III} N_{\rm IV} \tilde{a}_{0}$ & $(1,-1,-1)$ & $\frac{y_{\rm II}}{y_{\rm III}y_{\rm IV}}$ & 2 & sector-dep. & imaginary\\ \hline
$N_{\rm II} N_{\rm III} N_{\rm IV}\tilde{a}_{0}$ & $(-1,-1,-1)$  & $\frac{1}{y_{\rm II}y_{\rm III}y_{\rm IV}}$ & 2 & electric & real \\ \hline
\end{tabular}}
\renewcommand{\arraystretch}{1}
\caption{Properties of the charges that couple to the asymptotic graviphoton for $\mathrm{I} \to \mathrm{II} \to \mathrm{III} \to \mathrm{IV}$. The distinction between $N_{\rm II} \tilde{a}_{0}$ and $N_{\rm III} N_{\rm IV} \tilde{a}_{0}$ as electric or magnetic charge depends on the subsector of the growth sector that is being considered.}
\label{table:II-III-IV}
\end{table}

\textbf{Subsector 1:} $y^{\rm II}\ll y^{\rm III} y^{\rm IV}$. In this subsector we find that the charge $N_{\rm III} N_{\rm IV} \tilde{a}_{0}$ has a decreasing physical charge and is therefore electric, whereas the dual charge $N_{\rm II} \tilde{a}_{0}$ is magnetic. From the information in table \ref{table:II-III-IV} we then compute the radii via \eqref{eq:asymptoticradii} to be
\begin{align}
\gamma_1^{-2}&=\bigg(\frac{Q}{M}\bigg)^{-2}\bigg|_{N^{-}_{\mathrm{II}} N^{-}_{\mathrm{III}} N^{-}_{\mathrm{IV}}  \tilde{a}_{0}} \!\!\! =   \frac{1}{4} \, ,\\
\gamma_2^{-2}&=\bigg(\frac{Q}{M}\bigg)^{-2}\bigg|_{N^{-}_{\mathrm{II}} N^{-}_{\mathrm{III}} \tilde{a}_{0}} \!\!\!  +\bigg(\frac{Q}{M}\bigg)^{-2}\bigg|_{N^{-}_{\mathrm{II}} N^{-}_{\mathrm{IV}} \tilde{a}_{0}} \!\!\!  +  \bigg(\frac{Q}{M}\bigg)^{-2}\bigg|_{N^{-}_{\mathrm{III}} N^{-}_{\mathrm{IV}} \tilde{a}_{0}} \!\!\!  = \frac{1}{4}+ \frac{1}{4}+\frac{1}{4}=\frac{3}{4}\, . \nn
\end{align}
\textbf{Subsector 2:} $y^{\rm II}\gg y^{\rm III} y^{\rm IV}$. In this subsector we find that the charge $N_{\rm II} \tilde{a}_{0}$ has a decreasing physical charge instead and is therefore electric, whereas the dual charge $N_{\rm III} N_{\rm IV} \tilde{a}_{0}$ is now magnetic. From the information in table \ref{table:II-III-IV} we then compute the radii via \eqref{eq:asymptoticradii} to be
\begin{equation}
\begin{aligned}
\gamma_1^{-2}&=\bigg(\frac{Q}{M}\bigg)^{-2}\bigg|_{N^{-}_{\mathrm{II}} N^{-}_{\mathrm{III}} N^{-}_{\mathrm{IV}}  \tilde{a}_{0}}\!\!\!  +   \bigg(\frac{Q}{M}\bigg)^{-2}\bigg|_{N^{-}_{\mathrm{II}} \tilde{a}_{0}} \!\!\! =   \frac{1}{4}+\frac{1}{4} = \frac{1}{2} \, ,\\
\gamma_2^{-2}&=\bigg(\frac{Q}{M}\bigg)^{-2}\bigg|_{N^{-}_{\mathrm{II}} N^{-}_{\mathrm{III}} \tilde{a}_{0}} \!\!\!  +\bigg(\frac{Q}{M}\bigg)^{-2}\bigg|_{N^{-}_{\mathrm{II}} N^{-}_{\mathrm{IV}} \tilde{a}_{0}} \!\!\!  = \frac{1}{4}+ \frac{1}{4}=\frac{1}{2}\, .
\end{aligned}
\end{equation}

\end{subappendices}

\chapter{Moduli stabilization in asymptotic regimes}\label{chap:modstab}
In this chapter we investigate the stabilization of complex structure moduli after introducing background fluxes. Recall from sections \ref{ssec:IIBN=1} and \ref{ssec:Ftheory} that compactifications of Type IIB on Calabi-Yau orientifolds and F-theory on Calabi-Yau fourfolds yield $\mathcal{N}=1$ supergravity theories. The for us relevant part of the superpotential $W$ and K\"ahler potential $K$ can be evaluated by determining how the holomorphic $(D,0)$-form $\Omega$ varies with a change of the complex structure via the periods. Alternatively, the resulting scalar potential can be written down in terms of the Hodge star operator on $Y_D$ acting on three- or four-form fluxes in the middle cohomology $H^D(Y_D, \mathbb{Z})$. This allows for two equivalent approaches to moduli stabilization:
\begin{itemize}
\item demanding vanishing F-terms: $D_i W=\partial_i W + (\partial_i K) W=0$,
\item solving an (imaginary) self-duality condition: $\ast G_3 = i G_3$ or $\ast G_4 = G_4$.
\end{itemize} 
These complementary strategies allow us to highlight different aspects of how asymptotic Hodge theory captures the dependence on the complex structure moduli. To be precise, the self-duality condition exploits the sl(2)-approximation of the Hodge star operator to the fullest, while the F-terms emphasize the importance of essential exponential corrections in the periods. Below we utilize the sl(2)-approximation to set up a systematic algorithm for finding flux vacua, and the essential instantons to construct vacua with a small superpotential.

\subsubsection*{Approach using the self-duality condition}
In the first part of this chapter we develop a systematic algorithm for performing moduli stabilization near any boundary of moduli space by using the self-duality condition. The moduli dependence of the Hodge star is, in general, given by very complicated transcendental functions that depend on many details of the compactification geometry. However, in the asymptotic regime there exists an approximation scheme to extract the moduli dependence in essentially three steps:
\tikzstyle{block} = [draw,, rectangle, 
    minimum height=3em, minimum width=2em, align=center]
\begin{equation}
\begin{tikzpicture}[baseline = -0.4ex]
\node[block, text width=3.6cm] (a) at (0,0) {\small (1) sl(2)-approximation};
\node[block, text width=2.9cm] (b) at (4.3,0) {\small (2) Nilpotent orbit approximation};
\node[block, text width=3.7cm] (c) at (8.5,0) {\small (3) Full series of \\
exponential corrections };

\draw [ thick, ->] (a) -- (b);
\draw [ thick, ->] (b) -- (c);
\end{tikzpicture} \nn
\end{equation}
It is important to stress that it is a very non-trivial fact that such approximations exist in all asymptotic regimes. In particular, this is remarkable because from the analysis in chapter \ref{chap:models} we observed that in almost all asymptotic regimes exponential corrections are essential when deriving the Hodge star from the period integrals of the unique $(3,0)$-form or $(4,0)$-form of the Calabi-Yau manifold. The determination of the approximated Hodge star is non-trivial, but can be done explicitly for any given example following the methods we laid out in Part \ref{part1} of this thesis. 

The different degrees of approximation given break down the moduli dependence of the scalar potential. By using the sl(2)-approximation as a first step we reduce to polynomial terms in the complex structure moduli. This allows us to solve the relevant extremization conditions for the vevs of the moduli straightforwardly. Subsequently, we can include corrections to this polynomial behavior by switching to the nilpotent orbit approximation. Roughly speaking, this amounts to dropping all exponential corrections in the scalar potential,\footnote{We should stress that one has to be careful where these exponential corrections are dropped. While we can drop corrections at these orders in the scalar potential, essential exponential corrections have to be included in its defining K\"ahler potential and flux superpotential. These terms are required to be present for consistency of the period vector following the discussion in chapter \ref{chap:models}, and will also have to be included in the F-terms later in this chapter.} but including corrections to the simple polynomial terms of the sl(2)-approximation, yielding algebraic equations in the moduli. Taking the flux vacua found in the sl(2)-approximation as input, we then iterate a numerical program to extremize the scalar potential in the nilpotent orbit approximation. Finally, one can include exponentially small terms in the saxions in the scalar potential. In this work we aim to stabilize all moduli at the level of the nilpotent orbit, but in principle one could use these corrections to lift flat directions in the first two steps. 

We demonstrate our moduli stabilization algorithm on a set of Type IIB and F-theory examples. For the Calabi-Yau threefold examples we consider not only the large complex structure regime, but also so-called conifold-large complex structure boundaries that are not straightforwardly accessible via large complex structure/large volume mirror symmetry. We compare the sl(2)-approximation with the nilpotent orbit approximation, and find that the former actually provides a rather good approximation, even for a mild hierarchy among the moduli. For the Calabi-Yau fourfold examples we consider the large complex structure regime, and to be more precise the so-called linear scenario recently proposed in \cite{Marchesano:2021gyv}. This proves to be an interesting case where the sl(2)-approximation possesses a flat direction that is subsequently lifted by corrections in the nilpotent orbit approximation. 

\subsubsection*{Approach using the F-terms}
The second part of this chapter concerns the F-term approach, where for definiteness we restrict our attention to Calabi-Yau threefolds. The dependence of the F-terms on the complex structure moduli is encoded in the unique $(3,0)$-form $\Omega$, which can be conveniently captured by the period integrals of $\Omega$ over a basis of three-cycles of $Y_3$. There are a number of techniques available to derive these period integrals. While most of them require the explicit construction of the Calabi-Yau manifold, we build upon the general models for the asymptotic periods we constructed in chapter \ref{chap:models} using asymptotic Hodge theory. Our goal is to use the essential exponential corrections in these periods to engineer vacua with a small flux superpotential, which are important in certain phenomenological constructions.

One of the most prominent moduli stabilization scenarios, the KKLT scenario \cite{Kachru:2003aw}, requires to find vacua in complex structure moduli space, such that the vacuum superpotential is taking a very small, non-zero value.\footnote{In the mathematical literature, the vacua with exponentially small superpotential are a special case of the so-called extended locus of Hodge classes \cite{schnell2014extended}.}  It was recently suggested in \cite{Demirtas:2019sip} and further explored in \cite{Demirtas:2020ffz,Blumenhagen:2020ire,Honma:2021klo,Demirtas:2021nlu, Demirtas:2021ote,Broeckel:2021uty,Carta:2021kpk,Carta:2022oex} that, in fact, exponentially small vacuum superpotentials can be found near large complex structure and conifold-large complex structure boundaries.\footnote{See \cite{Denef:2004ze, Eguchi:2005eh} for statistical arguments about the abundance of vacua in these regions.} It was argued that the construction proceeds by introducing fluxes that preserve a continuous version of the monodromy symmetry at some leading order with a vanishing superpotential, while then including instanton corrections generates an exponentially small superpotential. We will explain how this construction is understood in asymptotic Hodge theory and how it can be generalized to other boundaries that have not been consider in the literature before. This will be done both in a hands-on way by considering explicit examples as well as by explaining how the abstract mathematical methods are useful. We will then see that our powerful techniques also allow us to control the moduli masses and identify alternative scenarios to the ones of \cite{Demirtas:2019sip,Demirtas:2020ffz}: in our new constructions the scaling of the masses is polynomial, rather than exponential, in the vacuum expectation values of the moduli, so these are parametrically heavier compared to the masses of the K\"ahler moduli.

Using the classification of asymptotic regimes \cite{Kerr2017,Grimm:2018cpv,Grimm:2019bey} we are able to identify a specific class of regions that generally admits essential instanton corrections that can set the scale of the vacuum superpotential. To be precise, our vacua arise near Type II points according to table \ref{table:HDclass} (signalling the presence of a K3 fibration in the mirror manifold) away from the large complex structure regime. Among the set of essential instantons in the flux superpotential we then identify a subset that enters the scalar potential already at polynomial order. These are precisely the metric-essential instanton corrections that are required to ensure a non-degenerate moduli metric, and induce the polynomial behavior of the moduli masses in the vacuum. We exemplify this discussion with one and two-moduli examples from chapter \ref{chap:models}, allowing us to work out the moduli stabilization explicitly. We should mention that the period models used in this thesis are given in a real rather than a rational symplectic basis. Our vacua do not require any fine-tuning of the flux quanta in contrast to e.g.~a racetrack potential, so we do not expect any technical issues to arise here. Nevertheless it would be interesting to work out the quantization of the fluxes in the future.



This chapter is organized as follows. In section \ref{sec:stabselfdual} we reduce the self-duality condition for the fluxes to the nilpotent orbit and sl(2)-approximation, the two regimes of interest for our moduli stabilization scheme. In section \ref{sec:IIBexamples} and \ref{sec:Ftheoryexamples} we then demonstrate this algorithmic approach on explicit Type IIB and F-theory examples. In section \ref{ssec:strategy} we outline our strategy for constructing vacua with a small flux superpotential by using the F-terms, for which in section \ref{sec:examples} we include some one- and two-moduli examples.

\section{Moduli stabilization using the self-duality condition}\label{sec:stabselfdual}
In this section we show how the self-duality condition simplifies when moving to asymptotic regimes in the moduli space. For simplicity we stick to the Type IIB setting here, although one can easily generalize to the F-theory setting by replacing $G_3$ with $G_4$ and removing factors of $i$ for the imaginary self-duality condition.

\subsubsection*{Sl(2)-approximation}
As a starting point, we study the extremization conditions for flux vacua in the sl(2)-approximation \eqref{eq:growthsector} for the saxions, i.e.~$y^1 \gg y^2 \gg \ldots \gg y^n$. Recall that the Hodge star in this regime can be approximated by the operator $C_{\rm sl(2)}$ given by \eqref{eq:Csl2ToCInf}. By writing the self-duality condition with $C_{\rm sl(2)}$ we then obtain
\begin{equation}\label{self_dual}
C_{\rm sl(2)}(x,y) \ G_3
= i G_3 \, .
\end{equation}
This approximation drastically simplifies how the Hodge star operator depends on the moduli. For the axions the dependence enters only through factors $e^{- x^k N_k^-} $. It is therefore useful to absorb these into $G_3$ by defining flux axion polynomials
\begin{equation}\label{axion_polynomials}
\rho_{\rm sl(2)}(x) \equiv  e^{- x^k N_k^-} G_3  \, .
\end{equation}
Note that $\rho_{\rm sl(2)}$ is invariant under monodromy transformations $x^i \to x^i +1$ when we shift the fluxes according to $G_3 \to e^{N_i^-} G_3$.

For the saxions we can make the dependence explicit by decomposing into eigenspaces of the weight operators $N_i^0$ of the $\mathfrak{sl}(2, \mathbb{R})$-triples. By applying this decomposition for $\rho_{\rm sl(2)}$ we obtain
\begin{equation}\label{rho_sl2}
\qquad \rho_{\rm sl(2)}(x^i)  = \sum_{\ell \in \cE}  \rho_\ell (x^i)\, , \qquad N_i^0 \rho_\ell(x^i) = \ell_i \rho_\ell(x^i)\, ,
\end{equation}
where $\ell = (\ell_1, \ldots , \ell_n)$, and $\cE$ denotes the index set that specifies which values the $\ell_i$ can take.  Generally one has bounds $-d \leq \ell_i \leq d$, with $d$ the complex dimension of the underlying Calabi-Yau manifold, which here is $d=3$. This decomposition \eqref{rho_sl2} allows us to write out the self-duality condition \eqref{self_dual} in terms of the sl(2)-approximated Hodge star \eqref{eq:Csl2ToCInf} componentwise as
\begin{equation}\label{self_dual_sl2}
y_1^{\ell_1} \cdots y_n^{\ell_n} \, C_\infty \, \rho_\ell (x^i) =     \rho_{-\ell} (x^i) \, , \qquad \ell \in \cE \, ,
\end{equation}
where $C_\infty$ is the boundary Hodge star operator independent of the limiting coordinates.\footnote{The operator \eqref{Cinfty} is also constructed as a product of the sl(2)-approximation. It can be understood as a Hodge star operator attached to the boundary for the Hodge structure \eqref{Fpinfty} arising in the asymptotic limit.} With \eqref{self_dual_sl2} we now have obtained a simple set of polynomial equations that one can solve straightforwardly.

\subsubsection*{Nilpotent orbit approximation}
For the next step in our approximation scheme we proceed to the nilpotent orbit, which amounts to dropping some exponential corrections. Generally speaking, the corresponding Hodge star operator $C_{\rm nil}$ then depends on these coordinates through algebraic functions, in contrast to more general transcendental functions when exponential corrections are included. On the other hand, in comparison to the sl(2)-approximation one finds that $C_{\rm nil}$ contains series of corrections to $C_{\rm sl(2)}$ when expanding in $y_1/y_2, \ldots, y_{n-1}/y_n, 1/y_n \ll 1$. For now, let us simply state the self-duality condition at the level of the nilpotent orbit as
\begin{equation}\label{self_dual_nil}
C_{\rm nil} \, G_3 = i G_3\, .
\end{equation}
Next, we write out the dependence of $C_{\rm nil}$ on the coordinates in a similar manner as we did for the sl(2)-approximation. The dependence on the axions again factorizes as
\begin{equation}
C_{\rm nil} (x^i, y^i)= e^{x^k N_k} \, \hat{C}_{\rm nil} (y^i)\,  e^{-x^k N_k} \, , \qquad \hat{C}_{\rm nil}  (y^i)= C_{\rm nil}(0, y^i) \, ,
\end{equation}
where here the log-monodromy matrices $N_i$ appear instead of the lowering operators $N_i^-$ of the $\mathfrak{sl}(2,\mathbb{R})$-triples in \eqref{eq:Csl2ToCInf}. In analogy to \eqref{axion_polynomials} we then absorb the axion dependence into flux axion polynomials as
\begin{equation}
\rho_{\rm nil} (x^i) = e^{-x^k N_k} G_3\, ,
\end{equation}
which are invariant under monodromy transformations $x^i \to x^i + 1$ if we redefine the fluxes according to $G_3 \to e^{N_i}G_3$. The dependence of $\hat{C}_{\rm nil}$ on the saxions $y^i$ is considerably more complicated than that of the sl(2)-approximated Hodge star operator $C_{\rm sl(2)}$, so we do not write this out further.\footnote{A systematic approach to incorporate these corrections in $y_1/y_2, \ldots, y_{n-1}/y_n, 1/y_n $ to the Hodge star operator was put forward in the holographic perspective of \cite{Grimm:2020cda, Grimm:2021ikg}.} The self-duality condition \eqref{self_dual_nil} can then be rewritten as
\begin{equation}\label{self_dual_nil_expanded}
\hat{C}_{\rm nil}(y^i)\,  \rho_{\rm nil}(x^i) = i \rho_{\rm nil}(x^i)\, .
\end{equation}
This extremization condition can be expanded in any basis of choice for the fluxes by writing out $\rho_{\rm nil}$ componentwise, yielding a system of algebraic equations in the moduli. Our approach in this work is to take solutions of \eqref{self_dual_sl2} as input, and then slowly vary the moduli vevs to find a solution to \eqref{self_dual_nil_expanded}.

In the following two sections we exemplify the proposed moduli stabilization scheme in Type IIB and F-theory. In Type IIB we consider examples where we stabilize all moduli, and compare the accuracy of our approximations by looking at the displacement of the vacua. In F-theory we consider the so-called linear scenario and show how flat directions in the sl(2)-approximations are lifted by corrections in the nilpotent orbit. We also make some comments about these vacuum loci regarding swampland conjectures, with a particular focus on the tadpole conjecture for (the Type IIB version of) the linear scenario.

\section{Self-duality: Type IIB examples}\label{sec:IIBexamples}
We first demonstrate our strategy on a set of Type IIB examples. We consider an example at large complex structure with $h^{2,1}=2$ and another where one modulus is sent close to a conifold locus with $h^{2,1}=3$.

\subsection{Large complex structure limit for $h^{2,1} = 2$}\label{ssec:LCS2}
Let us consider the example discussed in section~\ref{sec_aht_example}. 
We recall the triple intersection numbers of the mirror threefold  from 
\eqref{data_ex_001} as 
\begin{equation}
\kappa_{122}  = 4\, , \hspace{40pt} \kappa_{222} = 8\, ,
\end{equation}
which specify the  K\"ahler metric in the complex structure sector 
via the periods \eqref{eq:periodsexample}. The sl(2)-approximated 
Hodge-star operator in the regime
 $y^1 \gg y^2 \gg 1$
has been shown in \eqref{data_ex_csl2},
where we note that the full polynomial periods contain more detailed information 
about the moduli-space geometry 
than the sl(2)-approximation. In order to 
compare these two approaches, we define the relative difference 
\eq{
\label{diff_01}
  \Delta = \frac{|| \phi^{\rm nil} - \phi^{\rm sl(2)}||}{|| \phi^{\rm nil}||} \,,
}
where $\phi=(y^i,x^i,s,c)$ and where for simplicity we used the Euclidean norm.
To illustrate this point, let us give an explicit example:
\eq{
  \renewcommand{\arraystretch}{1.2}
  \arraycolsep10pt
  \begin{array}{||
  l@{\hspace{2pt}}c@{\hspace{2pt}}l
  ||
  l@{\hspace{2pt}}c@{\hspace{2pt}}l
  ||
  l@{\hspace{2pt}}c@{\hspace{2pt}}l
  ||}
  \hline\hline
  \multicolumn{3}{||c||}{\mbox{fluxes}}
  &
  \multicolumn{3}{c||}{\mbox{sl(2)-approximation}}  
  &
  \multicolumn{3}{c||}{\mbox{large complex structure}}  
  \\ \hline
  h^I &=& (0,0,1) 
  & y^i &=& (9.72,2.79) 
  & y^i &=& (7.85,2.79)   
  \\
  h_I &=& (-162,0,0) 
  & x^i &=& (0.00,0.00) 
  & x^i &=& (0.00,0.00)   
  \\
  f^I &=& (1,0,0) 
  &  s &=& 0.93
  &  s &=& 0.93  
  \\
  f_I &=& (0,1,37) 
  & c &=& 0.00
  & c &=& 0.00  
  \\
  \hline
  N_{\rm flux}&=&-199 & \multicolumn{6}{c||}{\Delta=0.22}
  \\
  \hline\hline
  \end{array}
\nn}
The choice of $H_3$ and $F_3$ fluxes together with their tadpole contribution (c.f.~ equation \eqref{tad_01})
is shown in the first column, in the second column we show the location of the minimum 
in the sl(2)-approximation, and in the third column the location of the minimum 
determined using the full polynomial periods (i.e.~the nilpotent orbit) is shown.
These two loci agree reasonably-well even for the small hierarchy of three, and their relative 
difference $\Delta$ is $22\%$.

Next, we want to investigate how well the sl(2)-approximation to the Hodge-star operator 
agrees with the large complex structure result depending on the hierarchy of the 
saxions. We implement the hierarchy through a parameter $\lambda$ as follows
\eq{\label{scaling_two_moduli}
  y^1=\lambda^2\,, \hspace{40pt}
  y^2=\lambda\,, 
}  
and for larger  $\lambda$ we expect a better agreement between the two approaches. 
We have considered three different families of fluxes characterized by 
different initial choices for $h^I$, $f^I$ and the axions $x^i$. 
The dependence of the relative difference $\Delta$ on the hierarchy parameter $\lambda$ 
is shown  in figure~\ref{fig_lcs2_01}, and
in figure~\ref{fig_lcs2_02} we shown the dependence of the (absolute value) of the tadpole
contribution $N_{\rm flux}$ on $\lambda$. 
We  make the following observations:
\begin{itemize}

\item As it is expected, for a large hierarchy $\lambda$ the sl(2)-approximation 
agrees better with the large complex structure result. Somewhat unexpected 
is however how quickly a reasonable agreement is achieved, for instance, for a 
hierarchy of $\lambda=6$ the two approaches agree up to a difference of $15\%$. 

\item Furthermore, it is also expected that when approaching the boundary
in moduli space the tadpole contribution increases, however, the rate 
with which the tadpole increases is higher than naively expected.
For the family corresponding to the green curve in figures~\ref{plot_lcs2_all}
the tadpole dependence can be fitted as
\eq{  N_{\rm flux} = 8.27\, \lambda^{3.99}+34.69 \,,}
which is in good agreement with the data for $\lambda\geq 2$. Thus, for 
these examples, when approaching the large complex structure limit the tadpole
contribution increases rapidly. The scaling in $\lambda$ of the tadpole can also be understood from the weights of the fluxes under the sl(2)-triples. The heaviest charge is $(1,0,0,0,0,0)$ with weights $(\ell_1,\ell_2)=(1,2)$ under $N^0_1,N^0_2$. Using \eqref{eq:Csl2ToCInf} for the asymptotic behavior of the Hodge star and plugging in \eqref{scaling_two_moduli} for the saxions we find that
\begin{equation}
N_{\rm flux} \sim (y^1)^{\ell_1} (y^2)^{\ell_2} = \lambda^4\, .
\end{equation}
\end{itemize}
\vspace*{-0.5cm}
\begin{figure}[h!]
\centering
\begin{subfigure}{0.45\textwidth}
\centering
\includegraphics[width=160pt]{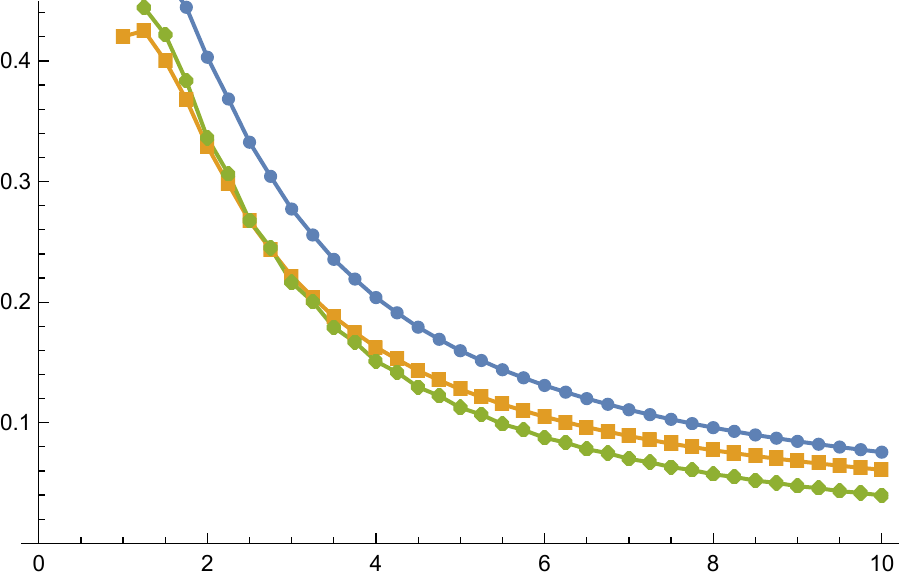}%
\begin{picture}(0,0)
\put(-6,-9){\footnotesize$\lambda$}
\put(-165,97){\footnotesize$\Delta$}
\end{picture}
\caption{Dependence of $\Delta$ on $\lambda$.\label{fig_lcs2_01}}
\end{subfigure}
\hspace{20pt}
\begin{subfigure}{0.45\textwidth}
\centering
\includegraphics[width=160pt]{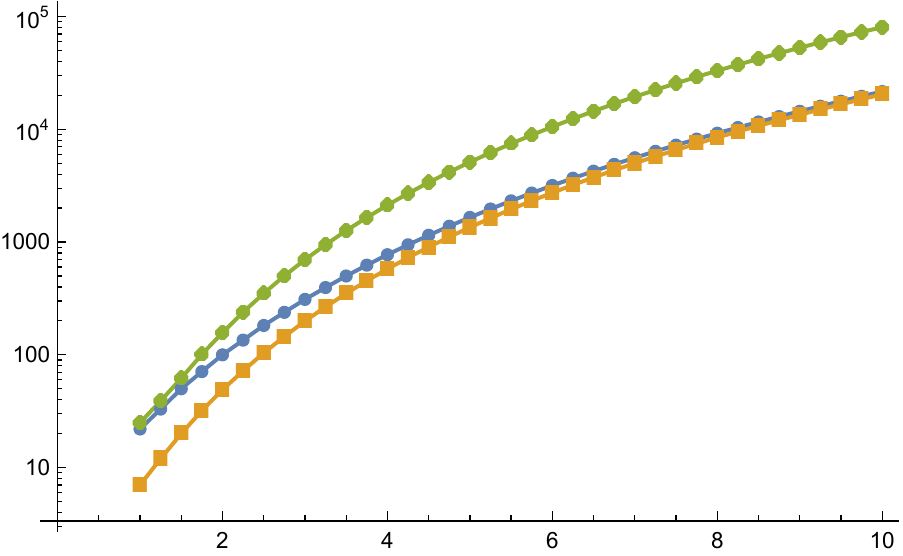}%
\begin{picture}(0,0)
\put(-6,-9){\footnotesize$\lambda$}
\put(-181,97){\footnotesize$N_{\rm flux}$}
\end{picture}
\caption{Log-dependence of $N_{\rm flux}$ on $\lambda$.\label{fig_lcs2_02}}
\end{subfigure}
\caption{Large complex structure limit for $h^{2,1}=2$ for three families: 
dependence of the relative difference $\Delta$ and of the tadpole contribution 
$N_{\rm flux}$ on the hierarchy parameter $\lambda$. \label{plot_lcs2_all}}
\end{figure}
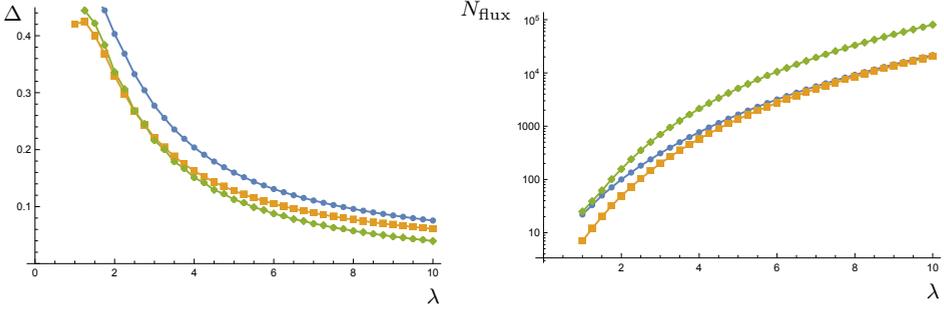



\subsection{Conifold--large complex structure limit for $h^{2,1} = 3$}\label{ssec:coniLCS}

As a third example we consider a combined conifold and large complex structure limit. 
We choose  $h^{2,1}=3$, and send one saxion to a conifold locus in moduli 
space and the remaining two saxions to the large complex structure point. 
This example has been considered before in \cite{Blumenhagen:2020ire}, but here we neglect some of the 
instanton contributions to the prepotential. In particular, for our purposes it
is sufficient to consider the following  prepotential  
\begin{small}
\eq{
\label{prepot_002}
  \mathcal F =  -\frac{1}{3!} \frac{\cK_{ijk} X^i X^j X^k}{X^0}  + \frac{1}{2}\op a_{ij} \op X^i X^j + b_i\op X^i X^0 
  + c\op \bigl(X^0\bigr)^2 - \frac{\bigl(X^3\bigr)^2 }{2\pi \op i} \log\left[ \frac{X^3}{X^0} \right] ,
}
\end{small}
where $i,j,k,=1,\ldots,3$. The non-trivial triple intersection numbers $\cK_{ijk}$ and 
 constants $a_{ij}$ and $b_i$ are given by
\eq{
  \begin{array}{lcl}
  \cK_{111} &=& 8\,, \\
  \cK_{112} &=& 2\,, \\
  \cK_{113} &=& 4\,, \\
  \cK_{123} &=& 1\,, \\
  \cK_{133} &=& 2\,, 
  \end{array}
  \hspace{40pt}
  a_{33} = \textstyle{ \frac{1}{2} + \frac{3-2\log[2\pi]}{2\pi\op i}\,,}
  \hspace{40pt}
  \begin{array}{lcl}
  b_{1} &=& \frac{23}{6}\,, \\
  b_{2} &=& 1\,, \\
  b_{3} &=& \frac{23}{12}\,.
  \end{array}
}
The constant $c$ can be set to zero for the limit we are interested in for simplicity. After computing 
the periods $\partial_I \mathcal F$ and the matrix $\partial_I\partial_J\mathcal F$, we 
set $t^i = X^i/X^0$ and $X^0=1$, and we 
perform a further field redefinition of the form 
\eq{
  t^1 \to \tilde t^1  \,,\qquad
  t^2 \to \tilde t^2+ 4\op \tilde t^3  \,,\qquad
  t^3 \to e^{2\pi \op i\op \tilde t^3} \,,
}
where the tildes will be omitted in the following. 

Let us now consider the sl(2)-approximation in the regime $y^3 \gg y^2 \gg y^1 \gg 1$. Using the periods following from the prepotential \eqref{prepot_002} and the algorithm of section \ref{sec_tech_details} we construct the sl(2)-approximated Hodge star \eqref{eq:Csl2ToCInf}. The relevant building blocks are the sl$(2,\mathbb{R})$-triples

\begin{footnotesize}
\begin{align}
N_1^- &= \scalebox{0.8}{$ \left(
\begin{array}{cccccccc}
 0 & 0 & 0 & 0 & 0 & 0 & 0 & 0 \\
 1 & 0 & 0 & 0 & 0 & 0 & 0 & 0 \\
 -\frac{4}{3} & 0 & 0 & 0 & 0 & 0 & 0 & 0 \\
 0 & 0 & 0 & 0 & 0 & 0 & 0 & 0 \\
 5 & 0 & 0 & 0 & 0 & -1 & \frac{4}{3} & 0 \\
 0 & -\frac{16}{3} & -2 & -\frac{8}{3} & 0 & 0 & 0 & 0 \\
 0 & -2 & 0 & -1 & 0 & 0 & 0 & 0 \\
 0 & -\frac{8}{3} & -1 & -\frac{4}{3} & 0 & 0 & 0 & 0 \\
\end{array}
\right)$} \,,  &N^0_1 &=\scalebox{0.8}{$ \left(
\begin{array}{cccccccc}
 2 & 0 & 0 & 0 & 0 & 0 & 0 & 0 \\
 0 & 0 & 0 & 0 & 0 & 0 & 0 & 0 \\
 0 & \frac{8}{3} & 2 & \frac{4}{3} & 0 & 0 & 0 & 0 \\
 0 & 0 & 0 & 0 & 0 & 0 & 0 & 0 \\
 0 & \frac{31}{3} & 4 & \frac{31}{6} & -2 & 0 & 0 & 0 \\
 \frac{31}{3} & 0 & 0 & 0 & 0 & 0 & -\frac{8}{3} & 0 \\
 4 & 0 & 0 & 0 & 0 & 0 & -2 & 0 \\
 \frac{31}{6} & 0 & 0 & 0 & 0 & 0 & -\frac{4}{3} & 0 \\
\end{array}
\right)$}  \,, \nn 
\end{align}
\end{footnotesize}

\begin{footnotesize}
\begin{align}
N_2^- &= \scalebox{0.8}{$\left(
\begin{array}{cccccccc}
 0 & 0 & 0 & 0 & 0 & 0 & 0 & 0 \\
 0 & 0 & 0 & 0 & 0 & 0 & 0 & 0 \\
 0 & 0 & 0 & 0 & 0 & 0 & 0 & 0 \\
 0 & 0 & 0 & 0 & 0 & 0 & 0 & 0 \\
 0 & 0 & 0 & 0 & 0 & 0 & 0 & 0 \\
 0 & 0 & 0 & 0 & 0 & 0 & 0 & 0 \\
 0 & 0 & 0 & 0 & 0 & 0 & 0 & 0 \\
 0 & 0 & 0 & \frac{1}{2} & 0 & 0 & 0 & 0 \\
\end{array}
\right)$}  \,,  \quad &N^0_2 &=\scalebox{0.8}{$\left(
\begin{array}{cccccccc}
 0 & 0 & 0 & 0 & 0 & 0 & 0 & 0 \\
 0 & 0 & 0 & -\frac{1}{2} & 0 & 0 & 0 & 0 \\
 0 & 0 & 0 & 0 & 0 & 0 & 0 & 0 \\
 0 & 0 & 0 & 1 & 0 & 0 & 0 & 0 \\
 0 & 0 & 0 & 0 & 0 & 0 & 0 & 0 \\
 0 & 0 & 0 & 0 & 0 & 0 & 0 & 0 \\
 0 & 0 & 0 & 0 & 0 & 0 & 0 & 0 \\
 0 & 0 & 0 & 1 & 0 & \frac{1}{2} & 0 & -1 \\
\end{array}
\right)$}\, , \\
N_3^- &= \scalebox{0.8}{$\left(
\begin{array}{cccccccc}
 0 & 0 & 0 & 0 & 0 & 0 & 0 & 0 \\
 0 & 0 & 0 & 0 & 0 & 0 & 0 & 0 \\
 4 & 0 & 0 & 0 & 0 & 0 & 0 & 0 \\
 0 & 0 & 0 & 0 & 0 & 0 & 0 & 0 \\
 8 & 0 & 0 & 0 & 0 & 0 & -4 & 0 \\
 0 & -8 & 0 & -4 & 0 & 0 & 0 & 0 \\
 0 & 0 & 0 & 0 & 0 & 0 & 0 & 0 \\
 0 & -4 & 0 & -2 & 0 & 0 & 0 & 0 \\
\end{array}
\right)$}  \,, &N^0_3 &= \scalebox{0.8}{$ \left(
\begin{array}{cccccccc}
 1 & 0 & 0 & 0 & 0 & 0 & 0 & 0 \\
 0 & 1 & 0 & \frac{1}{2} & 0 & 0 & 0 & 0 \\
 0 & -\frac{8}{3} & -1 & -\frac{4}{3} & 0 & 0 & 0 & 0 \\
 0 & 0 & 0 & 0 & 0 & 0 & 0 & 0 \\
 0 & 5 & 0 & \frac{5}{2} & -1 & 0 & 0 & 0 \\
 5 & 0 & 0 & 0 & 0 & -1 & \frac{8}{3} & 0 \\
 0 & 0 & 0 & 0 & 0 & 0 & 1 & 0 \\
 \frac{5}{2} & 0 & 0 & 0 & 0 & -\frac{1}{2} & \frac{4}{3} & 0 \\
\end{array}
\right)$}  \,, \nn
\end{align}
\end{footnotesize}
\noindent and the boundary Hodge star
\begin{equation}
C_\infty =\scalebox{0.85}{$  \left(
\begin{array}{cccccccc}
 0 & -\frac{23}{24} & -\frac{1}{4} & -\frac{23}{48} & \frac{1}{4} & 0 & 0 & 0 \\
 -\frac{5}{16} & 0 & 0 & \frac{1}{2} & 0 & \frac{5}{8} & -\frac{1}{6} & -1 \\
 -\frac{43}{12} & 0 & 0 & 0 & 0 & -\frac{1}{6} & \frac{38}{9} & 0 \\
 0 & 0 & 0 & -1 & 0 & -1 & 0 & 2 \\
 -\frac{281}{32} & 0 & 0 & 0 & 0 & \frac{5}{16} & \frac{43}{12} & 0 \\
 0 & -\frac{1745}{144} & -\frac{31}{24} & -\frac{1745}{288} & \frac{23}{24} & 0 & 0 & 0 \\
 0 & -\frac{31}{24} & -\frac{1}{2} & -\frac{31}{48} & \frac{1}{4} & 0 & 0 & 0 \\
 0 & -\frac{1745}{288} & -\frac{31}{48} & -\frac{2321}{576} & \frac{23}{48} & -\frac{1}{2} & 0 & 1 \\
\end{array}
\right)$}\, .
\end{equation}
To compare moduli stabilization within the sl(2)-approximation and the nilpotent orbit approximation
in the conifold-large complex structure limit (coni-LCS), we consider
first the following example
\begin{small}
\eq{
  \renewcommand{\arraystretch}{1.2}
  \arraycolsep10pt
  \begin{array}{||
  l@{\hspace{2pt}}c@{\hspace{2pt}}l
  ||
  l@{\hspace{2pt}}c@{\hspace{2pt}}l
  ||
  l@{\hspace{2pt}}c@{\hspace{2pt}}l
  ||}
  \hline\hline
  \multicolumn{3}{||c||}{\mbox{fluxes}}
  &
  \multicolumn{3}{c||}{\mbox{sl(2)-approximation}}  
  &
  \multicolumn{3}{c||}{\mbox{coni-LCS}}  
  \\ \hline
  h^I &=& (1,0,0,1) 
  & y^i &=& (1.88,3.78,8.20) 
  & y^i &=& (1.88,2.61,6.92)   
  \\
  h_I &=& (3,-60,1,-30) 
  & x^i &=& (-0.14,-1.27,0.99) 
  & x^i &=& (-0.14,-1.27,1.35)   
  \\
  f^I &=& (0,1,0,0) 
  &  s &=& 1.05
  &  s &=& 1.05  
  \\
  f_I &=& (132,32,0,18) 
  & c &=& 0.07
  & c &=& 0.07  
  \\
  \hline
  N&=&-210 & \multicolumn{6}{c||}{\Delta=0.22}
  \\
  \hline\hline
  \end{array}\nn
}
\end{small}
The hierarchy in the sl(2)-approximation has been chosen as a factor of two, 
the relative difference to the moduli stabilized via the coni-LCS prepotential 
\eqref{prepot_002} for this example is $22\%$.

Next, we want to study how well the sl(2)-approximation of the Hodge-star operator
captures moduli stabilization via the coni-LCS prepotential. We follow a strategy similar 
to the previous example and implement 
a hierarchy for the saxions through a parameter $\lambda$ as
\eq{\label{scaling_coniLCS}
  y^1=\lambda\,, \hspace{40pt}
  y^2=\lambda^2\,, \hspace{40pt}
  y^3=\lambda^3\,. 
}  
As before, we expect that  for larger  $\lambda$ the agreement between the two approaches
improves. 
We have again considered three different families of fluxes characterized by 
different initial choices for $h^I$, $f^I$ and the axions $x^i$. 
The dependence of the relative difference $\Delta$ on the hierarchy parameter $\lambda$ 
is shown  in figure~\ref{fig_clcs_01}, and
in figure~\ref{fig_clcs_02} we shown the dependence of the (absolute value) of the tadpole
contribution $N_{\rm flux}$ on $\lambda$. 
Our observations agree with the previous example, in particular:
\begin{itemize}

\item we see that even for a small hierarchy of $\lambda=4$ the relative difference between 
the stabilized moduli is only around $10\%$.

\item The tadpole contribution \eqref{tad_01} grows rapidly with the hierarchy parameter 
$\lambda$. For instance, for the orange curve in figure~\ref{fig_clcs_02} we obtain a fit 
\eq{
  N_{\rm flux} = 4.29\, \lambda^{4.97}+96.71 \,,
}
which is in good agreement with the data for $\lambda\geq3$. Therefore, also for
this example the tadpole contribution increases rapidly when approaching the boundary
in moduli space. Again this scaling is understood from the growth of the Hodge star for the heaviest charge $(1, 0, 0, 0, 0, 23/6, 1, 23/12)$. Using \eqref{eq:Csl2ToCInf} we find that
\begin{equation}
N_{\rm flux} \sim (y^1)^{\ell_1} (y^2)^{\ell_2} (y^3)^{\ell_3} = \lambda^5\, ,
\end{equation}
where we used that its weights under the $N^0_i$ are $(\ell_1,\ell_2,\ell_3)=(2,0,1)$, and plugged in the scaling \eqref{scaling_coniLCS} of the saxions in $\lambda$.

\end{itemize}

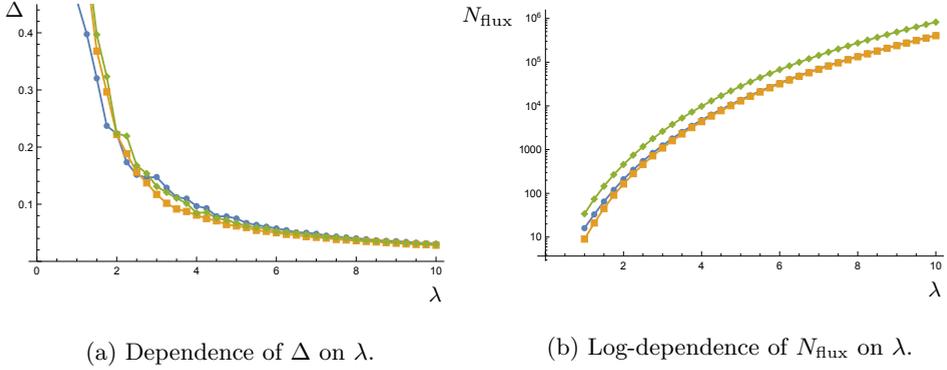
\begin{figure}
\centering
\vspace*{15pt}
\begin{subfigure}{0.45\textwidth}
\centering
\includegraphics[width=160pt]{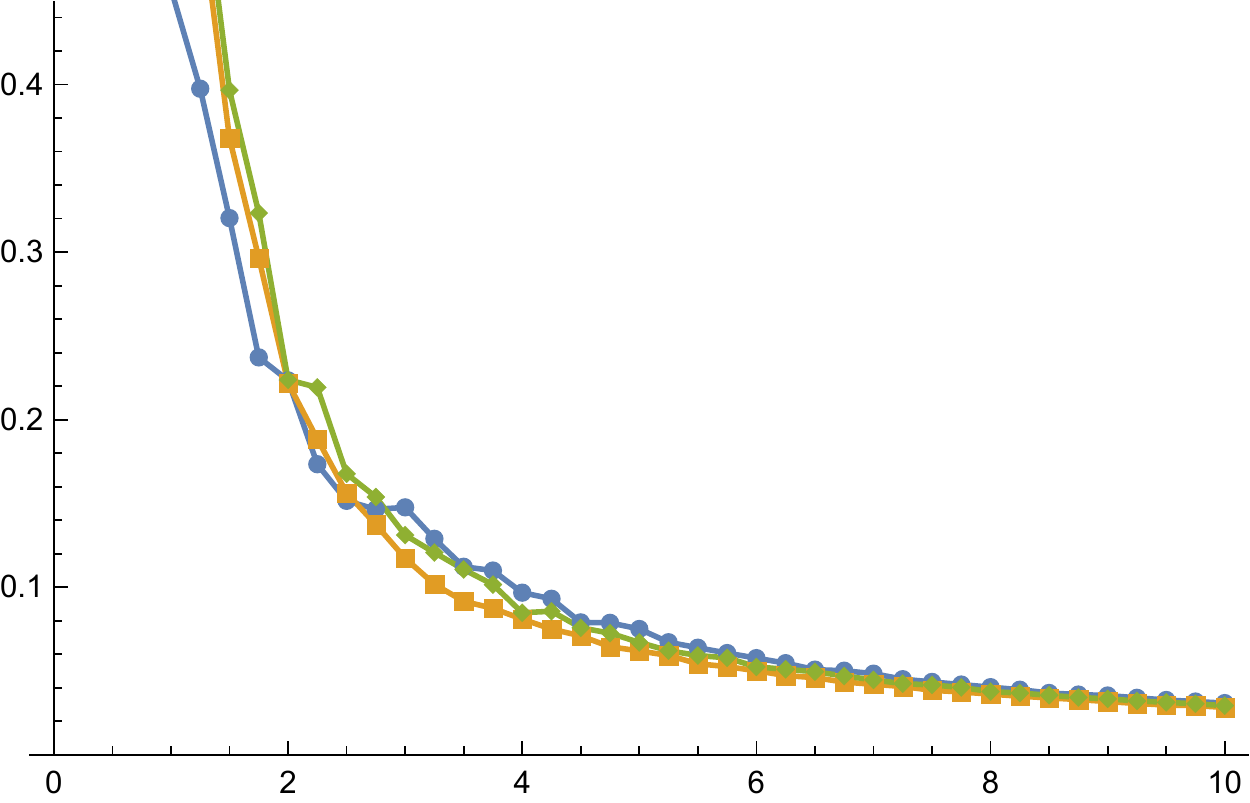}%
\begin{picture}(0,0)
\put(-6,-9){\footnotesize$\lambda$}
\put(-165,98){\footnotesize$\Delta$}
\end{picture}
\caption{Dependence of $\Delta$ on $\lambda$.\label{fig_clcs_01}}
\end{subfigure}
\hspace{20pt}
\begin{subfigure}{0.45\textwidth}
\centering
\includegraphics[width=160pt]{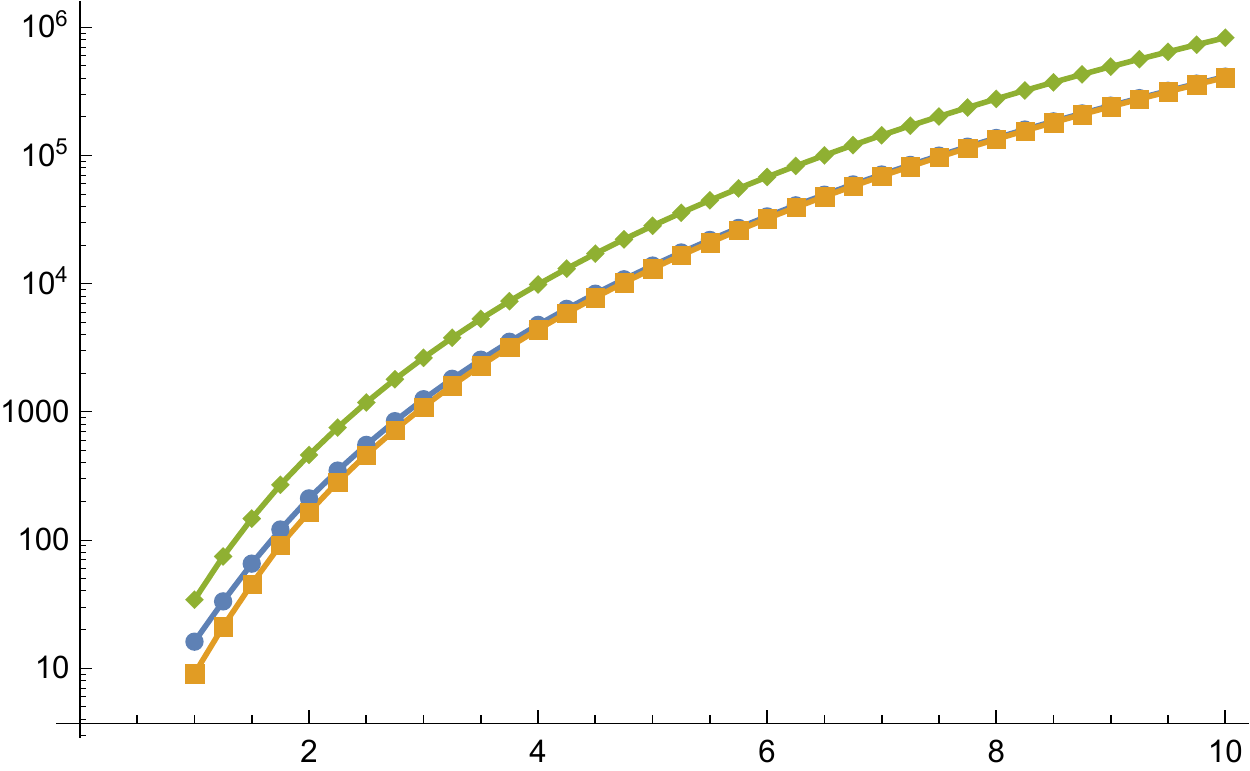}%
\begin{picture}(0,0)
\put(-6,-9){\footnotesize$\lambda$}
\put(-181,94){\footnotesize$N_{\rm flux}$}
\end{picture}
\caption{Log-dependence of $N_{\rm flux}$ on $\lambda$.\label{fig_clcs_02}}
\end{subfigure}
\caption{Conifold--large complex structure limit: dependence of the relative difference $\Delta$ and of the tadpole contribution 
$N_{\rm flux}$ on the hierarchy parameter $\lambda$. The plots show three different families, 
which all show a similar behaviour. \label{plot_clcs_all}}
\end{figure}

\section{Self-duality: F-theory examples}\label{sec:Ftheoryexamples}
We now want to apply the techniques discussed in the above sections to F-theory compactifications on elliptically fibered Calabi-Yau fourfolds. We study one example in great detail, where we show how flat directions in the sl(2)-approximation are lifted by corrections from the nilpotent orbit. We also comment on some relations of these vacua to swampland conjectures, with a particular focus on the tadpole conjecture.



\subsection{Large complex structure regime}\label{ssec:lcs}

To make our discussion here more explicit, we specialize to a particular region in complex structure moduli space. 
More concretely, we consider the large complex structure regime  of a Calabi-Yau fourfold $Y_4$. We first review the details of these asymptotic regimes here, and in the next subsection move on to the study of moduli stabilization in explicit examples.


\subsubsection*{Moduli space geometry}

Using homological mirror symmetry the superpotential can be expressed  in terms of the central charges of B-branes wrapping even-degree cycles of the mirror-fourfold $X_4$, and we refer to the references \cite{Grimm:2009ef,CaboBizet:2014ovf, Gerhardus:2016iot, Cota:2017aal} for a more detailed discussion. Following the conventions of \cite{Marchesano:2021gyv}, we expand the periods of the holomorphic four-form $\Omega$ around the large complex structure point as
\begin{equation}\label{fourfold_periods}
\Pi = \begin{pmatrix}
1 \\
-t^I \\
\frac{1}{2} \eta_{\mu \nu} \zeta^\nu_{IJ} t^I t^J \\
-\frac{1}{6} \cK_{IJKL} t^J t^K t^L +i K_I^{(3)}\\
\frac{1}{24} \cK_{IJKL} t^I t^J t^K t^L -i K_I^{(3)} t^I
\end{pmatrix},
\end{equation}
where $\cK_{IJKL}$ are the quadruple intersection numbers of $X_4$ and the coefficients $K_I^{(3)}$ arise from integrating the third Chern class. In formulas this reads
\begin{equation}\label{eq:quadruple}
\cK_{IJKL} = D_I \cdot D_J \cdot D_K \cdot D_L\, , \hspace{50pt} K_I^{(3)} = \frac{\zeta(3)}{8\pi^3}\  \int_{X_4} c_3(X_4) \wedge J_I \, ,
\end{equation}
where  $D_I$ denotes a basis of divisor classes for $X_4$ that generate its K\"ahler cone\footnote{We assume the K\"ahler cone to be simplicial, i.e.~the number of generators is equal to $h^{1,1}(X_4)$.} and $J_I \in H^2(X_4, \mathbb{Z})$ denote their Poincar\'e dual two-forms. Under mirror symmetry the $h^{1,1}(X_4)$ K\"ahler moduli $t^i$ of $X_4$ are identified with the $h^{3,1}(Y_4)$ complex structure moduli of $Y_4$, and the coefficients $K_I^{(3)}$ can be interpreted as perturbative corrections to the periods, similar to the correction involving the Euler characteristic  in the periods \eqref{eq:periodsexample} for the threefold case. Furthermore, we introduced a tensor $\zeta^\mu_{IJ}$ to expand all intersections of divisor classes $D_I \cdot D_J$ into a basis of four-cycles $H_\mu$ as
\begin{equation}\label{zeta_tensor}
D_I \cdot D_J = \zeta^\mu_{IJ} H_\mu \, .
\end{equation}
The intersection of two four-cycles $H_\mu \cdot H_\nu$ is  denoted by
\begin{equation}
\eta_{\mu \nu} =  H_\mu \cdot H_\nu\,,
\end{equation} 
and comparing with \eqref{eq:quadruple} leads to the following relation for the intersection numbers 
\begin{equation}
\cK_{IJKL} = \zeta^\mu_{IJ} \op \eta_{\mu\nu} \op \zeta^\nu_{KL} \, .
\end{equation}
The  superpotential  \eqref{fourfold_potentials} shown above can be expressed as $W = G_4 \op\Sigma\op \Pi$, where $\Sigma$
is the matrix coupling the periods to the flux quanta which is  given by
\begin{equation}
\Sigma = \begin{pmatrix}
0 & 0 & 0 & 0 & 1 \\
0 & 0 & 0 & -\delta^I_J & 0 \\
0 & 0 & \eta^{\mu \nu} & 0 & 0 \\
0 & -\delta_I^J & 0 & 0 & 0 \\
1 & 0 & 0 & 0 & 0 \\
\end{pmatrix}.
\end{equation}
Note that we used $I,\mu$ to label the rows and $J,\nu$ for columns. The $G_4$-flux written in these conventions takes the form 
\begin{equation}\label{fluxquanta}
G_4 = \left( \, m\,,  \hspace{4pt} m^I \,,  \hspace{4pt} \hat{m}_\mu \,,  \hspace{4pt}  e_I \,, \hspace{4pt}  e\, \right),
\end{equation}
where the flux-quanta are (half-)integer quantized ,
and the contribution of the fluxes to the tadpole cancellation condition is  given by 
\eq{
  N_{\rm flux} = \frac{1}{2} \int_{Y_4} G_4 \wedge G_4 = \frac{1}{2} \op G_4 \Sigma \op G_4\,.
}  
Let us stress that the period vector \eqref{fourfold_periods} is not expanded in an integral basis, which means that special care has to be taken with the quantized fluxes coupling to these periods in the superpotential \eqref{fourfold_potentials}. This quantization was worked out in \cite{Gerhardus:2016iot}, and has been reformulated as a rational shift of the flux quanta in \cite{Marchesano:2021gyv}. For our purposes implementing this shift will not be important, however, we still require the flux quanta in \eqref{fluxquanta} to take integer values. From the perspective of moduli stabilization these rational shifts typically only affect the moduli vevs and their masses slightly, while we merely want to demonstrate the effect of corrections to the sl(2)-approximation on aspects such as flat directions. Also for the minimal tadpole these shifts only add a small rational term, while we are interesting in the scaling in large $h^{3,1}$.

Finally, it is instructive to decompose the periods in \eqref{fourfold_periods} according to the nilpotent orbit form similar to
the threefold case discussed above. In particular, we can read off the log-monodromy matrices $N_A$ and leading term 
$a_0$ of the periods as
\begin{equation}\label{fourfold_logs}
\Pi = e^{t^I N_I} a_0 \, , \quad N_A = \scalebox{0.85}{$\begin{pmatrix}
0 & 0& 0 & 0 & 0 \\
-\delta_{AI} & 0 & 0 & 0 & 0 \\
0 & -\eta_{\mu \rho} \zeta^\rho_{AJ} & 0 & 0 & 0 \\
0 & 0 & -\zeta^\nu_{AI} & 0 & 0 \\
0 & 0 & 0 & -\delta_{AJ} & 0 
\end{pmatrix}$}, \quad a_0  = \scalebox{0.85}{$\begin{pmatrix}
1 \\
0 \\
0 \\
K^{(3)}_I \\
0
\end{pmatrix}$}\, .
\end{equation}


\subsubsection*{Periods for an elliptically fibered mirror}

Let us now specialize to an elliptically fibered mirror fourfold $X_4$, which corresponds to the example considered in the following section. For fourfolds with this fibration structure the topological data of $X_4$ is determined by the base $B_3$, and one can explicitly construct a basis for the four-cycles $H_\mu$ as discussed in \cite{Cota:2017aal}. In the following we work in the conventions of \cite{Marchesano:2021gyv}. 
We start by constructing a basis for the two- and six-cycles of $X_4$ from the base $B_3$. For the generators of the Mori cone of $B_3$ corresponding to two-cycles we write $C'^a$, while the dual four-cycles that generate the K\"ahler cone are denoted by divisors $D'_a$. The index runs as $a = 1, \ldots , h^{1,1}(B_3)$ with $h^{1,1}(B_3) = h^{1,1}(X_4)-1$. Denoting the projection of the fibration by $\pi$ and the divisor class of the section by $E$, we can then generate the Mori cone of $X_4$ by
\begin{equation}
C^0, \qquad C^a = E \cdot \pi^{-1} C'^a \, ,
\end{equation}
where $C^0$ corresponds to the class of the fiber. We generate the K\"ahler cone of $X_4$ by the dual basis of divisor classes
\begin{equation}\label{divisor_basis}
D_0 = E + \pi^* c_1(B_3)\, , \qquad D_a = \pi^* D'_a\, ,
\end{equation}
where $c_1(B_3) = c_1^a D_a'$ denotes the first Chern class of the base $B_3$. We can recover the intersection numbers of $B_3$ from those of $X_4$ as
\begin{equation}
\cK_{abc} = D'_a \cdot D'_b \cdot D'_c = D_0 \cdot D_a \cdot D_b \cdot D_c = \cK_{0abc}\, .
\end{equation}
Having constructed a basis for the two- and six-cycles, let us next consider the four-cycles $H_{\mu}$. We can generate these four-cycles via the divisors $D'_a$ and curves $C'^a$ of $B_3$ as
\begin{equation}\label{fourfold_fourcycles}
H_a = D_0 \cdot \pi^{-1}(D'_a) = D_0 \cdot D_a \, , \quad H_{\hat{a}} = \pi^{-1} (C'^a)\, , \quad a,\hat{a} = 1, \ldots , h^{1,1}(B_3) \,,
\end{equation}
where we split $\mu = (a, \hat{a})$. As a final task we construct the tensors $\zeta^\mu_{ij}$ and $\eta_{\mu\nu}$ appearing in the periods \eqref{fourfold_periods}. The tensor $\zeta^\mu_{IJ}$ relating intersections of two-cycles $D_I \cdot D_J$ to the four-cycle basis $H_\mu$ by \eqref{zeta_tensor} is found to be 
\begin{equation}
\zeta_{0b}^a =\delta_{ab}\, , \qquad \zeta^{\hat{a}}_{bc} = \cK_{abc} \, , \qquad \zeta^a_{00} = c_1^a\, ,
\end{equation}
and all other components either vanish or are fixed by symmetry. The intersections $\eta_{\mu \nu}$ between the four-cycles $H_\mu$ are given by
\begin{equation}
 \eta_{ab} = \cK_{abc}c_1^c \, , \qquad \eta_{a\hat{b}} = \delta_{ab} \, , \qquad \eta_{\hat{a} \hat{b}} = 0\, .
 \end{equation}


\subsection{The linear scenario -- construction} \label{ssec:linear}

With the expressions introduced above, we can now discuss a more concrete setting.
We consider the triple fibration $T^2 \to \mathbb{P}^1 \to \mathbb{P}^1 \to \mathbb{P}^1 $ as the mirror fourfold, and 
for  the toric construction  we refer to \cite{Mayr:1996sh}. In \cite{Marchesano:2021gyv} this geometry was used to realize a particular moduli stabilization scheme called the linear scenario. Here we discuss this setup from the perspective of the sl(2)-approximation,
and we comment on the scaling of the corresponding tadpole-contribution of the fluxes.


\subsubsection*{Moduli space geometry}

The relevant topological data of the above fourfold $X_4$ can be summarized by the following intersection numbers 
\begin{align}\label{fourfold_volume}
\cK_{ABCD} y^A y^B y^D y^D &= y_L( 32 y_0^3 + 24 y_0 y_1 y_2 +12 y_0 y_2^2 + 24 y_0^2 y_1 + 26 y_0^2 y_2 ) \nn \\
& \ \ \ + 64 y_0^2 y_1 y_2 + 24 y_0 y_1^2 y_2  + 8 y_0 y_2^3
+ 24 y_0 y_1 y_2^2 + 64 y_0^3 y_1 \nn \\
& \ \ \  + 24 y_0^2 y_1^2 + 36 y_0^2 y_2^2 + 72 y_0^3 y_2 + 52 y_0^4 \nn \\
& = 4\op y_L\op \cK_L + f(y^\alpha)\, ,
\end{align}
and by the integrated third Chern class
\begin{equation}
K_A^{(3)}  y^A =-  \frac{\zeta(3)}{8\pi^3} \Big(3136 y^0  + 480 y^L +  960 y^1 + 1080 y^2 \Big)\, ,
\end{equation}
and the base $B_3$ of the elliptic fibration has first Chern class
\begin{equation}
c_1(B_3) =  D_1 + 2D_2\, .
\end{equation}
The divisor $D_0$ corresponds to the zero section of the elliptic fibration as given in  the basis \eqref{divisor_basis}, while $D_L$ corresponds to the class of the Calabi-Yau threefold fiber over $\mathbb{P}^1$. In the last line of \eqref{fourfold_volume} we singled out the terms dependent on the saxion $y^L$  by writing 
\begin{equation}\label{fourfold_functions}
 \cK_L = \sum_{\alpha,\beta,\gamma} \cK_{Labc} y^\alpha y^\beta y^\gamma\, , \qquad f(y^a) = \sum_{\alpha,\beta,\gamma,\delta} \cK_{\alpha\beta\gamma\delta}   y^\alpha y^\beta y^\gamma y^\delta \, ,
\end{equation}
where $f(y^a)$ contains the remaining terms only depending on $y^\alpha=y^0,y^1,y^2$. Let us also note that we sort our complex structure moduli in the order $(t^0, t^L, t^1 , t^2)$ for the construction of the log-monodromy matrices as described by \eqref{fourfold_logs}, and split the indices as $A=(L, \alpha)$ with $\alpha=0,1,2$. Furthermore, we use the four-cycle basis as described by \eqref{fourfold_fourcycles}, where we let the indices run as $a, \hat{a} = L, 1, 2$.\footnote{Note that this four-cycle basis differs from the one used in \cite{Marchesano:2021gyv}, where another basis was taken using the fibration structure of the threefold fiber rather than the elliptic fiber.} 


\subsubsection*{Saxion hierarchies and the sl(2)-approximation}

Let us now introduce particular scalings for the saxions $y_A$. For the linear scenario of \cite{Marchesano:2021gyv} there is a hierarchy $y_L \gg y_\alpha $ which we realize by
\begin{equation}\label{scaling_linear_scenario}
y_L = \lambda^3\, , \hspace{50pt} y_\alpha = \lambda \, \hat{y}_\alpha\, .
\end{equation}
In order to understand this regime from the perspective of the sl(2)-approximation we need to consider a further hierarchy $y_L \gg y_1 \gg y_2 \gg y_0$, which we implement via an additional scaling parameter $\rho$ as
\begin{equation}\label{scaling_linear_sl2}
y_L = \lambda^3 \rho \, , \hspace{50pt} 
y_0 = \lambda  \, , \hspace{50pt}
y_1 = \lambda \rho^2 \, , \hspace{50pt}
y_2 = \lambda \rho\, .
\end{equation}
Setting for instance $\rho = \lambda$ we find scalings $(y_L, y_0,y_1,y_2) = (\lambda^4 , \lambda, \lambda^3, \lambda^2)$ as needed for the hierarchy of the sl(2)-approximation, while for $\rho = 1$ we reduce to a scaling of the form of \eqref{scaling_linear_scenario}. Typically we will keep both the scaling in $\rho$ and $\lambda$ explicit rather than making a choice for $\rho$. We can then use the scaling in $\lambda$ found in the sl(2)-approximation to make statements about the linear scenario. In order to keep the discussion here concise we included the data specifying the sl(2)-approximated Hodge star in appendix \ref{app:linearscenario}. It is, however, instructive to give the eigenspaces of the weight operators $N^{0}_i$, which have been summarized in table \ref{linear_scenario_eigenspaces}.

\begin{table}[t]
\centering
\renewcommand*{\arraystretch}{1.8}
\scalebox{0.8}{
\begin{tabular}{| c | c |}
\hline weights  & charges \\ \hline  \hline
$(-1,-3)$ & $(0, 0, 0, 0, 0, 0, 0, 0, 0, 0, 0, 0, 0, 0, 0, 1) $ \\ \hline
$(-1,-1)$ & \begin{minipage}{0.5\textwidth}\centering\vspace*{0.2cm}
$(0, 0, 0, 0, 0, 0, 0, 0, 0, 0, 0, 1, 0, 0, 0, 0)$\\
$ (0, 0, 0, 0, 0, 0, 0, 0, 0, 0, 0, 0, 0, 1, 0, 0)$ \\
$(0, 0, 0, 0, 0, 0, 0, 0, 0, 0, 0, 0, 0, 0, 1, 0 ) $ \vspace*{0.2cm}
\end{minipage} \\ \hline
$(-1,1)$ & \begin{minipage}{0.5\textwidth}\centering\vspace*{0.2cm}
$(0, 0, 0, 0, 0, 0, 1, 0, 0, 0, 0, 0, 0, 0, 0, 0)$ \\
$(0, 0, 0, 0, 0, 0, 0, 1, 0, 0, 0, 0, 0, 0, 0, 0)$ \\
$(0, 0, 0, 0, 0, 0, 0, 0, 1, 0, 0, 0, 0, 0, 0, 0)$ \vspace*{0.2cm}
\end{minipage} \\ \hline
$(-1,3)$ & $( 0, 0, 1, 0, 0, 0, 0, 0, 0, 0, 0, 0, 0, 0, 0, 0)$ \\ \hline
$(1,-3)$ & $(0, 0, 0, 0, 0, 0, 0, 0, 0, 0, 0, 1, 1, 1, \tfrac{1}{2}, 0)$ \\ \hline
$(1,-1)$ & \begin{minipage}{0.5\textwidth}\centering\vspace*{0.2cm}
$(0, 0, 0, 0, 0, 1, 0, 0, -\tfrac{1}{2}, 1, -1, 0, 0, 0, 0, 0)$\\
$(0, 0, 0, 0, 0, 0, 1, 0, 1, -\tfrac{3}{2}, 1, 0, 0, 0, 0, 0)$\\
$(0, 0, 0, 0, 0, 0, 0, 1, -1, 1, 0, 0, 0, 0, 0, 0)$ \vspace*{0.2cm}
\end{minipage} \\ \hline
$(1,1)$ & \begin{minipage}{0.5\textwidth}\centering\vspace*{0.2cm}
$(0, 1, 0, 0, -2, 0, 0, 0, 0, 0, 0, 0, 0, 0, 0, 0)$\\
$(0, 0, 1, 0, -2, 0, 0, 0, 0, 0, 0, 0, 0, 0, 0, 0)$ \\
$(0, 0, 0, 1, -2, 0, 0, 0, 0, 0, 0, 0, 0, 0, 0, 0)$\vspace*{0.2cm}
\end{minipage} \\ \hline
 $(1,3)$ & $(1, 0, 0, 0, 0, 0, 0, 0, 0, 0, 0, 0, 0, 0, 0, 0)$ \\ \hline
\end{tabular}}
\caption{\label{linear_scenario_eigenspaces} Eigenspaces of the weight operators given in \eqref{linear_operators_weights}. In order to connect to the scaling of the linear scenario \eqref{scaling_linear_scenario} we grouped the weights $\ell_\alpha$ under $N^0_\alpha$ together as $(\ell_L, \ell_0+\ell_1+\ell_2)$. }
\end{table}


\subsubsection*{Parametric scaling of the tadpole}

Regarding this model, the aim in  \cite{Marchesano:2021gyv} was to make a flux-choice which keeps the tadpole contribution 
$N_{\rm flux}$ finite
asymptotically. The parametric growth of the tadpole due to a flux is specified by its weights under the $\mathfrak{sl}(2, \mathbb{R})$-triples and we recall that for a charge in the eigenspace  $q \in V_{\ell_1 \ell_2 \ell_3 \ell_4}$ its Hodge norm \eqref{eq:Csl2ToCInf} grows as
\begin{equation}
\| q \|^2 \sim y_L^{\ell_L}\op y_0^{\ell_0} \op y_1^{\ell_1}\op y_2^{\ell_2} \, .
\end{equation}
Using the scalings of the saxions shown in \eqref{scaling_linear_sl2}, we find for large $\lambda$ that $3\ell_L+\ell_0+\ell_1+\ell_2 \leq 0$ must hold for our flux quanta. From \eqref{linear_operators_weights} we can compute that $\ell_A = \pm 1$ for all weights, so the fluxes with weights $\ell_L=1$ and $\ell_0+\ell_1+\ell_2 =-1,1,3$ should be turned off. Inspecting table \ref{linear_scenario_eigenspaces} we then find that this corresponds to the flux choice 
\begin{equation}\label{linear_fluxes}
m=0\,,  \qquad m_i = (0,m_L,0,0), \qquad \hat{m}^\mu = (0, \hat{m}^1, \hat{m}^2, \hat{m}^{\hat{L}},0,0)\, , 
\end{equation}
and all other fluxes unconstrained. This matches precisely with the fluxes that were turned off in \cite{Marchesano:2021gyv},  motivated from the sl(2)-approximation.
(Note that we chose a different four-cycle basis as compared to \cite{Marchesano:2021gyv} so the choice of $\hat{m}^\mu $ 
takes a slightly different form.)


\subsubsection*{Flat directions}

Given the above choice of fluxes, we can now investigate the stabilization of the saxions within various approximation schemes.  
Via the self-duality condition of the $G_4$-flux, the axions are fixed as
as \cite{Marchesano:2021gyv}
\begin{equation}\label{linear_axions}
\begin{aligned}
x^\alpha &= - \frac{\hat{m}^\alpha}{m^L} \, ,  \qquad &x^L &=  - \frac{e}{e_L} - \frac{1}{3 e_L (m^L)^2 } \Big( \cK_{Labc} \hat{m}^a \hat{m}^b \hat{m}^c - 3 e_a \hat{m}^a m^L \Big)\, .
\end{aligned}
\end{equation}
However, for simplicity we  set these axions to zero in the following, that is $x^L=x^0=x^1=x^2=0$, 
which means we impose $e=\hat{m}^1=\hat{m}^2= \hat{m}^{\hat{L}}=0$ on the flux quanta in addition to \eqref{linear_fluxes}.
\begin{itemize}

\item \textit{sl(2)-approximation}: For the specified choice of fluxes \eqref{linear_fluxes} with vanishing axions we find $(1, -3)$ and $(-1, 3)$ as allowed weights for $(\ell_L, \ell_0+\ell_1+\ell_2)$. 
Using the sl(2)-approximated Hodge star \eqref{eq:Csl2ToCInf} in the self-duality condition \eqref{fourfold_extremization}, this 
leads to
\begin{equation}
e_L = - \frac{y_0 \op y_1 \op y_2}{2\op y_L} m_L \, , \qquad e_\alpha = 2 e_L\, .
\end{equation}
Notice that the saxions only appear as the combination $y_0 y_1 y_2 / y_L$, so there are three saxionic directions left unconstrained. In particular, when plugging in the scaling of the linear scenario \eqref{scaling_linear_scenario} we see that $\lambda$ drops out completely. We can stabilize these seemingly flat directions by including corrections to the sl(2)-approximation.

\item \textit{Nilpotent orbit approximation}: We next include the full set of polynomial terms in the periods, in  particular, we investigate the difference between including the corrections $K^{(3)}_i$ or not. The relevant extremization conditions for the saxions now read \cite{Marchesano:2021gyv}
\begin{equation}\label{linear_solve}
\begin{aligned}
e_L &= -\frac{\cK}{6} g_{LL} m^L = \Big( -\frac{ \cK_L}{t_L}+ \frac{ f }{24 t_L^2 }  -\frac{ 
K_L^{(3)} \zeta(3)}{64 \pi^3 t_L}\Big) m^L+ \cO\Big(\frac{1}{\lambda^4}\Big) \, , \\
e_\alpha &= \frac{m^L}{6}  \cK_L \partial_\alpha \Big( \frac{f}{4 \cK_L} \Big) -  \frac{ 9 K^{(3)}_L  \cK_{L\alpha} m^L }{8  \cK_L }+ \cO\Big(\frac{1}{\lambda^2}\Big) \, ,
\end{aligned}
\end{equation}
where $\cK_{L \alpha} = \sum_{\beta, \gamma} \cK_{L \alpha \beta \gamma} y^\beta y^\gamma$. We also specified the order in $\lambda$ at which corrections to these equations enter under the scaling \eqref{scaling_linear_scenario}. Note that only $K^{(3)}_L$ appears here, so it dominates over the other corrections $K^{(3)}_\alpha$ in the expansion in $\lambda$. 
Since it is rather difficult to solve \eqref{linear_solve} explicitly for the saxions, we will instead consider some specific flux quanta $e_L, e_\alpha$ to exemplify that the saxions can indeed be stabilized. 
As flux quanta we take
\begin{equation}
m_L =  623\, , \qquad e_A = (-4698, -3072, -2760, -1566 )\, ,
\end{equation}
for which the saxions \eqref{scaling_linear_scenario} are  stabilized at
\begin{equation}
\lambda = 6 \, , \qquad \hat{y}^\alpha = 1\, .
\end{equation}

\item \textit{Nilpotent orbit approximation without $K^{(3)}_L$}: We also compute the eigenvalues of the mass matrix $K^{ab} \partial_a \partial_b V$ while formally setting $K^{(3)}_L = 0$ and  $K^{(3)}_\alpha = 0$. This yields the canonically-normalized masses
\begin{equation}
\begin{aligned}\label{masses_without}
m^2  = \bigl(&3.4 \cdot 10^{18}, \ 1.7 \cdot 10^{13}, \ 1.7 \cdot 10^{13}, \ 1.7 \cdot 10^{13}, \ \\
&1.6 \cdot 10^{13}, \ 3.8 \cdot 10^{11}, \ 2.7 \cdot 10^{11}, \ 0\bigr)\, ,
\end{aligned}
\end{equation}
and hence for this setting there is one flat direction.

\item \textit{Nilpotent orbit approximation with $K^{(3)}_L$}: We then include the correction $K^{(3)}_L $ but still set $K^{(3)}_\alpha = 0$ and find that the flat direction now acquires a mass
\begin{equation}\label{masses_with}
\begin{aligned}
m^2 = \bigl(&3.4 \cdot 10^{18}, \ 1.7 \cdot 10^{13}, \ 1.7 \cdot 10^{13}, \ 1.7 \cdot 10^{13}, \\
&1.6 \cdot 10^{13}, \ 3.6 \cdot 10^{11}, \ 2.5 \cdot 10^{11}, \ 2.3 \cdot 10^7 \bigr) ,
\end{aligned}
\end{equation}
while the other masses are only affected slightly. The now non-zero mass is significantly smaller as compared to the other moduli, since its mass scale is set by the correction $K_L^{(3)}$ rather than the leading polynomial terms.

\end{itemize}


\subsubsection*{Type IIB2 case}

In order to understand the above observations better, let us reduce the F-theory setting to Type IIB string theory. This brings us to the IIB2 moduli stabilization scheme of \cite{Marchesano:2021gyv}, which was originally considered in \cite{Palti:2008mg}. In order to  match 
the two configurations we set $f=0$ and $K^{(3)}_\alpha = 0$, and identify $K_{L\alpha \beta \gamma}$ with the intersection numbers of the Calabi-Yau threefold and $K^{(3)}_L$ as the Euler characteristic correction. Furthermore, the dilaton $t^L$ is interpreted as the axio-dilaton $\tau$. We will keep the notation of the F-theory setting, and simply drop the terms that are absent in the Type IIB setup, for which the self-duality condition \eqref{linear_solve}  reduces to
\begin{equation}\label{IIB_solve}
\begin{aligned}
e_L & = \Big( -\frac{1}{6} \frac{\cK_L}{y_L}  -\frac{ 
K_L^{(3)} \zeta(3)}{64 \pi^3 y_L}\Big) m^L\, , \qquad &e_\alpha &=  -  \frac{ 9 K^{(3)}_L  \cK_{L\alpha} m^L }{8  \cK_L }\, .
\end{aligned}
\end{equation}
When we drop the correction $K_L^{(3)}$ and plug in the saxion scaling \eqref{scaling_linear_scenario} we see that $\lambda$ cancels out of these equations. However, subsequently including the correction fixes $\lambda$ as
\begin{equation}\label{lambda_vev}
\lambda^3 = \frac{64 \pi ^3}{K_L^{(3)} \zeta(3)} \Big( - \frac{1}{6} \cK_{L\alpha \beta \gamma}\hat{y}^\alpha \hat{y}^\beta \hat{y}^\gamma - \frac{e_L}{m^L}  \Big)\, .
\end{equation}
Thus in the IIB2 scenario the modulus $\lambda$ parametrizes precisely the flat direction we encountered before, which is stabilized by including perturbative corrections. In the general F-theory setup this symmetry was more difficult to spot in \eqref{scaling_linear_scenario} due to the presence of the function $f(\hat{y}^\alpha)$ which alters this parametrization. Nevertheless, by studying the effect of $K_L^{(3)}$ on the masses in \eqref{masses_without} and \eqref{masses_with} we were able to observe this feature. We should also point out that, in contrast to the general F-theory setting, there is more than one flat saxionic direction present when $K_L^{(3)}$ is dropped here. In this case \eqref{IIB_solve} only imposes a single constraint on the saxions, so in addition to $\lambda$ there are $h^{2,1}-1$ other saxionic directions that remain flat. This shows how $\lambda$ is stabilized by corrections to the sl(2)-approximation, but we will not look further into these other flat directions.


\subsection{The linear scenario -- discussion} 

We now want to discuss the linear scenario in view of relations to swampland conjectures, 
to the finiteness theorem and to the tadpole conjecture.


\subsubsection*{Relations to swampland conjectures and finiteness theorem}
The above hierarchy in the masses of the linear scenario is quite interesting in light of some swampland conjectures. Say we place a cutoff scale between the moduli masses stabilized in \eqref{masses_without} and the field direction stabilized by the $K^{(3)}_L$ effect in \eqref{masses_with}. From the mass hierarchy it then follows that we can integrate out all axion fields (and three saxion fields), while one runaway direction remains. In the IIB2 case this direction is parametrized via the saxionic modulus $\lambda$ by \eqref{scaling_linear_scenario}, while in the more general F-theory setup this parametrization is slightly more involved. Either way, by restricting to this valley of the scalar potential we obtain a pseudomoduli space containing a single saxion field and no axions. In turn, we can send $\lambda \to \infty$, resulting in an infinite distance limit. Generically this path does not lift to a geodesic in the original higher-dimensional moduli space, so this provides us with an interesting class of examples for the Convex Hull Distance Conjecture of \cite{Calderon-Infante:2020dhm}. On the other hand, this infinite distance limit is rather intriguing from the perspective of the Distant Axionic String Conjecture \cite{Lanza:2020qmt, Lanza:2021qsu}. It predicts the emergence of asymptotically tensionless strings coupled to axion fields, but in this effective field theory all axions have already been integrated out due to the cutoff. It would be interesting to return to this puzzle in the near future, and see if a more detailed study of axion strings in the background of such scalar potentials elucidates this matter.

Let us also comment on the linear scenario in the context of the general finiteness theorems for flux vacua satisfying 
the self-duality condition \cite{Grimm:2020cda, Bakker:2021uqw,Grimm:2021vpn}. A first observation is 
that the tadpole $N_{\rm flux} = - e_L m^L$ seems to be independent of the flux $e_\alpha$. However, it is not hard to 
see that $e_\alpha$ cannot be chosen freely and 
there are only finitely many choices allowed in this setting. The second equation in \eqref{IIB_solve} implies that if we 
want to increase $e_\alpha$, we either have to increase $m^L$ or decrease the moduli. In the former case, we immediately 
see that the tadpole grows, while in the latter case we reach a point where our use of the asymptotic results are no longer applicable. 
Furthermore, we check that the possible vevs for the saxions are bounded from above, which is another necessary condition for the finiteness of solutions. 
Inserting the scaling \eqref{scaling_linear_scenario} into the first equation of \eqref{IIB_solve} we see that the $\hat{y}^\alpha$ are bounded from above. Namely, otherwise $e_L$ grows as we increase the volume $\cK_{L\alpha \beta \gamma} \hat{y}^\alpha \hat{y}^\beta \hat{y}^\gamma $, resulting in a diverging tadpole $N_{\rm flux} = - e_L m^L$. Using similar reasoning we find that $\lambda$ is bounded through the condition \eqref{lambda_vev}. 

Finally, let us also have a closer look at the flux vacuum loci themselves. In particular, consider the case $K_L^{(3)}=0$. 
For such geometries the modulus $\lambda$ is unfixed due to the absence of corrections in the nilpotent orbit approximation. 
This means the resulting flux vacua are not point-like, but rather infinite lines stretching to the boundary of moduli space.\footnote{It could be the case that the 
inclusion of exponential corrections stabilizes this flat direction. There are, however, examples where such corrections are absent \cite{Bakker:2021uqw}. 
In that regard it is interesting to point out the work of \cite{Palti:2020qlc}, where the absence of instantons in special cases was related to higher-supersymmetric settings.} 
It is interesting to point out that vacuum loci of this type do not need to be algebraic. This implies that if one aims to describe the structure of all flux vacua one has to 
leave the world of algebraic geometry. Remarkably, a delicate and powerful extension of algebraic geometry that can be use to describe flux vacua is provided by using 
tame topology and o-minimality \cite{Bakker:2021uqw}. The resulting tame form of geometry manifest the notion of finiteness and removes many pathologies that 
are allowed in geometric settings based on ordinary topology.


\subsubsection*{Tadpole conjecture I -- without axions}

We now focus on the tadpole contribution of the fluxes and its scaling with the number of 
moduli \cite{Bena:2020xrh,Bena:2021wyr}. For simplicity we restrict our attention to the IIB2 case, 
and our discussion follows closely the line of arguments first presented in \cite{Plauschinn:2021hkp}.
In particular, we will be using statistical data for Calabi-Yau threefolds obtained  in \cite{Demirtas:2018akl}
to show that the linear scenario is, under certain genericity assumptions, compatible with the tadpole conjecture.
Our main argument can then be summarized  as follows:
\begin{itemize}

\item Since we restrict our discussion to the large complex structure limit and ignore instanton corrections, 
we need to ensure that the latter are sufficiently suppressed. To implement this constraint 
on the mirror-dual side, we require that all holomorphic curves $\cC$ have a volume greater than a 
constant $c$. Similarly we require that all divisors $D$ and the mirror threefold $\tilde Y_3$ itself to have volumes greater than $c$,
and we use these conditions to define the stretched K\"ahler 
cone \cite{Demirtas:2018akl}, i.e.~we consider those $J$ that satisfy 
\beq \label{Kahler-cone}
  \int_C J > c\ , \qquad  \frac{1}{2}\int_D J^2 > c\ , \qquad  \frac{1}{6}\int_{\tilde Y_3} J^3 > c\ . 
\eeq
Note that (after applying mirror-symmetry) the asymptotic regime introduced at the beginning of section \ref{sec_tech_details} lies in the stretched K\"ahler cone with $c=1$. 

\item In \cite{Demirtas:2018akl} the authors analyzed the Kreuzer-Skarke list \cite{Kreuzer:2000xy} 
and determined properties associated with the stretched K\"ahler cone. Via mirror symmetry, these results on the K\"ahler-moduli 
side then translate to the large complex structure limit we are interested in.

\item  We note that the K\"ahler form $J$ can be expanded in any basis of the second cohomology.  
For example, we can use the generators $\omega_\alpha$ of a simplicial subcone 
as discussed in more detail in \cite{Grimm:2019bey} and 
write $J=y^\alpha\omega_\alpha$. Setting $c=1$ in \eqref{Kahler-cone} then implies $y^\alpha>1$ and we obtain the 
above parametrization of the asymptotic regime. However, in \cite{Demirtas:2018akl} the 
analysis was carried out in a basis naturally arising in the explicit construction of Calabi-Yau threefold examples. Denoting this
basis by $[D_\alpha]$ we expand $J= v^\alpha [D_\alpha]$. We note that the $v^\alpha$ now obey certain non-trivial inequalities 
for $J$ to be an element of the stretched K\"ahler cone \eqref{Kahler-cone}. Clearly, also the intersection numbers $\cK_{\alpha \beta \gamma} = D_\alpha \cdot D_\beta \cdot D_\gamma$ have to be 
evaluated in this different basis and the statistical statements of \cite{Demirtas:2018akl} concerning their behaviour is generically true in this `special' basis.

\item Of interest to us are the number of non-zero entries in the triple intersection numbers $\kappa_{\alpha\beta\gamma}$ 
and a measure for how stretched the K\"ahler cone is at large $h^{1,1}$. To associate a number to the latter, we introduce 
the distance $|v| = \sqrt{\sum_a(v^a)^2}$ from the origin of the K\"ahler cone to a K\"ahler form $J= v^a [D_a]$, and 
denote by $d_{\rm min}$ the minimal distance between the origin of the K\"ahler cone and the tip of the stretched K\"ahler cone. 
In  \cite{Demirtas:2018akl}  these have been determined for the Kreuzer-Skarke database and their dependence 
on $h^{1,1}$ have been determined. Under mirror symmetry this dependence translates into\footnote{
When using the basis $\omega_{\alpha}$ to expand the K\"ahler form then the dependence of the quantities in \eqref{rel_9379274}
on $h^{2,1}$ will be different. Since this information is not readily available, we use the basis employed in 
\cite{Demirtas:2018akl}.}
\eq{
  \label{rel_9379274}
   \#(\kappa_{\alpha\beta\gamma}\neq 0) \gtrsim 6.5\op h^{2,1} + 25\,,
   \hspace{40pt}
    d_{\rm min} \simeq 10^{-1.4} \op (h^{2,1})^{2.5}  \,.
}
Furthermore,  for large $h^{2,1}$ the size of the entries of $\kappa_{\alpha\beta\gamma}$ is of order $\mathcal O(10)$ 
which can be inferred from figure 2 of \cite{Demirtas:2018akl}.  For generic situations and in the large $h^{2,1}$ limit 
we then obtain the following rough scaling behavior (see \cite{Plauschinn:2021hkp} for details on the derivation) 
\eq{
  \label{fits_001}
  \cK_L \sim  (h^{2,1})^{-1/2} \op | v |^3 \,,
  \hspace{30pt}
  \cK_{L\alpha} \sim (h^{2,1})^{-1} \op | v  |^2,
  \hspace{30pt}
  \cK_L^{(3)} \sim h^{2,1} \,.
}
The scaling behaviour of $\cK_{L\alpha} $ can be determined using \eqref{rel_9379274} together with 
$v^{\alpha}=\lvert v\rvert\op \mathsf e^{\alpha}$ and $\mathsf e^{\alpha} \sim (h^{2,1})^{-1/2}$, and 
we included here the quantity $K_L^{(3)}$ which contains the Euler characteristic.
Note that this scaling behaviour matches roughly the statistical analysis presented in \cite{Demirtas:2018akl} for the minimal 
total volume and divisor volumes in the stretched K\"ahler cone when replacing $|v|$ with $d_{\rm min}$. 
  
\end{itemize}
Let us now use \eqref{fits_001} and determine the moduli-dependent expression 
in the second relation in equation \eqref{IIB_solve}. 
For generic situations and for large $h^{2,1}$ we can make the following estimate 
\eq{
  \label{rel_8394}
   \frac{ 9K^{(3)}_L  \cK_{L\alpha}}{8\cK_L } \sim (h^{2,1})^{1/2} \op | v |^{-1} \lesssim (h^{2,1})^{-2} \,,
}
where in the second step we  applied the bound 
 $| v | \geq d_{\rm min}$ and where we ignored numerical factors. 
This scaling can now be used in the self-duality condition  \eqref{IIB_solve}, where the second condition 
together with \eqref{rel_8394} translates to 
\eq{
e_\alpha \lesssim (h^{2,1})^{-2} \op m^L\,.
}
However, 
since the fluxes $e_{\alpha}$ and $m^L$ are integer quantized
it follows that for non-zero $e_{\alpha}$ and $m^L$ the flux $m^L$ has to scale at least as $(h^{2,1})^2$.
For the tadpole contribution this implies 
\eq{
  N_{\rm flux} = - e_L m^L  \sim (h^{2,1})^2 \,,
}
which is in agreement with the tadpole conjecture.

While this result is non-trivial, let us emphasize that we have used fitted statistical data to make this estimate. 
Some of our steps, most notably \eqref{rel_8394}, are true only approximately and are made under the assumption that 
conspicuous cancellations are absent. 
Comparing \eqref{fits_001} with statistical results on the divisor volumes obtained in \cite{Demirtas:2018akl} shows 
agreement, but ideally we would like to have available results on the minimal value that $\cK_{L}/\cK_{L\alpha} $ can take in the stretched K\"ahler cone directly. 
Moreover, it is conceivable that certain families of compactification spaces exist that show a slightly different scaling behaviour. 
Hence, we do not claim that the IIB2 moduli stabilization scheme of \cite{Marchesano:2021gyv} cannot produce examples 
that violate the tadpole conjecture -- however, under the assumptions stated above the 
scheme of \cite{Marchesano:2021gyv} generically satisfies the tadpole conjecture
(in the case of vanishing axions). 
Note that a similar observation has  been made independently in \cite{Lust:2021xds}.


\subsubsection*{Tadpole conjecture II -- including axions}

We finally want to discuss the case of non-vanishing axions.
Having non-zero fluxes $\hat{m}^\mu$ introduces additional terms, and the modified self-duality equation in F-theory reads
\begin{equation}
m^L e_\alpha -\frac{1}{2} \cK_{L \alpha \beta \gamma} \hat{m}^\beta \hat{m}^\gamma = (m^L)^2 \bigg( \frac{1}{6}  \cK_L \partial_\alpha \Big( \frac{f}{4 \cK_L} \Big)  -  \frac{ 9 K^{(3)}_L  \cK_{L\alpha} }{8  \cK_L } \bigg)\, .
\end{equation} 
For the type IIB limit we set  $f=0$, and together with the first relation in \eqref{IIB_solve} the self-duality condition 
can be expressed as
\eq{
\begin{aligned}
\label{rel_0987}
e_L & = \Big( -\frac{1}{6} \frac{\cK_L}{y_L}  -\frac{ 
K_L^{(3)} \zeta(3)}{64 \pi^3 y_L}\Big) m^L\, , \\
m^L e_\alpha -\frac{1}{2} \cK_{L \alpha \beta \gamma} \hat{m}^\beta \hat{m}^\gamma &= -  \frac{ 9 K^{(3)}_L  \cK_{L\alpha} (m^L)^2 }{8  \cK_L }\,.
\end{aligned}
}
We now want to repeat our reasoning from above, for which we have to distinguish two cases:
\begin{itemize}

\item We first consider the case in which  the combination 
$m^L e_\alpha -\frac{1}{2} \cK_{L \alpha \beta \gamma} \hat{m}^\beta \hat{m}^\gamma $
appearing in the second relation of \eqref{rel_0987} is generic, in particular, for large $m^L$ this combination 
is a (half-)integer of order $m^L$ (or larger). Using then the scaling \eqref{rel_8394}, we conclude 
again that 
\eq{
  m^L \sim (h^{2,1})^2 
  \hspace{40pt}\Longrightarrow\hspace{40pt}
  N_{\rm flux} \sim (h^{2,1})^2 \,,
}
as in our discussion for vanishing axions. Hence, also in this situation the tadpole conjecture is satisfied.

\item As a second case we consider a situation where the combination
$m^L e_\alpha -\frac{1}{2} \cK_{L \alpha \beta \gamma} \hat{m}^\beta \hat{m}^\gamma $
is fine-tuned to a non-vanishing $\mathcal O(1)$ half-integer. Here we only need to require 
a linear scaling of $m^L$ with $h^{2,1}$, which leads to 
\eq{
  m^L \sim h^{2,1}
  \hspace{40pt}\Longrightarrow\hspace{40pt}
  N_{\rm flux} \sim h^{2,1} \,.
}
The scaling behaviour of the tadpole conjecture is therefore satisfied also in this non-generic situation.

\end{itemize}


\section{Moduli stabilization using the F-terms}\label{ssec:strategy}
We now lay out our strategy for engineering vacua with a small flux superpotential. This construction relies on first searching for solutions to the F-term equations coming from the polynomial periods $\Pi_{\rm pol}$ and supplementing them accordingly with constraints involving essential instantons in order to stabilize all moduli. As a result we will have $W_{\rm pol}=0$, which by essential exponential corrections will be lifted to a small vacuum superpotential $|W_0|$. This procedure is then explicitly applied to the one- and two-moduli type $\rm II$ models in sections \ref{ssec:II0} and \ref{ssec:II1} respectively.

Let us begin by writing down the relevant extremization conditions at polynomial level for our flux vacua. As we want to engineer vacua with an exponentially small superpotential, we require the polynomial part in the expansion of the superpotential \eqref{eq:Wexpansion} to vanish
\begin{equation}
W_{\rm pol} \big|_* = 0\, ,
\end{equation}
where we used a star to denote the evaluation of moduli at their vevs. At the polynomial level of the superpotential the vanishing F-term constraints \eqref{eq:Fterms} can then be written as
\begin{equation}\label{eq:polextr}
\begin{aligned}
  \langle F_3 - \tau_* H_3 , \,  \,  e^{t^i_* N_i} N_a a_0 \rangle = 0\, , \quad \langle H_3 , \, e^{t^i_* N_i} a_0 \rangle = 0\, ,  \quad \langle F_3 , \, e^{t^i_* N_i} a_0 \rangle = 0\, .\\
\end{aligned}
\end{equation}
The first set of equations follows from $\partial_a W_{\rm pol} \big|_* = 0$, while the latter two follow from $\partial_\tau W_{\rm pol}  \big|_*= 0$. Note that we replaced the covariant derivative by a partial derivative for all constraints since $W_{\rm pol}\big|_*  = 0$ for the flux vacua we are interested in.

It is important to stress that these polynomial level conditions \eqref{eq:polextr} do not suffice to obtain the vacua of 
the scalar potential. The vectors $N_a a_0$ can be parallel or even vanishing resulting in a insufficient set of F-term conditions at the polynomial level. Another way to observe this is by looking at the scalar potential $V_{\rm pol}$ obtained from the complete $V$ defined in \eqref{eq:potential} 
by dropping all exponentially suppressed corrections in the final expression. We now look for a combination $\phi$ of moduli that is not constrained by 
the condition  \eqref{eq:polextr}. If such an unfixed direction $\phi$ exists, two possibilities can occur:
\begin{itemize}
\item[(1)] The direction $\phi$ is a flat direction of the polynomial scalar potential $V_{\rm pol}$. This implies that $\phi$ is massless at polynomial order, but might obtain an exponentially small mass upon including either essential or non-essential instanton corrections. 
\item[(2)]  The direction $\phi$  is \textit{not} a flat direction of the polynomial scalar potential $V_{\rm pol}$, but rather has a mass term already at polynomial order. This implies that their mass term must arise from \textit{metric-essential instantons} correcting the periods. 
\end{itemize}
Let us comment on these two cases in turn. First, note that for case (1) the field $t_*$ not appearing in $V_{\rm pol}$ might still be stabilized after including instanton corrections. If non-essential instantons are used in order to ensure the stabilization one needs 
to check if such corrections are actually present for a given Calabi-Yau geometry and implement an appropriate stabilization scheme, such as a racetrack potential. To make this concrete in explicit examples was one of the successes of \cite{Demirtas:2019sip,Blumenhagen:2020ire,Demirtas:2020ffz}. In particular, it was shown that this can be done at large complex structure point in \cite{Demirtas:2019sip}. While there are no essential instantons in this asymptotic limit, it is well-known that generically instanton corrections arise and contribute to the superpotential. Second, we note that realization of case (2) is primarily dependent only on the type of asymptotic regime, since we have a systematic classification in which asymptotic regime certain essential instanton corrections have to be present. One has thus a stabilization mechanism that is independent of the actual Calabi-Yau geometry that realizes this limit. In the following we will focus on case (2) and explain in detail how such situations can be engineered. 

To gain a better understanding of what happens in case (2), let us note that such vacua exactly arise if 
the polynomial part of the K\"ahler potential $\cK_{\rm pol}$ yields a degenerate K\"ahler metric, i.e.~when the full K\"ahler metric has exponential eigenvalues. To see this we 
recall from the discussion below \eqref{eq:Wexpansion} that such a K\"ahler metric allows exponentially suppressed contributions in the F-terms to enter the scalar potential \eqref{eq:potential} at polynomial order. Hence, the stabilization of the moduli at polynomial order 
requires a more careful treatment of the complete F-terms supplementing \eqref{eq:polextr} by additional constraints coming from metric-essential instantons. 

Let us outline an approach to read off these additional constraints by a more careful treatment of the F-terms. 
The exponentially suppressed contributions in the F-terms appear in the polynomial scalar potential through exponential eigenvalues of the K\"ahler metric, so it is convenient to expand $D_I W$ in an eigenbasis for the K\"ahler metric. The eigenvectors relevant for the supplementary constraints then have a vanishing eigenvalue under the polynomial part of the K\"ahler metric 
\begin{equation}
(\partial_i \bar{\partial}_{\bar{\jmath}}\log \cK_{\rm pol}) V^j = 0\, ,
\end{equation}
For boundaries with a linear $\cK_{\rm pol}$ such as in section \ref{ssec:II1} these eigenvectors $V^i$ will always be independent of the saxions, but for more general boundaries this need not be the case. Letting $a_{r_1 \cdots r_n}$ denote the essential instanton term corresponding to a given eigenvector $V^i$, then the relevant F-term constraint can be written as
\begin{equation}\label{eq:expextr}
V^i D_i W =V^i \partial_i W_{\rm inst} =  e^{2\pi i r_i t^i} \langle F_3 - \tau_* H_3, \, V^i (2\pi i  r_i + N_i) e^{t^i_* N_i} a_{r_1 \cdots r_n} \rangle = 0 \, ,
\end{equation}
where we used that $V^i \partial_i W_{\rm pol}=0$, and replaced the covariant derivative by a partial derivative because $V^i K^{\rm cs}_i \, W_{\rm inst}$ is subleading in the instanton expansion. Furthermore, note that we dropped correction terms subleading compared to the essential instanton $a_{r_1 \cdots r_n}$. The complete set of extremization conditions to stabilize the moduli at polynomial order of the scalar potential is then given by \eqref{eq:polextr} and as many equations of the form \eqref{eq:expextr} as there are metric-essential instantons. Since the exponential factor in \eqref{eq:expextr} is simply an overall factor, this yields a system of polynomial equations in the moduli.\footnote{This polynomial structure of the extremization conditions is natural from the perspective of the self-duality condition \eqref{eq:selfduality}. The dependence on the complex structure moduli then enters through the Hodge star operator, whose polynomial part is already non-degenerate as discussed below \eqref{eq:Wexpansion}. }

An interesting feature of the vacua constructed according to the above scheme is a natural hierarchy of mass scales. All moduli are stabilized by the polynomial part of the scalar potential, so the eigenvalues of the hessian $\partial_I \partial_J V$ will be polynomial as well. Recall that the inverse K\"ahler metric either has polynomial or exponentially large eigenvalues, and that the canonically normalized masses are then computed as the eigenvalues of $K^{IC} \partial_C \partial_J V$. This means that the moduli masses take polynomial or exponentially large values in the vevs of the saxions. Either way, this separates the mass scale of the complex structure moduli and the axio-dilaton from the exponentially small scale of $|W_0|^2$. This hierarchy makes our flux vacua particularly attractive from the perspective of the KKLT scenario \cite{Kachru:2003aw}, since it allows one to consistently integrate out the complex structure and axio-dilaton sector before dealing with the K\"ahler moduli. This point separates our construction from other findings \cite{Demirtas:2019sip,Blumenhagen:2020ire,Demirtas:2020ffz}, where the mass of the lightest complex structure modulus was of the same scale then $|W_0|^2$, requiring a more refined analysis. 

We end this general discussion by commenting on the application of our procedure near the different types of boundary points. Given a type $\rm I$ point our method does not work. Roughly speaking, this is due to the fact that there is too little information contained in the polynomial piece of the period vector and as a consequence there is not enough freedom in the fluxes to stabilize all moduli at the polynomial level of the scalar potential as a result. We elaborated on this point in \cite{Bastian:2021hpc}. Near type $\rm II$ points, we have a polynomial K\"ahler potential that is at most linear in the complex structure moduli, which makes it the simplest and most natural candidate for our procedure. In section \ref{ssec:II0} and \ref{ssec:II1}, we explicitly apply our method to the one and two moduli cases respectively. A type $\rm III$ point does not occur in complex structure moduli spaces of dimension less than three. As the models developed in \cite{Bastian:2021eom} only had a maximum of two moduli, we did not have an explicit realization of this scenario at our disposal. There is, however, no obvious reason for which the construction should not go through in that case. For a type $\rm IV$ that is not at LCS, the simplest scenario occurs for two moduli. In that case there is a technical obstruction to the success of our method, but this problem should be solvable given more than two moduli. These last two types deserve more concrete investigations in the future.

\section{F-terms: Type IIB examples with small $W_0$}\label{sec:examples}
In this section we illustrate our method for finding flux vacua with exponentially small superpotentials by studying one- and two-moduli boundaries in complex structure moduli space. Our search uses asymptotic models for the periods near these boundaries constructed in chapter \ref{chap:models}. We emphasize that, since all possible one- and two-moduli boundaries have been classified \cite{Kerr2017,Grimm:2018cpv,Bastian:2021eom}, this construction yielded an exhaustive set of asymptotic periods. 

\subsection{One-modulus  asymptotic region near a type $\mathrm{II}$ point}\label{ssec:II0}
We start with the simplest case, namely the one-modulus asymptotic regions that are near type $\mathrm{II}$ boundaries. 
These have already been systematically studied in the math literature \cite{Tyurin:2003}, and such boundaries correspond to so-called Tyurin degenerations. The period vectors near these boundaries have been computed for an explicit example in the physics literature in \cite{Joshi:2019nzi}, and general models for these periods including essential instantons have been constructed in chapter \ref{chap:models}. Using the results from the latter, we recall that the K\"ahler potential \eqref{eq:kpII0}
\begin{equation}\label{eq:kpII02}
\cK_{\rm pol}=4  y \, , \qquad \cK_{\rm inst} =  \frac{4  a^2(1+\pi y)}{\pi} e^{-4 \pi y} \, , 
\end{equation}
where $a \in \mathbb{R}_{\neq 0}$ is some model-dependent coefficient that controls the essential instanton term and we have dropped all the non-essential instanton corrections. The polynomial and exponential parts of the flux superpotential \eqref{eq:superpotII0} take the respective forms
\begin{equation}\label{eq:WII0}
\begin{aligned}
W_{\rm pol}&=-g_3-ig_4+(g_1+ig_2)t \, ,\\
W_{\rm inst}&=a \, e^{2\pi i t}(t-\frac{1}{\pi i})(g_1-ig_2)-a \, e^{2\pi i t}(g_3-ig_4)\, ,
\end{aligned} \,
\end{equation}
where the $g_i = f_i - \tau h_i$ are the components of the three-form fluxes with $\tau = c+is$ being the axio-dilaton. We now proceed to stabilizing the moduli at the polynomial level. The vanishing F-term conditions for the polynomial part of the flux superpotential can conveniently be written as
\begin{equation}\label{eq:DWII0}
\begin{aligned}
y D_t W_{\rm pol} - s D_\tau W_{\rm pol} &= (f_1+i f_2) y - (h_3+i h_4) s=0\, , \\
y D_t W_{\rm pol} + s D_\tau W_{\rm pol} &= -i(f_3+i f_4) -i  (h_1+i h_2) sy =0\, , 
\end{aligned}
\end{equation}
where we dropped exponentially suppressed terms and set the axions to zero, i.e.~$c=x=0$. We chose not to impose $W_{\rm pol}=0$ at first, leaving us with the full covariant derivatives at polynomial level in the above extremization conditions. The solution to these constraints is given by
\begin{equation}\label{eq:vevsII0}
s = \sqrt{-\frac{f_1 f_3}{h_1 h_3}}\, , \qquad y = \sqrt{-\frac{ f_3 h_3}{f_1 h_1}}\, , \qquad f_2 h_3 = f_1 h_4\, , \qquad f_4 h_1 = f_3 h_2\, , 
\end{equation}
where we require $f_1/h_3 >0$ and $f_3/h_1<0$ in order to get positive values for the saxions. Plugging this solution back into the superpotential \eqref{eq:WII0} we find that $W_{\rm pol} = 0$ holds when
\begin{equation}
h_2 =- f_1 \sqrt{-\frac{h_1 h_3}{f_1 f_3}}\, , \qquad h_4 = - f_3 \sqrt{-\frac{h_1 h_3}{f_1 f_3}}\, . \label{eq:II0Const2}
\end{equation}
Implementing the constraints \eqref{eq:vevsII0} and \eqref{eq:II0Const2}, we obtain flux vacua that have $W_{\rm pol}=0$. Thus, the scale of the superpotential is set by 
\begin{align}
|W|=|W_{\rm inst}| \sim a \, e^{- 2 \pi y} = a  \exp\Big(-2\pi \sqrt{-\frac{ f_3 h_3}{f_1 h_1}}\, \Big)\, .
\end{align}
We want to emphasize that the exponentially small scale of the superpotential is set by an essential instanton term that is required by consistency of the theory, i.e.~the coefficient $a$ cannot vanish for any geometric example. Furthermore, we can now relate the saxion masses and their vevs to the tadpole. The latter takes the form
\begin{equation}
Q_{\rm D3}=f_1 h_3 - f_3 h_1 \,.
\end{equation}
Computing the masses from the scalar potential as the eigenvalues of $K^{ac} \partial_c \partial_b V$ we find that they are given by the compact expression
\begin{align}
m_{t}^2=m_{\tau}^2=\frac{1}{\cV^2}(f_1 h_3 - f_3 h_1) =  \frac{Q_{\rm D3}}{\cV^2} \,,
\end{align}
where $m_t^2, m_\tau^2$ denote the canonically normalized moduli masses associated with the complex structure modulus and the axio-dilaton respectively. Let us also stress that this one-modulus case is rather non-generic as there are no metric-essential instantons required, which have to be present in the multi-moduli case as we will see below. Furthermore, the saxion vev $y$ is bounded from above by the tadpole as
\begin{align}
y \leq  2(f_1 h_3-f_3 h_1)= 2 Q_{\rm D3} \,,
\end{align}
which is saturated for $f_3=h_3$ and $f_1=h_1=1$. As the scale of $|W_0|$ is set by $e^{-2 \pi y}$, we cannot make it arbitrarily small by tuning the vev of the saxion $y$, i.e.
\begin{align}\label{eq:W0boundII0}
|W_0| \gtrsim e^{-2 \pi Q_{\rm D3}}\, .
\end{align}
 To be more concrete, we consider a specific set of fluxes that satisfy the above equations. Given $H_3 = (-1,-2,4,-2)$ and $F_3=(8,-4,8,16)$, we find that
\begin{equation}
s = 4\, , \qquad y=2\,, \qquad |W_0 | \sim a \, e^{-4\pi}\, ,\qquad m_t^2 = m_\tau^2 = Q_{\rm D3} =40\, ,
\end{equation}
where we dropped the volume factor $1/\cV^2$ in the masses for convenience.

\subsection{Two-moduli asymptotic region near a type II point}\label{ssec:II1}
Next we consider the more involved class of two-moduli  boundaries that intersect on a type II point. Such asymptotic regions are realized, for example, in the moduli space of the Calabi-Yau threefold in $\mathbb{P}^{1,1,2,2,6}_4[12]$ near a Seiberg-Witten point \cite{Kachru:1995fv, Curio:2000sc,Eguchi:2007iw,Lee:2019wij}. Using the asymptotic model from chapter \ref{chap:models}, we find that the K\"ahler potential \eqref{eq:II1kp} near these boundaries takes the form
\begin{align}\label{eq:kpII1}
\cK_{\rm pol} &= 4(y_1+n_2 y_2) \, ,  \nn \\
\cK_{\rm inst} &= -2 a^2 e^{-4\pi y_2} \Big(  n_1 y_1+y_2 + \frac{1-n_1 n_2}{2\pi} \Big) \\
& \ \ \ \, -2 n_2 b^2 e^{-4\pi y_1} \Big( n_2(n_1y_1+y_2) -\frac{1-n_1 n_2}{2\pi} \Big) \nn \\
& \ \ \ \, -4 ab e^{-2\pi y_1-2\pi y_2}\Big(  n_2 ( n_1 y_1+y_2) + \frac{(n_2-1)(1-n_2 n_2)}{4\pi} \Big) \cos(2\pi(x_1-x_2)) \, ,\nn 
\end{align}
where $a,b \in \mathbb{R}_{\neq 0}$ and $n_1,n_2 \in  \mathbb{Q}_{\geq 0}$ are model-dependent parameters that control the essential instanton terms. The polynomial and exponential parts of the flux superpotential \eqref{eq:superpotentialII1} take the respective forms
\begin{equation}\label{eq:WII1}
\begin{aligned}
W_{\rm pol} &=  ( g_1+ig_2) (t_1+n_{2}  t_2) +( g_1+i g_2)- (g_4+i g_5)\, ,\\
W_{\rm inst} &=   a \, e^{2\pi i t_2} \Big(ig_3\frac{1-n_1 n_2}{2 \pi } -g_3 n_{1} n_{2}  t_1 - g_3 n_{2} t_2-     g_6 n_{2}\Big) \\
  & \ \ \ +  b \, e^{2\pi i t_1} \Big(ig_3 \frac{n_{1} n_{2}-1}{2\pi}  -g_3 n_{1} t_1 -g_3 t_2 -
   g_6 \Big) +\mathcal{O}(e^{-4\pi y})\,  .
\end{aligned}
\end{equation}
 Let us already point out that the polynomial part of the K\"ahler potential and flux superpotential is very similar to those of the one-modulus $\mathrm{II}_0$ boundaries given in \eqref{eq:kpII02} and \eqref{eq:WII0}, for which we only need to replace $t$ by $t_1+n_2 t_2$ and relabel some of the fluxes. By computing the K\"ahler metric for \eqref{eq:kpII1} one sees that $\cK_{\rm pol}$ yields a degenerate polynomial metric, as elaborated upon below \eqref{eq:example_metric} . This degeneracy is cured by the exponentially suppressed term involving $a^2$ in $\cK_{\rm inst}$, which therefore is a metric-essential instanton. Note that the essential instanton term involving $b$ can also cure this degeneracy unless $n_2=0$. 

Now let us turn to the extremization conditions that have to be solved. The vanishing F-term conditions can conveniently be written as
\begin{equation}\label{eq:DWII1}
\begin{aligned}
(y_1 + n_2 y_2) D_{t_1} W - s D_\tau W &= (y_1+n_2 y_2)(f_1+if_2) -s(h_4+i h_5) = 0\, , \\
(y_1 + n_2 y_2) D_{t_1} W + s D_\tau W &=  -i f_4 +f_5 + s(y_1+n_2 y_2)(-ih_1+h_2)= 0\, , \\
\frac{n_2  D_{t_1} W - D_{t_2} W}{2\pi (a e^{-2\pi y_2}-n_2^2 b e^{-2\pi y_1} ) } &= i f_6 +s h_6 -(n_1 y_1 +y_2) (f_3 - i sh_3) = 0\, .
\end{aligned}
\end{equation}
The first two conditions take a similar form as \eqref{eq:DWII0} found near one-modulus $\mathrm{II}$ boundaries, where the saxion $y$ is now replaced by the linear combination $y_1+n_2 y_2$ and some fluxes are relabeled. Proceeding in the same fashion as the one-modulus setup, these two equations fix the saxions $y_1+n_2 y_2$ and $s$, and furthermore impose two relations on the fluxes that ensure their axionic partners have vanishing vevs
\begin{equation}\label{eq:solve1II1}
y_1+n_2 y_2 =\sqrt{-\frac{ f_4 h_4}{f_1 h_1}} \, , \quad s =\sqrt{-\frac{f_1 f_4}{h_1 h_4}} \, , \quad f_2 h_4 = f_1 h_5\, , \quad f_5 h_1 = f_4 h_2\, ,
\end{equation}
where we require $f_1/h_4 >0$ and $f_4/h_1<0$ in order to get sensible values for the saxions. 

The relevance of the third equation in \eqref{eq:DWII1} is a bit more subtle. While this F-term is exponentially suppressed, it does contribute to the polynomial part of the scalar potential. This can be seen by carefully inspecting the K\"ahler potential \eqref{eq:kpII1}, which yields a K\"ahler metric with an exponentially small eigenvalue for the eigenvector $(n_2, -1)$. This scaling then cancels against the scaling of the F-term when computing the scalar potential through \eqref{eq:potential}, hence contributing at polynomial order. Therefore we must require the third F-term given in \eqref{eq:DWII1} to vanish as well, which is solved by
\begin{equation}\label{eq:solve2II1}
n_1 y_1+y_2 =  \sqrt{-\frac{f_6 h_6}{f_3 h_3}}  \, , \qquad f_3 f_6 h_1 h_4 = f_1 f_4 h_3 h_6\, .
\end{equation}
where we must require $f_3 / h_6 >0$ and $f_6 / h_3 <0$ to have positive saxion vevs compatible with the positivity conditions given below \eqref{eq:solve1II1}.

Together \eqref{eq:solve1II1} and \eqref{eq:solve2II1} specify all vacua (with vanishing axion vevs) of the polynomial scalar potential arising from the flux superpotential \eqref{eq:WII1}. We are interested in constructing vacua with $W_{\rm pol}=0$, which amounts to additionally imposing
\begin{equation}
h_2 = -f_1 \sqrt{-\frac{h_1 h_4}{f_1 f_4}}\, , \qquad h_5 = -f_4 \sqrt{-\frac{h_1 h_4}{f_1 f_4}}\, .
\end{equation}
It is now instructive to compare the scale of the moduli masses and the vacuum superpotential again with the D3-brane tadpole. The tadpole contribution due to the fluxes can be split into two parts as
\begin{equation}
Q_{\rm D3}  = \underbrace{ f_1 h_4-  f_4 h_1}_{Q^{\rm pol}_{\rm D3} }   +\underbrace{\frac{1}{2}( f_3 h_6 -f_6 h_3)}_{Q^{\rm inst}_{\rm D3} }\, , 
\end{equation}
where we separated the fluxes that appear in the superpotential \eqref{eq:WII1} at polynomial order from the fluxes that enter at exponential order. From the positivity conditions on the fluxes given below \eqref{eq:solve1II1} and \eqref{eq:solve2II1} we see that all terms in the contributions $Q^{\rm pol}_{\rm D3}$ and $Q^{\rm inst}_{\rm D3}$ are positive. For the canonically normalized masses of the moduli we then compute the eigenvalues of $K^{ac}\partial_c \partial_b V$ and find that
\begin{equation}
m_\tau^2 = m_{t_1+n_2 t_2}^2 =  \frac{f_1 h_4-  f_4 h_1}{\cV^2} = \frac{Q^{\rm pol}_{\rm D3} }{\cV^2} \leq \frac{Q_{\rm D3}}{\cV^2}  \, ,
\end{equation}
where we dropped the overall volume factor $1/\cV^2$ for convenience. We do not write down the mass associated with the modulus $n_2 t_1 - t_2$ here, but note that it takes an exponentially large value in the fluxes, since this field corresponds to an exponential eigenvalue of the K\"ahler metric. 

For the scale of the vacuum superpotential we need to study the saxion vevs. By using \eqref{eq:solve1II1} and \eqref{eq:solve2II1} we find the following two bounds through the D3-brane tadpole
\begin{equation}
y_1+n_2 y_2 \leq \frac{1}{2}Q^{\rm pol}_{\rm D3}\, ,\qquad n_1 y_1+y_2 \leq Q^{\rm inst}_{\rm D3} 
\end{equation}
which are saturated for $f_1=h_1=1$, $f_4=-h_4$ and $f_3=h_3=1$, $f_6=-h_6$ respectively. The vacuum superpotential is then set by the smallest saxion vev as
\begin{equation}
|W_0| \sim e^{-2\pi \min(y_1,y_2)} \geq e^{-2\pi Q_{\rm D3}}\, .
\end{equation}
Given both this bound on the vacuum superpotential and \eqref{eq:W0boundII0} in the one-modulus setup, it is tempting to speculate that $|W_0|$ can be bounded from below by the D3-brane tadpole near any type II point, also in higher-dimensional moduli spaces. In light of the tadpole conjecture \cite{Bena:2020xrh,Bena:2021wyr} this hints at an interesting tradeoff. Recall from the discussion below \eqref{eq:fourfoldtadpole} that the allowed tadpole charge $Q_{\rm D3}$ grows linearly with the number of moduli, so for a geometry with large $h^{2,1}$ one is able to achieve smaller values for $|W_0|$. In turn, the tadpole conjecture indicates that for such geometries one cannot stabilize all moduli while satisfying the tadpole bound, so the catch is that there remain some flat directions. 

To make the above story more concrete, let us consider the Seiberg-Witten point in the moduli space of the Calabi-Yau threefold in $\mathbb{P}^{1,1,2,2,6}_4[12]$ as an example. Following \cite{Bastian:2021eom}, we find that this boundary corresponds to the monodromy data $n_1=0$, $n_2=1/4$. As flux quanta we pick
\begin{equation}
F_3 = (-4,-8,10,16,-8,-8)\, , \qquad  H_3 = (-2,1,1,-2,-4,5)\, .
\end{equation}
The scalar potential then has a minimum at
\begin{equation}
s = 4\, , \qquad y_1 = 1.5 \, , \qquad y_2 = 2 \, , \qquad |W_0| = 7.3\cdot 10^{-5}\, ,
\end{equation}
where we used that $b=-4 \Gamma(3/4)^4/(\sqrt{3} \pi^2)$, and ignored the coefficient $a$ in the expansion of the superpotential \eqref{eq:WII1} since $e^{-2\pi y_1} \gg e^{-2\pi y_2} $. The canonically normalized moduli masses are then computed to be
\begin{equation}
 m_{n_2 t_1-t_2}^2 = 1.3 \cdot 10^{10} \, , \qquad m_{t_1+n_2 t_2}^2  = m_{\tau}^2 = 40 = Q^{\rm pol}_{\rm D3} \, ,
\end{equation}
where we dropped the volume factor $1/\cV^2$ in the moduli masses for convenience.
Note in particular that there is one exponentially large mass, which corresponds to the eigenvector of the K\"ahler metric with an exponential eigenvalue. Furthermore, all moduli masses are orders of magnitude larger than the exponentially small $|W_0|^2$.

\paragraph{Summary.} In this chapter we have studied moduli stabilization near boundaries in complex structure moduli space. On the one hand, for the self-duality condition we found that the nilpotent orbit and sl(2)-approximations reduce the extremization conditions to a simple set of algebraic equations. On the other, for the F-terms metric-essential corrections to the superpotential had to be taken into account. The equivalence of the two approaches became apparent upon noting that the inverse K\"ahler metric in the scalar potential \eqref{eq:potential} cancels the exponential scaling of these corrections in the F-terms, thereby producing polynomial terms in the Hodge star formulation. Both these methods allowed for interesting constructions from a phenomenological point of view:
\begin{itemize}
\item For the self-duality condition we found that asymptotic Hodge theory turns the original very complicated problem of solving the flux vacuum conditions into a tractable polynomial task. We explicitly carried out its construction for a number of examples, and showed that the vacua found with the sl(2)-approximation resemble the vacua found in the nilpotent orbit approximation rather well --- even if one imposes only a moderate hierarchy of the saxions. We expect that the step-wise approach reduces the computational complexity (see \cite{Denef:2006ad, Halverson:2018cio} for an in-depth discussion of the arising challenges). Furthermore, it would be exciting to combine our approach with recent efforts to use machine learning algorithms in studying the string landscape \cite{Abel:2014xta, Ruehle:2017mzq, Halverson:2019tkf, Betzler:2019kon, Cole:2019enn, AbdusSalam:2020ywo,CaboBizet:2020cse, Bena:2021wyr, Krippendorf:2021uxu}.

\item For the F-term approach we proposed a method for constructing Type IIB flux vacua away from large complex structure with an exponentially small vacuum superpotential. These constructions relied on the insights of chapter \ref{chap:models} regarding essential exponential corrections to the $(3,0)$-form periods near such boundaries. The takeaway message was that there is a conceptual difference in whether the exponentially small value of the superpotential is induced by essential or non-essential instantons. Non-essential instantons are well-known to arise in most explicit geometric examples, but lead to mass matrices with exponentially small eigenvalues \cite{Demirtas:2019sip,Demirtas:2020ffz,Blumenhagen:2020ire}. In contrast, essential instantons that induce an exponentially small $W_0$ combine in the superpotential and K\"ahler potential such that they enter at polynomial level in the scalar potential. Consequently, the masses of the moduli stabilized by these terms are at least of $\cO(1)$, thus making it possible to study moduli stabilization at polynomial level while keeping the vacuum superpotential exponentially small.

\end{itemize}

\begin{subappendices}
\section{Sl(2)-approximation for the F-theory example}\label{app:linearscenario}
In this section we summarize the relevant building blocks for the sl(2)-approximated Hodge star \eqref{eq:Csl2ToCInf}. This consists of the weight and lowering operators $N^0_i, N^-_i$ of the sl(2)-triples, and the boundary Hodge star $C_\infty$.
The weight operators are given by
\begingroup
\def\myFigureScale{0.60}
\allowdisplaybreaks
\eq{
\label{linear_operators_weights}
\scalebox{\myFigureScale}{$N^0_0$} &\scalebox{\myFigureScale}{$= \left(
\begin{array}{cccccccccccccccc}
 1 & 0 & 0 & 0 & 0 & 0 & 0 & 0 & 0 & 0 & 0 & 0 & 0 & 0 & 0 & 0 \\
 0 & -1 & 0 & 0 & 0 & 0 & 0 & 0 & 0 & 0 & 0 & 0 & 0 & 0 & 0 & 0 \\
 0 & 0 & 1 & 0 & 0 & 0 & 0 & 0 & 0 & 0 & 0 & 0 & 0 & 0 & 0 & 0 \\
 0 & 1 & 0 & 1 & 0 & 0 & 0 & 0 & 0 & 0 & 0 & 0 & 0 & 0 & 0 & 0 \\
 0 & 2 & 0 & 0 & 1 & 0 & 0 & 0 & 0 & 0 & 0 & 0 & 0 & 0 & 0 & 0 \\
 0 & 0 & 0 & 0 & 0 & 1 & 0 & 0 & 0 & -2 & -3 & 0 & 0 & 0 & 0 & 0 \\
 0 & 0 & 0 & 0 & 0 & 0 & 1 & 0 & -2 & -4 & -6 & 0 & 0 & 0 & 0 & 0 \\
 0 & 0 & 0 & 0 & 0 & 0 & 0 & 1 & -3 & -6 & -6 & 0 & 0 & 0 & 0 & 0 \\
 0 & 0 & 0 & 0 & 0 & 0 & 0 & 0 & -1 & 0 & 0 & 0 & 0 & 0 & 0 & 0 \\
 0 & 0 & 0 & 0 & 0 & 0 & 0 & 0 & 0 & -1 & 0 & 0 & 0 & 0 & 0 & 0 \\
 0 & 0 & 0 & 0 & 0 & 0 & 0 & 0 & 0 & 0 & -1 & 0 & 0 & 0 & 0 & 0 \\
 0 & 0 & 0 & 0 & 0 & 0 & 0 & 0 & 0 & 0 & 0 & 1 & 0 & -1 & -2 & 0 \\
 0 & 0 & 0 & 0 & 0 & 0 & 0 & 0 & 0 & 0 & 0 & 0 & -1 & 0 & 0 & 0 \\
 0 & 0 & 0 & 0 & 0 & 0 & 0 & 0 & 0 & 0 & 0 & 0 & 0 & -1 & 0 & 0 \\
 0 & 0 & 0 & 0 & 0 & 0 & 0 & 0 & 0 & 0 & 0 & 0 & 0 & 0 & -1 & 0 \\
 0 & 0 & 0 & 0 & 0 & 0 & 0 & 0 & 0 & 0 & 0 & 0 & 0 & 0 & 0 & -1 \\
\end{array}
\right) , $}
\\
\scalebox{\myFigureScale}{$N^0_L$} &\scalebox{\myFigureScale}{$= \left(
\begin{array}{cccccccccccccccc}
 1 & 0 & 0 & 0 & 0 & 0 & 0 & 0 & 0 & 0 & 0 & 0 & 0 & 0 & 0 & 0 \\
 0 & 1 & 0 & 0 & 0 & 0 & 0 & 0 & 0 & 0 & 0 & 0 & 0 & 0 & 0 & 0 \\
 0 & -2 & -1 & -2 & -1 & 0 & 0 & 0 & 0 & 0 & 0 & 0 & 0 & 0 & 0 & 0 \\
 0 & 0 & 0 & 1 & 0 & 0 & 0 & 0 & 0 & 0 & 0 & 0 & 0 & 0 & 0 & 0 \\
 0 & 0 & 0 & 0 & 1 & 0 & 0 & 0 & 0 & 0 & 0 & 0 & 0 & 0 & 0 & 0 \\
 0 & 0 & 0 & 0 & 0 & 1 & 0 & 0 & 0 & 0 & 0 & 0 & 0 & 0 & 0 & 0 \\
 0 & 0 & 0 & 0 & 0 & 2 & -1 & 0 & 0 & 0 & 2 & 0 & 0 & 0 & 0 & 0 \\
 0 & 0 & 0 & 0 & 0 & 1 & 0 & -1 & 0 & 2 & 3 & 0 & 0 & 0 & 0 & 0 \\
 0 & 0 & 0 & 0 & 0 & 0 & 0 & 0 & -1 & -2 & -1 & 0 & 0 & 0 & 0 & 0 \\
 0 & 0 & 0 & 0 & 0 & 0 & 0 & 0 & 0 & 1 & 0 & 0 & 0 & 0 & 0 & 0 \\
 0 & 0 & 0 & 0 & 0 & 0 & 0 & 0 & 0 & 0 & 1 & 0 & 0 & 0 & 0 & 0 \\
 0 & 0 & 0 & 0 & 0 & 0 & 0 & 0 & 0 & 0 & 0 & -1 & 2 & 0 & 0 & 0 \\
 0 & 0 & 0 & 0 & 0 & 0 & 0 & 0 & 0 & 0 & 0 & 0 & 1 & 0 & 0 & 0 \\
 0 & 0 & 0 & 0 & 0 & 0 & 0 & 0 & 0 & 0 & 0 & 0 & 2 & -1 & 0 & 0 \\
 0 & 0 & 0 & 0 & 0 & 0 & 0 & 0 & 0 & 0 & 0 & 0 & 1 & 0 & -1 & 0 \\
 0 & 0 & 0 & 0 & 0 & 0 & 0 & 0 & 0 & 0 & 0 & 0 & 0 & 0 & 0 & -1 \\
\end{array}
\right), $}\\
\scalebox{\myFigureScale}{$N^0_1$} &\scalebox{\myFigureScale}{$= \left(
\begin{array}{cccccccccccccccc}
 1 & 0 & 0 & 0 & 0 & 0 & 0 & 0 & 0 & 0 & 0 & 0 & 0 & 0 & 0 & 0 \\
 0 & 1 & 0 & 0 & 0 & 0 & 0 & 0 & 0 & 0 & 0 & 0 & 0 & 0 & 0 & 0 \\
 0 & 2 & 1 & 2 & 1 & 0 & 0 & 0 & 0 & 0 & 0 & 0 & 0 & 0 & 0 & 0 \\
 0 & -2 & 0 & -1 & -1 & 0 & 0 & 0 & 0 & 0 & 0 & 0 & 0 & 0 & 0 & 0 \\
 0 & 0 & 0 & 0 & 1 & 0 & 0 & 0 & 0 & 0 & 0 & 0 & 0 & 0 & 0 & 0 \\
 0 & 0 & 0 & 0 & 0 & -1 & 0 & 0 & 0 & 0 & 2 & 0 & 0 & 0 & 0 & 0 \\
 0 & 0 & 0 & 0 & 0 & -2 & 1 & 0 & 0 & 0 & 2 & 0 & 0 & 0 & 0 & 0 \\
 0 & 0 & 0 & 0 & 0 & -1 & 1 & -1 & 2 & 2 & 3 & 0 & 0 & 0 & 0 & 0 \\
 0 & 0 & 0 & 0 & 0 & 0 & 0 & 0 & 1 & 2 & 1 & 0 & 0 & 0 & 0 & 0 \\
 0 & 0 & 0 & 0 & 0 & 0 & 0 & 0 & 0 & -1 & -1 & 0 & 0 & 0 & 0 & 0 \\
 0 & 0 & 0 & 0 & 0 & 0 & 0 & 0 & 0 & 0 & 1 & 0 & 0 & 0 & 0 & 0 \\
 0 & 0 & 0 & 0 & 0 & 0 & 0 & 0 & 0 & 0 & 0 & -1 & -2 & 2 & 0 & 0 \\
 0 & 0 & 0 & 0 & 0 & 0 & 0 & 0 & 0 & 0 & 0 & 0 & -1 & 0 & 0 & 0 \\
 0 & 0 & 0 & 0 & 0 & 0 & 0 & 0 & 0 & 0 & 0 & 0 & -2 & 1 & 0 & 0 \\
 0 & 0 & 0 & 0 & 0 & 0 & 0 & 0 & 0 & 0 & 0 & 0 & -1 & 1 & -1 & 0 \\
 0 & 0 & 0 & 0 & 0 & 0 & 0 & 0 & 0 & 0 & 0 & 0 & 0 & 0 & 0 & -1 \\
\end{array}
\right),$} \\
\scalebox{\myFigureScale}{$N^0_2$} &\scalebox{\myFigureScale}{$= \left(
\begin{array}{cccccccccccccccc}
 1 & 0 & 0 & 0 & 0 & 0 & 0 & 0 & 0 & 0 & 0 & 0 & 0 & 0 & 0 & 0 \\
 0 & 1 & 0 & 0 & 0 & 0 & 0 & 0 & 0 & 0 & 0 & 0 & 0 & 0 & 0 & 0 \\
 0 & 0 & 1 & 0 & 0 & 0 & 0 & 0 & 0 & 0 & 0 & 0 & 0 & 0 & 0 & 0 \\
 0 & 1 & 0 & 1 & 1 & 0 & 0 & 0 & 0 & 0 & 0 & 0 & 0 & 0 & 0 & 0 \\
 0 & -2 & 0 & 0 & -1 & 0 & 0 & 0 & 0 & 0 & 0 & 0 & 0 & 0 & 0 & 0 \\
 0 & 0 & 0 & 0 & 0 & -1 & 0 & 0 & 0 & 2 & 1 & 0 & 0 & 0 & 0 & 0 \\
 0 & 0 & 0 & 0 & 0 & 0 & -1 & 0 & 2 & 4 & 2 & 0 & 0 & 0 & 0 & 0 \\
 0 & 0 & 0 & 0 & 0 & 0 & -1 & 1 & 1 & 2 & 0 & 0 & 0 & 0 & 0 & 0 \\
 0 & 0 & 0 & 0 & 0 & 0 & 0 & 0 & 1 & 0 & 0 & 0 & 0 & 0 & 0 & 0 \\
 0 & 0 & 0 & 0 & 0 & 0 & 0 & 0 & 0 & 1 & 1 & 0 & 0 & 0 & 0 & 0 \\
 0 & 0 & 0 & 0 & 0 & 0 & 0 & 0 & 0 & 0 & -1 & 0 & 0 & 0 & 0 & 0 \\
 0 & 0 & 0 & 0 & 0 & 0 & 0 & 0 & 0 & 0 & 0 & -1 & 0 & -1 & 2 & 0 \\
 0 & 0 & 0 & 0 & 0 & 0 & 0 & 0 & 0 & 0 & 0 & 0 & -1 & 0 & 0 & 0 \\
 0 & 0 & 0 & 0 & 0 & 0 & 0 & 0 & 0 & 0 & 0 & 0 & 0 & -1 & 0 & 0 \\
 0 & 0 & 0 & 0 & 0 & 0 & 0 & 0 & 0 & 0 & 0 & 0 & 0 & -1 & 1 & 0 \\
 0 & 0 & 0 & 0 & 0 & 0 & 0 & 0 & 0 & 0 & 0 & 0 & 0 & 0 & 0 & -1 \\
\end{array}\right).$}}
The lowering operators are given by
\eq{
\scalebox{\myFigureScale}{$N_0^-$} &\scalebox{\myFigureScale}{$= \left(
\begin{array}{cccccccccccccccc}
 0 & 0 & 0 & 0 & 0 & 0 & 0 & 0 & 0 & 0 & 0 & 0 & 0 & 0 & 0 & 0 \\
 -1 & 0 & 0 & 0 & 0 & 0 & 0 & 0 & 0 & 0 & 0 & 0 & 0 & 0 & 0 & 0 \\
 0 & 0 & 0 & 0 & 0 & 0 & 0 & 0 & 0 & 0 & 0 & 0 & 0 & 0 & 0 & 0 \\
 \frac{1}{2} & 0 & 0 & 0 & 0 & 0 & 0 & 0 & 0 & 0 & 0 & 0 & 0 & 0 & 0 & 0 \\
 1 & 0 & 0 & 0 & 0 & 0 & 0 & 0 & 0 & 0 & 0 & 0 & 0 & 0 & 0 & 0 \\
 0 & -2 & 0 & -1 & -\frac{3}{2} & 0 & 0 & 0 & 0 & 0 & 0 & 0 & 0 & 0 & 0 & 0 \\
 0 & -4 & -1 & -2 & -3 & 0 & 0 & 0 & 0 & 0 & 0 & 0 & 0 & 0 & 0 & 0 \\
 0 & -\frac{9}{2} & -\frac{3}{2} & -3 & -3 & 0 & 0 & 0 & 0 & 0 & 0 & 0 & 0 & 0 & 0 & 0 \\
 0 & 0 & -1 & 0 & 0 & 0 & 0 & 0 & 0 & 0 & 0 & 0 & 0 & 0 & 0 & 0 \\
 0 & -\frac{1}{2} & 0 & -1 & 0 & 0 & 0 & 0 & 0 & 0 & 0 & 0 & 0 & 0 & 0 & 0 \\
 0 & -1 & 0 & 0 & -1 & 0 & 0 & 0 & 0 & 0 & 0 & 0 & 0 & 0 & 0 & 0 \\
 0 & 0 & 0 & 0 & 0 & 0 & -\frac{1}{2} & -1 & 2 & 4 & \frac{9}{2} & 0 & 0 & 0 & 0 & 0 \\
 0 & 0 & 0 & 0 & 0 & -1 & 0 & 0 & 0 & 1 & \frac{3}{2} & 0 & 0 & 0 & 0 & 0 \\
 0 & 0 & 0 & 0 & 0 & 0 & -1 & 0 & 1 & 2 & 3 & 0 & 0 & 0 & 0 & 0 \\
 0 & 0 & 0 & 0 & 0 & 0 & 0 & -1 & \frac{3}{2} & 3 & 3 & 0 & 0 & 0 & 0 & 0 \\
 0 & 0 & 0 & 0 & 0 & 0 & 0 & 0 & 0 & 0 & 0 & -1 & 0 & \frac{1}{2} & 1 & 0 \\
\end{array}
\right) ,$}\\
\scalebox{\myFigureScale}{$N_L^- $} &\scalebox{\myFigureScale}{$= \left(
\begin{array}{cccccccccccccccc}
 0 & 0 & 0 & 0 & 0 & 0 & 0 & 0 & 0 & 0 & 0 & 0 & 0 & 0 & 0 & 0 \\
 0 & 0 & 0 & 0 & 0 & 0 & 0 & 0 & 0 & 0 & 0 & 0 & 0 & 0 & 0 & 0 \\
 -1 & 0 & 0 & 0 & 0 & 0 & 0 & 0 & 0 & 0 & 0 & 0 & 0 & 0 & 0 & 0 \\
 0 & 0 & 0 & 0 & 0 & 0 & 0 & 0 & 0 & 0 & 0 & 0 & 0 & 0 & 0 & 0 \\
 0 & 0 & 0 & 0 & 0 & 0 & 0 & 0 & 0 & 0 & 0 & 0 & 0 & 0 & 0 & 0 \\
 0 & 0 & 0 & 0 & 0 & 0 & 0 & 0 & 0 & 0 & 0 & 0 & 0 & 0 & 0 & 0 \\
 0 & -2 & 0 & 0 & -1 & 0 & 0 & 0 & 0 & 0 & 0 & 0 & 0 & 0 & 0 & 0 \\
 0 & -3 & 0 & -1 & -1 & 0 & 0 & 0 & 0 & 0 & 0 & 0 & 0 & 0 & 0 & 0 \\
 0 & -1 & 0 & 0 & 0 & 0 & 0 & 0 & 0 & 0 & 0 & 0 & 0 & 0 & 0 & 0 \\
 0 & 0 & 0 & 0 & 0 & 0 & 0 & 0 & 0 & 0 & 0 & 0 & 0 & 0 & 0 & 0 \\
 0 & 0 & 0 & 0 & 0 & 0 & 0 & 0 & 0 & 0 & 0 & 0 & 0 & 0 & 0 & 0 \\
 0 & 0 & 0 & 0 & 0 & -1 & 0 & 0 & 0 & 0 & 0 & 0 & 0 & 0 & 0 & 0 \\
 0 & 0 & 0 & 0 & 0 & 0 & 0 & 0 & 0 & 0 & 0 & 0 & 0 & 0 & 0 & 0 \\
 0 & 0 & 0 & 0 & 0 & 0 & 0 & 0 & 0 & 0 & -1 & 0 & 0 & 0 & 0 & 0 \\
 0 & 0 & 0 & 0 & 0 & 0 & 0 & 0 & 0 & -1 & -1 & 0 & 0 & 0 & 0 & 0 \\
 0 & 0 & 0 & 0 & 0 & 0 & 0 & 0 & 0 & 0 & 0 & 0 & -1 & 0 & 0 & 0 \\
\end{array}
\right),$} \\
\scalebox{\myFigureScale}{$N_1^- $} &\scalebox{\myFigureScale}{$= \left(
\begin{array}{cccccccccccccccc}
 0 & 0 & 0 & 0 & 0 & 0 & 0 & 0 & 0 & 0 & 0 & 0 & 0 & 0 & 0 & 0 \\
 0 & 0 & 0 & 0 & 0 & 0 & 0 & 0 & 0 & 0 & 0 & 0 & 0 & 0 & 0 & 0 \\
 1 & 0 & 0 & 0 & 0 & 0 & 0 & 0 & 0 & 0 & 0 & 0 & 0 & 0 & 0 & 0 \\
 -1 & 0 & 0 & 0 & 0 & 0 & 0 & 0 & 0 & 0 & 0 & 0 & 0 & 0 & 0 & 0 \\
 0 & 0 & 0 & 0 & 0 & 0 & 0 & 0 & 0 & 0 & 0 & 0 & 0 & 0 & 0 & 0 \\
 0 & -2 & 0 & 0 & -1 & 0 & 0 & 0 & 0 & 0 & 0 & 0 & 0 & 0 & 0 & 0 \\
 0 & -2 & 0 & 0 & -1 & 0 & 0 & 0 & 0 & 0 & 0 & 0 & 0 & 0 & 0 & 0 \\
 0 & -3 & -1 & -1 & -1 & 0 & 0 & 0 & 0 & 0 & 0 & 0 & 0 & 0 & 0 & 0 \\
 0 & 1 & 0 & 0 & 0 & 0 & 0 & 0 & 0 & 0 & 0 & 0 & 0 & 0 & 0 & 0 \\
 0 & -1 & 0 & 0 & 0 & 0 & 0 & 0 & 0 & 0 & 0 & 0 & 0 & 0 & 0 & 0 \\
 0 & 0 & 0 & 0 & 0 & 0 & 0 & 0 & 0 & 0 & 0 & 0 & 0 & 0 & 0 & 0 \\
 0 & 0 & 0 & 0 & 0 & 1 & -1 & 0 & 0 & 0 & 0 & 0 & 0 & 0 & 0 & 0 \\
 0 & 0 & 0 & 0 & 0 & 0 & 0 & 0 & 0 & 0 & -1 & 0 & 0 & 0 & 0 & 0 \\
 0 & 0 & 0 & 0 & 0 & 0 & 0 & 0 & 0 & 0 & -1 & 0 & 0 & 0 & 0 & 0 \\
 0 & 0 & 0 & 0 & 0 & 0 & 0 & 0 & -1 & -1 & -1 & 0 & 0 & 0 & 0 & 0 \\
 0 & 0 & 0 & 0 & 0 & 0 & 0 & 0 & 0 & 0 & 0 & 0 & 1 & -1 & 0 & 0 \\
\end{array}
\right),$} \\
\scalebox{\myFigureScale}{$N_2^- $} &\scalebox{\myFigureScale}{$=\left(
\begin{array}{cccccccccccccccc}
 0 & 0 & 0 & 0 & 0 & 0 & 0 & 0 & 0 & 0 & 0 & 0 & 0 & 0 & 0 & 0 \\
 0 & 0 & 0 & 0 & 0 & 0 & 0 & 0 & 0 & 0 & 0 & 0 & 0 & 0 & 0 & 0 \\
 0 & 0 & 0 & 0 & 0 & 0 & 0 & 0 & 0 & 0 & 0 & 0 & 0 & 0 & 0 & 0 \\
 \frac{1}{2} & 0 & 0 & 0 & 0 & 0 & 0 & 0 & 0 & 0 & 0 & 0 & 0 & 0 & 0 & 0 \\
 -1 & 0 & 0 & 0 & 0 & 0 & 0 & 0 & 0 & 0 & 0 & 0 & 0 & 0 & 0 & 0 \\
 0 & -2 & 0 & -1 & -\frac{1}{2} & 0 & 0 & 0 & 0 & 0 & 0 & 0 & 0 & 0 & 0 & 0 \\
 0 & -4 & -1 & -2 & -1 & 0 & 0 & 0 & 0 & 0 & 0 & 0 & 0 & 0 & 0 & 0 \\
 0 & -\frac{5}{2} & -\frac{1}{2} & -1 & -\frac{1}{2} & 0 & 0 & 0 & 0 & 0 & 0 & 0 & 0 & 0 & 0 & 0 \\
 0 & 0 & 0 & 0 & 0 & 0 & 0 & 0 & 0 & 0 & 0 & 0 & 0 & 0 & 0 & 0 \\
 0 & \frac{1}{2} & 0 & 0 & 0 & 0 & 0 & 0 & 0 & 0 & 0 & 0 & 0 & 0 & 0 & 0 \\
 0 & -1 & 0 & 0 & 0 & 0 & 0 & 0 & 0 & 0 & 0 & 0 & 0 & 0 & 0 & 0 \\
 0 & 0 & 0 & 0 & 0 & 0 & \frac{1}{2} & -1 & 0 & 0 & \frac{1}{2} & 0 & 0 & 0 & 0 & 0 \\
 0 & 0 & 0 & 0 & 0 & 0 & 0 & 0 & 0 & -1 & -\frac{1}{2} & 0 & 0 & 0 & 0 & 0 \\
 0 & 0 & 0 & 0 & 0 & 0 & 0 & 0 & -1 & -2 & -1 & 0 & 0 & 0 & 0 & 0 \\
 0 & 0 & 0 & 0 & 0 & 0 & 0 & 0 & -\frac{1}{2} & -1 & -\frac{1}{2} & 0 & 0 & 0 & 0 & 0 \\
 0 & 0 & 0 & 0 & 0 & 0 & 0 & 0 & 0 & 0 & 0 & 0 & 0 & \frac{1}{2} & -1 & 0 \\
\end{array}
\right).$}
}
Finally, the Hodge star operator of the boundary Hodge structure is given by
\eq{
\scalebox{\myFigureScale}{$
C_\infty = \left(
\begin{array}{cccccccccccccccc}
 1 & 0 & 0 & 0 & 0 & 0 & 0 & 0 & 0 & 0 & 0 & 0 & 0 & 0 & 0 & 0 \\
 0 & 1 & 0 & 0 & 0 & 0 & 0 & 0 & 0 & 0 & 0 & 0 & 0 & 0 & 0 & 0 \\
 0 & 0 & 1 & 0 & 0 & 0 & 0 & 0 & 0 & 0 & 0 & 0 & 0 & 0 & 0 & 0 \\
 0 & 1 & 0 & 1 & 1 & 0 & 0 & 0 & 0 & 0 & 0 & 0 & 0 & 0 & 0 & 0 \\
 0 & -2 & 0 & 0 & -1 & 0 & 0 & 0 & 0 & 0 & 0 & 0 & 0 & 0 & 0 & 0 \\
 0 & 0 & 0 & 0 & 0 & -1 & 0 & 0 & 0 & 2 & 1 & 0 & 0 & 0 & 0 & 0 \\
 0 & 0 & 0 & 0 & 0 & 0 & -1 & 0 & 2 & 4 & 2 & 0 & 0 & 0 & 0 & 0 \\
 0 & 0 & 0 & 0 & 0 & 0 & -1 & 1 & 1 & 2 & 0 & 0 & 0 & 0 & 0 & 0 \\
 0 & 0 & 0 & 0 & 0 & 0 & 0 & 0 & 1 & 0 & 0 & 0 & 0 & 0 & 0 & 0 \\
 0 & 0 & 0 & 0 & 0 & 0 & 0 & 0 & 0 & 1 & 1 & 0 & 0 & 0 & 0 & 0 \\
 0 & 0 & 0 & 0 & 0 & 0 & 0 & 0 & 0 & 0 & -1 & 0 & 0 & 0 & 0 & 0 \\
 0 & 0 & 0 & 0 & 0 & 0 & 0 & 0 & 0 & 0 & 0 & -1 & 0 & -1 & 2 & 0 \\
 0 & 0 & 0 & 0 & 0 & 0 & 0 & 0 & 0 & 0 & 0 & 0 & -1 & 0 & 0 & 0 \\
 0 & 0 & 0 & 0 & 0 & 0 & 0 & 0 & 0 & 0 & 0 & 0 & 0 & -1 & 0 & 0 \\
 0 & 0 & 0 & 0 & 0 & 0 & 0 & 0 & 0 & 0 & 0 & 0 & 0 & -1 & 1 & 0 \\
 0 & 0 & 0 & 0 & 0 & 0 & 0 & 0 & 0 & 0 & 0 & 0 & 0 & 0 & 0 & -1 \\
\end{array}
\right).$}
}
\endgroup
\end{subappendices}





\addchap{Summary}
In this thesis we have studied various applications of asymptotic Hodge theory in string compactifications. This mathematical framework captures how physical couplings of the resulting effective theories behave near boundaries in the scalar field space where the internal Calabi-Yau manifold degenerates. Here we conclude by giving a summary of each of the three parts in which this thesis is divided.

Part \ref{part1} introduced the techniques from asymptotic Hodge theory we used throughout this thesis. We reviewed the results of the nilpotent orbit theorem of Schmid and the multi-variable sl(2)-orbit theorem of Cattani, Kaplan and Schmid. This discussion was tailored to applications in the study of string compactifications, explaining how to describe important physical couplings such as K\"ahler potentials and flux superpotentials near boundaries in moduli spaces. 
The nilpotent orbit approximation yields an asymptotic expansion divided into leading polynomial terms with exponential corrections. In turn, the sl(2)-orbit approximation implements a hierarchy in the large field limit towards this boundary, enabling the use of algebraic structures such as sl(2,$\mathbb{R}$)-triples to describe asymptotic behavior.

Part \ref{part2} discussed a geometrical application of asymptotic Hodge theory with the construction of general models for asymptotic periods. To be precise, we studied the $(3,0)$-form periods of the Calabi-Yau manifold near boundaries in complex structure moduli space. These periods determine for instance the $\mathcal{N}=2$ vector sector of Type IIB compactifications and encode part of the $\mathcal{N}=1$ K\"ahler potential and flux superpotential in Type IIB orientifolds. Taking the constraints imposed by asymptotic Hodge theory as consistency principles, we developed new methods for constructing these periods. We explicitly carried out our program for all possible one- and two-moduli boundaries and constructed general models for their asymptotic periods. One key finding was that near almost all boundaries exponential corrections played an essential role: for instance, such corrections can be needed to render the K\"ahler metric on the field space non-degenerate -- a surprising feature compared to well-studied corners such as the large complex structure regime.


Part \ref{part3} discussed two applications of asymptotic Hodge theory in string compactifications. Chapter \ref{chap:WGC} focused on bounds put by the Weak Gravity Conjecture, which predicts the existence of states whose charge must larger than or equal to its mass compared to the black hole extremality bound. More concretely, we studied the charge-to-mass ratios of BPS black holes in four-dimensional $\mathcal{N}=2$ supergravity theories arising from Type IIB Calabi-Yau threefold compactifications. Geometrically these states arise from D3-branes wrapped on certain three-cycles of the internal geometry, and the asymptotic behavior of their physical charges and masses can be described precisely by means of the sl(2)-approximation. We applied these techniques to compute asymptotic charge-to-mass ratios for a particular set of sl(2)-elementary states that couple to the asymptotic graviphoton. In turn, we determined the radii of the ellipsoid that forms the extremality region of electric BPS black holes, thereby giving us precise order one bounds.

Chapter \ref{chap:modstab} studied moduli stabilization in asymptotic regimes in complex structure moduli space. The setting was given by Calabi-Yau orientifold compactifications of Type IIB string theory and Calabi-Yau fourfold compactifications of F-theory with fluxes. We compared two equivalent sets of extremization conditions: a self-duality for the fluxes or demanding vanishing F-terms. The former is formulated via the Hodge star of the underlying geometry, while the latter are computed via the $(D,0)$-form periods. By the nilpotent orbit approximation the self-duality condition (or equivalently the scalar potential) then admits a leading polynomial approximation where all exponential corrections can be consistently dropped. In contrast, the F-terms (or equivalently the superpotential) required us to take metric-essential exponential corrections into account
. These complementary perspectives allowed for two concrete phenomenological applications:
\begin{itemize}
\item Using the self-duality condition we set up a three-step approximation scheme for finding flux vacua: (1) the sl(2)-approximation, (2) the nilpotent orbit approximation, and (3) the fully corrected result. The first two approximations turn moduli stabilization into a straightforward algebraic problem, where we can gradually incorporate corrections into the extremization conditions. It also gives a precise understanding of how flat directions in one approximation are lifted by corrections coming from another.

\item For the F-terms we 
used the fact that essential exponential corrections are controlled by asymptotic Hodge theory to set up a method for engineering vacua with a small flux superpotential. Our vacua feature $\cO(1)$ moduli masses because the scalar fields are stabilized by the polynomial scalar potential, making them particularly interesting for the KKLT scenario. We demonstrated our method on the one and two-parameter models constructed in part \ref{part2}.
\end{itemize}

\addchap{Samenvatting}
In dit proefschrift hebben we verschillende toepassingen van asymptotische Hodge-theorie in snaartheorie compactificaties bestudeerd. Dit wiskundige raamwerk legt vast hoe koppelingen van effectieve theorieën zich gedragen nabij randen in de scalaire veldruimte waar de Calabi-Yau variëteit degenereert. Hier sluiten we af met een samenvatting van de drie delen waarin dit proefschrift is verdeeld.

Deel \ref{part1} introduceerde de technieken uit asymptotische Hodge-theorie die we in dit proefschrift hebben gebruikt. We hebben een overzicht gegeven van de nilpotente baanstelling van Schmid en de sl(2)-baanstelling van Cattani, Kaplan en Schmid, toegespitst op toepassingen in snaartheorie compactificaties, waarbij werd uitgelegd hoe belangrijke koppelingen zoals de K\"ahler potentiaal en flux superpotentiaal zich gedragen nabij randen in moduliruimte. De nilpotente baanbenadering levert een asymptotische expansie op, verdeeld in leidende polynomiale termen met exponentiële correcties. Vervolgens implementeert de sl(2)-baanbenadering een hiërarchie in de limiet naar deze rand, waardoor algebraïsche structuren zoals sl(2,$\mathbb{R}$)-algebras het mogelijk maken om asymptotisch gedrag te beschrijven.

Deel \ref{part2} besprak een geometrische toepassing van asymptotische Hodge-theorie met de constructie van modellen voor periodes. We bestudeerden de $(3,0)$-vormperiodes van Calabi-Yau variëteiten nabij randen in de complexe-structuur moduliruimte. Deze periodes bepalen bijvoorbeeld de $\mathcal{N}=2$ vector sector van Type IIB compactificaties en een deel van de $\mathcal{N}=1$ K\"ahler potentiaal en flux superpotentiaal in Type IIB orientifolds. Door de beperkingen opgelegd door asymptotische Hodge-theory op te vatten als consistentieprincipes ontwikkelden we nieuwe methoden om deze periodes te bepalen. We voerden ons programma uit voor alle mogelijke randen met één of twee moduli en bouwden algemene modellen voor hun asymptotische periodes. Een belangrijke bevinding was dat bij bijna alle randen exponentiële termen een essentiële rol speelden: dergelijke correcties kunnen bijvoorbeeld nodig zijn voor de K\"ahler-metriek op de veldenruimte -- een verrassend kenmerk vergeleken met goed bestudeerde randen zoals het grote complex-structuur regime.


Deel \ref{part3} besprak twee toepassingen van asymptotische Hodge-theorie in snaartheorie compactificaties. Hoofdstuk \ref{chap:WGC} concentreerde zich op de grenzen die gesteld worden door het zwakke zwaartekrachtsvermoeden, dat het bestaan voorspelt van toestanden met lading groter of gelijk aan massa in vergelijking met de extremaliteitsgrens van zwarte gaten. We hebben de lading-tot-massaverhoudingen van BPS zwarte gaten bestudeerd in vierdimensionale $\mathcal{N}=2$ superzwaartekrachttheorieën die voortkomen uit Type IIB Calabi-Yau compactificaties. Geometrisch komen deze tot stand als D3-branen gewikkeld op bepaalde drie-cycli van de interne geometrie, en het asymptotische gedrag van hun fysieke ladingen en massa's kan nauwkeurig worden beschreven door middel van de sl(2)-benadering. We hebben de asymptotische lading-tot-massaverhoudingen berekend voor sl(2)-elementaire toestanden die gekoppeld zijn aan het asymptotische gravifoton. Vervolgens bepaalden we de stralen van de ellipsoïde die het extremaliteitsgebied van elektrische BPS zwarte gaten vormt, waardoor we een precieze extremaliteitsgrens kregen.

Hoofdstuk \ref{chap:modstab} bestudeerde moduli-stabilisatie nabij randen in complex-structuur moduliruimte. De setting werd gegeven door Calabi-Yau orientifold compactificaties van Type IIB snaartheorie en Calabi-Yau viervariëteit compactificaties van F-theorie. We vergeleken twee equivalente sets aan extremisatievergelijkingen: een zelfdualiteitseis voor de fluxen of de F-termen gelijk aan nul zetten. De eerste wordt geformuleerd via de Hodge-operator van de onderliggende geometrie, terwijl de laatste wordt berekend via de $(D,0)$-vormperiodes. Door de nilpotente baanbenadering laat de zelfdualiteitseis (of equivalent de potentiaal) dan een polynoombenadering toe waarbij alle exponentiële correcties consequent kunnen worden weggelaten. Daarentegen eisten de F-termen (of equivalent de superpotentiaal) dat we metriek-essentiële exponentiële correcties meenamen. Deze complementaire perspectieven maakten twee concrete fenomenologische toepassingen mogelijk:
\begin{itemize}
\item Met behulp van de zelf-dualiteitsvoorwaarde hebben we een driestaps benaderingsschema opgesteld voor het vinden van flux vacua: (1) de sl(2)-benadering, (2) de nilpotente baanbenadering, en (3) het volledig gecorrigeerde resultaat. De eerste twee benaderingen maken van moduli-stabilisatie een algebraïsch probleem, waarbij we geleidelijk correcties in de extremisatiecondities kunnen opnemen. Het geeft ook een nauwkeurig begrip van hoe vlakke richtingen in de ene benadering worden opgeheven door correcties uit een andere.

\item Voor de F-termen gebruikten we de controle van asymptotische Hodge-theorie over essentiële exponentiële correcties om vacua te vinden met een kleine superpotentiaal. Onze vacua hebben $\cO(1)$ moduli-massa's omdat de scalaire velden worden gestabiliseerd door de polynome potentiaal, wat ze bijzonder interessant maakt voor het KKLT-scenario. We hebben onze methode getest op de één en twee parameter-modellen die in deel \ref{part2} waren geconstrueerd.

\end{itemize}

\appendix

\addchap{Acknowledgements}
This thesis marks the end of four years of physics research in Utrecht, which has been a truly inspiring, sometimes challenging and overall wonderful time. Let me now take a moment to show some well-deserved appreciation to the people that have made it into such a great experience.

First and foremost, I would like to thank Thomas for his invaluable support throughout this time; I greatly enjoyed the (physics) discussions we have had, 
appreciated the many insights you shared with me, and always valued your encouragements in starting up (and finishing) projects. It has been a true pleasure working with you all these years, from the very first days as a master thesis student until now at end of the PhD and hopefully beyond. Erik, you came to Utrecht halfway during my PhD, and you've been a great addition to the group here; thank you a lot for the enjoyable collaborations and engaging discussions. Let me also thank Stefan here, for all the pleasant conversations and discussions during lunches and coffee breaks.

One of the things I learned quickly during my PhD is that doing research together is much more enjoyable than working on projects alone. Throughout these years I have had the chance to work with many excellent people, each of which I would like to thank here. Let me begin with Koen, one of my paranymphs; it has been a great joy to go through this experience together, starting all the way back in our masters. You made sure we never left a milestone uncelebrated, be it a finished project or any other important moment. I enjoyed the many coffee breaks we had playing chess, sparring about our separate (and joint) projects and, to top it off, you even set me up with Olga -- thanks for everything. Brice, we have collaborated together through most of my PhD, and I liked working with you a lot. In some periods we wrote more often a paper together than we were able to meet in-person, but with the dinners and wine-tastings we have tried to make up for that. Chris, thanks for sharing your enthusiasm with us, from random paper suggestions in the journal club to mini-lectures related to our projects and everything in-between; it has been great to have you around here in Utrecht. Eric, I admire your broad range of interests and ability to discuss about anything; thanks for the work together, engaging conversations and all the other good times. 

Fabian, as my first collaborator (apart from Thomas), thanks for the pleasant cooperation, and I look forward to seeing you around in Boston. Jeroen, it was a pleasure assisting in supervising your master thesis, and even more to follow up with the first project of your PhD. Lorenz, thanks for the nice discussions and sharing all your insights on the computation of periods. Alvaro, I thoroughly enjoyed our online discussions that somehow always seemed to overrun; thanks for your comments on the draft of this thesis, and let's play some table tennis again in real life soon.

I would also like to thank the rest of the members of our institute. Huibert and Kilian, thanks for the enjoyable coffee breaks, lunches, discussions, and all the fun nights out. Govert, it has been great to have you around as a fellow physicist in our mutual group of friends, and I am happy to be joining you soon in Boston. I would also like to thank all the other people for the good company during lunches and coffee breaks: Alexander, Anna, Arno, Camilo, Chongchuo, Enis, Guoen, Joren, Nava, Nick, Nico, Markus, Matti, Pierre, Ronnie, Stefano, Tycho, Umut, and Wilke. 

I would also like to thank all the people I had the chance of meeting outside of Utrecht during conferences and workshops abroad, too many of which to name all individually. Staying within the Netherlands, thanks to Antonio, Carlos, Dora, Evita, Gabriele, Greg, Johannes, John, and Joren for brightening up the holography meetings and DRSTP schools.

Outside of physics I want to thank all the friends that have made my time in Utrecht so great. To the guys in my jaarclub for the many Tuesday evenings together providing the necessary distractions of PhD life. A special thanks here goes to my other paranymph Noud, especially for enduring my explanations of the quantum-wiggling strings I've been working on, I hope at least some parts came across. Also I would like to thank the large group of friends originating from USBC, making the occasional board game night, race cycling route, holiday or any other time so enjoyable. And finally, many thanks to my housemates for the cozy tea times after work, pleasant home office at our dining table during lockdowns, and all the other fun moments we have had throughout these years.


%

Olga, it has been great to have you along my side for the last steps of this PhD; thanks for sharing all your enthusiasm and optimism with me, making me already look forward to the adventures that lie ahead of us. And last but certainly not least, my family, especially my parents and my brother: you stimulated me to always follow my own interests, made the time for quick calls to catch up -- which luckily always lasted much longer -- and you were always there for me; thanks for all your support throughout the years.


\addchap{About the author}
\begin{wrapfigure}{r}{0.35\linewidth}
\begin{center}
\vspace*{-20pt}
\includegraphics[width=0.35\textwidth]{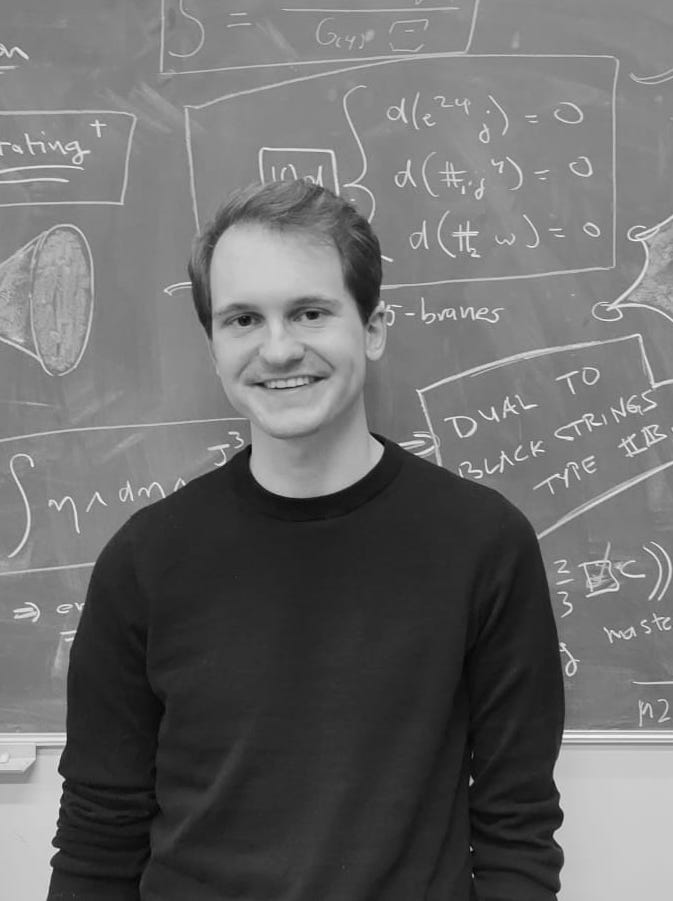}
\vspace*{-30pt}
\end{center}
\end{wrapfigure}
Damian van de Heisteeg was born on the 30th of July 1995 in Ede. He obtained his high school diploma from the Marnix College in Ede in 2013. In that same year he started with a double bachelor program in mathematics and physics at Utrecht University, which he completed cum laude in 2016. His bachelor thesis on phase transitions in antiferromagnets was supervised by Rembert Duine. He then continued his studies with the master's program in theoretical physics, graduating cum laude in 2018. The research of his master thesis focused on string theory, specifically backreaction of fluxes and branes in F-theory, under the supervision of Thomas Grimm. 

After obtaining his master degree he began as a PhD candidate advised by Thomas Grimm (and later joined by Erik Plauschinn as copromotor) at the Institute for Theoretical Physics in Utrecht under an NWO start-up grant. His research focused on applications of asymptotic Hodge theory in the setting of string compactifications, the contents of which are presented in this thesis. Independent of his advisors, he has investigated near-horizon geometries of black holes in string theory with other collaborators in the high-energy theoretical physics group. Apart from his research, he has also written popular science articles on the Dutch website `The Quantum Universe'. He will continue as a postdoctoral fellow at the Center of Mathematical Sciences and Applications at Harvard University.

\bibliographystyle{jhep}
\bibliography{references.bib}

\backmatter

\end{document}